\documentclass{svjour3}                     
\journalname{Astronomy Astrophysics Review}

\smartqed  
\usepackage{graphicx}
\usepackage{color}
\usepackage[colorlinks=true,linkcolor=blue,citecolor=black,urlcolor=blue]{hyperref}

\usepackage[]{natbib}

\usepackage{txfonts}
\usepackage{lscape}

\def\apj{ApJ}%
\def\apjl{ApJLett}%
\def\apjs{ApJSupp.}%
\def\mnras{MNRAS}%
\def\araa{ARA\&A}
\def\aj{AJ}
\def\apss{Ap\&SS}  
\def\aap{A\&A}   
\def\pasa{PASA}
\def\pasp{PASP} 
\def\pasj{PASJ}
\def\nat{Nature}
\def\memsai{Mem.~Soc.~Astron.~Italiana}
\def\rmxaa{Rev. Mex. Astron. Astrof.}

\newcommand{\Te}{$T_e$}     
\newcommand{\Ne}{$n_e$}     
\newcommand{\msun}{M$_\odot$}  

\newcommand{\ha}{H$\alpha$}
\newcommand{\hb}{H$\beta$}
\newcommand{\hc}{H$\gamma$}
\newcommand{\oiii}[1]{[OIII]$\lambda$#1}
\newcommand{\oii}[1]{[OII]$\lambda$#1}
\newcommand{\nii}[1]{[NII]$\lambda$#1}
\newcommand{\niii}[1]{[NIII]$\lambda$#1}
\newcommand{\neii}[1]{[NeII]$\lambda$#1}
\newcommand{\neiii}[1]{[NeIII]$\lambda$#1}
\newcommand{\sii}[1]{[SII]$\lambda$#1}
\newcommand{\siii}[1]{[SIII]$\lambda$#1}
\newcommand{\cii}[1]{[CII]$\lambda$#1}

\begin{document}

\title{De Re Metallica: The cosmic chemical evolution of galaxies}



\titlerunning{Cosmic chemical evolution of galaxies}        

\author{
Maiolino R.         \and
Mannucci F. 
}


\institute{R. Maiolino \at
              Cavendish Laboratory and Kavli Institute for Cosmology,
              University of Cambridge \\
              Madingley Rise, Cambridge, CB3 0HA, United Kingdom\\
				\email{r.maiolino@mrao.cam.ac.uk}\\
                ORCID: 0000-0002-4985-3819 \\
           \and
           F. Mannucci \at
              INAF - Osservatorio Astrofisico di Arcetri\\
              Largo E. Fermi 5, 50125 Firenze, Italia\\
              \email{filippo.mannucci@inaf.it}\\
              ORCID: 0000-0002-4803-2381
}

\date{Received: date / Accepted: date}

\maketitle

\tableofcontents

\newpage

\begin{abstract}
The evolution of the content of heavy elements in galaxies,
the relative chemical abundances, their spatial distribution,
and how these scale with various galactic properties, provide
unique information on the galactic evolutionary processes
across the cosmic epochs.
In recent years major progress has been made in constraining the
chemical evolution of galaxies and inferring key information relevant
to our understanding of the main mechanisms involved in galaxy evolution.
In this review we provide an overview of these various areas. After an overview of the methods used to constrain the  chemical enrichment in galaxies and their environment, we discuss the observed scaling relations between metallicity
and galaxy properties, the observed relative chemical abundances,  how
the chemical  elements are distributed within galaxies, and how these properties
evolve across the cosmic epochs. We discuss how the various observational
findings compare with the predictions from theoretical models
and numerical cosmological simulations. Finally, we briefly discuss the open
problems the prospects for progress in this field in the nearby future.

\end{abstract}

\section{Introduction}  
\label{sec:intro}

The evolution of the chemical properties of stellar populations and of the interstellar and intergalactic medium across the cosmic epochs provides unique information on the evolutionary processes driving the formation and evolution of galaxies. Theory and cosmological simulations give a relatively simple scenario on how dark matter evolve from the primeval perturbations, forming dark matter halos and large scale structures 
\citep[e.g.,][]{Springel18}. 
However, the evolution of the baryonic component is much more complex as baryons interact with radiation and are subject to dissipative processes. The evolution of the baryonic component and how this results into the formation of stars and in the properties of galaxies as we see them in the local universe and across the cosmic epochs, has been subject to numerous models and cosmological simulations, which use different prescriptions and assumptions. 

The investigation of the evolution of the content of chemical elements provides tight constraints on such models. Indeed, the content of metals gives a measure of the 
integrated star formation in galaxies, but also on the fraction of
metals lost through outflows and stripping. The metallicity, i.e. the content of metals relative to hydrogen and helium, is also sensitive to dilution resulting from inflow of pristine gas.
Therefore, the investigation of the metals content and of the metallicity in galaxies provide
truly crucial information on the key mechanisms involved in galaxy evolution.
In addition, different chemical elements are enriched on different timescales by different populations of stars, therefore the relative abundance of chemical elements enables us to obtain unique constraints on the 
star formation history.

The analysis of the metallicity and chemical abundances on spatially resolved
scales (gradients) give additional information on the processes that have regulated
the growth and assembly of galaxies (e.g. inside-out or outside-in formation and/or quenching),
as well as information on other internal phenomena such as galactic
fountains, stellar migration and radial gas inflows.

Major advances of several observing techniques and the advent of new major observing facilities
have recently enabled astronomers to investigate the chemical evolution in the early cosmic epochs,
by directly probing the early enrichment process of galaxies and of the
intergalactic/circumgalactic medium, hence setting tight constraints on the models of early galaxy formation.\\

In this review we primarily provide an overview of the chemical properties of galaxies in the local universe and
at high redshift from an observational
perspective, but we also discuss how such properties give important constraints on the galaxy evolutionary
processes, including extensive comparisons with theoretical models and numerical simulations.

We will generally discuss the overall statistical properties of galaxies 
rather then focusing on
individual galaxies, although we will unavoidably open some themes by quickly discussing the Milky Way (MW) or some
nearby well studied targets.

In the first part we also give a more technical overview of the methods, definitions and techniques adopted to measure the metallicity and chemical enrichment in stellar populations and in the interstellar medium (ISM), 
also discussing the strengths and weaknesses of the various methods. A large fraction of the review will be dedicated to the metallicity scaling relations, i.e. the
trends of the stellar/gas metallicity with other galaxy properties, such as mass, star formation rate,
gas content, environment, etc. We will then discuss these properties on resolved scales, i.e. the investigation of metallicity
gradients, which has recently been a steadily growing field. 
The relative chemical abundances of various elements is another major topic that, as mentioned above, provide key information, and to which we dedicate an entire major section; however, we cannot realistically review all chemical elements, hence we will mostly focus
on some specific abundances ratios that are particularly useful to constrain the star formation history
and galaxy evolution, and which have been measured across large samples of galaxies. In all of these major topics we discuss both the finding in the local universe and the evolution of these properties at high redshift. 
We will dedicate a section to the current constraints of the metallicity in the host galaxies of Active Galactic Nuclei (AGN). We only provide very brief discussions about resolved stellar populations and absorption systems associated with the intergalactic medium (IGM).
Finally we give an overview of the current understanding of the global metal content of galaxies and discuss
their metal budget.

\subsection{Expressing metallicity and chemical abundances}
\label{sec:definitions}

Different definitions are adopted to measure the abundance of metals and of individual chemical elements.
In contrast to other scientific disciplines \citep[e.g.][]{Agricola1556}, in astrophysics the term ``metals'' refers to the all elements heavier than helium. The ``metallicity'' (Z) indicates the  mass of all metals relative to the total mass of baryons (dominated by hydrogen and helium):

\begin{equation}
\rm Z \equiv M_{metals}/M_{baryons}
\end{equation}

The relative abundance of two generic chemical elements X and Y is generally expressed in
terms of relative number densities N, relative to the Solar value, with the following notation:
\begin{equation}
\rm [X/Y] \equiv \log{(N_X/N_Y)} - \log{(N_X/N_Y)_\odot}
\end{equation}

When expressing the abundance of chemical elements relative to hydrogen, the following expression is also often used:
\begin{equation}
\rm 12+\log{(X/H)} \equiv 12+ \log{(N_X/N_H)} \,,
\label{eq:12log}
\end{equation}
where the value 12 was introduced so that any element, even
the most rare, has a positive Solar values in Eq.~\ref{eq:12log}.

Since oxygen is generally the most abundant heavy element in mass, often the ``metallicity'' is expressed in terms of oxygen abundance, generally under the implicit assumption that the abundance of all other chemical elements scale proportionally maintaining the solar abundance ratios. Therefore, often the metallicity is indicated as
\begin{equation}
\rm 12+\log{(O/H)} \equiv 12+ \log{(N_O/N_H)}
\label{eq:12logO}
\end{equation}
The Solar (photospheric) reference is still not completely settled. The review by \cite{Asplund09} gives
	$\rm 12+\log{(O/H)_\odot} = 8.69 \pm 0.05$ based on
    three-dimensional (3D), time-dependent hydrodynamical model of the Solar atmosphere. However, for instance,
    \cite{Delahaye06} give
	$\rm 12+\log{(O/H)_\odot} = 8.86 \pm 0.05$
    based on helioseismology.
However, it should be clear that in the context of this review the Solar metallicity (or Solar chemical abundances) do not have a
particular meaning, at most being representative of the chemical enrichment of the (pre-)Solar Neighborhood, i.e. a specific region
of the Milky Way disc. The Solar metallicity/abundances
should only be considered as
reference values. What is important is that when comparing different studies they should be consinstently scaled to the same Solar reference value.

However, it should be clear that using the O/H abundance is only an approximation of the real metallicity of the gas, as the relative abundance of the chemical elements can vary in a drastic way with respect to the solar value.

\subsection{The origin of the elements}
\label{sec:origin}

While primordial nucleosynthesis accounts for the origin of
hydrogen, deuterium, the majority helium and a small fraction of lithium
\citep[e.g.,][]{Cyburt16},
all other elements are produced by stellar nucleosynthesis
or by the explosive burning and photodisintegration associated
with the late stages of stellar evolution.
Boron, beryllium and a small fraction of lithium are  exceptions because they are
produced by cosmic rays spallation of heavier elements.

Extensive reviews have been published on the origin of chemical
elements through stellar processes
\citep[e.g.,][]{Rauscher11,Matteucci12,Nomoto13}, therefore
in this section we only provide a quick overview of the basic processes
associated with the production of heavy elements and a short summary
of the primary sources of some key element, as well as the associated
production timescales. 

Stars on the main sequence burn hydrogen atoms, producing $\rm ^{4}He$ atoms,
either through
the proton-proton nuclear reaction chain (pp-chain), or through the 
CNO-cycle; the relative role of these two processes depends on the
stellar mass (the latter dominates at masses larger than about 
$\rm 1.3~M_{\odot}$) and, of course, metallicity.

At later stages,
when hydrogen is exhausted in the core and the this becomes hotter,
helium burning starts, which
produces $\rm ^{12}C$ through the triple-alpha reaction,
and $\rm ^{16}O$ is also produced through the capture of an additional
helium nucleus. During this phase hydrogen burning continues around the
helium burning core.

If the initial stellar mass is less than
$\rm 8~M_{\odot}$ no additional burning phases take place.

If the stellar mass is less than about $\rm 2~M_{\odot}$ the star
undergoes the so-called core helium flash, a runaway process which
results into an explosive expansion of the outer core. This results
into a white dwarf surrounded by a planetary nebula.

Stars between $\rm 2~M_{\odot}$ and $\rm 8~M_{\odot}$ after leaving
the main sequence enter the red giant phase and, once the
helium burning core is exhausted, they go trough the so-called
Asymptotic Giant Branch (AGB), characterized by both a hydrogen- 
and a helium-burning shell. Thermonuclear runaway of the latter results
into a sequence of several He-shell flashes, with increases mass losses and
strong stellar winds, which eject a large fraction of the previous burning
products (in particular carbon and nitrogen). Such flashes are also responsible
for producing large convection zones (which mix the product of nuclear
reactions) and for the production of s-process nuclei.
Eventually these stars also end their life as white dwarfs
surrounded by planetary nebulae.

It should be noted that during the CNO-cycle, hence for stars more
massive than about $\rm 1.3~M_{\odot}$, while the primary reaction chain
leaves unaffected the ``catalytic'' nuclei, secondary branches of
the reaction produce $\rm ^{14}N$ at expenses of $\rm ^{16}O$ and
$\rm ^{12}C$. The result is a strong enrichment of $\rm ^{14}N$ and also
a drastic reduction of the $\rm ^{12}C/^{13}C$ ratio.
These effects are obviously strongly dependent on the stellar metallicity,
which boosts the role of the CNO cycle and explains the ``secondary'' nature
of nitrogen, i.e.,
whose production is greatly enhanced in metal rich environments.

In stars more massive than $\rm 8~M_{\odot}$, after helium burning
the core shrinks and increases the temperature to
ignite carbon burning into $\rm ^{20}Ne$ or $\rm ^{23}Na$. Then 
heavier elements are subsequently
burned as the lighter elements in the core are 
exhausted and, consequently, the core temperature increases.
This process in particular results into the production and burning
of elements such as oxygen, magnesium and silicon.
During this
sequential process occurring in the stellar core, the burning of lighter
elements occurs in stratified shells around the core that,
in the meantime, have reached the adequate temperature for igniting
the associated nuclear reactions.
The process ends when the core is primarily composed of iron and nickel.
At this stage no more energy is gained by further nuclear reactions, the
energy released can no longer sustain the hydrostatic equilibrium with
the weight of the outer layer and collapse begins, yielding to a core-collapse
supernova (CC SN, type~II, or Ib-Ic). The outward propagating shock wave
results in the stellar material undergoing
shock heating and explosive nucleosynthesis. The latter
affects the abundance pattern of the outer layers, and especially
the distribution of $\alpha$\ elements and the production of Fe-peak
elements. Supernovae also produce 
elements heavier than iron through neutron rapid capture of iron-seed nuclei
(r-process), followed by (slower) $\beta$-decay.
The resulting yield of the various chemical elements implies complex
calculations \citep[e.g.][and references therein]{Nomoto13} and, in particular, depends on the location of
the ``mass-cut'', i.e., the
boundary between the core remnant that retain metals and the envelope
that is expelled.

\begin{figure}

\centerline{\includegraphics[width=10cm]{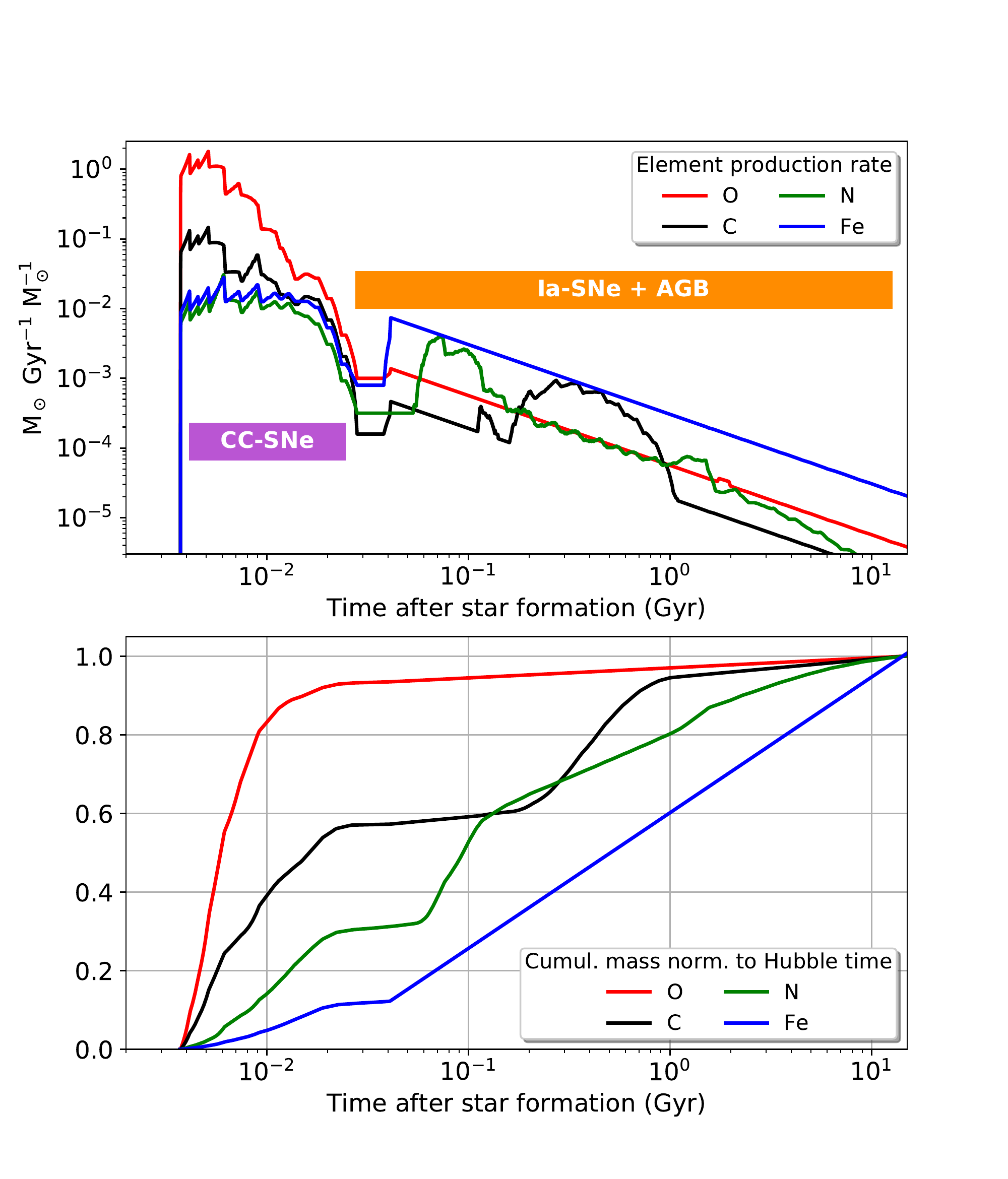}}
\caption{Timescales of production of various elements after a single episode of star formation (a single stellar population, SSP) of solar metallicity, based on the model by Vincenzo et al. (in prep.), see text for details. The upper panel shows the production rate in \msun/Gyr normalized to 1~\msun\ of formed stars. The lower panel shows the cumulative mass produced, normalized to the amount after one Hubble time. Oxygen (red line) in mainly produced by CC SNe and therefore has the shortest formation timescales. Iron (blue line) is dominated by type Ia SNe, Carbon (black) has contributions from both kinds of SNe and from AGB stars. The production of Nitrogen (green) is dominated by AGB stars. In this plot, the production of elements before 30Myr is due to CC SNe, type Ia SNe are described by a power-law $t^{-1}$ after 40Myr and up to the Hubble time, and AGB stars give additional contributions above this power-law at intermediate ages of $\sim$0.04-5 Gyr.
}
\label{fig:metal_production}
\end{figure}

Type Ia supernovae are an additional important source of elements. SNIa are
thermonuclear explosions of C-O white dwarfs that accrete mass
either as a consequence of mass exchange
in a close binary system with a non-degenerate star  (``single
degenerate scenario'') or as a consequence of the merging of two white
dwarfs
(``double degenerate scenario'') \citep[see][for a review]{Maoz14}.
SNIa explosions are thought to arise from the ignition of carbon burning
in the C-O core, which results in the total disruption of the white
dwarf \citep[as the nuclear energy released by the dwarf is higher than its
gravitational binding energy,][]{Leibundgut01}. The resulting nucleosynthesis
produces elements primarily around the iron peak, but also silicon, argon, sulfur, and calcium.
Clearly, since SNe Ia require first the formation a white dwarf
from a low-mass (less than 8\msun) star, and then a significant mass transfer from a companion star,
the production of the SNIa is delayed with
respect to the onset of star formation. Specifically, the first SNe
require a minimum timescale of 30~Myr \citep{Greggio83}, although
the bulk of SNIa explode on longer timescales, 
due to the longer timescales associated to lower-mass stars and mass transfer, see \cite{Maoz12a} and Fig.~\ref{fig:metal_production}.

Finally, the merging of binary neutron stars
is an additional source of elements beyond the iron peak, driven
by the r-process.\\

Clearly, each of these processes, and the associated
release of chemical elements into the interstellar medium, is linked
to the terminal stages of stars with a specific stellar mass on the
main sequence and, therefore, with different timescales.

The amount of metals injected into the interstellar medium (ISM) by each star at the end
of its lifetime is quantified by the so-called {\it stellar yield}, $p_i$,
which is defined as the fractional mass of newly formed $i^{th}$ chemical
element, injected into the ISM, relative to the mass of the stellar
progenitor on the main sequence.
The computation of the stellar yields is quite complex and often subject to large
uncertainties, as they also depend on the assumed mass loss and stellar
rotation. The yields also depend of the progenitor metallicity, and in some
cases the dependence is very strong (such as in the case of nitrogen, as
discussed above). Compilations of stellar yields are given
in \cite{Romano10} and \cite{Nomoto13}. Here we only qualitatively summarize
the main production channels of some of the key elements.

Fig~\ref{fig:metal_production} summarizes the production timescales of a few critical elements (O, C, N, and Fe) after a single episode of star formation, i.e., for a single population of stars (SSP) created together along a given stellar initial mass function (IMF).
Most of the $\alpha$
elements (e.g. O, Ne, Mg) 
are thought to be produced by stars more massive than $\rm
8~M_{\odot}$ and by the associated core-collapse SNe, and therefore
released into the ISM promptly, soon after the beginning
of the star formation. Iron peak elements (e.g. Fe, Ni)
are partly produced by massive stars,
but most of them are produced by type Ia SNe, hence are released into the
ISM with a delay ranging from $\sim$40--50 Myr to a few Gyr, depending on the
stellar IMF and on the star formation history. It  should be noted that zinc is often
referred to as an iron-peak element, but it also seems to have an $\alpha$-like enrichment component and not
always closely follow iron \citep[e.g.][]{Berg16b}. Elements such as carbon and nitrogen
are partly produced by massive stars, but most of the production is
due to 
intermediate mass stars ($\rm 2M_{\odot}<M_{star}<8M_{\odot}$), primarily
through their AGB winds (or Wolf Rayet stars), and therefore these elements are also subject to a
delayed enrichment.
The results in ~Fig.~\ref{fig:metal_production} are obtained by Vincenzo et al. (in prep.) for stars of solar metallicity,
with the IMF from \cite{Kroupa93}, 
stellar lifetimes from  \cite{Kobayashi04},
stellar yields from \cite{Nomoto13},
SNe Ia yields from \cite{Iwamoto99},
and a $t^{-1}$ delay-time distribution for the type Ia SNe \citep{Maoz12a}.

In Sect.~\ref{sec:abund_ratios}, we will see that the different enrichment mechanisms and
timescales of these different elements provide a powerful tool to constraint
the evolution of the star formation history in galaxies.

\section{Measuring metallicities of stellar populations}
\label{sec:measmet_stars}

UV, optical and infrared spectra of galaxies contain a wealth of information about their stellar populations. Except for a number of galaxies in the local group where single stars can be resolved,
galaxy spectra consist of the integrated light of the stellar population which is virtually always a composition of different generations.  The broad-band colors and the continuum of the spectra are dominated by the distribution of stellar ages and metallicity, modulo dust reddening. The emission lines reflect the ionizing properties of the most-massive stars (and of the AGN, when present), the absorption features reveal the properties of stellar photospheres, stellar winds, and interstellar gas.

Several methods have been developed during the years to derive the chemical abundances from galaxy spectra. Two are the components that are usually considered,  stars and the interstellar medium (ISM). This section deals with the former component, the next section with the latter one.

\begin{figure}
\centerline{\includegraphics[width=12cm]{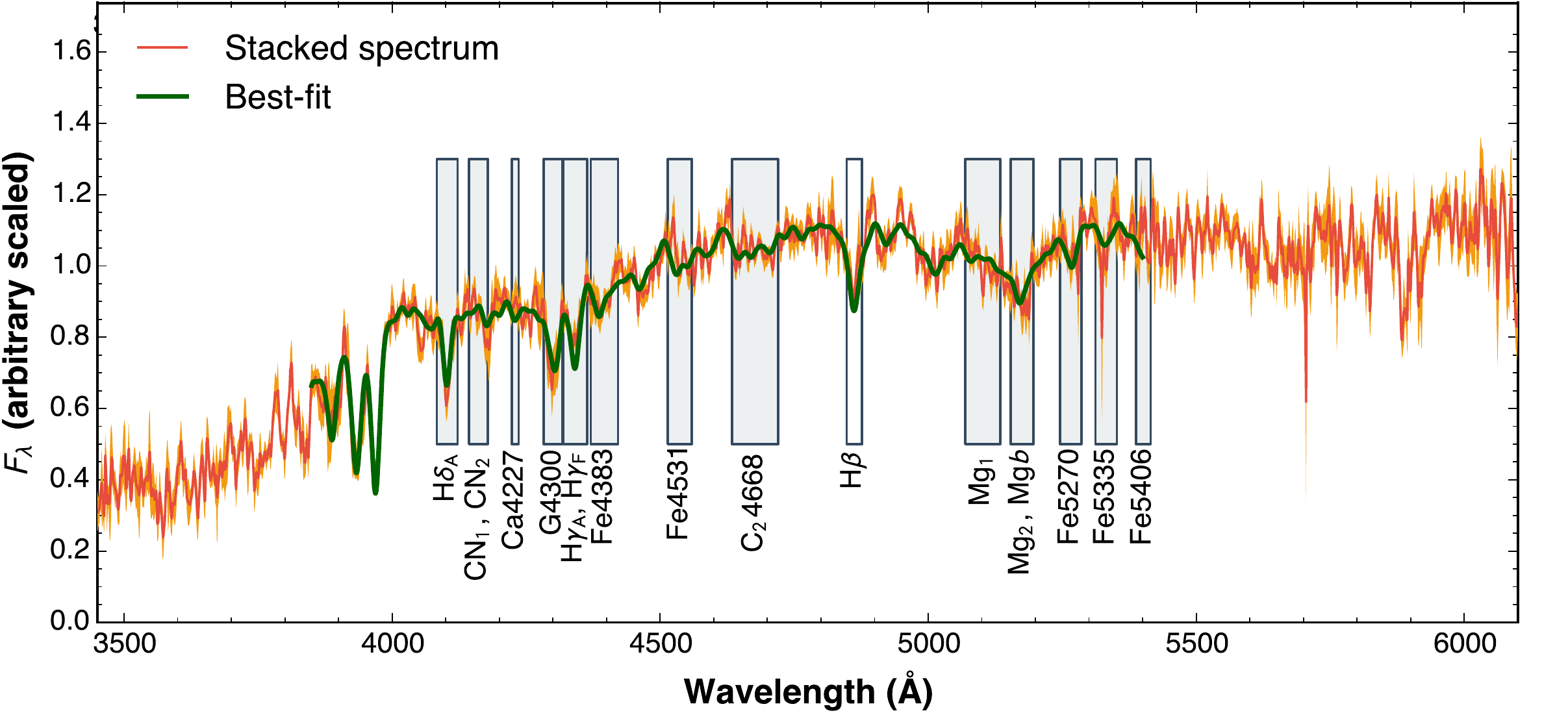}}
\caption{The wavelength ranges covered by the set of Lick indices (grey rectangles) used by \cite{Onodera15} are overplotted on a stacked spectrum of a sample of quenched galaxies at $z\sim1.6$. The red line is the stacked spectrum, with orange lines showing the $\pm1\sigma$ uncertainties. The green line is the best-fitting model stellar spectrum.
}
\label{fig:measuremet_onodera15}
\end{figure}
\subsection{Rest-frame optical spectra}
\label{sec:measmet_stars_optical}

Extracting information about the metallicity of stellar populations is subject to  ongoing efforts. A complete discussion of this topic is far beyond the scope of this review. Here we summarize the two main techniques that are often used to ``invert'' the spectra and derive the physical properties of the stellar population. Both are based on comparing observed spectra with synthetic spectra.

The first tool to be developed was a set of standardized indices, calibrated on model spectra, aimed at 
maximizing the sensitivity to some parameters (e.g., age, or metallicity) while minimizing the dependence on the other parameters.
These indeces, including the ``Lick indeces'' (see, e.g., Fig~\ref{fig:measuremet_onodera15}), 
were introduced in the 1970s and later refined and re-calibrated, and are still widely used, 
especially when only low-resolution spectra are available
\citep[e.g.,][]{Faber73, Worthey94a, Worthey97, Trager98,Trager00a, Thomas03a, Schiavon07, Thomas11d}. 
Each index is defined either by a central bandpass whose flux is  compared with those of two adjacent wavelength ranges intended to estimate a ``pseudo-continuum'', or by the flux ratio in two nearby bandpasses to measure a break. 
The Lick indices use relatively large bandpasses, of the order of 50\AA, this is useful to increase the S/N when needed,
but it is not optimal to derive individual abundances.
Some indices, such as the depth of the 4000 \AA\ break, $D_n(4000)$ \citep{Hamilton85,Balogh99},
and the depth of the Balmer absorption lines, are mainly sensitive to age and to the fraction of young stars relative to old ones. Other indices are defined to be sensitive to abundances of particular elements, such as Fe and Mg in case of the [MgFe]' and [Mg$_2$Fe] indices defined by \cite{Thomas03a} and \cite{Bruzual03}.
The results often depend on the particular set on indices available or used, but these indices are at the base of most studies about unresolved stellar populations in nearby galaxies (see Sect.~\ref{sec:MZR_stellar}).

More recently, spectro-phometric models of the stellar population have been developed aimed at reproducing the full observed spectra with a combination of input stellar populations with different properties
\citep[e.g.,][among many others]{Bruzual03,Conroy09,Leitherer14b,Chevallard16,Wilkinson17,Cappellari17},
see \cite{Conroy13} for a recent review, and \url{http://www.sedfitting.org} (by T.~Budavari, D.~Dale, B.~Groves and J.~Walcher) for a complete and updated list of the available tools.
These methods are in principle very powerful because they use all the information contained in the spectra, often also including broad-band photometry to simultaneously derive chemical abundance, age distribution (i.e., the star-formation history), dust extinction, and possibly other parameters such as the IMF. The weak point is that often the solution is not unique and there are strong degeneracies among the parameters. In particular, 
the problem is strongly non-linear in stellar mass and age, with the youngest and more massive stars often completely outshining the older, less massive stars that constitute most of the mass 
\citep[see, e.g.,][]{Maraston10}.
Uncertainties in the spectra of the input stellar populations, for example due to the presence of stellar rotation and binary stars \citep[e.g.,][]{Levesque12a, Leitherer14, Stanway16, Choi17} also affect the results. 
Despite such uncertainties, these methods are becoming the standard tool to analyze composite stellar population, in particular when spectra with enough spectral resolution and S/N are available.

\subsection{UV spectra}

At high redshift ($z>1$), the UV part of the spectrum becomes more easily accessible at optical wavelengths and often constitutes the most important piece of information about the properties of stellar populations. At these wavelengths spectra are dominated by the photospheric properties of young, massive stars, allowing us to derive the properties of the on-going star formation activity. 

A metallicity dependence of the depth of many UV absorption features is expected on the base of theoretical models \citep[e.g.,][]{Leitherer95,Leitherer10,Eldridge12} and has been actually observed in the spectra of  local starburst galaxies \citep[e.g.,][]{Heckman98, Heckman05,Leitherer11}. 

In the UV, correlations between metallicity and equivalent widths (EWs) of many absorption features are a consequence of several physical causes.
First, the contribution from the photospheres of the O and B, which depend on metallicity. 
Second, the strong, metal-dependent winds from hot stars dominate the spectra close to high-ionization lines, such as CIV and SiIV 
\citep[e.g.,][]{Walborn95}.
Third, the ISM produces the absorption of interstellar lines whose column density is related to metallicity. 
The interstellar lines are usually heavily saturated \citep[e.g.,][see \citealt{Savage96} for a review]{Pettini02a}, and the observed EWs are more sensitive to velocity dispersion than column density \citep{Gonzalez-Delgado98a,Leitherer11}.

\begin{figure}

\centerline{\includegraphics[width=12cm]{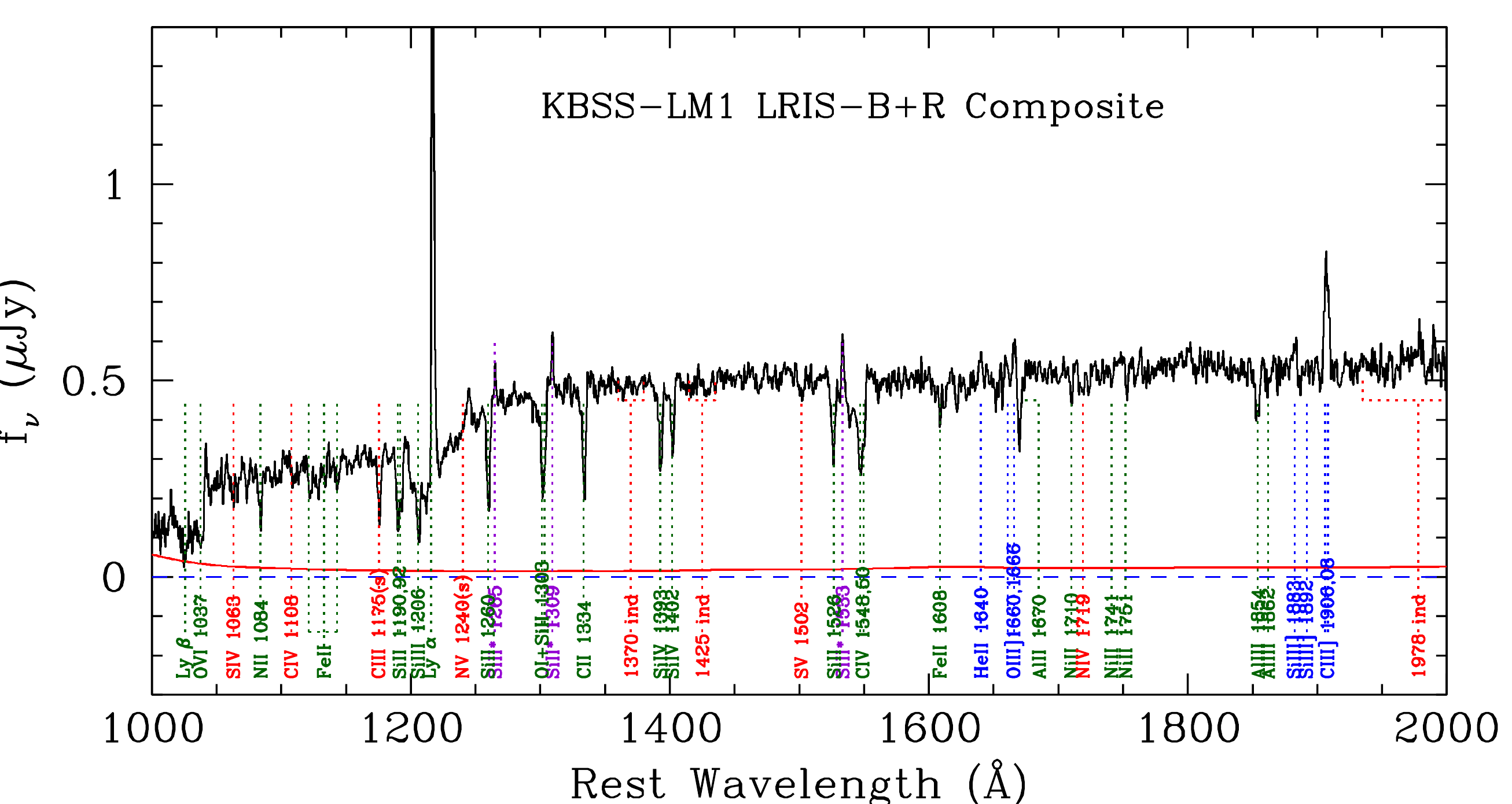}}
\caption{Stacked rest-frame UV spectrum of star-forming galaxies at $\langle z \rangle \sim2.4$, from \cite{Steidel16}. The most important absorption and emission features are shown, color-coded according to their nature: red, stellar absorption features; green, interstellar absorption features; blue, nebular emission lines; violet, fine structure lines. 
}
\label{fig:UVstellar_steidel16}
\end{figure}

A complete list of features with their physical origin can be found in 
Tab.~1 of \cite{Leitherer11}.  Here we only mention that the features that have usually attracted the highest interest
are stellar wind lines such as 
NV$\lambda$1238,1242, 
SiIV$\lambda$1393,1402,
CIV$\lambda$1548,1550, and 
HeII$\lambda$1640,
interstellar lines such as 
SiII$\lambda$1190,1260,1304,1526,
SiIII$\lambda$1206, 
OI$\lambda$1302,  
CII$\lambda$1334, 
SiIV$\lambda$1393,1402,
FeII$\lambda$1608, 
AlIII$\lambda$1670,1854,1862,
NiII$\lambda$1710,1741,1751, and
MgII$\lambda$2796,2803,
and  photospheric lines such as
FeV$\lambda$1360-1380, 
OV$\lambda$1371, 
SiIII$\lambda$1417, 
CIII$\lambda$1427, 
FeV$\lambda$1430, 
SiII$\lambda$1533, and 
CIII$\lambda$2300 
\citep{Leitherer01,Pettini02a,Leitherer14b}. Some of these 
are shown in the rest-frame UV composite spectrum of distant galaxies in
 Fig.~\ref{fig:UVstellar_steidel16}.
 Weaker interstellar lines, which are not saturated, are  sometimes also detected  \citep{Pettini00,Pettini02a}, specifically:
FeII$\lambda$1144, SII$\lambda$1250,1259, NiII$\lambda$1317,1370,1703,1709,1741,1751, and SiII$\lambda$1808.

These features usually have complex dependences on age, metallicity and IMF. 
For these reasons, similar to the optical case, several authors have identified a number of indices optimized to be sensitive mainly on metallicity 
\citep{Heckman98, Leitherer01, Rix04,  Maraston09, Leitherer11, Sommariva12, Zetterlund15, Faisst16, Byler18}.
These works are based either on empirical spectral libraries or on theoretical stellar spectra. Both approaches have strengths and weaknesses. Empirical libraries automatically take into account the uncertain effects of stellar winds, but usually include only a narrow range of metallicities, limited by the metallicity distribution of young stars in the local universe, and are affected by the unknown contribution of interstellar lines \citep[e.g.,][]{Kudritzki16}. In contrast, theoretical libraries can be build for a large range of different metallicities, but are affected by model uncertainties.
The resulting indeces depend on a number of assumptions, such as the age of the star-forming episode and the evolution of the SFR, usually assumed to be constant, instantaneous, or exponentially declining.

The IMF of the stellar population 
also affects the results. While this introduces yet another source of uncertainty, it also opens a way to test the dependence of the IMF on galaxy type, environment, star-formation activity and cosmic age, as discussed, for instance, by
\cite{Lemasle14}, \cite{La-Barbera16}, \cite{Fontanot17b}, \cite{Sarzi18}, and \cite{De-Masi18}.

\section{Measuring metallicity and chemical abundances of the gaseous phase}
\label{sec:measmet_ism}

In this section we provide a brief overview of the methods
used to infer the metallicity and chemical abundances of the gaseous phase focusing on the ISM, but also discussing
the techniques exploited for the circum-galactic
medium (CGM) and intergalactic medium (IGM), as well
as for the intra-cluster medium (ICM).
For each method we critically assess advantages and
limitations. 
A more detailed, recent review can be found in \cite{Peimbert17}.

\subsection{Direct method based on electron temperature}
\label{sec:measmet_ism_Te}

The spectra of ionized gas in astrophysical conditions are usually  rich of collisionally excited emission lines (CEL). The flux of each metal line is proportional to the abundance
of the element (specifically the observed ionic species)
times its emissivity. If the latter can be measured, then the abundance can be constrained accurately \citep{Aller84}.

The emissivity of these lines depends on both electron temperature \Te\  and on electron density \Ne. Once these two parameters are measured, ionic abundances follow from relations based only on atomic physics. In a two-levels atom, the rate of collisional de-excitation of the transition $2 \to 1$ of an ion is given by $n_2 n_0 C_{21}$,
where $n_2$ is the density of ions whose level 2 is populated, $n_0$ is the density of colliding particles (typically electrons),  and $C_{21}$ is the coefficient for collisional de-excitation given by
\begin{equation}
C_{21} = \left(\frac{2\pi}{kT_e}\right)^{0.5} \frac{\hbar^2}{m^{3/2}}
  \frac{\Omega(1,2)}{\omega_2} \,,
\end{equation}
where $\Omega(1,2)$ is the ``collision strength'' of the transition, $\omega_2$ is the statistical weight of the upper level $2$. Note that there is only a mild dependence on \Te\ as $\sqrt{T_e}$.

The rate of collisional excitation is similarly given by $n_1 n_0 C_{12}$, where
\begin{equation}
C_{12}=\frac{\omega_2}{\omega_1} e^{-E_21/kT_e}C_{21}
\end{equation}
that depends exponentially on \Te.

A CEL is produced by the collisional excitation of the upper level followed by a radiative de-excitation. Neglecting stimulated emission (usually not important in diffuse nebulae), the population $n_2$ of the upper level is given by
\begin{equation}
\frac{dn_2}{dt} = -n_2(A_{21}+n_eC_{21}) + n_1n_eC_{12} \,,
\end{equation}
where $A_{21}$ is the Einstein coefficient for spontaneous
(radiative) transition.

At equilibrium ($dn_2/dt=0$):
\begin{equation}
\frac{n_2}{n_1} = \frac{n_e C_{12}}{A_{21}+n_e C_{21}}
\end{equation}
The critical density $n_c$ is defined as the density at which the rate of collisional de-excitations equals the rate of spontaneous radiative transitions
\begin{equation}
n_c = A_{21}/C_{21} 
\end{equation}
and therefore:
\begin{equation}
\frac{n_2}{n_1} = \frac{n_e/n_c\exp^{-E/kT_e}}{1+n_e/n_c}
\end{equation}

When the medium has a density much lower than the critical density of 
the transition ($n_e\ll n_c$), $n_2$ depends exponentially on $T_e$ and $n_1\sim n_X$, where $n_X$ is the density of the ion $X$. The volumetric emissivity $J_{21}$ of a line is therefore given by
\begin{equation}
J_{21} = h\nu_{21}\frac{n_2A_{21}}{4\pi} \sim n_e n_x e^{-E_{21}/kT_e} \,.
\end{equation}

Once $n_e$ and \Te\ are measured, ionic abundance can be obtained by comparing the flux of the CEL to the hydrogen recombination lines. 

Adding up the abundances of the observed ions, and assuming an ionization correction for the unobserved ones, the total elemental abundance is derived  \citep[e.g.,][]{Aller54,Dinerstein90,Pilyugin05b,Pilyugin06b,Pilyugin09,Bresolin09,Pilyugin10a,Perez-Martinez14,Perez-Montero14,Perez16}.

Electron density is usually measured by density-sensitive doublets, i.e., doublets that
have critical densities not far from the gas density, hence whose flux ratio depends strongly on the density,
such as  \oii3726,3729 and \sii6717,6731. However,
since in order to infer the abundance the flux of the CEL lines has to be compared with the flux of the hydrogen recombination lines, and since the emissivity of the hydrogen recombination lines scales as $n_e n_p$ (where $n_p$ is the density of protons), when comparing the emissivities the dependence on density cancels out (as long as the density is below the critical density), and therefore only the dependence on temperature is really critical to determine the abundance of the ion.

Electron temperatures can be measured through ``auroral'' lines, i.e.,
lines coming from high quantum levels, whose excitation is
very sensitive to temperature.
The most commonly used of such lines are OIII]1661,1666, \oiii4363, \oii7320,7330, \sii4069, \nii5755, and \siii6312 \citep[e.g.,][]{Castellanos02}, 
whose flux are compared to other brighter lines from the same species but from a very different energy levels.  A more complete list of the line ratios used can be found in \cite{Perez-Montero17b}. The most accurate results are only obtained when several of these lines are used because they trace different regions of the emitting nebula, in particular regions of different ionization levels. 
The \siii6312 is an interesting line, although not often used. In fact, it can be observed to higher metallicity than, for example, \oiii4363, but the corresponding nebular lines needed to measure the abundance are \siii9069,9532, which are in a part of the spectrum which is often not observed.
Specialized routines are now available that derive temperature from line ratios, such as PYNEB
\citep{Luridiana12,Luridiana15}, a PYTHON version of the STSDAS NEBULAR routines. A list of sources for atomic
data can be found, for example, in \cite{Garcia-Rojas09} and \cite{Fang15}. A tutorial on the use on this
method was recently published by \cite{Perez-Montero17b}.

The practical use of this method is often limited by the intrinsic faintness of the auroral lines, which typically are 10-100 times fainter than the corresponding Balmer lines. While the auroral lines are routinely observed in local or low redshift, metal-poor, star forming galaxies \citep{Kennicutt03,Izotov06a,Izotov06b,Izotov07a,Izotov11a,Izotov12,Kreckel15,Haurberg15,Amorin15,Pilyugin15,Lagos16,Ly16,Sanchez-Almeida16,Izotov18},  at higher redshift ($z>1$) only a few detection claims exists. While some of these measurements are based on UV lines such as OIII]$\lambda$1661,1666 \citep{Villar-Martin04,Erb10},  most of the claims are based on optical lines observed in the near-IR and are often affected by very low signal-to-noise  ratios  (often S/N$<$2), are based on uncertain identifications (due, for example, to the nearby \hc), and would produce unrealistically high \oiii4363/\oiii5007 ratios
\citep{Yuan09, Christensen12b,Stark13b,James14a,Maseda14,Sanders16b}.
In contrast, auroral lines are often detected in stacked spectra of both local and distant galaxies \citep{Andrews13,Trainor16,Curti17,Bian18}. 
Average relations between the temperatures of regions of different ionizations in HII regions and galaxies are also computed
\citep{Perez-Montero03,Izotov06a,Pilyugin07a,Pilyugin06b,Pilyugin09,Pilyugin10a,Curti17}. These relations can be used when some auroral lines are not observed. Direct relations between ``strong'' lines and auroral line fluxes, the so-called {\it f-f relations}, have been proposed by \cite{Pilyugin05b}, \cite{Pilyugin06a}, and \cite{Pilyugin09}. Recently, \cite{Curti17} and \cite{Curti19a}  have extended these relations and applied them to SDSS galaxies, obtaining very tight relations between auroral and strong line fluxes.

A source of uncertainty of this method is due to the thermal and density structure of the emitting nebulae. The exponential dependence on \Te \ means that emission is dominated by the regions of higher \Te \ and the derived values of the parameters can be biased \citep[e.g.,][]{Peimbert67,Stasinska02,Liu02,Liu03,Esteban04,Garcia-Rojas07,Peimbert07}.
This  problem is typically addressed by introducing a temperature-fluctuation parameter $t^2$ that can be estimated by comparing different measurements of \Te\ \citep[e.g.,][]{Peimbert67,Peimbert04,Peimbert12,Peimbert17}).

The problem of temperature fluctuations can be alleviated by exploiting coronal lines from various ionization stages of the same elements. These lines are emitted by different parts of the HII regions, including the partially ionized regions, and the thermal structure of the HII region can be better sampled \citep[e.g.,][]{Campbell86,Garnett92,Kennicutt03,Bresolin05,Pilyugin09, Curti17}. This point is further discussed in Sect.~\ref{sec:measmet_comparison}.\\

\subsection{Abundances from metal recombination lines}
\label{sec:measmet_ism_rec}

The permitted, recombination lines (RL) of metal ions are in principle the most direct way to derive the chemical abundances. In the usual conditions of HII nebulae (no stimulated emission), the volumetric emissivity of a permitted line of the ion $X$ due to a transition between the levels $i$ and $j$ is
\begin{equation}
J_{ij} = \frac{h\nu_{ij}}{4\pi} n_{X^{+i+1}} n_e \alpha^{\rm eff}_{ij} \left(X^{+i},T_e\right) ,
\label{eq:JRL}
\end{equation}
where $\alpha^{eff}_{ij}$ has only a weak dependence on \Te. The ionic abundance is computed by comparison with hydrogen recombination lines, which have the same dependence on density, hence the estimated abundances are nearly insensitive to the gas density.

The total element abundance is measured after assuming a ionization correction for the unobserved ions
\citep{Peimbert03,Tsamis03,Esteban04,Lopez-Sanchez07, Peimbert07,Peimbert14a, Peimbert14b, Esteban14,Toribio-San-Cipriano17}. In practice the d

The RLs most commonly used to measure metallicity are  OI~8446,8447, OII~4639,4642,4649, OIII~3265, OIV~4631, NII~4237,4242, NIII~4379, CII~4267, CIII~4647, and CIV~4657.

The mild dependence on density and temperature reduces the impact of clumping and temperature fluctuation that can affect the CEL-\Te\ method described above.
As the line emissivity in Eq.~(\ref{eq:JRL}) is proportional to the abundance of each element, recombination lines from metallic species are very faint when compared to the H recombination lines, of the order of $10^{-3}$--$10^{-4}$ with respect to Balmer lines even for the most abundant elements like C and O. The detection of RL from metals is practically limited only to bright HII regions, planetary nebulae (PNe) and supernova remnants (SNRs), with spectra of high resolution and high signal-to-noise ratio \citep[e.g.,][]{Peimbert04,Garcia-Rojas07}.

\subsection{Photoionization models} 
\label{sec:measmet_ism_mod}

The widely adopted alternative to the direct methods
consists in using photoionization models to predict or interpret
the relative strength of some of the main nebular lines to constrain
the gas-phase metallicity. This approach has high potential, but also a number of limitations,
as unavoidably only a small number of parameters involved in the
photoionization calculations can be
realistically explored, with simplified geometrical configuration, generally not
properly reflecting the complexity and distribution of real HII regions.
However, the advantage is that in principle there is no limit on the
possible properties of star forming regions that can be explored,
especially in terms of metallicity range and properties of the ionizing
spectrum. This flexibility enables also the potential exploration to systems
at high redshift, even in extremely metal poor environment or extreme
ionizing spectra, which do not have local counterparts
\citep[e.g.,][]{Schaerer03,Kewley13b,Jaskot16,Xiao18}. The additional advantage is that
such models can constrain, together with metallicity, other properties
of the ionized gas, such as the ionization parameter. Finally, as
discussed in the next sections, photoionization
models allow to calibrate strong-line diagnostic associated with lines that
are not possible to calibrate empirically through the direct method because
the required data are not available.

Generally, the classical approach is to use detailed photoionization
codes, such as CLOUDY \citep{Ferland13}
or Mappings \citep{Binette85,Sutherland93,Dopita13} to generate a grid
of models out of which a number of line ratios are extracted and proposed
as diagnostics of the gas metallicity 
\citep[e.g.,][]{Kewley02a,Nagao11,Dopita13,Jaskot16,Gutkin16,Chevallard16,Feltre16,Dopita16}. Generally, with the exception of a few cases \citep{Jaskot16}, models assume a simple plane-parallel
geometry. Most of them assume that chemical abundances scale proportionally
to solar, except generally for nitrogen whose abundance is assumed to scale
with the global metallicity assuming a fixed relationship 
\citep[see, e.g., the discussion in][]{Nicholls17}.
The effect of dust is generally included, both in terms of dust extinction
and in terms of
dust depletion of chemical elements, and the assumptions on dust distribution affect the resulting structure of the HII region \citep[e.g.][]{Stasinska01}. Dust depletion is
generally inferred from Galactic studies and assuming 
that the dust-to-metal ratio remains constant with metallicity, which however
may not be the case at low metallicities \citep{DeCia16,DeCia18a}. 
Another important assumption is that most ionized clouds are ionization bounded 
(i.e. the ionized zone is not truncated by the dimension of the cloud), but
this assumption may not apply in number of galaxies, especially in some young,
strongly star forming systems \citep{Nakajima14}. Some models have incorporated
this possibility to investigate metallicity diagnostics, although
restricted to the UV \citep{Jaskot16}.
The primary quantities
that are varied in the grid of parameters are the gas metallicity and ionization parameter, defined as the dimensionless ratio of the incoming photon flux density and gas density
at the cloud surface, normalized by the speed of light, 
$U=q/c=Q_{ion}/(4\pi r^2n_ec$) where $Q_{ion}$ is the number of ionizing photon emitted per unit time by the source, and $r$ is the distance to the emitting cloud . 
Metallicity and ionization parameter are, for a given shape of the ionizing flux,
the two  most important parameters in affecting the flux ratios
of the main nebular lines, and are often subject to degeneracies, in the
sense that most emission line ratios depend on both parameters.

Some authors, especially in early models, adopted a constant gas density.
Later \cite{Dopita14} have pointed out that radiation pressure on dust is probably
the primary physical mechanism regulating the physical properties of 
gaseous clouds and that, therefore,
it is physically more sensible to adopt a constant pressure and derive the density
distribution accordingly. In either case the assumed density or pressure  can be another
parameter that is varied to construct the grid of models.

While original models assumed only a single representative
shape of the ionizing
spectrum, more recent models typically explore a broad range of ionizing
stellar continua from stellar population synthesis models,
lately even expanding the range to include the contribution of stellar binarity (especially at low luminosity) and rotation \citep{Levesque12a, Leitherer14, Stanway16, Choi17,Xiao18}.
The possible presence of turbulence introduces further uncertainties \citep{Gray17}.

While such photoionization models are widely used, we  warn about
some important caveats and issues.
Despite extensive efforts, these photoionization models are still quite
simplistic, and are still far from capturing the complexity of the HII regions and their distribution
in galaxies. An indication of this is that, while models tend to statistically reproduce fairly well
many diagnostic diagrams, when individual systems are considered it is often very difficult to 
find a single model that simultaneously reproduces all observed nebular line ratios.
The basic issue is likely to be that both within each HII region and among the several HII regions
typically sampled by the large projected aperture of extragalactic surveys, the ionization parameter 
is not characterized by a single value, but by a broad distribution. The same issue  applies to other parameters
such as density, temperature and metallicity. Future generation models will hopefully
incorporate this additional feature, although it is likely demanding to implement.

Another issue is that assuming a fixed relationship between N/H and O/H, typically of the steep form
$\rm N/H\propto (O/H)^2$ in the intermediate/high-metallicity regime,
makes the flux of the nitrogen nebular lines (and in particular \nii6584) hypersensitive to
metallicity. However, one should take into account that nitrogen nebular lines
are probing directly the abundance of nitrogen; if a system deviates from the assumed relation (which, for
instance seems
to be the case in galaxies of different masses and other sub-galactic regions) then the nitrogen emission
line may provide deceiving information. A similar problem applies to carbon nebular lines when using
UV spectra.

More recent models are introducing a Bayesian (or multi-features fit) approach
in which the information  coming from multiple nebular lines (and sometimes also from the continuum)
is combined to identify the best model among those provided by the grid \citep[e.g.,][]{Tremonti04,Chevallard16,
Blanc15,Perez-Montero16,Perez-Montero17,Vale-Asari16}. These codes are certainly a step forward. However, some of them still make
the assumption of a fixed relationship between N/H, C/H and O/H given a priori to the code. As mentioned,
this risks the determination of the metallicity to be dominated by the flux of a single
nitrogen or carbon line and, in addition, it is not really possible to distinguish between effect of
differential chemical abundances and global metallicity. However, some of the most recent codes
do consistently derive the global metal abundance and, separately, the abundance of nitrogen and carbon,  without making  assumptions on their abundance scaling with metallicity
\citep{Blanc15,Perez-Montero16,Perez-Montero17,Vale-Asari16}.

As  will be discussed in Sect.~\ref{sec:measmet_comparison}, 
photoionization models tend to overestimate
gas metallicity 
with respect to the \Te\ method
from $\sim0.2$ up to even $\sim0.6$~dex.
The discrepancy is particularly strong at high
metallicities. The origin of the discrepancy is not totally clear. Dust depletion is
certainly a factor contributing for at least 0.1~dex, indeed photoionization models
do account for dust depletion, while the direct methods simply give the actual gas metallicity.
Another potential problem is how nitrogen is included in the photoionization models;
its assumed quadratic dependence on the metallicity may artificially boost the inferred
metallicity. Alternatively, \Te-based metallicities may be biased low as a consequence
of low metallicity regions being characterized by brighter auroral lines. An interesting additional possibility is that the basic assumption that electrons
in HII regions follow a Maxwell-Boltzmann distribution may be incorrect and that
instead they follow a distribution similar to that observed in astrophysical
plasmas in and beyond the solar system, i.e., a so-called $\kappa$-distribution
(a generalized Lorenzian distribution),
which is characterized by a more extended tail towards high energies than
the Maxwell--Boltzmann distribution. This possibility has been investigated recently
by \cite{Binette12}, \cite{Nicholls12} and \cite{Dopita13}, who have pointed out that, very
interestingly, the introduction of a
$\kappa$-distribution in models of HII regions solves some of the outstanding problems,
such as the discrepancies in the temperatures measured by some auroral lines (see Sect.~\ref{sec:measmet_comparison}), and makes
the photo-ionization derived metallicities in better agreement with those inferred from
the auroral lines. This is certainly an area of study worth  developing further.
A revision of some of the nebular line diagnostics by including
the effects of a $\kappa$ distribution has already been presented by \cite{Dopita13}.
It is also possible that the introduction of the $\kappa$-distribution is actually a way to reproduce the effect of gradients in the properties of the HII regions or of a collection of HII regions with different conditions, for example different temperatures.

\subsection{Comparison among the different methods} 
\label{sec:measmet_comparison}

In general the three methods described above give different results. The ratio of the abundances obtained from the more direct methods, i.e., from RLs and \Te-CELs,  is often referred to as ``Abundance Discrepancy'' \citep[AD,][]{Tsamis03,Garcia-Rojas07,Garcia-Rojas09,Garcia-Rojas13,Peimbert17,Toribio-San-Cipriano17} and can be very high, up to a factor of $\sim 5$ \citep{Tsamis03}. This discrepancy is a serious limitation to our understanding of ionized nebulae and its origin is debated. As already mentioned, temperature fluctuations in HII regions are definitely present \citep{Garcia-Rojas07} but their real effect on abundance determination is not clear and there also are indications, based on the comparison with fine-structure infrared lines, that they are not the source of the AD \citep{Tsamis03}. The presence of small-scale, chemical inhomogeneities due to a clumpy, not well mixed gas distribution, possibly related to the presence of SN remnants, has been proposed by \cite{Tsamis05} and \cite{Stasinska07b} as a way to explain the different results from the RLs (dominated by the metal-rich clumps) and the CELs (dominated by the more diffuse medium),  and by \cite{Binette12} to explain the difference in temperature between [OIII] and [SIII].
Flourescence excitations via continuum or resonance lines have also been proposed as causing unrealistic RL flux ratios and therefore affect RL-derived metallicities \citep{Grandi76,Liu01,Tsamis03}.

\begin{figure}
\centerline{\includegraphics[width=12cm]{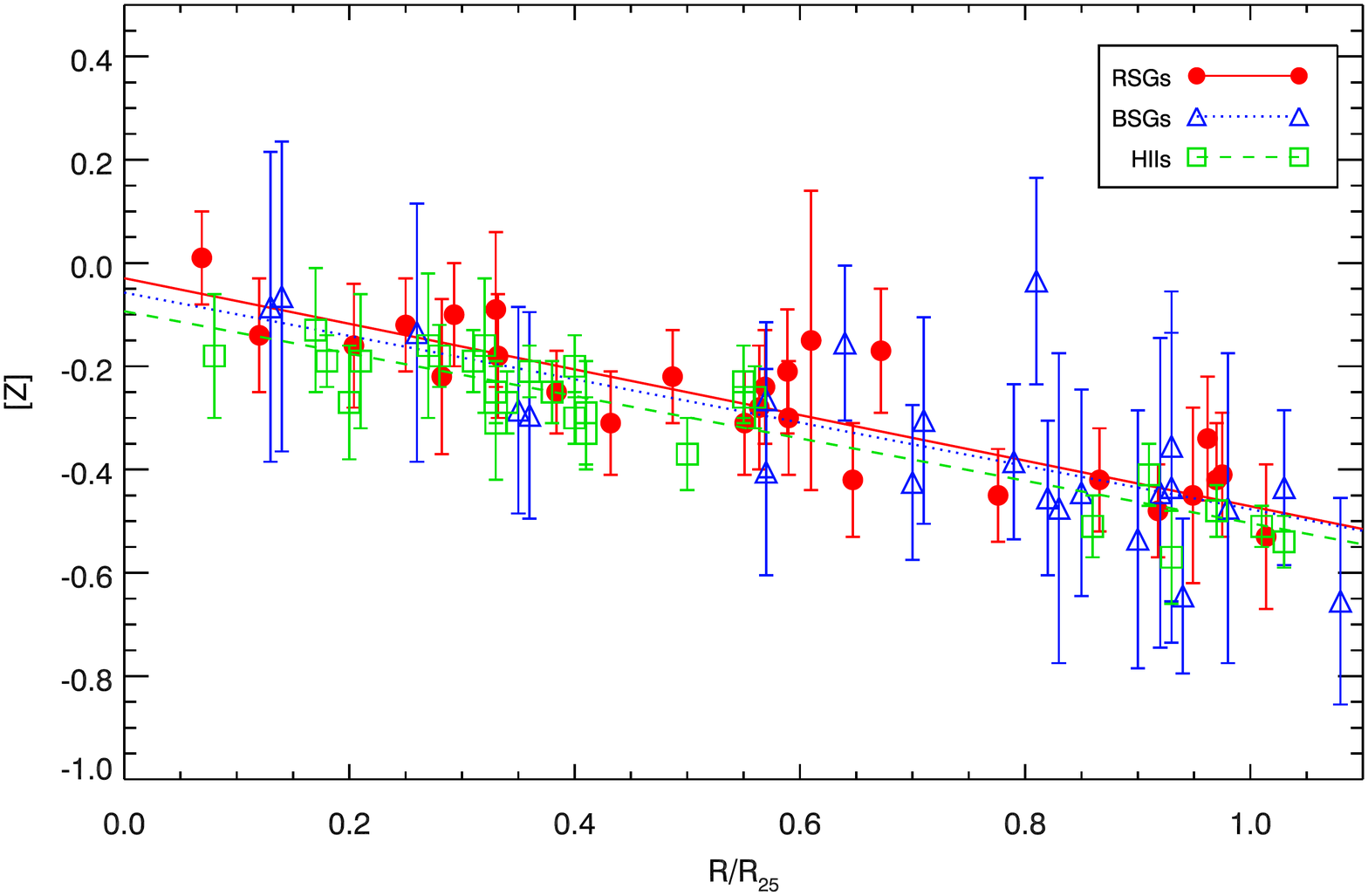}}
\caption{The radial metallicity gradient in M300 measured by three different methods: blue supergiants (BSG, blue triangles), red supergiants (RSG, red dots), HII region metallicities with the ``\Te'' method (green squares). The agreement is remarkable. From \cite{Davies15}.
}
\label{fig:METALS_davies15}
\end{figure}

As mentioned in the previous section,
even higher differences exists between the results of CELs and RLs and those of the photoionization models,
systematic discrepancies of 0.6--0.7 dex are not uncommon \citep[e.g.,][]{Kewley08,Moustakas10,Lopez-Sanchez12}, although some photoionization models
tend to be in better agreement with the \Te\ method \citep{Perez-Montero14}. Generally, direct \Te \ metallicities tend to be lower than those derived from photoionization models, with RLs usually in between these two values.

As discussed in Sect.~\ref{sec:measmet_ism_mod}, when making these comparisons one should also take into account that the direct \Te \ and RL methods estimate the metallicity of the gas phase only, while photoionization models generally take into account dust depletion and provide results in terms of {\it total} metallicity of the ISM (i.e. gas and dust). Therefore, \Te \ and RL metallicities should be increased by about 0.1 dex in order to have a fair comparison with the photoionization models.

The problem of understanding these differences and obtain the best estimates of the gas-phase metallicities can be addressed by comparing the values obtained from CELs, RLs, and photoionization methods with those of the young stars formed in the same region. The use of young stars is needed to guarantee that the stellar metallicity is similar to gas-phase metallicity, as young stars have been recently created from the same ISM that is observed in the HII regions, with the possible exception of some CNO reprocessing.

Independent estimates of metallicities can be obtained for blue supergiants (BSG) and red supergiants (RSG). 
Spectral type A and B BSG are massive (12--40 \msun), young (less than 10 Myr) stars in the short evolutionary stage leading to RSG. They are very bright and therefore their spectra can be obtained with the high-S/N needed to accurately measure chemical abundances. Their spectra show absorption lines from many elements (e.g., C, N, O, Mg, Al, S, Si, Ti and Fe). High resolution spectra can provide abundances with uncertainties of less then 0.05~dex \citep[e.g.][]{Kudritzki12},
while lower resolution spectra can also be used to obtain uncertainties of $\sim$0.1dex \citep{Kudritzki08,Kudritzki16}.

\begin{figure}[t]

\centerline{\includegraphics[width=12cm]{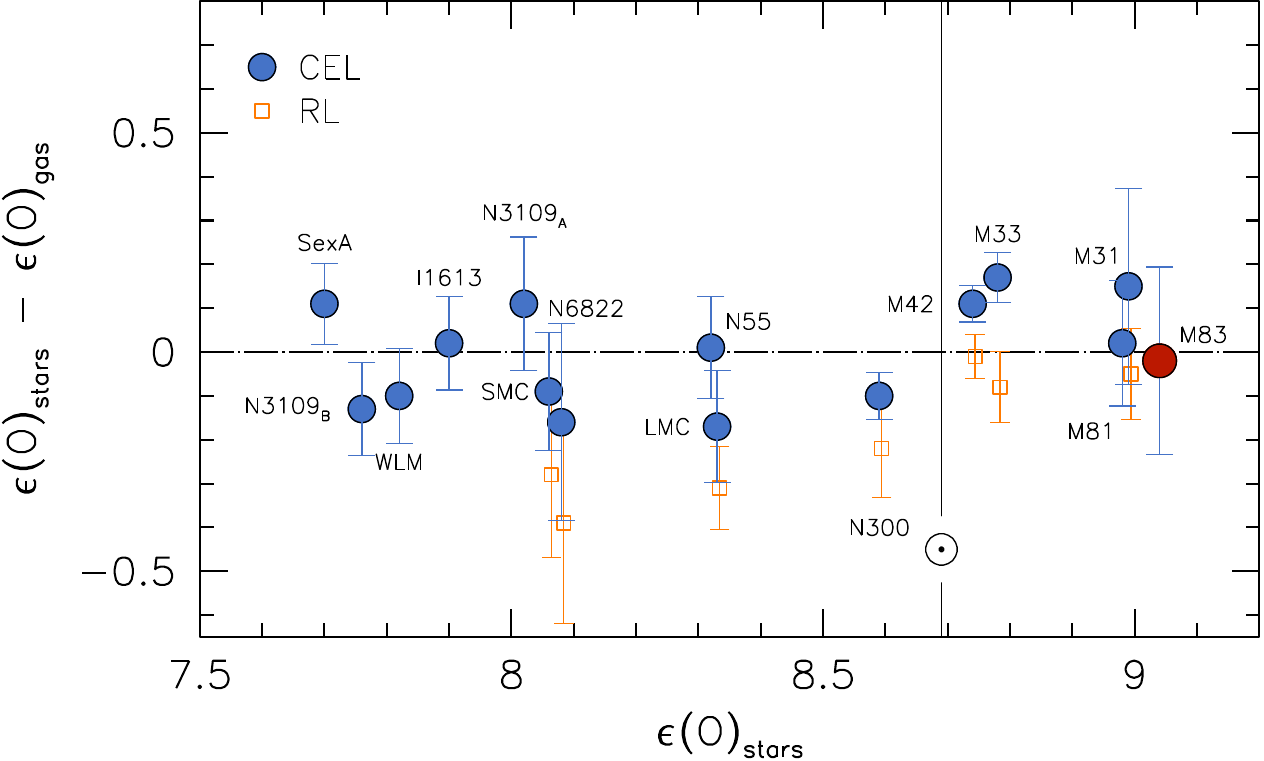}}
\caption{
Difference between stellar and gaseous metallicity as inferred ``directly'' from the \Te\  method (blue circles) and from the Recombination Lines (orange squares) in a sample of local galaxies and star-forming regions, as a function of stellar metallicity. The vertical line shows the adopted Solar value. From \cite{Bresolin16}.
}

\label{fig:METALS_bresolin16}
\end{figure}

RSGs have initial masses between 8 and 35 \msun, high bolometric luminosities peaking in the near-IR, and young ages (less than 50~Myr). 
Abundances of RSG can also be reliably measured if good spectroscopy is available \citep{Cunha07,Davies10,Patrick16,Davies17a}. 

The metallicity estimates derived for RSG and BSG are largely independent because the physical conditions of the atmosphere of the two types of stars are very different. 
The BSG method exploits hot atmospheres, and obtain metallicities from optical lines of singly and doubly ionized ions, while for RSG much colder atmospheres are used and near-IR lines from neutral metals are measured.  The agreement between these two methods is excellent \citep[e.g.][]{Gazak15,Davies17a}, and \cite{Zahid17} recently showed that the derived values are also in excellent agreement with those derived by spectral fitting of stacked spectra of many galaxies.\\

The stellar values can be compared to those derived for the ISM with various methods. This comparison has been subject of a number of recent studies \citep{Bresolin07a,Kudritzki12,Hosek14,Lardo15,Gazak15,Bresolin16,Davies17a,Toribio-San-Cipriano17} in which metallicity gradients of individual, local galaxies with different metallicities are derived with all the available methods to allow  a comparison on local scales (see Fig.~\ref{fig:METALS_davies15}). In most cases there is a good agreement between stellar metallicity and the results of the \Te-CEL method for the ISM. The status is summarized by \cite{Bresolin16} (see Fig.~\ref{fig:METALS_bresolin16} and \citet{Davies17a}). 
Photoionization models \citep{Kewley02a, Tremonti04, Maiolino08} usually provide higher metallicities, with differences of the order of 0.2--0.3~dex.

It is still not clear whether the stellar metallicities agree better with RLs or \Te-CEL derived values. According to the summary in \cite{Bresolin16} (Fig.~\ref{fig:METALS_bresolin16}), a good agreement is found with the \Te-CEL at any metallicity, while the RL method seems to overestimate metallicity with respect to the stellar values for sub-solar metallicities. The situation is not clear yet and better agreement with RL metallicities are also found in the Orion nebula \citep{Simon-Diaz10,Simon-Diaz11}.\\

{\bf Summarizing}, \Te\ direct gas-phase metallicities are in better agreement with the metallicities of young stars and therefore are now considered to be a more solid base for the ``strong line'' methods described in the next section\footnote{As explained above, RL lines are too faint and detected in too few galaxies to be used to define a strong line calibration.}
\citep{Denicolo02,Pettini04,Andrews13,Curti17}.

\subsection{Strong line calibrations} 
\label{sec:measmet_strong}

The emission lines required to directly measure the gas metallicity are very weak
and challenging even with large telescopes. This has generally confined the use
of the direct method to a few tens/hundreds local galaxies and HII regions, or resorting
to the use of stacking of large number of spectra \citep[e.g.,][]{Andrews13,Curti17}.
At high redshift the auroral lines
needed to apply the \Te\ method have been detected (or marginally detected)
only in a handful of sources.
This issue has prompted astronomers to calibrate alternative diagnostics of the
metallicity that exploit relatively strong nebular emission lines, which can be detected more easily,
even in low S/N spectra. This technique is often referred to as the ``strong line method''.

It is important to note that the strong line method is not a primary technique to derive metallicity but is a way to allow for an easier, albeit less precise, application of one of the primary methods introduced above.
 
These strong line calibrations have been performed empirically, through the direct methods \citep[e.g.,][]{Pettini04,Pilyugin05b,Pilyugin10b,Pilyugin16,Curti17},
through photoionization models \citep[e.g.,][]{Zaritsky94,McGaugh91,Kewley02a,Kobulnicky04,
Tremonti04,Nagao11,Dopita16},
or a combination of the two \citep[e.g.,][]{Denicolo02,Nagao06,Maiolino08}. 

An important warning is that a number of the ``strong line'' metallicity diagnostics
are highly degenerate with other parameters (ionization parameter,
density, pressure,\dots) or even exploit indirect correlations between metallicity
and other parameters, such as the correlation between metallicity and ionization
parameter \citep{Nagao06,Dopita06}
and the correlation between metallicity and nitrogen abundance.
It is important to be aware of these issues and using multiple diagnostics
is strongly encouraged. Within this context, for
what concerns the direct calibrations, \cite{Pilyugin10b} and \cite{Pilyugin16} have
developed methods which use strong line ratios to consistently
provide  metallicity, nitrogen abundance, and mitigate the effects of ionization
parameter.

In the context of photoionization models, the calibration of strong line ratios
is rapidly making the way, as mentioned above, to codes which simultaneously
fit multiple line ratios \citep{Blanc15,Chevallard16,Perez-Montero16,Perez-Montero17},
and, since most of these codes are publicly available, having
calibrations for individual ``strong line'' diagnostics is becoming less compelling.
However, the use of individual strong line diagnostics is still popular and handy, especially
for high redshift studies, where generally only a few nebular lines are measured.

The use of hybrid calibrations (i.e., both empirical, through the direct methods,
and exploiting photoionization models) was in the past needed to properly cover
the metallicity range \citep{Nagao06,Maiolino08}.
Indeed, while the low-metallicity range is fairly populated in terms of
galaxies and HII regions with auroral lines detections for the $T_e$ method,
at high metallicities the weakness of the auroral lines has resulted in
poor statistics and sparse sampling; this, together with concerns
of biases of the auroral-line detected samples in the high metallicity
range, has led to efforts to complement the empirical calibrations (at low metallicities)
with photoionization models-based calibrations at high metallicities.
However, more recently extensive stacking of SDSS galaxies \citep{Curti17}
has mitigated these issues and fully empirical ($T_e$-based) calibrations 
are available up to high metallicities.

We finally point out that all these strong line calibrations, either empirical
or through photoionization models, are based on HII regions and/or star forming galaxies. 
This is an important caveat for various reasons. The  presence of diffuse ionized gas (DIG, see Sect.~\ref{sec:measmet_strong_BPT}) or of contamination
by other excitation mechanisms, such as photoionization by shocks or harder
sources, such as AGNs or evolved post-AGB stars, can strongly
affect the nebular line ratios in a way that is independent of the metallicity and can vary from galaxy to galaxy.
The selection of HII regions/star forming galaxies is typically based
on the so-called ``Baldwin--Phillips--Terlevich'' (BPT)  diagnostic diagrams
which attempt to isolate
HII regions from regions excited by other mechanisms and sources 
(see Sect.~\ref{sec:measmet_strong_BPT}). However, the simple
demarcation reported by some authors \citep{Kauffmann03c,Kewley06b} is identified
in an empirical way or by using considerations based on theoretical models. In reality
the transition between different excitation mechanisms is certainly much smoother
and mixed. It is therefore very likely that the datasets used for calibrations
include contamination from AGN, shocks and post-AGB stars and, vice-versa,
samples of star forming regions are missed from the calibrations, especially
in the high metallicity regime. The same contamination issue is certainly
present when applying these calibrations to galactic regions which are marginally resolved,
or to the integrated spectra of galaxies in which different contributions
are likely mixed. We will discuss these issues in the following.

\subsubsection{Strong line calibrations: Optical lines} 
\label{sec:measmet_strong_opt}

Optical nebular line ratios are among the most widely exploited
to constrain the metallicity in galaxies, both because some of the
strongest nebular lines are in this wavelength range, and because
this spectral range is easily accessible from ground with a variety
of facilities and huge amount of data have been delivered by several
surveys.

Table \ref{tab:strong_line_calib} provides a list of the main strong line
diagnostic, and their definition, that have been proposed and calibrated 
either empirically or theoretically by several teams
\citep{McCall85,Skillman89c,McGaugh91,Zaritsky94,
Denicolo02,Kewley02a,Kobulnicky04,Tremonti04,Pettini04,Pilyugin05b,Nagao06,
Maiolino08,Pilyugin10b,Nagao11,Pilyugin16,
Brown16,Dopita16,Curti17}. 
In some of these
studies authors have suggested a combination of them to account for
secondary (or primary!) dependences, such as the dependence on ionization parameter
or nitrogen abundance \citep[e.g.,][]{Kobulnicky04,Pilyugin05b,Pilyugin10b,Pilyugin16,Curti17}.

\begin{figure}[t]

\centerline{\includegraphics[width=12cm]{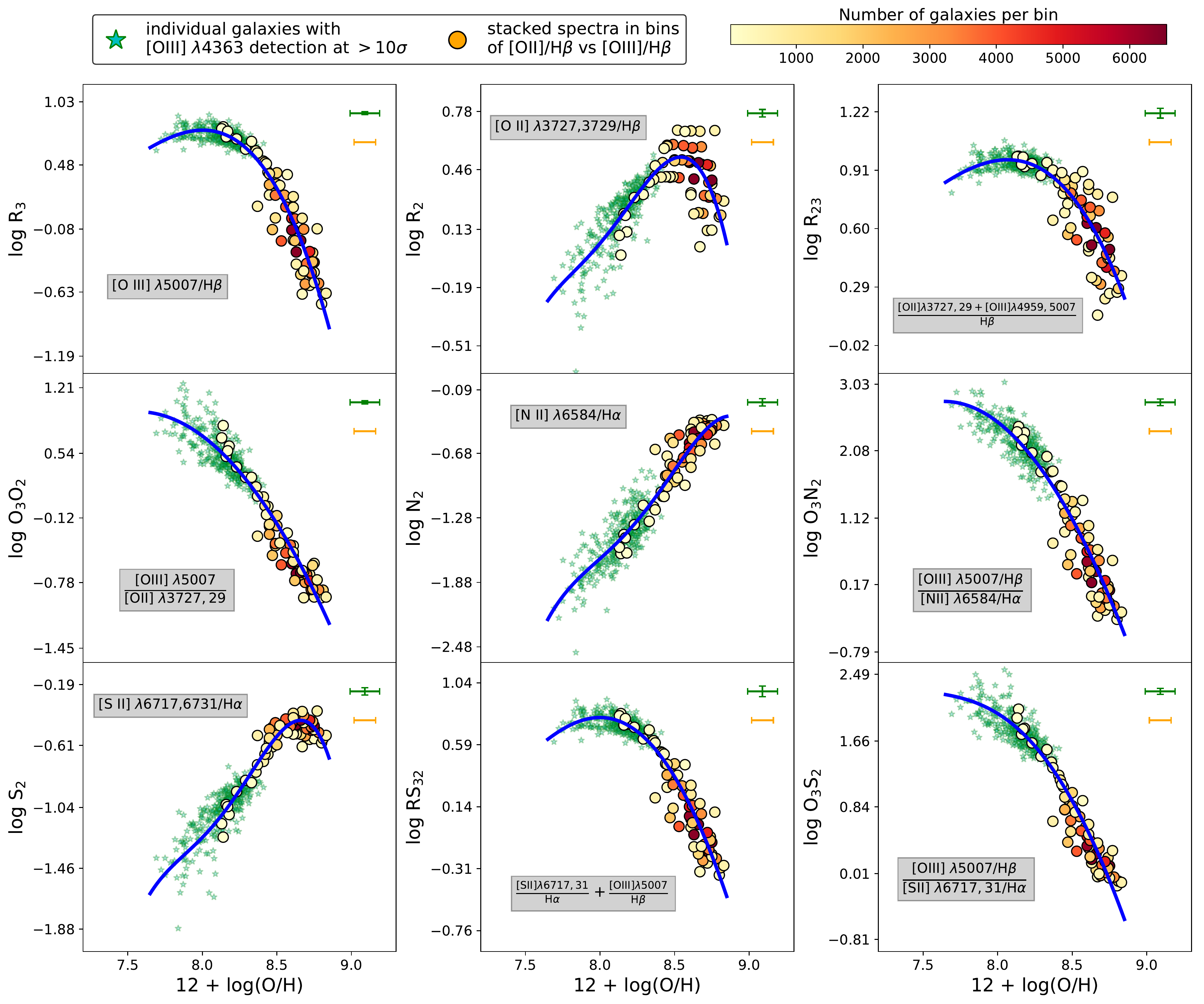}}
\caption{Optical strong line diagnostics as a function of oxygen abundance 
obtained by \cite{Curti17} and \cite{Curti19a}. For the definition of the strong line diagnostics on the Y-axes please refer to Table~\ref{tab:strong_line_calib}. Green symbols are individual 
galaxies and circles are measurements from
stacks of galaxies from the SDSS, color coded by the number of galaxies in each
stack (top color bar). These calibrations are based on the most recent realization of \Te\ direct method.}
\label{fig:calib_Curti}
\end{figure}

In the third column we provide the references for some of the calibrations
that have been proposed. We strongly suggest to use the empirical calibrations ($T_e$-based)
obtained by \cite{Curti17} and \cite{Curti19a}, and shown in Fig.~\ref{fig:calib_Curti}, as they
are based both on individual HII 
galaxies and stacks of SDSS galaxies
in which also low-ionization coronal lines are detected and which, therefore, enable
the empirical calibration to extend to high metallicities, while also mitigating
potential biases.

As mentioned, given that each of these line ratios is generally degenerate with
other parameters, it is advised to combine multiple of these diagrams to disentangle
dependences (a publicly available routine to combine these calibrations can be found at \url{http://www.arcetri.astro.it/metallicity/}).
However, if only some of these
diagnostics are available it is important to be aware of a number of potential caveats,
strengths and weaknesses, which are summarized in columns 4 and 5 of
Table \ref{tab:strong_line_calib}, and discussed in the following.

R23 ($\equiv$log((\oii3727 + \oiii5007,4958)/\hb)) has been one of the parameters
most widely adopted, as it involves emission lines of both the main ionization
stages of oxygen, O$^+$ and O$^{+2}$, hence it is less affected by the
ionization structure of the HII regions. However, it is subject to a significant
dependence on the ionization parameter, which some authors attempt to correct by including additional 
transitions such as O32, as discussed below \citep{Kobulnicky04,Pilyugin05b,Nagao06}. 
The additional problem of this
diagnostic is that it is double-branched (i.e., a R23 value can be associated
with two very different metallicities), hence identifying which of the
two branches applies requires the use of another diagnostics. 
Moreover, it has a very weak dependence on metallicity at the turnaround between the two branches 
(roughly between 12+log(O/H)=8.0 and 8.5, Fig.~\ref{fig:calib_Curti}), 
therefore being little sensitive over a significant range of metallicities. Finally,
the use of \oii3727 line
(far in wavelength from H$\beta$ and \oiii5007)
implies that this diagnostic is significantly sensitive to dust reddening, hence
requiring correction.

R2 ($\equiv$log(\oii3727/\hb)) and 
R3 ($\equiv$log(\oiii5007/\hb)) are individually strongly dependent on the ionization
parameter and on the ionization state of the gas, hence are not really
meant to be used in isolation, unless each of them is really the only indicator available.
Nevertheless, Fig.~\ref{fig:calib_Curti} shows that, at least for the stacked spectra, a tight, well-behaved relation exist between R3 and metallicity. This is mainly driven by the relation between metallicity and ionization parameter, but seems to work well across a large range of metallicity. R3 is also insensitive to dust reddening. In contrast, R2 is affected by the same issues as R23 in terms of being double-valued and being affected by dust reddening.

N2 ($\equiv$log(\nii6584/\ha)) is very convenient to use, especially at high-$z$,
as the two lines are very close-by hence requiring a very small spectral coverage
and being completely free from dust reddening (unless differential dust extinction is
present  inside the galaxies). However, this diagnostic is actually primarily tracing the abundance of
nitrogen, hence if galaxies follow a different N/O--O/H relation than the average sample
used for the calibration then this indicator can be deceiving. This diagnostic
is also well known to be strongly sensitive to ionization parameter \citep[e.g.,][]{Nagao06}. 
This ratio is also one of the parameters used to select star forming
galaxies in the BPT diagram, hence its calibration and application is highly sensitive to
the detailed BPT-boundary adopted to select HII regions, implying that the resulting
metallicity distribution of galaxies is also quite subject on the assumed BPT boundary.
Finally, given its quadratic dependence on metallicity, N2 can be very low in low metallicity galaxies, down to $-2$, 
and therefore the  \nii6584 line can be vary faint. This can introduce severe selection biases through the undetected galaxies if this effect is not properly taken into account.

S2 ($\equiv$log(\sii6717,31/\ha)) is very similar to N2, but it has the advantage
of not being linked to the nitrogen abundance. Sulfur is an $\alpha$-element like O, 
therefore S/O is expected to evolve much less than N/O.

O32 ($\equiv$log(\oiii5007/\oii3727)) is mostly a tracer of the ionization parameter
\citep{Diaz00,Kewley02a,Nagao06}.
The strong dependence on metallicity (Fig.~\ref{fig:calib_Curti}) is mostly secondary and is mostly due to the correlation between ionization parameter and metallicity.
Therefore, this diagnostic should really be used only
if no other tracers are available and always bearing in mind that it is a
very indirect tracer, through the U-Z relation \citep[which may evolve with
redshift, see][]{Kewley13a}, and subject to large scatter. Both O32 and R3 can be used to distinguish 
between the two branches of R23.

Ne3O2 ($\equiv$log(\neiii3870/\oii3727)), proposed by \cite{Nagao06} and \cite{Perez-Montero07}, is equivalent to O32, i.e. is primarily a tracer of the ionization parameter, however it has the advantage that \neiii3870 and \oii3727 are very close in wavelength, hence they require a very small wavelength range to be observed
simultaneously (quite convenient at high redshift), and their ratio is essentially unaffected by dust reddening. 
 Also, it is the only ratio available in the near-IR bands above $z\sim3.6$ when \hb\ and \oiii5007 exit the near-IR K band.
The drawback is that the \neiii3870 line is typically
ten times fainter than \oiii5007.

O3N2 ($\equiv$log(\oiii5007/\hb)-log(\nii6584/\ha)) is used quite extensively as it is a monotonic function of metallicity and unaffected by dust extinction (unless significant differential extinction is present). However,
it is very sensitive to the ionization parameter and to the N/O abundance ratio,
hence with the same major caveats discussed above for the other diagnostics
affected by the same issues.

O3S2 ($\equiv$log(\oiii5007/\hb +\sii6717,31/\ha)) is a very promising diagnostics,
proposed only recently by \cite{Curti19a} and Kumari (in prep.). It is essentially
equivalent to R23, but where \oii3727/\hb\ is replaced by \sii6717,31/\ha. 
Hence, as R23,
it probes both the high and low ionization stages of the gas,
but it has the additional advantage of being
nearly insensitive to dust reddening. However, it shares some of the same issues as R23, in
the sense that it has a secondary dependence on the ionization parameter and is double valued, although the inversion point is conveniently shifted to lower metallicities
relative to R23.

N2S2H$\alpha$\ ($\equiv$N2/S2+0.264$\cdot$N2) was developed by \cite{Dopita16} and
is expected to be less sensitive to the ionization parameter with respect to N2 and S2,
and share with them the same requirement for the small wavelength range and
the feature of being little affected by extinction.
The calibration provided by \citep{Dopita16}  through photoionization models
is strongly dependent on the assumed N/O--O/H relation.
This diagnostic has been re-calibrated empirically by \cite{Curti19a}, but the
inclusion of the nitrogen line still preserves its dependence on the nitrogen abundance.

N2O2 ($\equiv$log(\nii6584/\oii3727)) and N2S2 ($\equiv$log(\nii6584/\sii6717,31))
are both very good tracers of the abundance of nitrogen relative to $\alpha$\ elements
\citep[e.g.,][]{Perez-Montero09}; however, they are often used also as tracers
of metallicity thanks to the correlation of the nitrogen to $\alpha$-elements abundance ratio (N/$\alpha$, such as N/O and N/S) with O/H at high metallicities. Yet,
one should bear in mind that the dependence on metallicity is therefore indirect
and that these diagnostics can be deceiving for systems deviating from
the N/$\alpha$--O/H average relation. Both diagnostics fail to be a sensitive
metallicity diagnostics
at low metallicities where the N/$\alpha$--O/H flattens.
N2S2 has the advantage, relative to N2O2, of being much less sensitive to dust
reddening and requiring a small wavelength range.

\begin{landscape}
\begin{table}
\caption{List of optical strong line ratio metallicity diagnostics that
have been proposed in the literature. }
\begin{tabular}{lllcc}
\hline
\hline
Name & Definition & Calibrations$^a$     & Strengths & Weaknesses and Caveats\\
   &            &   & & \\
\hline
\hline
R23 & log((\oii3727 + \oiii5007,4958)/\hb) & 1--9,13--15 & Accounts for both ionized phases of oxygen &
	Secondary dependence on U \\
 & & & & Double valued \\
 \hline
R2 & log(\oii3727/\hb) & 6--9,13 & & Strongly dependent on U\\
 & & & & Double valued \\
 \hline
R3 & log(\oiii5007/\hb) & 6--9,13,14 & Unaffected by dust reddening & Strongly dependent on U\\
 & & & Requires small wavelength range & Double valued \\
\hline
N2 & log(\nii6584/\ha) & 6--10,12,14,17 & Unaffected by dust reddening & Strongly dependent on U\\
	& & & Requires small spectral range & Strongly dependent on N/H \\
\hline
S2 & log(\sii6717,31/\ha) & 9 & Little affected by dust reddening & Strongly dependent on U\\
	& & & Requires small spectral range &  \\
\hline
O32 & log(\oiii5007/\oii3727) & 6--9,13,14 & Sometimes last resort at high-$z$ & Traces primarily U\\
\hline
Ne3O2 & log(\neiii3870/\oii3727) & 6--8,13,14, 16 & Unaffected by dust reddening & Traces primarily U\\
	 &  & & Requires small spectral range & \\
	 &  & & Available from ground out to  z$<$5.2  & \\

\hline
O3N2 & R3 - N2 & 6--10,12,14,17 & Unaffected by dust reddening & Strongly dependent on U\\
	  &  & & & Strongly dependent on N/O \\
\hline
O3S2 & log(\oiii5007/\hb + \sii6717,31/\ha) & 9 & Little affected by dust reddening & Secondary dependence on U\\
 &   & & Traces both high and low ionization stages &  Double valued \\
 \hline
N2S2H$\alpha$ & N2/S2+0.264$\cdot$N2 & 9,11 & Little affected by dust reddening & Strongly dependent on N/H and N/S\\
	 &  & & Requires small spectral range & \\
	&  & & Mitigates U dependence & \\
\hline
N2O2 & log(\nii6584/\oii3727) & 4,6,7,9,12 & Little affected by U & Traces primarily N/O\\
	& & & & Requires dust reddening correction \\
    	&  &  &  & Insensitive at low metallicities\\
\hline
N2S2 & N2-S2 & 4,6,7,9,12,16 & Little affected by dust reddening & Traces primarily N/S\\
	&  &  & Requires small spectral range & Insensitive at low metallicities\\
     &	 &  & Little affected by U & \\ 
\hline
\hline
\end{tabular}
$^a$References for the some of the calibrations: 1) \cite{McCall85}, 2) \cite{Skillman89b},
3) \cite{McGaugh91}, 4) \cite{Kewley02a}, 5) \cite{Kobulnicky04}, 6) \cite{Nagao06},
7) \cite{Maiolino08}, 8) \cite{Curti17}, 9) \cite{Curti19a}, 10) \cite{Pettini04}, 11) \cite{Dopita16},
12) \cite{Perez-Montero09}, 
13) \cite{Jones15b}, 14) \cite{Bian18}, 15) \cite{Strom17b}, 16) \cite{Perez-Montero07}, 17) \cite{Marino13}
\label{tab:strong_line_calib}
\end{table}

\end{landscape}

Even more indicators, based a different combinations of several lines or on other elements such as Argon, have been proposed, among others, by
\cite{Perez-Montero05},
\cite{Stasinska06a}, \cite{Perez-Montero07}, \cite{Kobulnicky99}, \cite{Dopita16}, and \cite{Pilyugin16}. Most importantly, some of these authors provides
interlaced combination of multiple line diagnostics that attempt to explicitly take into account
(and therefore mitigate) the effects of other physical parameters, such as the ionization parameter.

It is important to recall that, when attempting to investigate the nitrogen abundance N/O as a function of global metallicity O/H, obviously none of the strong line metallicity diagnostics involving the nitrogen lines should be used to measure the
global metallicity O/H, as in this case both axes
would essentially measure more or less directly the nitrogen abundance; for this
kind of studies the global metallicity O/H should be measured by using only nitrogen-free
diagnostics. Calibrations giving N/O (or N/S) in terms of N2O2 and N2S2 are given, for instance, by
\cite{Perez-Montero09} (based on empirical calibration) and \cite{Strom17b}  (through photoionization modeling).\\

{\bf Summarizing}, the results of strong line methods based on optical lines critically depend on the method used to calibrate their line ratios, and calibrations based on the \Te\ method should be preferred.  When lines of other non-$\alpha$ elements, such as nitrogen, are used to measure oxygen abundance, variations of the relative element abundances add another important source of uncertainty.

\subsubsection{Strong line calibrations: UV lines}
\label{sec:measmet_strong_UV}

Although extensive models and codes have been developed
to infer the metallicity from the UV lines \citep{Fosbury03,Jaskot16,Gutkin16,Feltre16,Chevallard16,Perez-Montero17,Byler18,Nakajima18},
and some works also  measure  metallicities directly using UV auroral lines
\citep[e.g.,][]{Erb10,Berg18}, no clear, simple calibration has been proposed
to derive the gas metallicity from ratios of UV strong lines. Ratios between
CIII]1908, CIV1549 and HeII1640 (often some of the brightest line detected in the UV spectra of star forming galaxies)  are all strongly sensitive to the ionization
parameter and shape of the ionizing continuum, and they are also produced/contributed by different physical processes: CIII]1908 is a collisional excited nebular line, CIV1549 is a resonance line which is generally blended with interstellar absorption and stellar P-Cygni profile, HeII 1640 is an interstellar recombination line in the highly ionized part of HII regions, but is also produced by the atmospheres of Wolf-Rayet stars. \cite{Perez-Montero17}
 also investigated the ratio of UV carbon lines with the
optical hydrogen recombination lines; however, besides the problem of these
ratios being extremely sensitive to dust reddening, no clear trend
was found with metallicity, likely also because of the strong dependence 
on the ionization parameter.

However, the UV range contains the optimal nebular lines
for the measurement of the C/O abundance ratio. 
Indeed, based on empirical calibrations, \cite{Perez-Montero17}
have shown that the flux ratio

$$\rm C3O3\equiv \log{\left( \frac{CIII]1908+CIV1549}{OIII]1664}\right) }$$

\noindent
is primarily sensitive only to the C/O ratio, and they provide an empirical
relation in the form $\rm \log{(C/O)} = -1.069+0.796\cdot C3O3$.
One however has to be aware of the caveats on the different production mechanisms of these lines discussed above, especially for what concerns CIV1549. If information on the electron temperature is available from auroral lines, then a more accurate determination of the C/O abundance (or at least of $\rm C^{+2}/O^{+2}$) can be obtained from the CIII]1908/OIII]1664 ratio \citep{Garnett95,Garnett97}, hence not having to rely on CIV1549 and the  issues associated with this transition.

\subsubsection{Strong line calibrations: far-infrared lines}
\label{sec:measmet_strong_IR}

The advent of IR space observatories such as Spitzer \citep{Werner04} and Herschel \citep{Pilbratt10},
the prospect of new major spectroscopic IR surveys with the next
generation satellites \citep[SPICA,][]{Roelfsema18}, and the possibility of observing the rest-frame
far-IR wavelength range in distant galaxies with ground based
sub-mm/mm observatories, such as ALMA and NOEMA, has fostered the
investigation of mid/far-IR transitions as metallicity tracers.

Far-IR, fine-structure lines are expected to become a main coolant of HII regions at moderately high metallicities and low temperatures \citep[e.g.][]{Stasinska02}.
One problem of these transitions is that the most useful of them,
within this context, are sparsely distributed across the broad IR wavelength
range, often posing observational challenges. The additional problem is
that these transitions tend to have low critical densities, making
the gas density an additional important parameter affecting these
diagnostics. Moreover, some of these transitions come from multiple gas
components, making their modeling difficult and subject to large uncertainties
and assumptions; for instance [CII]158$\mu$m is emitted partly in the
HII regions and partly in the Photo-Dissociation-Regions (PDR).

However, despite these caveats, IR transitions offer the possibility
of investigating the metallicity of galaxies virtually without suffering from
any dust extinction effects, therefore are worth exploring and using, whenever
possible 
\citep{Moorwood80a,Moorwood80b,Lester87,Rubin88,Tsamis03}. 
For example, the metallicities of the central, obscured regions of starburst galaxies are only accessible via these far-IR lines, while metallicities derived from optical lines are likely to apply only to the outer, less dust-extincted part of these galaxies \citep[e.g.][]{Puglisi17,Calabro18}.

\begin{figure}
\centerline{
\includegraphics[width=5.8cm]{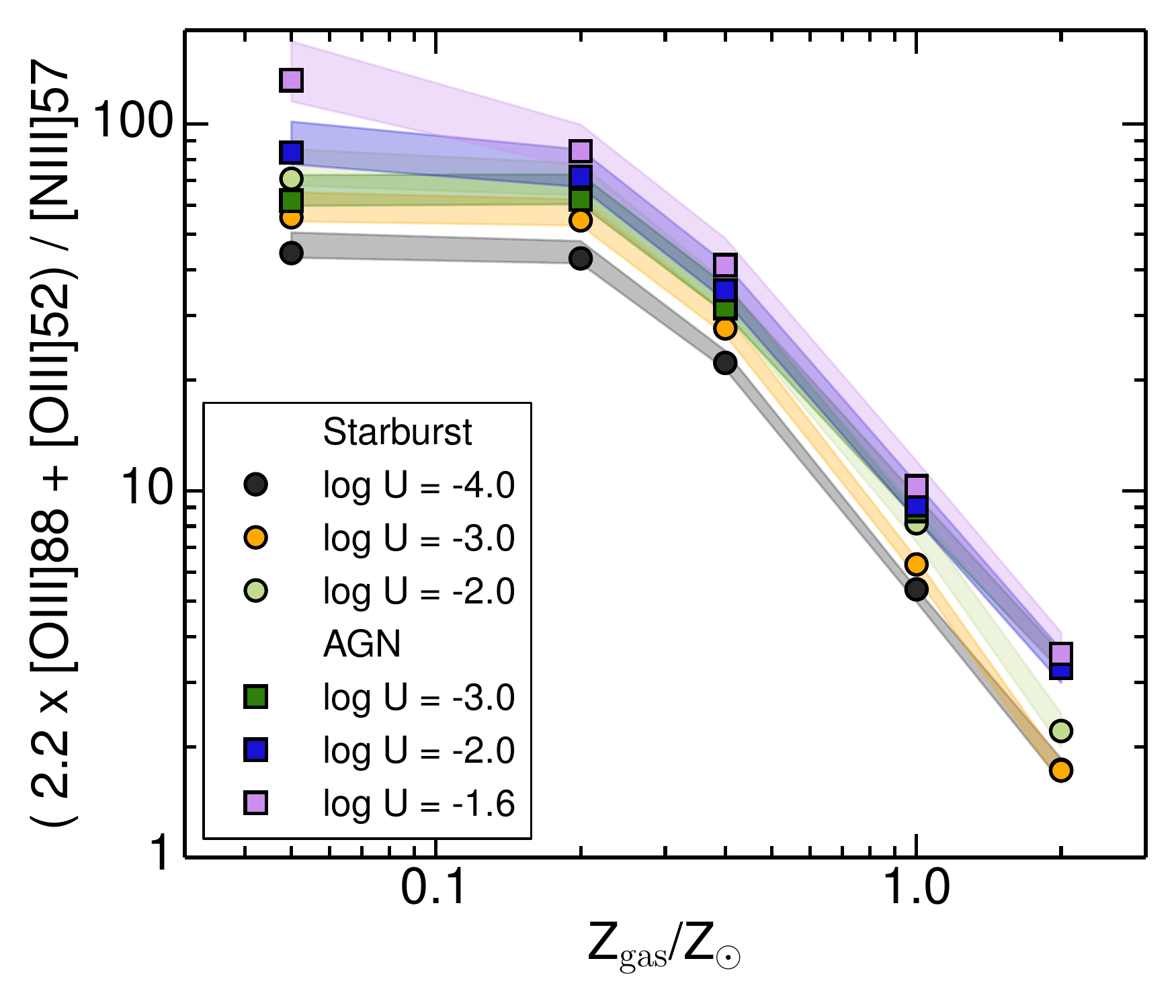}
\includegraphics[width=6.2cm]{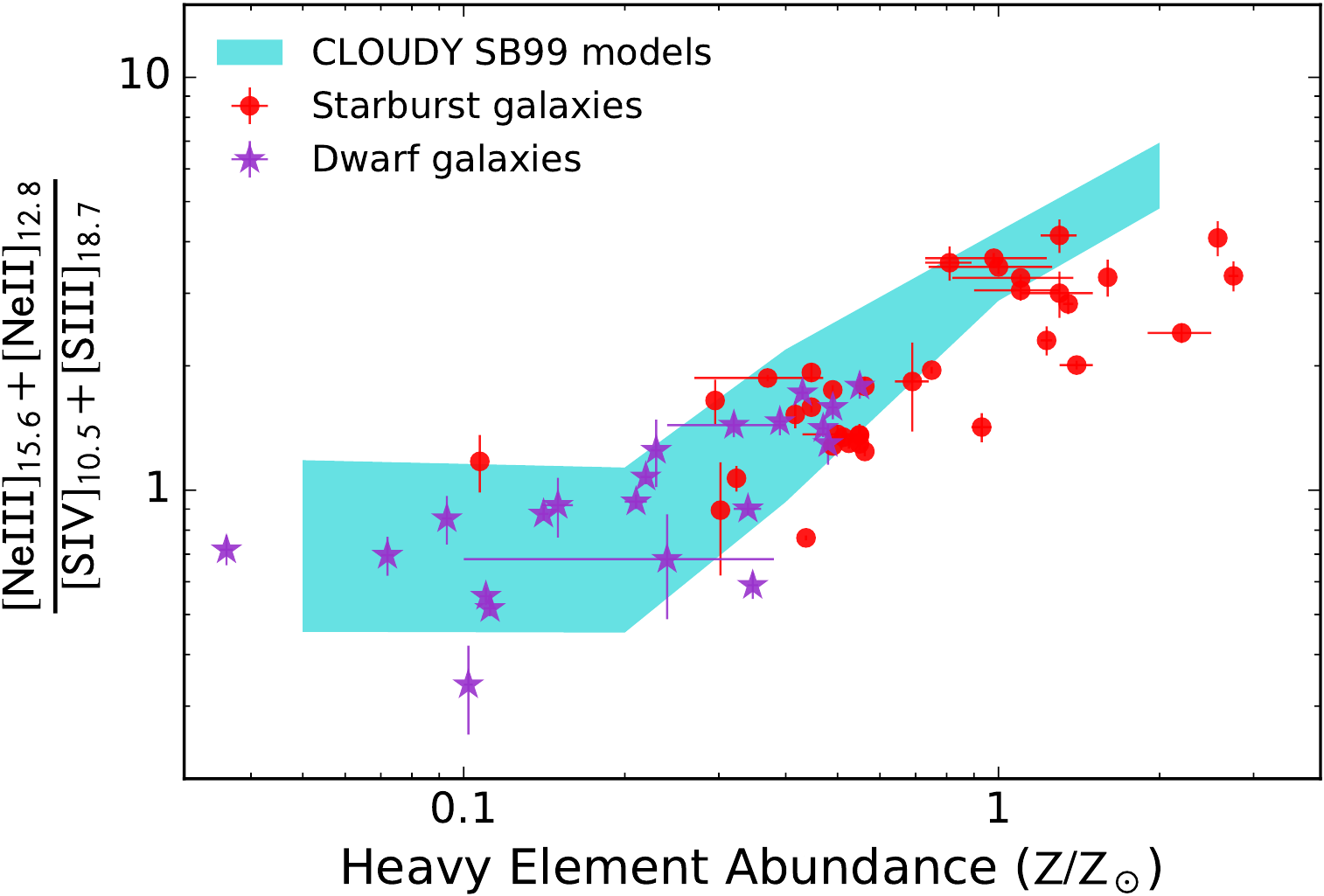}
}
\caption{Two metallicity calibrations based on far-IR lines.
Left:
(2.2$\times$\oiii88$\mu$m+\oiii52$\mu$m)/[NIII]57$\mu$m flux ratio
as a function
of the gas metallicity, for different ionization parameters and sources
of ionization, according to the photoionization models developed
by \cite{Pereira-Santaella17}.
Right: ([NeIIII]16$\mu$m+[NeII]13$\mu$m)/([SIV]11$\mu$m+[SIII]16$\mu$m)
as a function of metallicity along with a range of photoionization models, 
from \cite{Fernandez-Ontiveros17}.
}
\label{fig:diagIR}
\end{figure}

Most efforts have used photoionization modeling and primarily using
nebular lines coming only from HII regions, hence reducing model
uncertainties and assumptions.
After investigating various possible line ratios,
\cite{Nagao11} identified the line flux ratio (\oiii52$\mu$m+88$\mu$m)/[NIII]57$\mu$m
as a good tracer of the gas metallicity, which is little dependent on
gas density, ionization parameter and source of ionizing continuum.
This has been confirmed later
on by \cite{Pereira-Santaella17} by adding a correction factor
that minimizes 
the effects of some scaling relations not directly associated with metallicity
(Fig.~\ref{fig:diagIR}-left).
Note however, that the metallicity sensitivity of this
diagnostic is primarily due to the relation N/O--O/H assumed
in the photoionization models adopted to calibrate it (hence the steep
dependence at high metallicities and flattening at low metallicities). So
one should be aware that this diagnostic is primarily tracing N/O.

In absence of one of the two [OIII] transitions, the
\oiii52$\mu$m/\niii57$\mu$m and \oiii88$\mu$m)/\niii57$\mu$m ratios can
still be used individually, but they are much more sensitive to the gas density.
In fact, the line ratio \oiii88$\mu$m/52$\mu$m in sensitive to density and can be used to measure it \citep{Lester83}.

\cite{Fernandez-Ontiveros17} have also proposed the ratio\\
([NeIIII]16$\mu$m+[NeII]13$\mu$m)/([SIV]11$\mu$m+[SIII]16$\mu$m) as a good
metallicity diagnostics, by computing both an empirical calibration
and a range of photoionization models (Fig.~\ref{fig:diagIR}-right),
and \cite{Croxall13} have proposed a calibration based on the ratios \oii88$\mu$m/\ha\ (although strongly affected by dust extinction) and 
 \neii12.8$\mu$m/\neiii15.5$\mu$m.

Finally, \cite{Nagao12}  proposed that \nii205$\mu$m/\cii158$\mu$m could be a
viable metallicity tracer for observations of high redshift galaxies with
ALMA (at $z>4$, where both lines are detectable from ground), again
exploiting the N/O--O/H, although subject to significant model uncertainties.

\subsubsection{Excitation source and BPT diagrams}
\label{sec:measmet_strong_BPT}

Understanding the source of excitation of the nebular emission lines is
important as the calibration of the metallicity diagnostics is typically
restricted to a class of ionizing/excitation sources, generally young hot
stars in star forming regions. This section
gives a quick overview of some of the diagnostics used to classify galaxies and
galactic regions, as well as of some of the main issues.:w

Diagrams comparing two or more emission line flux ratios contain a wealth of information about the conditions of the emission line nebulae. Different pieces of information can be obtained from different line ratios. Usually the influence of dust extinction
on these diagrams
is minimized either by using pairs of lines close in wavelength, or by plotting extinction-corrected line ratios.

The BPT diagrams, originally proposed by \cite{Baldwin81}, are among the most widely used.
In these diagrams
the \oiii5007/\hb\ ratio is used for its dependence on ionization parameter, and compared either to \nii6584/\ha\ (N2-BPT), 
[SII]/\ha\ (S2-BPT), or [OI]/\ha\ (OI-BPT) \citep{Veilleux87}.
Diagnostic diagrams similar to the BPT plots have been discussed also considering lines at  wavelengths other than optical, such as mid- and far-IR \citep{Dopita13} and UV \citep{Byler18}.

In general, line flux ratios depend on many quantities such as total metallicity, chemical abundance ratios, ionization parameter and hardness of the ionizing continuum. For this reason 
BPT diagrams are widely used to compare observational and theoretical results on emission lines \citep[e.g.,][]{Lehnert96,Kewley01b,Kewley02a,Dopita06}. These are the first diagrams that any photoionization model must reproduce, at least at the statistical level.
Diagrams have also been introduced that are optimized to 
reveal the effect of single parameters.
For example, the distribution of galaxies in the diagram O32 vs.\ R23 is sensitive to both the ionization parameter and metallicity, and can be used to break the degeneracies between  these two quantities both in the local universe and at high redshift \citep{Lilly03,Hainline09,Richard11, Nakajima13}. 

The presence of diffuse ionized gas (DIG) in local and distant galaxies \citep[e.g.,][]{Oey07} can also affect the position of galaxies on the BPT diagrams and bias the measurements of metallicity, especially in galaxies with low sSFR and low surface brightness of \ha\ emission \citep{Sanders17a,Zhang17,Lacerda18}. 

Note that, since \neiii3870 and \oiii5007 are emitted by essentially the same region and scale nearly proportionally to each other,  the [NeIII]/[OII] and O32 have the same meaning and both depend mainly on ionization parameter. This is the reason why the [NeIII]/[OII] vs. O32 plots show small scatters \citep{Steidel16}. The two lines can therefore be used equivalently.

\begin{figure}
\centerline{
\includegraphics[width=12cm]{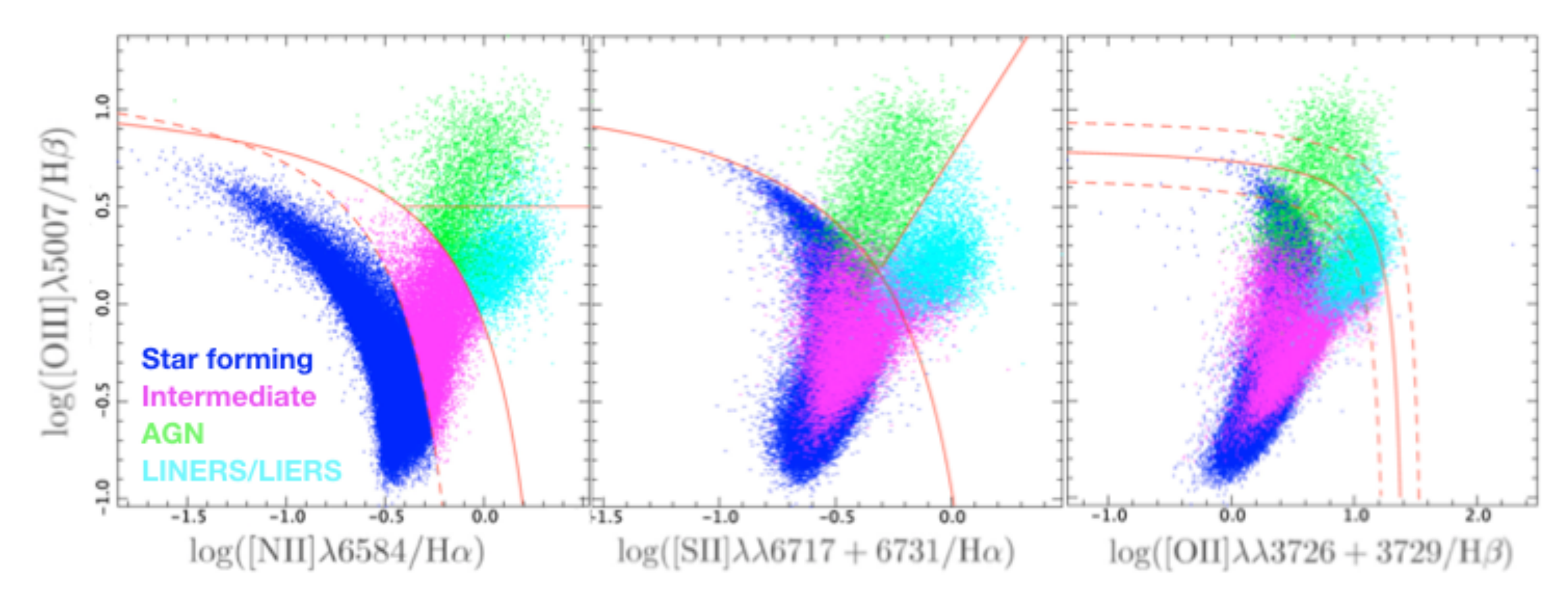}
}
\caption{The three classical diagnostic diagrams N2-BPT, S2-BPT, and R3 vs. R2 for SDSS galaxies, from \cite{Lamareille10}. Galaxies are color-coded according to the N2-BPT (blue and purple) and S2-BPT (green and cyan) diagrams. Blue are star forming galaxies, purple are composite galaxies, green are Seyfert galaxies and cyan are LINERs/LIERs. 
}
\label{fig:BPT_lamareille10}
\end{figure}

Local galaxies show a typical bimodal distribution in the classic N2-BPT ([OIII]/\hb\ \ vs [NII]/\ha), S2-BPT ([OIII]/\hb\ \ vs [SII]/\ha) and O1-BPT ([OIII]/\hb\ vs [OI]/\ha) diagrams, which are not sensitive to dust extinction. HII regions and galaxies whose emission is dominated by star formation, i.e., whose ionizing continuum is due to young, hot stars, follow a well defined sequence, where high values of [OIII]/\hb\ correspond to low values in [NII]/\ha, [SII]/\ha\ and [OI]/\ha, as shown in 
Fig.~\ref{fig:BPT_lamareille10} for the former two
\citep{Veilleux87,Kewley01a,Kewley01b,Kauffmann03c,Bamford08,Lamareille10}. 
The position of the galaxies along the star-forming sequence is dominated by the luminosity-weighted ionization parameter. This is linked to metallicity \citep{Nagao06,Dopita06}, therefore this is also a metallicity sequence, increasing toward higher values of [NII]/\ha\ \citep[e.g.,][]{Andrews13,Curti17}.

Nebulae excited by a harder, AGN continuum are located in a different part of the diagram, with high values of all these diagnostic ratios: [NII]/\ha, [SII]/\ha, [OI]/\ha\  and [OIII]/\hb; therefore, the BPT diagrams are routinely used as a classification tool to separate the two populations of star-forming and AGN-dominated galaxies. 
 The reason why AGN are located in these regions of the BPT diagrams is primarily the combination of two effects: high energy photons produced by AGN increase the fraction of highly ionized oxygen and also penetrate deeper into the gaseous clouds producing extended, partially ionized regions where singly ionized nitrogen and sulfur, as well as neutral oxygen, are abundant and where their collisionally excited transitions are the main coolant; moreover, the intense radiation field produced by AGNs also often results into high ionization parameter, which further boosts the [OIII]5007 emission.
Composite galaxies with contributions from both star formation and AGN occupy an intermediated position.

The region of the BPT diagrams with high values of [NII]/\ha,  [SII]/\ha, and [OI]/\ha\  and relatively low values of [OIII]/\hb\ are occupied by the so-called ``Low Ionization Nuclear Emission Line Regions'' \citep[LINERS,][]{Heckman80}. The bimodality of
the population in this region is even more evident in the S2-BPT diagram \citep{Kewley06}. Recent works have shown that this kind of emission is not confined to the nuclear
region and is actually extended on kilo-parsec scales \citep{Sarzi10,Singh13,Belfiore16a}, 
it has therefore been suggested to rename these regions as ``LIER'' (i.e., dropping the ``N''
that stands for ``Nuclei'' in the original acronym). It has been shown that LINER/LIER are a composite population that can be ionized by weak AGN, by post-AGB stars (associated
with evolved stellar populations) or shocks
\citep[e.g.,][see Sect.~\ref{sec:measmet_strong}]{Shull79,Chevalier80,Lehnert96,Allen08,Rich10,Binette12,Newman14}.\\

The most commonly used separation boundaries between these classes of galaxies are from \cite{Kewley01a}, \cite{Kewley01b}, \cite{Kauffmann03c} and \cite{Kewley06}, but other classifications have been proposed by \cite{Veilleux87}, \cite{Tresse96}, \cite{Dopita00}, \cite{Stasinska06b}, \cite{Cid-Fernandes10a}, and \cite{Lamareille10}. In all cases,  galaxy classification based on different BPT diagrams (either N2, S2, or OI) can be very different (see Fig.\ref{fig:BPT_lamareille10}). To obtain more robust and stable results \cite{Vogt14} defined multiparameteric, 3D BPT diagrams, simultaneously using more than two line ratios.\\

Recently, a number of large size, integral field unit (IFU) spectrographs have been used to map a large number of emission line galaxies. These data have been used to obtain spatially-resolved BPT diagrams studying the ionization properties of galactic sub-regions, both in  local galaxies
\citep[e.g.,][]{Belfiore15,Belfiore16a,Davies16,Davies17b} 
and at high redshift
\cite[e.g.,][]{Newman14,Curti18a}.
In particular, the use of the VLT IFU spectrograph MUSE 
\citep{Bacon10}
on nearby galaxies has allowed astronomers to obtain spatially resolved spectroscopy with very high spatial resolution on entire galaxies
\cite[e.g.,][]{Cresci17,Venturi17,Venturi18}
distinguishing different ionization conditions in different parts of  galaxies.\\

\begin{figure}[t]
\centerline{
\includegraphics[width=12.0cm]{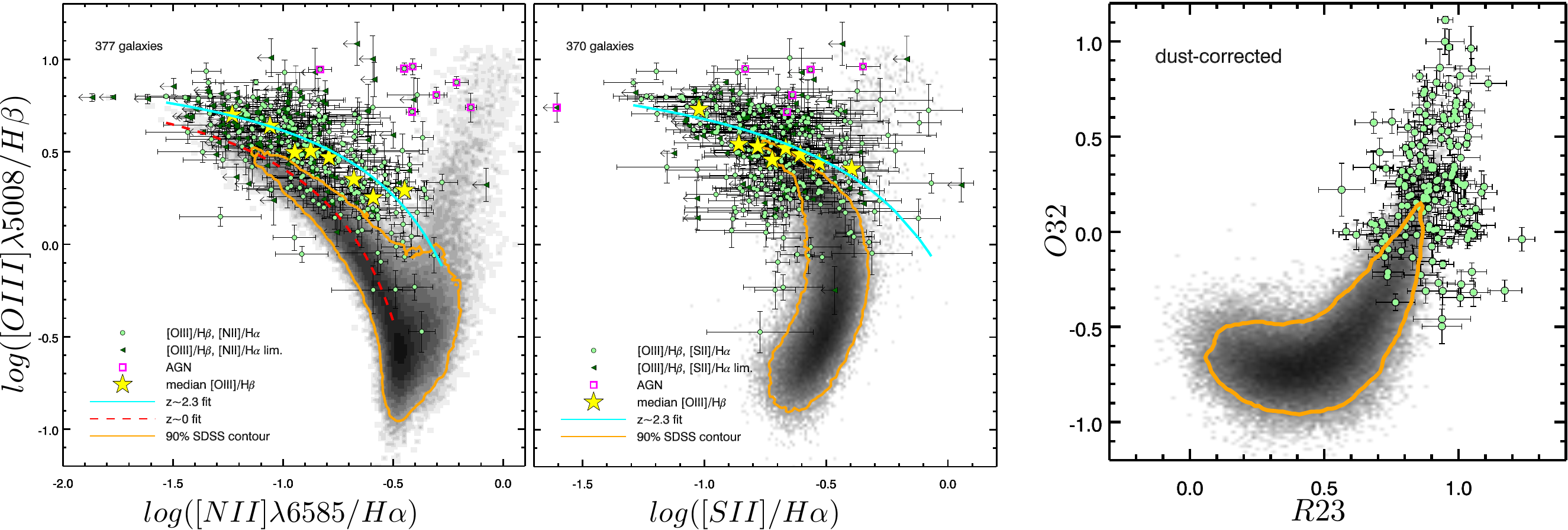}
}
\caption{Three classical diagnostic diagrams for high-$z$ galaxies
from\cite{Strom17b}. The N2-BPT, S2-BPT, and O32 vs.\ R23 diagrams are shown. Grey areas show the location of local SDSS galaxies, while green dots are star forming galaxies at $z\sim2.3$ observed with MOSDEF at Keck. High-$z$ galaxies show a clear offset from the local HII-sequence on the N2-BPT diagram and more limited offsets (if any, see \cite{Shapley15})
on S2-BPT diagram and on the local O32 vs R23 diagram. 
The cyan curve in the left panel reflects the ridge-line of the $z\sim2$ KBSS sample from \cite{Strom17a}. Yellow stars are the median [OIII]/H$\beta$ ratio in bins of the X-axis.
}
\label{fig:BPT_strom17}
\end{figure}

The new, near-IR, multi-object spectrographs on 8-m class telescopes have allowed the observation of the near-IR spectra of a large number of intermediate redshift galaxies, obtaining high S/N ratios and reliable fluxes for the rest-frame optical lines. These studies have confirmed and clarified earlier results, based either on few galaxies or stacked spectra,   showing the presence of a shift between the BPT distribution of high-$z$ galaxies with respect to local forming galaxies \citep[][among others]{Shapley05a, Erb06a, Kriek07, Liu08a, Hainline09, Finkelstein09, Erb10, Yabe12, Cullen14, Yabe14a, Yabe15, Newman14, Masters14, Hayashi15, Salim15, Shapley15, Sanders16a, Steidel16,Trainor16,Strom17a,Kashino17a,Strom17b}.
High redshift galaxies appear to have higher [OIII]/\hb\ \ for a given [NII]/\ha, or higher [NII]/\ha\  for a given [OIII]/\hb, or both, see Fig.~\ref{fig:BPT_strom17}. 
Studies comparing the redshift evolution of only one line ratio with respect to other properties of the galaxies such as mass or sSFR also find evidences for evolution, but in this case the results are more ambiguous because it is not easy to distinguish an evolution of the  ionizing properties of the gas from a simple metallicity evolution \citep{Cullen16, Holden16}

\begin{figure}

\begin{center}
\centerline{\includegraphics[width=8cm]{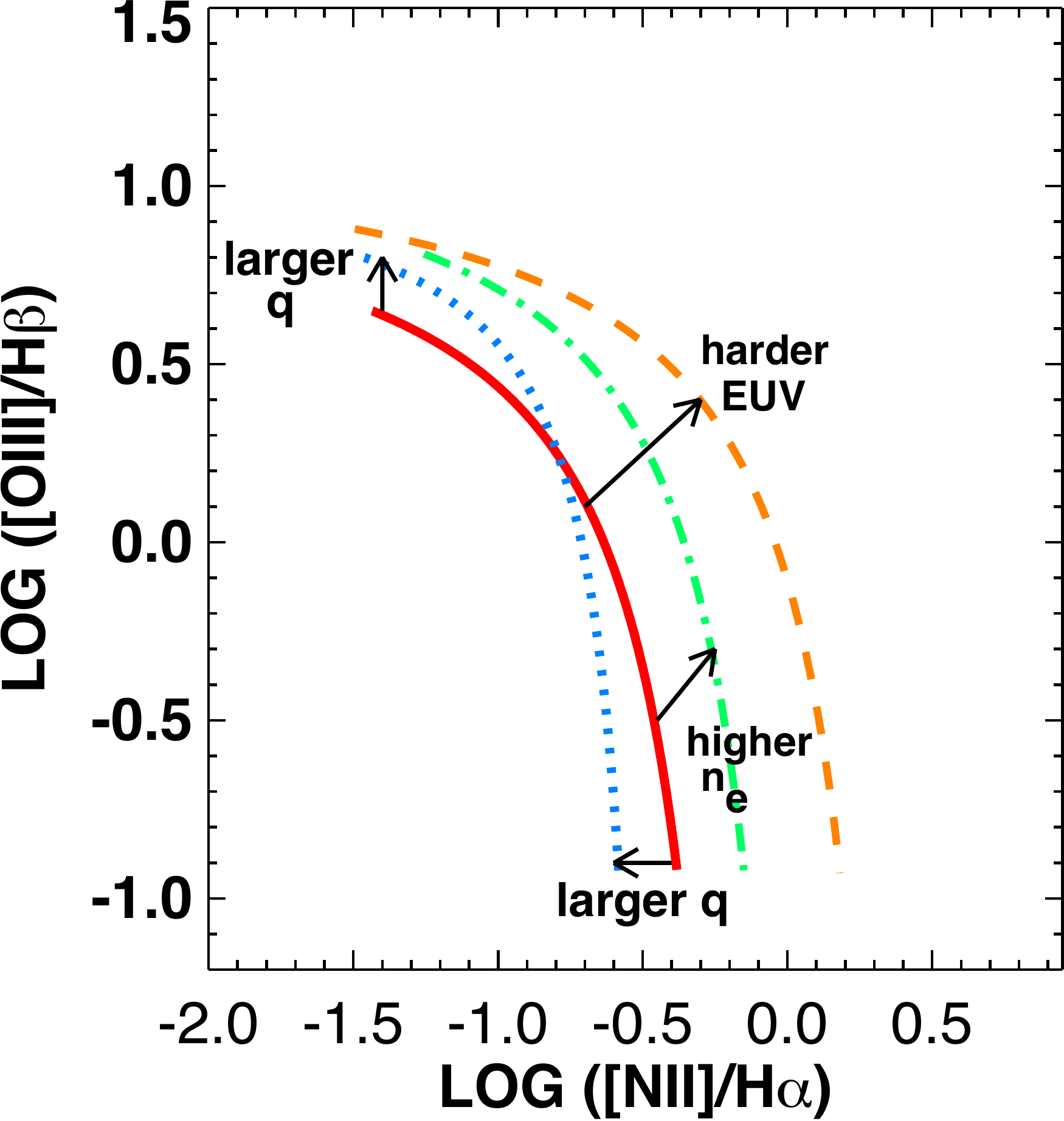}}
\end{center}
\caption{Illustration of the effect of the variation of several parameters on the offset from the HII sequence on the N2-BPT diagram, from \cite{Kewley13b}. It shows how the SDSS star-forming galaxy sequence (in red) is modified by a harder ionizing field into the orange line, by an increase of electron density into the green line, by an increase of the ionization parameter $q$ into the blue line, and by an increase of N/O into the purple line.
The effects of most of these parameters are highly degenerate.
}
\label{fig:BPT_kweley13}
\end{figure}

Many possible explanations have been proposed. 
\cite{Kewley13b} ascribe this shift to an evolution of the ISM conditions in high-$z$ starbursts, such as higher densities, harder ionizing radiation, or higher ionization parameter (see Fig.~\ref{fig:BPT_kweley13}).
Several authors explain the effect as the consequence of a high ionization parameter due either to higher SFR densities or to a top-heavy IMF \citep{Brinchmann08,Hayashi15,Shimakawa15b,Kewley15,Cullen16,Kashino17a,Hirschmann17,Kaasinen18}. Higher densities or higher ISM pressures are suggested by \cite{Shirazi14}, but this possibility is not supported by \cite{Strom17b} who found no correlation between density and offset from the local BPT. Higher densities and higher ionization parameters are also seen in the rare local galaxies that fall into the same, offseted part of the N2-BPT diagram \citep{Bian16, Bian17, Bian18}. These galaxies have low mass and high sSFR, but other SDSS galaxies with similar mass and sSFR do not necessarily have the same properties of the ISM, pointing towards other explanations for their peculiar properties.  Also \cite{Poetrodjojo18} do not find a relation between ionization parameter and sSFR in their sample of local, face-on spiral galaxies.
Higher N/O abundance ratio have been proposed, as an explanation of the offset, by \cite{Perez-Montero09}, \cite{Masters14},  \cite{Shapley15}, \cite{Yabe15} and \cite{Masters16}. The argument is based on the observed evolution in the N2-BPT and not in the S2-BPT, on the absence of offset in the O32 vs.\ R23 diagram (see Fig.~\ref{fig:BPT_strom17}), and on some N/O ratios measured at high redshift \citep{Kojima17}. \cite{Jones15b} reached the same conclusion on the base of a metallicity calibration at z=0.8 through the ``direct'' method.  The difference in N/O could be related to the presence of hot ($T_e \sim 80,000$ K)  Wolf--Rayet stars enriching the ISM with N, or to
selection effects at high redshift \citep{Masters16}.
\cite{Cowie16} proposed a combined effect of increasing N/O and increasing ionization parameters. 
However, a variable N/O has been questioned by \cite{Kashino17a} and \cite{Strom17b}, as these authors find evolution also in the S2-BPT diagram, and by the results of the models proposed by \cite{Hirschmann17}.  
\cite{Steidel14}, \cite{Nakajima16}, and \cite{Strom17b} proposed the effect of harder stellar ionizing radiation as the dominant contribution to the BPT evolution, but this is not supported by the lack of correlation between the ionizing photon production efficiencies and the offset from the local BPT diagram found by \cite{Shivaei18}. 
In addition to the hardness of the spectra, \cite{Nakajima16} suggested that matter-bounded HII regions could boost the value of the O32 ratio. Weak AGNs and shocks \citep{Wright10,Newman14} could also have a role, albeit \cite{Steidel14} found no evidence for high ionization lines from AGNs, and possibly selection effects could be present \citep{Juneau14}. \\

In conclusions, it is not clear if there is one dominant effect and, actually, it is likely that many of the effects discussed above contribute to the observed BPT ``evolution'' with redshift.
This large range of possible explanations ultimately derives from the intrinsic degeneracies of the photoionization models with respect to many parameters, as shown in fig.~\ref{fig:BPT_kweley13}.\\

Most importantly, in the context of this review, it is  not clear how the strong-line calibrations derived for local galaxies are affected by this evolution of the excitation mechanisms/processes or of the relative abundances. Definitely the shift from the local distribution on the BPT diagrams results in an internal inconsistencies of the different methods \citep[e.g.,][]{Kewley02a}. The effect on metallicity can be severe when using N-based indicators \citep{Newman14,Cullen16} or relatively minor \citep{Salim15}, especially when using many line ratios \citep{Brinchmann08}.
The problem of the evolution of the metallicity diagnostics could be potentially solved if direct detections of \oiii4363 and other auroral lines in a significant number of high redshift galaxies is obtained. Using observations at $z\sim 0.8$, \cite{Jones15b} obtained that the calibrations remain stable within 0.01~dex with respect to the local ones, revealing a small effect of the shift in the BPT diagrams on metallicity determinations. Similarly, by exploiting the currently limited number of high redshift ($z\sim2$) galaxies for which ``direct'' \Te\ metallicity measurements are available (mostly lensed galaxies), \cite{Patricio18} have shown that diagnostics such as R23 and R3 appear to provide still reliable metallicity measurements, consistent with the local calibrations, while diagnostics involving nitrogen seem to be affected by a systemic offset, suggesting that a redshift evolution of the N/O abundance ratio, or selection effects, are probably affecting the reliability of these diagnostics. Diagnostics that are primarily tracing the ionization parameter and tracing
metallicity only indirectly (or through a secondary dependence), such as O32 and Ne3O2, are those that show
the largest dispersion and deviations in terms of `true' metallicity with respect to that inferred from applying the local calibrations, confirming that enhanced ionization parameter in distant galaxies is heavily affecting these diagnostics. Certainly more direct
\Te\ based measurements at high redshift are needed to
further assess these issues.\\

{\bf Summarizing}, the various flavors of the BPT and other diagnostic diagrams are critical tools  to study the conditions of the ionized ISM in local and distant galaxies and determine the main contribution to the ionization, allowing for a classification of  galaxies. Clear signs of evolution are seen between local and high-redshift ($z\sim2$) galaxies, and the origin of this evolution is debated. It is likely that the values of metallicity derived for distant galaxies with some of the strong line methods are affected by this evolution, but the importance of this effect is still not clear.

\subsection{Interstellar and Intergalactic absorption lines}
\label{sec:measmet_ism_abs}

The ultraviolet spectral region is rich of resonant transitions
of metal lines, associated with different kind of ions.
If a clump of gas is observed against a source
of radiation, the absorption introduced by such transitions
provide a measure of the column density of the associated
ionic species \citep[through the curve of growth, see][for a review]{Savage96} .
More specifically, the depth of the absorption produced by the transition between levels $\rm m$
and $\rm n$ of an ion $\rm i$ along the line of sight is given by $\rm e^{-\tau (\nu )}$, where the optical depth is defined as
$$ \rm \tau (\nu ) = \frac{\pi e^2}{m_e c} f N^i \phi (\nu ,b)$$
where $\rm N^i$
is the column density of the ion $\rm i$, $\rm f$ is the
oscillator strength and $\rm \phi (\nu ,b)$ describes the line
profile as a function of frequency, which is also dependent on the Doppler
parameter $\rm b$.

If the column of hydrogen is also measured through one of the
Lyman absorption lines, and constraints on the ionization status of
the gas can be obtained, then accurate measurements of the cloud
metallicity, or of its relative chemical abundances, can be inferred.
Provided that the spectrum has high enough signal-to-noise, and adequate
resolution, this techniques provides some of the most accurate
measurements of gas metallicity and chemical abundances. Indeed,
by measuring directly the column of metals/ions the technique
is often (nearly) model independent and subject to little uncertainties.

Most metal absorption lines used to investigate the chemical abundances
are resonant lines in the UV. These
have been extensively used to investigate the
chemical enrichment 
of the circumgalactic/intergalactic medium
and in the interstellar medium of galaxies
 \citep[e.g.][]{Le-Brun97,Rauch98,Pettini02c,Noterdaeme08,Lehnert09,Steidel10,Rafelski12,DeCia18b}.
In most cases quasars have been used as background light, 
but sometimes also other bright sources such as SNe and 
Gamma-ray Bursts (GRBs) constituted the beacon
\citep[e.g]{Prochaska07, Ledoux09, Berger12, Vreeswijk14}.
The advent of sensitive spectrometers in space has enabled
to extend these studies to local systems where stellar clusters or star forming regions are used as background sources
\citep[e.g.,][]{James14b,Tumlinson11,Werk14,Tumlinson17}.

The fundamental difference with respect to luminosity-selected galaxies is that absorption lines, sensitive to column-density cross-section, probe the most numerous objects which are expected to be low-mass galaxies with low metallicity, a population whose emission is not easily accessible at high-redshift. Continuum or line emission metallicity studies instead tend to probe preferentially more massive (hence more metal rich) systems. Moreover,
while metallicities measured
through emission lines or stellar photospheric features are
luminosity weighted (hence provide a  view of the metal
content in galaxies biased towards the most active or massive regions), absorption systems
provide a totally unbiased view of the metal content from this point of view.

Damped Lyman Alpha systems (DLAs) are by far the class of systems most widely exploited to trace metal enrichment and chemical abundances through absorption features at cosmological distances. These are defined as systems detected in absorption with column densities of neutral hydrogen $\rm N(HI)>2\times 10^{22}~cm^{-2}$. The several metal transitions observed in absorption together with Ly$\alpha$ often enable a detailed characterization of their metallicity and abundance pattern. Good overviews of DLAs can be found in \cite{Pettini04b} and \cite{Wolfe05}, although much work has obviously been done since these reviews.

An important aspect to bear in mind about DLAs and other absorption systems, especially in the context of this review,
is that it is not really clear which kind 
of systems they are probing and what is their
relation to galaxies. Even in those cases in which the optical counterpart
of the DLA is identified, it is not clear whether the absorption system is tracing
the outer parts of the galactic disc, or gas in the circumgalactic medium that has
been ejected or is in an accretion phase, or a satellite galaxy/clump, or simply clumps in the nearby
ICM/CGM  \citep[e.g.][]{Fynbo10,Krogager12,Fumagalli15}, possibly characterized by a complex multiphase status and a poor chemical mixing 
\citep[e.g.][]{Zahedy18}.
The fact that DLAs are not directly tracing the bulk of the active and luminous part of galaxies has been discussed for instance by \cite{Pettini99} and \cite{Pettini04b}, based on the weak redshift evolution of the metallicity in DLAs. 

\cite{Prochaska03b} found evidence for a significant redshift evolution of the DLA metallicities, subsequently confirmed by various
authors \citep[e.g.][]{Rafelski12,Rafelski14,DeCia18b,Poudel18}, however the extrapolation of these evolutions to z=0 is around $\rm log(Z/Z_{\odot})\sim 0.7-0.8$, i.e. well below the metallicity of the bulk of local galaxies (at least within their effective radius).
For the same reasons, and in particular due to the fact that absorption systems probe different regions than the bright parts of galaxies, it is difficult to compare the metallicity inferred from absorption systems with those inferred from emission lines or stellar features in galaxies. While early studies claimed a strong discrepancy (i.e. that absorption systems were systematically more metal poor than the metallicity inferred from emission line systems), these claims have been significantly revisited by studies that have compared absorption-derived metallicities with close-by HII regions and found
that they are indeed consistent with each other \citep[e.g.][]{Bowen05}. Statistical studies have also found consistency between metallicities inferred from absorption systems and the luminous part of their counterpart once  metallicity gradients are taken into account \citep[e.g.][]{Krogager17}.

Even with the difficulties in associating DLAs with the galactic component, absorption
systems still provide the most accurate measurements of the relative abundances on
an extremely wide range of chemical elements. By probing both refractive and volatile
species, absorption systems also provide some of the best determination of the
relative depletion of chemical elements in dust grains in different environments and as
a function of metallicity, as it will be discussed in sect.~\ref{sec:dust}.

DLAs will be recalled and discussed in several parts of this review.
It is however  important here to recall some caveats and difficulties of this technique, in particular the saturation of absorption
lines, which in some cases only provides a lower limit on the column
density of the ionic species, and the need to correct
for the ionic fraction. The latter
is often the main source of uncertainty as, unless multiple
transitions from different ionization stages of the same element are
observed, it may imply some modeling
to account for the fraction of the unobserved ionic species. 
However, in many cases either several transitions
are available to tightly constrain the ionization stage, or the column density
is large enough that one can safely assume that the bulk of the 
cloud is self-shielded and therefore mostly neutral or
in low ionization stages \citep{Wolfe05}.

In the case of absorption lines used to
probe the ISM in galaxies, the clump(s) covering factor is another
potential source of uncertainty, but it can often be constrained either
through the depth of saturated lines, or by using doublets whose
relative equivalent widths is tied to the relative oscillator strengths.

The fact that transitions of different elements can be traced at different redshifts,
depending on the available wavelength band, and at different column densities (depending on the oscillator strength of different transitions), may introduce difficulties in properly comparing the metallicities observed at different cosmic epochs and different systems (e.g., \citealt{Rafelski12,Rafelski14,Wolfe05}, and reference therein).

At very high redshift ($z>5$) the Ly$\alpha$ optical depth of the IGM
is so high to make the continuum blueward of the Ly$\alpha$ wavelength
at the redshift of the background source nearly totally
absorbed. Therefore, at these redshifts
it generally becomes nearly impossible to obtain
constraints on the column of hydrogen. However, at very high
redshifts absorption systems still provide extremely precious information on the relative chemical abundances 
\citep[modulo ionization corrections][]{Becker12,DOdorico13}.\\

{\bf Summarizing}, absorption lines probe a different population of galaxies (or different galactic regions) with respect to those investigated by luminosity-selected samples, because they are most sensitive to the gas column density distribution. Due to their high column density, accurate measurements of the abundance of several elements can be obtained for the DLAs, allowing for a detailed study of the evolution of the abundance ratios and of dust depletion.

\subsection{Chemical abundances from X-ray spectroscopy}
\label{sec:measmet_ism_X}

X-ray spectroscopy has been extensively used to measure the
chemical abundances in hot plasmas ($T\sim 10^6$--$10^8$~K),
especially in the case of the hot ICM
\citep{Mushotzky96, Mushotzky97,de-Plaa07,Sato07,Mernier16,
Mernier18,Simionescu18}, but also in the galactic
winds \citep[e.g.,][]{Ranalli08,Nardini13,Veilleux14}.

Chemical abundances and metallicity are inferred through thermal
collisional excitation and ionization models. This
is  relatively straightforward, as
these plasmas are generally optically thin and in collisional
ionization equilibrium \citep[although accounting for
the charge exchange X-ray emission has complicated the interpretation
of some low-energy spectra, ][]{Liu12,Zhang14a}. The X-ray spectrum of hot
plasma is rich of transitions from nearly all elements, from carbon to nickel,
hence rich of information that can be used to infer chemical abundances.
Of course, no transitions from hydrogen are detectable in the X-rays,
hence the absolute metallicity can only be constrained by modeling the
metal lines emission relative to the underlying free-free emission, which is mostly due to H and He.

Until recently, many studies had mostly used CCD spectra resulting in
very low spectral resolution that produces blending of many metal
emission lines and, therefore, significant uncertainties in the resulting
abundances. The use of X-ray dispersion grating spectrometers
has enabled astronomers to greatly improve the determination of the chemical abundances
in plasmas by disentangling the transitions of several chemical elements
\citep[e.g.,][]{Ranalli08,Pinto13,Mernier16,Mernier18}. However, the dispersive nature
of this technique has often limited the use of this method to the systems with
high surface brightness, such as galaxy cluster cores.
The use of the innovative array of micro-calorimeters on board of the Hitomi
space observatory
has delivered a fantastic combination of
very high spectral resolution and high sensitivity,
therefore enabling an unprecedented, extremely detailed determination of
the abundances in the Perseus cluster 
\citep{Hitomi17,Simionescu18}.
Unfortunately, the limited lifetime of the Hitomi observatory has
prevented the extension of such studies to larger samples.

Grating X-ray spectroscopy has been used also to detect highly ionized metal
absorption lines in Warm-Hot systems along the line of sight of bright background
quasars \citep{Buote09,Zappacosta10,Fang10,Zappacosta12,Nicastro18}.
These observations are very challenging, and do not really provide constraints
on the gas metallicity (which is generally assumed or constrained from
other tracers), however they can provide key information on the content
of baryons in these intervening systems.

\subsection{Dust depletion}
\label{sec:dust}

In the interstellar medium a significant fraction of metals
is locked into dust grains, and this is a major source of uncertainty
for the determination of the metallicity and chemical abundances
of the ISM. The ``direct'' methods only probe the gas phase
metallicity, while photoionization models assume, {\it a-priori},
a fixed depletion pattern of the various chemical elements, hence
assume that the dust-to-metals ratio does not depend on metallicity
or other environmental effects.

As mentioned in sect.~\ref{sec:measmet_ism_abs}, absorption line studies are the most effective
in tracing the depletion pattern, as they can both trace
elements that are heavily depleted onto dust (e.g. iron)
and elements of the same group that are little depleted. 

Extensive studies and reviews on the dust
depletion patterns and dust properties have been published
\citep[e.g.,][]{Savage96,Draine03,Jenkins09,DeCia16,Galliano18} and an extensive discussion
goes beyond this paper. However, in the following we provide some
basic information that is particularly useful when investigating
the metallicity of the ISM.

The amount of depletion is heavily variable and depends on environment
\citep[e.g.][]{Jenkins14}.
Typically, of the order of 30\%-50\% of the metals is locked in dust grains.
The depletion fraction is quite different from element to element,
depending on its ``refractory'' nature. Typically, in the ISM
nearly all ($\sim90-99\%$) of iron is locked in dust, 
together with most of the silicon ($\sim30-97\%$, making silicate grains). 
About $\sim0-40\%$
of oxygen and carbon are in grains, while
other elements such as nitrogen, are almost totally in the gas phase \citep{Jenkins09}
even in the densest environments \citep[e.g.,][]{Caselli02}. 
Zinc is often assumed to be free from dust depletion and therefore to be a good tracer of the abundance of the iron peak elements, but detailed studies have shown that it can suffers of significant depletion as well \citep{Jenkins09,Berg15b}.
The resulting structure of dust is complex, with cores, mantles and intrusions of the different elements.
These are all important aspects
when measuring the metallicities by using specific metal lines.

The  assumption that the depletion pattern and, more generally,
the dust-to-metal ratio is independent of metallicity seems to
be reasonable only around solar metallicities, while recent studies
(mostly based on the DLA observations) have shown that the
dust-to-metal ratio decreases significantly with metallicity
\cite[e.g., about 50\% lower at $\rm Z\sim 0.1~Z_{\odot}$,][]
{Vladilo11,Wiseman17,De-Cia13,DeCia16}, suggesting that grain-growth in the ISM 
is a dominant
mechanism of dust formation. The variation of dust depletion
with metallicity introduces a level of complexity that has not been
incorporated yet in photoionization models.

\section{The landscape of galaxy chemical evolution models} 
\label{sec:models}

Each galaxy is subject to a number of  processes acting together that determine its chemical evolution. Metal poor gas is accreted from the IGM; gas is used inside the galaxy to form stars, but stars can also be acquired via major or minor mergers; stellar evolution provides chemically enriched gas to the ISM via SN explosions and stellar winds; part of this enriched gas can leave the galaxy through galactic winds to enrich the CGM or the IGM; the gas inside galaxies is recycled several times and goes through a number of stellar generations; dust is created and destroyed, locking and releasing metals; the presence of a central AGN can affect the properties of the ISM and of the CGM, either by heating the gas or by removing it through AGN-driven
winds, and both these effects can suppress star formation; dynamical interactions with nearby galaxies and with the ICM can alter the properties of the ISM and affect the star formation activity. All these effects can depend on several parameters such as dark matter halo mass, cosmic time, and environment. The study of the chemical enrichment of galaxies can give information on all these effects.

Different kinds of models have been developed to take into account all the effects discussed above in a consistent way and inside a cosmological framework. Extensive reviews of the methods used to develop galaxy formation models and of the current status in this area can be found in \cite{Silk12}, \cite{Silk14}, \cite{Somerville15}, \cite{Naab17} and \cite{Dayal18}.
In the context of galaxy formation models, \cite{Matteucci12} provides an extensive overview on the theoretical approaches to model and reproduce  chemical enrichment of galaxies.

In this section we give a brief summary of the main classes of existing models of galaxy formation and evolution, which will serve as a reference for the discussions presented together with the observational results. As emphasized by \cite{Somerville15}, all these models are based on a common set of basic physical processes and obtain similar, although not at all identical, results \citep[see, e.g.,][]{Mitchell18}.

One critical problem that all models must solve is how to keep star formation efficiency low, as in all galaxies only a small fraction of baryons are converted into stars. 
Galaxies contain a smaller fraction of baryons with respect to the cosmic average, and this fraction increases with mass up to halo mass of the order of $10^{12}$\msun\ \citep[e.g.,][]{Baldry08, Papastergis12}. This requires the existence of mechanisms either to prevent accretion or to remove baryons from galaxies.
Star formation must also be limited. For this, the gas can be put into some state unsuitable for star formation (e.g. hot or highly turbulent) rather than removed from the galaxy, or further accretion of gas can be prevented therefore depriving the galaxy of the necessary fuel for a sustained and prolonged star formation. Sources of these kinds of feedback have been proposed to be, among others, SNe, radiation from young stars, AGN, and ram-pressure stripping due to a intracluster medium.\\

All  models contain a few main, critical parameters that are described via  mathematical formulae in the analytical and SAM models or with physical prescription in the numerical ones:

\begin{itemize}
\item The star formation ``efficiency'', defined as the number of stars formed per unit time per unit gas mass
\begin{equation}
\rm \epsilon = SFR/M_{\rm gas} \,,
\label{eq:sf_efficiency}
\end{equation}
often referred to as the inverse of the gas ``depletion time'', i.e.,
the time required for star formation to completely use up all gas if there is no
further gas accretion and the star formation rate remains constant. Note, however, that
in case of no accretion the SFR cannot remain constant, must decline exponentially, as simply obtained by solving the ``closed-box'' differential equation
\begin{equation}
\rm SFR = -dM_{\rm gas}/dt=\epsilon M_{\rm gas} \,,
\end{equation}
hence
the term ``depletion time'' is not totally appropriate, as it is actually linked to the e-folding time of the SFR. The star formation efficiency is often
taken as constant, independent of the gas mass or surface density; this is equivalent of
assuming that the Schmidt--Kennicutt relation is linear, in contrast to the original
formulation, which has a slope of 1.4 \citep[see ][]{Kennicutt12}. Whether the slope of the relation is linear or superlinear is still a debated issue, which goes beyond the scope of this review. More detailed models tend to distinguish between the star formation efficiency associated with the sole molecular gas, i.e. the phase out of which stars form, and the global star formation efficiency, i.e. including the atomic component of the cold gas, which is typically
inactive and mostly a reservoir on larger scales.

\item The outflow mass loading factor, i.e., the ratio between mass outflow rate and
star formation rate, 
\begin{equation}
\rm \eta =  \frac{\dot{M}_{\rm outfl}}{SFR} \,.
\label{eq:outflow}
\end{equation}
For star forming galaxies, in which
the outflow is primarily driven by SNe and radiation pressure from stars, the loading
factor is typically observed to be around unity \citep[e.g.,][]{Steidel10,Heckman15,Fluetsch18}, but it is expected to anti-correlate with the mass of the
galaxy \citep[as a consequence of the deeper gravitational potential well,][]{Somerville12},
and it can be boosted by a large factor by the presence of an AGN \citep{Cicone14,Fluetsch18}. Often models leave the outflow loading factor as a free parameter that is constrained by fitting the observational data.
When talking about outflow, it is often important
to discriminate between gas that does not escape the galaxy (falling back, hence being
recycled, on relatively
short timescales) and gas that leaves the halo (or which is potentially re-accreted
only very long timescale, close to the Hubble time). Some models introduce the concept
of ``effective'' outflow loading factor by considering only the fraction of gas that
is expelled and not recycled \citep{Peng15}. Most models assume that the metallicity of the outflowing gas is the same as the average metallicity of the ISM in the galaxy (or of the galactic sub-region), while some models also consider the possibility of differential outflows in which the species produced by core collapse SNe are expelled more efficiently than other elements.

\item The gas inflow rate. This is one of the most critical parameters, as it
regulates most of the galaxy evolution. Many models {\it assume that the gas inflow rate
is proportional to the SFR in the galaxy} as this enables a simpler analytical solution
of the differential equations that describe galaxy evolution and chemical enrichment. However, although mathematically convenient, 
{\it we warn that this assumption is unphysical}, as there is no physical reason why
the inflowing gas should know about the SFR in the galaxy. Moreover, assuming that the accretion rate is proportional to the SFR, when combined with Eqs.~(\ref{eq:sf_efficiency})--(\ref{eq:outflow}), automatically implies that the gas mass and the SFR must decline
exponentially with time, hence implying that the galaxy cannot follow the 
observed tight relation between mass and SFR dubbed Main Sequence of Star Formation 
\citep[MSSF, e.g.][]{Brinchmann04,Daddi07}.
The only case in which the inflow rate can be physically assumed to be proportional
to the SFR is in the condition of perfect equilibrium in which
the inflow rate is exactly compensated by the SFR and outflow rate.
More physically sound models assume the inflow rate proportional to the halo
mass. Generally the inflowing gas is assumed to be chemically pristine; this assumption may not
be appropriate in several cases; indeed, as we will see in Sect.~\ref{sec:MZR_DLAs},
the circumgalactic gas can be significantly enriched, also at high redshift.

\end{itemize}

It is important to emphasize that a meaningful comparison between the results of the models, especially those based on numerical techniques, and the observation must take into account all the selection effects on both sides. In fact, generally large populations of observed galaxies are compared to a large number of model galaxies derived from 
the simulated evolution of baryons inside a distribution of dark matter halos, and the selection of the objects in the models can be different to those affecting the actual observations. 
While the former are usually selected on stellar mass or halo mass, the latter are selected based on luminosity or gas column density. 
In other words, simulations must be ``observed'' using the same selections used for the telescope observations, producing ``synthetic observations'' of the simulation outputs.
This is not always done, but the result of the comparison and the values of the derived physical quantities can depend critically on these effects \citep[e.g.][]{Governato09,Scannapieco10,Guidi16,Guidi18}.

\subsection{Numerical simulations}
\label{sec:models_hydro}

These models start from $N$-body simulations of the effect of gravity on dark and baryonic matter and the hierarchical growing of  structures via a continuous process of merging and accretion. Several implementations exist that can be based either on particles, or on a geometric mesh, or hybrid. Critical parameters of these simulations are the spatial and mass resolution, i.e., the finest detail that can be resolved and reproduced, and the total size of the simulation, i.e., the number of particles used and the size of the universe volume that is simulated. Being limited by the total computational time available, usually high resolutions corresponds to small volume sizes, and vice-versa. Often several simulations are made with the same physics and different choices of resolution/size \cite[e.g.,][]{McAlpine16}.

In hydrodynamical simulations baryonic physics is included by solving the hydrodynamical equations albeit in some simplified form 
\citep[e.g.][]{Finlator08}.
Complex processes on small scales such as star formation and the effects of SN explosions on the ISM would require sub-parsec resolution, which is never achieved, and sub-grid recipes are still used. The impressive advances in computational power of the last decade, and a better understanding of the description of the basic physical processes have allowed these codes to reach the level where
reproducing most of the observational constraints is possible.
Several large simulations are now available and have released their databases to the public to allow for a better and more extensive analysis of the results. Among the best know, 
ILLUSTRIS \citep{Vogelsberger14,Genel14,Genel16},
ILLUSTRIS-TNG \citep{Springel18,Pillepich18,Torrey17},
EAGLE \citep{Schaye15, Crain15,McAlpine16,De-Rossi17},
HORIZON-AGN \citep{Dubois14, Dubois16},
FIRE \citep{Hopkins14, Ma16},
and MUFASA \citep{Dave16, Dave17b}. 
The review by \cite{Somerville15} provides an extensive description of these models. Some hydro codes are optimized to reach the best possible resolution, albeit on single galaxies \citep[e.g.,][]{Pallottini14a,Pallottini17a,Katz15,Costa15}.

\subsection{Semi-analytic models (SAMs)}
\label{sec:models_SAM}

These models start from considering the merging trees, obtained from N-body dark matter simulations or Monte-Carlo techniques, that give rise to the formation of cosmic structures.
The resolution is given by the mass of dark matter particles used.
Galaxies are associated with dark matter halos, and evolve inside them by describing the important physics with
analytic formulae, usually based on a number of free parameters whose values are varied to reproduce the observations, in a forward-modeling approach.
As mentioned above, crucial quantities and processes are, among others, gas cooling time, star formation efficiency, feedback from SNe, feedback from AGN, morphological transformation, and metals distribution 
\citep[e.g.,][]{Kauffmann93,Lacey93,Cole00,Kauffmann04,Croton06,De-Lucia07,Benson12,Collacchioni18}.

The advantage of SAM models is that they are less computationally expensive and 
allow for a faster exploration of the effects of changing the description of the physical processes. The role of these processes can therefore be better understood. The weak points are that the physics is controlled ``by hand'' (which is however an issue also for many simulations), i.e., the assumptions on the physical processes are approximate and not necessary realistic,
the evolution of baryons and dark matter may not be self-consistent because baryons evolve inside pre-determined dark matter halos, and
a large freedom in the choice of the parameters is allowed so that it is not guaranteed that a unique solution can be obtained.
Examples of SAM models, discussing in particular the chemical properties of galaxies, can be found in  \cite{Thomas99a}, \cite{De-Lucia04}, \cite{Somerville08}, \cite{De-Lucia09}, \cite{Hirschmann13b}, \cite{Fu13}, \cite{Yates14}, \cite{Porter14}, \cite{Cousin16}, \cite{Zoldan17}, and \cite{De-Lucia17}.

\begin{figure}[t]
\centerline{\includegraphics[width=8cm]{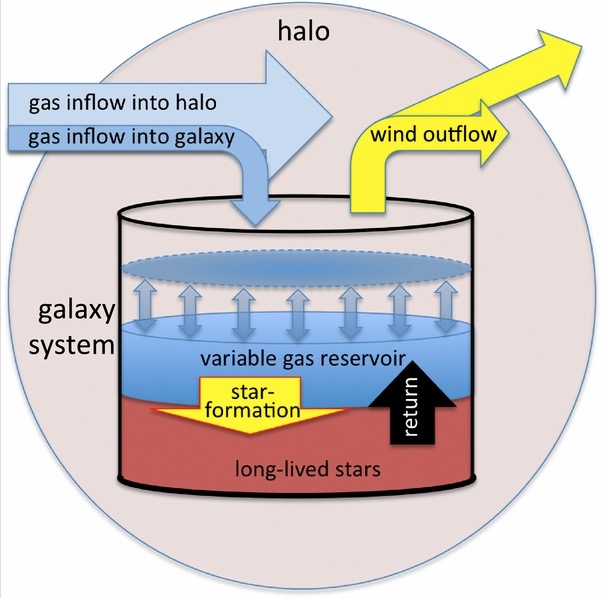}}
\caption{Sketch illustrating the basic processes involved in the galaxy equilibrium models. From \cite{Lilly13}.
}
\label{fig:MODELS_lilly13}
\end{figure}

\subsection{Analytical models}
\label{sec:models_analytical}

For the MW and the galaxies in the Local Group with resolved stellar populations, a set of detailed chemical evolution models have been developed to reproduce the wealth of data available in terms of total metallicity, chemical abundance ratios, and radial gradients of these quantities.
Besides the galaxy-wide processes listed above,  the (generally unknown) star-formation history of a galaxy, its dependence on radius, and many physical effects related to stellar evolution (stellar evolutionary sequences for each initial mass, SN explosions timescale, chemical and dust yields, radial re-distribution of matter, among others) are to be taken into account. All these quantities are often expressed as a function of some parameters whose value is to be obtained by comparison with the observations
\citep[e.g.,][]{Matteucci86a,Matteucci89,Matteucci90,Matteucci94,Chiappini97,Chiappini01, Matteucci01b, Naab06, Pipino06, Prantzos08, Prantzos09, Calura09, Cescutti15, Ryde16, Vincenzo16, Grisoni17, Grisoni18}.

The treatment of dust production and destruction is an important ingredient that can affect the results especially for the elements most depleted into dust, and change important aspects of the chemical evolution models \citep{Dwek98,Calura08a,Valiante09,Gioannini17a,Gioannini17b}.

A class of models for star-forming galaxies are named ``equilibrium'' models
(or ``gas-regulator'' models) because they consider how the interplay of all the on-going processes discussed above affects the gas reservoir of the galaxies and, therefore,
their star formation, giving a slowly-evolving, quasi-steady state, in which gas inflow
is compensated by star formation and outflows, yielding a nearly constant, or slowly
evolving, gas content
\citep[][see Fig.~\ref{fig:MODELS_lilly13}, ]{Bouche10,Peeples11,Dayal13, Lilly13, Forbes14,Pipino14,Peng14b,Feldmann15, Yabe15a,Harwit15,Kacprzak16,Hunt16b}.

From the point of view of chemical evolution, there are still many chemical elements whose evolution is not satisfactorily reproduced in the solar neighborhood. This is very probably due to failures in the stellar yields suggesting that nucleosynthesis calculations should be revised. Recently, \cite{Matteucci14} have shown that a possibility of reproducing, for instance, the solar abundance of the r-process element Europium, as well as [Eu/Fe] versus [Fe/H] in the Galaxy, is to assume that this element is mainly produced during merging of neutron stars. Such an event has been witnessed for the first time in connection to the gravitational waves event GW170817 \citep{Pian17}. The other channel of Eu production is represented by core-collapse SNe, although their Eu production is not enough to reproduce the solar Eu. New nucleosynthesis calculations are necessary also for elements such as Mn, Cr, K, Ti \citep[see ][]{Romano10}. Improved dust production and destruction prescriptions are expected to better reproduce the galactic chemical evolution in the presence of dust. \\

{\bf Summarizing}, models of galaxy formation and evolution have been developed by using very different analytical and numerical methods, each of them with strengths and weaknesses. All these models use different techniques to study the effect of the same set of critical mechanisms, in particular the processes fostering and limiting star formation in galaxies. Total metallicity and element abundance ratios are very sensitive to these processes and to the timescales of star formation, therefore they are among the main observables used to constrain the models.

\section{Metallicity scaling relations in galaxies}
\label{sec:scaling_relations}

Metallicity, both of gas and stars, shows clear scaling relations with several integrated properties of the galaxies. These relations are present both in star-forming and quiescent galaxies, and constitute one of the most revealing pieces of information about the evolution of galaxies, as discussed in Sect.~\ref{sec:models}. The processes of galaxy formation and evolution depend critically on several parameters, at least halo mass and environment; metallicity, being a consequence of all the history of star formation, gas accretion, merging, and gas outflow, is critically dependent on these parameters in the form of scaling relations.

\subsection{The Mass-metallicity relation (MZR)}
\label{sec:MZR}

The primary scaling relation of metallicity is observed to be with galaxy stellar mass and, as a consequence, with the quantities that scale with mass, such as luminosity in bands dominated by relatively old stars. The MZR exists for both gas-phase and stellar metallicities.

\begin{figure}[t]
\centerline{\includegraphics[width=9cm]{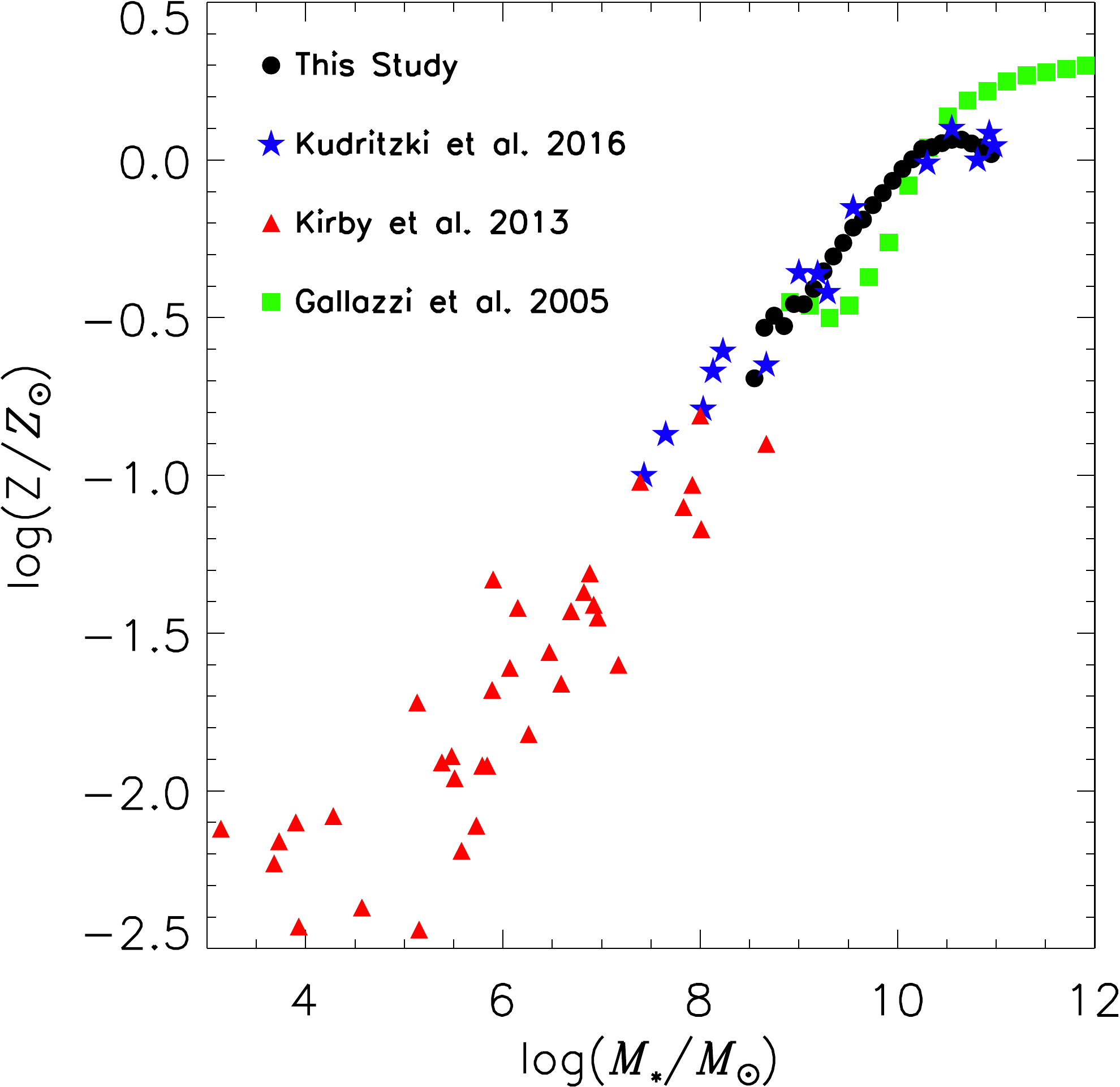}}
\caption{
Stellar MZRs of local galaxies derived by various authors, from \cite{Zahid17}.
Green squares are the median values of local SDSS galaxies with metallicities derived from Lick indices \citep{Gallazzi05}; black dots are also SDSS galaxies but with metallicities derived from spectral fitting of stacked spectra \citep{Zahid17}; 
Blue stars are nearby, single galaxies whose metallicity is based on spectroscopy of individual supergiant stars \citep{Bresolin16,Kudritzki16,Davies17b}. Red triangles are local dwarf galaxies \citep{Kirby13}.
}
\label{fig:Stellar_MZR}
\end{figure}
\subsubsection{Stellar metallicity}
\label{sec:MZR_stellar}

The stellar MZR was first discovered in local ellipticals by studying color-magnitude diagrams and stellar spectroscopy \citep{McClure68, Sandage72, Mould83, Buonanno85} and soon interpreted as the result of chemically-enriched SN winds preferentially ejecting metals from low mass galaxies, due to their shallower gravitational potential well \citep{Tinsley74, Tinsley78, Mould84}.

In more recent times, several authors have applied the methods explained in Sect.~\ref{sec:measmet_stars} to optical spectra of local galaxies, and in particular to the extensive SDSS spectroscopic database, to derive the stellar MZR of local galaxies, 
both quiescent and star forming
\citep{Trager00b,Kuntschner01,Gallazzi05,Thomas05,Gallazzi06,Gallazzi08,Panter08, Graves09a,Thomas10a, Harrison11, Petropoulou11, Kirby13, Conroy14, Gonzalez-Delgado14, Fitzpatrick15, Sybilska17,Zahid17, Lian18a, Zhang18b}.
A clear MZR is seen, with metallicity increasing with mass, as shown in Fig.~\ref{fig:Stellar_MZR}.

As detailed in Sect.~\ref{sec:measmet_stars}, the stellar populations of local and distant galaxies are studied using not only rest-frame optical spectra, but also the UV spectra, especially in starburst galaxies.  The UV stellar MZR have been derived using samples of galaxies selected in different ways in the local universe \citep{Heckman98,Gonzalez-Delgado98a, Leitherer11, Zetterlund15}. When UV spectra are used, only the young, massive stars are sampled, and the results are expected to be similar to those derived for the ISM (out of which young stars have recently formed, see Sect.~\ref{sec:MZR_gas}).

By deconvolving the integrated spectra it is also possible to derive stellar metallicity as a function of stellar age in the same galaxy \citep[e.g.,][]{Panter08}. 
With this technique evidence for temporal evolution is found, but the actual significance of the effect is debated because the process of deconvolving spectra is often model-dependent and always associated with significant uncertainties.

Usually, the derived metallicities are luminosity-weighted and can be significantly different from the mass-weighted values provided by  models of galaxy evolution \citep{Zahid17, Lian18a}. The observed scatter is often greater than the observational uncertainties, showing that  other parameters besides the stellar mass affect the chemical evolution of galaxies.

As detailed in Sect.~\ref{sec:alpha_over_iron}, the same authors cited above also find a systematic increase of the $\alpha$-elements-to-iron ratio with stellar mass. The $\alpha$-elements/Fe abundance ratio is a powerful tool to constrain the relative contribution of SNIa and core-collapse SNe, hence of the star formation history timescale,
and an enhanced $\alpha$/Fe abundance ratio is usually explained as a consequence of shorter formation timescales in massive galaxies, the so-called downsizing \citep[e.g.,][]{Thomas10a,Onodera15}.\\

\begin{figure}
\centerline{\includegraphics[width=9cm]{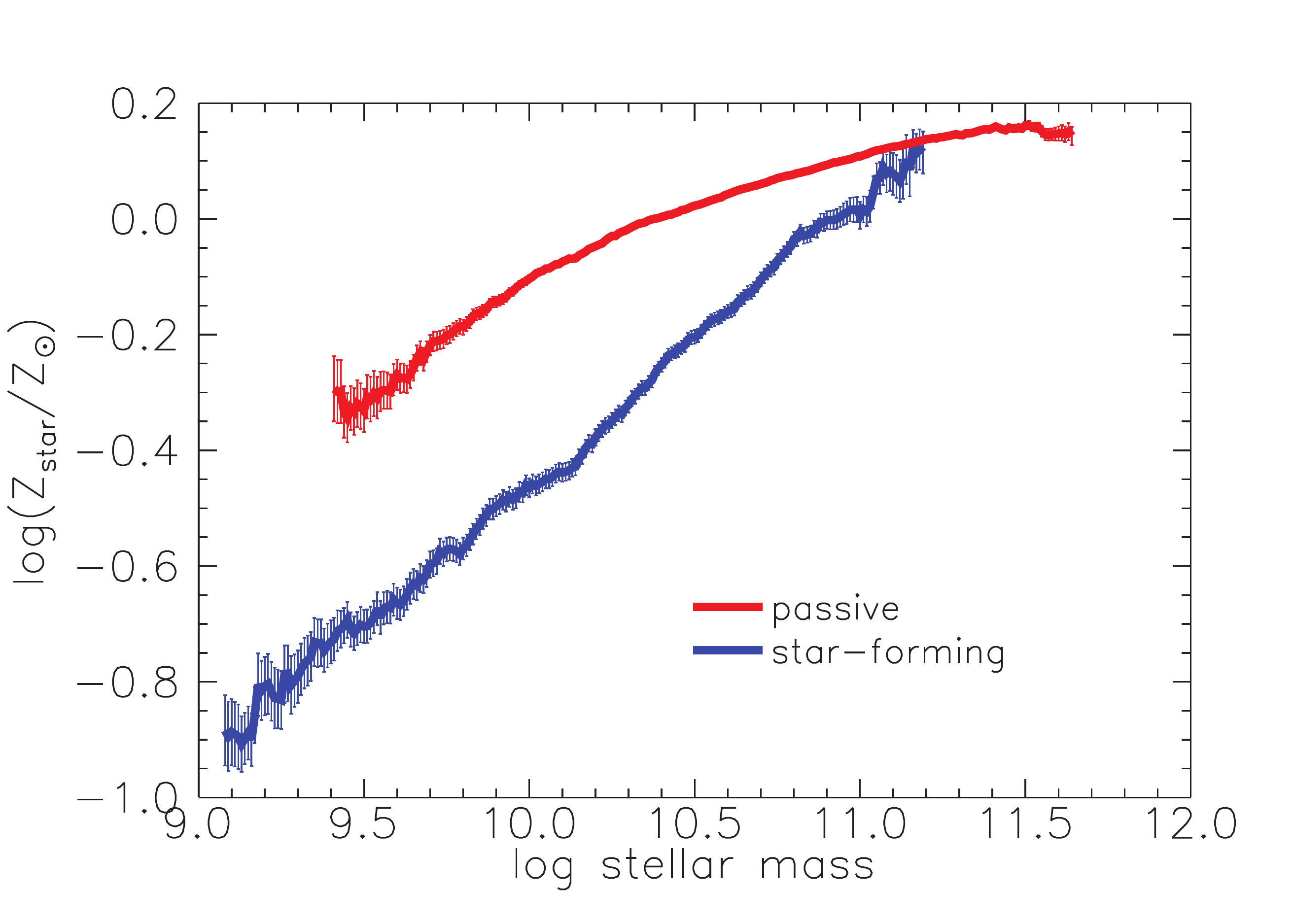}}
\caption{Stellar MZR for quiescent (red) and star-forming (blue)  galaxies in the local universe, from \cite{Peng15}. The difference is ascribed to a sudden reduction of gas infall, producing a reduction of star formation and a rapid increase in metallicity during the ``starvation'' phase, due to
the lack of fresh supply of external gas diluting the metallicity. 
}
\label{fig:MZR_peng15}
\end{figure}

Splitting the SDSS sample into the two classes of quiescent and star-forming galaxies, \cite{Peng15} found evidence of the central role of ``strangulation'' or, more generally, ``starvation''  to explain the stellar metallicity properties of galaxies (Fig.~\ref{fig:MZR_peng15}). Strangulation is the suppression of gas accretion due to dynamical or physical processes \citep{Larson80, Balogh00, Keres05}. When the infall is halted, the galaxy evolves as a closed box
\citep{Searle72,Tinsley80a},
reducing gas fraction, enriching gas with the residual, decreasing activity of star formation, and producing stars with rapidly increasing metallicities, primarily as a consequence of the lack of inflowing gas that dilutes the metallicity. As observed by \cite{Peng15}, in such a scenario higher metallicities are expected in quiescent (starved) galaxies with respect to star forming galaxies which are still experiencing infall (hence dilution) of metal-poor gas.
A similar interpretation is given by \cite{Spitoni17} in their detailed analytical models,
where the metal enrichment of the passive population relative to the star forming
population is achieved through a faster decline (relative to normal discs) of
the inflow rate, which is similar to the simple ``starvation'' scenario but expressed
through a smoother, more physically plausible declining function. 
 Also the recent results of the hydrodynamical code EAGLE are consistent with this scenario \citep{De-Rossi18}.
Whatever is its origin, the dominant feedback effect in these galaxy does not remove gas but prevents further accretions.
Very rencently, \cite{Trussler18} have used mass-weighted metallicities (on a larger sample of galaxies) to show that the metallicity difference between quiescent and star forming galaxies also holds for massive galaxies ($M_{star}\sim 10^{11} M_{\odot}$), especially if also taking into account that the high-z star forming progenitors of local passive galaxies were even more metal poor than local star forming galaxies. This finding indicates that even massive passive galaxies must have quenched through a starvation phase, though \cite{Trussler18} point out that in order to explain their metallicity properties a final ejective or heating phase must also have played a role (else their metallicity would be much higher than observed).
\\

The role of environment in shaping the stellar MZR has been the subject of considerable efforts because it is considered, together with mass, one of the main, independent variables driving galaxy evolution
\citep{Trager00b,Kuntschner01,Thomas05, Sheth06, Sanchez-Blazquez06, Thomas10a, Pasquali10}; \cite{Zhang18b}.
In clusters, galaxies on the red sequence show metallicities that do not depend on age or morphology: old, quiescent ellipticals and younger lenticular follow the same MZR with similar dispersion \citep{Nelan05, Mouhcine11}.
It appears that while environment has a strong effect in defining the morphology, age, and the overall level of activity of the galaxies \citep{Pasquali10, Peng10, Peng12}, the direct (i.e., not mediated by mass) effect on the MZR is modest \citep{Thomas10a,Mouhcine11,Fitzpatrick15,Sybilska17}. 
The effect of environment is larger for dwarf satellite galaxies in high overdensities, in the sense that they tend to be more metal rich than in low density environment \citep{Peng15,Trussler18}, an effect ascribed to starvation as these systems plunge in the hot environment of massive overdensities. In the Illustris simulations \cite{Engler18} do find a dependence of stellar metallicity of dwarf galaxies in clusters on the time elapse from the infall into the cluster; however, they explain this effect with stellar stripping which reduces the stellar mass of infalling galaxies.\\

{\bf Summarizing}, increasing amounts of data are available on the stellar metallicities of nearby galaxies and their dependence on the other main parameters of galaxies. Clear MZRs are present in all galaxy samples. Models can be tested by investigating whether they can reproduce these relations in various classes of galaxies, for stars of different ages, and in different environments. The difference in MZR between passive and star-forming galaxies has been interpreted as evidence for quenching by starvation. 

\begin{figure}
\centerline{\includegraphics[width=9cm]{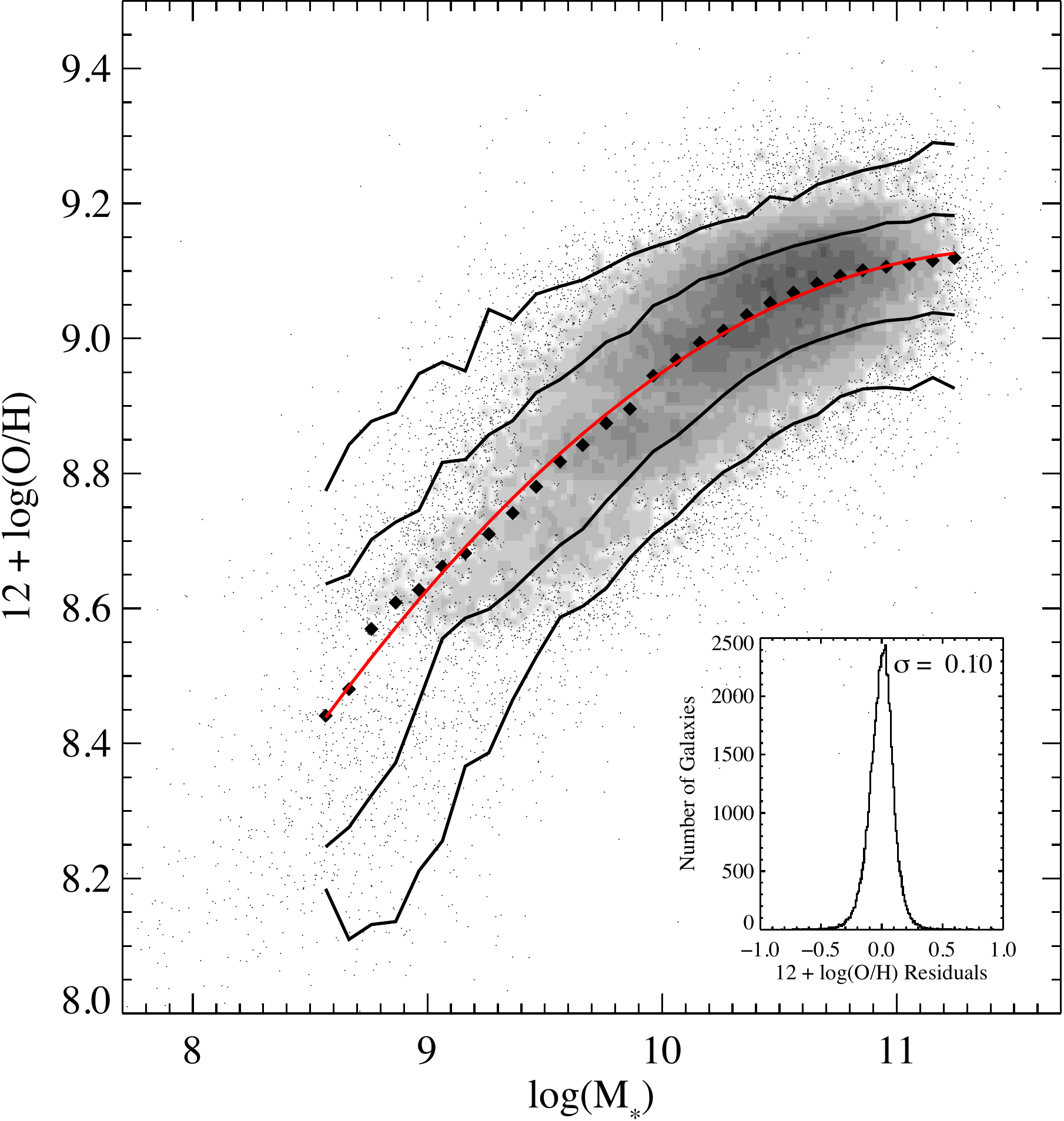}}
\caption{Gas-phase MZR in the local universe.  Black dots are the median of the distribution of the SDSS galaxies (shown as gray dots), from \cite{Tremonti04}. The black lines enclose 68\% and 95\% of the distribution, and the red line is a polynomial fit to the data. The inset histogram is the distribution of residuals of the fit.  This MZR is derived using a metallicity calibration based on photoionization models, and this is likely to produce abundances higher than the actual values, as discussed in Sect~\ref{sec:measmet_comparison}. 
}
\label{fig:MZR_tremonti}
\end{figure}
\subsubsection{Gas-phase metallicity}
\label{sec:MZR_gas}

This dependence of ISM metallicity on mass was observed, for the first time, in galaxies of the Local Group \citep{Peimbert70}.
The first MZR was inferred for a small sample of local, star-forming galaxies (irregulars and blue-compact dwarfs) in an early work by \cite{Lequeux79} in the form of a dependence of chemical abundance on total (dynamical) mass, and the correlation was soon after confirmed by \cite{Talent81} and \cite{Kinman81} (see \citet{Pagel81} for an early review). The same authors found a clear anti-correlation of metallicity with gas fraction, which is mainly driven by the anti-correlation between mass and gas-fraction \citep[e.g.,][]{Rodrigues12,Peeples14}. Total and stellar mass are difficult to obtain for a large number of galaxies, and this was even more true in the '80s. As a consequence luminosity was often used as a proxy for mass when studying these scaling relations \citep{Garnett87,Skillman89b,Garnett02a,Lamareille04}.

The quality of the MZR observed in the local universe improved significantly with the use of SDSS spectra that allowed astronomers to measure the flux ratios of the main optical emission lines for more than $100,000$ galaxies 
\citep[e.g.,][]{Tremonti04,Mannucci10,Perez-Montero13,Lian15}, 
as shown in Fig.~\ref{fig:MZR_tremonti}.  
The observed scatter around the relation in SDSS galaxies is of the order of 0.1dex \citep{Tremonti04,Mannucci10}, somewhat larger than the metallicity measurement uncertainties.
The MZR was also extended towards low mass galaxies, rare in the SDSS sample, 
albeit with a larger scatter and possible biases due to selection effects \citep{Skillman88,Lee06,
van-Zee06,Haurberg13,Pilyugin13,Haurberg15}.
The existence of the MZR is nowadays established from $\sim10^7$\msun\ to $\sim10^{12}$\msun, with a steep dependence at low masses, up to $\rm M_*\sim 10^{10}$\msun, that then flattens out at higher masses.\\

\begin{figure}
\centerline{\includegraphics[width=11cm]{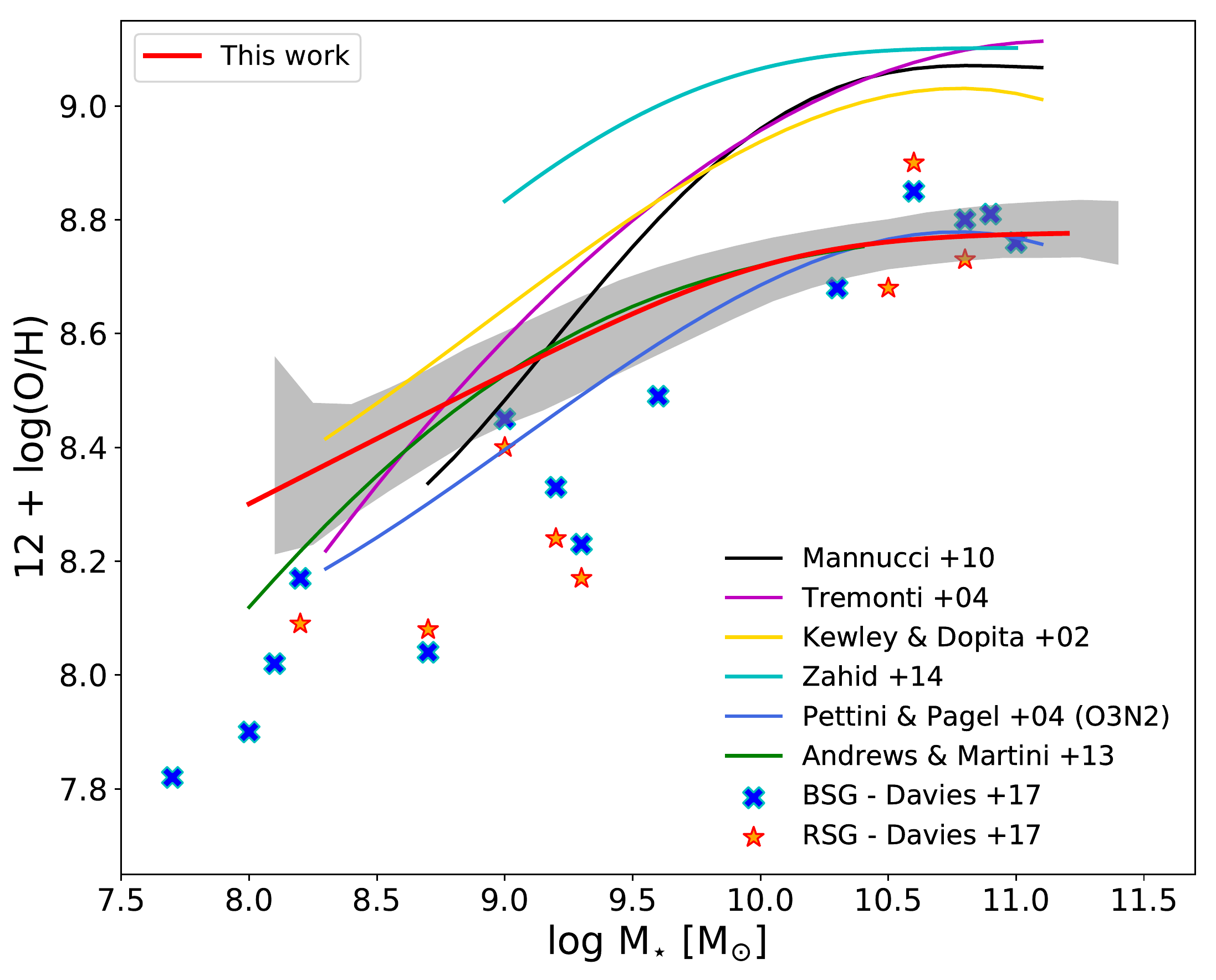}}
\caption{Comparison of the MZRs obtained in the local universe (SDSS data, $<z>\sim0.08$) using different metallicity calibrations, from \cite{Curti19a}. The grey shaded area gives the average metallicity of galaxies using the \Te-based calibration in \cite{Curti17}, in good agreement with \cite{Andrews13} and \cite{Pettini04}. 
Some of the differences, especially at low masses, are due to the different galaxy selection criteria and are linked to the dependence of metallicity on SFR (see below). The curves with higher normalizations are obtained when using calibrations based on photoionization models, while lower curves are based on ``direct'' \Te-based metallicities.
Stars and crosses and RSG and BSG in local galaxies from \cite{Davies17a}, in reasonable agreement with the \Te\ metallicities. 
The relative normalization between stellar and gas-phase metallicities depends on the assumed solar abundance.
}
\label{fig:MZR_curti18}
\end{figure}

Most of these studies are based on metallicities derived with the strong-line method, the only technique that can applied to large numbers of galaxies. It should be noted that both the shape and, even more, the overall normalization of the gas-phase MZR depend on the
calibration used. In particular, the MZR based on photoionization models, like \cite{Tremonti04}, and \cite{Mannucci10} at high metallicities, provide high normalizations that, for example, do not fit the positions of the MW, the LMC and the SMC which are all significant more metal-poor than these MZR. 
For this reasons,
of particular interest are the MZRs derived by using ``direct'' metallicities (see Sect.~\ref{sec:measmet_ism_Te})
\citep{Berg12,Andrews13,Ly16,Curti19a}. These studies confirm the overall shape of the relations found with the strong line methods but with a significant lower normalization, as shown in Fig.~\ref{fig:MZR_curti18}.

The contribution from the DIG to the line ratios and, therefore, to the computation of metallicities can also significantly affect the measured shape of the MZR and therefore is another source of uncertainty \citep{Sanders17a}.

The comparison between stellar and gas-phase MZR can be used to obtain scientific insights on several aspects of galaxies. For example, the small amount of ISM present in early-type galaxies (ETG) shows metallicities similar to that of the old stellar population \citep{Griffith18}. This reveals that the ISM in these galaxies is little affected by infalling gas (expected to be less metal rich than the stars) and mostly due to internal production.

\subsubsection{Interpreting the MZR}
\label{sec:MZR_models}

Various possible driving mechanisms have been proposed to explain the existence and the shape of the MZR. 

First, MZR could be shaped by outflows produced by feedback \citep[e.g.][]{Garnett02a,Brooks07}.
Outflows mainly due to SNe are very common in starburst galaxies, both in the local universe and at high redshift  \citep[e.g.,][]{Heckman02,Law07,Weiner09,Steidel10,Martin12b,Heckman17}.
These outflows are observed to have metallicities higher than the ISM of the parent galaxies \citep{Chisholm18}, and  are expected to be much more effective in small galaxies, where the potential well is shallower, hence removing a larger fraction of metal-enriched gas from low-mass systems towards the CGM and the IGM 
\citep{Tremonti04, Tumlinson11,Chisholm18}.

Second, it is known that high-mass galaxies evolve more rapidly and at higher redshifts than low-mass ones, the so-called ``downsizing'' \citep[e.g.,][see also Sect.~\ref{sec:alpha_over_iron}]{Cowie96, Gavazzi96, Somerville15}, therefore at the present stage they are expected to have converted a larger fraction of their gas into stars and metals, reaching a higher metallicity \citep{Maiolino08,Zahid11}. Under this interpretation the MZR is a sequence of evolutionary stages.

Third, the earlier evolutionary stage of  smaller galaxies and their larger gas fraction \citep[e.g.,][]{Erb06c, Rodrigues12, Lagos16b} could be linked to the on-going infall of metal-poor gas, which, once mixed with the existing ISM, contributes to reduce metallicity and to the build up of the stellar population through star formation. 

Fourth, the shape of the high-mass end of the IMF could depend on galaxy mass, introducing a systematic change in the average stellar yields and in the rate of metal enrichment 
\citep{Trager00b,Koppen07,Molla15,Vincenzo16b,Lian18c}.

Fifth, the metallicity of the accreted gas, recycled from previous episodes of star formation, may be larger for larger mass galaxies \citep{Brook14, Ma16}.\\

The equilibrium models introduced in Sect.~\ref{sec:models} are explicitly built to reproduce the MZR and therefore explain this relation as the consequence of the interplay of many on-going processes. 
The simple ``bathtub'' models explain the MZR without having to invoke any direct effect of the gravitational potential in terms of capability of
retaining metals (i.e., no mass-dependent outflow rate), although mass is indirectly included through the DM framework, which accelerates the evolution of more massive systems
\citep[e.g.][]{Bouche10,Lilly13,Peng14b,Dekel13,Dekel14}. 

SAM and hydro numerical codes  (see Sect.~\ref{sec:models}) have also been tuned to reproduce the MZR and its evolution \citep{De-Lucia04, Croton06, Finlator08, Oppenheimer08, Oppenheimer10, Dutton10, Dave11c, Dave12,Somerville12, Dayal13, Forbes14, Lu15c, Pipino14, Torrey14, Zahid14, Feldmann15, Harwit15, Kacprzak16, Christensen16, Hirschmann16, Rodriguez-Puebla16, Torrey17,Torrey18}.
These models often produce an anti-correlation between metallicity and gas fraction \citep{Schaye15,De-Rossi15a,Lagos16b,Segers16,De-Rossi17} which is actually observed, see Sect.~\ref{sec:FMR_gas}. \\

A critical role in shaping the MZR is played by the properties of stellar and AGN feedback, the chemical yields,  the actual metallicity of the outflowing wind with respect to the parent ISM, the fraction of metals that are re-accreted, and the evolution of the star formation efficiency. Many of these parameters are degenerate, and in some cases measurements of the metallicity help to break these degeneracies. For example, by comparing the stellar and gas mass-metallicity relations, \cite{Lian18a} concluded that only two scenarios can reproduce both relations as well as the MSSF: either strong outflows remove most of the metals, or a steeper IMF characterizes the early stages of galaxy formation. As it will be discussed in Sect.~\ref{sec:abund_ratios}, the study of the abundance ratios between different elements is potentially capable of breaking degeneracies because they are sensitive to the timescales of star formation.

An important issue when comparing observations with theory is that of normalization which drives the total amount of metals observed in the ISM. As shown in Fig.~\ref{fig:MZR_curti18} and discussed in Sect.~\ref{sec:measmet_comparison}, metallicity calibrations totally or partially based on photoionization models provide a significantly higher normalization of the MZR than the direct '\Te' methods. The amount of metals that the models must produce and disperse into the IGM depends critically on this problem. The direct method is now considered to be more reliable and the low normalizations should be preferred.\\

{\bf Summarizing}, the MZR could be a sequence of metals removal by outflows in low mass galaxies, dilution, or different evolutionary stages of galaxies with different masses, and all these effects could be simultaneously present. Information on the relative importance of these effects can be obtained by studying the effective yields (see Sect.~\ref{sec:effective_yields}).

\begin{figure}

\centerline{\includegraphics[width=9cm]{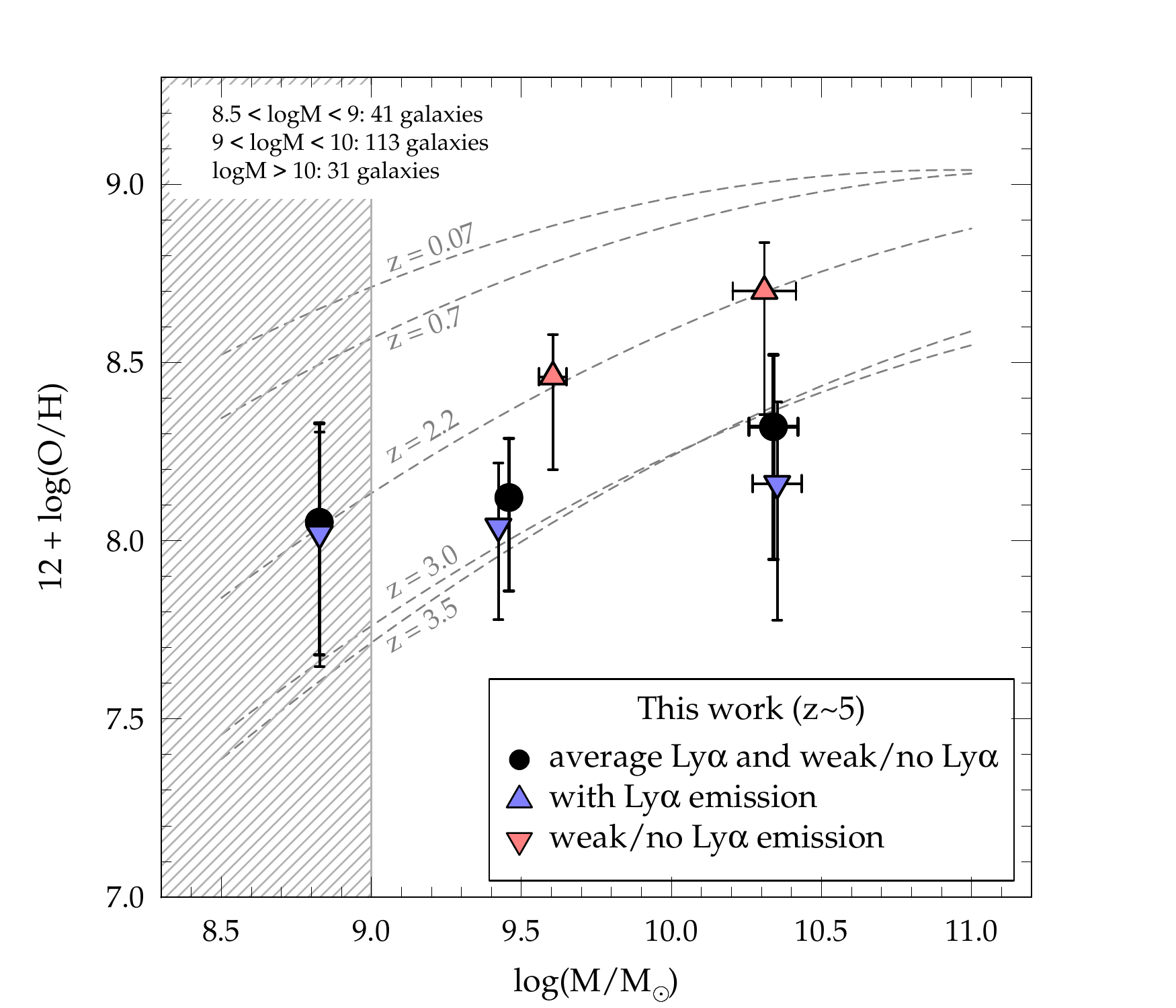}}
\caption{Stellar MZR at $z\sim5$ for averages of galaxies with (blue) and without (red) Ly$\alpha$\ emission, from \cite{Faisst16}. The stellar MZR in compared with the gas-phase MZR in \cite{Mannucci09b}. 
}
\label{fig:MZR_faisst16}
\end{figure}

\subsubsection{Redshift evolution of the MZR}
\label{sec:MZR_evolution}

Measuring the evolution of the MZR with redshift is not an easy task. A large number of spectra are needed, and 
measuring metallicities generally require high S/N ratio spectra, especially to measure stellar metallicity but also when measuring the abundances in the gas phase with the ``strong-line'' method.\\

The evolution of the {\it stellar MZR} has been studied with optical spectroscopy to sample the stellar populations dominating the total stellar mass up to intermediate redshifts in massive galaxies, both in clusters and in the field, finding a modest evolution, consistent with the passive evolution of the stellar population \citep{Kelson06,Ferreras09,Choi14,Gallazzi14,Onodera15,Leethochawalit18}. 
A significant number of rest-frame UV spectra of high redshifts galaxies have been obtained \citep{Mehlert02, Fosbury03, Shapley03, Steidel04, Rix04, Savaglio04, Halliday08, Quider09, Dessauges-Zavadsky10, Erb10, Mouhcine11, Sommariva12, Steidel16, Faisst16}. In particular,
a number of bright, usually lensed galaxies have been studied in considerable detail in the UV, obtaining a wealth of information about their level of metal enrichment and about the chemical abundance ratios \citep[e.g.,][]{Pettini01,Villar-Martin04,Dessauges-Zavadsky10}.
Despite large uncertainties, as expected the stellar MZR derived from rest-frame UV spectra (hence probing young stars) is similar to the gas-phase MZR, has a significant redshift evolution (see Fig.~\ref{fig:MZR_faisst16}), and shows an inverse relation between metallicity and SFR \citep{Faisst16}.\\

\begin{figure}
\centerline{\includegraphics[width=9cm]{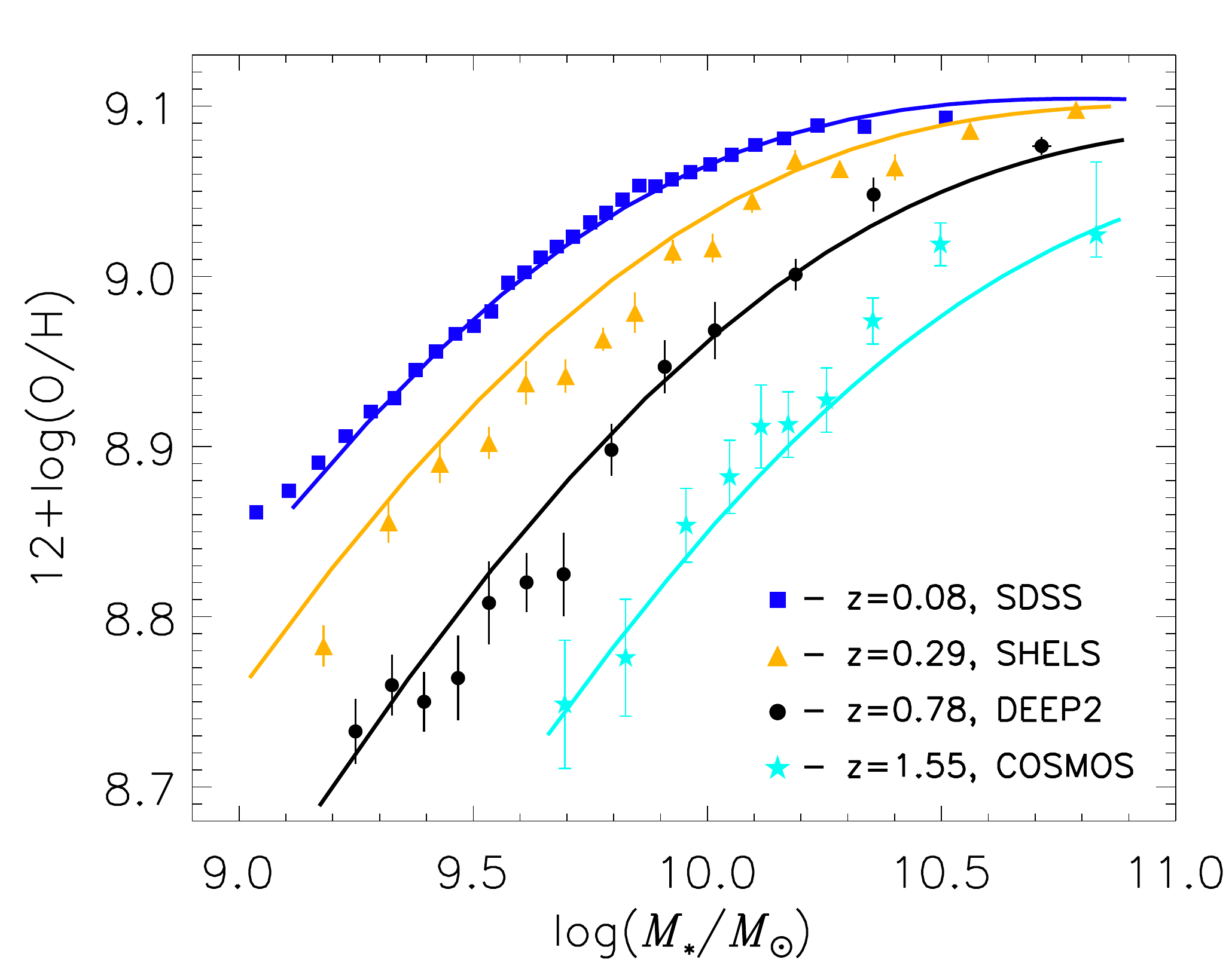}}
\caption{Redshift evolution of the MZR up to $z=1.55$ from \cite{Zahid14a}. Blue, yellow, black and cyan points refer to $z=0.08$, 0.29, 0.78, and 1.55, respectively. These metallicities have the ``high'' normalization related to the use of photoionization models (see Fig.~\ref{fig:MZR_curti18}).
}
\label{fig:MZR_zahid14}
\end{figure}

Significant observational efforts by many groups have provided a clear picture of the evolution of the {\it gas-phase MZR} up to z$\sim$3.5, i.e., out to the maximum redshift where the main optical lines are still in the near-IR bands. This MZR is found to evolve monotonically with redshift, with metallicity declining with redshift at a given mass. At low redshift the evolution is faster at lower mass (see Fig.~\ref{fig:MZR_zahid14}), while high mass galaxies 
have already reached their current metallicity by $z\sim 1$, in what appears to be the chemical version of downsizing. 

This results is based on many observational works at intermediate ($z\leq1.5$) redshifts, 
mainly using optical spectroscopy 
\citep[e.g.][]{Contini02, Kobulnicky03, Kobulnicky04, Maier04, Maier05, Savaglio05, Maier06, Cowie08, Zahid11, Moustakas11, Cresci12, Foster12, Zahid13a, Perez-Montero13, Yuan13, Nakajima13, Guo16b, Perez16, Suzuki17}.
The evolution at higher redshifts requires near-IR spectroscopy. The much brighter sky background, the poorer atmospheric transmission, the technological limitations of near-IR spectrographs with respect to optical ones, and the fainter apparent brightness of  high-redshift objects produce a lower number of useful spectra, usually with lower S/N ratio and for galaxies with higher SFR. Nevertheless a number of authors have produced significant databases at $z\sim2$ 
\citep[e.g.,][]{Erb06a, Finkelstein11a,Wuyts12,Erb10, Cullen14, Zahid14b,Wuyts14b,Steidel14,Sanders15, Onodera15, Sanders16a, Bian17,Sanders18} 
and $z\sim3$ \citep{Maiolino08, Mannucci09b,Belli13a,Maier14a,Troncoso14, Onodera16}.

In many cases, the detection of the faintest lines needed to measure metallicities is only possible when many galaxy spectra are stacked together. This procedure is not free from uncertainties because it involves choices on which galaxies to stack, how to do the stacking, and is prone to non-linear effects.  Lensed and single bright galaxies have been extensively observed to obtain higher S/N ratio on the spectra of single galaxies, especially in the low mass range \citep{Teplitz00a, Kobulnicky00,Richard11,Troncoso14,Wuyts14a,Jones15a,Perna18}.

While the resulting big picture outlined above is clear, the details are debated and depend on the method used. 
First, metallicity indicators play a critical role, especially because at high redshift each
observing program typically has access to a restricted number of diagnostics, hence introducing scatter and systematics among different surveys \citep[][see Sect.~\ref{sec:measmet_strong}]{Kewley08}. This is even more important when taking into account that 
different results can be obtained also when using different line ratios of the same calibration
\citep[e.g.,][]{Brown16}. 

Moreover, the possibility of a significant evolution with redshift of the calibrations due to different conditions of the star-forming regions, for example in terms of ionization parameter, ionizing spectra, density, pressure, and N/O ratio is always present (see Sect.~\ref{sec:measmet_strong_BPT}), and could also be differential for the various methods
\citep{Kewley13a, Kewley13b,Steidel14, Kewley15,Shapley15,Strom17b,Kashino17a}. 

Second, the method used to measure stellar mass (and SFR) affects the results, as discussed, e.g., in \cite{Yates12} and \cite{Cresci19}. 

Third, how galaxies are selected affects the final results. For example, when galaxies are selected according to the flux of a metallicity-sensitive emission line, such as \oiii5007 
\citep[e.g.,][]{Izotov11a,Xia12,Izotov15} 
the results can be biased toward lower or higher metallicities. 

Fourth, similar but more subtle effects are introduced by the need of putting a S/N threshold on the flux the line to be used to measure metallicity. If a relatively high (3-5) minimum S/N is used on all the lines \citep[e.g.,][]{Yates12},
metallicity-dependent selection effects can be introduced near the detection threshold. For example, 
the flux of \oiii5007 is about 1/10 of \ha\ at solar metallicities (and low extinction) and about as bright as \ha\ at $Z\sim 0.2 Z_\odot$. 
If a threshold of S/N=3 is used for \oiii5007, 
at each SFR (i.e., roughly at each \ha\ luminosity), a larger fraction of low metallicity galaxies are included while high metallicity galaxies are preferentially excluded, altering the cosmic average and the resulting MZR.  
Opposite biases are introduced by \nii6584, whose flux increases with metallicities. The result is the introduction of biases that are difficult to trace \citep[][]{Salim14, Cresci19}.  The opposite approach consists in using a high S/N threshold only on the emission line more directly related to SFR and less dependent on metallicity, such as \ha\ and \hb, obtaining a more SFR-selected sample \citep[e.g.,][]{Mannucci10}. In both cases the resulting sample is usually not mass- or volume-selected.

Fifth, often spectra are obtained inside a fixed aperture irrespective to galaxy distance. For example, SDSS spectra are obtained with a 3~arcsec circular fiber placed on the galaxy center. This, together with the existence of radial metallicity gradients, can introduce spurious correlations of metallicity with distance and galaxy size that are not easy to estimate and correct.

Finally, the presence of a dependence of metallicity on SFR and other galactic properties, such as size and surface density (see Sect.~\ref{sec:FMR}), means that the shape of the observed MZR and its redshift evolution depend critically on how the targets are selected in terms of luminosities and redshift range.  In other words, the MZR is only defined for a given average SFR at each mass, using intrinsically fainter or brighter galaxies for a given mass and redshift affects the shape of the MZR. Marked differences between published results (e.g., \citealt{Steidel14} vs. \citealt{Wuyts14b}) can be explained by this effect, see the discussion in \cite{Cresci19}.

Despite all these uncertainties, there is a general agreement on the fact that the observed MZR evolves with redshift (see Fig.~\ref{fig:MZR_zahid14}), especially at z$>$1, i.e., the observed metallicity at a given stellar mass decreases with redshift at a rate which depends on redshift and mass. The evolution of the MZR can be parametrized in different ways
\citep[e.g.][]{Moustakas11}.
\cite{Maiolino08} and \cite{Mannucci09b} used a simple 3rd-order polynomial fit, while \cite{Zahid14} introduced a different parametrization named ``Universal Metallicity Relation''. The variation of only one of the parameters used in this analytic formula, the mass at which the mass dependence on metallicity starts to flatten out, which is often enough to reproduce the observed evolution. \cite{Curti19a} proposed a similar parametrization but with one more parameter to better match the observations.\\

Most galaxy evolution models cited in the previous section reproduce the evolution of the MZR. The steady decrease of metallicity with redshift at constant mass is ascribed to several reasons, including higher efficiency in ejecting gas and reduced stellar yields. Based on the IllustrisTNG simulations, \cite{Torrey17} identify one of the main drivers of the MZR evolution in the increasing gas fraction with redshift, in agreement with the anti-correlation between metallicity and gas fraction and SFR observed in the local universe (see Sect.~\ref{sec:FMR_gas}), while \cite{Yabe15a} attribute the evolution of the MZR to higher infalls and outflows at high redshifts, and \cite{Lian18c} propose either higher metal loading factors or a steeper IMF at high redshifts. As discussed in Sect.~\ref{sec:models}, a meaningful comparison with the models should consider all the selection- and observational- effects listed above, and this is not always done.\\

{\bf Summarizing}, there is a clear evidence that the gas-phase MZR and the stellar MZR based on UV observations (i.e., related to young stars) are evolving with redshift, showing lower metallicities at earlier cosmic times. The details of the evolution depend on a number of possible selection effects and issues regarding how metallicities are estimated. The increase of metallicity with time for a given stellar mass is a feature which is commonly reproduced by the models of galaxy formation.

\begin{figure}
\centerline{\includegraphics[width=9cm]{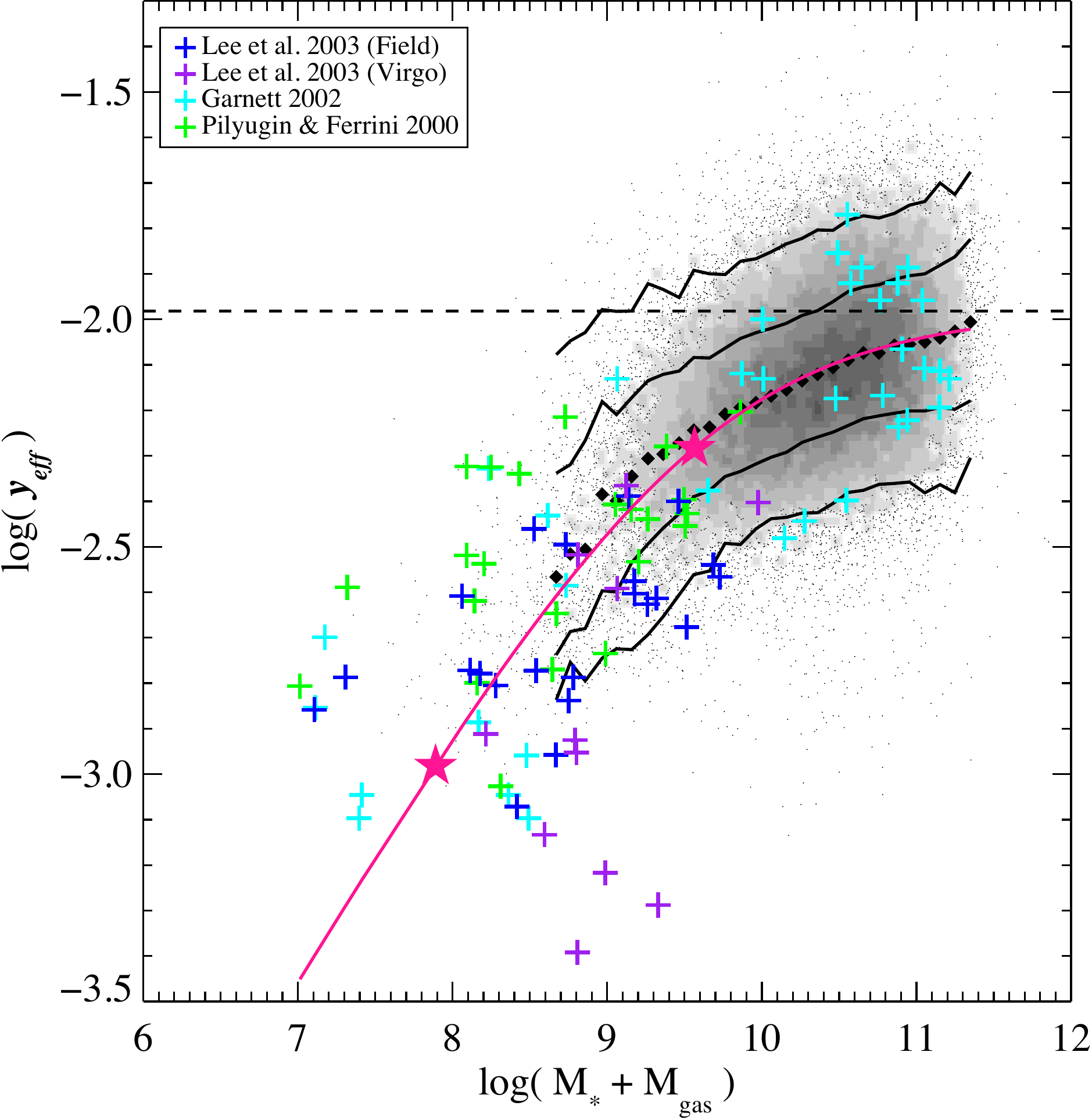}}
\caption{Effective yields as a function of total baryonic mass for SDSS galaxies (gray) and other low-mass galaxies, from \cite{Tremonti04}.
The black diamonds are the median of the data distribution. The dashed line indicates true yield $y$ if no metals are lost. The pink line is a fit with a physically-motivated analytic formula, and the pink stars denote galaxies that have lost 50\% and 90\% of their metals. 
}
\label{fig:Yields_tremonti04}
\end{figure}

\subsubsection{Effective yields}
\label{sec:effective_yields}

The effective yields are a way to measure the influence of infall and outflows on the chemical evolution of a galaxy \citep{Matteucci01b,Garnett02a,Dalcanton07}. 
The differential equations describing a closed-box system, i.e., with no infalling gas and not outflowing winds, with the assumptions of instantaneous recycling and instantaneous mixing,  can be solved \citep{Edmunds90} to express gas metallicity $Z$ as a function of the gas fraction $f_{\rm gas}$ and of the true constant stellar yield $y$ (assumed to be constant):

\begin{equation}
Z = y\ \ln (1/f_{gas})
\end{equation}

The true yields $y$  can be expressed in terms of the solar yield $y_\odot$=0.0142, i.e., the fractional contribution of metals to the solar mass \citep{Asplund09}. Using the observed quantities for $Z$ and $f_{gas}$, the effective yields is defined as:

\begin{equation}
y_{\rm eff} =Z/\ln (1/f_{\rm gas})
\end{equation}

In general the measured values of $y_{\rm eff}$  differ from the true stellar yields $y$ if the system is not a closed box. 
In other words, comparing the metallicity with the gas fraction gives information on the gas flow from and into  galaxies.
An outflow removes gas, hence reduces $f_{\rm gas}$ and therefore reduces also $y_{\rm eff}$. If the outflow preferentially eject metals, then it also
reduces the metallicity, further reducing $y_{\rm eff}$. Inflow of metal-poor gas
decreases the metallicity and, although it increases the gas fraction, it can be shown
that the net effect is to reduce $y_{\rm eff}$.
 Both effects therefore tend to reduce $y_{\rm eff}$ with respect to $y$, and the difference is a measure of how much these gas flows affect the chemical evolution of the galaxies.

\begin{figure}[t]
\includegraphics[width=6.0cm]{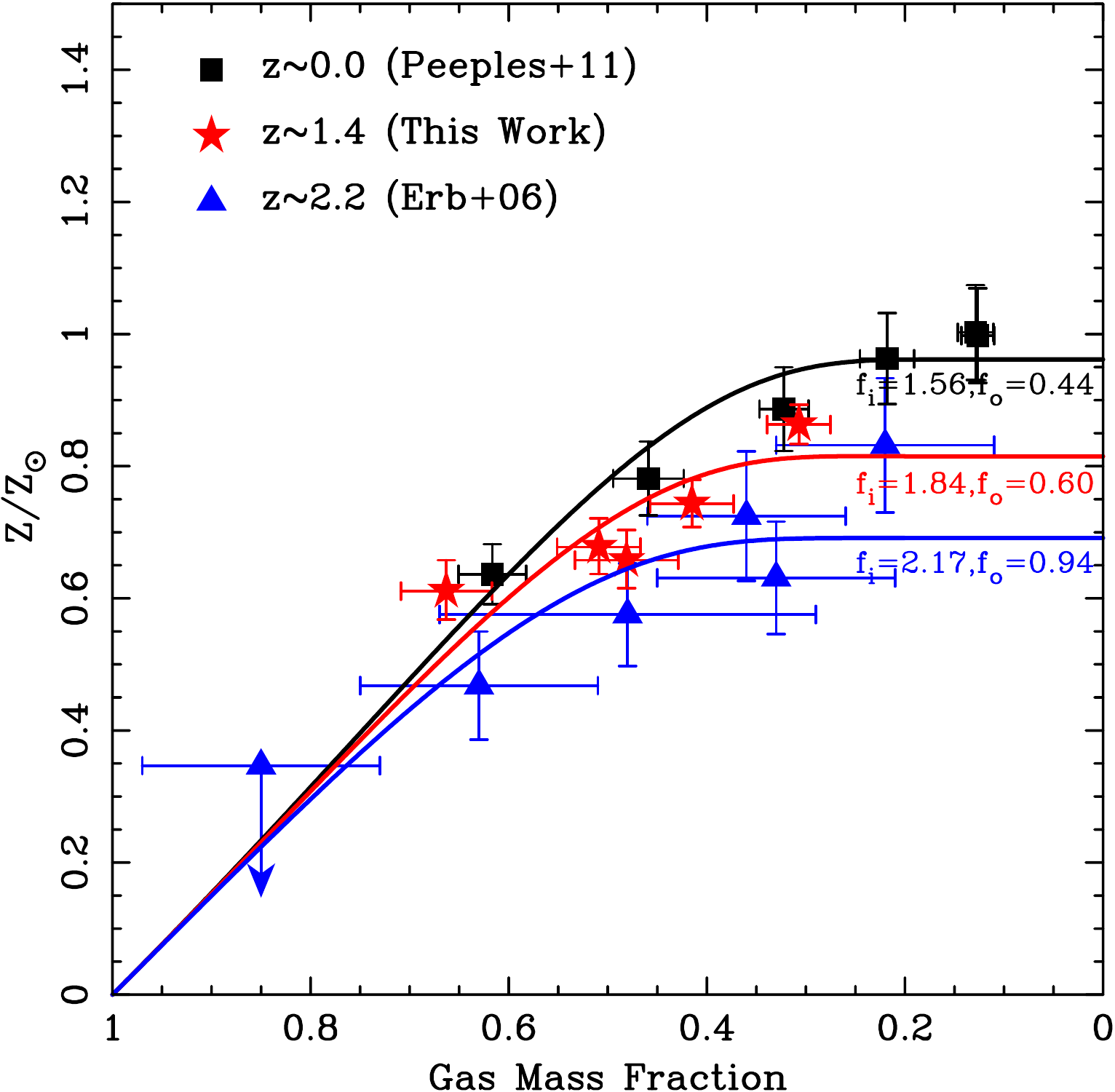}
\includegraphics[width=6.0cm]{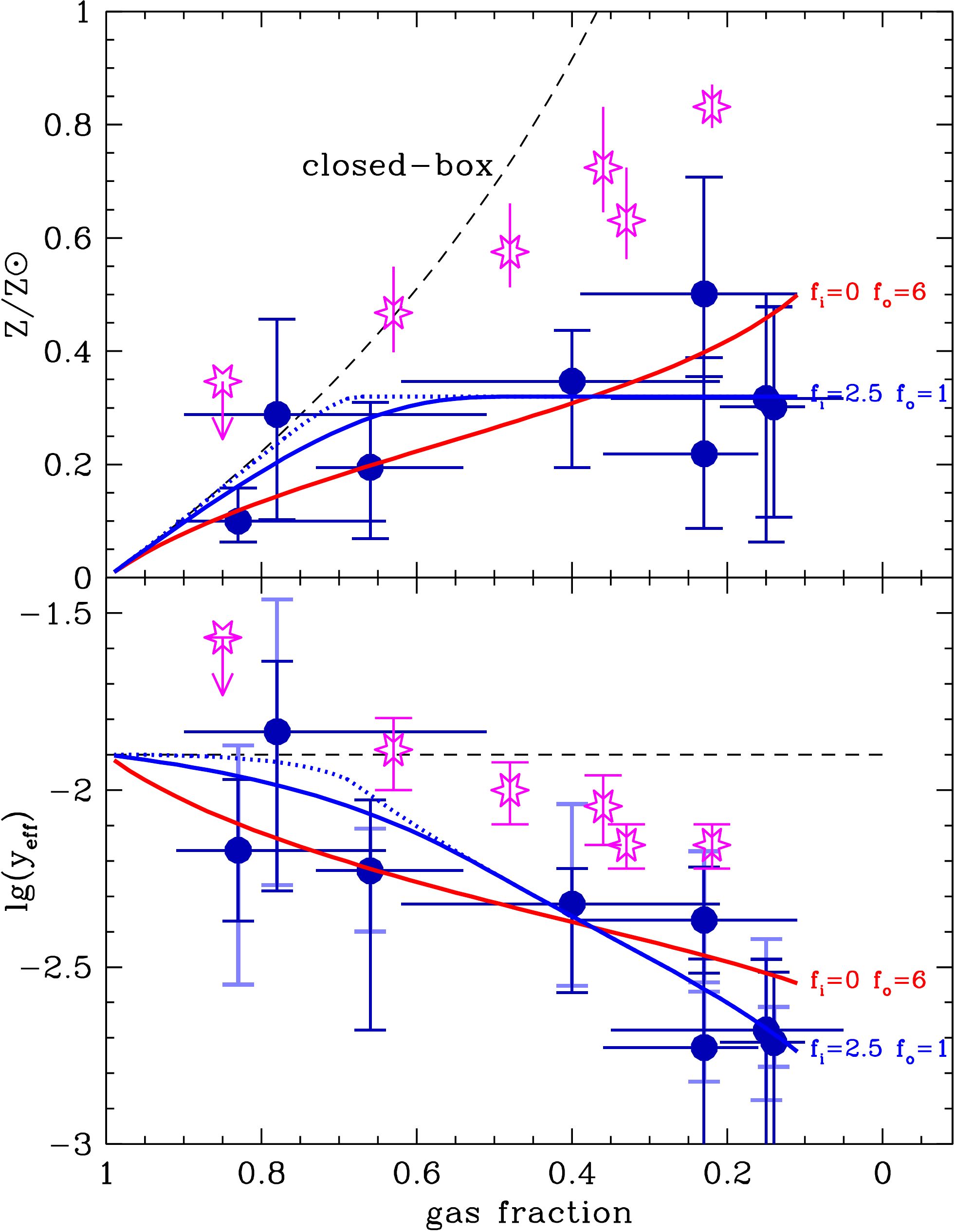}
\caption{
Left: Metallicity as a function of $f_{gas}$ for three sample of galaxies at various redshifts, from \cite{Yabe15a}. The properties of each galaxy sample are reproduced by models with gas infalls and outflows proportional to the SFR, where $f_i$ and $f_o$ are the relative loading factors.
Right: Metallicity (upper panel) and effective yields (lower panel) as a function of $f_{gas}$ in the sample of $z\sim2$ by \cite{Erb08} (magenta stars) and $z\sim3$ galaxies from \cite{Mannucci09b} (blue dots). The black dashed line are the expectation of a close-box model \citep{Edmunds90,Matteucci01b,Matteucci08a}. Blue and red lines are 
two different fits to the blue data points and emphasize the different contribution of outflows ($f_o$) and infalls ($f_i$). In this model, infalls are more effective in changing $y_{\rm eff}$ for the gas-poor galaxies, while outflows are more effective in gas-rich galaxies.
}
\label{fig:Yields_mannucci09}
\end{figure}

Estimating gas masses via the Schmidt-Kennicutt relation, \cite{Tremonti04} found a significant dependence of $y_{\rm eff}$ on mass for the local SDSS galaxies (Fig.~\ref{fig:Yields_tremonti04}), with lower mass galaxies having lower $y_{\rm eff}$. At variance with this result, \citep{Yabe15a} used the gas masses directly measured by \cite{Peeples11} and metallicities
estimated from the MZR , obtaining $y_{\rm eff}$ decreasing with mass,
see Fig.~\ref{fig:Yields_mannucci09}-left.  
This shows the dependence of the results on how gas fractions are measured, but in either cases it proves that galaxies are not closed boxes, that infall and/or outflows of gas have an important effects on metallicity, and that earlier evolutionary stages  in low mass galaxies cannot be the only explanation for the MZR. Either outflows of metal-rich gas, or infall of metal-poor gas, or both, must have measurable consequences on metallicity. \\

At high redshift  $y_{\rm eff}$ is found increase with gas fraction and, as a consequence, to decrease with stellar mass. This is observed at $z\sim1$ \citep{Rodrigues12,Yabe15a}, at $z\sim2$ \citep{Erb06a, Erb08,Wuyts12}, and at $z\sim3$ \citep{Mannucci09b,Troncoso14}, see Fig.~\ref{fig:Yields_mannucci09}, and mass loading factors of infall and outflows significantly larger than 1 are often obtained.

\subsubsection{The mass-metallicity relation of DLAs and GRB host galaxies}
\label{sec:MZR_DLAs}

At even higher redshifts ($z>3.5$), all main optical lines leave the near-IR bands and metallicity can only be obtained with different techniques and on galaxy samples selected in very different ways with respect to the common magnitude- or mass-based selections. 

A MZR relation up to $z\sim5$ has been derived for gamma-ray burst (GRB) host galaxies by \cite{Laskar11} by measuring the metallicity through ISM absorption lines. 
To what degree this is representative of the high-redshift universe is debated because it is still not clear if long-duration GRBs select an unbiased sample of star-forming galaxies. A discussion of this complex issue is beyond the scope of this review. Here we only mention that evidence for a strong metallicity bias has been proposed and negated several times, 
\citep[see, e.g.,][and references therein]{Fynbo06a,Mannucci11a,Arabsalmani14,Piranomonte15,Trenti15,Kruhler15,Vergani17,Arabsalmani18}.\\ 

As explained in Sect.~\ref{sec:measmet_ism_abs}, DLAs \citep[e.g.,][]{Wolfe86} provide a unique opportunity to obtain accurate measure of the metallicity of the CGM (and the ISM of the outer disc) in high-z galaxies \citep[see, e.g.,][]{Pettini02c,Pettini06}.
Comparing this information to that derived from luminosity-selected galaxies is not straightforward \citep[e.g.,][]{Fynbo08, Christensen14}, for the various  reasons already discussed in sect.~\ref{sec:measmet_ism_abs}.

In particular, DLAs are selected on the gas cross-section, and the information derived 
only applies to the part of the galaxy projected on the background source, whose distance from the center is usually not known. As a consequence, the same galaxy can show very different metallicity properties if, for example, two lines of sight at different radial distances are studied. 
We recall that it is not even clear whether DLA probe the extended discs of galaxies, or clumps in the circumgalactic medium, or both (see the discussion in \citealt{DeCia18b} and \citealt{Krogager17}).

Moreover, different elements are used to measure the metallicity of the ISM in emission and absorption: typically O and N abundances are estimated for emission-selected galaxies (see Sect.~\ref{sec:measmet_ism}),  while in DLAs various elements can be traced (Zn, S, Fe, Si, O, C, Mg, \dots) depending on the observed band and column density, etc.

Finally, the integrated properties of the galaxy associated with the DLA, such as stellar mass and SFR, are often not measured as the galaxy itself is often not even detected, although
a small number of DLA galaxies are actually identified in emission \citep[e.g.,][]{Rhodin18,Kanekar18,Krogager17,Noterdaeme12,Peroux11}.  At $z\sim0.7$ these galaxies follow a MZR similar to the emission-selected galaxies when a somewhat uncertain correction for metallicity gradients is included \citep{Rhodin18}.
Moreover, \cite{Krogager17} identified the optical counterparts of a small sample of DLAs at z$\sim$2 and,
by combining these with some additional previous detections, suggest that the DLA host galaxies follow a luminosity-metallicity relation.

However, more generally, as a consequence of the issues discussed above,
most studies investigate the mass-metallicity relation by using the velocity dispersion of
the DLAs as a proxy for the mass \citep{Haehnelt98, Ledoux06, Pontzen08}.
A 2D-linear (planar) correlation between velocity dispersion (mass), metallicity and redshift have been proposed by \cite{Neeleman13}, which is capable of reducing the scatter about the relations involving only two quantities. The correlation between velocity dispersion and metallicity is
found to be in place already at $z=4$ \citep{Ledoux06}, and \cite{Moller13} explored the relation out to $z=5$ by  using a sample of ~100 DLAs  at $0.1<z<5$ \citep[see also ][for a confirmation that the same relation applies to DLA  along the line of sight of both QSOs and GRBs]{Arabsalmani15}.

\begin{figure}
\centerline{\includegraphics[width=7cm]{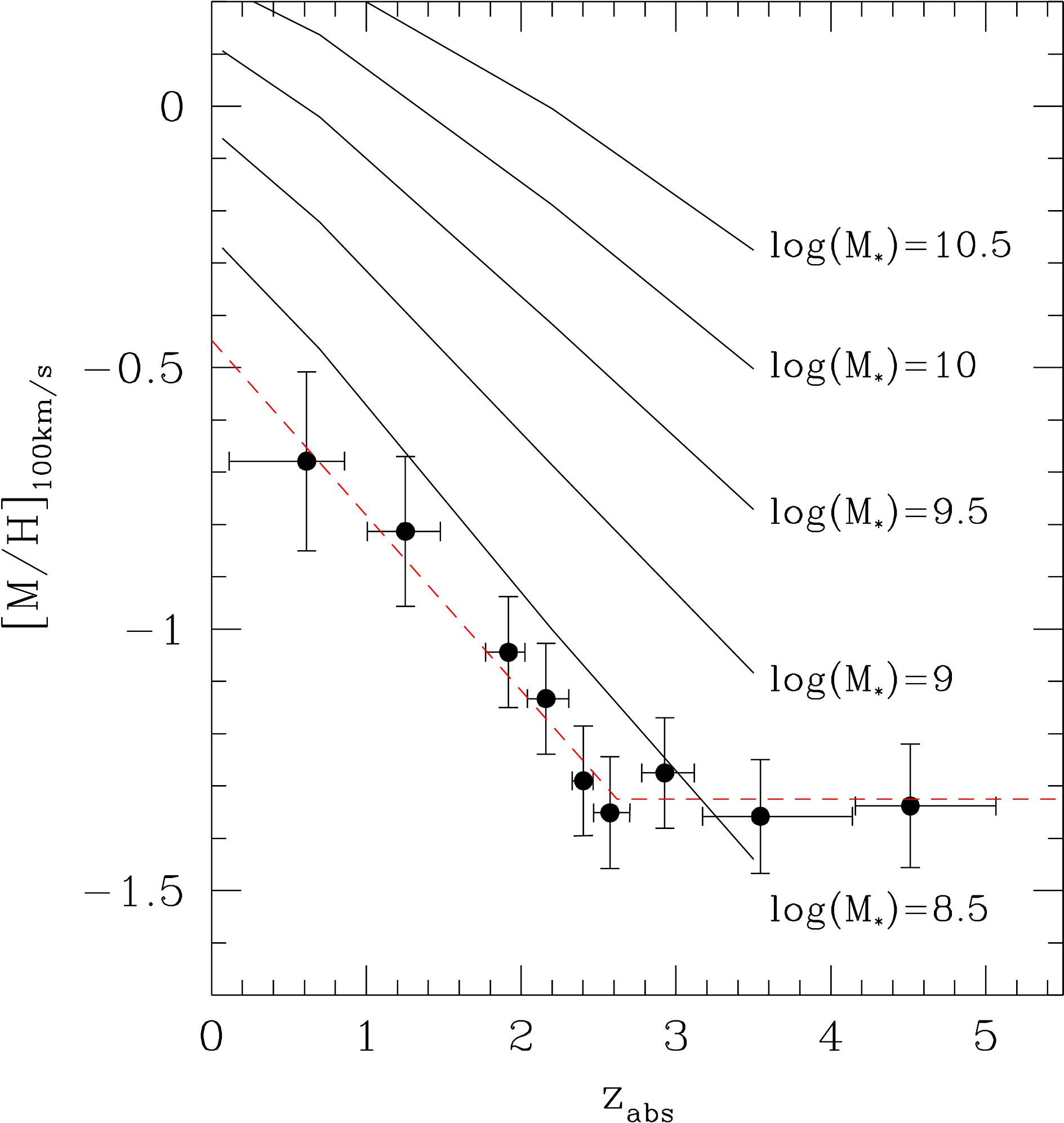}}
\caption{Redshift evolution of the normalization of the MZR for DLAs. Dots show average of DLAs, from \cite{Moller13}, compared to the evolution of luminosity-selected galaxies galaxies from \cite{Maiolino08}, \cite{Mannucci09b}, and \cite{Troncoso14}. At low masses the MZR of these works is based on ``direct'' \Te\ metallicities, therefore the comparison is not strongly affacted by the use of more modern calibrations (see Sect.~\ref{sec:measmet_comparison}).
}
\label{fig:DLA_moller13}
\end{figure}

Clearly the absolute normalization of the MZR for DLA and luminosity-selected galaxies cannot be easily compared, as this implies translating the DLA velocity dispersion (at the location of the DLA impact parameter) into a host galaxy stellar mass. However, the relative redshift evolution
of the two relations can be compared, and this can give information on the nature of the galaxies associated with the DLAs. 
Figure~\ref{fig:DLA_moller13} shows the evolution of the DLA MZR from $z=5$ to $z=0$ \citep[from ][]{Moller13} compared to the evolution of the MZR of emission-selected galaxies \citep{Maiolino08, Mannucci09b, Troncoso14}. Interestingly, the DLA MZR follow the same
rapid evolution as low mass galaxies. However, clear evolution is only seen out to $z=2.6$, beyond
this redshift the normalization of the DLA MZR remains constant.
The cosmic epoch of this break (z$\sim$2.6) is interesting for various reasons: it is close to the peak of galaxy formation \citep{Madau14} and, intriguingly, this is also the redshift where the FMR (see Sect.~\ref{sec:FMR}) starts to show clear signs of evolution \citep{Mannucci10}, and
it is also the epoch when the comoving HI mass density of galaxies change from rapidly increasing (at z$>$2) to remaining constant (at z$<$2) \citep{Prochaska09}.
This change has been attributed to several possible effects: 
a variation of the IMF \citep{Bailin10}, 
variations in the properties of the IGM and of the equilibrium between accretion from
the IGM and gas processing in galaxies \citep{Dixon09,Prochaska09,Prochaska10}, or
the switch from ``in-situ' star formation, due to infalling gas, to ''ex-situ'' formation followed by accretion of stellar systems already formed
\citep[as expected at these redshifts by some cosmological simulations][]{Oser10}.
In some of these scenarios,
some of the hypotheses on which the gas-equilibrium models are based
\citep[e.g.,][see Sect.~\ref{sec:models}]{Dave11c,Lilly13} are no longer valid at $z>2.6$.\\

{\bf Summarizing}, DLAs follow a mass-metallicity relation which is also evolving with redshift, sign of ongoing process of chemical enrichment in these systems from high redshifts. The link with the evolution of the MZR in emission-selected galaxies is not well established, but there is similarity  between DLA and low-mass galaxies.

\subsubsection{Other classes of galaxies}
\label{sec:MZR_merging}

Special classes of galaxies have also been studied to understand the key processes affecting  chemical evolution.

Merging and interacting galaxies generally show lower metallicities than the MZR \citep{Kewley06,Michel-Dansac08,Ellison08b,Reichard09,Rupke10a,Morales-Luis11,Mouhcine11,Ellison13,Torres-Flores13,Chung13,Cortijo-Ferrero17}.
This difference is interpreted as the effect of the tidally-driven gas infalls that also produce the increase of SFR \citep{Reichard09,Rupke10b,Perez11,Torrey12,Ellison13}. 
This results can be used to study the nature of peculiar galaxies. Fox example, the agreement  with the local MZR, together with the study of galaxy dynamics and metallicity gradients, allowed  \cite{Bournaud08} to conclude that a chain-like, clumpy galaxy is not an on-going merger but is actually a clumpy disk.\\

Starburst galaxies, Ultra-Luminous Infrared Galaxies (ULIRGs) and low-redshift analogues of Lyman-Break Galaxies (LBG) are usually found to have metallicity lower than expected for their mass \citep{Liang04, Rupke08, Roseboom12, Lian15}, and this can be explained by the effect of higher SFR (see Sect.~\ref{sec:FMR}).\\

Green peas (GP) are compact, star-forming, galaxies selected for the presence of a bright \oiii5007 line \citep{Cardamone09}. As the \oiii5007 flux rises with decreasing metallicity (see Sect.~\ref{sec:measmet_strong}), these galaxies, as other line-selected samples, are preferentially selected to have low metallicities and, as a consequence, are usually below the MZR  \citep{Amorin10,Izotov11a,Xia12,Amorin12b,Ly15,Lofthouse17,Senchyna18}. Similar to the GPs, the extremely metal poor (XMP) galaxies are rare objects selected to have extreme line ratios and very low metallicities, down to a few percent solar in the local universe. They usually have low masses, high sSFR, and disturbed morphologies \citep{Izotov06a,Izotov07a,Morales-Luis11,Izotov12,Sanchez-Almeida15,Sanchez-Almeida16,Izotov18a,Izotov18b}. By definition they fall below the MZR.\\

Lyman-alpha galaxies (LAG), i.e., galaxies selected from their Ly$\alpha$\ emission, are found to have low metallicities. The brightness of the Ly$\alpha$ line in these galaxies is therefore interpreted in terms of low column density of dust associated with the low chemical abundance, which makes it easier to the Ly$\alpha$\ photons to escape \citep{Finkelstein11b, Finkelstein11a, Nakajima13, Song14,Trainor16}.\\

\begin{figure}
\centerline{
\includegraphics[width=6.0cm]{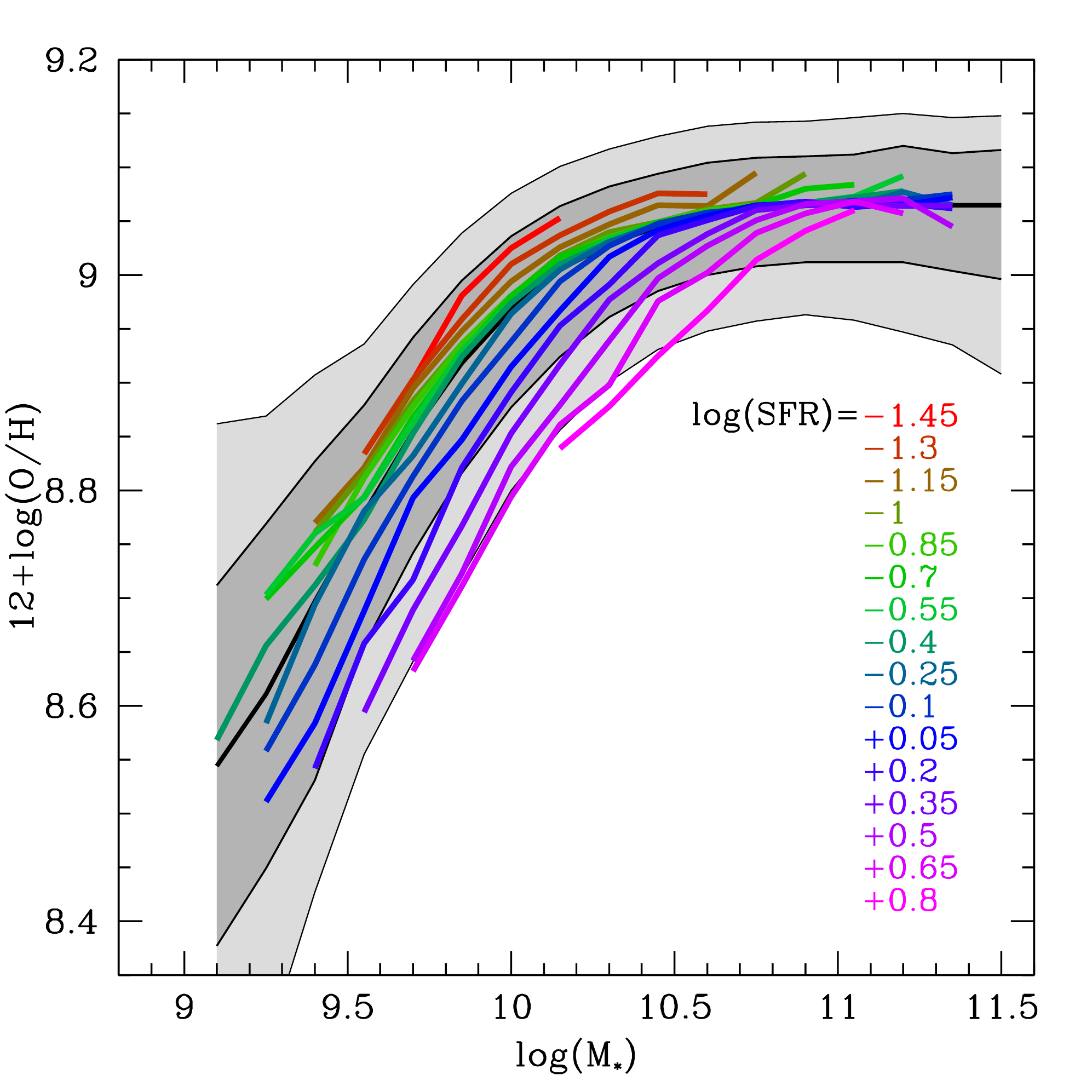}
\includegraphics[width=6.0cm]{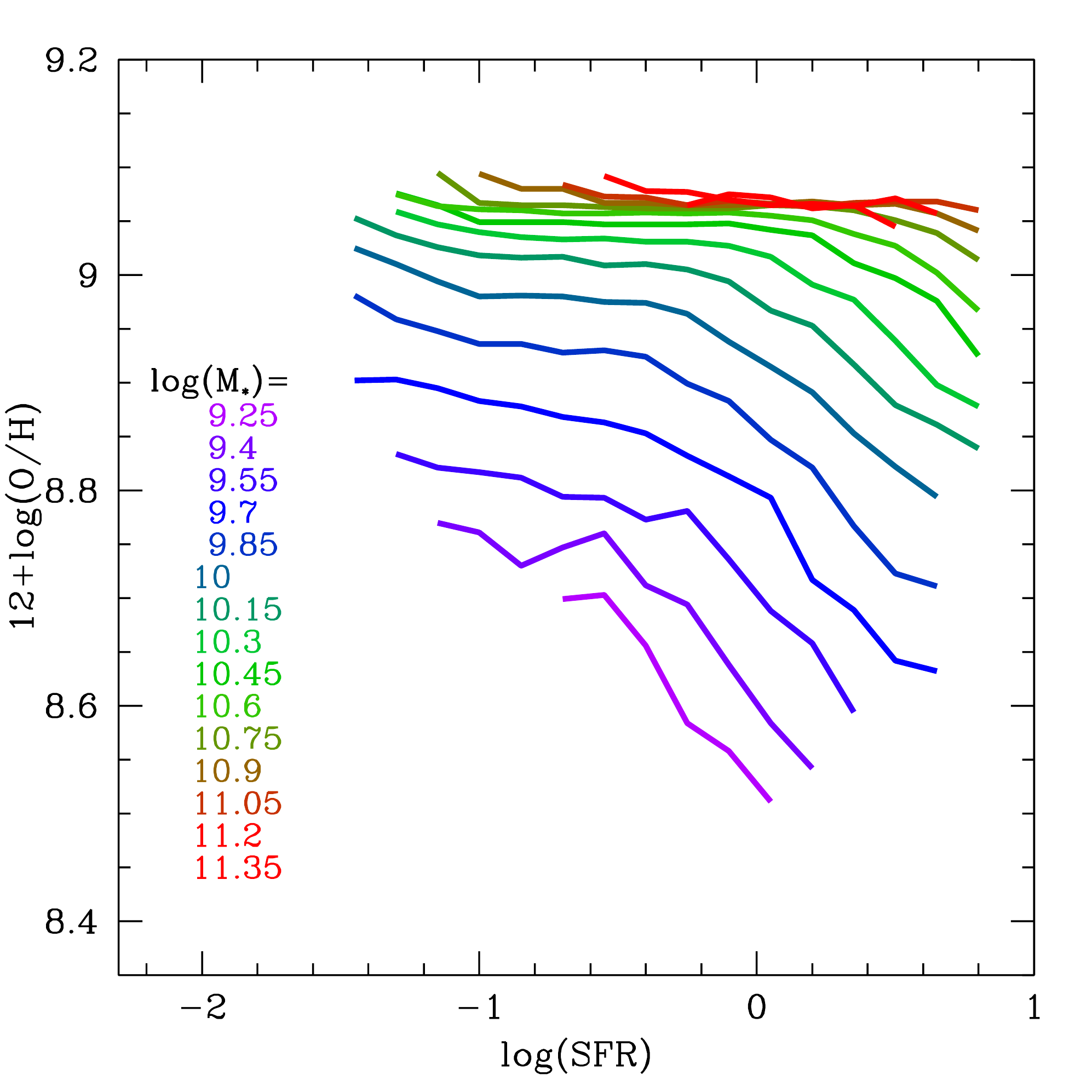}
}
\caption{The fundamental metallicity relation from \cite{Mannucci10}. Left: dependence of the gas metallicity on mass, in bins of SFR. The gray shaded areas contains 68\% and 95\% of the full, unbinned galaxy sample. Right: dependence of metallicity on SFR in  bins of stellar mass. 
}
\label{fig:FMR_mannucci10}
\end{figure}

\subsection {The dependence of metallicity on SFR and gas fraction}
\label{sec:FMR}

As soon as a sufficient precision was reached, evidences for further dependences of metallicity on other galactic properties started to appear. 
\cite{Tremonti04} reported a correlation of the metallicity residuals from the MZR with galaxy color, ellipticity, and central mass density. Soon after, \cite{Hoopes07} pointed out a dependence of the shape of the MZR on galaxy size, and \cite{Ellison08a} was the first to explicit state the dependence of metallicity on SFR for a given mass.
\cite{Mannucci10} introduced a 3D relation between mass, metallicity and SFR, named Fundamental Metallicity Relation (FMR), so that
in the local universe  the residual metallicity scatter across the median relation is reduced and becomes very low, $\sim$0.05 dex (i.e., $\sim 12\%$), 
consistent with the uncertainties of the measurements.
For a given stellar mass, metallicity decreases with SFR and sSFR, i.e., more actively star-forming galaxies have lower metallicities than more quiescent galaxies (Fig.~\ref{fig:FMR_mannucci10}). 
As introduced in Sect.~\ref{sec:MZR_stellar}, recently a dependence on SFR was also found for stellar metallicities by \cite{Faisst16}.\\

\begin{figure}[t]
\centerline{\includegraphics[width=8cm]{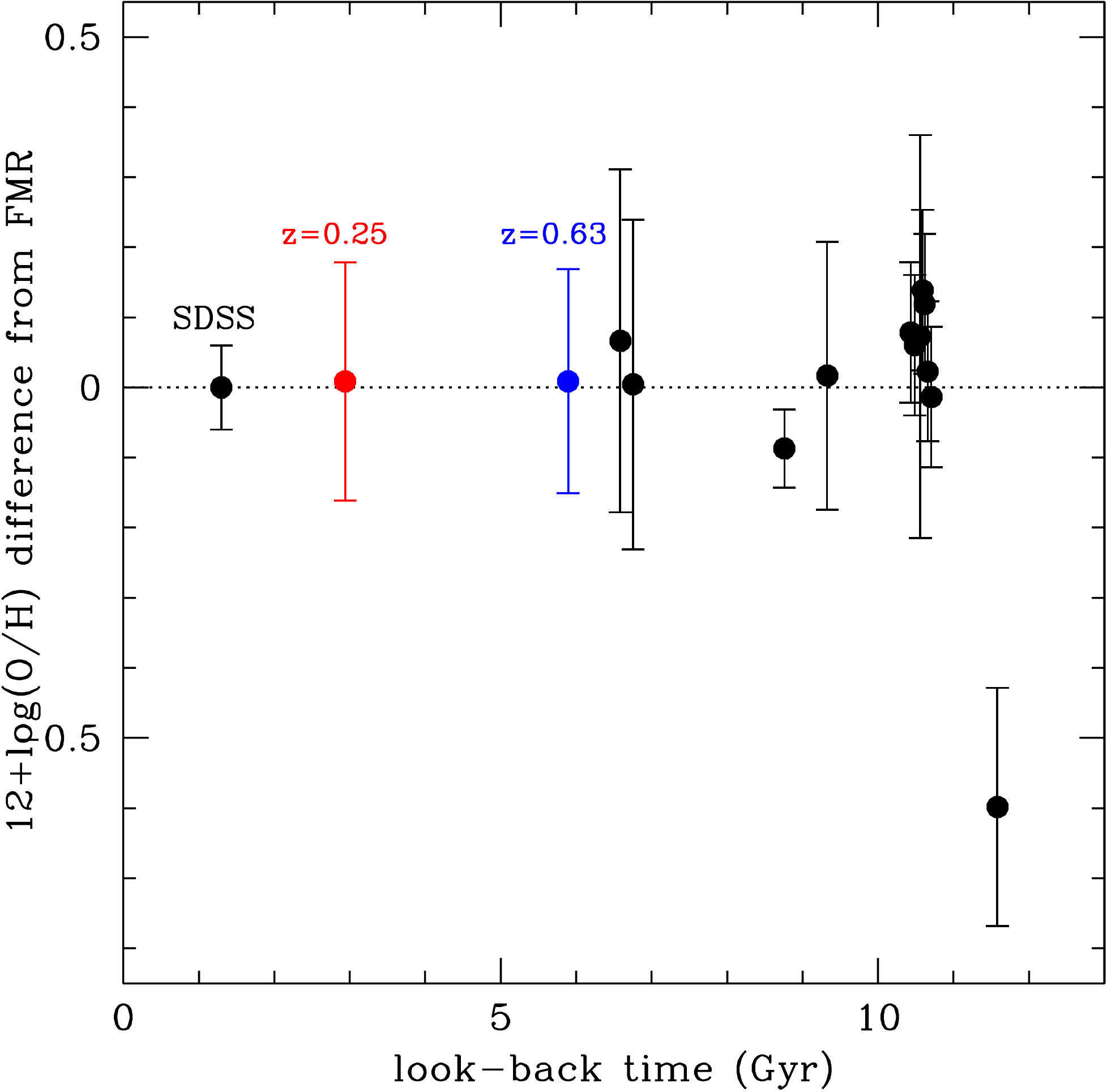}}
\caption{Difference between the metallicity predicted by the local FMR in \cite{Mannucci10} and the metallicity actually observed in a number of galaxy samples at various look-back times (lower scale) or redshifts (upper scale). This diagram shows that no evolution is observed up to $z\sim2.5$. The evolution of the MZR can be explained by the selection of galaxies with progressively higher SFR at higher redshifts. From \cite{Cresci12}.
}
\label{fig:FMR_cresci12}
\end{figure}

Apparently at odds with the clear evolution of the MZR, \cite{Mannucci10} also showed that the FMR  {\it does not evolve} with redshift up to $z=2.5$ (Fig.~\ref{fig:FMR_cresci12}). All the data at $z<2.5$ available to \cite{Mannucci10} 
\citep{Savaglio05,Shapley05a,Erb06a,Liu08a,Epinat09a,Wright09,Forster-Schreiber09,Law09,Lehnert09}
closely follow the surface defined by the local SDSS. In contrast, data at higher redshifts \citep{Maiolino08, Mannucci09b} show much lower metallicities, with a difference of about 0.6dex. This difference is discussed in Sect.~\ref{sec:FMR_evol}.

The actual shape of the local FMR depends on several factors, such as how galaxies are selected and how  mass, SFR, and metallicities are measured. As a result, relations with different shapes have been published \citep[e.g.][]{Lara-Lopez10b,Hunt12,Yates12}, while other authors \citep{Brisbin12,Nakajima14} obtained a shape consistent with that derived by \cite{Mannucci10}.
A FMR using ``direct'' metallicities, based on the \Te\ method, was derived by  
\cite{Andrews13}, who found a significantly stronger dependence on SFR than in \cite{Mannucci10}. To detect the faint auroral lines needed to measure \Te, these authors stacked SDSS spectra according to 
mass and SFR, i.e., assume that metallicity depends only (or mainly) on these two parameters.
As discussed in Sect.~\ref{sec:measmet_strong}, \cite{Curti17} derived a new \Te - based calibration using a different stacking scheme, based on similarities of the spectra 
(same \oii3727/\hb\ and \oiii5007/\hb ratios)
rather then on the galaxy parameters.  Using these calibrations, \citep{Curti19a} derived the corresponding MZR and FMR
(see Fig.~\ref{fig:MZR_curti18})
finding, as expected, metallicities lower that in \cite{Tremonti04} and \cite{Mannucci10}, a MZR similar to \cite{Andrews13}, and a strong dependence of metallicity on SFR.

\cite{Salim14}, \cite{Telford16} and \cite{Cresci19} critically re-analyzed the FMR using different metallicity indicators, various ways to measure SFR, studying the further dependence on galaxy size, and considering the SFR distance of the galaxies from the MSSF as basic parameter, finding similar results to \cite{Mannucci10}.

A number of authors have studied the dependence of metallicity on mass and SFR looking for linear relations or through the use of principal component analysis (PCA), i.e., using a technique that rotates the axes in a multi-dimensional space to minimize the scatter. Being a simple linear transformation, the PCA technique cannot really account
for more complex, non-linear correlations, such as the MZR and the FMR.
 Within the same context, using the SDSS sample \cite{Lara-Lopez10b} proposed the existence of a ``fundamental plane'' between mass, SFR, and metallicities.  In contrast to virtually all other authors, \cite{Lara-Lopez10b} use a different approach in which they derive the mass as a function of SFR and metallicity, deriving a plane with a relatively large scatter, 0.16~dex. The FMR, which has a curved shape at high masses and different dependence on SFR for different masses, can be approximated by a plane by reducing the upper limit in redshift which has the effect of lowering the number of high-mass galaxies present in the sample. The resulting plane is quantitatively very different from the FMR in \cite{Mannucci10}. This Fundamental Plane was later revised and extended toward higher masses by \cite{Lara-Lopez13}, confirming the correlation between the three quantities albeit with a larger scatter than the FMR.
\cite{Hunt12} used a composite sample of metal poor, starburst galaxies at $0<z<3.4$, including the many GPs, with metallicities measured in various ways, finding a planar correlation of metallicities with mass and SFR at any redshift.
A new PCA  was computed by \cite{Hunt16a} using a different and larger sample including a number of high-$z$ objects, testing various metallicity estimators, and comparing with the local SDSS galaxies.
The linear relation derived by \cite{Hunt16a} does not evolve with redshift. As discussed by \cite{Cresci19} it does not reproduce the local dependence of metallicity on mass and the detailed MZR at high redshifts.\\

\begin{figure}
\centerline{\includegraphics[width=9cm]{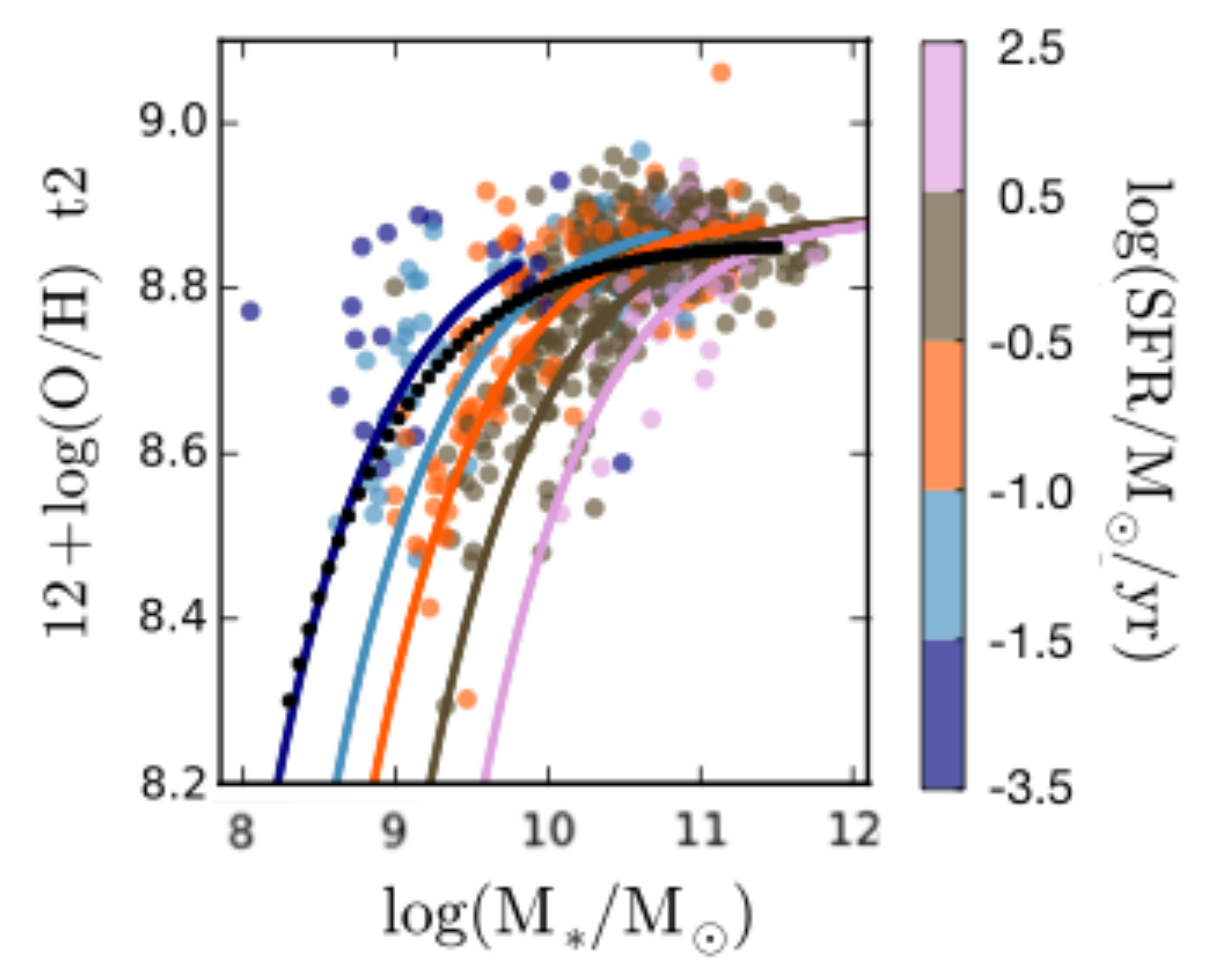}}
\caption{Dependence of metallicity on mass and SFR for the 612 CALIFA galaxies in \cite{Sanchez17}. Galaxies are color-coded with SFR as indicated in the colorbar. Lines are the best-fitting MZRs in different SFR bins.
}
\label{fig:FMR_Sanchez17}
\end{figure}

Surveys of local galaxies based on the use of two large, IFU spectrographs, CALIFA \citep{Sanchez12a} and MaNGA \citep{Bundy15}, were used to investigate the existence and properties of the FMR in these nearby, well-resolved galaxies. 
A series of papers based on these data
\citep{Sanchez13a,Hughes13,de-los-Reyes15,Barrera-Ballesteros17,Sanchez17}
have questioned the existence of the relation in this sample, and hypothesized that the dependence on SFR observed by many authors on SDSS and other data is due to aperture effects. In these works a metallicity is computed for each HII region detected, a radial gradient is computed, and the value of metallicity at the effective radius is used as representative metallicity of the galaxy.
\cite{Salim14} re-analyzed the same data finding opposite results, i.e., the presence on an anti-correlation between metallicity and sSFR, and \cite{Pilyugin13} excluded the importance of the aperture effects. 
As discussed in \cite{Cresci19} and shown in Fig.~\ref{fig:FMR_Sanchez17}, 
a clear, monotonic dependence of the metallicity on SFR is actually present in the CALIFA and MaNGA data at the expected level.
More recently also Belfiore et al. (2018, in prep.) have re-analyzed
the MaNGA data, finding a clear dependence on SFR, also on resolved
scales, as discussed more extensively in Sect.\ref{sec:gradients}.\\

In the local universe, the FMR was found to hold also for Ly$\alpha$\ local analogs \citep{Lian15} and Herschel-selected starburst galaxies \citep{Roseboom12}, with properties similar to the SDSS galaxies. This is somewhat surprising as the conditions in these galaxies are expected to be quite different from the more common galaxies dominating the objects in the SDSS.
In contrast, the central regions of barred galaxies observed in the SDSS do not follow the FMR. The radial motions of the gas produced by the bar creates a temporary central increase of SFR which is not linked to a decrease of metallicity. The metallicity is actually observed to increase, even if the amount of this increase, of the order of 0.02-0.06 dex, is debated and depends on the indicator used and on the element considered \citep{Ellison11,Cacho14}. This metallicity increase is explained with accretion of metal-rich gas toward the central regions of the galaxies \citep{Martel13,Martel18}.
As noted in Sect,~\ref{sec:MZR_merging}, interacting galaxies show lower metallicities and higher SFR, in qualitative agreement with the FMR \citep[e.g.][]{Ellison13}. Nevertheless they tend to have large metallicity dispersion and some offset from the FMR toward lower abundances \citep{Gronnow15,Bustamante18}.\\

{\bf Summarizing}, the gas-phase metallicity  not only correlates with mass but also anti-correlates with SFR, i.e. more star forming galaxies show lower metallicities. The actual shape of this relation depends on how galaxies are selected and mass, SFR and metallicities are measured. This relation seems not to evolve up to z=2.5, as detailed in the next section.

\subsubsection{Redshift evolution of the FMR}
\label{sec:FMR_evol}

The issue of the evolution of the FMR, or lack thereof, with redshift has been the subject of much work, both observational and theoretical. The dependence of metallicity on both mass and SFR was measured in many galaxy samples at various redshifts, and the results compared with the predictions based on the local FMR in \cite{Mannucci10}.

Many works have found good agreement between the local FMR and the metallicities observed
at $z\sim1$ \citep{Yabe12,Cresci12,Stott13,Henry13a,Henry13b,Yabe14a,Maier15a,Maier15b,Calabro17} 
and at $z\sim2$ \citep{Nakajima12,Belli13a,Nakajima14,Maier14a,Stott14,Song14,Yabe15,Salim15,Kacprzak16,Wuyts16,Sanders18,Hirschauer18}, i.e. no redshift evolution of the FMR.
Lensed galaxies also played an important role in studying the evolution of the FMR over a larger redshift range and toward lower masses and lower SFR, usually finding good agreement with the FMR, at least for the average metallicity level
\citep{Richard11,Christensen12b,Wuyts12}.
Some authors confirm the anti-correlation between metallicity and SFR, but with a weak dependence or with some evolution with respect to the local FMR \citep{Niino12b,Perez-Montero13,Grasshorn-Gebhardt16,Cullen14,Zahid14b,Salim15,Wu16}.
Other works do not have the sensitivity investigate the evolution of the FMR \citep{Divoy14,de-los-Reyes15}.

\begin{figure}[t]

\centerline{\includegraphics[width=10cm]{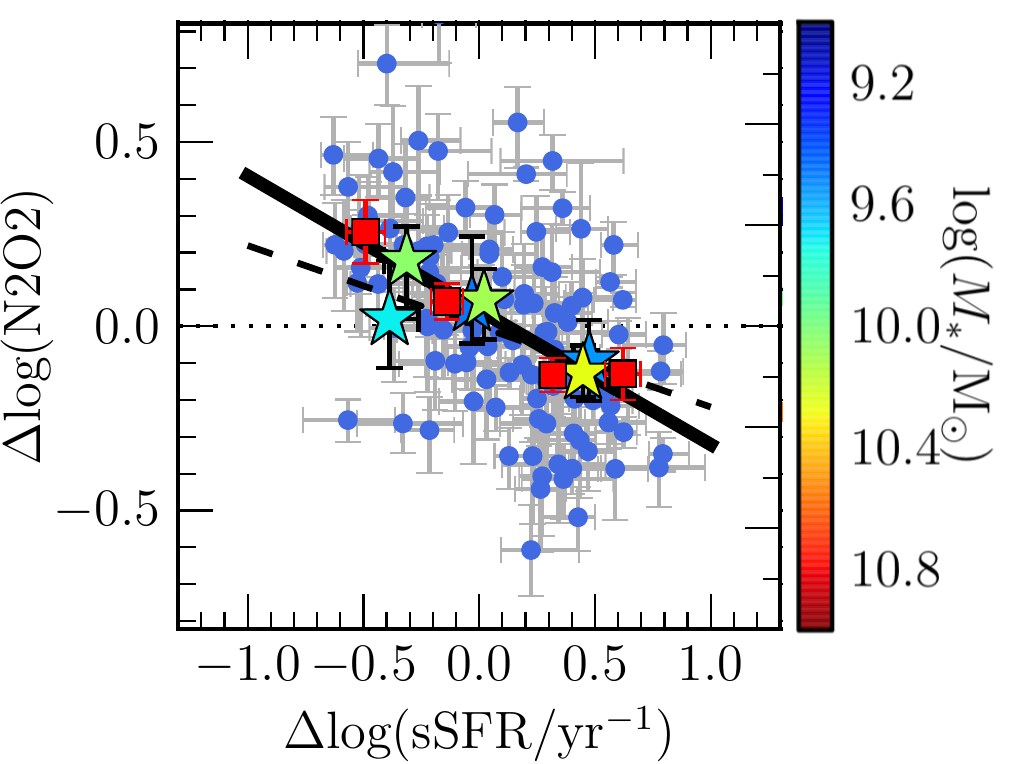}}
\caption{Observed dependence of metallicity on the sSFR at $z=2.3$ from \cite{Sanders18}.
The plot shows the  metallicity difference $\Delta$log(O/H) from the mean MZR as a function of the sSFR difference $\Delta$sSFR from the MSSF. The metallicity shown is computed with the N2O2 strong-line indicator, and the original paper also analyzes other indicators, finding similar results.  
Blue points are individual galaxies, red squares show medians in bins of $\Delta$log(sSFR),  stars display the results from stacked spectra, color-coded by stellar mass. The solid line shows the best fit to the medians, and the dashed line shows the prediction from the cosmological simulations of \cite{Dave17b}. A clear dependence of the residuals with sSFR is seen. 
}
\label{fig:FMR_sanders18}
\end{figure}

Several works on high redshift galaxies \citep{Wuyts12, Steidel14, Yabe14a, Sanders15, Yabe15, Wuyts14b, Onodera15,Wuyts16} do not find a clear dependence of metallicity on SFR within their data. 
It should be considered that
the comparison of the  metallicity properties of distant galaxies with the FMR of the local (SDSS) galaxies must take into account two separate issues. The first issue is whether the local FMR is able to predict the {\it average} metallicity of a galaxy sample with a given average mass and average SFR, i.e, whether  high-redshift galaxies on average follow the local FMR or not. The second issue is whether within the high-redshift galaxy sample alone it is possible to detect the metallicity dependence on the SFR as in the local universe.  
The latter issue is considerably more difficult to tackle than the former because it requires accurate mass, SFRs and metallicities for each galaxy, with final uncertainties smaller than the expected dependence of metallicity on SFR, and a large dynamic range of SFR.
Measurements of the SFR at high-$z$ are often quite uncertain due the effect of dust extinction. Similarly, accurate metallicities are often difficult to obtain at high-$z$.
Moreover, at high-$z$ it is difficult to achieve a
large dynamical range in mass and SFR, especially at low-masses where the dependence of metallicity on SFR is larger. 

About the first issue, \cite{Cresci19} has shown that the papers quoted above that do not find an internal dependence of metallicity with SFR are actually confirming the predictions of the FMR in terms of average metallicity. This is a remarkable result given that the FMR is only based on local galaxies.
About the second issue, a few works report the detection of the dependence of metallicity on SFR also within the high redshift data alone.
\cite{Zahid14b}, \cite{Salim15}, \cite{Kacprzak16} and \cite{Kashino17a} find lower metallicities in galaxies with higher SFR at $z\sim2$, albeit with uncertainties due to the low S/N, the presence of galaxies with no detections in the faint lines, and the large metallicity scatter. Finally, in the context of the MOSDEF survey \cite{Sanders18} investigated the evolution of the FMR using a large sample (260 objects) of star-forming galaxies at z$\sim2.3$  using several emission-line ratios. In contrast to earlier works by the same group based on smaller data sets \citep{Sanders15}, this study detected the explicit dependence of metallicity with SFR for a given stellar mass within the data sample itself without the need of a comparison with lower redshift samples, see Fig.~\ref{fig:FMR_sanders18}. The level of metal enrichment of galaxies with a given stellar mass and SFR is similar to that in the local universe, being only $\sim0.1$ dex lower, consistent with the no-evolution upper limit derived by \cite{Cresci19}. \\

As mentioned, at $z>2.5$ various works have found that galaxies deviate from the FMR by being more metal poor than expected
\citep{Mannucci10,Troncoso14,Onodera16}, at least in the redshift range $2.8<z<3.6$ where optical metallicity diagnostics are still detectable in the K-band. 
The observed difference, albeit large ($\sim$0.6dex) should be taken with care: only a few galaxies have reliable metallicities measured at these redshifts, these galaxies often have very high SFRs and could be in special phases of their evolution, metallicity is usually measured using only one indicator (often R23), and the N2 indicator, which is most often used at lower redshift,  is no longer available.
Nevertheless is interesting to note that this redshift range $z\sim3.5$ where the FMR appears to break down is the same range where the DLA scaling relations seem to change behavior \citep{Moller13} and where the cosmic density of HI start to evolve rapidly
with respect to the non-evolution at lower redshifts \citep{Prochaska09}. As mentioned, this is the epoch beyond which galaxies may no longer be in equilibrium, i.e., at $z>2.5$ probably
the infall of (pristine) gas on most galaxies is too fast for them to
efficiently transform it into stars, hence most of the accreting gas is being accumulated and resulting into a larger dilution of the metals in the host galaxy.\\

It would be interesting to extend these studies to even higher redshifts. The use of (rest-frame) optical diagnostics shall await JWST. However, the UV nebular lines of CIII]1909, CIV1549 and OIII]1808, although weak, offer an alternative potential tool to trace the gas metallicity out to very high redshift. Interestingly, the few current detections of these lines at high redshift indicate moderately low metallicities ($z\sim0.1$) even out to $z\sim7$ \citep{Stark17}, while at such high redshift one would expect lower levels of metal enrichment, at least for `normal' galaxies. Yet, one must take into account that these early detections may be biased toward higher metallicities.

It is interesting that far-IR fine structure atomic transitions, such as [CII]158$\mu$m and [OIII]88$\mu$m, redshifted into the millimeter bands of atmospheric transmission, are increasingly being used to trace
the metal enrichment in `normal' galaxies even at $z>7$ 
\citep{Maiolino15,Pentericci16,Inoue16,Carniani17,Carniani18,Hashimoto18a,Tamura18}
and out to $z\sim9$ \citep{Hashimoto18a}. In some of the galaxies the very high [OIII]88$\mu$m--to--[CII]158$\mu$m ratio is interpreted as indication of low metallicity \citep[$z\sim0.1$,][]{Inoue16}, although this interpretation is subject to degeneracies, as the same high ratio can also be interpreted in terms of ionization parameter \citep{Carniani17,Katz17}.
Very interestingly, for some of these galaxies the thermal dust continuum is also detected \citep{Hashimoto18b,Tamura18} which implies a significant amount of dust, and therefore of metals, already in place at such early epochs. Some of these authors infer metallicities as `high' as Z$\sim$0.2 at $z\sim8$, which requires a very fast and efficient enrichment process in these low mass ($\rm M_{star}$ a few times $\rm 10^9~M_{\odot}$) primeval galaxies.\\

{\bf Summarizing}, most of the works confirm or are consistent with the existence of the FMR at high redshift, and find no or a very limited evolution of this relation with redshift up to $z=2.5$. The situation at even higher redshift is unclear but there are several indications of evolution during the first $\sim$2 Gyrs after the Big Bang. This stability of the FMR is an important observation that models must reproduce, as discussed in the next section.


\begin{figure}
\centerline{\includegraphics[width=10cm]{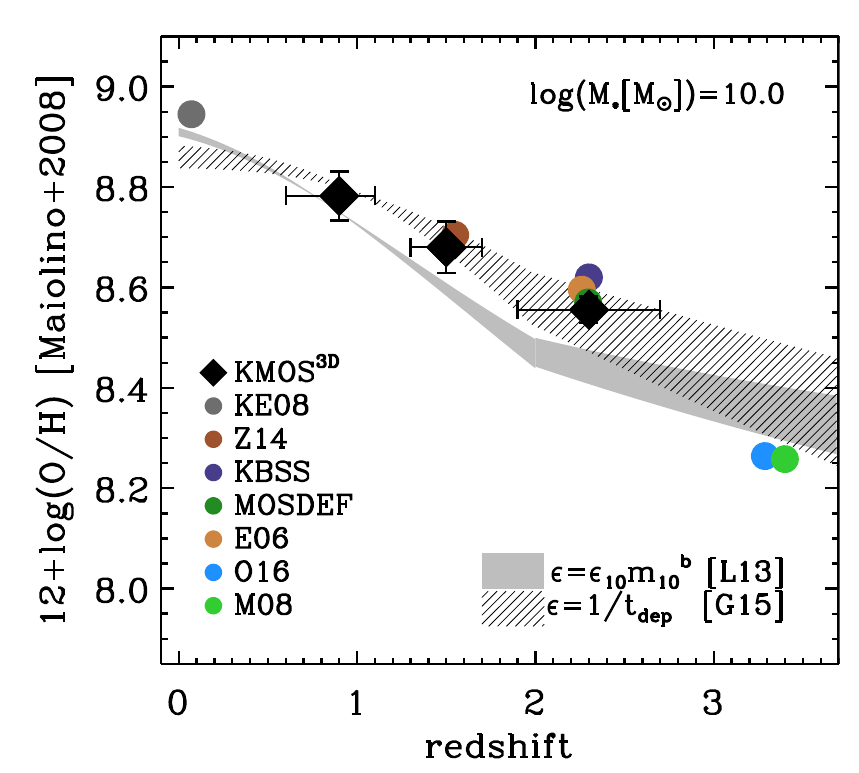}}
\caption{Redshift evolution of the MZR at a stellar mass $M=10^{10}$\msun from \cite{Wuyts16}. The gray shaded areas are the predictions of the \cite{Lilly13} model with the variable depletion time by \cite{Genzel15}, calibrated on the local FMR by \cite{Mannucci10}. 
}
\label{fig:MODELS_wuyts16}
\end{figure}

\begin{figure}[t]

\centerline{\includegraphics[width=8cm]{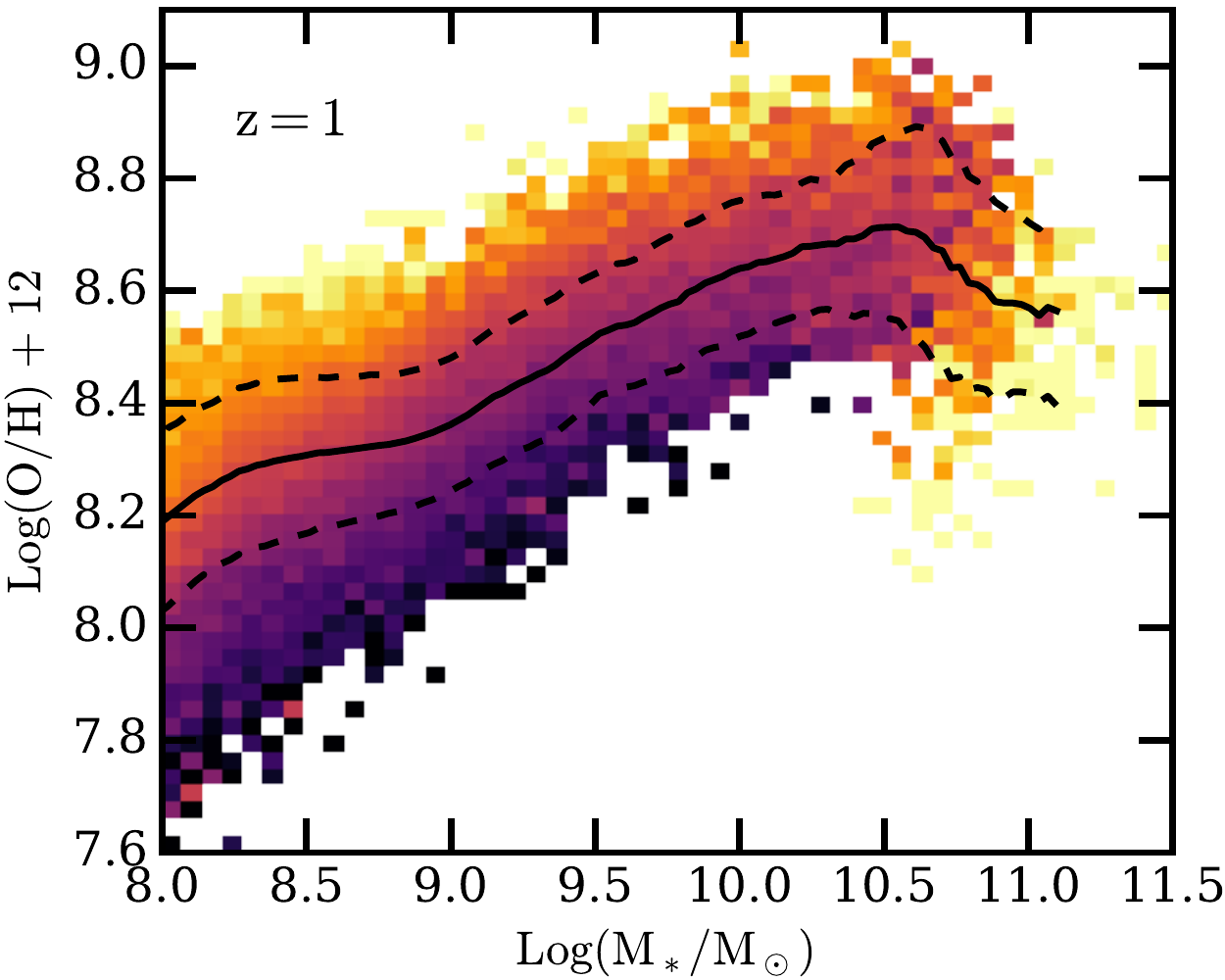}}
\caption{Dependence of metallicity on mass and sSFR produced at z=1 by the IllustrisTNG simulation from \cite{Torrey18}. Solid and dashed black lines indicate the median MZR and one sigma scatter, respectively. Colors show how the residuals about the MZR are correlated with SFR. This model produces a low gas retention in the ISM in the local universe, where $\sim 85\%$ of the metals are outside the ISM. 
}
\label{fig:MODELS_torrey18}
\end{figure}

\subsubsection{Origin of the FMR}
\label{sec:FMR_origin}

The origin of the dependence of metallicity not only on mass but also on SFR is debated. \cite{Ellison08a} proposed a varying star formation efficiency as the most probable explanation for the dependence of metallicity on SFR and galaxy size. \cite{Mannucci10} proposed that the interplay of infall of metal poor gas and star formation
may play a central role in shaping the FMR: on the one hand, infall provides chemically poor gas, lowering metallicity; on the other hand gas accretion delivers additional fuel for
star formation, hence enhancing the SFR.  As such, the FMR is yet another piece of evidence for the importance and ubiquity of cold gas accretion as a dominant driver of galaxy evolution.
As presented in Sect.~\ref{sec:MZR_merging}, the same kind of explanation had been previously proposed to explain the deviations from the MZR of several classes of galaxies, such as interacting systems, ULIRG and Green Peas. The spatially-resolved inverse dependence between SFR and metallicity observed within  galaxies, both in the local universe  \citep{Sanchez-Almeida13,Sanchez-Almeida14,Sanchez-Almeida18a,Belfiore18} and at high redshift \citep{Cresci10} is also naturally interpreted as the effect of the infall of chemically poor gas, as discussed in Sect.~\ref{sec:gradients}.

In a more consistent and quantitative way, the FMR is predicted or explained by
the gas-equilibrium models described in Sect.~\ref{sec:MZR_models}. 
This is in agreement with the observations that local galaxies that are expected to be far from equilibrium, such as interacting pairs and barred galaxies, show quantitative disagreement wit the FMR.
Some models \citep[e.g.,][]{Lilly13} predict no evolution of the FMR because the equilibrium is based on basic physical processes that can remain stable with cosmic time, while others \cite[e.g.,][]{Dave11c} expect a mild evolution with redshift due to the progressive enrichment of the CGM. The observational results are not precise enough to distinguish between these two scenarios in a robust way \citep{Sanders18, Cresci19}. 
Gas infall, with a super-linear dependence of SFR on infalling mass, is the explanation proposed by \cite{Brisbin12}.
In the context of the equilibrium models, \cite{Wuyts16} modified the model by \cite{Lilly13} by introducing the evolving gas depletion time obtained by \cite{Genzel15}, obtaining a good fit to the evolution of the MZR starting from the local FMR, see Fig.~\ref{fig:MODELS_wuyts16}.
The modest evolution of $\sim0.1$ dex of the FMR found by \cite{Sanders18} at $z\sim2.3$  can be interpreted either as a limited but systematic increase in the mass loading factors of the outflowing gas for a given stellar mass, or with the decrease with redshift of the average metal abundance of the inflowing gas \citep{Dave11c}.

\cite{Lagos16}, \cite{De-Rossi17}  and \cite{De-Rossi18} find a very good match between the observations of the FMR and their hydrodynamical codes up to high redshift, and propose gas fraction as the parameter that is better correlated with metallicity (see next section). Critical parameters to reproduce the observations are the properties of stellar and AGN feedback and the dependence of star formation on metallicity. In the IllustrisTNG simulation \cite{Torrey18} find the FMR up to high redshift 
(see Fig.~\ref{fig:MODELS_torrey18})
and ascribe its origin to the similar timescales of the evolution of SFR and metallicity. This opens up the interesting possibility of using the observed scatter across the FMR to test the existence of variations of the properties of high redshift galaxies on timescales shorter than cosmic evolutionary times.

Differences among the various chemical elements can also have a role in defining and explaining the shape of the FMR. Recently, \cite{Matthee18} computed the dependence of the abundance of various elements on mass and SFR in the results of the EAGLE simulation. 
They found a FMR for the abundance of every element, and also a 3D relation between mass, SFR and $\alpha$/Fe.

They found a FMR for the abundance of every element, with a a smaller dispersion for the $\alpha$-elements, and they also find a 3D relation between mass, SFR and [O/Fe].\\

Interestingly, \cite{Perez-Montero13} and \cite{Kashino16} showed that the dependence of metallicity on mass seen in SDSS galaxies when using R23 or N2, as in \cite{Mannucci10}, is not present when the N2S2 or N2S2H$\alpha$\ indicators \citep{Dopita16} is used. 
\cite{Telford16} made a critical analysis of these studies, finding that, in contrast, the FMR is also present in the N2S2 indicator, albeit with a weaker dependence on SFR.
If confirmed, this effect would be in agreement with the explanation of the FMR as due to infall of metal-poor gas: N2S2 is based on the ratio between two metal lines, therefore is not sensitive to dilution due to metal-free gas as both abundances go down simultaneously. In contrast, N2 and R23 are based on ratios between metal, collisionally-excited lines and H recombination line, therefore  sensitive to dilution. \\

\begin{figure}

\centerline{\includegraphics[width=9cm]{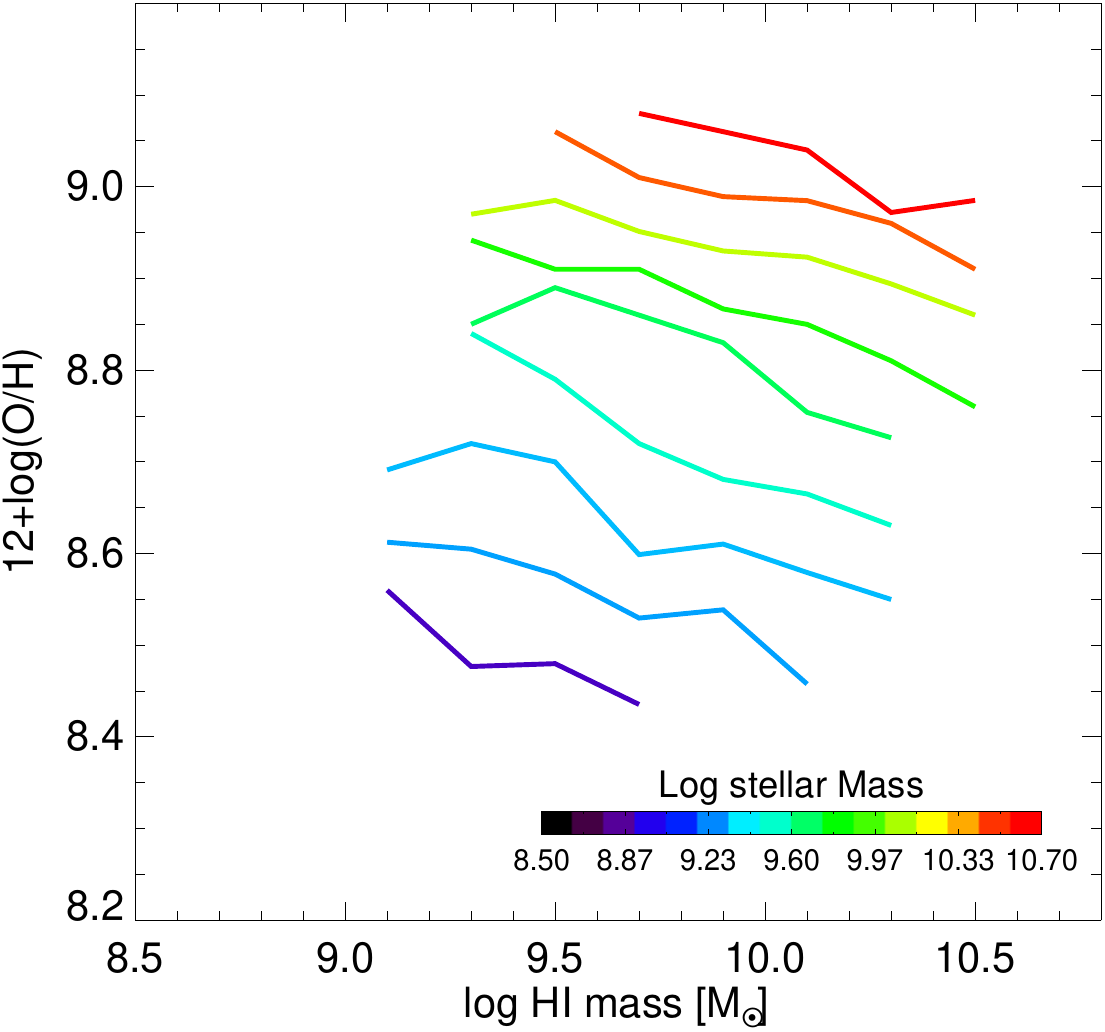}}
\caption{Observed dependence of gas-phase metallicity on atomic (HI) gas mass, in bins of stellar mass, for the SDSS galaxies, from \cite{Bothwell13}
}
\label{fig:FMR_bothwell13}
\end{figure}

\subsubsection{The relation between metallicity and gas content}
\label{sec:FMR_gas}

According to the explanations of the FMR given by the gas-equilibrium models, the link between metallicity and SFR is actually due to two relations: the increase of SFR with gas content (or, equivalently, gas fraction, $f_{gas}$), and  the decrease of metallicity with gas content (or $f_{gas}$) due to dilution. If this is true, a direct dependence of metallicity on gas content ( or $f_{gas}$) is expected, and several models obtain this relation as the most fundamental \citep{Lagos16b,Segers16,De-Rossi17}.

\cite{Peeples08} and \cite{Peeples09} showed that
such a correlation is present in the local universe, in the sense that SDSS galaxies with low gas-fractions also have metallicities above the MZR. 
In the formalism of the FMR, \cite{Bothwell13}, \cite{Hughes13}, \cite{Lara-Lopez13}, \cite{Jimmy15}, and \cite{Brown17} showed that metallicities anti-correlate with the atomic gas mass, and that this correlation is tighter than with SFR. Interestingly, the dependence of metallicity on HI gas fraction persists at high stellar mass, where metallicity does not depend on SFR any more \citep[see Fig.~\ref{fig:FMR_bothwell13}][]{Bothwell13}. \cite{Bothwell16} and \cite{Bothwell16b} showed that an even tighter correlation can be present with molecular hydrogen, more directly related to the SF activity, although the statistics for the molecular gas in galaxies (for which information on the metallicity is
available) are still poor, hence the result needs to be confirmed
with larger sample of galaxies having measurements for both the
gas metallicity 
and for the molecular gas content.

\subsection {Metallicity dependence on environment}
\label{sec:MZR_env}

The gas-phase MZR shows also some dependence on environment. In the local universe gas-phase metallicities are observed to correlate weakly with environment, with higher values in denser environments 
 \citep{Shields91,Skillman96,Mouhcine07,Cooper08b,Ellison09,Petropoulou11, Petropoulou12, Kulas13, Hughes13, Pilyugin17a, Wu17}. 
 However, the environmental effects are clearly seen only if satellite and central galaxies are considered separately. Indeed,
central galaxies do not show significant correlation
with the environmental density, while satellite galaxies do show, at a given stellar
mass, a significant metallicity dependence on environment, i.e.
satellite galaxies in denser environments are characterized by higher metallicities \citep{Pasquali12,Peng14,Peng15,Trussler18,Maier18}. 
In galaxy clusters the metallicity
of satellite galaxies seems to also correlate with the stage of accretion  of
galaxies into the cluster \citep{Maier16}.
These trends have been interpreted as a combination of ``strangulation'' (i.e. satellite galaxies being prevented from  accreting cold, near-pristine gas as they plunge into the hot halo of massive environments, see Sect.~\ref{sec:MZR_stellar}), ram-pressure stripping, higher metallicities in the infalling gas, 
and external pressure reducing the amount of metals lost to the CGM due to gas re-accretion \citep{Pasquali10, Peng14, Peng15, Spitoni15b, Segers16, Pilyugin17a, Trussler18}. The correlation is probably also mediated by the gas fraction, which is higher in galaxies in low-density environments \citep{Wu17}.
Based on the results of the Illustris simulation, \cite{Genel16} attributes the metallicity difference partly to the different SF history, partly to the smaller size of forming disks, biasing star formation toward the inner, more metal rich parts. In contrast, \cite{Bahe17} see the effect in the EAGLE simulation and identify gas stripping and suppression of metal-poor infalls as the main drivers of the effect.

At high redshift the situation if even more uncertain, with contrasting results and, in general, little evidence for the existence of any environmental dependence \citep{Magrini12a,Kulas13,Williams14,Shimakawa15a,Kacprzak15,Valentino15}.
However, the statistics in these high-$z$ studies are still poor, while
it is well known that large statistics are required to identify
the role of the environment and disentangle it from mass segregation effects.

{\bf Summarizing,}
the galaxy metallicity (and the metallicity scaling relations) does not depend significantly
on environment for central galaxies, while satellite galaxies tend to have
 higher metallicities in denser environments. The latter phenomenon is possibly
 associated with the ``strangulation/starvation'' of galaxies as they plunge into hot halo of
 massive environments, but other phenomena may also play a role.

\begin{figure}[t]

\centerline{\includegraphics[width=9cm]{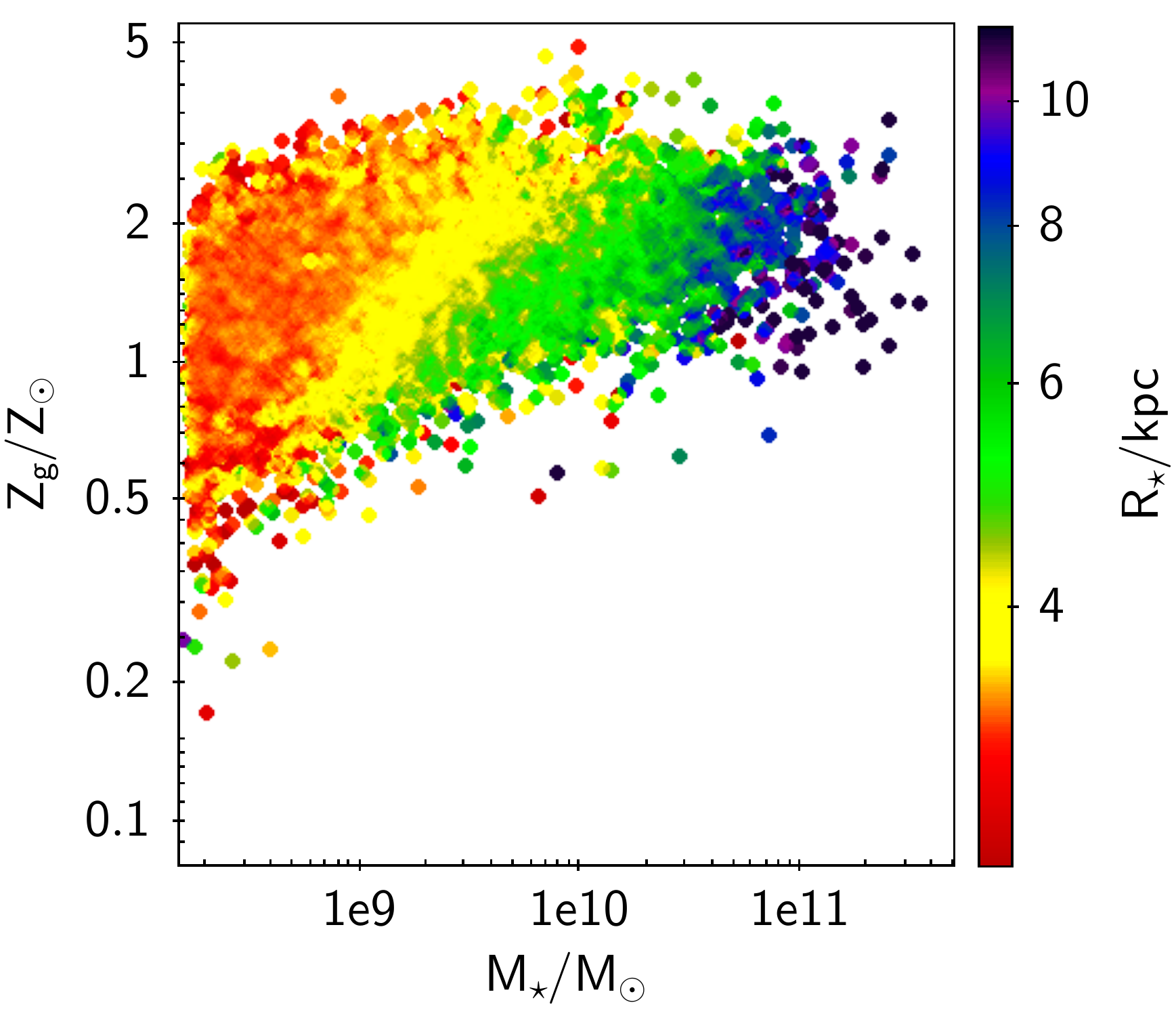}}
\caption{Predicted dependence of gas-phase metallicity on stellar mass for galaxies of different size from the EAGLE cosmological simulations, from \cite{Sanchez-Almeida18b}.}
\label{fig:FMR_sanchezalmeida18}
\end{figure}

\subsection{Metallicity dependence on other physical properties}
\label{sec:FMR_furtherdep}

Besides SFR and gas fraction, other dependencies have been proposed to further reduce the scatter of the MZR or explain the evolution of metallicity with respect to the FMR at $z>2.5$. 

\cite{Hoopes07} and \cite{Ellison08a} found a dependence of gas metallicity on size, according to which, at fixed mass, more compact galaxies are more metal rich.  The actual difference between the more compact and more diffuse galaxies depends on mass and is about 0.05--0.2dex. This relation was later analyzed by \cite{Brisbin12}.
\cite{Wu15} and \cite{Hashimoto18c} found similar dependence of metallicity
on quantities related to size, i.e., respectively, galaxy surface brightness and stellar surface density.
(see also Sect.~\ref{sec:gradients}).
\cite{Yabe12} found a similar relation at $z\sim1.4$ albeit with a large scatter.

\cite{Sanchez-Almeida18b} investigated the origin of this effect by using the EAGLE cosmological simulations (see Sect.~\ref{sec:models}). The relation between mass, metallicity and size is present in the results of the simulation (see Fig.~\ref{fig:FMR_sanchezalmeida18}), together with the FMR. They found that galaxies whose last major episode of gas accretion happened earlier are now both smaller and more metal rich. Also \citep{Ellison08a}, \cite{Brisbin12}, and \cite{Wu15} relate the difference in metallicity to the different consequences of gas accretion in galaxies of different sizes.


The dependence on both mass and radius was further discussed by \cite{DEugenio18} by studying the relation of metallicity with average gravitational potential $(\Phi \equiv M_*/R_e)$ and 
average surface mass density $(\Sigma \equiv M_*/R^2_e)$. They found a more direct relation of metallicity with $\Phi$, in agreement with the equilibrium model and, in particular, with the expected dependence of metallicity on escape velocity, giving a central role to metal losses due to galactic winds. The dependence on observing aperture is discussed in Sect.~\ref{sec:gradients_scalingrel}.\\

As discussed in Sect.~\ref{sec:measmet_strong_BPT}, the ionization parameter $q$ is an important quantity to understand the nebular spectra of galaxies. This parameter shows some correlation with stellar mass and SFR \cite[e.g.,][]{Dopita06,Brinchmann08,Nagao06}.
Its influence is often studied placing the galaxies in a plane defined by R23, mainly sensitive to metallicity (but with secondary dependence on $q$), and O32, sensitive to $q$ (but with a secondary dependence on metallicity 
\citep[e.g.,][see Sect.~\ref{sec:measmet_strong} for the definitions]{Kewley02a,Nakajima13,Shapley15}.
Based on a sample of local and high-$z$ galaxies, \cite{Nakajima14} analyzed the relation between the residuals from the MZR and the ionization parameters  as measured by O32 and by photoionization models. On the base of a correlation between $q$ and the main integrated parameters of the galaxies, i.e., mass, SFR and metallicity, they extended the FMR into a four-dimensional relation including $q$, the ``Fundamental Ionization Relation'' (FIR). This relation is able to account for the lower metallicities at $z>2.5$.

\section{Metallicity gradients in galaxies} 
\label{sec:gradients}

\subsection{Overall properties of metallicity gradients in galaxy discs}
\label{sec:gradients_general}

Since the first seminal works by \cite{Aller42}, \cite{Searle71},
and \cite{Pagel81}, 
the investigation of gradients of metallicity in galactic discs
has been subject to continuously growing interest, especially
in recent years thanks to the developments of new facilities and
extensive surveys that have enabled the measurement of
gradients in detail using multiple tracers and also over large
samples of galaxies.

\begin{figure}
\centerline{
\includegraphics[width=10cm]{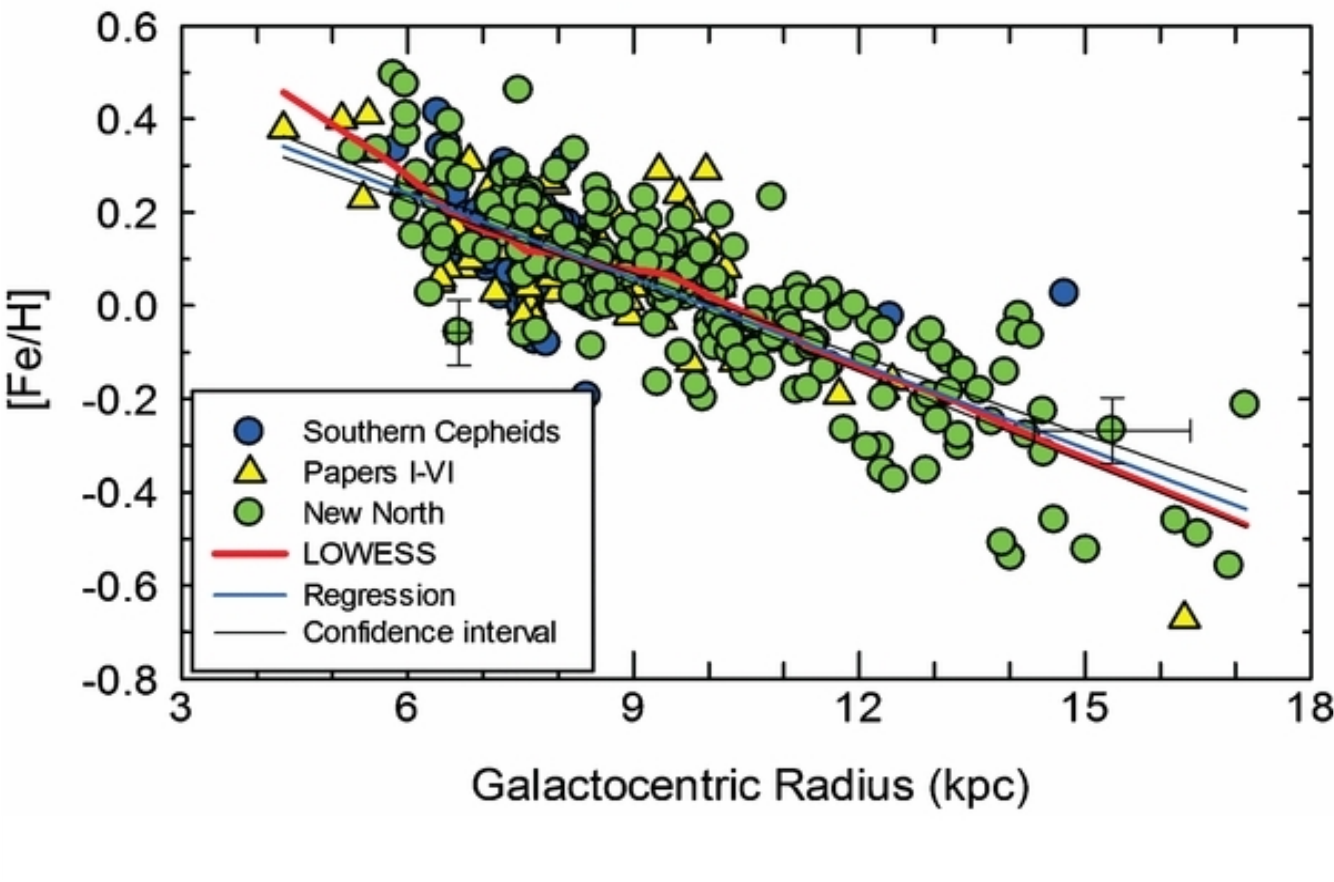}
}
\caption{Metallicity gradient of the MW based on Cepheids measurements
\citep{Luck11}.}
\label{fig:grad_Ceph_Luck2011}
\end{figure}

In the local Universe the general finding is that, at least within
the optical radius, in most spiral galaxies the metallicity
decreases exponentially with galactocentric radius.
As a consequence, gradient of the metallicity in
its logarithmic form (e.g. 12+log(O/H))
is linear with radius, and can be expressed in units of $\rm dex~kpc^{-1}$.
In the case of the Milky Way, typically radial metallicity gradients
range between $-0.06$ and $-0.01\ \rm dex~kpc^{-1}$, depending on the metallicity
tracer adopted, as discussed in the following.

Indeed, different metallicity tracers have been used to investigate the metallicity
gradients, providing different kind of information, especially because
they trace the enrichment at different cosmic times, hence their comparison
can provide precious information on the evolution of the gradients with time.

HII regions are used to trace the current metallicity of
the gaseous component of galactic discs. The preferred method is
obviously through the ``direct''-T$_e$ method (see Sect.~\ref{sec:measmet_ism_Te}), 
which however, due
to the weakness of the auroral lines, can
be properly mapped only in the MW \citep{Deharveng00,Esteban05,Rudolph06,Balser11}
and in a few nearby galaxies \citep[e.g.,][]{Bresolin07a,Bresolin09,
Bresolin12,Werk11,Berg12,Berg13,Berg15a}. The strong line method
has been extended to probe much larger samples of galaxies, as discussed later
on, but with the uncertainties associated with the latter method, as discussed
in Sect.~\ref{sec:measmet_strong}.

Massive stars are also an alternative, accurate method to probe the metallicity
of the gas out of which these stars have recently formed. Also in this case,
the difficulty of obtaining the high quality and (high resolution) spectra
required to measure the abundances has limited this method to the MW
and a few other nearby galaxies \citep[e.g.,][]{Daflon04,Bresolin07b,Davies15}, see the discussion in Sect.~\ref{sec:measmet_comparison}.

\begin{figure}
\centerline{\includegraphics[width=9cm]{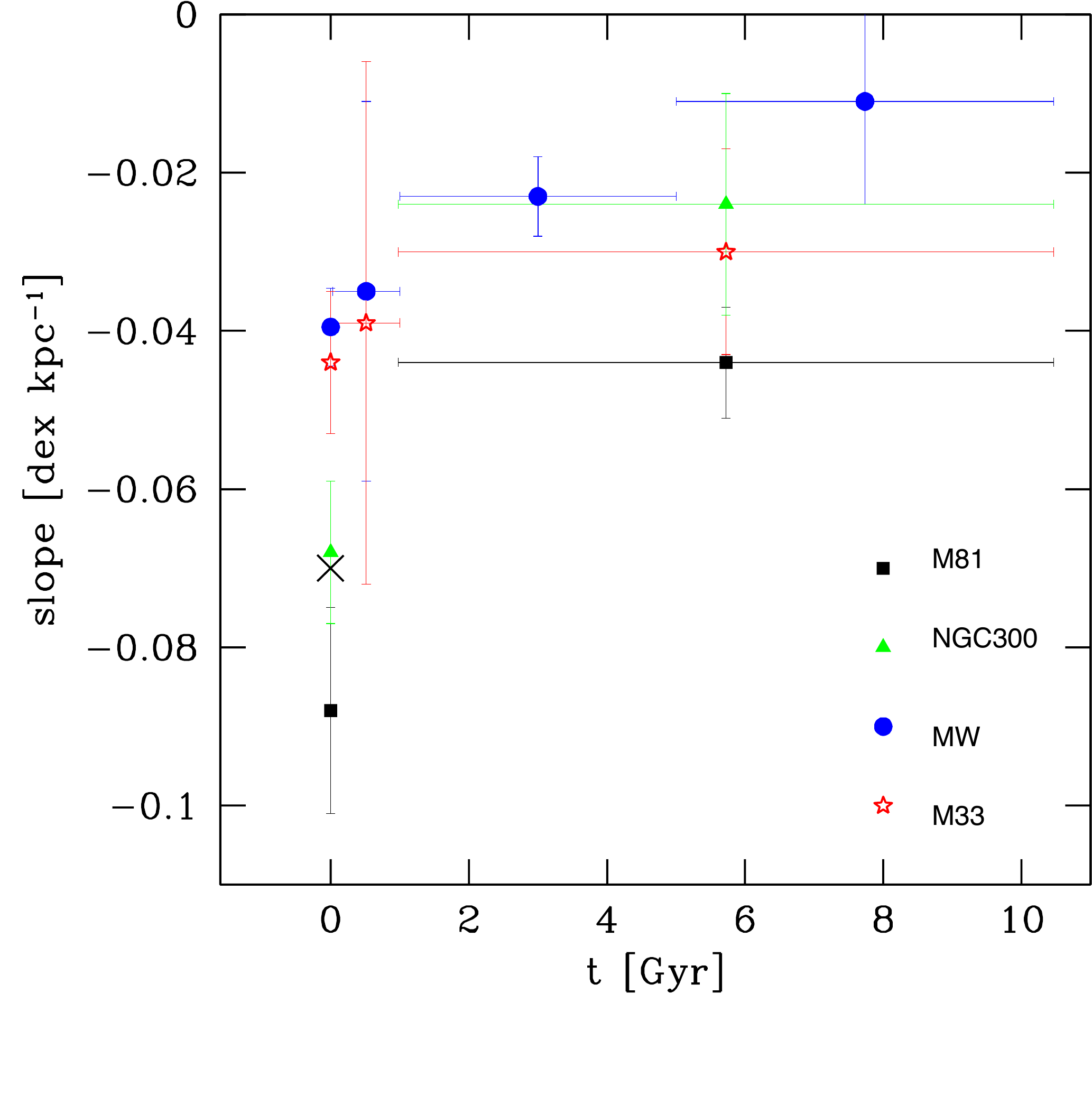}}
\caption{Evolution of the metallicity gradient in the MW and in a few additional galaxies
as a function of lookback time, based on different metallicity tracers probing the gas phase at different epochs
\citep{Stanghellini14}.
This diagram shows that gradients become steeper, more negative, as time flows.
}
\label{fig:grad_compil_Stanghellini2014}
\end{figure}

Cepheids have also been extensively used as a tool to investigate the current metallicity
gradient, especially in the MW, as in their case, besides being very luminous, the
distance is determined with high accuracy \citep{Luck11,Luck06,Andrievsky04}. As a result,
they provide probably the most accurate metallicity gradient measurements in the
Milky Way (Fig.~\ref{fig:grad_Ceph_Luck2011}). The MW radial gradient of [Fe/H] obtained by \cite{Luck11} with this method is $\rm d[Fe/H]/dR_G = -0.062 \pm 0.002~dex~kpc^{-1}$. The gradient of [O/H] has nearly the same slope within uncertainties: $\rm d[O/H]/dR_G = -0.056 \pm 0.003~dex~kpc^{-1}$.

Open clusters have instead been used to probe the metallicity of the
gas when they formed, typically probing ages of a few Gyr
\citep{Friel95,Chen03,Sestito06,Magrini09a,Lepine11}. By using the extensive
ESO-Gaia survey \cite{Magrini17b} has provided the most extensive mapping
of the metallicity of open clusters (and field stars) in the MW, by also
differentiating between clusters younger and older than 2~Gyr.

Planetary Nebulae  instead probe the enrichment of the gas back to the
time when their
progenitors formed, on timescales ranging from about 3--4~Gyr to potentially
the oldest epochs ($\sim 13$~Gyr), and they have been extensively used
to investigate gradients in nearby galaxies and in the MW
\citep[e.g.,][]{Maciel99,Maciel03,
Henry10,Stanghellini10,Stanghellini14}, in some
works even differentiating among the ages of the planetary
nebulae \citep{Stanghellini18}.

In principle the different tracers discussed above enable us to determine the evolution
of the chemical gradient in galaxies as a function of lookback time.
The general result is that diagnostics tracing metal enrichment on longer time scales tend to
give gradients that are flatter than those inferred
from HII regions, suggesting that metallicity gradients have become steeper (more negative)
with time \citep{Magrini16}.
This is  shown in Fig.~\ref{fig:grad_compil_Stanghellini2014},
from \cite{Stanghellini14}, where the evolution of the metallicity gradients
is shown as a function of lookback time, for a few galaxies for which this information
can be extracted.

This is somewhat in contrast with simple expectations of inside-out galaxy formation (inferred
by other tracers), in which the inner regions would be expected to start forming at earlier
times, hence having more times to produce metals, than the outer galactic regions \citep[e.g.,][]{Dave11c,
Gibson13,Prantzos00,Pilkington12}.
This steepening requires some heavy redistribution of metals in the early galaxy formation,
such as powerful feedback effects, prominent radial flows (for instance induced by gravitational
instabilities or galaxy interactions) or early stochastic accretion/dilution  (e.g. from intensive
accretion and minor merging events in the early universe)
\citep{Dekel13,Dekel14,Tissera18,Grisoni18}.

However, one should also be aware that the tracers probing the metallicity gradients across
long lookback
times, such as PNe or older open clusters, are potentially affected by the effect of radial stellar
migration. Indeed, stars may potentially migrate significantly from their original
birth site, washing out an originally steep gradient, as a consequence of stellar bars, galaxy
interactions or other secular processes. In this scenario the flatter gradients observed
at later cosmic times would simply be a consequence of the longer timescale during which
stellar migration has been mixing the older stellar populations.
Yet, as discussed later, other results seem to independently support the scenario in which
metallicity gradients have become steeper with time.
Moreover, some detailed modelling of the prominence of stellar migration across the lifetime
of galaxies have indicated that this effect is minor and unlikely to substantially affect the
slope of the metallicity gradient in galaxies
\citep{Spitoni15a}.

\begin{figure}
\centerline{\includegraphics[width=10cm]{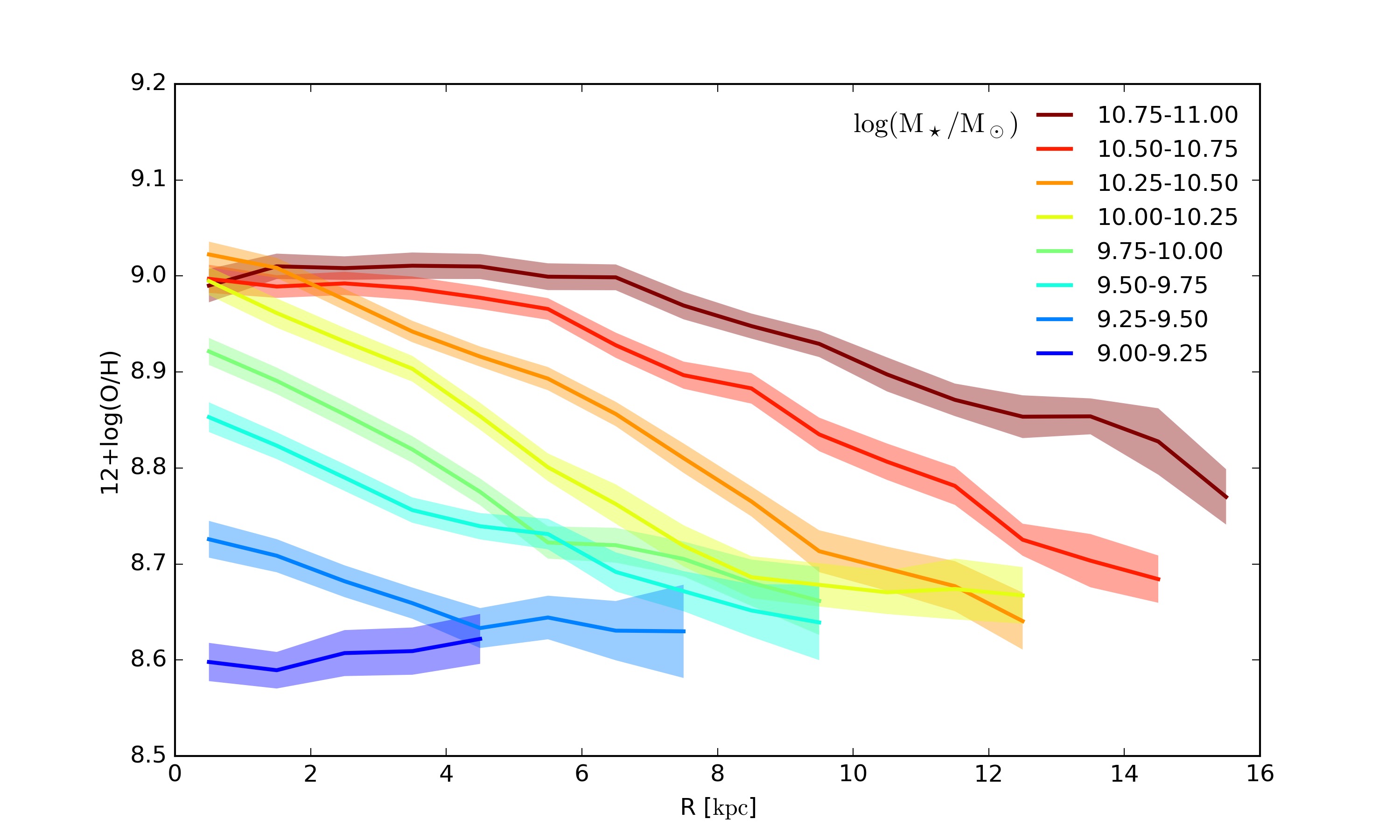}}
\caption{Average gaseous metallicity profile in bins of stellar mass for $\sim$2,800
galaxies from the MaNGA survey \citep{Belfiore17a}.
Metallicities are derived from O3N2 using the calibrations in \cite{Pettini04}.
}
\label{fig:gradients_mass_Belfiore2017}
\end{figure}

\subsection{Statistical properties of galactic discs metallicity gradients}
\label{sec:gradients_statistics}

The advent of large integral field spectroscopic surveys has made it possible to investigate
gradients more systematically for large sample of galaxies, especially exploiting HII regions,
whose nebular lines are easier to detect and map in galaxies. By using spatially resolved spectroscopic
data from a sample of 306 star forming discs from the CALIFA survey, \cite{Sanchez14}  found
a large spread of metallicity gradients. However, both \cite{Sanchez14} and \cite{Ho15}
have pointed out that the spread is
greatly reduced, and metallicity gradients become comparable,
if the radii are normalized to the galaxy effective radius ($\rm R_e$),
suggesting that the chemical evolution of galaxies with different masses
is governed by the same
enrichment processes occurring on local scales.

\cite{Ho15} and \cite{Sanchez-Menguiano16} found no evidence that the metallicity gradient (normalized to $\rm R_{25}$) depends
on galaxy mass, based on  data of star forming galaxies from the SAMI and CALIFA
integral field surveys, 
However, the size and/or the mass range of these samples may not be sufficient to identify clear trends with mass. Indeed,
by using the second MaNGA--SDSS4 data release, comprising integral field spectroscopic
data for about 2,800 galaxies spanning two orders of magnitudes in stellar mass, \cite{Belfiore17a}   later revealed that the metallicity gradients 
of star forming galaxies  (normalized
to $\rm R_e$) actually strongly depend on the stellar mass of the galaxy. 
Indeed, as illustrated
in Fig.~\ref{fig:gradients_mass_Belfiore2017}, the metallicity gradient is nearly flat
for low mass galaxies ($\rm M_{star}\sim 10^9~M_{\odot}$) and becomes progressively
steeper (more negative) for more massive galaxies. If one considers that the sequence in mass
somehow reflects an evolutionary sequence, the mass-gradient relationship is at least qualitatively
in agreement with the indication that the metallicity gradient was shallower at earlier epochs as
inferred by the different tracers of individual galaxies.

A steepening of the metallicity gradient with
stellar mass has later been confirmed by
\cite{Poetrodjojo18} by using data from the SAMI Survey \citep{Bryant15}.
The dependence seems
weaker than observed by \cite{Belfiore17}, but the sample
of galaxies is also much smaller.

\begin{figure}
\centerline{
\includegraphics[width=4cm]{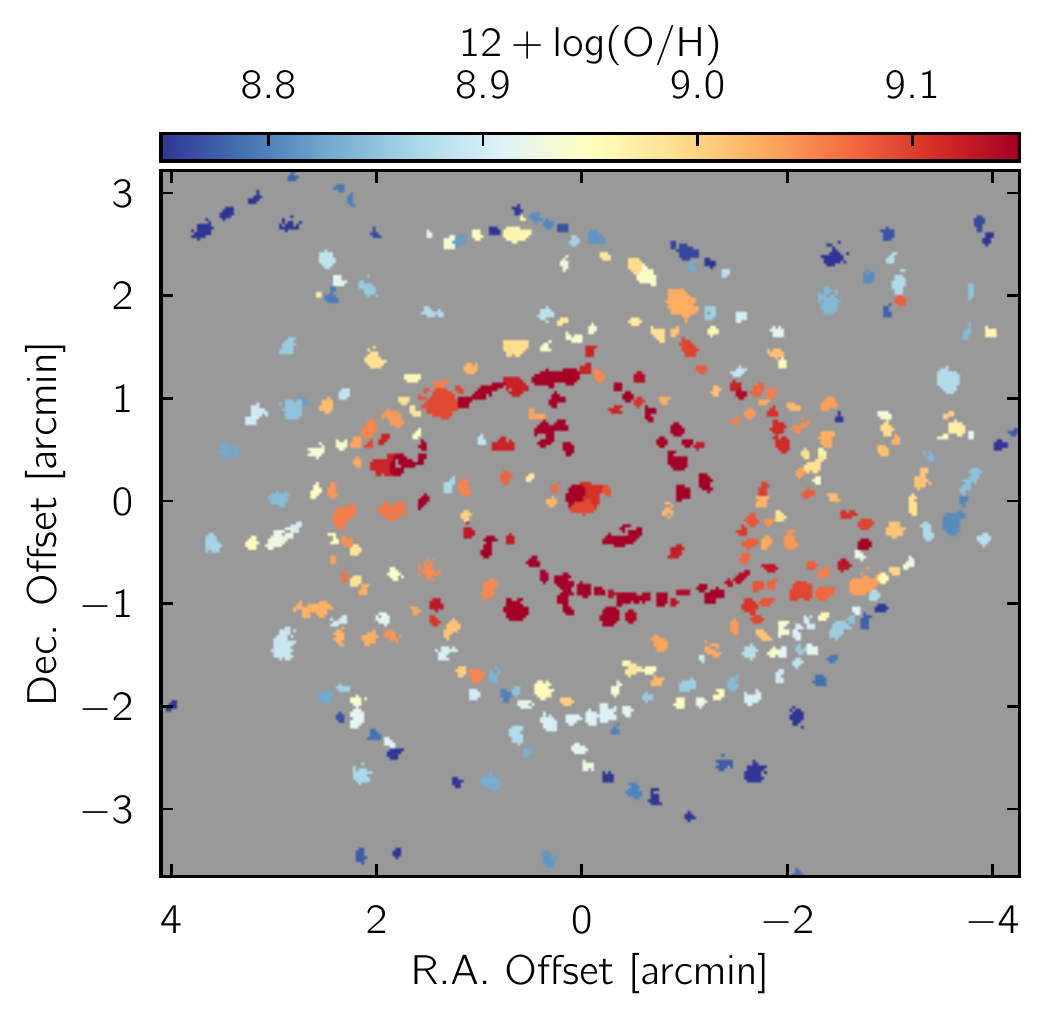}
\includegraphics[width=8cm]{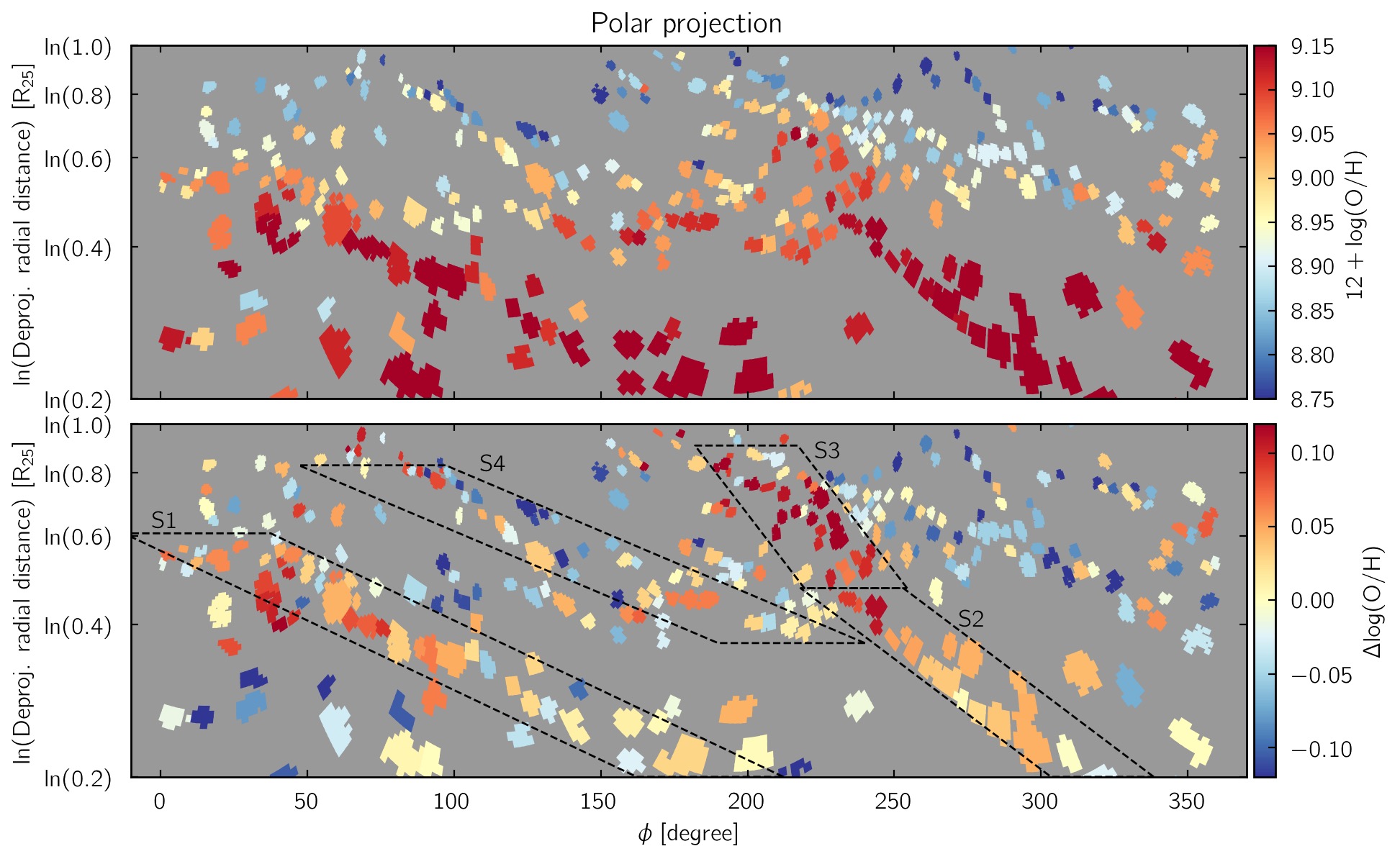}
}
\caption{
Left: Metallicity distribution of HII regions in the galaxy NGC2997 from \cite{Ho18}.. Right-top: Metallicity distribution for the same galaxy in polar projection. Right-bottom: metallicity difference with the average radial gradient. The positions of the spiral arms are indicated. 
}
\label{fig:azim_grad_Ho18}
\end{figure}

It should be noted that these results on metallicity gradients of large samples of
galaxies, based on the extensive MaNGA, CALIFA and SAMI surveys, adopt the strong line methods
for measuring the metallicity, with all caveats discussed in Sect.~\ref{sec:measmet_comparison}.
However,
 the absolute calibration scale offset potentially plaguing the strong line method is partly
mitigated in the case of metallicity gradients as these imply differential measurements.
It is also important to recall, that all these studies confine the measurement of the metallicity
gradient to the galactic regions showing evidence for star formation; extensive bulge or inter-arm
regions with LIER-like emission or with nebular emission too weak to be probed are excluded
from the determination of the metallicity gradients, which may result in either potential bias or
may be missing some key information associated with the metal evolution in these specific regions.

The flattening of the metallicity gradient in the central region of the most
massive spiral galaxies (Fig.~\ref{fig:gradients_mass_Belfiore2017}, see also \citealt{Zinchenko16})
is likely a consequence
of the metallicity saturating, and approaching the yield, in the central most metal rich regions.
However, it is also possible that, despite the attempt to confine the measurement of the
metallicity gradient to star forming regions, some contamination from the central LIER-like emission
in massive galaxies may affect the strong line diagnostics.

\begin{figure}
\centerline{\includegraphics[width=10cm]{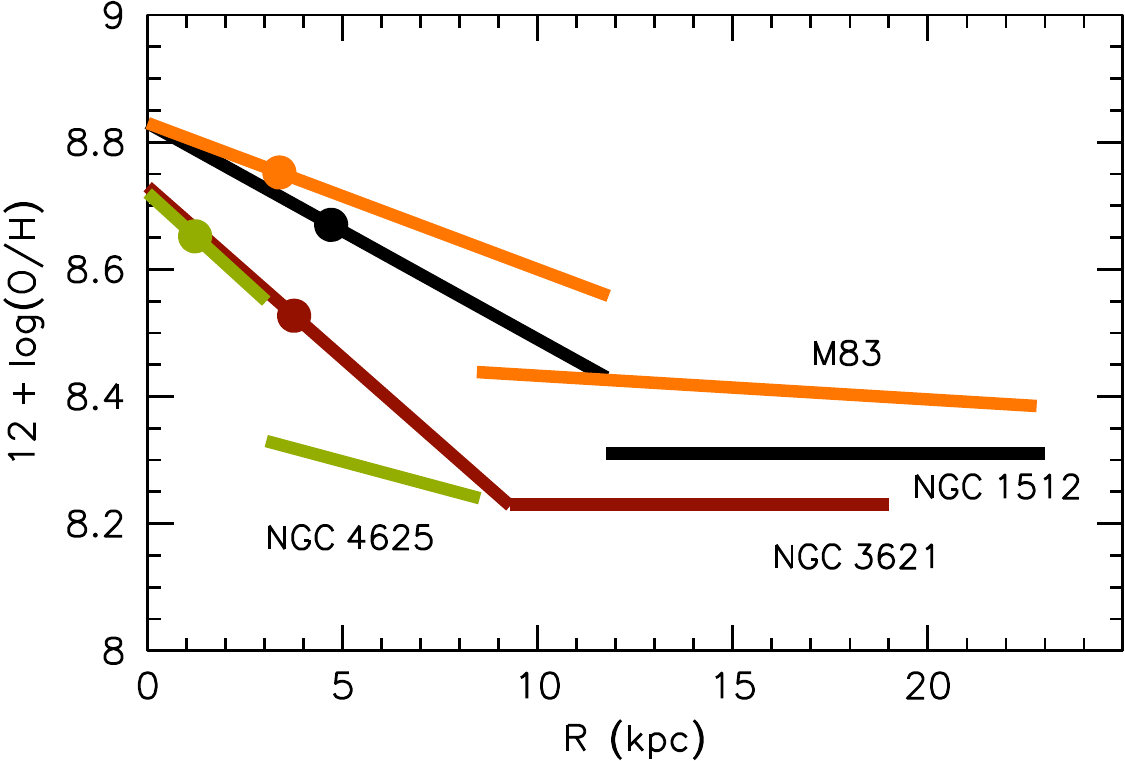}}

\caption{Schematic representation of the abundance gradients for a sample of local galaxies, from cite{Bresolin12}.
Dots are shown at $R = 0.4 R_{25}$ for each galaxy to represent the characteristic abundances of their inner disks. 
}
\label{fig:gradients_out_Bresolin12}
\end{figure}

Some attempts have been made to explore azimuthal variations of the metallicity (i.e., at
fixed galactocentric distance) in galactic discs,
both by using the direct $T_e$ method \citep{Li13,Berg15a} 
and the strong line method \citep{Zinchenko16,Sanchez-Menguiano17a,Ho17,Ho18}. Generally,
the azimuthal variations, if anything, are found to be
small (less than 0.1 dex), but significant in
a growing number of systems, as shown for
example in Fig.~\ref{fig:azim_grad_Ho18} \citep{Ho18}. There are also claims that, in particular, metallicity varies between the arm and inter-arm regions in some galaxies \citep{Ho17,Ho18}, which
may reveal fast enrichment by the enhanced star
formation along the spiral arms. However, these
properties do
not seem to be common to most galaxies \citep{Sanchez-Menguiano17a}. Moreover, one should take into account
that the azimuthal metallicity gradients, as well as the  arm/inter-arm variations, have so far been assessed primarily through strong line diagnostics and, despite the efforts to take the ionization parameter into account, the small variations observed
may still be affected by changes in the physical
conditions of the ISM and excitation conditions
between arm and inter-arm regions. Studies of azimuthal variations based on direct (T$_e$)
measurements are less conclusive in establishing
whether there are significant azimuthal metallicity
gradients in local galaxies, although the statistics of HII-regions is admittedly poor in this case. Certainly more studies
are needed both to expand the samples of galaxies
with angular resolution high enough to resolve azimuthal structures and sensitive enough to detect
auroral lines in multiple galactic HII regions.

{\bf In summary}, the study of the radial metallicity gradients is an important piece of information on the current status of the galaxies and on the process dominating their formation. As different populations of the MW  (HII regions, young massive stars, open clusters, PNe, etc.) reveal the chemical abundances at different ages and stages of the Galaxies, the evolution of chemical abundance gradient with time can be studies via stellar archeology. A steepening of the gradient with time towards more negative values is detected, and this is explained by the presence of radial redistribution of metals.

\subsection{Galactic discs outskirts}
\label{sec:gradients_outskirts}

While the slope of the radial gradients has generally been probed within the optical radius,
high sensitivity observations have made it possible to explore the radial gradients in the outer
discs. Observations have shown that at large galactocentric radii (typically at $R>2R_e$),
metallicity gradients tend to become very flat and settle to a relatively high value, typically 
around 0.3--0.5~$\rm Z_{\odot}$. This has been confirmed both through the direct method applied to individual galactic discs \citep{Bresolin09b,Goddard11,Bresolin12,Sanchez-Menguiano17b}, as
illustrated in Fig.~\ref{fig:gradients_out_Bresolin12},
and through the stacking of large sample of galaxies \citep{Sanchez14}.
It has been suggested that the flat slope of the metallicity gradient at large radii may be
associated with the low efficiency and discontinuous star formation typical of outer discs,
similar to what is observed in low-mass galactic discs.
However, the really puzzling
finding is the relatively high level of enrichment in these outer regions, where the formation of
stars has been very low and certainly not enough to bring the content of metals to the observed
values.
The only realistic explanation is that these outer regions have accreted pre-enriched material,
either as a consequence of cooling from the halo, gas stripping from the center due
to galaxy interactions, minor merging with enriched satellites,
galactic fountains or major outflows from the central (metal rich) regions
\citep{Bresolin12,Sanchez13b,Belfiore16b}.

It is also important to mention that there are some notable exceptions to these trends.
Indeed, \cite{Moran12} have found cases of spiral galaxies whose metallicity gradient
drops significantly at $\rm R>R_{90}$ (the radius enclosing 90\% of the r-band light)
and they find that this feature is linked to the amount
of atomic gas content HI. This may indicate that galaxies with prominent metallicity drop 
in their outskirts have been recently accreting pristine/low-metallicity gas
from the intergalactic medium that, for momentum conservation, is predominantly deposited
in the outer regions.

\subsection{Spatially resolved scaling relations}
\label{sec:gradients_scalingrel}

Within this context there has been recently an extensive effort in trying to depart from metallicity
gradient studies with the classical radial axisymmmetric (or azimuthal)  approach, and investigate
the variation of metallicities on local scales
focusing on the potential correlation with the other local galactic properties, such as
stellar surface density and star formation rate surface density. This approach is equivalent
to investigating whether scaling relations apply locally. 

\begin{figure}
\centerline{
\includegraphics[width=7.9cm]{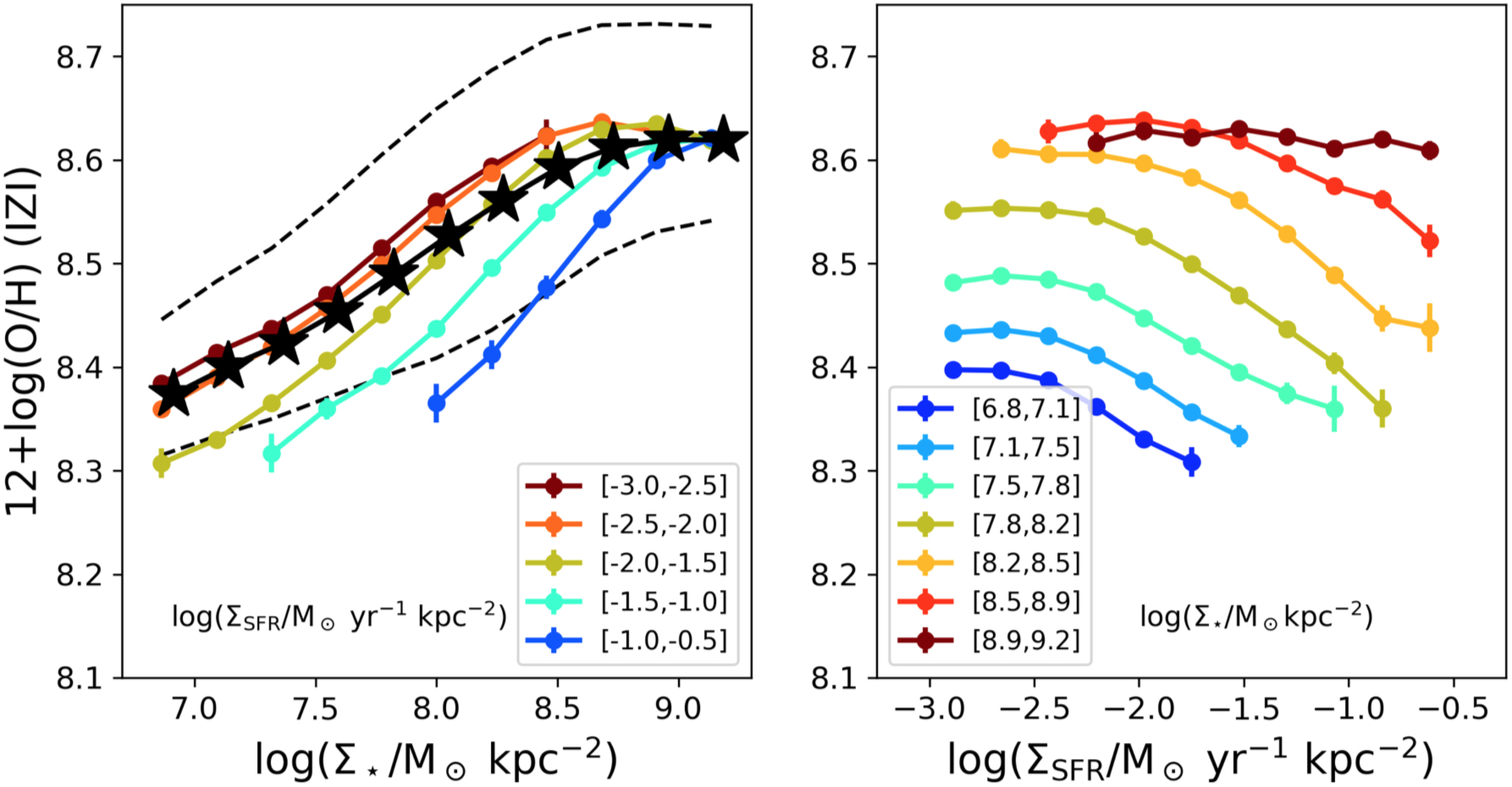}
\includegraphics[width=3.9cm]{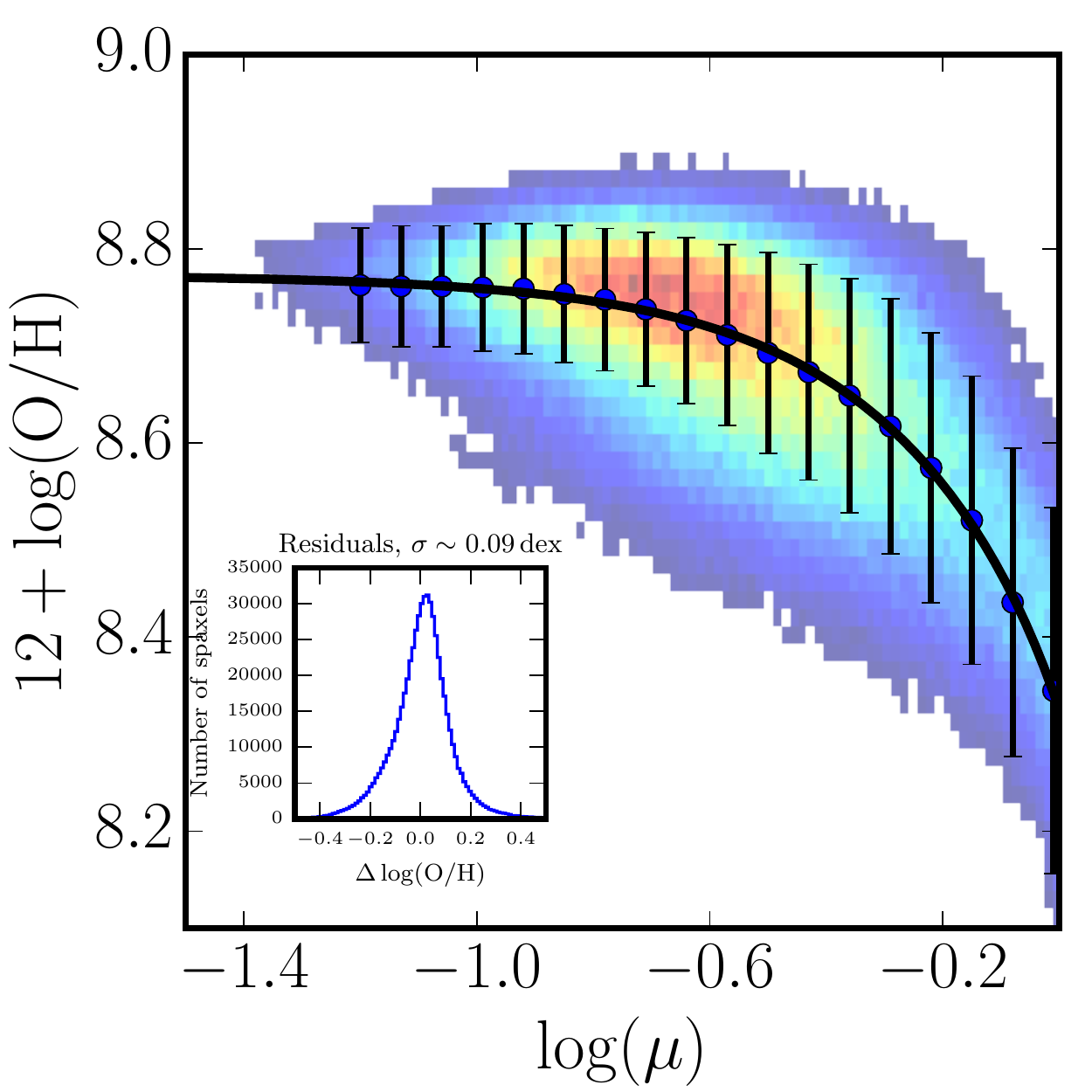}
}
\caption{Left: gas metallicity as a function of stellar mass surface density, averaged across the entire
MaNGA sample (black stars) and averaged in bins
of star formation rate surface density (coloured
lines), from \cite{Barrera-Ballesteros18}
Center: gas metallicity as a function of star formation
rate surface density, averaged in bins
of stellar mass surface density.
Courtesy of Belfiore et al. (in prep.).  Right: gas metallicity as a function of local gas fraction $\mu$, as
inferred from the Balmer decrement. 
}
\label{fig:sigma_star_Barrera}
\end{figure}


More specifically, by using CALIFA and MaNGA data, \cite{Gonzalez-Delgado14} and \cite{Barrera-Ballesteros16} investigated the spatially resolved dependence of the gas metallicity on local properties of galaxies.  They both find a clear 
correlation between metallicity and stellar mass surface density $\rm \Sigma_*$. \cite{Barrera-Ballesteros16}
investigate also the possible dependence of metallicity on local surface density of star formation rate, $\rm \Sigma_{SFR}$, but they claimed that a Z-$\rm \Sigma_{SFR}$ relation is not significant, although (at fixed $\rm \Sigma_*$) their highest $\rm \Sigma_{SFR}$ bin does show a clear drop in metallicity.
More recently, Belfiore et al. (in prep.) have further investigated in detail the spatially resolved metallicity scaling relations by using the latest MaNGA releases; their analysis confirmed the Z-$\rm \Sigma_*$ scaling relation and revealed also the existence of a clear dependence of metallicity on SFR surfance density, i.e. a Z-$\rm \Sigma_{SFR}$ anti-correlation, as illustrated in Fig.~\ref{fig:sigma_star_Barrera} (left and center).

\cite{Barrera-Ballesteros18} found also a strong anti-correlation between
local metallicity and gas surface density $\mu$ (inferred from the
dust extinction traced by the Balmer decrement) as illustrated in Fig.~\ref{fig:sigma_star_Barrera}-right. 
This finding is in good agreement with the Z-$\Sigma_{\rm SFR}$
anti-correlation, as $\Sigma_{\rm SFR}$ and gas surface
density $\mu$ are tightly linked by the Schmidt--Kennicutt relation. It is also
in agreement with the results on the gas-FMR by, e.g., \cite{Bothwell13}, see Sect.~\ref{sec:FMR_gas}.
 
 The (anti-)correlation between local metallicity
 and surface density of SFR (or, equivalently, on gas surface density) may really be driving the global FMR; in the scenario
 in which local metallicity dilution due to
 infalling gas both decreases the local metallicity
 and boosts the local SFR, the cumulative effect can
 scale up to induce the FMR over the entire galaxy.
 
Whether the global MZR stems from the local correlation
 between metallicity and $\Sigma _*$, or vice-versa, is more difficult
 to establish. Indeed, total stellar
 mass and $\Sigma _*$ are correlated, and therefore
 it is difficult to establish which of the two
 is the primary, driving correlation.
 \cite{Barrera-Ballesteros16} claim that
 the local Z--$\Sigma _*$ is the primary relation, which
 drives both the global MZR and the radial metallicity
 gradients in galaxies.
 However, this certainly cannot
be the case at large galactic radii, where the metallicity profile flattens and remains
at relatively high levels, while the stellar surface density keeps fading exponentially
or even more rapidly.
Moreover, recently \cite{DEugenio18} have shown that
the gas metallicity is more tightly correlated with the gravitational
potential ($\rm \Phi\sim M_*/R_e$) than with the galaxy stellar mass or the stellar
surface density, indicating that the gravitational potential is likely
the primary mechanism establishing the level of metal content (see also Sect.~\ref{sec:FMR_furtherdep}).
The weak correlation found by \cite{Barrera-Ballesteros18} between metallicity and local escape velocity suggests that metals lost by winds only play a minor role in shaping the local metallicity. 
In conclusions, these evidences show that the MZR might be shaped by global rather than local effects.\\

{\bf Summarizing}, the scaling relations observed at the global, galaxy-scale level are now also observed at the local level. While the global FMR is probably driven by the local one, the situation for the MZR is uncertain because conflicting evidences are present.

\subsection{Interacting galaxies}
\label{sec:gradients_interacting}

\begin{figure}
\centerline{
\includegraphics[width=9cm]{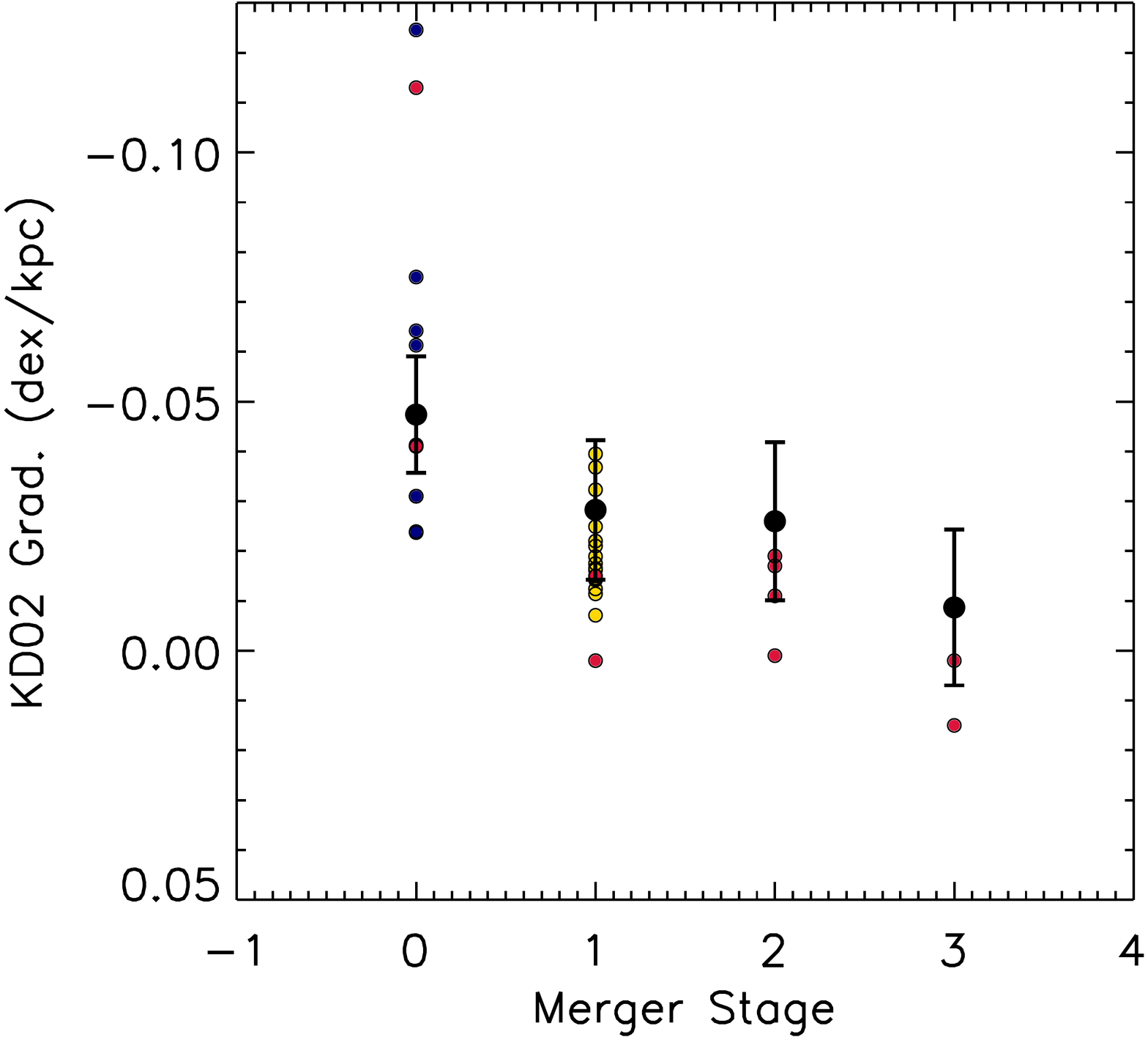}
}
\caption{
Gas phase metallicity gradient versus merger stage. Blue points are isolated systems
from \cite{Rupke10b}, yellow points are wide pairs, red points are
luminous infrared galaxies  and black points are from models \citep{Rich12a}.
}
\label{fig:grad_interct}
\end{figure}

While most of these studies have focused on regular, isolated galactic discs, 
a few
studies have investigated gradients in interacting/merging systems. It has been found
that interacting systems 
generally have significantly flatter gradients than isolated galaxies, and the effect
is stronger in systems that are in a more advanced stage of merging
(Fig.~\ref{fig:grad_interct})
\citep{Rupke10b,Rich12a,Torres-Flores14}. 
Moreover,
the extended tails resulting from galaxy interaction display remarkably flat gradients
out to 70 kpc \citep{Olave-Rojas15}.
 Recently \cite{Ellison18a} has used MaNGA data to study the spatial distribution of the excess of SFR and deficit of metallicity in interacting/merging galaxies, finding that both are more prominent in the inner parts of the galaxies.

The most widely accepted interpretation is that
interactions on the one hand make the outer low-metallicity gas lose angular momentum
and flow towards the central region of the galaxy causing dilution, while on the other hand
metal enriched gas is stripped into extended tails where the metallicity
imprint of the original location is rapidly lost and mixed up with gas from other regions
\citep{Torrey12,Rupke10a}. 
 This interpretation is consistent with the lower total metallicity observed in these galaxies as discussed in Sect.~\ref{sec:MZR_merging}, and with the observations that recently merged galaxies have larger amounts of atomic gas \citep{Ellison18b}.

\subsection{Stellar metallicity gradients}
\label{sec:gradients_stellar}

Several studies have been performed on the metallicity gradients of the stellar population
in early type galaxies \citep[e.g.,][]{Spolaor10,Bedregal11,Koleva11,Harrison11,La-Barbera12,Goddard17a}.
They are often more demanding than gas metallicity gradients,
as resolved stellar metallicities require much higher S/N data on the stellar continuum, 
which is especially difficult to achieve
in the outer parts of early type galaxies galaxy due to the rapidly declining surface brightness of the
stellar light with galactocentric radius. However, given typically the lack of dust
reddening, stellar colors have also been employed to map the metallicity in these
passive systems \citep[e.g.,][]{Tortora10}.
Fewer studies have been performed on the stellar metallicity gradients of star forming galaxies
\citep[e.g.,][]{Morelli15,Sanchez-Blazquez14,Gonzalez-Delgado15,Goddard17a,Li18}
both because disentangling age-metallicity degeneracies of the stellar population requires
even higher S/N and because the presence of nebular lines makes the analysis of the stellar
features more difficult. Moreover,  the simple analysis of the stellar
metallicity provides light-weighted stellar metallicities, typically dominated by
the most recent stellar population. A proper determination of the dominating stellar
population, especially in late type galaxies is difficult as it depends on the star formation history and on the stellar
indices or wavelength range used to extract the stellar metallicity. Mass-weighted stellar
metallicities are potentially more interesting to investigate the metallicity gradient of the
bulk of the stellar population, but, as discussed in Sect.~\ref{sec:measmet_stars}, mass-weighted stellar metallicities
are more difficult to infer, especially in the outer regions of galaxies where the S/N is not
as high as in the central galactic regions.

\begin{figure}
\centerline{
\includegraphics[width=12cm]{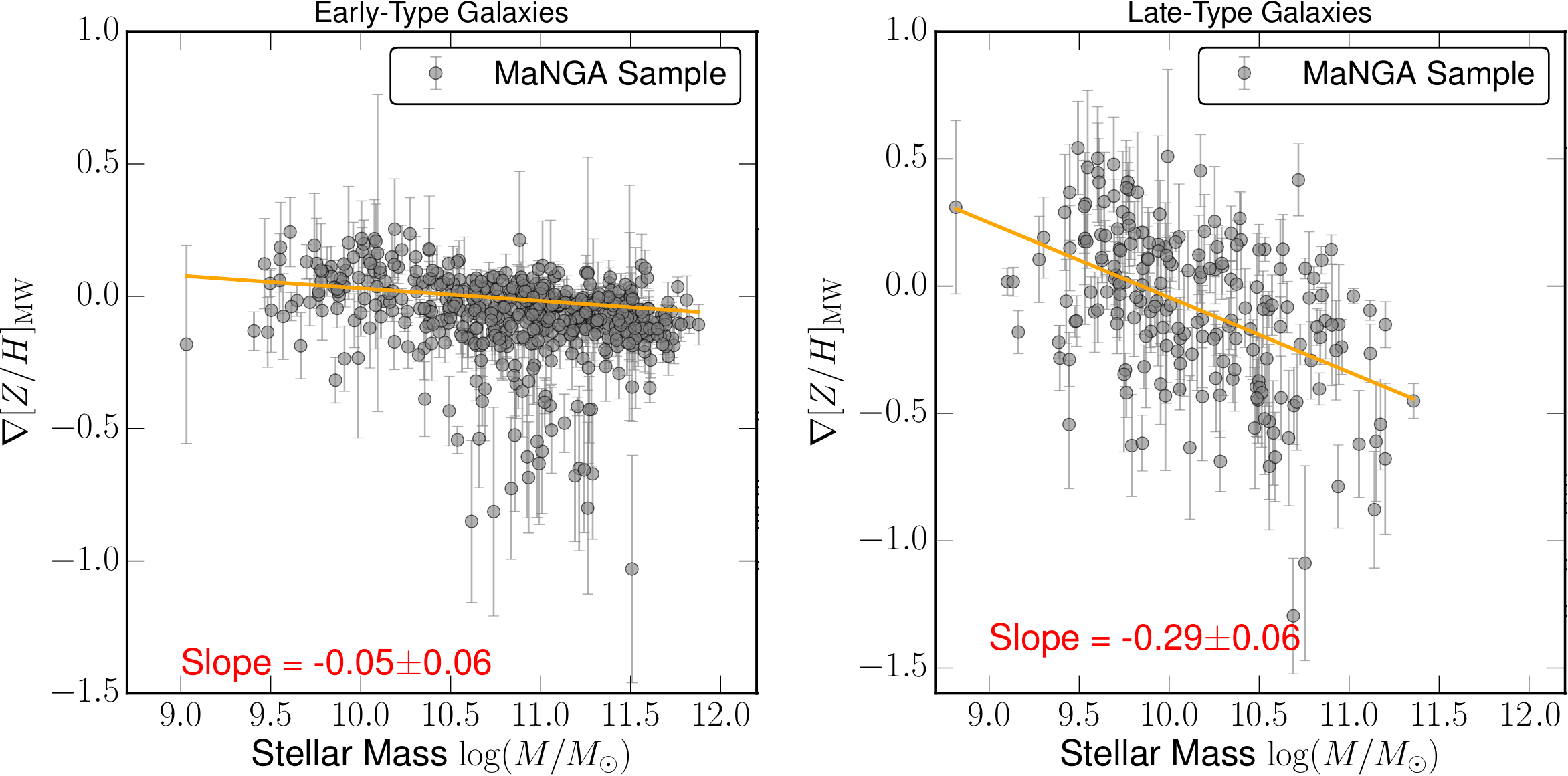}
}
\caption{
Mass-weighted stellar metallicity gradient as a function of stellar mass for
early-type (left) and late-type (right) local galaxies from the MaNGA survey
\citep{Goddard17a}.
}
\label{fig:grad_star_mass_Goddard2017a}
\end{figure}

Typically, stellar metallicity gradients are shallow. Although the dispersion is
large, works based both on CALIFA
\citep{Gonzalez-Delgado15} and on MaNGA  \citep{Goddard17a}
find a significant dependence of the stellar
metallicity gradient with stellar mass, with a stronger dependence
for late type galaxies than for early type galaxies
(Fig.~\ref{fig:grad_star_mass_Goddard2017a}). This is a trend similar to that
observed for the gas phase metallicity, however the important difference is that
for stellar metallicities the gradient is negative but shallower at high masses and becomes
positive (``inverted'') at low masses, indicating a change in the formation and accretion
histories between low mass and high mass galaxies, and also between late type and early type
galaxies.
A strong dependence of stellar metallicity gradients with
stellar mass for late type galaxies was also confirmed by \cite{Lian18b}; they point out that the mass dependence is significantly stronger than for the gaseous metallicity gradients, which is surprising, given that close to equilibrium the stellar metallicity should approach the gas metallicity \citep{Peng14b}; they interpret the strong difference  between stellar and ISM metallicity gradients invoking a variation of either the outflow loading factor, or of the IMF, both in time and radially.

Interestingly, by using MaNGA data,
\cite{Li18} have found that the stellar metallicity gradients depend on stellar
velocity dispersion and that they peak (becoming most negative) at intermediate
velocity dispersions of about 100~km/s.
This is interpreted as indicating a change in the evolutionary history in galaxies.
In particular, metallicity gradients becoming flat at very large velocity dispersions
is likely to indicate a growing role of mergers, which redistribute the metallicity
in very massive galaxies whose velocity dispersion has been enhanced by the merging history. 

Finally, very interestingly, \cite{Sanchez-Blazquez14} have investigated the
relation of stellar metallicity gradients on the presence and strength of stellar
bars, finding no evidence for any correlation \citep[in agreement with similar
studies tracing the gas metallicity gradients, see][]{Sanchez14}.
This result indicates that stellar
migration associated with stellar bars does not play a significant role in shaping
the metallicity gradient of galaxies, in agreement with the expectations of some models
\citep{Spitoni15a}. However, one should take into account that stellar bars are
a recurrent phenomenon in the life of galaxies \citep[about 40\% of spiral galaxies in the local universe
are barred, ][]{Sheth08}, therefore the lack of a correlation with the current strength of
bar in galaxies does not necessarily mean that past barred phases 
(even in galaxies currently non-barred)
have not played a role.

\subsection{Metallicity gradients at high redshift}
\label{sec:gradients_highz}

Measuring metallicity gradients at high redshift is obviously much more difficult
for various reasons.

Firstly, the steep cosmological dimming of the surface brightness
($\propto (1+z)^4$) makes it much more difficult to achieve the 
S/N required to measure metallicity gradients in the outer parts of galaxies; this
effectively translates into measuring high-$z$ metallicity gradients only for the gas phase
and only by using strong line methods.

\begin{figure}
\centerline{\includegraphics[width=12cm]{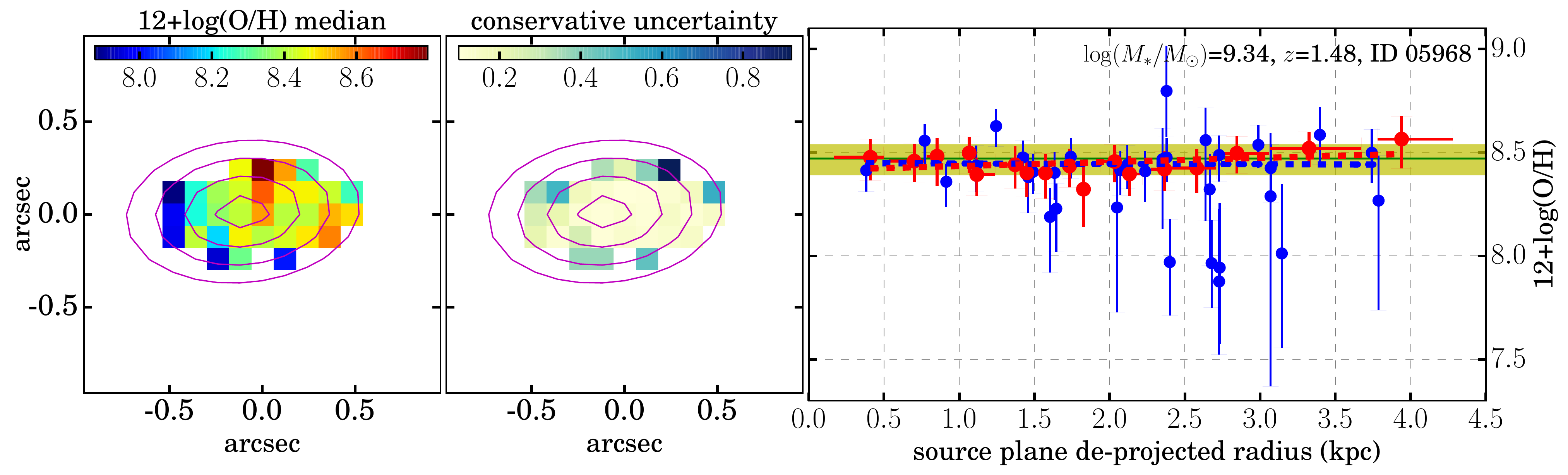}}
\caption{
Metallicity gradient in a lensed galaxy at $z=1.48$ obtained with HST grism data \citep{Wang17}.
Left: metallicity map. Center: uncertainty. Right: radial metallicity distribution (blue
points are individual pixels, red points show the average within annuli). 
}
\label{fig:grad_highz_hst}
\end{figure}

Secondly, it becomes increasingly difficult to spatially resolve
galaxies in integral field spectroscopy. In most studies the metallicity gradients
are only marginally resolved. Targeting lensed galaxies generally helps
\citep[e.g.,][]{Yuan11,Jones13,Jones15a,Leethochawalit16,Wang17}, however
at the cost of introducing the additional uncertainty associated with the lens modelling.
Moreover, even when the lens model is well constrained, gravitational magnification
is differential, hence the resulting lensed image and metallicity map are strongly weighted towards
the regions close to the lens caustic, therefore potentially resulting into distorted
metallicity maps. The use of adaptive optics certainly helps \citep[e.g.,][]{Leethochawalit16,Perna18,Forster-Schreiber18}
although the modest
Strehl-ratios cause the sensitivity to drop significantly, especially towards the outer low surface
brightness regions. HST grism spectroscopy has also been effectively used
to map the metallicities at high-$z$ with HST-like high angular resolution (Fig.~\ref{fig:grad_highz_hst})
\citep{Jones15a,Wang17}, the only problem being the low spectral resolution
of the spectra and small wavelength range, which often limit the use of diagnostics and
makes the subtraction of the stellar continuum more problematic. In other studies,
not exploiting gravitational lensing and from the ground without adaptive optics
\citep[e.g.,][]{Cresci10,Queyrel12,Swinbank12,Stott14,Troncoso14,Wuyts16},
the measurement of metallicity
gradients has generally been limited to the larger (hence typically more massive) galaxies
therefore with potential bias.

The additional problem
is that most nebular diagnostics are shifted to the near infrared bands.
The basic, bluest diagnostics enabling the use of the strong line method (e.g., the $\rm R_{23}$
parameter) can be traced in the optical band through optical integral field spectrometers
only out to $z\sim0.8$ \citep{Carton18}. While integral field spectrometers in the near-IR
bands are well developed and available on large telescopes to probe the strong line metallicity
diagnostics at $z>1$, the reduced sensitivity in these bands (primarily because of the higher
background, both thermal and from bright OH sky lines)
and discontinuous spectral coverage (because of the deep atmospheric absorption bands), makes
the measurement of metallicity gradients even more challenging.
Often, spectra are obtained in a single spectral band, resulting in limited information.
For instance, [NII]/H$\alpha$\ is used to effectively trace metallicity
gradients at high-$z$, as these two lines are conveniently observed in the same band
\citep[e.g.,][]{Wuyts16,Forster-Schreiber18}, but at the cost of introducing all uncertainties associated
with this single diagnostic (e.g., dependence on nitrogen enrichment, dependence on ionization parameter, etc...), see Sect.~\ref{sec:measmet_strong_BPT}.

\begin{figure}
\centerline{\includegraphics[width=11cm]{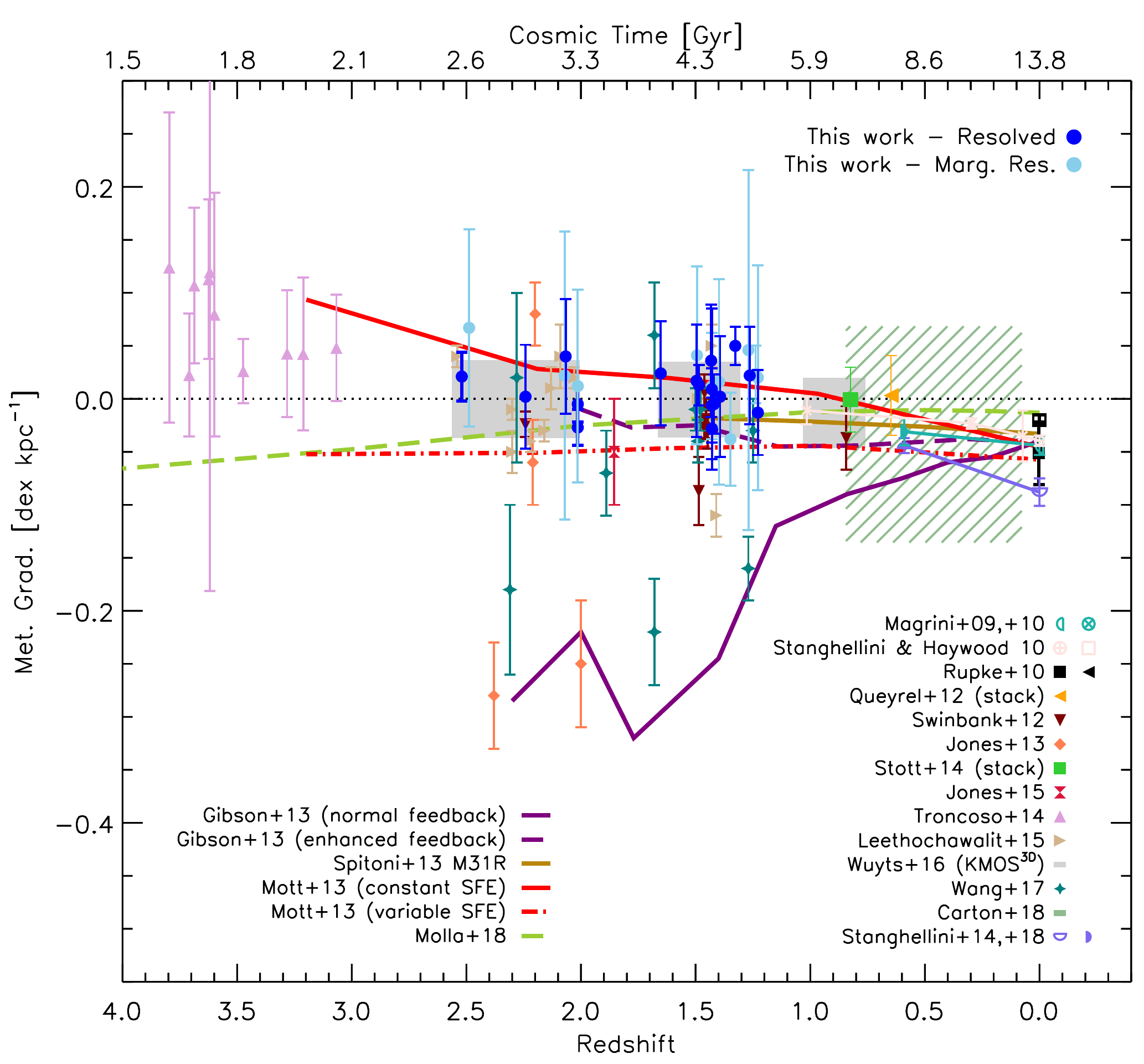}}
\caption{
Overview of the metallicity gradients measured at different redshifts, both in lensed
and unlensed galaxies, from different surveys, including local galaxies (whose gradient evolution has been inferred through tracers at different lookback
times). Evolutionary
tracks of different models are also show (see legend). Courtesy of Curti et al. (in prep.).}
\label{fig:highz_met_grad_curti18}
\end{figure}

With all these caveats, extensive studies have been undertaken to constrain the
evolution of metallicity gradients at high redshift
\citep[e.g.,][]{Cresci10,Queyrel12,Swinbank12,Stott14,Troncoso14,Wuyts16,
Yuan11,Jones13,Jones15a,Leethochawalit16,Wang17,Carton18,Forster-Schreiber18}.
A summary of the observed evolution of the metallicity gradients is shown in
Fig.~\ref{fig:highz_met_grad_curti18}  (from new observations and a compilation
courtesy of Curti et al., in prep.). There is clearly
a large dispersion, part of which is likely due to the observational uncertainties, 
but it is also likely reflecting a real dispersion of the metallicity gradients during the early
phases of galaxy evolution, when the accretion and merging processes were more stochastic
and resulting into a more irregular behavior \citep[e.g.][]{Ceverino16}. Despite the large scatter, and with the
exception of a few rare cases of very steep gradients \citep[][based on low S/N spectra]{Jones13},
most studies find that at high redshift the metallicity gradients are on average flatter
than observed locally. This is in  agreement with the flattening of the gradients in local
galaxies when using tracers that probe longer lookback times, i.e., primarily PNe
\citep{Stanghellini14,Stanghellini18}, as discussed in Sect.~\ref{sec:gradients_general}.

It is interesting to note that studies
comparing metallicities measured in DLA absorption
systems with their optical counterparts (whose metallicity is measured through nebular lines)
have also independently
inferred gradients that are, on average,
quite flat ($-0.022\ {\rm dex\ kpc}^{-1}$)
even out to galactocentric radii of several tens kpc,
in high-$z$ galaxies \citep{Christensen14,Rhodin18}.

However, it is becoming increasingly evident that at such early epochs the concept of
``radial'' (azimuthally averaged) gradient loses part of its meaning, 
as expected by some models \citep{Ceverino16}. Indeed, 
the metallicity distribution is often very irregular, with large local variations
{\citep[e.g.][and Fig.~\ref{fig:grad_highz_hst}]{Forster-Schreiber18}}, 
hence the apparently flat radial gradient simply
results from averaging large azimuthal variations into the same radial bins.
These large irregular metallicity variations in high-$z$ galaxies probably reflect the
chaotic accretion and formation processes during these early phases. Therefore, the comparison
with models should not be done simply in terms of radial gradients (which can be partly deceiving
of the chemical complexity of these systems), additional information should be considered
such as the metallicity scatter.

\begin{figure}

\centerline{\includegraphics[width=12cm]{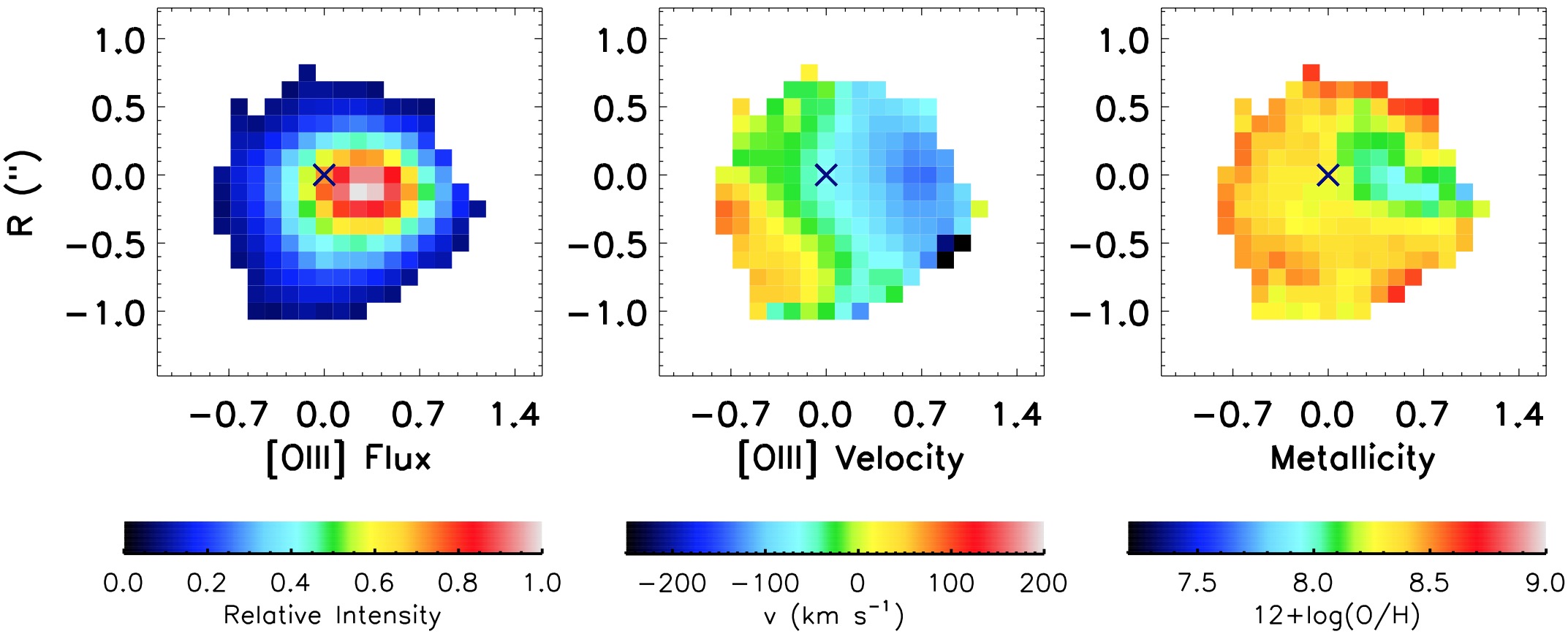}}
\caption{
Example of galaxy with inverted metallicity gradient at $z=3.06$, from \cite{Cresci10}.
Left: map of the [OIII]5007 line emission (approximately proportional to the
surface density of SFR). Center: Velocity field. Right: Metallicity map. 
}
\label{fig:grad_inv_Cresci2010}
\end{figure}

A population
of positive (i.e., ``inverted'') metallicity gradients has recently been discovered
at high redshift
\citep{Cresci10,Troncoso14,Carton18,Wang18}. These inverted gradients appear to be particularly
common at $z>3$ (Fig.~\ref{fig:grad_inv_Cresci2010}). However, it has been pointed
out that rather than simply being inverted, the central metallicity suppression
is mostly associated with the central enhancement of star formation rate. \cite{Cresci10}
and \cite{Troncoso14} suggest that this is a consequence of enhanced
inflow of metal-poor gas toward the central galaxies, in their early epoch of formation,
which results in both a local dilution of the metallicity and a local enhancement of the SFR
due to the larger amount of gas available. \cite{Stott14} suggest a similar scenario
to interpret the correlation that they find at $z\sim1$ between metallicity gradient and sSFR:
the enhanced inflow of near-pristine gas makes the central region metal poor and also boosts the sSFR. This is in agreement with the infall interpretation of the FMR, see Sect.~\ref{sec:FMR_origin}.

Whether the metallicity gradients of high-$z$ galaxies have a mass dependence similar
to that observed locally is not totally clear, mostly
because of the large scatter. \cite{Wuyts16} and
\cite{Stott14} do not find a clear mass dependence in their samples at $z\sim1$--$2$.
\cite{Carton18} find a weak mass dependence in their sample at $0.1<z<0.8$. Together with
the local findings
these results suggest that the mass dependence of the metallicity gradients is established
at late epochs.
Correlations of the gradients with sSFR have been also proposed \citep{Stott14,Wuyts16}, but in all cases only with low significance and in disagreement with other studies \citep{Leethochawalit16,Troncoso14}.

Finally, \cite{Carton18} find an interesting correlation at $0.1<z<0.8$ between
metallicity gradient and galaxy size, in which small galaxies show a large spread
in metallicity gradient (both positive and negative), while large galaxies present a much
 smaller spread and regular negative gradients.

\subsection{Metallicity gradients models}
\label{sec:gradients_models}

Several models  have been proposed to reproduce the metallicity gradients in galaxies, both for the gas and the stars
\citep{Molla97,Chiappini01,Naab06,Mott13,Spitoni13,Ho15,Kudritzki15,Ascasibar15,Spitoni15a,Schonrich17,Lian18b}.
Many models consist in the extension of the analytical
``gas regulator'' scenario applied to concentric galactic
rings, in which there is a radial dependence of the primary parameters, such as gas infall timescale and star formation efficiency (also introducing threshold for star formation). Some models, as discussed above, also introduce variable
outflow loading factors and even variable IMF.
 The main challenge of these models
is that they have also to reproduce the inside-out growth of galaxies, i.e., the finding that, based on both the stellar population age gradients and surface density gradients, the central parts of galaxies must have grown earlier and faster than
the outer parts. This requires models to have accelerated star formation and enrichment in the central regions, relative
to the outer regions, which results into a negative metallicity gradient, as observed in galaxies. 

\begin{figure}
\centerline{
\includegraphics[width=6cm]{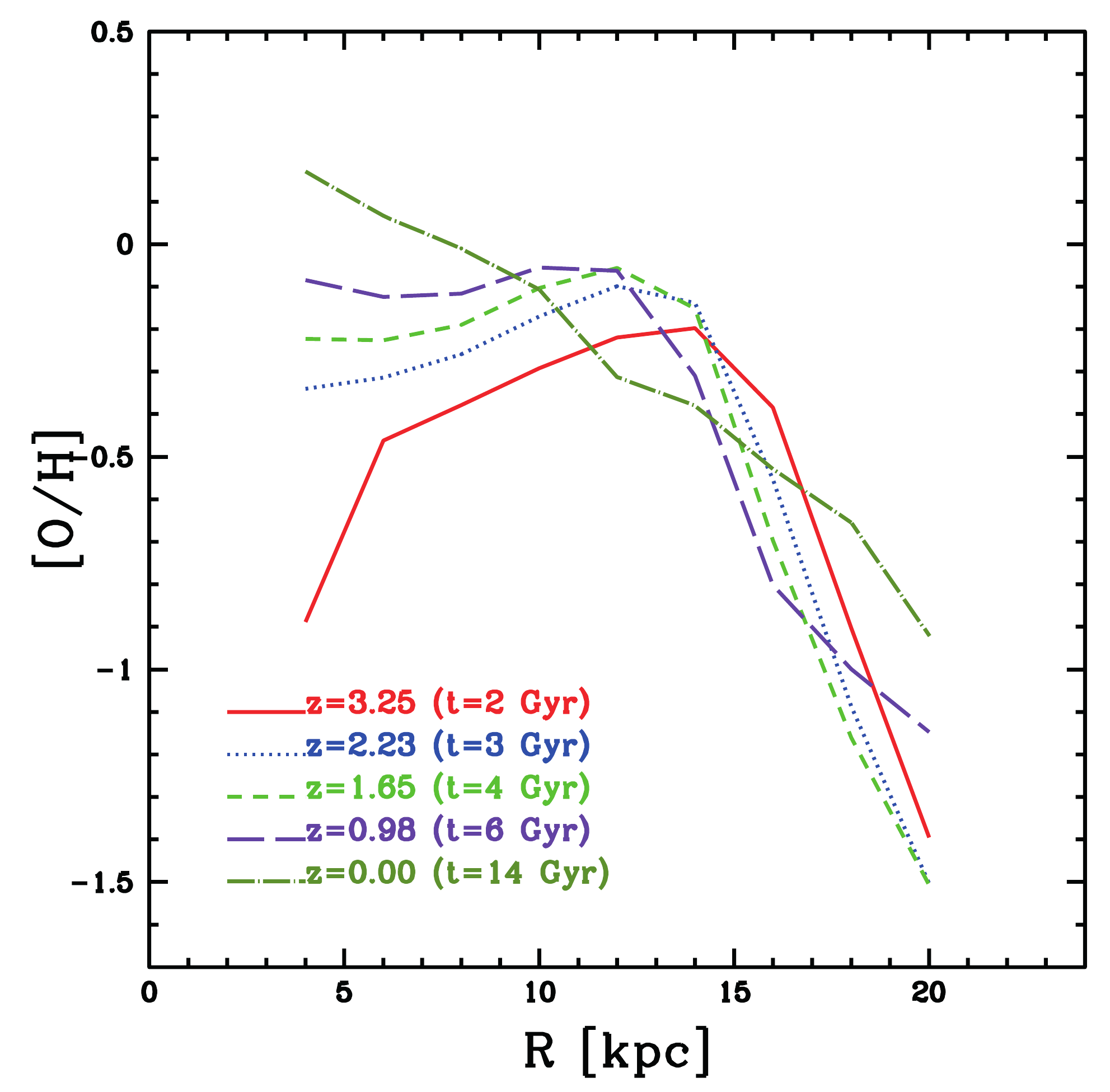}
\includegraphics[width=6cm]{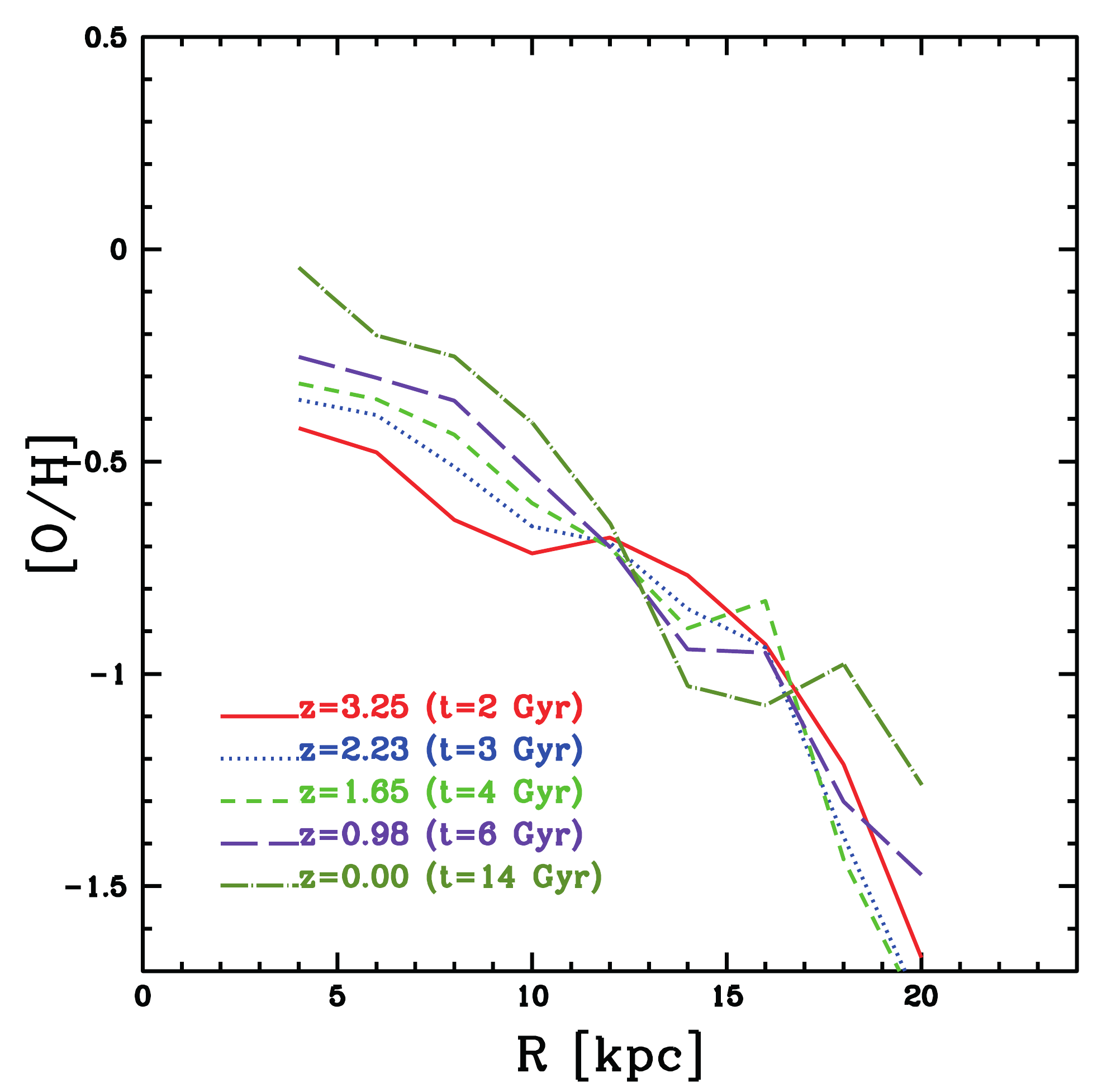}
}
\caption{
Examples of metallicity gradients predicted by the analytical
model of \cite{Mott13}, which include radial flows of gas. The model on the left assumes a threshold, inside-out formation, constant star formation efficiency, and radial flows. The model on the right uses a radially variable star formation efficiency. To be noticed that the metallicity gradient becomes more negative with time (at least within the central 10 kpc, which is the only part  observed at high redshift) and also that in the left-side model
at early times the metallicity gradient is centrally inverted as a consequence of strong central infall of pristine gas.
}
\label{fig:met_grad_model_Mott13}
\end{figure}

The main problem of this scenario is that, if different galactic annular
rings do not exchange metals (as it is the case in most models), then the unavoidable consequence
is that the metallicity gradient should flatten with cosmic time, which is opposite to what observed as inferred from the observation in local galaxies diagnostics that trace metallicity gradients at earlier
epochs (Fig.~\ref{fig:grad_compil_Stanghellini2014}), from
the observed evolution of metallicity gradients at high redshift
(Fig.~\ref{fig:highz_met_grad_curti18}) and even
based on the simple finding that the metallicity gradient
steepens with galaxy stellar mass (Fig.~\ref{fig:gradients_mass_Belfiore2017}) and assuming the stellar mass sequence of galaxies
somehow reflects their evolutionary pattern. Possible solutions to
comply with the inside-out growth of galaxies {\it and}
having metallicity gradients that {\it steepen} with time
is that stellar migration plays a role in mixing stellar metallicities and the production of metal at different radii,
from different generations of stars \citep{Spitoni15a,Schonrich17}, or prominent radial
flows of gas that can dilute the central regions at early
epochs \citep{Spitoni11,Mott13,Spitoni13}, or strong feedback (in the form
of outflows)
at early epochs has redistributed the metals produced
by the central active region in the 
circumgalactic medium and towards the external region of galactic discs.
This is not unreasonable given that
recent observations have revealed that as much as 40\%
of the metals produced by galaxies have been expelled in their halo (see Sect.~\ref{sec:metal_budget}) and that the circumgalactic medium
has already been significantly enriched ($\rm Z\sim 0.1~Z_{odot}$) by $z\sim 2$
\citep{Prochaska13}. 
Fig.~\ref{fig:met_grad_model_Mott13}  shows the effect of introducing radial flow in the analytical models proposed by \cite{Mott13}.
Interestingly, in this models the high inflow rate of pristine
gas towards the central regions even inverts the gradient
centrally, nicely reproducing the inverted gradients observed
at high redshift
\citep[Fig.~\ref{fig:grad_inv_Cresci2010}][]{Cresci10,Troncoso14}.

Zoom-in cosmological simulations have become increasingly
popular to investigate the evolution of metallicity
gradients in galaxies \citep{DiMatteo09,Torrey12,Pilkington12,Gibson13,Tissera16a,Tissera16b,Tissera17,Tissera18}, although these high resolution simulations are computationally expensive and, therefore, typically results are shown for relatively small samples of simulated galaxies. Simulations typically obtain reasonably metallicity gradients, although some of them struggle to reproduce
the observed mass dependence. A common outcome is that galaxy
interactions tend to flatten the metallicity gradients at any
epoch \citep{DiMatteo09,Ma17a}. There is typically a large scatter and rapid variations in the gradients observed
in simulations especially at high redshift; \cite{Ma17a} suggest
that the observed gaseous metallicity gradients reflect more
the current status of the galaxy rather than tracing its past
cosmological evolution.
Many simulations tend to have the same problem as analytical
models in reproducing the temporal steepening of metallicity
in regular, rotating galaxies. Within this context the
most successful models are those that introduce enhanced feedback
from star formation, which redistributes the metals across the
galaxy more effectively in the early, actives stages of galaxy formation \citep{Gibson13}.

\subsection{Summary of metallicity gradients in galaxies}

In this section we summarize some of the key results regarding
the spatially resolved distribution of metals in galaxies.

Radial metallicity gradients tend to steepen in more massive
galaxies. This is observed both for the gas phase metallicity and for the stellar
metallicity (although with larger scatter).

  However, metallicity gradients are not described by a single
slope, they generally flatten at high galactocentric radii (suggestive
that galactic outskirts have accreted accreted pre-enriched material)
and, in massive galaxies, they tend to flatten also in the central region (likely because of
metallicity saturation).

There is evidence that the metallicity is not simply a function of galactocentric radius,
but it depends on the local physical properties of the galactic disc. More specifically, the
gas metallicity is found to correlate with
the surface density of stellar mass and to anti-correlate with
the surface density of star formation rate and
with the surface density of gas. Whether these local scaling relations are responsible for
the global metallicity scaling relations has yet to be properly assessed.

The use of indicators that trace the metallicity in galaxies at different lookback times, as well as 
the direct observation of metallicity gradients at high redshift, indicate that metallicity gradients
were flatter in the past (which is also consistent with the local steepening of metallicity
gradients as a function of stellar mass, if one consider low mass galaxies progenitors of more massive ones).
The fact that radial metallicity gradients steepen with cosmic time is in contrast with the simplest
 expectations of the
chemical enrichment in the
scenarios in which galaxies grow inside-out (as confirmed by independent observations),
which would expect a flattening of the metallicity gradients with time. Solutions to this issue
involve either stellar radial migration, or prominent early radial inflows of low-metallicity gas
or feedback (outflow) effects that redistribute the metals produced in the central regions
in the circumgalactic medium and towards the outer galactic regions.

It is finally interesting to note the growing evidence for a population of high-z galaxies
with inverted (i.e. positive) radial gradients. These may be tracing systems in which
prominent inflows of low metallicity gas is taking place or which are associated
with galaxy merging/interaction which are effective in driving metal-poor gas
from the outskirts towards the central region, hence flattening or even inverting metallicity
gradients.

\section{Relative chemical abundances}
\label{sec:abund_ratios}

Since different elements are produced by different classes
of stars/SNe and released on different timescales (see Fig.~\ref{fig:metal_production}), the relative
abundances between different chemical elements provide precious
information on the star formation history and on the IMF.

The ratio between $\alpha$-elements, which are primarily produced
on short timescales by massive stars through core-collapse SNe,
and iron-peak elements, which are primarily produced by SN Ia on
longer timescales, is a classical example of tracer of star formation
history. Stellar populations characterized by ``enhanced''  $\alpha$/Fe must
have formed on short timescale, before that SN Ia had time to enrich
the ISM, while stellar populations characterized by ``low'' or solar-like 
 $\alpha$/Fe must have formed over a prolonged phase of star formation.
Other chemical elemental ratios, including CN/Fe \citep{Carretero04,Carretero07} have a similar potential of constraining
 the star formation history.

The detailed abundance pattern of several chemical elements can potentially
provide the fingerprints of the specific stellar progenitors that have
been responsible for enriching the ISM. Potentially they can even provide
the signature of the enrichment by the first population of stars
\citep{Caffau11,Frebel15,deBennassuti17}.

In this section, after a rapid overview of the metal abundances observed
in the Milky Way, we will focus primarily on the chemical abundances observed
across the galaxy populations, locally and at high redshift.
A detailed analysis of all chemical elements would require a full, dedicated
review. We will therefore focus on the analysis of the most commonly
used chemical elements and, in particular, $\alpha$/Fe, N/O and C/O,
although we will also mention some other chemical abundances ratios.

\subsection{The Milky Way}
\label{sec:abund_MW}

Extensive spectroscopic surveys have enabled astronomers
to map the relative chemical
elements of large numbers of stars in the Galaxy \citep[e.g., SEGUE, RAVE,
APOGEE, GAIA-ESO, HERMES-GALAH, ][]{Yanny09,Steinmetz06,Gilmore12,Freeman12,Majewski17}.
High spectral resolution observations have provided
unprecedented constraints on the relative abundances of most chemical
elements, although on smaller samples of stars. 
\citep[e.g. ][ and references therein]{Reddy06,Bensby10,Nissen10,Johnson14,Zoccali17}.
The advent of high resolution
near-IR spectroscopic surveys has further enabled astronomers to trace
the chemical abundances in the inner bulge and for metal-rich cool stars, whose
heavy metal lines blending and blanking at optical wavelengths
makes it difficult to measure chemical abundances with classical optical
spectroscopy \citep[e.g.][]{Rich12b,Onehag12,Lindgren16}.

\begin{figure}
\centerline{\includegraphics[width=12cm]{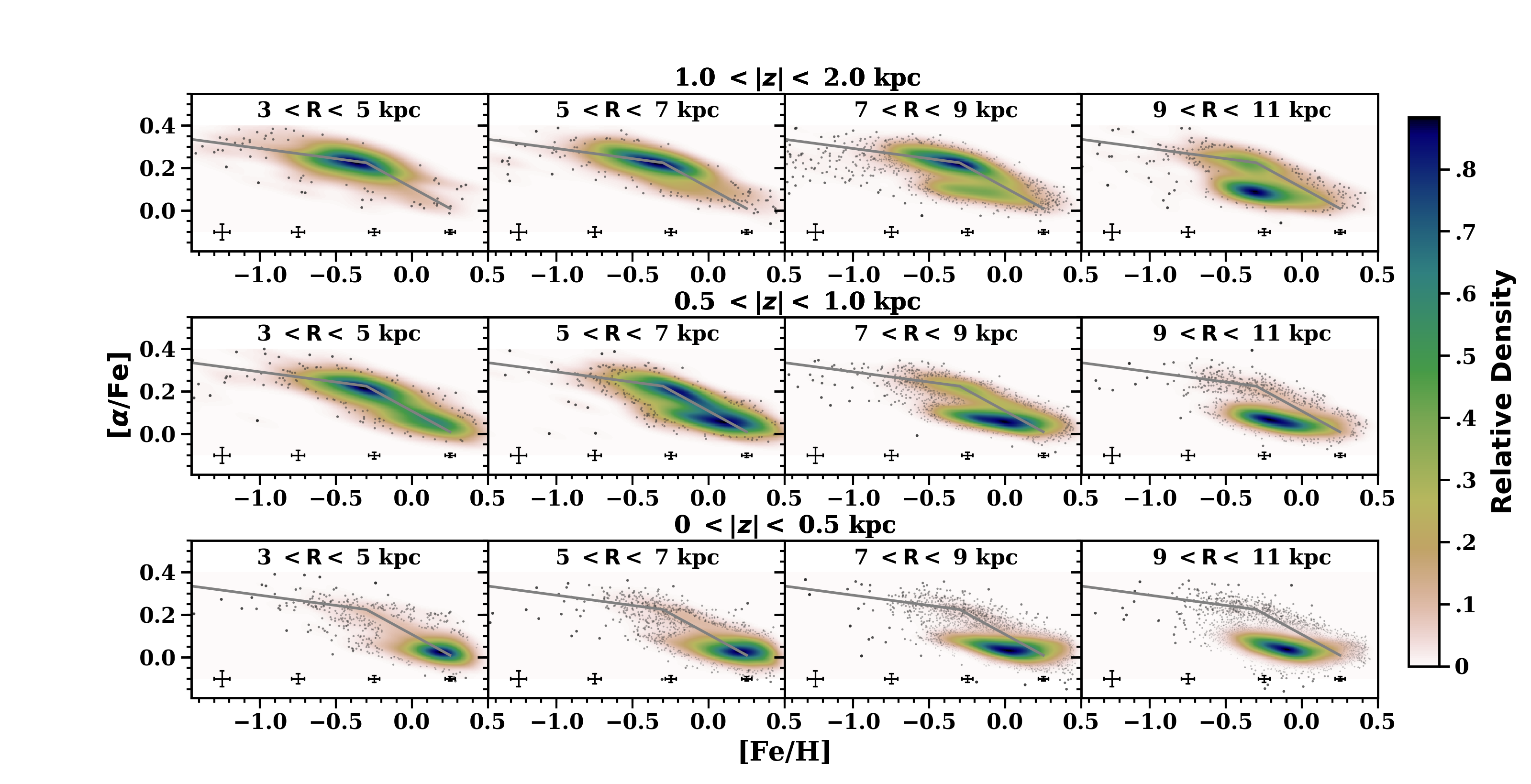}}
\caption{[$\alpha$/Fe] versus [Fe/H] in the Galactic disc stars in bins of galactocentric
distance and vertical distance from the Galactic plane from a sample of $\sim$70.000 red giants observed by the SDSS-III/APOGEE survey, from \cite{Hayden15}.
}
\label{fig:abund_MW_Hayden2015}
\end{figure}

An overview of the [$\alpha$/Fe] versus [Fe/H] for the Galactic disc
is given
in Fig.~\ref{fig:abund_MW_Hayden2015}, obtained with APOGEE near-IR, medium resolution
spectroscopic data of $\sim$70,000 red giants
\citep{Hayden15}. $[\alpha$/Fe]--[Fe/H]
diagrams are very useful in identifying stellar populations resulting
from different star formation histories. Indeed, as mentioned
above, the [$\alpha$/Fe] ratio works like a clock of the star formation history,
indicating  how rapidly star formation has occurred, while the [Fe/H] ratio
distributes stellar populations along their temporal evolutionary sequence (of
course, depending on the star formation efficiency stellar populations spread
more or less quickly along this axis).
In Fig.~\ref{fig:abund_MW_Hayden2015} different panels conveniently
show the [$\alpha$/Fe]--[Fe/H] distribution in bins of galactocentric
radius and distance
from the Galactic plane, illustrating that
the stellar populations in the disc
have a bimodal distribution, which has been identified with the
thin and thick disc. The thin disc, which dominates the stellar population
on the disc plane and at intermediate and large radii, is characterized
by a rather flat [$\alpha$/Fe]--[Fe/H] distribution indicative of a prolonged
star formation history, on timescales of about 7~Gyr, probably
also associated with normal star formation efficiency (typical of disc galaxies)
\citep{Matteucci86a,Micali13,Ryde16}. The thick disc population, dominating at higher vertical distances
from the Galactic plane and in the inner disc, is clearly $\alpha$-enhanced (and,
on average, more metal poor than the thin disc), indicating that the thick
disc was formed faster, on a timescale of about $\sim 2$ Gyr, and likely
with higher star formation efficiency. It is important to highlight that the existence of thick-thin disc bimodality is still debated. In particular, \cite{Bovy12} claim that there is not a real bimodality but a smooth, continuous distribution if one considers the mass-weighted scale-height distribution of stellar populations.\\

\begin{figure}
\centerline{\includegraphics[width=8cm]{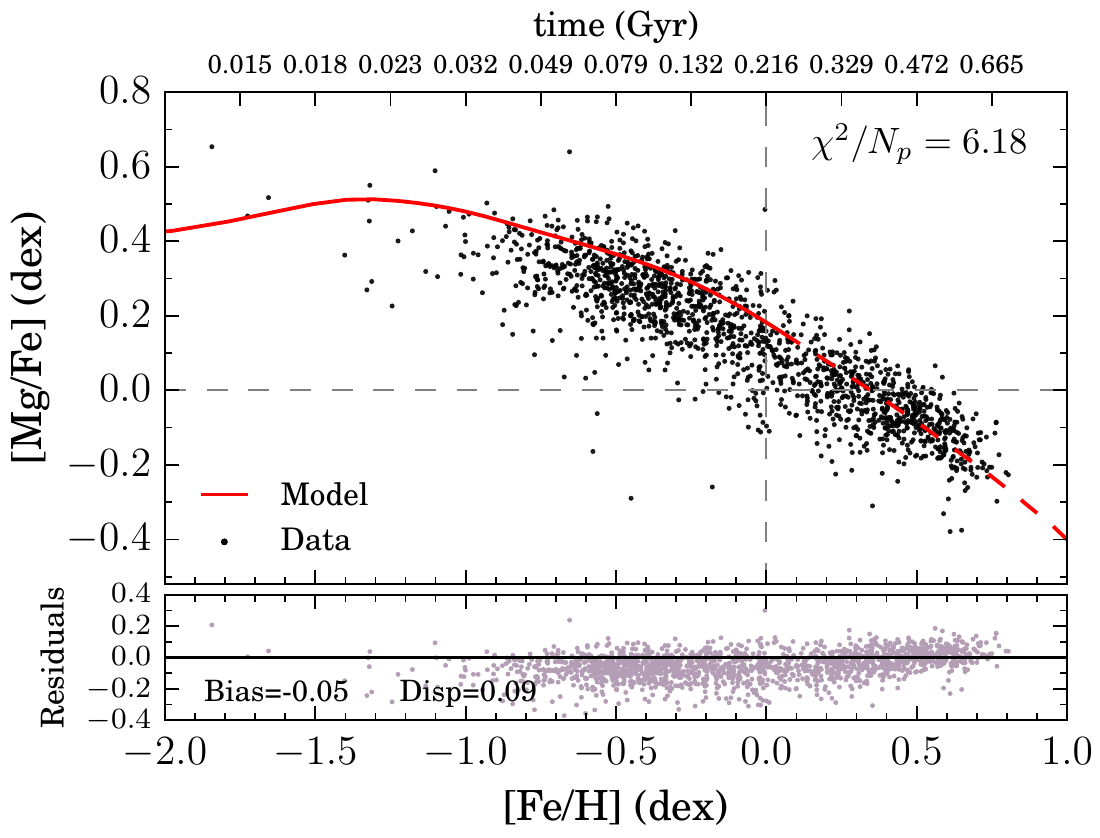}}
\centerline{\includegraphics[width=8cm]{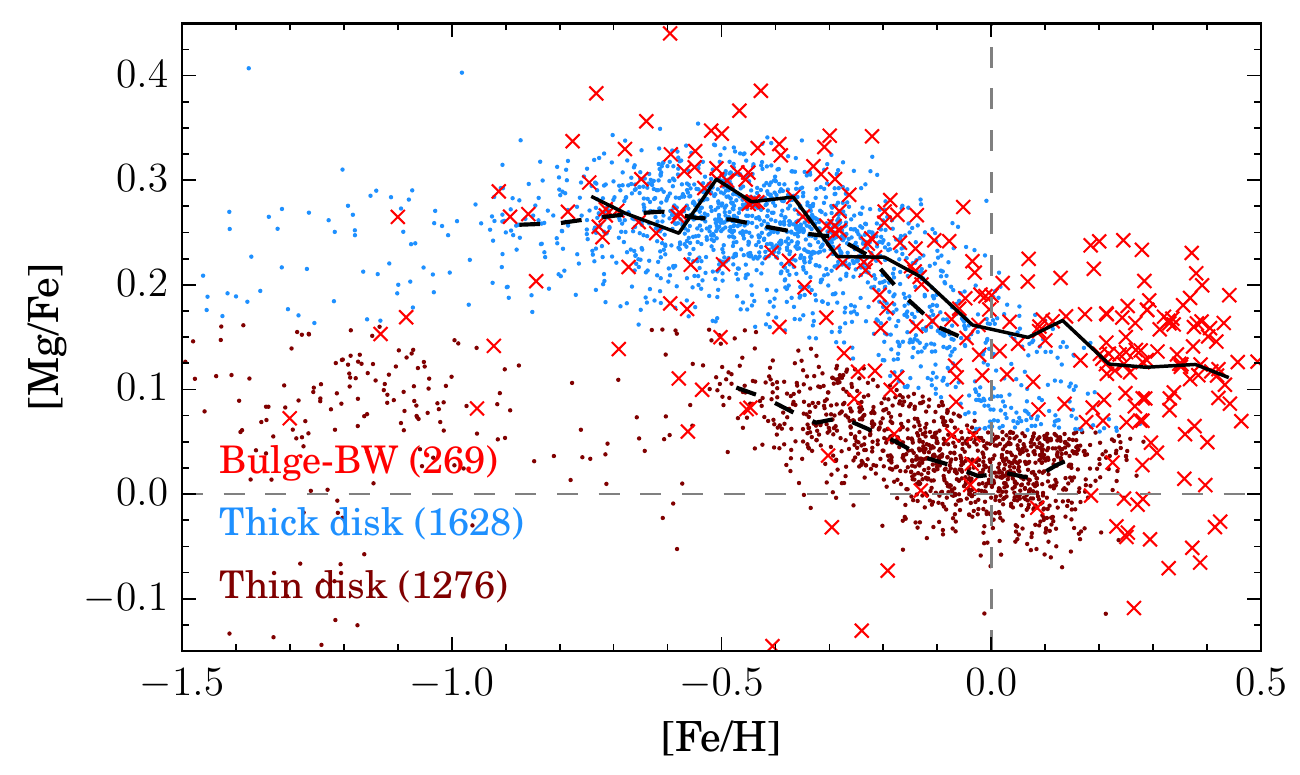}}
\caption{[$\alpha$/Fe] versus [Fe/H] in the Galactic bulge from the Gaia-ESO (optical) survey
\citep[top, ][]{Rojas-Arriagada17} and from the APOGEE (near-IR) survey \citep[bottom, ][]{Schultheis17}. 
}
\label{fig:abund_halo}
\end{figure}

There has been growing evidence that the bulge stellar population is bimodal in metallicity
\cite[e.g.,][]{Rojas-Arriagada17}, hosting both a sub-solar population and a supersolar population.
The two populations also have different [$\alpha$/Fe] enrichment levels, as illustrated in
Fig.~\ref{fig:abund_halo}-top. The low-metallicity component chemical properties are consistent
with those of the thick disc. The high metallicity component appeared similar to the thick-disk based
on the optical spectroscopic surveys, but when measured through near-IR data (i.e., consistent
with the disc data shown in Fig.~\ref{fig:abund_MW_Hayden2015}), more adequate to
probe the high metallicity component of the bulge, also the high metallicity component appear
clearly $\alpha$-enhanced relative to both the thick and thin discs, as illustrated
in Fig.~\ref{fig:abund_halo}-bottom \citep{Schultheis17}. The emerging picture is that the high metallicity component
was formed through the same secular process as the thick disc, and actually associated with the
inner Galactic bar, while the low metallicity, $\alpha$-enhanced component was formed quickly (within
less than about 0.5~Gyr), with high efficiency (which enabled quick enrichment), as a consequence
of the initial gravitational collapse of the galaxy.

The halo is also $\alpha$-enhanced but characterized by even more metal poor stars
\citep{Cayrel04,Frebel15} and it has been suggested that its chemical properties are also consistent
with the early, fast, and efficient collapse of the galaxy \citep{Micali13},
although more recent data point at two 
stellar populations, differentiated both chemically and kinematically \citep{Carollo07,Fernandez-Alvar18},
corresponding to faster and slower star formation histories, respectively, the former possibly also
associated with a top-heavier IMF.

\begin{figure}
\centerline{
\includegraphics[width=5.9cm]{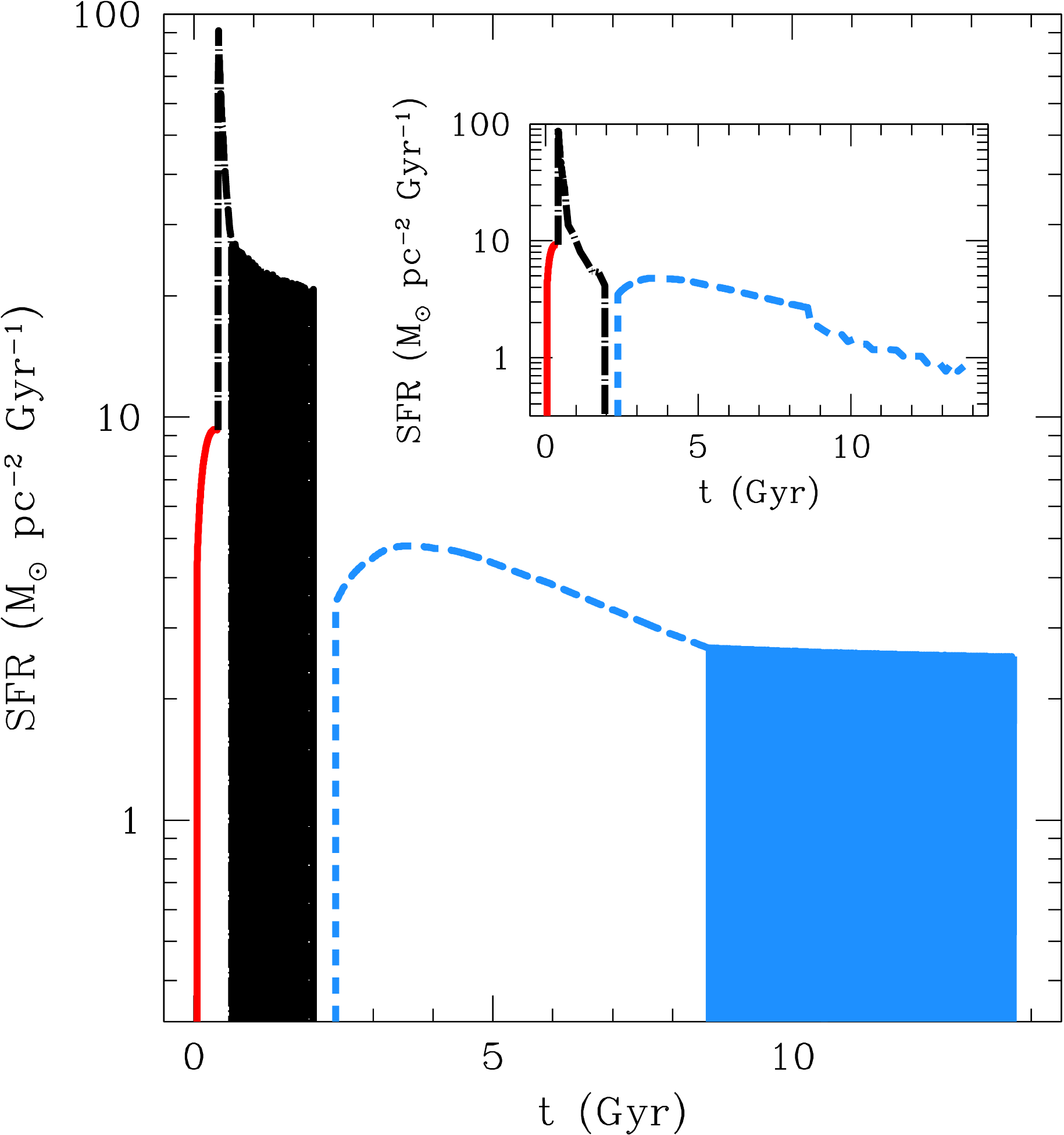}
\includegraphics[width=5.9cm]{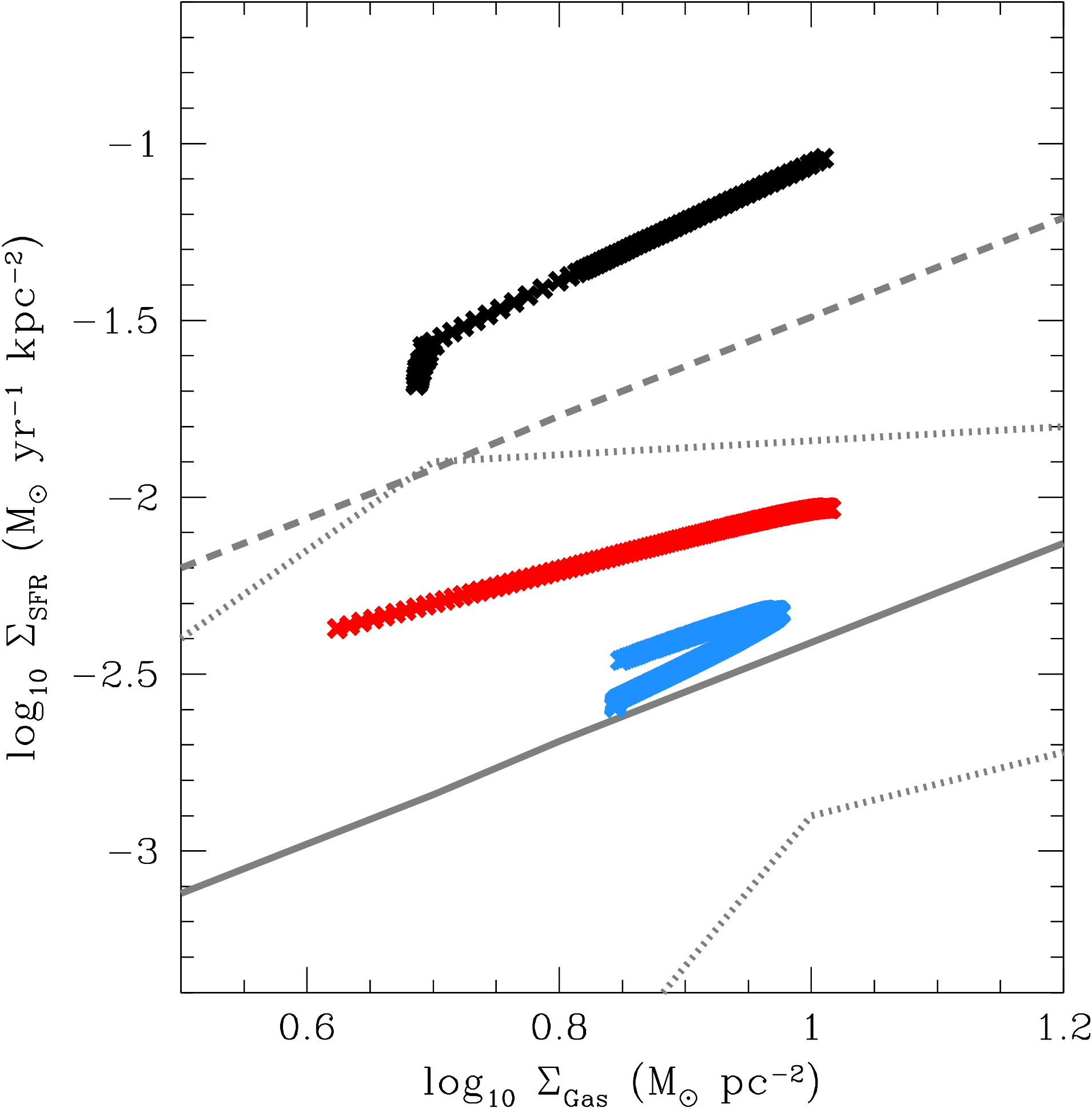}
}
\caption{Galaxy's three-phases infall model \citep[3IM,][]{Micali13}.
Left: temporal evolution of the SFR as predicted by the 3IM. The (red) solid portion of the curve refers to the halo
phase, the (black) dot-dashed one to the thick-disc phase, the (blue) dashed one to the thin-disc phase. The colored bands are due to a large number of subsequent, rapid variations.
Right: SFR density versus gas density in the halo (red crosses), thick disc (black crosses) and thin disc (blue crosses). Also
shown are the fit to the $\rm \Sigma _{\rm SFR}-\Sigma_{\rm gas}$
relation for local spirals and $z=1.5$ BzK galaxies (grey solid line), the extrapolation of the starburst sequence from the same authors (grey dashed line) and the region of the plot occupied by spiral galaxies data (delimited by the dotted grey lines). 
}
\label{fig:abund_Micali13}
\end{figure}

Overall, among analytical models, the abundances and metallicities in the various components of the Milky Way can be explained
well in the framework of the so-called two- or three-phase infall model \citep[Fig~\ref{fig:abund_Micali13}, ][]{Micali13,Chiappini97, Chiappini01,Chiappini05}
in which the halo and
old bulge component have formed very fast (within less than 0.5~Myr) and very efficiently (above the Schmidt--Kennicutt
relation), the thick disc (and lower metallicity component of the bulge) have formed in a second infall event
(on intermediate timescales of $\sim 2$~Gyr) and with even higher efficiency, while the thin disk has formed
on much longer timescale ($\sim 7$~Gyr), following the Schmidt--Kennicutt relation for normal star forming discs. Both
the thick and (especially) the thin disc must also have had 
a radially variable inflow rate.
Alternative models by \cite{Schonrich09a, Schonrich09b} reproduce the multiple components as the effect of different star formation conditions in different parts of the disc followed by radial mixing of stars.\\

{\bf Summarizing}, different morphological components of the Milky are characterized
by different chemical abundance patterns, especially for what concerns $\alpha$/Fe,
indicating different formation histories and different formation processes.
More specifically, halo, bulge, thick and thin disc are characterized by gradually
later ages of star formation, on gradually longer timescales, and likely
associated with different star formation efficiencies. Whether these were
physically distinct phases (yielding to distinct, different populations) or part of a more smoother, continuous evolution is not totally
clear.

\subsection{Local galaxies}
\label{sec:abund_local}

\subsubsection{$\alpha$/Fe}
\label{sec:alpha_over_iron}

As introduced above, [$\alpha$/Fe] is sensitive to the number ratio of CC to thermonuclear SNe and, therefore, is a powerful way to study the star formation timescales of galaxies. It is also sensitive to a number of parameters such as the IMF, the assumed stellar yields, the delay time distribution of type Ia SNe, and the differential ejection of metals into the CGM.
In simple, close-box, constant-IMF models, [$\alpha$/Fe] is expected to evolve from a high value when, at early times and low metallicities, (little) iron production is dominated by CC events, toward a lower value, when iron produced by Ia events becomes dominant.
The faster star formation occurs, the higher is the enrichment of $\alpha$-elements by CC SNe before that SNIa start polluting the ISM with iron.

Beyond the MW, chemical abundances of individual stars (giants/supergiants) has been determined only for a few galaxies of the Local Group, mostly dwarf satellites of the MW and Andromeda
\citep[e.g.][]{Bonifacio04,Monaco05,Sbordone07,Tolstoy09,Cohen10,Kirby11,Hill12,Starkenburg13,Hendricks14}.
Dwarf spheroidal galaxies
are characterized by a distribution
on the $\alpha$/Fe vs Fe/H diagram that
is below the plateau observed in MW halo and thick disc stars, joining the $\alpha$/Fe abundance ratio of MW halo stars only at [Fe/H]$<$-1.5. The presence of a knee in the distribution is still debated \citep{Tolstoy09,Kirby11,Hendricks14}. This distribution implies that SNIa have contributed to the enrichment of the stars in these systems at most metallicities, i.e. during most of the formation of these systems, except for the their earliest phases, suggesting a bursty evolution possibly resulting from a sequence of minor accretion or merging events. Dwarf irregulars (e.g. the Small Magellanic Cloud), which are still in the process of actively forming stars, are also characterized by low $\alpha$/Fe abundance ratio, indicating that they have been forming stars slowly, stochastically and/or inefficiently
\citep{Matteucci83,Recchi01}. As a consequence of their shallow gravitational potential, in dwarf galaxies supernova-driven winds are also expected to play a more important role,  with respect to more massive galaxies, in regulating the chemical enrichment history, by removing metals (likely in a differential way, i.e. preferentially $\alpha$-elements), by reducing the efficiency of star formation and by contributing to its stochasticity.

Except for the few galaxies in the Local Group, the bulk of the investigation of the chemical abundances
of the stellar population in galaxies has been based on spatially integrated spectra, unavoidably implying
larger uncertainties and degeneracies. However, despite these caveats, large spectroscopic
surveys have enabled us to investigate the relative chemical abundances across a broad range of galaxy
masses and environments.

\begin{figure}
\centerline{\includegraphics[width=12cm]{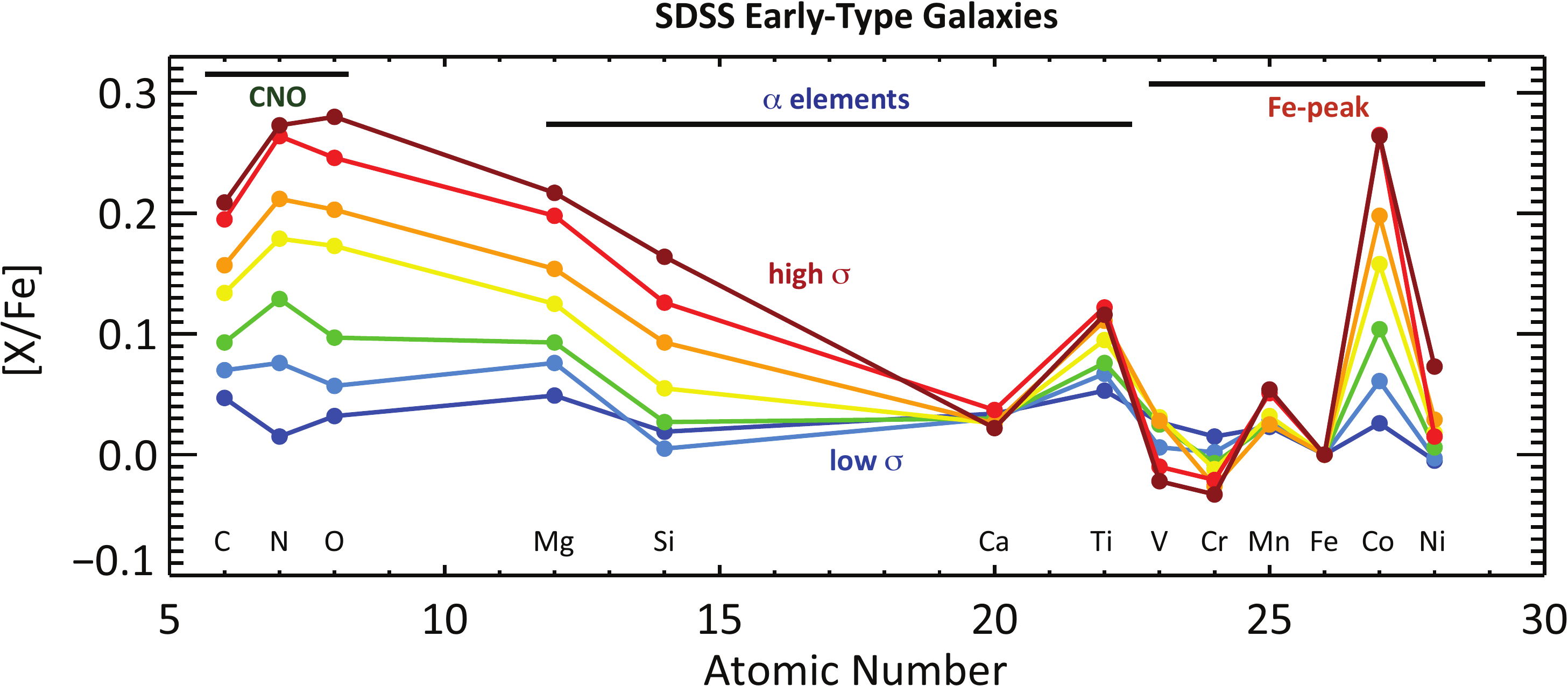}}
\caption{Elemental abundances relative to iron inferred 
by \cite{Conroy14}
from the stacked spectra of local early-type
galaxies in bins of velocity dispersion (hence dynamical mass), from $\sigma=90$ km/sec (blue) to $\sigma=270$ km/sec (red). 
}
\label{fig:abund_Conroy14}
\end{figure}

\begin{figure}
\centerline{\includegraphics[width=11cm]{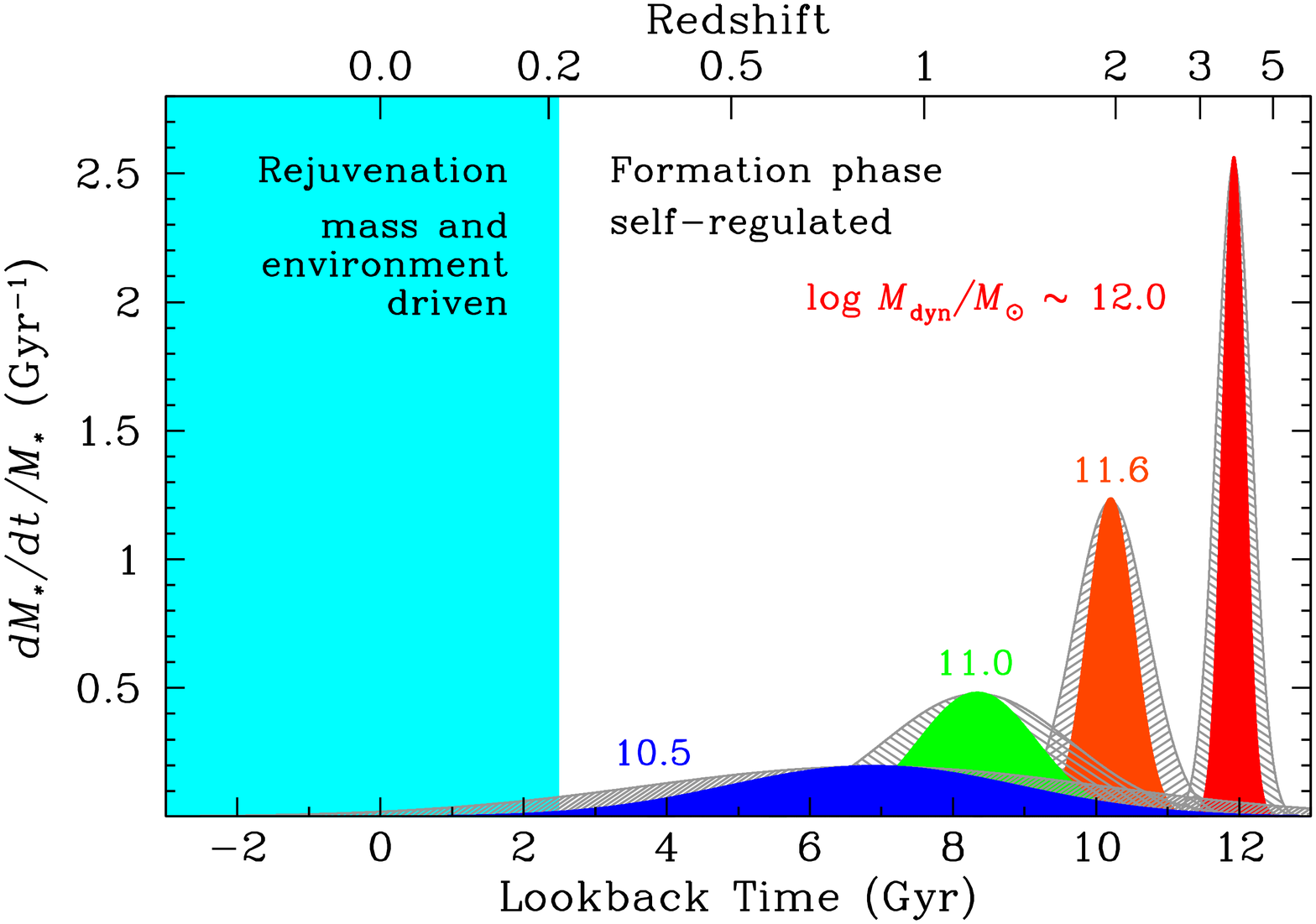}}
\caption{Representation of the specific star formation rate as a function of look-back time for galaxies of the different masses shown in the labels
\citep{Thomas10}.
}
\label{fig:abund_Thomas10}
\end{figure}

Initial works have investigated chemical abundances using primarily 
the Lick indices and focusing on early type galaxies to constrain their evolutionary
history. Such early works already identified that early-type galaxies have enhanced $\alpha$
elements compared to the abundance patterns of the stars in the Galactic disc \citep{Worthey94b,Thomas99a}.
It was further found that such $\alpha$-enhancement increases steadily as a function of stellar velocity
dispersion, which is a tracer of galaxy mass \citep{Trager00b,Thomas05}. Such a trend has been clearly confirmed
by the full-fitting (i.e., not limited to the Lick indices) of the spectra of local
galaxies \citep{Walcher09, Conroy12} as illustrated in Fig.~\ref{fig:abund_Conroy14}, where
the abundance of different chemical elements relative to iron is obtained from Sloan 
spectra stacked in bins of velocity dispersion \citep{Conroy14}. This trend indicates that more massive
galaxies have formed much more rapidly than low mass galaxies.
More specifically,
translating the velocity dispersion into galaxy mass,
and combining the $\alpha$-enhancement information with the age of the
stellar population (more massive galaxies are typically older) results into the scenario originally
proposed by \cite{Matteucci87} and \cite{Matteucci94}, further developed by later studies, and summarized  in
Fig.~\ref{fig:abund_Thomas10} from \cite{Thomas10} \citep[see also ][]{Thomas05},
in which more massive galaxies formed at earlier cosmic
epochs (a phenomenon often referred to as ``cosmic downsizing''), on shorter timescales and (based on models)
more efficiently. This scenario, at least in terms of timeline sequence, has been verified through the
evolution of the mass function of galaxies at high redshift, illustrating that most massive galaxies were already
in place at early cosmic epochs, while lower mass galaxies have evolved more slowly
\citep[e.g.,][]{Gavazzi96,Cowie96,Perez-Gonzalez08,Muzzin13,Santini15}.
Theoretical models and numerical simulations explain this phenomenon in terms of accelerated evolution
in the overdense regions of the Universe, where baryons collapse more rapidly in the deepest
gravitational potential wells of dark matter. The enhanced star formation efficiency in these dense regions facilitates the
rapid formation of stars and rapid gas consumption. Moreover, the strong negative feedback from the resulting supernovae
and rapid black-hole accretion (releasing large amount of energy through the luminous quasar phase)
result into rapid quenching of the star formation  \citep{Matteucci87,Matteucci94,Pipino11,Segers16,De-Lucia17}.
Additional theoretical explanations involve 
the effect of varying IMFs \citep{Fontanot17b}.


Early-type galaxies show rather flat radial gradients in terms of [$\alpha$/Fe], or slightly positive \citep[e.g.,][]{Greene13,Roig15}. 
If confirmed, the positive gradients of $\alpha$/Fe can be explained through models in which
the quenching in the outer parts of massive galaxies results primarily through explosion of supernovae, whose
cumulative injection of energy is more effective in ejecting gas in the shallower gravitational
potential of the galaxy outskirts \citep{Pipino08,Pipino10}.
However, such outside-in quenching effect would be in contrast with
numerical simulations that expect massive ellipticals to grow inside-out
through a sequence of minor (dry) mergers from z$\sim$2 \citep{Naab07,Naab09}, also
supported by observed size growth of elliptical as a function of redshift
\citep{vanDokkum10}, although the latter has also been interpreted
as an observational effect in terms of ``progenitors bias'' \citep{Lilly16}. However, if the positive radial gradients of $\alpha$/Fe in massive elliptical
galaxies are confirmed with high significance, then this would be problematic
to explain in the minor mergers scenarios for the size growth of ellipticals.

Reproducing the $\alpha$/Fe enhancement in massive galaxies
requires processes that enable the rapid production of stars and that then
quench star formation on relatively short timescales ($\sim 0.5-1$~Gyr).
In analytical models this is achieved by requiring a high star formation
efficiency (which makes the formation of stars and enrichment of $\alpha$ elements faster) and then a quenching effect that suppress star formation
\citep[either through SNe or AGN feedback, e.g.][]{Matteucci87,Matteucci98,Romano02,Pipino04,Pipino08}.
In cosmological simulations the introduction of quasar feedback seems to
reduce the lifetime of massive galaxies enough to reproduce the
relationship between $\alpha$/Fe enhancement and galaxy mass
\citep{Segers16}. However, more recently \cite{DeLucia17} have pointed out
that AGN feedback alone may not be capable to simultaneously reproduce
the $\alpha$/Fe enhancement and other galactic properties, leaving the problem open.
Within this context, we note that it is not necessary to invoke the ``ejective'' mode of quasar feedback (i.e. removal of gas through massive quasar-driven outflows) in order to achieve a rapid quenching of star formation; indeed,  a scenario in which
the galaxy is simply``starved'' (e.g. because its surrounding halo has been heated) 
would also result in a rapid cessation of star formation, as in distant powerful starburst galaxies (such as the Submillimeter Galaxies, which are often regarded as the progenitors of local massive ellipticals) the gas depletion times (by the simple effect of the highly efficient star formation) are as short as a few hundred million years.

\begin{figure}
\centerline{\includegraphics[width=12cm]{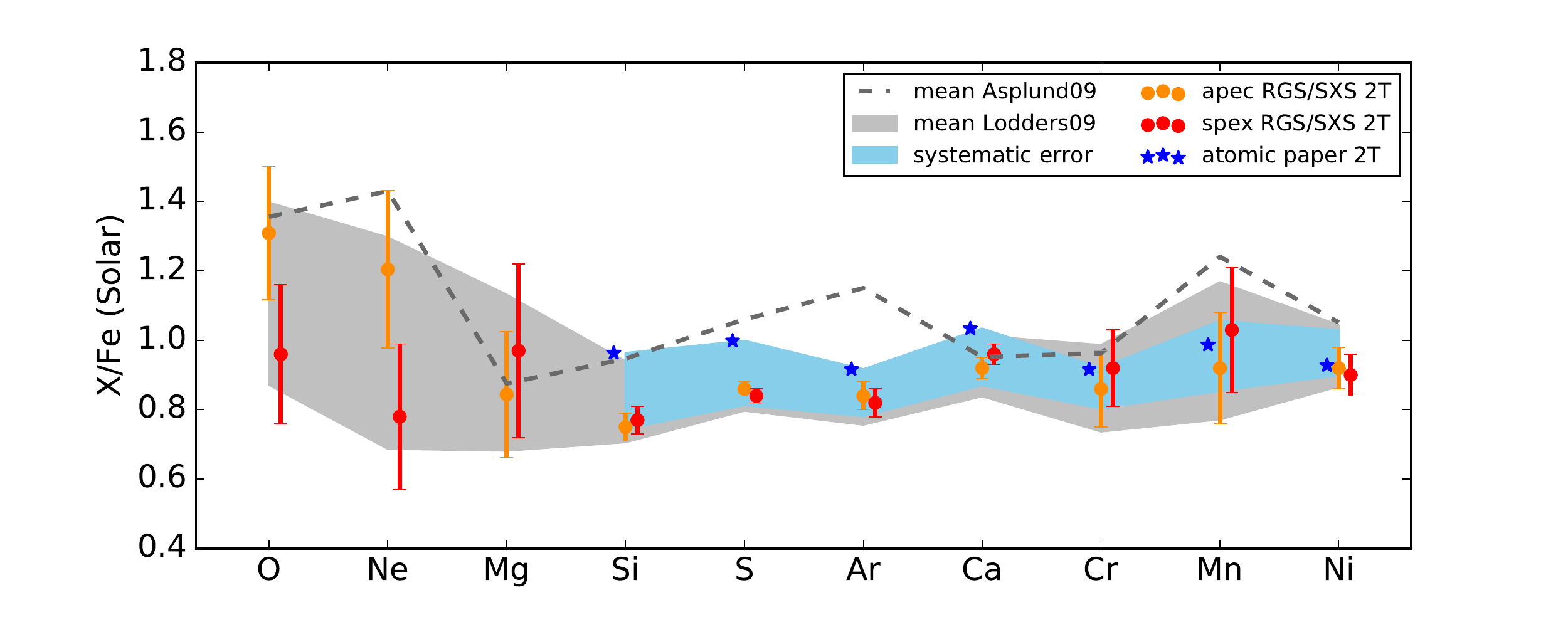}}
\caption{Chemical abundances relative to iron inferred for the ICM of the Perseus cluster, obtained with XMM-Newton RGS for O, Ne and MG, and with the Hitomi SXS for Si to Ni. From \cite{Simionescu18}.
}
\label{fig:Xray_Perseus}
\end{figure}

X-ray spectroscopy offers the possibility of measuring the abundance of several chemical elements, including iron and $\alpha$-elements, of  hot plasmas ($\rm >10^6$~K). Therefore, X-ray spectroscopy is extremely important to investigate, for instance, the ISM and CGM heated by SNe and galactic winds, as well as the hot gas in clusters and groups of galaxies, although sensitivity, spectral and angular resolution issues have often limited the exploitation of this technique.

X-ray spectroscopy of the hot phase of the galactic superwind of the prototypical starburst galaxy M82 has revealed a much higher metallicity in the outer parts of the outflow and with a $\alpha$/Fe abundance ratio significantly higher than in the host galaxy, confirming that the outflow is associated with hot plasma freshly enriched by recent generation of core-collapse supernovae produced by the starburst event \citep{Ranalli08}. These results provide the direct observational evidence that starburst superwinds eject metals with velocities of a few/several hundred km/s hence directly enriching the the CGM and IGM.

X-ray spectroscopy of the hot plasma in galaxy clusters and galaxy groups have generally revealed Solar-like chemical abundances (Fig.\ref{fig:Xray_Perseus}) and a surprisingly flat radial distribution of the relative abundances \citep{Simionescu10,Simionescu15,Mernier17,Simionescu18}.
One should not confuse these abundances with those of the stellar populations in the galaxies belonging to the clusters (which may be $\alpha$-enhanced if they formed rapidly), indeed even if star formation may have stopped in the cluster's galaxies, SNIa keep enriching the ICM (with their typical high Fe/$\alpha$\ pattern) over time. \cite{Mernier17} estimate that, on average, the fraction of SNIa with respect to the total number of SNe (i.e. SNIa+SNcc) that have contributed to the enrichment of the ICM must be in the range of 20\%--40\%. Interestingly, \cite{de-Plaa17} find that the O/Fe abundance ratio does not depend on ICM temperature, therefore suggesting that the enrichment of the ICM is not related to cluster mass and that most of the enrichment has occurred before the ICM was formed.\\

Within the context of X-ray spectroscopy, we mention that
excellent high angular resolution maps of the metal enrichment of some
individual clusters have been performed, revealing very interesting substructures. For instance, \cite{Sanders16}
has obtained a detailed map of the metal enrichment of the Centaurus cluster revealing high metallicity blobs on scales of 5-10 kpc, which are likely tracing material uplifted by the AGN hosted in the central galaxy. \\

{\bf Summarizing},
the $\alpha$/Fe abundance ratio is an important clock to investigate the
star formation history in galaxies. Massive galaxies are characterized by systematically higher $\alpha$/Fe ratio that,  together with information on their
stellar ages, indicates that more massive galaxies formed on shorter timescales and at earlier cosmic epochs than lower mass galaxies. The rapid star formation in massive systems is generally modelled in terms of a combination of enhanced
star formation efficiency and strong feedback effect that rapidly quench star formation. The intracluster medium typically shows solar abundances, with
no significant radial variation,  reflecting the additional, continuous ejection
of iron by SNIa over time also from passive galaxies.

\begin{figure}
\centerline{\includegraphics[width=10cm]{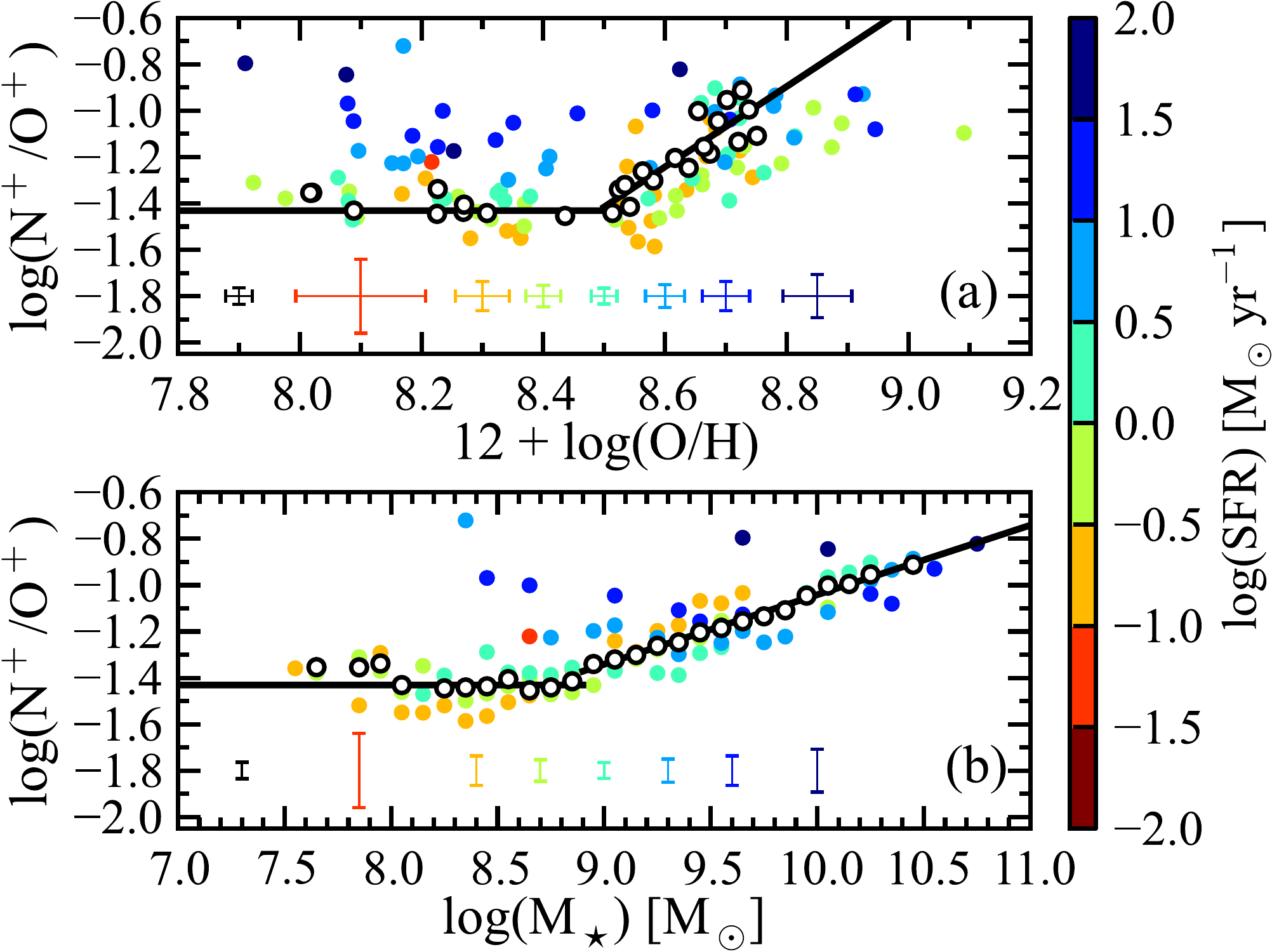}}
\caption{N$^+$/O$^+$ ratio as a function of direct-method oxygen abundance (a) and M$_*$ (b), from \cite{Andrews13}.
Open circles are obtained from SDSS stacks binned in stellar mass, colored symbols result from stacks in mass and SFR. 
}
\label{fig:NO_Andrews13}
\end{figure}

\subsubsection{N/O}
\label{sec:N/O}

In late-type galaxies, the analysis of chemical abundances has often focused on the gas phase, and on those elements
whose abundance can be inferred through the nebular emission lines (although DLA have also been extensively used to explore in detail the circumgalactic medium and outer parts of galactic discs, as discussed later on). Among these, nitrogen is one of the elements
that has been subject to many extensive studies. Indeed, its abundance can be inferred from the relative bright
doublet of [NII]$\lambda \lambda$6548,6584 next to H$\alpha$. The nitrogen abundance can be inferred `directly' through the T$_e$
method, through auroral lines tracing the temperature of the partially ionized zone
\citep[e.g.,][]{Andrews13,Pilyugin10a,Perez-Montero09,Berg11,Berg13,Berg15a}, see Sect.~\ref{sec:measmet_ism_Te}. As noted earlier, the latter method
requires the detection of very faint lines, hence can be applied only to limited samples of nearby galaxies/HII-regions
or stacked spectra of galaxies. However, at least in star forming galaxies,
the N/O abundance ratio is nearly proportional to the \nii6584/\oii3727 line flux
ratio, as these lines are emitted from nearly the same zone in HII-regions, hence their ratio is little dependent on other
factors such as the ionization parameter and shape of the ionizing continuum. Therefore, since these are both
relatively bright lines in most star forming galaxies, the \nii6584/\oii3727 ratio can be used to investigate
the N/O abundance in relatively large samples of galaxies, and calibrations of this diagnostic have
been provided \citep[e.g.,][]{Perez-Montero09}. In the absence of \oii3727 (which requires a relatively large
wavelength range to be observed together with \nii6584), the \sii6717,6730 doublet can be used as an alternative
proxy of $\alpha$\ elements, so to infer N/S from \nii6584/\sii6717,6730 \citep{Perez-Montero09}.

Nitrogen is a particularly interesting element to investigate in galaxies, because,
 in contrast to oxygen and other $\alpha$-elements, it is produced primarily by intermediate-mass stars and only with a smaller contribution by
 massive stars \citep[possibly enhanced in the presence of stellar rotation][and references therein]{Vangioni18,Vincenzo18}
Therefore, the N/O ratio provides precious information on the evolutionary stage of the galaxy. Moreover,
nitrogen has also a ``secondary'' component, whose production increases with metallicity \citep{Edmunds78}; indeed,
being a product of the CNO cycle, its abundance increases at expenses of the C and O abundances.
As a consequence, at high metallicities the nitrogen abundance is expected to evolve quadratically with
the metallicity, $\rm N/H \propto (O/H)^2$ or, equivalently, $\rm N/O \propto O/H$.

The nitrogen-to-oxygen abundance ratio in nearby galaxies has been investigated by multiple studies
\citep[e.g.,][]{Edmunds78,Vila-Costas92,Vila-Costas93,Thuan95,van-Zee98a,Perez-Montero09,Pilyugin10a,Pilyugin12,
Perez-Montero13,Andrews13,Berg11,Berg13,Berg15a,Belfiore15,Belfiore17a}.
Figure~\ref{fig:NO_Andrews13}-top shows the nitrogen and oxygen abundances inferred
from the direct-T$_e$ method from SDSS galaxy spectra stacked in bins of SFR and mass; hollow points are in bins
of stellar mass, while colored points are further split in bins of SFR \citep{Andrews13}. At low metallicities the N/O abundance
is relatively constant (if one ignores galaxies with high SFR, which will be discussed later); this is the region where nitrogen is thought to mostly have a `primary' contribution from
massive stars. At $\rm 12+log(O/H) > 8.3$--$8.4$, N/O increases steeply with metallicity; this is the region where secondary
nitrogen production is thought to take over and where intermediate mass stars start to contribute significantly.

\begin{figure}
\centerline{
\includegraphics[width=12cm]{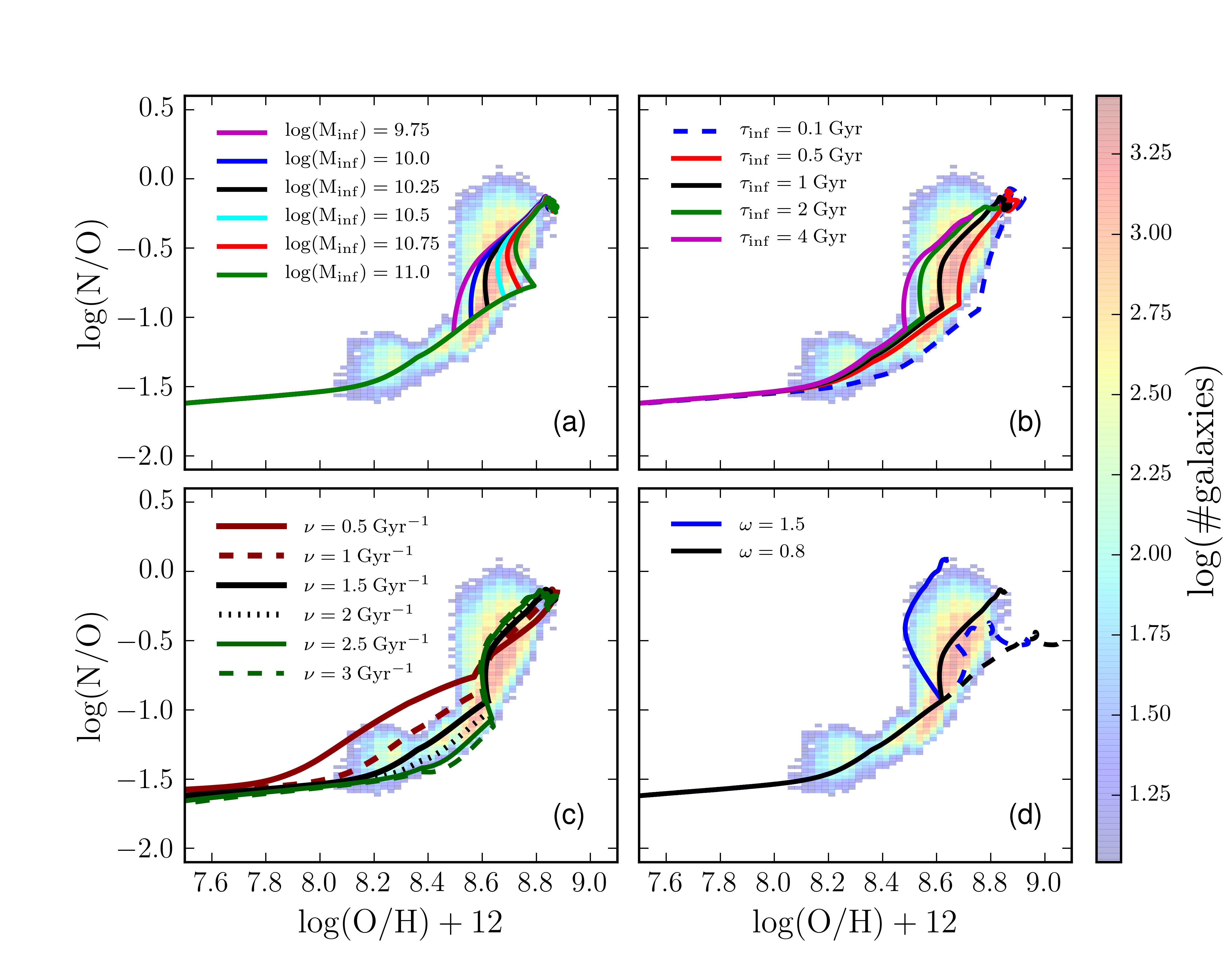}
}
\caption{Models of nitrogen enrichment in galaxies
overplotted onto the SDSS data (color shaded area), from \cite{Vincenzo16}.
(a) effect of varying the infall mass $M_{\rm inf}$; (b) effect of varying the infall time-scale $\tau _{\rm inf}$;
(c) effect of varying the star formation efficiency $\nu$; (d) effect of varying the outflow loading factor $\omega$.
In panel (d), the dashed lines correspond to a non-differential outflow (where both N and O are expelled with the same
efficiency) while the solid lines refer to the reference assumption of a differential outflow where N is not expelled (with
$\omega_{\rm N} = 0$). 
}
\label{fig:NO_Vincenzo16_2}
\end{figure}

A similar trend has been clearly observed also for the nitrogen abundance
of DLA, which  populate mostly the low metallicity plateau
\citep[e.g.][]{Pettini02c,Centurion03,Zafar14}.
However, for DLA there is evidence for a bimodal distribution of this plateau, with most
systems clustering at [N/$\alpha$]$\sim-$0.85 and a smaller fraction of them ($\sim$25\%)
clustering around [N/$\alpha$]$\sim-$1.4.
Such bimodal distribution may be related to the N/O spread observed in emission lines galaxies (especially the strongly star forming ones) at low metallicities (Fig.~\ref{fig:NO_Andrews13}).

For local galaxies most studies have focused mostly on the use of nebular emission lines in star forming galaxies.
We shall warn that a non-negligible number of studies investigate the distribution of galaxies on the
N/O versus O/H diagram by using a nitrogen-based strong line diagnostic as a tracer of O/H, such as
N2 or O3N2. This is a fundamental mistake that should be avoided by any means. Indeed, the use of the same information
(the [NII] line flux) on both axes unavoidably introduces artificial correlations between N/O and O/H. 
Moreover, strong line diagnostics based on theoretical photoionization models (and several Bayesian methods)
assume {\it a-priori} a relationship between N/O and O/H, therefore these strong line diagnostics cannot
be used to investigate the N/O vs O/H trends.

It has been pointed out by several authors that the N/O abundance ratio is a strong function
of the galaxy mass \citep{Perez-Montero09,Andrews13,Masters16}. This is, for instance, shown in the
stacked analysis by \cite{Andrews13} in Fig.~\ref{fig:NO_Andrews13}.
Such a trend with stellar mass is regarded as a  consequence of the fact
that massive galaxies are more evolved, hence the nitrogen enrichment contribution by intermediate mass stars
has been more prominent. However, it is also probably a secondary product of the mass-metallicity relation, indeed
the higher metallicity of massive galaxies likely boosts the production of secondary nitrogen.

Chemical evolutionary models have been proposed to interpret the evolution of the N/O abundance
\citep[e.g.,][]{Matteucci86b,Garnett90,Coziol99,Henry00,Chiappini05,Koppen05,Torres-Papaqui12}. The most recent effort in this area is from \cite{Vincenzo16} where
the observational data are compared with the different model predictions by varying different parameters
such as gas inflow properties, efficiency of star formation, outflow loading factor and also including
the scenario of differential outflow rates, in which oxygen is expelled more preferentially by the SN-driven winds
than nitrogen. Some of these models are shown in Fig.~\ref{fig:NO_Vincenzo16_2}, overplotted on the density distribution observed in several thousands galaxies from the SDSS survey (color shaded area). 
The broad distribution of
galaxies in the N/O versus O/H diagram implies that different galaxies have evolved through different paths.
However, models show that, on average, the global population of star forming galaxies require an initial
phase in which star formation in fueled by gas infall over a timescale of 1~Gyr, and that
outflows start being effective in quenching further enrichment when the galaxy has reached a metallicity
close to solar. The data seem to constrain the average past star formation efficiency to a value of about 
$\nu \sim 1.5$--$2\ {\rm Gyr}^{-1}$.

\begin{figure}
\centerline{\includegraphics[width=10cm]{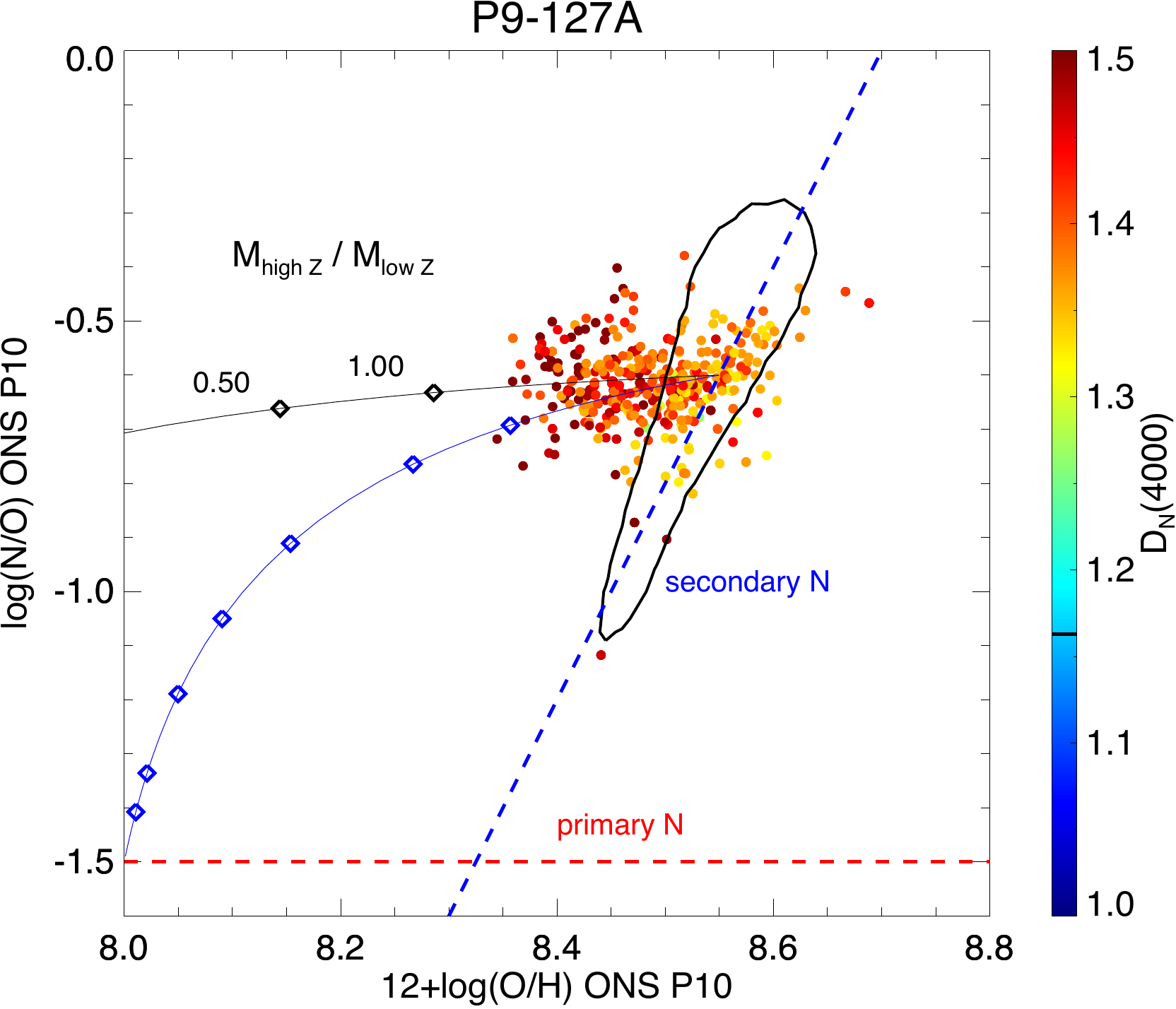}}
\caption{
Spatially resolved N/O vs O/H diagram for one galaxy in the SDSS4-MaNGA sample, \citep{Belfiore15}. Points are color-coded according to
their $D_{\rm n}(4000)$ parameter (strength of the 4000~\AA\ break), which is a tracer of the age of the stellar
population (redder points are older). The contours show the distribution in the SDSS sample (with the same calibrators
adopted for the galaxy). The blue and black solid lines with diamonds show the fountain mixing effect in which metal rich gas
expelled from the central region are mixed with metal poor gas from the outer regions, for two different values of the lower metallicity.
The deviation
from the main N/O-O/H main sequence observed in this galaxy can be reproduced fairly well through
such fountain mixing simple model.
The correlation of the deviation with the $D_{\rm n}(4000)$ parameter can
be explained with the fact that older regions are more gas poor hence the dilution effect is more effective. 
}
\label{fig:NO_Belfiore1}
\end{figure}

It is interesting to note that the characteristic shape of the N/O versus O/H diagram for star forming galaxies can be
very useful to identify secondary evolutionary effects at play when galaxies deviate from this sequence. In particular,
observations have shown that a significant fraction of galaxies and star forming regions tend to scatter
toward the high N/O region at fixed O/H \citep[e.g.,][]{Koppen05,Belfiore15}. One of such examples is illustrated
in Fig.~\ref{fig:NO_Belfiore1}, which shows the N/O versus O/H diagram for the spatially resolved star forming regions
of a galaxy in the SDSS4-MaNGA sample \citep{Belfiore15}. Such deviations can be explained through different scenarios:
1) a burst of star formation with increased star formation efficiency that, as shown in Fig.~\ref{fig:NO_Vincenzo16_2}c,
at later time boosts the N/O abundance relative to the sequence; 2) the infall of pristine/metal-poor gas at late
times; such an event dilutes the overall metallicity leaving unaffected the N/O abundance, hence moving the
galaxy/region horizontally on the diagram \citep{Koppen05}; 
3) a fountain scenario in which metal rich gas with high N/O abundance
is ejected from the central region and deposited on the outer galactic regions, which are more metal poor and have
lower N/O, resulting into a mixing sequence. Each of these scenarios is characterized by a different pattern
on the N/O versus O/H diagram. For instance, in the case of the galaxy shown in Fig.~\ref{fig:NO_Belfiore1} a fountain/mixing
scenario seems to describe well the deviations from the N/O main sequence in this system.
Another important mechanism that has been proposed to explain local enhanced nitrogen abundance in some star forming regions (especially thanks to IFU techniques) is the enrichment of nitrogen by Wolf-Rayet outflows
\citep{Walsh89,Lopez-Sanchez07,James09,Perez-Montero11,Monreal-Ibero12}.

\begin{figure}
\centerline{\includegraphics[width=9cm]{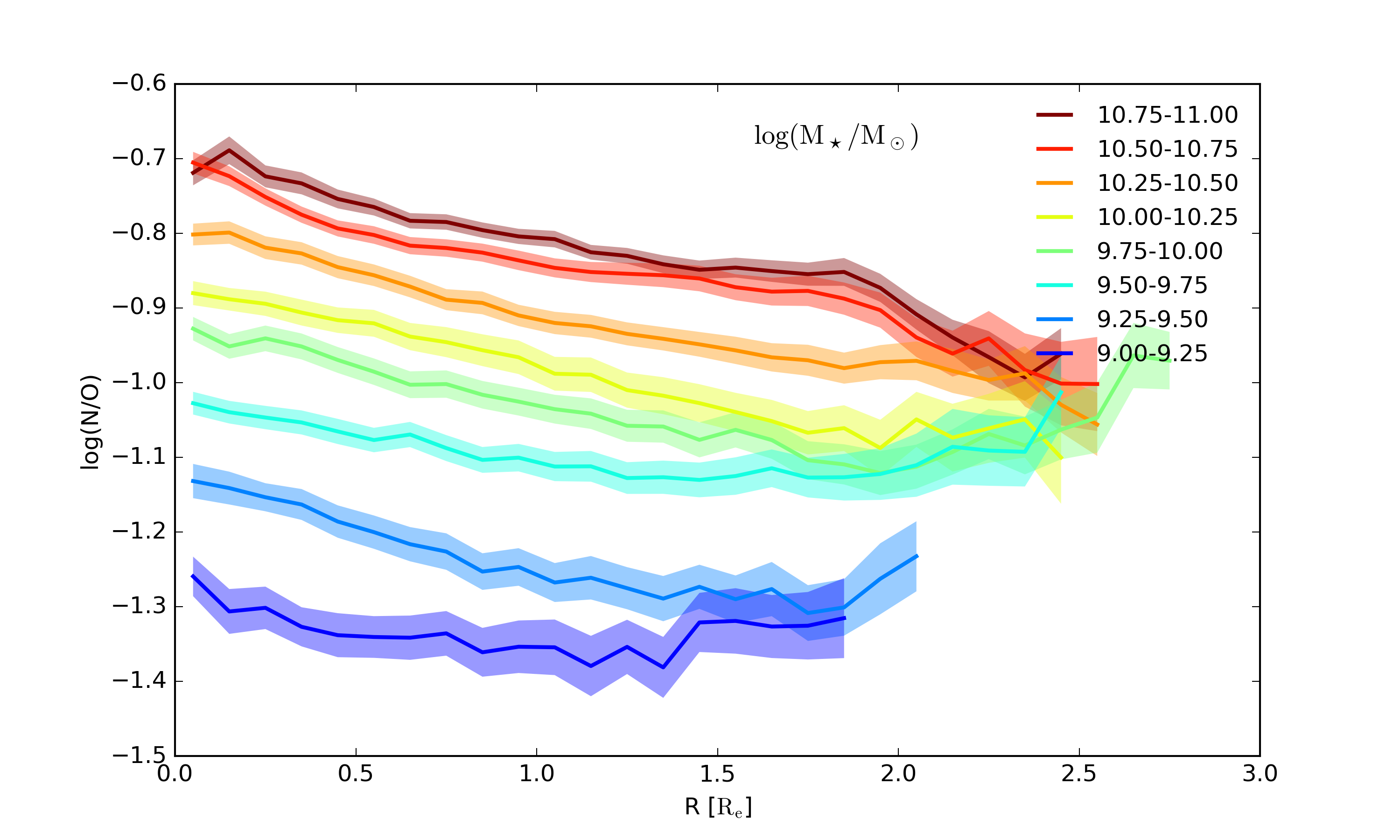}}
\caption{
N/O abundance ratio radial gradient for star forming galaxies from the SDSS4-MaNGA survey
in bins of stellar mass \citep{Belfiore17a}
}
\label{fig:NO_grad_Belfiore}
\end{figure}

The gradient of N/O abundance has also been investigated recently \citep{Berg13,Berg15a,Belfiore17a,Esteban18,James09,James13,Westmoquette13,Kumari18}.
Fig.~\ref{fig:NO_grad_Belfiore} shows the N/O abundance ratio radial gradient for star forming galaxies from the SDSS4-MaNGA survey
in bins of stellar mass \citep{Belfiore17a}. The systemically increasing nitrogen abundance with galaxy mass, the tendency
for the gradient to steepen with galaxy mass and to flatten in the outer parts are all trends similar to those
seen in the metallicity gradients (Fig.~\ref{fig:gradients_mass_Belfiore2017}). However, one important
difference is that the N/O gradient does not flatten in the central region as instead observed
for the O/H abundance. Within the context of the inside-out growth of late-type galaxies,
this finding indicates that the central regions of massive galaxies have locally evolved to an equilibrium metallicity
(saturated around the yield), while the nitrogen abundance continues to increase as a consequence of both the
delayed secondary nucleosynthetic production and the contribution from intermediate mass stars.\\

{\bf Summarizing}, the N/O vs O/H diagram of galaxies shows a dual behavior, with
a plateau at low metallicities and a steeply increasing trend at high metallicities,
which can be interpreted in terms of primary production of nitrogen at low metallicities, while at high metallicities can be interpreted as the delayed production by nitrogen by intermediate-mass stars
together with the ``secondary'' production channel, hence the N/O vs O/H diagram
can be interpreted as an evolutionary sequence.
The nitrogen abundance also correlates significantly with galaxy stellar mass,
which is interpreted in terms of more massive galaxies being more evolved, hence intermediate mass stars have had more time to enrich the ISM with nitrogen.
Individual galaxies or galactic
regions may show significant deviations from this trend (in particular by showing enhanced nitrogen abundance), which can be explained by models in terms of
different effects (such as metallicity dilution by accreting near-pristine gas,
variation of star formation efficiency, differential outflows effects, enhanced enrichment by Wolf-Rayet stars).

\begin{figure}
\centerline{\includegraphics[width=9cm]{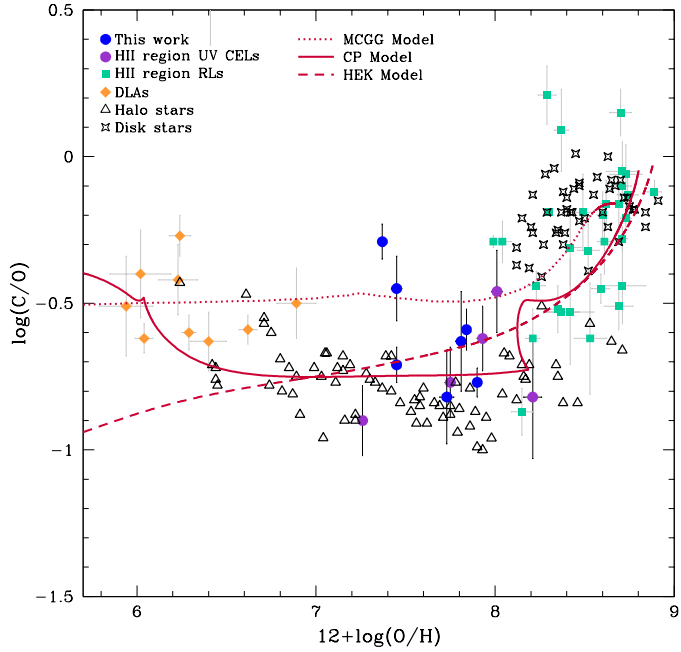}}
\caption{
C/O abundance ratio as a function of O/H for Galactic and extragalactic
systems, compared with various models \citep{Berg16a}.
Blue dots are HII regions in dwarf galaxies observed by \cite{Berg16a} with HST/COS. Purple dots are other objects with direct oxygen abundances and C/O abundances determined from UV CELs. Green filled squares are star forming galaxies with abundances based on RLs.  Triangles are MW halo stars, while 4-pointed stars are disk stars. Finally, orange diamonds are DLA systems, and lines are the results of three enrichment models, see \cite{Berg16a} for details.
}
\label{fig:CO_Berg16}
\end{figure}

\begin{figure}
\centerline{\includegraphics[width=9cm]{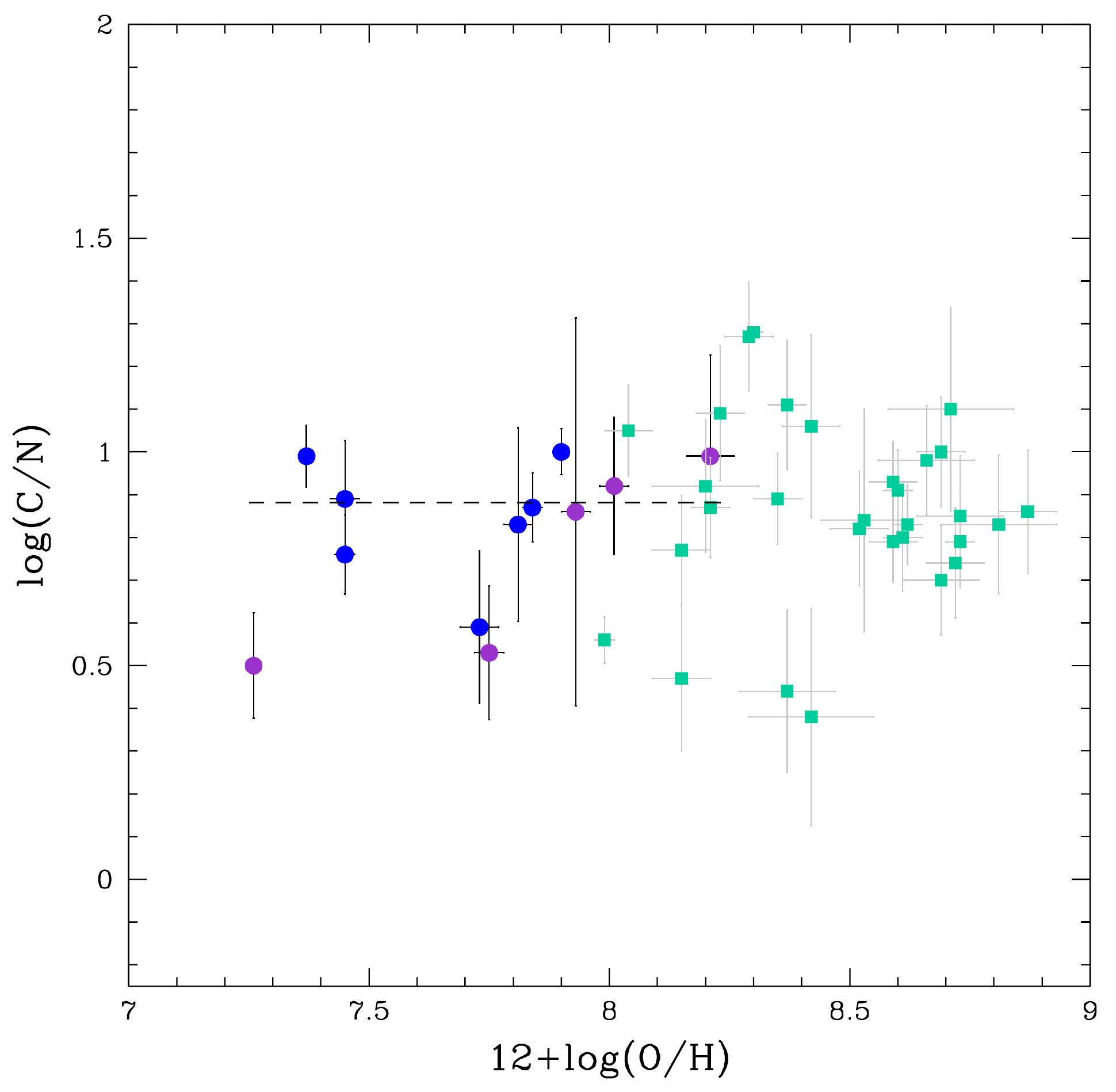}}
\caption{
C/N abundance ratio as a function of O/H for the same systems
as in Fig.~\ref{fig:CO_Berg16}, from \cite{Berg16a}. The dashed line is the 
weighted average of the significant detections based on CEL.
}
\label{fig:CN_Berg16}
\end{figure}
\subsubsection{C/O}
\label{sec:C/O}

The abundance of carbon provides additional important
information on the evolutionary stage of galaxies
as carbon is primarily released by intermediate mass stars
\citep[although Wolf-Rayet stars, whose progenitors are though to be high mass stars, are regarded as additional important contributors of carbon, ][]{Dray03a,Dray03b}. 
Therefore,
significant carbon enrichment is generally delayed
with respect to $\alpha$\ elements.

In the local universe the carbon abundance
has been investigated in galactic stars through medium/high resolution surveys
\citep{Gustafsson99, Bensby06,Akerman04,Spite05,Fabbian09,Nieva12,Nissen14,Tautvaisiene16}, although
one should be aware that when observed in evolved giant and supergiant stars
complex internal mixing and dredge up processes make it difficult to properly interpret
the observed abundances and to use them to trace galaxy evolution.
Carbon abundance in stellar spectroscopy has been extended
also to some nearby galaxies, and recently \cite{Conroy14} have extended the analysis
to thousands of galaxies from the SDSS, through 
fitting of stacked spectra (see fig.~\ref{fig:abund_Conroy14}).

The gas phase metallicity has been more difficult to determine as carbon does
not have strong transitions at optical wavelengths. However, deep observations
of bright HII regions, both in our galaxy and external galaxies, have enabled
the detection of CII recombination lines in the optical, enabling the
measurement of the carbon abundance for some of them
\citep{Esteban02,Peimbert05,Lopez-Sanchez07,Garcia-Rojas07,Esteban09,Esteban14}.
 HST has enabled
sensitive spectroscopy in the UV, where collisionally excited lines of carbon
are present, especially CIII]$\lambda \lambda 1907,1909$ (but also CIV1459, for
galaxies with harder ionizing spectrum, such as AGNs), which
have been used to estimate the carbon abundance in HII regions, although
often requiring photoionization modeling
\citep{Garnett97,Kobulnicky97,Kobulnicky98,Garnett99,Izotov99,Berg16a,Perez-Montero17,Pena-Guerrero17}.
As mentioned in sect.\ref{sec:measmet_strong_UV}, if information on the gas temperature is available then the carbon abundance can be inferred with higher accuracy \citep{Garnett95,Garnett99}. The UV spectral range also
contains absorption features from UV resonant lines which can further be used, through HST data,  to
constrain the carbon abundance of the ISM \citep[primarily using the CII1334 ISM absorption, ][]{James14b} and  of the young stellar population \citep{Leitherer11,Leitherer11b}.

Figure~\ref{fig:CO_Berg16}, from \cite{Berg16a},
summarizes some of the main findings on the C/O versus O/H diagram for
different galactic systems and components, specifically halo and disk MW stars,
HII regions measured either through recombination lines or through UV collisionally
excited lines (as well as high-$z$ DLA, which will be discussed in the next section).
The plot shows large dispersion. However, at least at metallicities higher than
12+log(O/H)$=$7, it resembles the trend observed in the N/O versus O/H diagram,
with a flat relation at low metallicity and a steeply increasing abundance
at high metallicities. The latter trend has led some authors to suggest
that carbon may also have a secondary production, i.e., yields that
are strongly metallicity-dependent; alternatively,  or in addition,
the delayed release of C may also mimic a secondary production effect
\citep{Garnett99,Henry00,Carigi00,Chiappini03}.
The similar behavior of carbon and nitrogen is further supported by the
observed C/N trend (Fig.~\ref{fig:CN_Berg16}), which is constant with metallicity
\citep{Berg16a} and which has strengthened the idea the carbon follows an
enrichment pattern similar to nitrogen.

The stacking analysis of \cite{Conroy14} has revealed that carbon is enhanced
relative to iron in more massive galaxies (i.e., with larger velocity dispersion,
Fig.~\ref{fig:abund_Conroy14}), an effect similar to the $\alpha$ enhancement in massive galaxies, although slightly less
extreme, indicating that carbon is capturing star formation on intermediate temporal scales.\\

{\bf Summarizing}, the C/O vs O/H diagram of galaxies shows a similar dual behavior as for nitrogen,
with
a plateau at low metallicities and a increasing trend at high metallicities.
These similarities with the N/O diagram
suggests that carbon share a common origin with nitrogen (as confirmed
by the constant C/N ratio). In particular,
at high metallicities carbon is enriched with delay by intermediate mass stars
and may also have a ``secondary'' component. Therefore, the C/O vs O/H diagram  describes a temporal sequence. In the next section we will discuss the
peculiar of the C/O abundance ratio at very low metallicities.


\begin{figure}
\centerline{
\includegraphics[width=12cm]{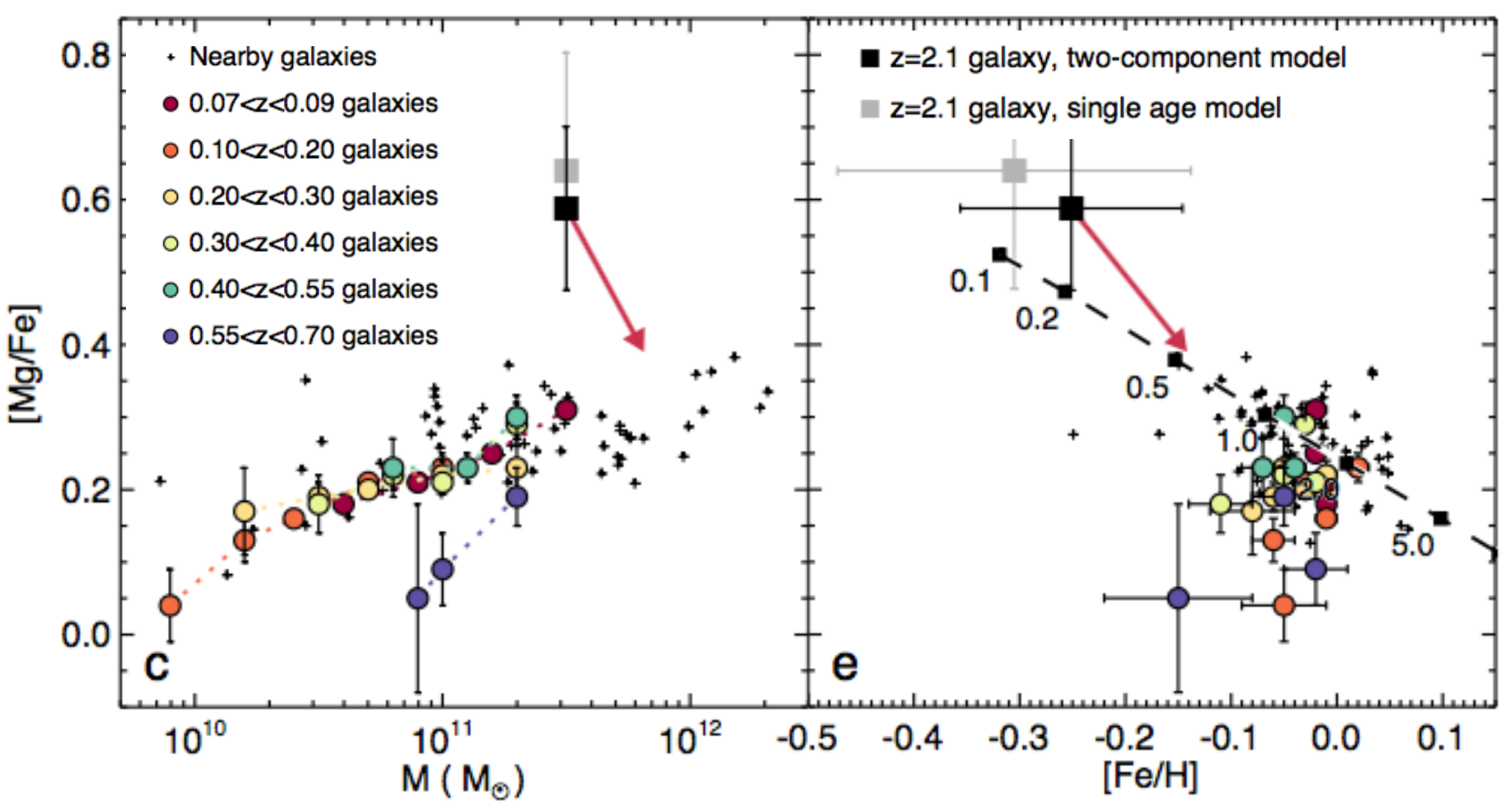}
}
\caption{
[Mg/Fe] abundance for a massive galaxy at $z=2.1$ as a function
of mass (left) and as a function of [Fe/H] (right), compared with
lower-redshift galaxies (see legend). 
The black dashed line and the red arrows are the results of two evolutionary models. From \cite{Kriek16}.
}
\label{fig:abund_hz_Kriek16}
\end{figure}

\subsection{High redshift and the very low metallicity regime}

At high redshift the measurement of the relative chemical abundances are
obviously much more difficult to obtain as the limited S/N of distant galaxies
often prevents detecting the required multiple spectral diagnostic features.
This is especially true for what concerns measuring
the [$\alpha$/Fe] ratio in the spectra of stellar populations of high-$z$ galaxies,
as this requires excellent S/N on the continuum.
Indeed, currently, the [$\alpha$/Fe] ratio has been measured
only in a few cases by
exploiting individual spectra \citep{Lonoce15,Kriek16} or  by using
stacking of star forming or quiescent galaxies \citep{Onodera15,Steidel16}.
The general result is that these distant galaxies are all $\alpha$-enhanced,
in some cases even relative to the lower redshift passive galaxies of the
same mass (e.g., Fig.~\ref{fig:abund_hz_Kriek16}), indicating that these
galaxies have formed quickly, with little iron pollution from SNIa,
generally inferring star formation timescales
of only 0.5--1~Gyr.

\begin{figure}
\centerline{\includegraphics[width=9cm]{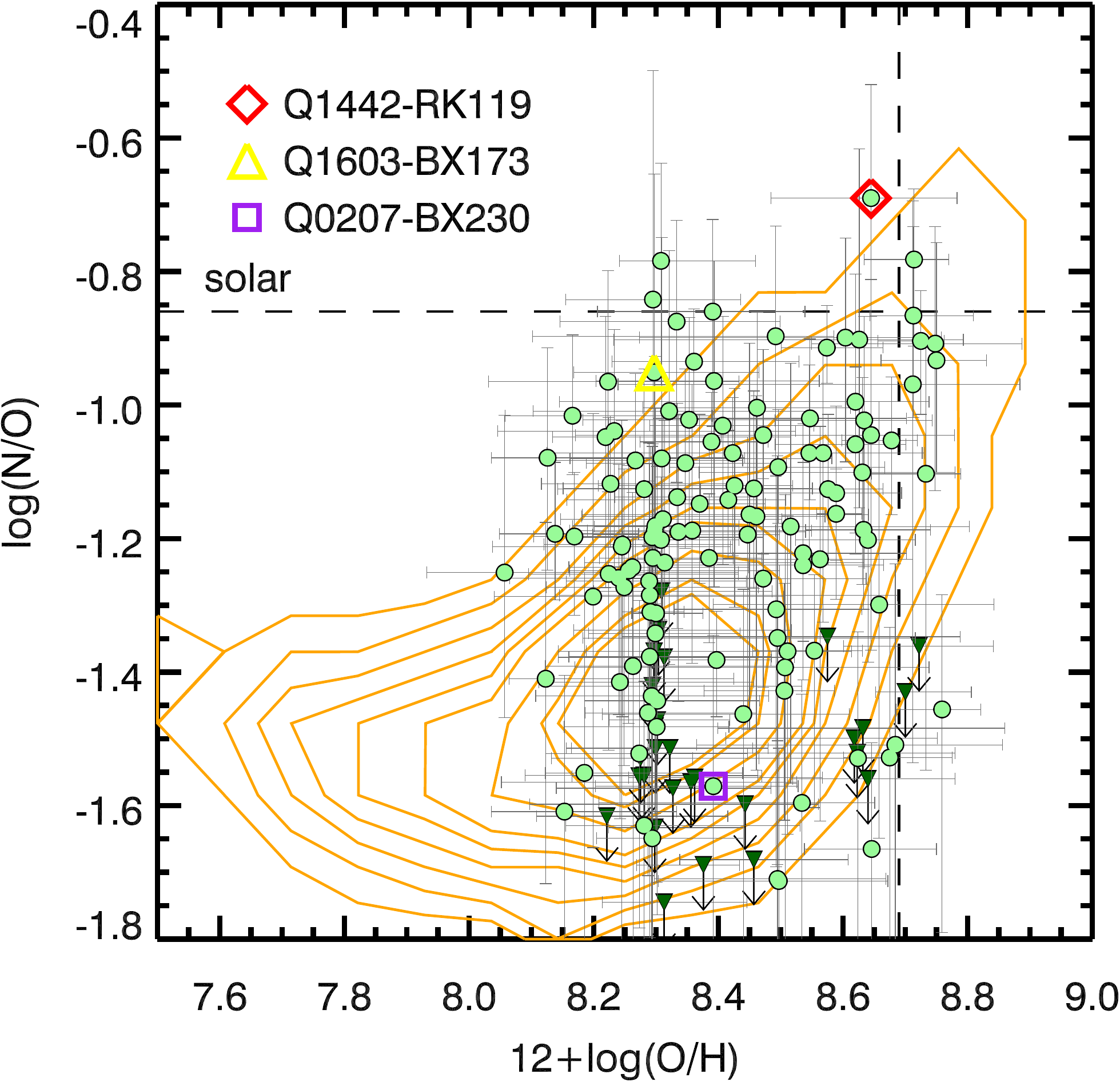}}
\caption{
N/O versus O/H for a sample of star forming galaxies at $z\sim2.3$ (symbols) compared with local star forming galaxies (contours), from \cite{Strom17b}.
}
\label{fig:NO_hz_Strom17}
\end{figure}

The N/O abundance ratio in high-$z$ star forming galaxies
has been investigated little so far, due to the need to sample a
broad range of nebular diagnostics (hence multi-band near-IR/optical
observations are needed,
which may be subject to differential slit losses if not performed
with IFU), both to measure N/O (which requires measuring [NII] and [OII])
and to measure the metallicity O/H  with diagnostics that do {\it not} use
nitrogen. One of such attempts is shown in Fig.~\ref{fig:NO_hz_Strom17}
based on a near-IR spectroscopic survey of galaxies at $z\sim2.3$ \citep{Steidel16,Strom17b}
Generally high-$z$ star forming galaxies (symbols) follow the same relation
as local galaxies (contours), although with somewhat larger scatter, some tendency of being more nitrogen-rich at a given O/H. This seems confirmed (especially in terms of larger scatter) by a smaller sample, but based on T$_e$ measurements, at lower metallicities, in the work by \cite{Kojima17}.
The larger scatter,
if confirmed with higher statistics in future surveys,
may be a consequence of enhanced star formation efficiency at such
early epochs \citep{Vincenzo18}, or may reflect frequent and prominent
inflows of near-pristine gas in high-$z$ galaxies (which dilute the metallicity,
hence O/H, but affect little N/O), as expected by many theoretical
models (see the discussion in Sect.~\ref{sec:N/O}).

The C/O ratio was studied by \cite{Steidel16} in the stacked spectrum of $\sim20$ galaxies at $z\sim2.4$ and by \cite{Amorin17} in individual galaxies
at $z\sim 3$. They found values in  agreement with the values in local stars and HII regions, although with large scatter.

\begin{figure}
\centerline{
\includegraphics[width=8cm]{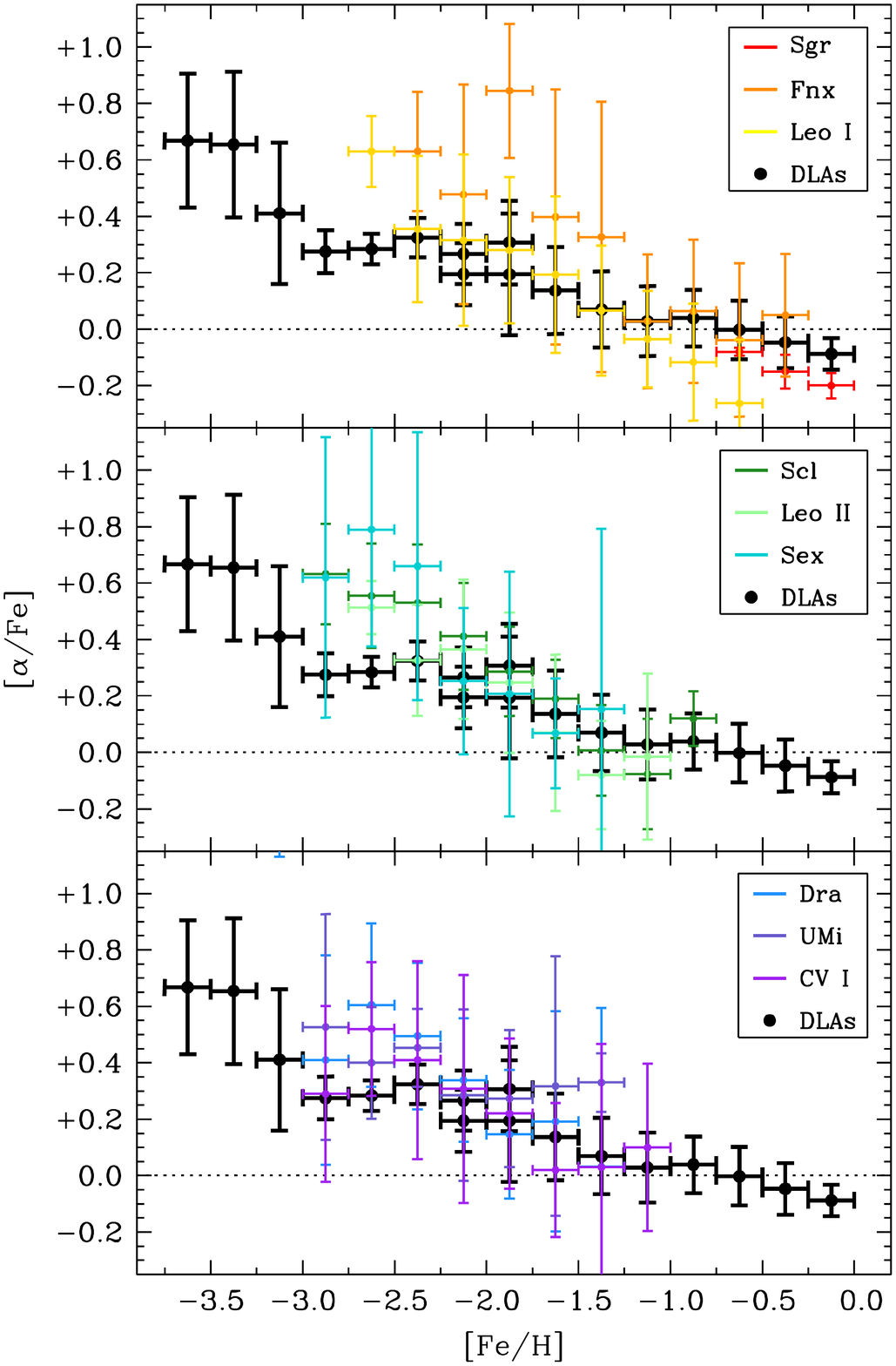}
}
\caption{
[$\alpha$/Fe] versus [Fe/H] for high redshift DLA (black
symbols) compared with with local
dwarf spheroids \citep[adapted from ][]{Cooke15}.
}
\label{fig:DLA_alpha_Fe_Cooke15}
\end{figure}

As discussed in Sect.~\ref{sec:measmet_ism_abs}, high column density absorption systems (DLA) generally provide the most accurate determination
of the relative (and often absolute) chemical abundances
\citep[e.g.,][]{Prochaska99,Berg15b,Fumagalli14,Pettini08,Henry07,Dessauges-Zavadsky06,Wolfe05,Prochaska03,
Rafelski12,Neeleman13,Jorgenson13,Berg16b}, although they probe a broad range of environments,
whose connection with galaxies is generally not fully clear (likely ranging from outskirts of galactic discs
to clumps in the intergalactic medium).
In contrast to early claims that DLA abundances may resemble the chemical pattern of stars
in the MW halo, hence that DLA may probe the formation of galactic haloes,
more recent studies have shown that the chemical and kinematic properties of DLA
are more similar of those seen in dwarf galaxies of the Local Group
(Fig.~\ref{fig:DLA_alpha_Fe_Cooke15}; \citealt{Cooke15,Berg15b,DeCia16}), hence DLA might be tracing
the early formation of dwarf satellite galaxies (also based on their similar velocity dispersion).

\begin{figure}
\centerline{\includegraphics[width=12cm]{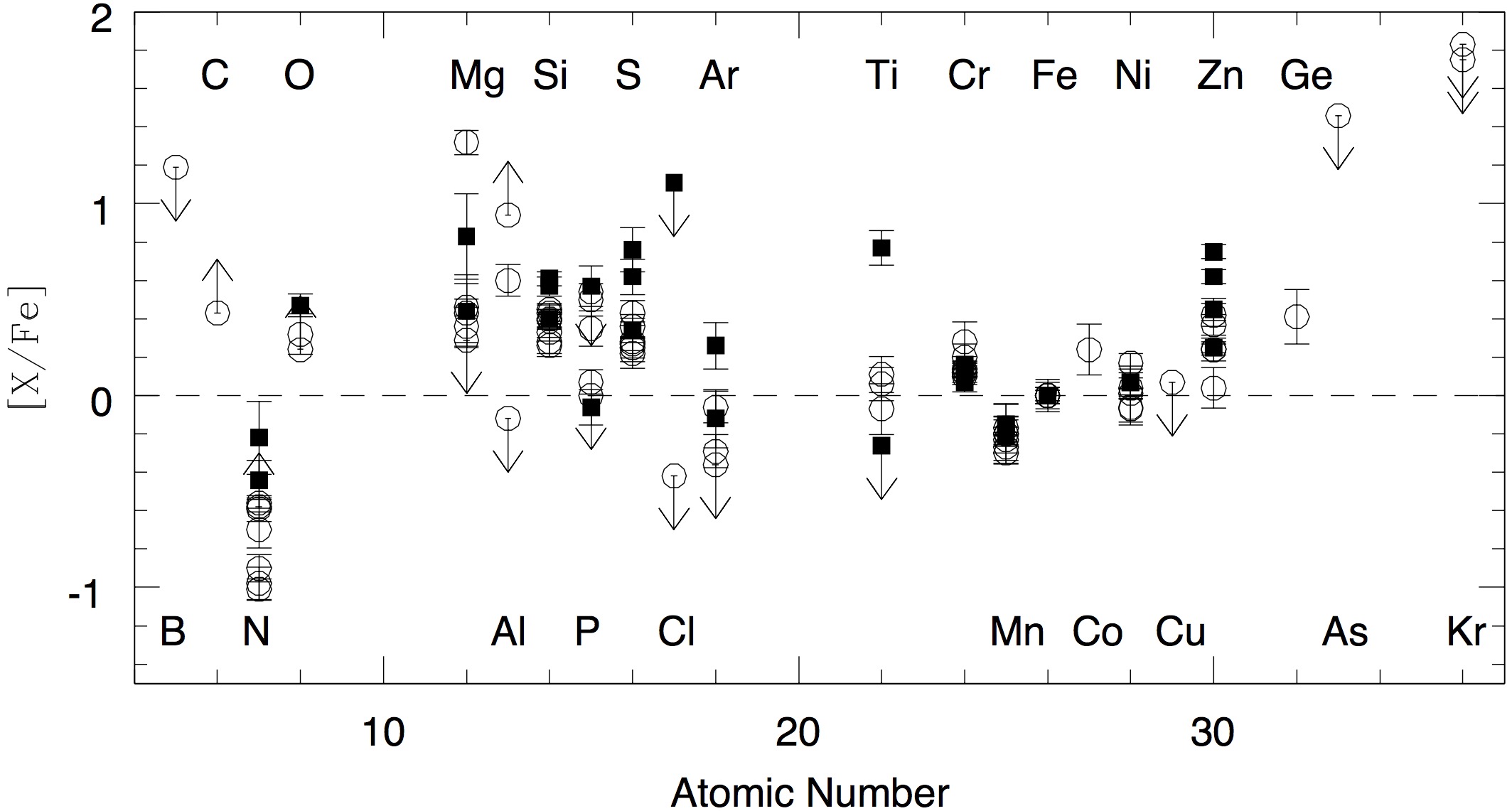}}
\caption{
Detailed chemical abundances inferred for the sample of 11 DLAs at z$\sim$2
in from \cite{Dessauges-Zavadsky06}.
}
\label{fig:DLA_elements_pattern_Dessauges06}
\end{figure}

In many DLA it is possible to trace
the detailed chemical enrichment pattern of several elements
(Fig.~\ref{fig:DLA_elements_pattern_Dessauges06}), in most cases confirming that
these result from the enrichment of core-collapse supernovae from massive stars
\citep{Prochaska03,Dessauges-Zavadsky06}.

\begin{figure}
\centerline{\includegraphics[width=12cm]{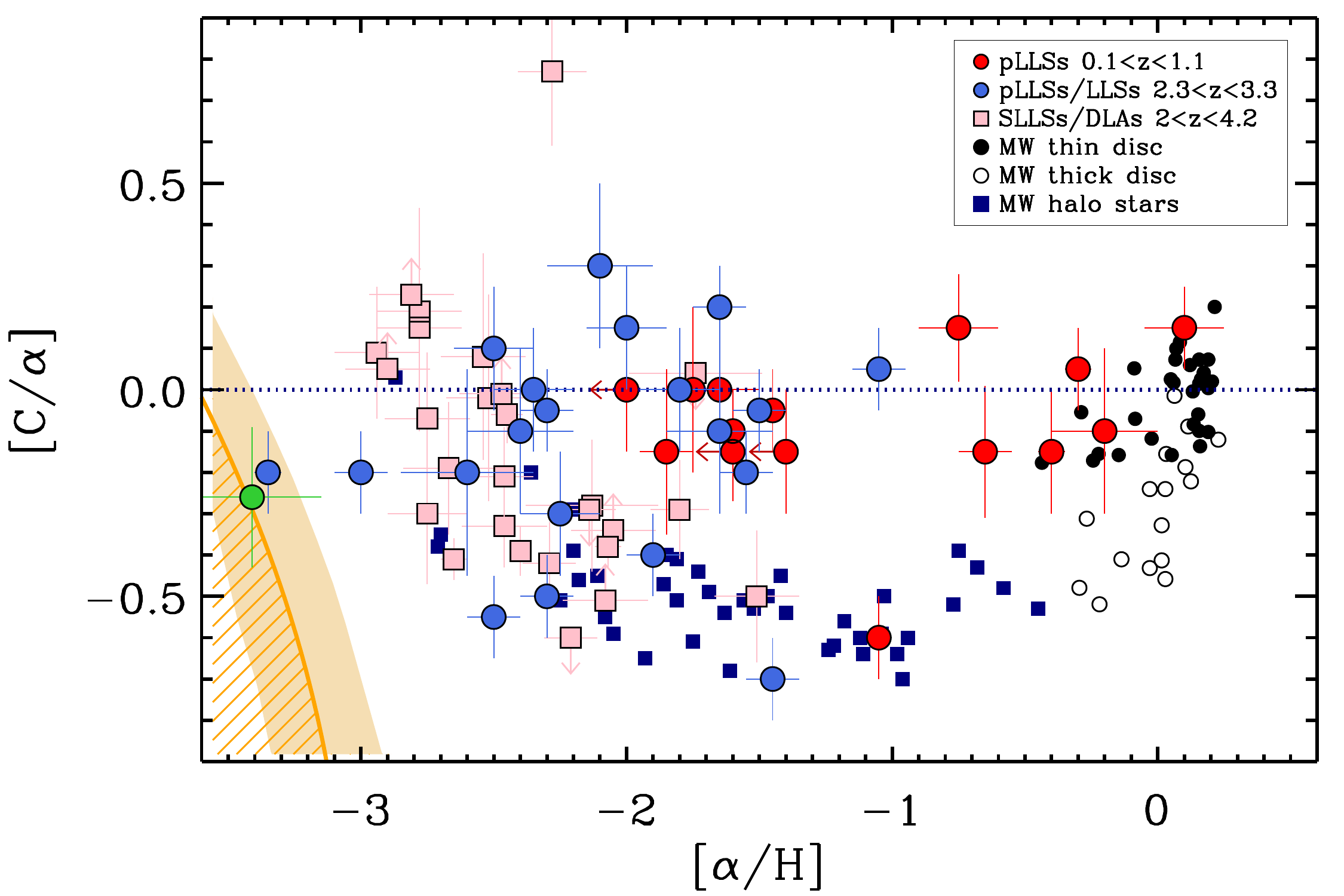}}
\caption{
[C/$\alpha$] vs [$\alpha$/H] for high-redshift DLAs and Lyman Limit Systems,
compared with Galactic stars. The hatched orange
region identifies the area within which gas
may have been primarily polluted by Pop~III stars
\citep{Frebel07}.
From \cite{Lehner16}.
}
\label{fig:CO_Lehner16}
\end{figure}

\begin{figure}
\centerline{\includegraphics[width=9cm]{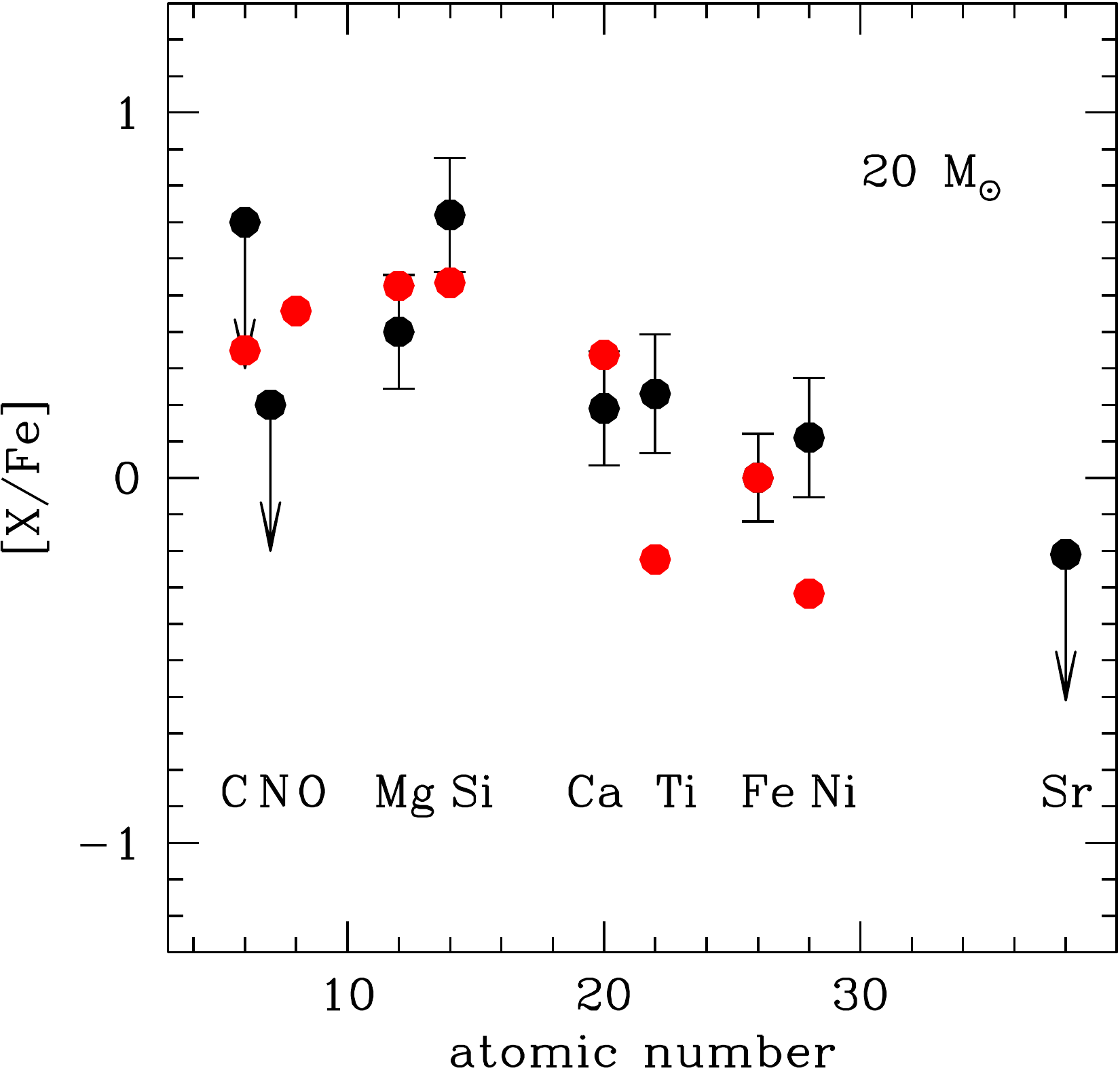}}
\caption{
Left: Abundance pattern observed in the most metal poor halo star \citep[black points,][]{Caffau11} compared
with yields expected from Pop~III stars of different masses. Right: difference between Pop~III yields and observations.
From \cite{Schneider12b}.
}
\label{fig:Schndeider12}
\end{figure}

It is interesting the finding that, for both DLAs at high-$z$ and for halo stars,
the carbon abundance relative to $\alpha$
elements increases systematically at very low metallicities \citep[12+log(O/H)$<$7,
e.g., Fig.~\ref{fig:CO_Berg16} and Fig.~\ref{fig:CO_Lehner16}][] {Cooke12,Berg16b,Lehner16}
which has been interpreted as possible signature of enrichment by
PopIII stars \citep{Carigi11}, although such high values of C/O at low
metallicities can also been explained by other models without invoking
the contribution of Pop~III-like yields, but simply different carbon
yields in the metal poor (Pop~II) regime \citep{Molla15}. A similar
enrichment has been observed in a population of metal poor halo stars collectively known
as carbon-enhanced metal-poor [CEMP] stars \citep{Beers05}. 

Lyman Limit Systems (LLS), which are characterized by lower absorbing column density relative to DLA,
appear to deviate from the trend observed in MW stars and DLA, by being carbon enhanced at metallicities 
$-2< [\alpha/{\rm H}] < -1$. \cite{Lehner16} suggest
that this indicates that LLS trace gas clouds enriched by
preferential ejection of carbon from low metallicity galaxies.

Even more interesting is the discovery of extremely
low metallicity DLA and LLS systems ($[\alpha/{\rm H}]<-3$) that are also carbon poor (Fig.~\ref{fig:CO_Lehner16})
\citep{Lehner16,Crighton16,Cooke17}. These are considered the best candidates as tracers of gas polluted by Pop~III stars. Overall, their chemical
enrichment pattern is well reproduced by
pollution from supernovae originating from PopIII stars, with progenitor
masses of about $\rm 20~M_{\odot}$.
The abundance pattern of these systems is partly reminiscent of some of the most metal poor halo stars.
The most metal poor of these stars has a metallicity
of $\rm 4.5 \times 10^{-5}~M_{\odot}$ \citep{Caffau11} and its chemical abundance pattern is well reproduced by a metal-free progenitor
with mass of about 20--30 $\rm M_{\odot}$ (Fig.~\ref{fig:Schndeider12}) \citep{Schneider12b}.
Interestingly,  both in the case of DLA/LLS and halo stars
the abundance pattern excludes the scenario of pollution
by very massive Pop~III progenitors, which would result in hypernovae. It is also very interesting that the extremely metal poor halo stars have sub-solar stellar masses, at metallicities well below the critical value
that, according to standard model, would allow cooling
and fragmentation of the gas that would enable the formation of low mass stars \citep[shaded region in
Fig.~\ref{fig:CO_Lehner16}, ][]{Frebel07}. While initially very puzzling, \cite{Schneider12a} and \cite{Schneider12b} pointed out that small amount
of dust formed in the ejecta of Pop~III SNe would be enough
to enable the cooling and fragmentation of the gas that
would result into the formation of the first generation of extremely metal poor low mass stars.

We conclude this section by highlighting
that millimeter/submillimeter observations of molecular transitions at high redshift are now sensitive enough
to provide valuable constraints not only on the
composition of molecular species in the ISM of primeval
galaxies, but also on the relative
abundance of different atomic isotopes, which can
provide precious information on the properties of the stellar populations responsible for the early chemical enrichment.
\cite{Zhang18a} have measured the $\rm ^{13}C/^{18}O$ abundance ratio in a sample of distant lensed, starburst
galaxies ($z\sim2-3$), by measuring multiple transitions
of the $\rm ^{13}CO$ and $\rm C^{18}O$ isotopologues of carbon monoxide. Since $\rm ^{13}C$ is mostly produced
by low/intermediate mass stars ($\rm M_*<8M_{\odot}$)
while $\rm ^{18}O$ is mostly produced by massive
stars ($\rm M_*>8M_{\odot}$), the $\rm ^{13}C/^{18}O$
ratio is sensitive to the shape of the IMF. From their
measurements \cite{Zhang18a} find that these early systems are
likely characterized by a top-heavy IMF.\\

{\bf Summarizing}, galaxies with $\alpha$-enhanced stellar populations
are already seen at high redshift ($\rm z\sim 1-2$), indicating that
these systems have formed rapidly, likely within 0.5--1~Gyr. The N/O abundance
ratio of distant galaxies generally follow the local relation, but with larger
dispersion and with a larger fraction of nitrogen-enhanced (relative to oxygen)
galaxies, which may indicate that these galaxies have experienced enhanced
star formation efficiency or absolute metallicity dilution by infalling
near-pristine gas. The $\alpha$/Fe properties of DLA (together with their
velocity dispersion) suggest that they may trace the early
formation of dwarf galaxies and generally be primarily enriched by massive stars
(also based more broadly by a wider set of chemical abundances). The properties of very metal poor DLA with sub-solar carbon abundances suggest that these
may trace the early enrichment by the first generation of stars (PopIII).

\section{Metallicity and chemical abundances in AGN}

Active galactic nuclei (AGN), powered by supermassive accreting black holes,
in their various manifestations, have been extensively investigated
to probe the metallicity in galactic nuclei, in their host
galaxies and even in the CGM.
Given that they can reach very high luminosities (quasar phase)
they have been effectively used to probe the metallicity of their circumnuclear region and of their host galaxies out to
very high redshift.

The nebular emission lines in AGN are primarily divided in two classes,
``broad'' lines and ``narrow'' lines.

Broad emission lines, with line widths up of a few to several thousands
km/s, are emitted by a nuclear region typically smaller than a fraction of a parsec
(the so-called Broad Line Region, BLR). The density of the clouds
in the BLR is so high ($\sim 10^{11}~{\rm cm}^{-3}$)
that ``forbidden'' transitions (e.g., [OIII]5007, [OII]3727, [SII]6730, etc\dots)
are not detected; indeed the critical densities of all these collisionally excited
transitions are well below
the typical density of the BLR, implying that in this regime their emissivity increases only
linearly with gas density, in contrast to the permitted lines (such as hydrogen Balmer recombination lines) whose
emissivity increases quadratically with density even in the extreme conditions of the BLR.

The narrow lines have widths more comparable to those typically observed
in the host galaxies (a few 100 km/s), although typically broader because
often associated with outflows, and extend on scales ranging from
a few 100~pc to several kpc (the so-called Narrow Line Region, NLR).

Metallicity determinations of the BLR and NLR have mostly relied
on photoionization models \citep[e.g.,][]{Hamann99,Nagao06b},
although attempts have been made to use the direct-$\rm T_e$ method
\citep{Dors15} but which have revealed the inadequacy of this method
for AGNs (the origin of such ``temperature problem'' in AGNs
is not yet clear).

Some of the broad line ratios from metal transitions
in the UV have been proposed as sensitive metallicity tracer of the BLR.
The nebular emission
ratio $\rm (SiIV\lambda 1397+OIV\lambda 1402)/CIV\lambda 1549)$ as been
proposed to be the most stable against distribution of gas densities and
ionization parameter in the BLR clouds, and also in terms of hardness
of the ionizing continuum \citep{Nagao06}.
The ratios $\rm NV\lambda 1240/CIV\lambda 1549$ and $\rm NV\lambda 1240/HeII\lambda 1640$
have also been proposed \citep[e.g.,][]{Hamann99,Dietrich03b,Nagao06b,Matsuoka11a,Wang12}, but they are
more sensitive to ionization parameter, shape of the ionizing continuum and
are primarily sensitive to the nitrogen abundance rather than metallicity.

The general finding is that the metallicity of the BLR in quasars
is very high, nearly always supersolar and up to several times
solar \citep{Hamann99,Dietrich03b,Nagao06,Jiang07,Juarez09,Simon10a,Matsuoka11a,Wang12,Shin13,Xu18}.
Such high values of the metallicity have posed questions on whether the
photoionization modeling of the extreme environment characterizing the BLR is
appropriate. However, very high nuclear metallicities (a few/several times solar) are also confirmed by the iron emission and absorption features observed in the X-ray
emission coming from the nuclear region \citep[e.g.,][]{Jiang18}.
Yet, such high metallicities in the nuclear region of AGN are not really unexpected. Indeed
the very high
densities and large amount of gas in the central region of AGN
likely foster rapid star formation and quick enrichment on the ISM.
Moreover, it is important to bear in mind that the mass of gas in the
BLR is very small, a few times $\rm 10^4~M_{\odot}$. As pointed out by
\cite{Juarez09}, such small mass can be quickly enriched with less than
a SN explosion every $10^4$ yrs.

\begin{figure}
\centerline{\includegraphics[width=8cm]{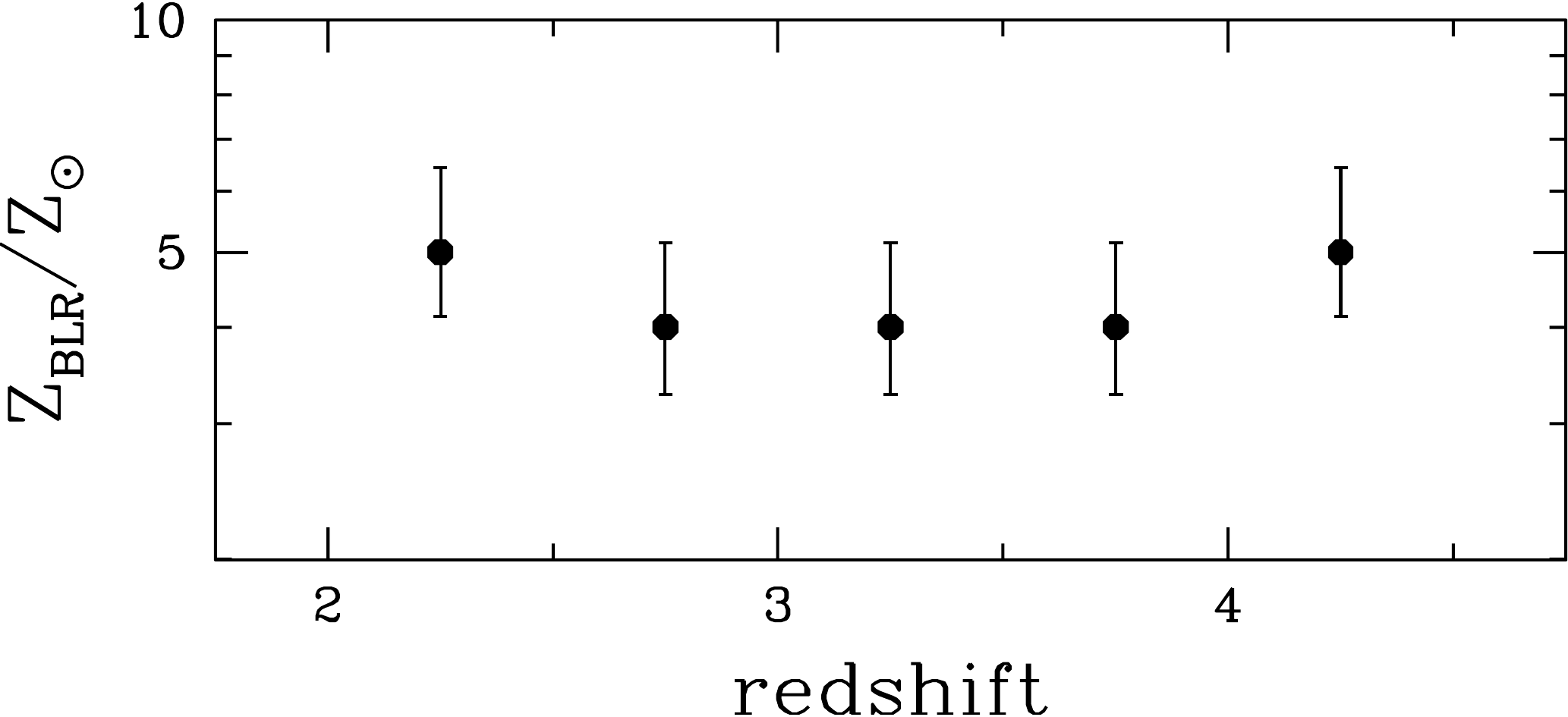}}
\centerline{\includegraphics[width=12cm]{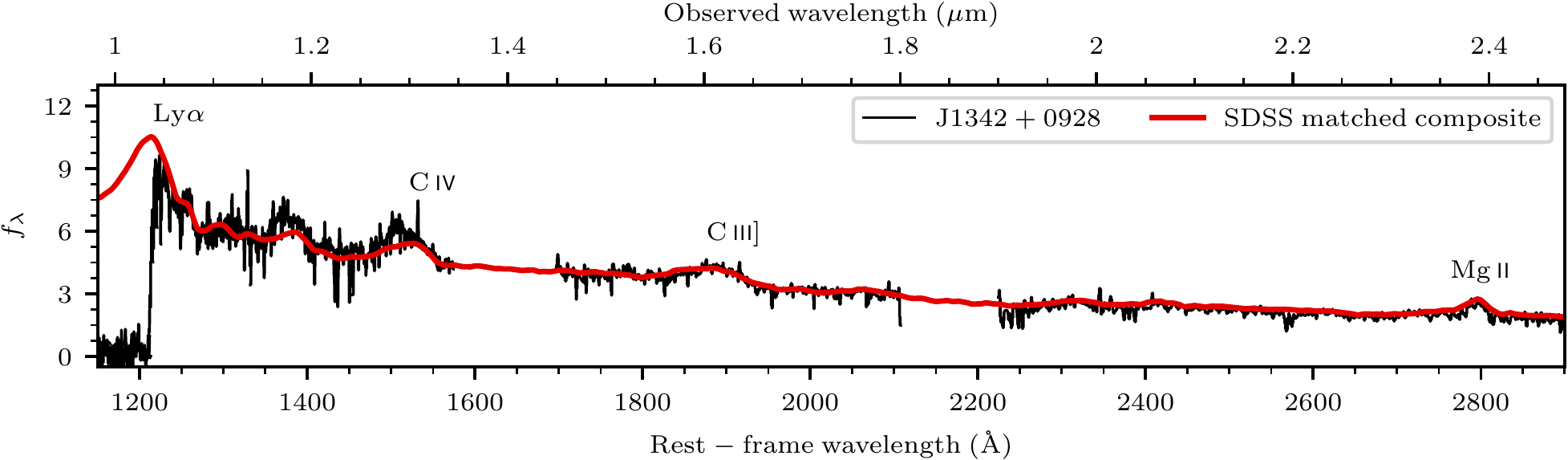}}

\caption{Top: average metallicity of the BLR in quasars (from stacked spectra)
as a function of redshift, from \cite{Nagao06b}.
Bottom: spectrum of the most distant quasar currently known ($z=7.5$) compared
with the average spectrum of intermediate redshift quasars from SDSS, illustrating
that the two are nearly identical suggesting similar chemical enrichments of the BLR \citep{Banados18}.
}
\label{fig:Z_BLR_z}
\end{figure}

The somewhat puzzling result is that the metallicity of the BLR does not 
seem to evolve with redshift (Fig.~\ref{fig:Z_BLR_z}-top) and such lack of evolution
appear to persist out to the most distant quasars known, at $z=7.5$ (Fig.~\ref{fig:Z_BLR_z}-bottom)
\citep{Nagao06b,Juarez09,Mortlock11,Banados18,Xu18}.
The lack of redshift evolution of metallicity (which is in contrast to what is observed for galaxies),
which seem to remain high at all redshifts,
is likely a consequence of the mass-metallicity relation combined with selection effects.
Indeed, in order to be selected in large scale  surveys  the quasar has to be luminous
enough to pass the sensitivity threshold of the survey; this generally implies that (even if
accreting at the Eddington limit) the black hole must have already become fairly
massive. If some form of black hole/galaxy relation is already in place at high redshift,
this implies that, at any redshift, the host galaxy must already be massive hence
typically display high metallicity when the quasar enters into the survey (at any epoch).
This combination of effects, explaining the lack of redshift evolution of the
BLR metallicities, was discussed more quantitatively in \cite{Juarez09}.

\begin{figure}
\centerline{
\includegraphics[width=8cm]{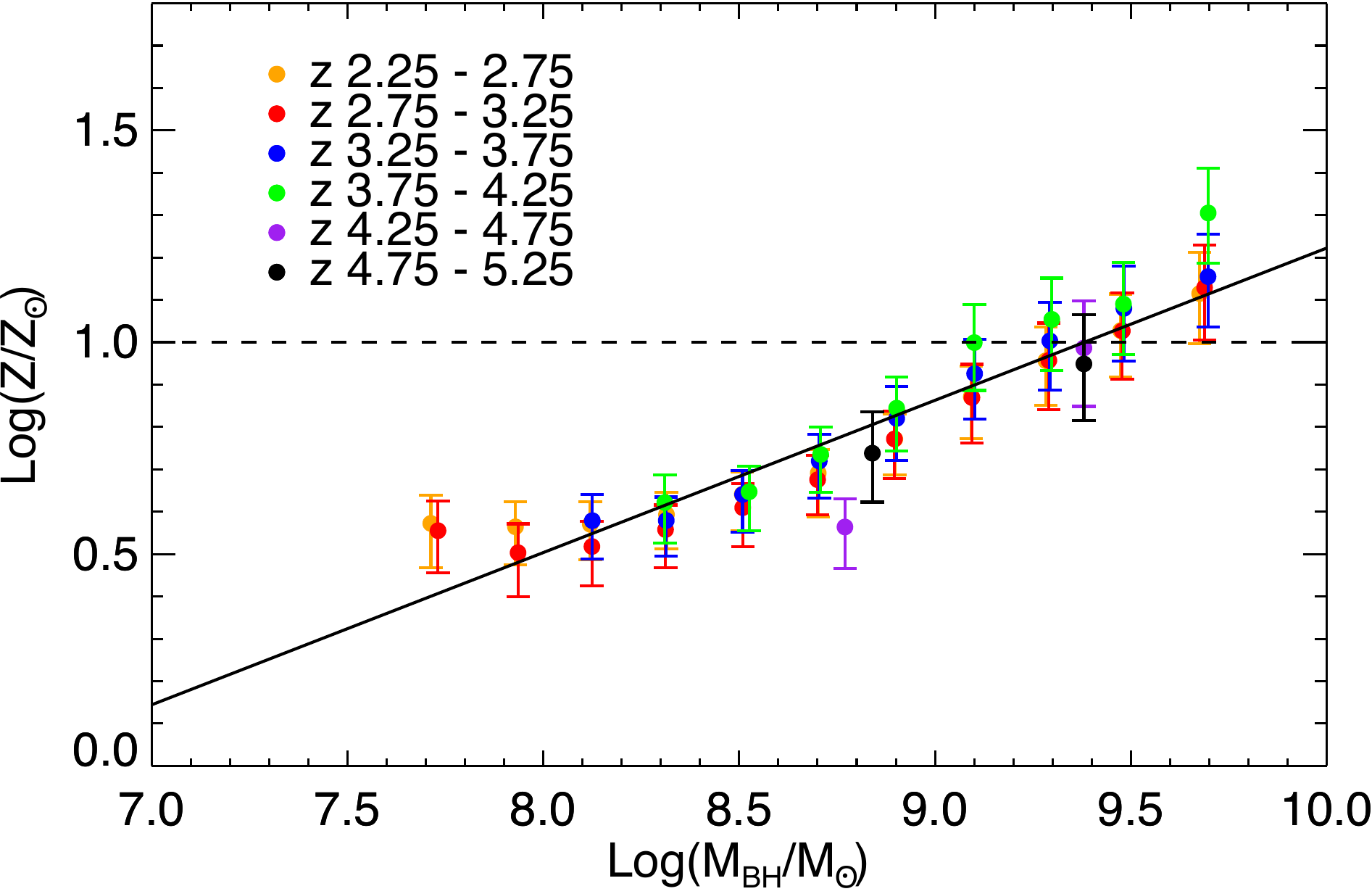}
}
\caption{
Metallicity of the BLR as a function of black hole mass
for quasars divided in bins of redshift (see legend).
From \cite{Xu18}
}
\label{fig:Z_BLR_BH}
\end{figure}

The fact that the metallicity in AGN is linked to the mass-metallicity relation of the
host galaxy was first hinted by the fact that the metallicity of the BLR scales
with AGN luminosity \citep{Hamann99,Nagao06b,Xu18}.
Indeed, if the AGN luminosity is a function of the black hole
mass (assuming an average L/L$_{\rm Edd}$ ratio) and the black hole mass is linked to the
host galaxy mass through BH-spheroid relation,
then one would expect the AGN nuclear metallicity to
scale with the AGN luminosity as a consequence of the mass-metallicity relation of the
host galaxy \citep{Juarez09}.
A more clear evidence of this, which by-passes the use of AGN luminosity, is
the more direct relationship between BLR metallicity and black hole mass
obtained by \cite{Matsuoka11a} and \cite{Xu18}, as illustrated in Fig.~\ref{fig:Z_BLR_BH} \citep[see also ][ for a potential extension of the relation to low masses.]{Ludwig12}

\cite{Matsuoka11a} report the lack of correlation between metallicity
and Eddington ratio L/L$_{\rm Edd}$.
A correlation with the Eddington ratio is seen only
when using the NV/CIV and NV/HeII ratios, which are primarily tracing the nitrogen enrichment.
Since nitrogen enrichment is delayed with respect to $\alpha$\  elements,
the latter correlation has been interpreted as indication that black hole accretion
is triggered with a delay of a few 100~Myr with respect to the onset of star formation.
This is delay that has been suggested also in local AGNs \citep{Davies07} and it has been interpreted
as a consequence of the initial strong turbulence induced by SNe, which may prevent effective accretion
onto the BH, while at later epochs the more gentle stellar winds may be effective
in removing angular momentum from the gas (hence enabling it to move towards
the centre) without introducing excessive turbulence or gas removal through SN-driven winds.

Many authors have used the flux ratio of the MgII~2798\AA \ doublet relative to the UV FeII ``bump'' (due to a blending of multiplets at
2200--3090 \AA), with the goal of
constraining the redshift evolution of the $\alpha$/Fe ratio in the Broad Liner Region \citep[e.g.][]{Dietrich03a,Maiolino03,Freudling03,Iwamuro04,Jiang07,DeRosa11,DeRosa14,Calderone17,Mazzucchelli17}. These various studies find no evidence for a redshift evolution of the MgII/FeII flux ratio out to most distant quasars at z$\sim$6.5. If one assumes that the flux ratio is, to a first order, a proxy  of the $\alpha$/Fe abundance ratio, then the lack of evolution would imply that the relative contribution of SNIa and core-collapse 
does not change.
However, ones has to take into account that both emission features, and especially the FeII ``bump'' are primary coolants of the BLR and, therefore, their flux does not really scale linearly with abundance, but it rather tend to adjust in order to keep the thermal equilibrium of the BLR clouds \citep{Verner04a,Verner04b}, therefore the lack of evolution of the MgII/FeII ratio may simply reflect the ``thermostatic'' role of the associated transitions.\\

Similar studies have been performed to investigate the metallicity
of the NLR, i.e., on much larger galactic scales in AGN hosts.
In this case studies have generally focused on type 2 AGNs, in which the BLR (whose strong,
broad lines would
otherwise prevent a proper disentangling of the flux of the narrow lines)
is obscured along the line of sight. Studies have both exploited optical narrow nebular
lines ratios (especially in local galaxies) and UV nebular lines (especially in distant
galaxies, whose UV lines are redshifted into the optical bands) and by 
using photoionization models to infer the metallicity from a combination of line
ratios.
Initial claims about highly supersolar metallicities in the NLR \citep{Groves06}
have been revised downwards, however it is still true that most NLR appear to be
metal rich, with metallicities around solar or super-solar
\citep{Nagao06c,Matsuoka09,Stern13,Matsuoka11b,Coil15,Dors15,Castro17,Dors17}.
\cite{Matsuoka09} suggest that in the redshift range 1--4 there is little cosmic
evolution of the NLR metallicity, although admittedly the statistics
are much poorer than for the BLR metallicities; however, \cite{Coil15} have pointed
out that, in their sample of type 2 AGNs at $z\sim2.3$, the NLR metallicity (inferred
from rest-frame optical diagnostics)
is lower than the metallicity in the NLR of local AGN. They also point out that the metallicity of the NLR in their sample of type 2 AGN is higher than in a matched sample of star forming galaxies.

The solar/super-solar metallicities in the NLR of AGNs already at high redshift,
can be partly explained in terms of dust destruction in the NLR, which releases
metals into the ISM \citep{Nagao06c,Matsuoka09,Dors14}, but probably is also partly
due to the fact that the NLR is often also associated with galactic (AGN-driven) outflows
originating from the central, metal-rich region of the galaxy.
Indeed, several studies have also directly investigated the outflowing gas in quasars
and AGNs, especially through absorption lines (and especially in Broad Absorption Line
Quasars, where prominent blueshifted absorption troughs probe powerful winds),
revealing high metallicity gas being expelled on kpc scales
\citep{Hamann99,Simon10a,Ganguly03,DOdorico04,Ganguly06,Gabel06,Arav07,Borguet12,Shin17}.

However, it has been suggested that the NLR also follows the (host galaxy) mass-metallicity
relation (although offset towards higher values),
either directly \citep{Matsuoka18} or
indirectly through the AGN luminosity or black hole mass
 as indirect tracers of the host galaxy mass \citep{Matsuoka09,Ludwig12}.

We finally mention that, while the NLR tend to be generally metal rich, the investigation of the outer, most extended region
of the NLR (sometimes referred to as Extended Narrow Line Region, ELR) does reveal low metallicity regions \citep[e.g.,][]{Fu09,Husemann11} indicating that the outer parts of the NLR
probe gas in the outer galaxy that are still poorly enriched.

Of growing interest is becoming the technique of probing
the metallicity of the halo of quasars through the analysis of associated absorption systems detected in the spectrum of a nearby (in projection) background quasar \citep{Prochaska13,Prochaska14}. This technique has revealed that that the circumgalactic medium of quasars at $z\sim2$ hosts significant quantities of cold gas ($\rm >10^{10}~M_{\odot}$) significantly metal enriched
($\rm Z>0.1~Z_{\odot}$ out to the virial radius ($r_{\rm vir}\sim 160~kpc$), implying that by $z\sim2$ feedback has already been quite effective in enriching the CGM of massive galaxies and also
implying that
the assumption of pristine gas accretion in many models may
be inappropriate.

{\bf In summary}, the BLR show very high metallicities at any redshift, with no indication of evolution with time. Selection effects are probably contributing to hide signs of redshift evolution, nevertheless this points out toward very early enrichments of the central regions of massive galaxies. 
The metallicity of the BLR correlates strongly with the black hole mass, which is likely a result of the mass-metallicity relation of the host galaxy combined with the black hole-host galaxy mass scaling relation.
The NLR also show high metallicities, though much lower than those observed in
the BLR. The NLR metallicity also show no evidence for redshift evolution. The interpretation is complex because contributions to the metallicity of the NLR
are expected to come from the AGN-driven outflows, from the effects of dust destruction, and also to be linked to
the mass-metallicity relation of the host galaxy.
The investigation of absorption systems in the proximity of quasars (by exploiting pairs of quasars that are close in projection) has revealed large amounts of metal enriched cold gas in their halos, suggesting that quasars activity have polluted significantly their circumgalactic medium through outflows, already at early cosmic times.

\section{Metal budget}
\label{sec:metal_budget}

Since metals are produced by the star formation activity across the cosmic epochs,
comparing the total amount of metals seen in the various phases to the cosmic evolution of stellar mass and SFR is useful to investigating consistency of the interpretation of
 these various independent observational results, and in particular
 the validity of the underlying assumptions and models about the production of
 chemical elements and their transfer among the various galactic and intergalactic
 phases.
In other words, the metal budget is fundamental information whose evolution should be in agreement with the other independent observations of galaxy evolution
and should be matched by models. For this reason it has been subject of considerable work
\citep{Pei95,Edmunds97,Pettini04b,Pettini06,Ferrara05,Bouche07a,Gallazzi08,Zahid12b,Peeples14,Madau14}.

Determining the total mass in metals is a challenging goal, as heavy elements are produced in galaxies and then dispersed in the universe in different forms, often in states that, as we have discussed in this review, are difficult to observe.
Moreover, in general observational studies measure the metallicity of the various
phases, i.e., the  abundance of metals {\it relative} to the total content of baryons. Therefore, even in those cases where the metallicity is well constrained, inferring
the {\it absolute} total content of metals implies having a good knowledge of the
total mass associated with the same phase (stars, ISM, CGM, ICM, WHIM, IGM), which
are all distributed on different scales and forms, hence adding to the problem
a additional level of complexity and uncertainty.

Attempts to infer the total mass in metals in the local universe
have implied combining the contribution to the metal budget
from all these components.
More specifically:

\begin{itemize}
\item Stars and ISM. Galaxies contain large amounts of metals locked into stars and star remnants and dispersed into the ISM. The mass of metals contained in galaxies can be obtained by integrating the mass function of galaxies convolved by the metallicity and gas fraction as a function of mass. As the chemical abundance ratios and the overall metallicities depend on galaxy types, it is necessary to make the computation dividing galaxies in bins of morphology or SF histories. Roughly  mass-weighted solar metallicities are obtained for this component when assuming a Salpeter IMF \citep{Calura04,Gallazzi08}.

\item CGM and IGM. The CGM is enriched by  galactic winds and plays a critical role in the current ``equilibrium models'' (see Sect.~\ref{sec:models}) because it constitutes a reservoir of metals 
extracted from the galaxy by winds and that could rain back onto the galaxy. Similarly, the IGM is thought to be enriched by galactic winds, especially those escaping from low-mass galaxies. As discussed in the previous sections, the chemical abundance of the CGM and IGM is obtained from the absorption features of several ionization species of various elements
and shows significant evolution with redshift, rising to about 0.1 solar in the local universe and containing about 10\% of the metals produced
\citep{Meiksin09,DOdorico10,Simcoe11b,Shull14,DOdorico16}. 
While the amount of metals in galaxies declines steadily with redshift, the amount of metals in the CGM/ISM as traced by DLAs seems to remain roughly constant out to $z\sim4$ \citep{Prochaska13,Rafelski14}, implying that the fraction of metals in the CGM/IGM is much larger at high redshift than locally. It had been claimed that the amount of metals in DLAs  shows a decline at $z>4$ \citep{Rafelski14}, however this has not been confirmed by more recent studies \cite{DeCia18b,Poudel18}.

\item The intra-cluster medium (ICM). As already mentioned, X-ray observations have revealed that the intracluster medium is highly enriched with all metals (generally with solar-like relative abundances, sect.\ref{sec:alpha_over_iron}). Metals in clusters are mainly produced in the evolved populations of early type  galaxies \citep{Matteucci12}, which are enriched in Fe by ongoing production of type Ia SNe, also long after the end of star formation \citep{Maoz12a}. The metal transfer toward the ICM is due either to AGNs, to SN explosion, or to ram stripping,  and the system as a whole evolves nearly as a closed box. As already mentioned, extensive works have been undertaken to estimate the content of metals in the ICM, especially with the advent of high resolution spectroscopy \citep{Mushotzky97, Balestra07,Blanc08b, De-Plaa13, Molendi16,Mernier16,Mernier18,Hitomi17,Simionescu18}. 
The metallicity of the ICM is generally
very high (supersolar), and therefore, despite the small contribution to the total baryonic mass ($\sim4\%$), it is expected to contribute significantly to the total metal budget (see the discussion later  in this section).

\item The warm-hot intergalactic medium (WHIM). The WHIM is the warmer/hotter phase of the IGM, thought
to result from the gravitational shock-heating of the intergalactic
medium in the local universe to temperatures of $\sim 10^5$--$10^7$~K, where
most of the baryons in the local universe are though to reside \citep{Nicastro17}.
It is thought to be the local (hotter) counterpart of the (cooler) IGM at high-z
observed through the Ly$\alpha$ Forest.
Its content of metals has been inferred through UV and X-ray absorption
spectroscopy \citep[e.g.,][]{Tripp00,Fang01,Prochaska04,Cooksey08,Fang10,Zappacosta10,Zappacosta12}.
Typically the inferred metallicity is of the order of 0.1~Z$_{\odot}$ and its
 contribution to the local, total metal budget is less than 5\%.

\end{itemize}

Amid these issues and uncertainties,
the published values for the total amount of metals show a significant scatter, with values around $\Omega_Z\sim 10^{-4}$ \citep{Pei95,Madau96a,Zepf96, Mushotzky97, Madau98a, Pagel02, Dunne03, Calura04, Gallazzi08} where $\Omega_Z$ is the density of metals normalized to the critical density of the Universe for $h=0.7$,  $\rho_c=1.36\times10^{11}M_\odot Mpc^{-3}$.

The relative distribution distribution of metals among these different components is clearly still subject to significant uncertainties.

Expectations
from the integrated cosmic production of stars can be achieved
by integrating the star formation rate density as a function of redshift
and convolving it with the yields per stellar generation \citep{Molla15,Vincenzo16b}. This is obviously subject to additional uncertainties, not only associated with the measurement of the evolution of the
star formation history \citep{Madau14}, but also with our knowledge
on the return fraction of metals to the gas phase and with our yet
limited knowledge of the IMF and its potential variations.
Bearing in mind all these uncertainties, the expected average metallicity
of the local universe is inferred to be $Z\sim0.09$ solar for a Salpeter IMF and to decrease by one order of magnitude by $z=2.5$ \citep{Madau98a,Pettini06,Madau14}.

There is reasonable agreement between the expected
and measured metal budget in the local universe. However, given
large uncertainties in both the measured and expected
content of metals the agreement is not too surprising and really
not very constraining of any of the underlying assumptions.

It is has been perhaps more instructive to investigate the metal budget
in individual systems, as this exercise may provide information on
processes resulting in loss of metals, or even provide constraints on the
yield of metals.

For instance, \cite{Renzini14} compare the amount of iron in the ICM
(as inferred from X-ray observations)
with the amount of iron expected to be produced by the stellar
populations of galaxies within the cluster, based on empirical yields
of iron. While they find a good agreement for intermediate
mass clusters ($\rm M_{500}\approx 10^{14}~M_{\odot}$), in more
massive clusters they reveal a clear tension, in the sense that the
ICM contain much more iron mass (up to a factor of $\sim6$)
than that produced by stars in galaxies, revealing 
higher rates of type Ia SNe in clusters \citep[e.g.][ and references therein]{Mannucci08,Friedmann18},
or issues either in the
metallicity measurements or with our knowledge of the yields.

{\cite{Calura04}, \cite{Bouche07a}, \cite{Peeples14} and \cite{Tumlinson17} present an extensive
analysis of the metal budget in galaxies and including their CGM, 
by exploiting the extensive results from the COS-Halos project
\citep{Tumlinson11,Prochaska17}. They show that, nearly independently
of mass, only about 20\%--25\% of metals produced in stars remain
in galaxies (in stars or in the ISM). They infer that, for $\rm L_*$ galaxies,
as much as 40\% of the metals produced by stars are  deposited
in the halo (CGM,
within a radius of $\sim$150~kpc), while the remaining must be lost
into the intergalactic medium.

Finally, it is interesting to note that the quality of the data
is becoming good enough to enable a spatially resolved metal
budget in galaxies. For instance, \cite{Belfiore16b} use spatially
resolved metallicities for stars and gas combined with spatially
resolved maps of the gas content and surface brightness to
illustrate that within the central 7~kpc ($\rm \sim 3~R_e$)
of the well studied galaxy NGC628
about 50\% of the metals have been lost (somewhat in tension with the result
obtained by \citet{Peeples14}, unless many more metals are lost at larger
radii). Interestingly,  \cite{Belfiore16b} also find that the fraction
of metals lost increases
 to about 70\% in the central kpc of the galaxy \citep[a similar result
 was found by ][]{Greggio11}, suggesting
that such metals were ejected either by the SNe associated with the early central
burst of star formation, associated with the formation of the
bulge, or by AGN/quasar driven winds, during the past evolution of the
central region of the galaxy.
Very recently, \cite{Telford18} have performed a very similar, extensive and detailed analysis of the spatially resolved metal budget in M31, finding very similar results, i.e. a higher loss of metals from the central region of M31. Very interestingly, they also find that during the past 1.5~Gyr some of the metal lost from the central region have been redistributed in the galactic disc outskirts.

\section{Conclusions}

In this paper, we have tried to review the measuring methods, the observational results, and the implications for  models of galaxy metallicity evolution. It is the result of many years of effort by many researchers, sometime using dedicated instruments, only part of this effort is reproduced here.  

\subsection{Summary}

The study of  chemical abundances in galaxies is a complex and extended field with many open problems and conflicting results.  Nevertheless a few clear points are emerging about methods, observations and models:

\begin{itemize}

\item Stellar metallicities are now routinely measured using UV and optical spectra. Spectrophotometric models of increasing precision, complexity and spectral resolution use the full information contained in the  spectra to derive the metallicity together with other parameters of the stellar population. Simplified methods exist that use particular features to derive metallicities, and these methods are more apt to study large samples of galaxies with lower resolution spectra.

\item The absolute scale of the gas-phase metallicity is still uncertain because of the difference among the three main  methods to measure it (Recombination Lines, \Te\ method, and photoionization models). Discrepancies have been reduced, but they still persist. The ``direct'' method based on measuring \Te\ is currently the most reliable and seems to be in agreement with the metallicity of young stars.

\item These methods have been used to calibrate a large range of strong line ratios diagnostics, which can be applied to faint, distant galaxies.  The difference among these secondary calibrations  are dominated by the different  method used for the primary calibration, photoionization model 
vs.\ direct '\Te' method.

\item Most heavy elements are generally largely depleted onto dust grains, therefore the evolutions of gas-phase metallicity and of dust are linked together. Dust depletion is a very important, often neglected source of uncertainty in the study of gas-phase abundances.

\item Both stellar and gas-phase metallicities follow a well defined mass-metallicity relation (MZR) in which the metallicity of galaxies increases with stellar mass. The observed MZR evolves with redshift, with metallicity decreasing at any mass, although more rapidly at lower masses. 

\item The gas metallicity of galaxies has also other secondary dependencies. The most important of which is the anti-correlation between gas metallicity and  SFR (or gas content, which
is related to the SFR), which is called (together with the mass dependence) the Fundamental Metallicity Relation (FMR). This relation has no or a very limited evolution with redshift up to $z\sim2.5$, and a possible strong evolution at $z\sim3.5$. Most authors describe the FMR as an effect of gas infall, providing further evidence for the ubiquity and importance of cold gas accretion in shaping galaxy evolution, and explain the absence of evolution as the effect of the same dominant physical processes at $z<2.5$.

\item Environment also has a secondary effect on the metallicity of galaxies, but only for satellites, whereby satellite galaxies in denser environments (e.g. group and clusters) tend to be more metal rich than galaxies in low density environments. This is probably a consequence of multiple different effects (such as ``strangulation'' and accretion of metal-enriched gas).

\item Understanding the redshift evolution of the MZR and of the FMR is made challenging and uncertain by the evident evolution of the ISM properties seen in the excitation BPT diagnostic diagrams. The dominant cause of this evolution is not clear, and  most likely a combination of different effects (higher pressure, harder ionizion continua, higher ionization parameter, and variation of the N/O abundance ratio relative to local galaxies). How this evolution affects the determination of metallicities is not yet clear.

\item The metallicity evolution of DLA systems provides independent information about the evolution of galaxies and of their CGM. If their velocity dispersion is taken as a proxy of their mass, then the metallicity of DLA systems follow a mass-metallicity relation as well.
Although it not straightforward to link absorption-selected systems to emission-selected galaxies,  both classes of objects identify the redshift range $2<z<3$ (coincident with the peak of star formation density) as a turning point in galaxy evolution, as this is the redshift range where the evolution of most scaling relations change significantly.

\item The metallicity distribution inside galaxies contains a wealth of information about the spatially-resolved processes of galaxy formation.
 The 
radial metallicity gradients of local galaxies steepen as a function of galaxy stellar mass (at least within the central $\rm \sim 2R_e$). However, the outskirts of galaxies show very flat metallicity gradients, which imply accretion of pre-enriched gas from the halo.

\item  The scaling relations between galaxy metallicity, stellar mass,
star formation rate and gas content are also found locally, on spatially resolved scales, in the form of correlations between metallicity and surface density
of stellar mass, surface density of SFR and surface density of gas. However, it is
not yet clear whether the local scaling relations are totally driving the global, galaxy-wide ones.

\item  Based both on diagnostics that trace the metallicity at different lookback times in local galaxies and the direct observation of metallicity
gradients at high redshift, there is a clear indication that the radial  metallicity gradients of galaxies have become steeper with cosmic time. This result is difficult to explain in the context of inside-out formation of galaxies and it requires radial migration of stars, radial inflows of low-metallicity gas or  radial redistribution of metals at the early stages of galaxy formation.

\item  There is growing evidence for the existence of inverted (i.e. positive) metallicity gradients in distant galaxies. These may trace accretion of near-pristine gas or radial inflow of metal-poor gas induced by galaxy mergers or interactions.

\item Different elements provide different information on the evolutionary stage and star formation history of galaxies. Oxygen, Nitrogen, Carbon and the $\alpha$-elements are particularly useful and are studied in detail to understand galaxy formation and evolution because they sample different time scales. 

\item The $\alpha$/Fe abundance ratio shows that the different components
of the MW (halo, bulge, thick and thin disc) have formed not only at different epochs, but also on different timescales and with different star formation efficiencies.

\item
Similarly, the $\alpha$/Fe abundance ratio in local galaxies indicates
that more massive galaxies formed faster (probably with higher star formation
efficiency) and at earlier times than lower mass galaxies, a phenomenon which known as galaxy ``downsizing''. 

\item The N/O and C/O abundance ratios, together with O/H, are used to obtain further information on galaxy evolution as nitrogen and carbon are characterized by  longer production timescales than $\alpha$-elements, and therefore provide important information on the evolutionary stage of galaxies and on other galaxy evolutionary processes (such as gas accretion and efficiency
of star formation).

\item X-ray observations have shown that the intracluster medium is highly enriched and with solar-like chemical abundances. The estimated global
content of metals in massive clusters exceed significantly the
amount of metals that are expected to be produced by the clusters' galaxies. This is still an unsolved, open problem.

\item AGN are currently one of the few ways to study metallicity in galaxies up to very high redshifts ($z>7$). The metallicity information can generally be extracted only for gas in the so called Broad Liner Region  (BLR, on sub-parsec nuclear scales) and for the Narrow Line Region (NLR, on galactic scales).

\item
The BLR metallicity scales with black hole mass at all redshifts, which is likely tracing a combination of mass-metallicity relation in the host galaxy and
$\rm M_{BH}-M_{gal}$ relation, both already in place at early cosmic epochs.

\item
The metallicity of the BLR is generally very high (often a few/several times super-solar) and does not evolve with redshift. The latter is probably a consequence of observational selection effects (quasars are detected only
when their black holes, and therefore their galaxies, are massive enough,
and therefore already highly enriched).

\item
The metallicity of the NLR is lower than in the BLR, but still higher
than in normal galaxies, probably as a consequence of dust destruction and
enrichment by  quasar-driven outflows. Also the NLR shows little evolution with redshift.

\item Different types of models have been developed to reproduce and explain all these and other observations. The main contribution of the metallicity and chemical abundance analysis is to put strong constraints on the role and the properties of gas infall, galactic winds, stellar and AGN feedback, stellar yields, amount of re-accretion of gas, and IMF. Both analytic and numerical models can now account for most of the observational results, but they often disagree on the dominant effects.

\item The distribution of metal mass in the various components of the universe is still not well constrained, nevertheless it appears that most of the metals have left the parent galaxies.

\end{itemize}

\subsection{Open issues}

Although impressive progress has been achieved in recent years,
it is clear that several outstanding problems and uncertainties have
yet to be faced by observations and theory.

The difficulty of accessing direct gas-phase metallicity tracers of the
gas phase for the vast majority of galaxies, especially at high
redshift, hence having to rely on strong-line diagnostics or photoionization
models remains one of the main issues. Indeed, the latter methods
are still prone to degeneracies with other galaxy parameters such as
ionization
parameter, shape of the ionizing continuum, relative chemical
abundances, geometry, pressure and density of the ionized clouds,
and the broad distribution of all these parameters inside galaxies;
these issues make the comparison between different classes of
galaxies and with models still difficult. 
Within this context, it is not yet clear how much the evolution of
the average excitation conditions of star forming regions at high redshift
(i.e. the evolution of the BPT diagrams) affects the metallicity determination
of the gas phase through models or strong line diagnostics calibrated locally.

Similarly, it is still not clear if and how much the FMR evolves
at high redshift and, if so, why. As discussed, galaxies out to $z\sim2.5$
follow the same FMR observed locally once
consistent calibrations and a consistent formalism are adopted. However,
the dispersion is larger at high redshift than locally and it is not
clear why, i.e., whether this is a consequence of observational uncertainties
or it  truly reflects an evolution of the galaxy evolutionary mechanisms. Even more
intriguing is the evolution of the FMR beyond $z\sim2.5$, which is 
accompanied by a similar regime change in other properties of galactic and intergalactic
properties around the same redshift (such as the evolution of the content of neutral atomic
gas in galaxies, and the evolution of the DLA's mass-metallicity relation); this change of regime at z$\sim$2.5
remains to be fully explained by models.

The redshift evolution of other scaling relations at high redshift,
such as with gas content (either molecular or atomic) or with environment,
cannot yet be investigated, due to the lack of data on these other properties,
jointly to metallicity determinations.

Mainly because of sensitivity limitations,
the measurement of the stellar metallicities at high redshift is still limited to
very small samples, preventing us to investigate the evolution of scaling relations
and the comparison with models at high redshift.

The investigation of metallicity gradients at high redshift is still largely
hampered by the lack of spatial resolution. Gravitational lensing, the use of
adaptive optics and HST have delivered nicely resolved metallicity maps for small
samples, but we are still far from achieving the statistics obtained locally, which
has made it possible to investigate metallicity gradients as a function of galaxy properties.
Moreover, it is becoming clear that at high redshift metallicity variations within
galaxies do not follow the same, simple radial behavior as in local galaxies, but
more complex non-radial variations, implying that a re-thinking of the metallicity
gradients characterization is needed.

Models and numerical simulations have made an excellent progress during recent years
and can nicely reproduce many of the observed metallicity properties in galaxies,
locally and at high redshift. However, all models still suffer from a number
of degeneracies and {\it a-priori} assumptions that are difficult to control
or verify. The IMF (both in terms of shape and cut-offs)
is one of the critical input parameters of models,
which drastically affects their results. Models obviously depend
even more critically on their particular choice of  yields
and enrichment delay times of stars with different  masses.
Result from models also depends on the assumed dependence of the outflow loading
factor and star formation efficiency as a function of galaxy mass, star formation rate
and AGN activity.

Numerical simulations are still limited by lack of high enough resolution to correctly model the subgrid physics associated to star formation and feedback by SNe and AGN.
It is observed that the shape of some of the metallicity scaling relations depends on
the adopted resolution. Higher resolution simulations can better
capture the baryonic physics, but the unavoidably smaller volumes sampled by these
simulations result both into a potential bias towards lower density environments
and a shortage of massive systems, which may affect the comparison with observations,
especially in terms of dependence on environment.

Another outstanding problem consists in the way models and observations are compared.
As the actual shape of MZR and FMR depends on how the galaxies are selected, it is necessary to select the model galaxies for the comparison in the same way. This is only rarely done, and usually biased observed samples are compared with volume-limited model samples.

However, major progress on these various fronts
is expected in the near future, thanks to the development of new observing facilities
and new generation of models, as discussed in the next section.

\subsection{Future prospects}

In order to properly constrain models, and to advance  our understanding of the
mechanisms driving galaxy evolution, additional and more accurate observational
data are needed.
The next generation of observing facilities and surveys will certainly enable a major
step forward in this area of research in the next decade.

The James Webb Space Telescope (JWST) holds some of the major expectations. Its unprecedented
sensitivity in the near/mid-IR bands, coupled with its high angular resolution,
and multiple spectroscopic modes (including a multi-object spectroscopic mode,
 integral field units and slitless spectroscopy) will enable astronomers to probe
 nebular emission lines that are metallicity diagnostics at high redshift,
 out to the re-ionization 
 epoch and beyond, for several thousands of galaxies, including very low mass systems.
 Most importantly, JWST will enable astronomers to directly detect auroral lines in hundreds
 of individual
 galaxies, hence directly measure the metallicity and recalibrate the strong line
 diagnostics at different epochs. JWST will also deliver high fidelity maps of the
distribution of metals in hundreds of galaxies. The expectation of JWST is to
trace the metallicity evolution of metals back to the first generation of stars,
i.e., the so-called Pop~III stars, formed out of  primordial pristine gas.

On a similar time-frame the Extremely Large Telescopes (such as the Giant Magellan
Telescope, GMT, the Thirty Meters Telescope, TMT, and the European Extremely
Large Telescope, E-ELT), with their huge collecting areas will deliver very high
signal-to-noise spectra at intermediate spectral resolution of the stellar continuum
in large samples of distant galaxies, therefore enabling a major leap forward in the characterization
of the stellar metallicities, relative chemical abundances (especially $\alpha$/Fe)
and the associated scaling relations at high redshift. The determination of metallicity
gradients will also greatly benefit from the leap in angular resolution delivered by
these telescopes, together with their adaptive optics systems.
High resolution spectroscopy, the technique that is most severely affected by
photon starving, will probably be the area that will most  benefit from
the huge collecting area of these telescopes. The number of new systems that will
be observable at high spectral resolution will increase by orders of magnitude (thanks to the steep luminosity function of quasars),
hence enabling an unprecedented mapping of chemical elements across the universe
through absorption systems, with the ultimate goal of finding the chemical
signatures of the first generation of stars.

On shorter timescales, the advent of the next generation of 
high-multiplexing, optical, multi-object
spectrographs on 4m class telescopes, such as WEAVE on WHT and 4MOST on ESO/VISTA, will allow to expand the number of observed galaxies to several millions, obtain spectra at higher resolution, and increase the redshift range sampled. Even more interesting, the new
near-IR multi-object spectrographs on 8m class telescopes, such as PFS at Subaru and MOONS at the VLT, will 
 deliver Sloan-like surveys at high redshift by providing near-IR spectra
 for millions of galaxies out to $z\sim2$ and gas metallicities for hundreds of
 thousands of them. This will enable astronomers to explore the redshift 
 evolution of the metallicity scaling relations with unprecedented statistics.
 The expected leap in statistics, volume and completeness for distant galaxies will
 make it possible to investigate, for the first time, the environmental effects
 on the metallicity scaling relations at high redshift.
 Stacking of hundreds/thousands of spectra is also expected to enable the
 investigation of the stellar metallicities and also to detect the auroral
 lines to recalibrate the strong line diagnostics at high redshift.
 
 The current generation of 8m-class telescopes with the next generation of adaptive-optics assisted spectrographs (like ERIS on VLT) will be used to obtain spatially-resolved metallicity for a large number of galaxies and with an higher level of accuracy.
 
For the warm/hot gas phase the short-lived Hitomi mission has given just a glimpse
of the wealth of information that can be obtained through high resolution-high sensitivity
X-ray spectroscopy. 
The X-ray Imaging and Spectroscopy Mission (XRISM), 
to be launched in 2021 by JAXA and NASA with a European Space Agency (ESA) participation,
will provide observing capabilities similar to the Hitomi satellite
and will therefore enable us to obtain detailed and accurate measurements of the
hot plasma in galaxy clusters, in galactic halos and in galactic winds. This
will finally enable astronomers to both derive a much more accurate budget of metals
and to directly witness the metal enrichment of the CGM
in different classes of galaxies. XRISM will pave the way to Athena, the large
X-ray observatory to be launched around 2028, which will trace the
metallicity and chemical abundance of the hot gas even in distant systems, thanks
to its unprecedented sensitivity.

The Atacama Large Millimeter Array (ALMA), which has recently entered in full operation is already delivering exceptional results. In the coming years it is expected to provide
a detailed census of the molecular gas in galaxies across the cosmic epochs. The capability
of tracing the transitions of multiple molecular species, involving several different
elements and associated with
different isotopes, will provide unique constraints on the star formation history
and on the IMF, both locally and at high redshift.

The Square Kilometer Array (SKA), among many expected ground-breaking results, 
will finally provide a census of the content
and distribution of atomic neutral gas in galaxies at high redshift, which is
the key ingredient, still largely unconstrained, to understand galaxy evolution
at high redshift and the role played by the HI gas reservoir in distant galaxies.

On longer timescales, the SPICA space mission, if selected by ESA,
will offer a sensitivity improvement by orders of magnitude in the mid- and far-infrared
spectroscopic ranges. By measuring fine structure transitions of several chemical elements,
SPICA will enable astronomers to trace the metal enrichment 
in several hundreds, or even thousands high-$z$ galaxies,
without being
affected by dust extinction, hence probing, in an unbiased way, also the heavily
obscured population of galaxies.

\begin{acknowledgements}

It is a pleasure to thank the referee for a careful reading of the manuscript and many comments and insights. We also thanks many colleagues for useful discussions and comments on the early manuscript, and suggestions in identifying the relevant results and the most important open problems: F.~Belfiore, G.~Cresci, M.~Curti., P.~Dayal, A.~De~Cia, G.~De Lucia, M.~E.~De Rossi, V.~D'Odorico, S.~Ellison, A.~Fabian, F.~Fontanot, A.~Gallazzi, B.~James, R.~Kennicutt, N.~Kumari, A.~Marconi, F.~Matteucci, L.~Origlia, Y.~Peng, E.~Perez-Montero, M.~Pettini,  J.~Shaye, E.~Spitoni, P.~Torrey,  F.~Vincenzo, and s.~Wuyts. We also thanks F.~Vincenzo for the data used for fig.~\ref{fig:metal_production}, and M.~Heyden and A.~Motta for providing modified versions of their figures.
FM acknowledges support by PRIN INAF ``ESKAPE-HI'', and RM acknowledges ERC Advanced Grant 695671 "QUENCH" and support by the Science and Technology Facilities Council (STFC).

\end{acknowledgements}

\appendix

\twocolumn[\section{List of acronyms}]

\setlength{\parindent}{-0.2cm}
\setlength{\columnsep}{2.0cm}

4MOST: 4-metre Multi-Object Spectroscopic Telescope \citep[ESO,][]{De-Jong16}

AAS: American Astronomical Society

AD: Abundance Discrepancy

AGB: Asymptotic Giant Branch

AGN: Active Galactic Nucleus

ALMA: Atacama Large Millimeter Array

APOGEE: APO Galactic Evolution Experiment \citep[SDSS,][]{Majewski17}

BLR: Broad-Line Region

BH: Black Hole

BPT: Baldwin--Phillips--Terlevich

BSG: Blue Supergiant Stars

CALIFA:  Calar Alto Legacy Integral Field spectroscopy Area survey \citep{Sanchez12a}

CC: core-collapse (supernovae)

CCD: charge-coupled device

CEL: Collisionally-Excited Lines

CEMP: Carbon-Enhanced Metal-Poor stars

CGM: Circumgalactic medium

DIG: Diffused Ionized Gas

DLA: Damped Lyman-$\alpha$ system

E-ELT: European Extremely Large Telescope

ELR: Extended Narrow Line Region

ESA: European Space Agency 

ESO: European Southern Observatory

ETG: Early-Type Galaxy

EW: Equivalent width

FMR: Fundamental Metallicity Relation

GP: Green Peas

GMT: Giant Magellan Telescope

GRB: Gamma-Ray Burst

GALAH:  Galactic Archeology with HERMES \citep[Anglo-Australian Telescope,][]{De-Silva15}

ICM: intracluster Medium

IFU: Integral Field Unit

IGM: intergalactic Medium

IMF: Initial Mass Function

IR: Infrared

ISM: Interstellar Medium

JAXA: Japan Aerospace Exploration Agency

JWST: James Webb Space Telescope

LAG: Lyman-$\alpha$ Galaxies

LIER: Low Ionization Emission Line Region

LINER: Low Ionization Nuclear Emission Line Region

LLS: Lyman-Limit Systems

LMC: Large Magellanic Cloud

MaNGA: Mapping Nearby Galaxies at APO \citep{Bundy15}

MOONS: Multi Object Optical and Near-infrared Spectrograph \citep[ESO-VLT,][]{Cirasuolo12}

MSSF: Main Sequence of Star Formation

MUSE: Multi Unit Spectroscopic Explorer \citep[ESO-VLT,][]{Bacon10}

MW: Milky Way

MZR: Mass-Metallicity Relation

NASA: National Aeronautics and Space Administration

NLR: Narrow-Line Region

NOEMA: NOrthern Extended Millimeter Array (IRAM)

PCA: principal component analysis 

PDR: Photo-Dissociation Regions

PFS: Prime-Focus Spectrograph \citep[Subaru,][]{Tamura16}

PN: Planetary Nebula

QSO: Quasi-Stellar Object

RAVE: Radial Velocity Experiment \citep[Australian Astronomical Obs.,][]{Steinmetz06}

RL: Recombination lines

RSG: Red Supergiant Stars

SAM: Semi-analytic Model

SAMI: Sydney-AAO Multi-object Integral field spectrograph (Galaxy Survey) \citep{Bryant15}

SDSS: Sloan Digital Sky Survey

SEGUE: Sloan Extension for Galactic Understanding and Exploration \citep{Yanny09}

SFR: Star Formation Rate

sSFR: Specific Star Formation Rate

SKA: Square-Kilometer Array

SN: Supernovae

S/N: signal-to-noise ratio

SNR: Supernova Remnant

SPICA: Space Infrared Telescope for Cosmology and Astrophysics \citep{Roelfsema18}

SSP: Single Stellar Population

ULIRG: Ultra Luminous Infrared Galaxy

UV: Ultraviolet

VISTA: Visible and Infrared Survey Telescope for Astronomy (ESO)

VLT: Very Large Telescope (ESO)

TMT: Thirty-Meters Telescope

WEAVE: Wide-field multi-object spectrograph
for WHT \citep{Dalton16}

WHIM: Warm-Hot Intergalactic Medium

WHT: William Herschel Telescope

XRISM: X-ray Imaging and Spectroscopy Mission 

XMP: Extremely Metal Poor galaxies

\setlength{\parindent}{0.5cm}
\onecolumn

\bibliographystyle{mnras}
\bibliography{bibliography}   

\begin{thebibliography}{}
\makeatletter
\relax
\def\mn@urlcharsother{\let\do\@makeother \do\$\do\&\do\#\do\^\do\_\do\%\do\~}
\def\mn@doi{\begingroup\mn@urlcharsother \@ifnextchar [ {\mn@doi@}
  {\mn@doi@[]}}
\def\mn@doi@[#1]#2{\def\@tempa{#1}\ifx\@tempa\@empty \href
  {http://dx.doi.org/#2} {doi:#2}\else \href {http://dx.doi.org/#2} {#1}\fi
  \endgroup}
\def\mn@eprint#1#2{\mn@eprint@#1:#2::\@nil}
\def\mn@eprint@arXiv#1{\href {http://arxiv.org/abs/#1} {{\tt arXiv:#1}}}
\def\mn@eprint@dblp#1{\href {http://dblp.uni-trier.de/rec/bibtex/#1.xml}
  {dblp:#1}}
\def\mn@eprint@#1:#2:#3:#4\@nil{\def\@tempa {#1}\def\@tempb {#2}\def\@tempc
  {#3}\ifx \@tempc \@empty \let \@tempc \@tempb \let \@tempb \@tempa \fi \ifx
  \@tempb \@empty \def\@tempb {arXiv}\fi \@ifundefined
  {mn@eprint@\@tempb}{\@tempb:\@tempc}{\expandafter \expandafter \csname
  mn@eprint@\@tempb\endcsname \expandafter{\@tempc}}}

\bibitem[\protect\citeauthoryear{{Agricola}}{{Agricola}}{1556}]{Agricola1556}
{Agricola} G.,  1556, {De Re Metallica}.
Hieronymus Froben and Nicolaus Episcopius, Basel

\bibitem[\protect\citeauthoryear{{Akerman}, {Carigi}, {Nissen}, {Pettini}  \&
  {Asplund}}{{Akerman} et~al.}{2004}]{Akerman04}
{Akerman} C.~J.,  {Carigi} L.,  {Nissen} P.~E.,  {Pettini} M.,   {Asplund} M.,
  2004, \mn@doi [\aap] {10.1051/0004-6361:20034188}, \href
  {http://adsabs.harvard.edu/abs/2004A%26A...414..931A} {414, 931}

\bibitem[\protect\citeauthoryear{{Allen}, {Groves}, {Dopita}, {Sutherland}  \&
  {Kewley}}{{Allen} et~al.}{2008}]{Allen08}
{Allen} M.~G.,  {Groves} B.~A.,  {Dopita} M.~A.,  {Sutherland} R.~S.,
  {Kewley} L.~J.,  2008, \mn@doi [\apjs] {10.1086/589652}, \href
  {http://adsabs.harvard.edu/abs/2008ApJS..178...20A} {178, 20}

\bibitem[\protect\citeauthoryear{{Aller}}{{Aller}}{1942}]{Aller42}
{Aller} L.~H.,  1942, \mn@doi [\apj] {10.1086/144372}, \href
  {http://adsabs.harvard.edu/abs/1942ApJ....95...52A} {95, 52}

\bibitem[\protect\citeauthoryear{{Aller}}{{Aller}}{1954}]{Aller54}
{Aller} L.~H.,  1954, \mn@doi [\apj] {10.1086/145931}, \href
  {http://adsabs.harvard.edu/abs/1954ApJ...120..401A} {120, 401}

\bibitem[\protect\citeauthoryear{{Aller}}{{Aller}}{1984}]{Aller84}
{Aller} L.~H.,  ed. 1984, {Physics of thermal gaseous nebulae}  Astrophysics
  and Space Science Library Vol. 112, \mn@doi{10.1007/978-94-010-9639-3.
}

\bibitem[\protect\citeauthoryear{{Amor{\'{\i}}n}, {P{\'e}rez-Montero}  \&
  {V{\'{\i}}lchez}}{{Amor{\'{\i}}n} et~al.}{2010}]{Amorin10}
{Amor{\'{\i}}n} R.~O.,  {P{\'e}rez-Montero} E.,   {V{\'{\i}}lchez} J.~M.,
  2010, \mn@doi [\apjl] {10.1088/2041-8205/715/2/L128}, \href
  {http://adsabs.harvard.edu/abs/2010ApJ...715L.128A} {715, L128}

\bibitem[\protect\citeauthoryear{{Amor{\'{\i}}n}, {P{\'e}rez-Montero},
  {V{\'{\i}}lchez}  \& {Papaderos}}{{Amor{\'{\i}}n} et~al.}{2012}]{Amorin12b}
{Amor{\'{\i}}n} R.,  {P{\'e}rez-Montero} E.,  {V{\'{\i}}lchez} J.~M.,
  {Papaderos} P.,  2012, \mn@doi [\apj] {10.1088/0004-637X/749/2/185}, \href
  {http://adsabs.harvard.edu/abs/2012ApJ...749..185A} {749, 185}

\bibitem[\protect\citeauthoryear{{Amor{\'{\i}}n} et~al.,}{{Amor{\'{\i}}n}
  et~al.}{2015}]{Amorin15}
{Amor{\'{\i}}n} R.,  et~al., 2015, \mn@doi [\aap]
  {10.1051/0004-6361/201322786}, \href
  {http://adsabs.harvard.edu/abs/2015A%26A...578A.105A} {578, A105}

\bibitem[\protect\citeauthoryear{{Amor{\'{\i}}n} et~al.,}{{Amor{\'{\i}}n}
  et~al.}{2017}]{Amorin17}
{Amor{\'{\i}}n} R.,  et~al., 2017, \mn@doi [Nature Astronomy]
  {10.1038/s41550-017-0052}, \href
  {http://adsabs.harvard.edu/abs/2017NatAs...1E..52A} {1, 0052}

\bibitem[\protect\citeauthoryear{{Andrews} \& {Martini}}{{Andrews} \&
  {Martini}}{2013}]{Andrews13}
{Andrews} B.~H.,  {Martini} P.,  2013, \mn@doi [\apj]
  {10.1088/0004-637X/765/2/140}, \href
  {http://adsabs.harvard.edu/abs/2013ApJ...765..140A} {765, 140}

\bibitem[\protect\citeauthoryear{{Andrievsky}, {Luck}, {Martin}  \&
  {L{\'e}pine}}{{Andrievsky} et~al.}{2004}]{Andrievsky04}
{Andrievsky} S.~M.,  {Luck} R.~E.,  {Martin} P.,   {L{\'e}pine} J.~R.~D.,
  2004, \mn@doi [\aap] {10.1051/0004-6361:20031528}, \href
  {http://adsabs.harvard.edu/abs/2004A%26A...413..159A} {413, 159}

\bibitem[\protect\citeauthoryear{{Arabsalmani}, {M{\o}ller}, {Fynbo},
  {Christensen}, {Freudling}, {Savaglio}  \& {Zafar}}{{Arabsalmani}
  et~al.}{2015a}]{Arabsalmani14}
{Arabsalmani} M.,  {M{\o}ller} P.,  {Fynbo} J.~P.~U.,  {Christensen} L.,
  {Freudling} W.,  {Savaglio} S.,   {Zafar} T.,  2015a, \mn@doi [\mnras]
  {10.1093/mnras/stu2138}, \href
  {http://adsabs.harvard.edu/abs/2015MNRAS.446..990A} {446, 990}

\bibitem[\protect\citeauthoryear{{Arabsalmani}, {M{\o}ller}, {Fynbo},
  {Christensen}, {Freudling}, {Savaglio}  \& {Zafar}}{{Arabsalmani}
  et~al.}{2015b}]{Arabsalmani15}
{Arabsalmani} M.,  {M{\o}ller} P.,  {Fynbo} J.~P.~U.,  {Christensen} L.,
  {Freudling} W.,  {Savaglio} S.,   {Zafar} T.,  2015b, \mn@doi [\mnras]
  {10.1093/mnras/stu2138}, \href
  {http://adsabs.harvard.edu/abs/2015MNRAS.446..990A} {446, 990}

\bibitem[\protect\citeauthoryear{{Arabsalmani} et~al.,}{{Arabsalmani}
  et~al.}{2018}]{Arabsalmani18}
{Arabsalmani} M.,  et~al., 2018, \mn@doi [\mnras] {10.1093/mnras/stx2451},
  \href {http://adsabs.harvard.edu/abs/2018MNRAS.473.3312A} {473, 3312}

\bibitem[\protect\citeauthoryear{{Arav} et~al.,}{{Arav} et~al.}{2007}]{Arav07}
{Arav} N.,  et~al., 2007, \mn@doi [\apj] {10.1086/511666}, \href
  {http://adsabs.harvard.edu/abs/2007ApJ...658..829A} {658, 829}

\bibitem[\protect\citeauthoryear{{Ascasibar}, {Gavil{\'a}n}, {Pinto}, {Casado},
  {Rosales-Ortega}  \& {D{\'{\i}}az}}{{Ascasibar} et~al.}{2015}]{Ascasibar15}
{Ascasibar} Y.,  {Gavil{\'a}n} M.,  {Pinto} N.,  {Casado} J.,  {Rosales-Ortega}
  F.,   {D{\'{\i}}az} A.~I.,  2015, \mn@doi [\mnras] {10.1093/mnras/stv098},
  \href {http://adsabs.harvard.edu/abs/2015MNRAS.448.2126A} {448, 2126}

\bibitem[\protect\citeauthoryear{{Asplund}, {Grevesse}, {Sauval}  \&
  {Scott}}{{Asplund} et~al.}{2009}]{Asplund09}
{Asplund} M.,  {Grevesse} N.,  {Sauval} A.~J.,   {Scott} P.,  2009, \mn@doi
  [\araa] {10.1146/annurev.astro.46.060407.145222}, \href
  {http://adsabs.harvard.edu/abs/2009ARA%26A..47..481A} {47, 481}

\bibitem[\protect\citeauthoryear{{Ba{\~n}ados} et~al.,}{{Ba{\~n}ados}
  et~al.}{2018}]{Banados18}
{Ba{\~n}ados} E.,  et~al., 2018, \mn@doi [\nat] {10.1038/nature25180}, \href
  {http://adsabs.harvard.edu/abs/2018Natur.553..473B} {553, 473}

\bibitem[\protect\citeauthoryear{{Bacon} et~al.,}{{Bacon}
  et~al.}{2010}]{Bacon10}
{Bacon} R.,  et~al., 2010, in Ground-based and Airborne Instrumentation for
  Astronomy III. p. 773508, \mn@doi{10.1117/12.856027}

\bibitem[\protect\citeauthoryear{{Bah{\'e}}, {Schaye}, {Crain}, {McCarthy},
  {Bower}, {Theuns}, {McGee}  \& {Trayford}}{{Bah{\'e}} et~al.}{2017}]{Bahe17}
{Bah{\'e}} Y.~M.,  {Schaye} J.,  {Crain} R.~A.,  {McCarthy} I.~G.,  {Bower}
  R.~G.,  {Theuns} T.,  {McGee} S.~L.,   {Trayford} J.~W.,  2017, \mn@doi
  [\mnras] {10.1093/mnras/stw2329}, \href
  {http://adsabs.harvard.edu/abs/2017MNRAS.464..508B} {464, 508}

\bibitem[\protect\citeauthoryear{{Bailin}, {Stinson}, {Couchman}, {Harris},
  {Wadsley}  \& {Shen}}{{Bailin} et~al.}{2010}]{Bailin10}
{Bailin} J.,  {Stinson} G.,  {Couchman} H.,  {Harris} W.~E.,  {Wadsley} J.,
  {Shen} S.,  2010, \mn@doi [\apj] {10.1088/0004-637X/715/1/194}, \href
  {http://adsabs.harvard.edu/abs/2010ApJ...715..194B} {715, 194}

\bibitem[\protect\citeauthoryear{{Baldry}, {Glazebrook}  \& {Driver}}{{Baldry}
  et~al.}{2008}]{Baldry08}
{Baldry} I.~K.,  {Glazebrook} K.,   {Driver} S.~P.,  2008, \mn@doi [\mnras]
  {10.1111/j.1365-2966.2008.13348.x}, \href
  {http://adsabs.harvard.edu/abs/2008MNRAS.388..945B} {388, 945}

\bibitem[\protect\citeauthoryear{{Baldwin}, {Phillips}  \&
  {Terlevich}}{{Baldwin} et~al.}{1981}]{Baldwin81}
{Baldwin} J.~A.,  {Phillips} M.~M.,   {Terlevich} R.,  1981, \mn@doi [\pasp]
  {10.1086/130766}, \href {http://adsabs.harvard.edu/abs/1981PASP...93....5B}
  {93, 5}

\bibitem[\protect\citeauthoryear{{Balestra}, {Tozzi}, {Ettori}, {Rosati},
  {Borgani}, {Mainieri}, {Norman}  \& {Viola}}{{Balestra}
  et~al.}{2007}]{Balestra07}
{Balestra} I.,  {Tozzi} P.,  {Ettori} S.,  {Rosati} P.,  {Borgani} S.,
  {Mainieri} V.,  {Norman} C.,   {Viola} M.,  2007, \mn@doi [\aap]
  {10.1051/0004-6361:20065568}, \href
  {http://adsabs.harvard.edu/abs/2007A%26A...462..429B} {462, 429}

\bibitem[\protect\citeauthoryear{{Balogh}, {Morris}, {Yee}, {Carlberg}  \&
  {Ellingson}}{{Balogh} et~al.}{1999}]{Balogh99}
{Balogh} M.~L.,  {Morris} S.~L.,  {Yee} H.~K.~C.,  {Carlberg} R.~G.,
  {Ellingson} E.,  1999, \mn@doi [\apj] {10.1086/308056}, 527, 54

\bibitem[\protect\citeauthoryear{{Balogh}, {Navarro}  \& {Morris}}{{Balogh}
  et~al.}{2000}]{Balogh00}
{Balogh} M.~L.,  {Navarro} J.~F.,   {Morris} S.~L.,  2000, \mn@doi [\apj]
  {10.1086/309323}, \href {http://adsabs.harvard.edu/abs/2000ApJ...540..113B}
  {540, 113}

\bibitem[\protect\citeauthoryear{{Balser}, {Rood}, {Bania}  \&
  {Anderson}}{{Balser} et~al.}{2011}]{Balser11}
{Balser} D.~S.,  {Rood} R.~T.,  {Bania} T.~M.,   {Anderson} L.~D.,  2011,
  \mn@doi [\apj] {10.1088/0004-637X/738/1/27}, \href
  {http://adsabs.harvard.edu/abs/2011ApJ...738...27B} {738, 27}

\bibitem[\protect\citeauthoryear{{Bamford}, {Rojas}, {Nichol}, {Miller},
  {Wasserman}, {Genovese}  \& {Freeman}}{{Bamford} et~al.}{2008}]{Bamford08}
{Bamford} S.~P.,  {Rojas} A.~L.,  {Nichol} R.~C.,  {Miller} C.~J.,  {Wasserman}
  L.,  {Genovese} C.~R.,   {Freeman} P.~E.,  2008, \mn@doi [\mnras]
  {10.1111/j.1365-2966.2008.13963.x}, \href
  {http://adsabs.harvard.edu/abs/2008MNRAS.391..607B} {391, 607}

\bibitem[\protect\citeauthoryear{{Barrera-Ballesteros}
  et~al.,}{{Barrera-Ballesteros} et~al.}{2016}]{Barrera-Ballesteros16}
{Barrera-Ballesteros} J.~K.,  et~al., 2016, \mn@doi [\mnras]
  {10.1093/mnras/stw1984}, \href
  {http://adsabs.harvard.edu/abs/2016MNRAS.463.2513B} {463, 2513}

\bibitem[\protect\citeauthoryear{{Barrera-Ballesteros}, {S{\'a}nchez},
  {Heckman}, {Blanc}  \& {The MaNGA Team}}{{Barrera-Ballesteros}
  et~al.}{2017}]{Barrera-Ballesteros17}
{Barrera-Ballesteros} J.~K.,  {S{\'a}nchez} S.~F.,  {Heckman} T.,  {Blanc}
  G.~A.,   {The MaNGA Team} 2017, \mn@doi [\apj] {10.3847/1538-4357/aa7aa9},
  \href {http://adsabs.harvard.edu/abs/2017ApJ...844...80B} {844, 80}

\bibitem[\protect\citeauthoryear{{Barrera-Ballesteros}
  et~al.,}{{Barrera-Ballesteros} et~al.}{2018}]{Barrera-Ballesteros18}
{Barrera-Ballesteros} J.~K.,  et~al., 2018, \mn@doi [\apj]
  {10.3847/1538-4357/aa9b31}, \href
  {http://adsabs.harvard.edu/abs/2018ApJ...852...74B} {852, 74}

\bibitem[\protect\citeauthoryear{{Becker}, {Sargent}, {Rauch}  \&
  {Carswell}}{{Becker} et~al.}{2012}]{Becker12}
{Becker} G.~D.,  {Sargent} W.~L.~W.,  {Rauch} M.,   {Carswell} R.~F.,  2012,
  \mn@doi [\apj] {10.1088/0004-637X/744/2/91}, \href
  {http://adsabs.harvard.edu/abs/2012ApJ...744...91B} {744, 91}

\bibitem[\protect\citeauthoryear{{Bedregal}, {Cardiel}, {Arag{\'o}n-Salamanca}
  \& {Merrifield}}{{Bedregal} et~al.}{2011}]{Bedregal11}
{Bedregal} A.~G.,  {Cardiel} N.,  {Arag{\'o}n-Salamanca} A.,   {Merrifield}
  M.~R.,  2011, \mn@doi [\mnras] {10.1111/j.1365-2966.2011.18752.x}, \href
  {http://adsabs.harvard.edu/abs/2011MNRAS.415.2063B} {415, 2063}

\bibitem[\protect\citeauthoryear{{Beers} \& {Christlieb}}{{Beers} \&
  {Christlieb}}{2005}]{Beers05}
{Beers} T.~C.,  {Christlieb} N.,  2005, \mn@doi [\araa]
  {10.1146/annurev.astro.42.053102.134057}, \href
  {http://adsabs.harvard.edu/abs/2005ARA%26A..43..531B} {43, 531}

\bibitem[\protect\citeauthoryear{{Belfiore} et~al.,}{{Belfiore}
  et~al.}{2015}]{Belfiore15}
{Belfiore} F.,  et~al., 2015, \mn@doi [\mnras] {10.1093/mnras/stv296}, \href
  {http://adsabs.harvard.edu/abs/2015MNRAS.449..867B} {449, 867}

\bibitem[\protect\citeauthoryear{{Belfiore}, {Maiolino}  \&
  {Bothwell}}{{Belfiore} et~al.}{2016a}]{Belfiore16b}
{Belfiore} F.,  {Maiolino} R.,   {Bothwell} M.,  2016a, \mn@doi [\mnras]
  {10.1093/mnras/stv2332}, \href
  {http://adsabs.harvard.edu/abs/2016MNRAS.455.1218B} {455, 1218}

\bibitem[\protect\citeauthoryear{{Belfiore} et~al.,}{{Belfiore}
  et~al.}{2016b}]{Belfiore16a}
{Belfiore} F.,  et~al., 2016b, \mn@doi [\mnras] {10.1093/mnras/stw1234}, \href
  {http://adsabs.harvard.edu/abs/2016MNRAS.461.3111B} {461, 3111}

\bibitem[\protect\citeauthoryear{{Belfiore} et~al.,}{{Belfiore}
  et~al.}{2017a}]{Belfiore17a}
{Belfiore} F.,  et~al., 2017a, \mn@doi [\mnras] {10.1093/mnras/stx789}, \href
  {http://adsabs.harvard.edu/abs/2017MNRAS.469..151B} {469, 151}

\bibitem[\protect\citeauthoryear{{Belfiore} et~al.,}{{Belfiore}
  et~al.}{2017b}]{Belfiore17}
{Belfiore} F.,  et~al., 2017b, \mn@doi [\mnras] {10.1093/mnras/stx789}, \href
  {http://adsabs.harvard.edu/abs/2017MNRAS.469..151B} {469, 151}

\bibitem[\protect\citeauthoryear{{Belfiore} et~al.,}{{Belfiore}
  et~al.}{2018}]{Belfiore18}
{Belfiore} F.,  et~al., 2018, \mn@doi [\mnras] {10.1093/mnras/sty768}, \href
  {http://adsabs.harvard.edu/abs/2018MNRAS.477.3014B} {477, 3014}

\bibitem[\protect\citeauthoryear{{Belli}, {Jones}, {Ellis}  \&
  {Richard}}{{Belli} et~al.}{2013}]{Belli13a}
{Belli} S.,  {Jones} T.,  {Ellis} R.~S.,   {Richard} J.,  2013, \mn@doi [\apj]
  {10.1088/0004-637X/772/2/141}, \href
  {http://adsabs.harvard.edu/abs/2013ApJ...772..141B} {772, 141}

\bibitem[\protect\citeauthoryear{{Bensby} \& {Feltzing}}{{Bensby} \&
  {Feltzing}}{2006}]{Bensby06}
{Bensby} T.,  {Feltzing} S.,  2006, \mn@doi [\mnras]
  {10.1111/j.1365-2966.2006.10037.x}, \href
  {http://adsabs.harvard.edu/abs/2006MNRAS.367.1181B} {367, 1181}

\bibitem[\protect\citeauthoryear{{Bensby}, {Alves-Brito}, {Oey}, {Yong}  \&
  {Mel{\'e}ndez}}{{Bensby} et~al.}{2010}]{Bensby10}
{Bensby} T.,  {Alves-Brito} A.,  {Oey} M.~S.,  {Yong} D.,   {Mel{\'e}ndez} J.,
  2010, \mn@doi [\aap] {10.1051/0004-6361/201014809}, \href
  {http://adsabs.harvard.edu/abs/2010A%26A...516L..13B} {516, L13}

\bibitem[\protect\citeauthoryear{{Benson}}{{Benson}}{2012}]{Benson12}
{Benson} A.~J.,  2012, \mn@doi [NewA] {10.1016/j.newast.2011.07.004}, \href
  {http://adsabs.harvard.edu/abs/2012NewA...17..175B} {17, 175}

\bibitem[\protect\citeauthoryear{{Berg}, {Skillman}  \& {Marble}}{{Berg}
  et~al.}{2011}]{Berg11}
{Berg} D.~A.,  {Skillman} E.~D.,   {Marble} A.~R.,  2011, \mn@doi [\apj]
  {10.1088/0004-637X/738/1/2}, \href
  {http://adsabs.harvard.edu/abs/2011ApJ...738....2B} {738, 2}

\bibitem[\protect\citeauthoryear{{Berg} et~al.,}{{Berg} et~al.}{2012}]{Berg12}
{Berg} D.~A.,  et~al., 2012, \mn@doi [\apj] {10.1088/0004-637X/754/2/98}, \href
  {http://adsabs.harvard.edu/abs/2012ApJ...754...98B} {754, 98}

\bibitem[\protect\citeauthoryear{{Berg}, {Skillman}, {Garnett}, {Croxall},
  {Marble}, {Smith}, {Gordon}  \& {Kennicutt}}{{Berg} et~al.}{2013}]{Berg13}
{Berg} D.~A.,  {Skillman} E.~D.,  {Garnett} D.~R.,  {Croxall} K.~V.,  {Marble}
  A.~R.,  {Smith} J.~D.,  {Gordon} K.,   {Kennicutt} Jr. R.~C.,  2013, \mn@doi
  [\apj] {10.1088/0004-637X/775/2/128}, 775, 128

\bibitem[\protect\citeauthoryear{{Berg}, {Ellison}, {Prochaska}, {Venn}  \&
  {Dessauges-Zavadsky}}{{Berg} et~al.}{2015a}]{Berg15b}
{Berg} T.~A.~M.,  {Ellison} S.~L.,  {Prochaska} J.~X.,  {Venn} K.~A.,
  {Dessauges-Zavadsky} M.,  2015a, \mn@doi [\mnras] {10.1093/mnras/stv1577},
  \href {http://adsabs.harvard.edu/abs/2015MNRAS.452.4326B} {452, 4326}

\bibitem[\protect\citeauthoryear{{Berg}, {Skillman}, {Croxall}, {Pogge},
  {Moustakas}  \& {Johnson-Groh}}{{Berg} et~al.}{2015b}]{Berg15a}
{Berg} D.~A.,  {Skillman} E.~D.,  {Croxall} K.~V.,  {Pogge} R.~W.,  {Moustakas}
  J.,   {Johnson-Groh} M.,  2015b, \mn@doi [\apj] {10.1088/0004-637X/806/1/16},
  \href {http://adsabs.harvard.edu/abs/2015ApJ...806...16B} {806, 16}

\bibitem[\protect\citeauthoryear{{Berg} et~al.,}{{Berg}
  et~al.}{2016a}]{Berg16b}
{Berg} T.~A.~M.,  et~al., 2016a, \mn@doi [\mnras] {10.1093/mnras/stw2232},
  \href {http://adsabs.harvard.edu/abs/2016MNRAS.463.3021B} {463, 3021}

\bibitem[\protect\citeauthoryear{{Berg}, {Skillman}, {Henry}, {Erb}  \&
  {Carigi}}{{Berg} et~al.}{2016b}]{Berg16a}
{Berg} D.~A.,  {Skillman} E.~D.,  {Henry} R.~B.~C.,  {Erb} D.~K.,   {Carigi}
  L.,  2016b, \mn@doi [\apj] {10.3847/0004-637X/827/2/126}, \href
  {http://adsabs.harvard.edu/abs/2016ApJ...827..126B} {827, 126}

\bibitem[\protect\citeauthoryear{{Berg}, {Erb}, {Auger}, {Pettini}  \&
  {Brammer}}{{Berg} et~al.}{2018}]{Berg18}
{Berg} D.~A.,  {Erb} D.~K.,  {Auger} M.~W.,  {Pettini} M.,   {Brammer} G.~B.,
  2018, \mn@doi [\apj] {10.3847/1538-4357/aab7fa}, \href
  {http://adsabs.harvard.edu/abs/2018ApJ...859..164B} {859, 164}

\bibitem[\protect\citeauthoryear{{Berger} et~al.,}{{Berger}
  et~al.}{2012}]{Berger12}
{Berger} E.,  et~al., 2012, \mn@doi [\apjl] {10.1088/2041-8205/755/2/L29},
  \href {http://adsabs.harvard.edu/abs/2012ApJ...755L..29B} {755, L29}

\bibitem[\protect\citeauthoryear{{Bian}, {Kewley}, {Dopita}  \&
  {Juneau}}{{Bian} et~al.}{2016}]{Bian16}
{Bian} F.,  {Kewley} L.~J.,  {Dopita} M.~A.,   {Juneau} S.,  2016, \mn@doi
  [\apj] {10.3847/0004-637X/822/2/62}, \href
  {http://adsabs.harvard.edu/abs/2016ApJ...822...62B} {822, 62}

\bibitem[\protect\citeauthoryear{{Bian}, {Kewley}, {Dopita}  \& {Blanc}}{{Bian}
  et~al.}{2017}]{Bian17}
{Bian} F.,  {Kewley} L.~J.,  {Dopita} M.~A.,   {Blanc} G.~A.,  2017, \mn@doi
  [\apj] {10.3847/1538-4357/834/1/51}, \href
  {http://adsabs.harvard.edu/abs/2017ApJ...834...51B} {834, 51}

\bibitem[\protect\citeauthoryear{{Bian}, {Kewley}  \& {Dopita}}{{Bian}
  et~al.}{2018}]{Bian18}
{Bian} F.,  {Kewley} L.~J.,   {Dopita} M.~A.,  2018, \mn@doi [\apj]
  {10.3847/1538-4357/aabd74}, \href
  {http://adsabs.harvard.edu/abs/2018ApJ...859..175B} {859, 175}

\bibitem[\protect\citeauthoryear{{Binette}}{{Binette}}{1985}]{Binette85}
{Binette} L.,  1985, \aap, \href
  {http://adsabs.harvard.edu/abs/1985A%26A...143..334B} {143, 334}

\bibitem[\protect\citeauthoryear{{Binette}, {Matadamas}, {H{\"a}gele},
  {Nicholls}, {Magris C.}, {Pe{\~n}a-Guerrero}, {Morisset}  \&
  {Rodr{\'{\i}}guez-Gonz{\'a}lez}}{{Binette} et~al.}{2012}]{Binette12}
{Binette} L.,  {Matadamas} R.,  {H{\"a}gele} G.~F.,  {Nicholls} D.~C.,  {Magris
  C.} G.,  {Pe{\~n}a-Guerrero} M.~{\'A}.,  {Morisset} C.,
  {Rodr{\'{\i}}guez-Gonz{\'a}lez} A.,  2012, \mn@doi [\aap]
  {10.1051/0004-6361/201219515}, \href
  {http://adsabs.harvard.edu/abs/2012A%26A...547A..29B} {547, A29}

\bibitem[\protect\citeauthoryear{{Blanc} \& {Greggio}}{{Blanc} \&
  {Greggio}}{2008}]{Blanc08b}
{Blanc} G.,  {Greggio} L.,  2008, \mn@doi [New Astronomy]
  {10.1016/j.newast.2008.03.010}, \href
  {http://adsabs.harvard.edu/abs/2008NewA...13..606B} {13, 606}

\bibitem[\protect\citeauthoryear{{Blanc}, {Kewley}, {Vogt}  \&
  {Dopita}}{{Blanc} et~al.}{2015}]{Blanc15}
{Blanc} G.~A.,  {Kewley} L.,  {Vogt} F.~P.~A.,   {Dopita} M.~A.,  2015, \mn@doi
  [\apj] {10.1088/0004-637X/798/2/99}, \href
  {http://adsabs.harvard.edu/abs/2015ApJ...798...99B} {798, 99}

\bibitem[\protect\citeauthoryear{{Bonifacio}, {Sbordone}, {Marconi}, {Pasquini}
   \& {Hill}}{{Bonifacio} et~al.}{2004}]{Bonifacio04}
{Bonifacio} P.,  {Sbordone} L.,  {Marconi} G.,  {Pasquini} L.,   {Hill} V.,
  2004, \mn@doi [\aap] {10.1051/0004-6361:20031692}, \href
  {http://adsabs.harvard.edu/abs/2004A%26A...414..503B} {414, 503}

\bibitem[\protect\citeauthoryear{{Borguet}, {Edmonds}, {Arav}, {Dunn}  \&
  {Kriss}}{{Borguet} et~al.}{2012}]{Borguet12}
{Borguet} B.~C.~J.,  {Edmonds} D.,  {Arav} N.,  {Dunn} J.,   {Kriss} G.~A.,
  2012, \mn@doi [\apj] {10.1088/0004-637X/751/2/107}, \href
  {http://adsabs.harvard.edu/abs/2012ApJ...751..107B} {751, 107}

\bibitem[\protect\citeauthoryear{{Bothwell}, {Maiolino}, {Kennicutt}, {Cresci},
  {Mannucci}, {Marconi}  \& {Cicone}}{{Bothwell} et~al.}{2013}]{Bothwell13}
{Bothwell} M.~S.,  {Maiolino} R.,  {Kennicutt} R.,  {Cresci} G.,  {Mannucci}
  F.,  {Marconi} A.,   {Cicone} C.,  2013, \mn@doi [\mnras]
  {10.1093/mnras/stt817}, \href
  {http://adsabs.harvard.edu/abs/2013MNRAS.433.1425B} {433, 1425}

\bibitem[\protect\citeauthoryear{{Bothwell}, {Maiolino}, {Peng}, {Cicone},
  {Griffith}  \& {Wagg}}{{Bothwell} et~al.}{2016a}]{Bothwell16}
{Bothwell} M.~S.,  {Maiolino} R.,  {Peng} Y.,  {Cicone} C.,  {Griffith} H.,
  {Wagg} J.,  2016a, \mn@doi [\mnras] {10.1093/mnras/stv2121}, \href
  {http://adsabs.harvard.edu/abs/2016MNRAS.455.1156B} {455, 1156}

\bibitem[\protect\citeauthoryear{{Bothwell}, {Maiolino}, {Cicone}, {Peng}  \&
  {Wagg}}{{Bothwell} et~al.}{2016b}]{Bothwell16b}
{Bothwell} M.~S.,  {Maiolino} R.,  {Cicone} C.,  {Peng} Y.,   {Wagg} J.,
  2016b, \mn@doi [\aap] {10.1051/0004-6361/201527918}, \href
  {http://adsabs.harvard.edu/abs/2016A%26A...595A..48B} {595, A48}

\bibitem[\protect\citeauthoryear{{Bouch{\'e}}, {Lehnert}, {Aguirre},
  {P{\'e}roux}  \& {Bergeron}}{{Bouch{\'e}} et~al.}{2007}]{Bouche07a}
{Bouch{\'e}} N.,  {Lehnert} M.~D.,  {Aguirre} A.,  {P{\'e}roux} C.,
  {Bergeron} J.,  2007, \mn@doi [\mnras] {10.1111/j.1365-2966.2007.11740.x},
  \href {http://adsabs.harvard.edu/abs/2007MNRAS.378..525B} {378, 525}

\bibitem[\protect\citeauthoryear{{Bouch{\'e}} et~al.,}{{Bouch{\'e}}
  et~al.}{2010}]{Bouche10}
{Bouch{\'e}} N.,  et~al., 2010, \mn@doi [\apj] {10.1088/0004-637X/718/2/1001},
  \href {http://adsabs.harvard.edu/abs/2010ApJ...718.1001B} {718, 1001}

\bibitem[\protect\citeauthoryear{{Bournaud} et~al.,}{{Bournaud}
  et~al.}{2008}]{Bournaud08}
{Bournaud} F.,  et~al., 2008, \mn@doi [\aap] {10.1051/0004-6361:20079250},
  \href {http://adsabs.harvard.edu/abs/2008A%26A...486..741B} {486, 741}

\bibitem[\protect\citeauthoryear{{Bovy}, {Rix}  \& {Hogg}}{{Bovy}
  et~al.}{2012}]{Bovy12}
{Bovy} J.,  {Rix} H.-W.,   {Hogg} D.~W.,  2012, \mn@doi [\apj]
  {10.1088/0004-637X/751/2/131}, \href
  {http://adsabs.harvard.edu/abs/2012ApJ...751..131B} {751, 131}

\bibitem[\protect\citeauthoryear{{Bowen}, {Jenkins}, {Pettini}  \&
  {Tripp}}{{Bowen} et~al.}{2005}]{Bowen05}
{Bowen} D.~V.,  {Jenkins} E.~B.,  {Pettini} M.,   {Tripp} T.~M.,  2005, \mn@doi
  [\apj] {10.1086/497617}, \href
  {http://adsabs.harvard.edu/abs/2005ApJ...635..880B} {635, 880}

\bibitem[\protect\citeauthoryear{{Bresolin}}{{Bresolin}}{2007a}]{Bresolin07a}
{Bresolin} F.,  2007a, \mn@doi [\apj] {10.1086/510380}, \href
  {http://adsabs.harvard.edu/abs/2007ApJ...656..186B} {656, 186}

\bibitem[\protect\citeauthoryear{{Bresolin}}{{Bresolin}}{2007b}]{Bresolin07b}
{Bresolin} F.,  2007b, \mn@doi [\apj] {10.1086/510380}, \href
  {http://adsabs.harvard.edu/abs/2007ApJ...656..186B} {656, 186}

\bibitem[\protect\citeauthoryear{{Bresolin}, {Schaerer}, {Gonz{\'a}lez Delgado}
   \& {Stasi{\'n}ska}}{{Bresolin} et~al.}{2005}]{Bresolin05}
{Bresolin} F.,  {Schaerer} D.,  {Gonz{\'a}lez Delgado} R.~M.,   {Stasi{\'n}ska}
  G.,  2005, \mn@doi [\aap] {10.1051/0004-6361:20053369}, \href
  {http://adsabs.harvard.edu/abs/2005A%26A...441..981B} {441, 981}

\bibitem[\protect\citeauthoryear{{Bresolin}, {Ryan-Weber}, {Kennicutt}  \&
  {Goddard}}{{Bresolin} et~al.}{2009a}]{Bresolin09b}
{Bresolin} F.,  {Ryan-Weber} E.,  {Kennicutt} R.~C.,   {Goddard} Q.,  2009a,
  \mn@doi [\apj] {10.1088/0004-637X/695/1/580}, \href
  {http://adsabs.harvard.edu/abs/2009ApJ...695..580B} {695, 580}

\bibitem[\protect\citeauthoryear{{Bresolin}, {Gieren}, {Kudritzki},
  {Pietrzy{\'n}ski}, {Urbaneja}  \& {Carraro}}{{Bresolin}
  et~al.}{2009b}]{Bresolin09}
{Bresolin} F.,  {Gieren} W.,  {Kudritzki} R.-P.,  {Pietrzy{\'n}ski} G.,
  {Urbaneja} M.~A.,   {Carraro} G.,  2009b, \mn@doi [\apj]
  {10.1088/0004-637X/700/1/309}, \href
  {http://adsabs.harvard.edu/abs/2009ApJ...700..309B} {700, 309}

\bibitem[\protect\citeauthoryear{{Bresolin}, {Kennicutt}  \&
  {Ryan-Weber}}{{Bresolin} et~al.}{2012}]{Bresolin12}
{Bresolin} F.,  {Kennicutt} R.~C.,   {Ryan-Weber} E.,  2012, \mn@doi [\apj]
  {10.1088/0004-637X/750/2/122}, \href
  {http://adsabs.harvard.edu/abs/2012ApJ...750..122B} {750, 122}

\bibitem[\protect\citeauthoryear{{Bresolin}, {Kudritzki}, {Urbaneja}, {Gieren},
  {Ho}  \& {Pietrzy{\'n}ski}}{{Bresolin} et~al.}{2016}]{Bresolin16}
{Bresolin} F.,  {Kudritzki} R.-P.,  {Urbaneja} M.~A.,  {Gieren} W.,  {Ho}
  I.-T.,   {Pietrzy{\'n}ski} G.,  2016, \mn@doi [\apj]
  {10.3847/0004-637X/830/2/64}, \href
  {http://adsabs.harvard.edu/abs/2016ApJ...830...64B} {830, 64}

\bibitem[\protect\citeauthoryear{{Brinchmann}, {Charlot}, {White}, {Tremonti},
  {Kauffmann}, {Heckman}  \& {Brinkmann}}{{Brinchmann}
  et~al.}{2004}]{Brinchmann04}
{Brinchmann} J.,  {Charlot} S.,  {White} S.~D.~M.,  {Tremonti} C.,  {Kauffmann}
  G.,  {Heckman} T.,   {Brinkmann} J.,  2004, \mn@doi [\mnras]
  {10.1111/j.1365-2966.2004.07881.x}, \href
  {http://adsabs.harvard.edu/abs/2004MNRAS.351.1151B} {351, 1151}

\bibitem[\protect\citeauthoryear{{Brinchmann}, {Pettini}  \&
  {Charlot}}{{Brinchmann} et~al.}{2008}]{Brinchmann08}
{Brinchmann} J.,  {Pettini} M.,   {Charlot} S.,  2008, \mn@doi [\mnras]
  {10.1111/j.1365-2966.2008.12914.x}, \href
  {http://adsabs.harvard.edu/abs/2008MNRAS.385..769B} {385, 769}

\bibitem[\protect\citeauthoryear{{Brisbin} \& {Harwit}}{{Brisbin} \&
  {Harwit}}{2012}]{Brisbin12}
{Brisbin} D.,  {Harwit} M.,  2012, \mn@doi [\apj]
  {10.1088/0004-637X/750/2/142}, \href
  {http://adsabs.harvard.edu/abs/2012ApJ...750..142B} {750, 142}

\bibitem[\protect\citeauthoryear{{Brook}, {Stinson}, {Gibson}, {Shen},
  {Macci{\`o}}, {Obreja}, {Wadsley}  \& {Quinn}}{{Brook}
  et~al.}{2014}]{Brook14}
{Brook} C.~B.,  {Stinson} G.,  {Gibson} B.~K.,  {Shen} S.,  {Macci{\`o}} A.~V.,
   {Obreja} A.,  {Wadsley} J.,   {Quinn} T.,  2014, \mn@doi [\mnras]
  {10.1093/mnras/stu1406}, \href
  {http://adsabs.harvard.edu/abs/2014MNRAS.443.3809B} {443, 3809}

\bibitem[\protect\citeauthoryear{{Brooks}, {Governato}, {Booth}, {Willman},
  {Gardner}, {Wadsley}, {Stinson}  \& {Quinn}}{{Brooks}
  et~al.}{2007}]{Brooks07}
{Brooks} A.~M.,  {Governato} F.,  {Booth} C.~M.,  {Willman} B.,  {Gardner}
  J.~P.,  {Wadsley} J.,  {Stinson} G.,   {Quinn} T.,  2007, \mn@doi [\apjl]
  {10.1086/511765}, \href {http://adsabs.harvard.edu/abs/2007ApJ...655L..17B}
  {655, L17}

\bibitem[\protect\citeauthoryear{{Brown}, {Martini}  \& {Andrews}}{{Brown}
  et~al.}{2016}]{Brown16}
{Brown} J.~S.,  {Martini} P.,   {Andrews} B.~H.,  2016, \mn@doi [\mnras]
  {10.1093/mnras/stw392}, \href
  {http://adsabs.harvard.edu/abs/2016MNRAS.458.1529B} {458, 1529}

\bibitem[\protect\citeauthoryear{{Brown}, {Cortese}, {Catinella}  \&
  {Kilborn}}{{Brown} et~al.}{2018}]{Brown17}
{Brown} T.,  {Cortese} L.,  {Catinella} B.,   {Kilborn} V.,  2018, \mn@doi
  [\mnras] {10.1093/mnras/stx2452}, \href
  {http://adsabs.harvard.edu/abs/2018MNRAS.473.1868B} {473, 1868}

\bibitem[\protect\citeauthoryear{{Bruzual} \& {Charlot}}{{Bruzual} \&
  {Charlot}}{2003}]{Bruzual03}
{Bruzual} G.,  {Charlot} S.,  2003, \mn@doi [\mnras]
  {10.1046/j.1365-8711.2003.06897.x}, \href
  {http://adsabs.harvard.edu/abs/2003MNRAS.344.1000B} {344, 1000}

\bibitem[\protect\citeauthoryear{{Bryant} et~al.,}{{Bryant}
  et~al.}{2015}]{Bryant15}
{Bryant} J.~J.,  et~al., 2015, \mn@doi [\mnras] {10.1093/mnras/stu2635}, 447,
  2857

\bibitem[\protect\citeauthoryear{{Bundy} et~al.,}{{Bundy}
  et~al.}{2015}]{Bundy15}
{Bundy} K.,  et~al., 2015, \mn@doi [\apj] {10.1088/0004-637X/798/1/7}, \href
  {http://adsabs.harvard.edu/abs/2015ApJ...798....7B} {798, 7}

\bibitem[\protect\citeauthoryear{{Buonanno}, {Corsi}, {Fusi Pecci}, {Hardy}  \&
  {Zinn}}{{Buonanno} et~al.}{1985}]{Buonanno85}
{Buonanno} R.,  {Corsi} C.~E.,  {Fusi Pecci} F.,  {Hardy} E.,   {Zinn} R.,
  1985, \aap, \href {http://adsabs.harvard.edu/abs/1985A%26A...152...65B} {152,
  65}

\bibitem[\protect\citeauthoryear{{Buote}, {Zappacosta}, {Fang}, {Humphrey},
  {Gastaldello}  \& {Tagliaferri}}{{Buote} et~al.}{2009}]{Buote09}
{Buote} D.~A.,  {Zappacosta} L.,  {Fang} T.,  {Humphrey} P.~J.,  {Gastaldello}
  F.,   {Tagliaferri} G.,  2009, \mn@doi [\apj] {10.1088/0004-637X/695/2/1351},
  \href {http://adsabs.harvard.edu/abs/2009ApJ...695.1351B} {695, 1351}

\bibitem[\protect\citeauthoryear{{Bustamante}, {Sparre}, {Springel}  \&
  {Grand}}{{Bustamante} et~al.}{2018}]{Bustamante18}
{Bustamante} S.,  {Sparre} M.,  {Springel} V.,   {Grand} R.~J.~J.,  2018,
  \mn@doi [\mnras] {10.1093/mnras/sty1692}, \href
  {http://adsabs.harvard.edu/abs/2018MNRAS.479.3381B} {479, 3381}

\bibitem[\protect\citeauthoryear{{Byler}, {Dalcanton}, {Conroy}, {Johnson},
  {Levesque}  \& {Berg}}{{Byler} et~al.}{2018}]{Byler18}
{Byler} N.,  {Dalcanton} J.~J.,  {Conroy} C.,  {Johnson} B.~D.,  {Levesque}
  E.~M.,   {Berg} D.~A.,  2018, \mn@doi [\apj] {10.3847/1538-4357/aacd50},
  \href {http://adsabs.harvard.edu/abs/2018ApJ...863...14B} {863, 14}

\bibitem[\protect\citeauthoryear{{Cacho}, {S{\'a}nchez-Bl{\'a}zquez}, {Gorgas}
  \& {P{\'e}rez}}{{Cacho} et~al.}{2014}]{Cacho14}
{Cacho} R.,  {S{\'a}nchez-Bl{\'a}zquez} P.,  {Gorgas} J.,   {P{\'e}rez} I.,
  2014, \mn@doi [\mnras] {10.1093/mnras/stu935}, \href
  {http://adsabs.harvard.edu/abs/2014MNRAS.442.2496C} {442, 2496}

\bibitem[\protect\citeauthoryear{{Caffau} et~al.,}{{Caffau}
  et~al.}{2011}]{Caffau11}
{Caffau} E.,  et~al., 2011, \mn@doi [\nat] {10.1038/nature10377}, \href
  {http://adsabs.harvard.edu/abs/2011Natur.477...67C} {477, 67}

\bibitem[\protect\citeauthoryear{{Calabr{\`o}} et~al.,}{{Calabr{\`o}}
  et~al.}{2017}]{Calabro17}
{Calabr{\`o}} A.,  et~al., 2017, \mn@doi [\aap] {10.1051/0004-6361/201629762},
  \href {http://adsabs.harvard.edu/abs/2017A%26A...601A..95C} {601, A95}

\bibitem[\protect\citeauthoryear{{Calabr{\`o}} et~al.,}{{Calabr{\`o}}
  et~al.}{2018}]{Calabro18}
{Calabr{\`o}} A.,  et~al., 2018, \mn@doi [\apjl] {10.3847/2041-8213/aad33e},
  \href {http://adsabs.harvard.edu/abs/2018ApJ...862L..22C} {862, L22}

\bibitem[\protect\citeauthoryear{{Calderone}, {Nicastro}, {Ghisellini},
  {Dotti}, {Sbarrato}, {Shankar}  \& {Colpi}}{{Calderone}
  et~al.}{2017}]{Calderone17}
{Calderone} G.,  {Nicastro} L.,  {Ghisellini} G.,  {Dotti} M.,  {Sbarrato} T.,
  {Shankar} F.,   {Colpi} M.,  2017, \mn@doi [\mnras] {10.1093/mnras/stx2239},
  \href {http://adsabs.harvard.edu/abs/2017MNRAS.472.4051C} {472, 4051}

\bibitem[\protect\citeauthoryear{{Calura} \& {Matteucci}}{{Calura} \&
  {Matteucci}}{2004}]{Calura04}
{Calura} F.,  {Matteucci} F.,  2004, \mn@doi [\mnras]
  {10.1111/j.1365-2966.2004.07648.x}, \href
  {http://adsabs.harvard.edu/abs/2004MNRAS.350..351C} {350, 351}

\bibitem[\protect\citeauthoryear{{Calura} \& {Menci}}{{Calura} \&
  {Menci}}{2009}]{Calura09}
{Calura} F.,  {Menci} N.,  2009, \mn@doi [\mnras]
  {10.1111/j.1365-2966.2009.15440.x}, \href
  {http://adsabs.harvard.edu/abs/2009MNRAS.400.1347C} {400, 1347}

\bibitem[\protect\citeauthoryear{{Calura}, {Pipino}  \& {Matteucci}}{{Calura}
  et~al.}{2008}]{Calura08a}
{Calura} F.,  {Pipino} A.,   {Matteucci} F.,  2008, \mn@doi [\aap]
  {10.1051/0004-6361:20078090}, \href
  {http://adsabs.harvard.edu/abs/2008A%26A...479..669C} {479, 669}

\bibitem[\protect\citeauthoryear{{Campbell}, {Terlevich}  \&
  {Melnick}}{{Campbell} et~al.}{1986}]{Campbell86}
{Campbell} A.,  {Terlevich} R.,   {Melnick} J.,  1986, \mn@doi [\mnras]
  {10.1093/mnras/223.4.811}, \href
  {http://adsabs.harvard.edu/abs/1986MNRAS.223..811C} {223, 811}

\bibitem[\protect\citeauthoryear{{Cappellari}}{{Cappellari}}{2017}]{Cappellari17}
{Cappellari} M.,  2017, \mn@doi [\mnras] {10.1093/mnras/stw3020}, \href
  {http://adsabs.harvard.edu/abs/2017MNRAS.466..798C} {466, 798}

\bibitem[\protect\citeauthoryear{{Cardamone} et~al.,}{{Cardamone}
  et~al.}{2009}]{Cardamone09}
{Cardamone} C.,  et~al., 2009, \mn@doi [\mnras]
  {10.1111/j.1365-2966.2009.15383.x}, \href
  {http://adsabs.harvard.edu/abs/2009MNRAS.tmp.1256C} {pp 1256--+}

\bibitem[\protect\citeauthoryear{{Carigi}}{{Carigi}}{2000}]{Carigi00}
{Carigi} L.,  2000, \rmxaa, \href
  {http://adsabs.harvard.edu/abs/2000RMxAA..36..171C} {36, 171}

\bibitem[\protect\citeauthoryear{{Carigi} \& {Peimbert}}{{Carigi} \&
  {Peimbert}}{2011}]{Carigi11}
{Carigi} L.,  {Peimbert} M.,  2011, \rmxaa, \href
  {http://adsabs.harvard.edu/abs/2011RMxAA..47..139C} {47, 139}

\bibitem[\protect\citeauthoryear{{Carniani} et~al.,}{{Carniani}
  et~al.}{2017}]{Carniani17}
{Carniani} S.,  et~al., 2017, \mn@doi [\aap] {10.1051/0004-6361/201630366},
  \href {http://adsabs.harvard.edu/abs/2017A%26A...605A..42C} {605, A42}

\bibitem[\protect\citeauthoryear{{Carniani} et~al.,}{{Carniani}
  et~al.}{2018}]{Carniani18}
{Carniani} S.,  et~al., 2018, \mn@doi [\mnras] {10.1093/mnras/sty1088}, \href
  {http://adsabs.harvard.edu/abs/2018MNRAS.478.1170C} {478, 1170}

\bibitem[\protect\citeauthoryear{{Carollo} et~al.,}{{Carollo}
  et~al.}{2007}]{Carollo07}
{Carollo} D.,  et~al., 2007, \mn@doi [\nat] {10.1038/nature06460}, \href
  {http://adsabs.harvard.edu/abs/2007Natur.450.1020C} {450, 1020}

\bibitem[\protect\citeauthoryear{{Carretero}, {Vazdekis}, {Beckman},
  {S{\'a}nchez-Bl{\'a}zquez}  \& {Gorgas}}{{Carretero}
  et~al.}{2004}]{Carretero04}
{Carretero} C.,  {Vazdekis} A.,  {Beckman} J.~E.,  {S{\'a}nchez-Bl{\'a}zquez}
  P.,   {Gorgas} J.,  2004, \mn@doi [\apjl] {10.1086/422749}, \href
  {http://adsabs.harvard.edu/abs/2004ApJ...609L..45C} {609, L45}

\bibitem[\protect\citeauthoryear{{Carretero}, {Vazdekis}  \&
  {Beckman}}{{Carretero} et~al.}{2007}]{Carretero07}
{Carretero} C.,  {Vazdekis} A.,   {Beckman} J.~E.,  2007, \mn@doi [\mnras]
  {10.1111/j.1365-2966.2006.11370.x}, \href
  {http://adsabs.harvard.edu/abs/2007MNRAS.375.1025C} {375, 1025}

\bibitem[\protect\citeauthoryear{{Carton} et~al.,}{{Carton}
  et~al.}{2018}]{Carton18}
{Carton} D.,  et~al., 2018, \mn@doi [\mnras] {10.1093/mnras/sty1343}, \href
  {http://adsabs.harvard.edu/abs/2018MNRAS.478.4293C} {478, 4293}

\bibitem[\protect\citeauthoryear{{Caselli}, {Benson}, {Myers}  \&
  {Tafalla}}{{Caselli} et~al.}{2002}]{Caselli02}
{Caselli} P.,  {Benson} P.~J.,  {Myers} P.~C.,   {Tafalla} M.,  2002, \mn@doi
  [\apj] {10.1086/340195}, \href
  {http://adsabs.harvard.edu/abs/2002ApJ...572..238C} {572, 238}

\bibitem[\protect\citeauthoryear{{Castellanos}, {D{\'{\i}}az}  \&
  {Terlevich}}{{Castellanos} et~al.}{2002}]{Castellanos02}
{Castellanos} M.,  {D{\'{\i}}az} A.~I.,   {Terlevich} E.,  2002, \mn@doi
  [\mnras] {10.1046/j.1365-8711.2002.04987.x}, \href
  {http://adsabs.harvard.edu/abs/2002MNRAS.329..315C} {329, 315}

\bibitem[\protect\citeauthoryear{{Castro}, {Dors}, {Cardaci}  \&
  {H{\"a}gele}}{{Castro} et~al.}{2017}]{Castro17}
{Castro} C.~S.,  {Dors} O.~L.,  {Cardaci} M.~V.,   {H{\"a}gele} G.~F.,  2017,
  \mn@doi [\mnras] {10.1093/mnras/stx150}, \href
  {http://adsabs.harvard.edu/abs/2017MNRAS.467.1507C} {467, 1507}

\bibitem[\protect\citeauthoryear{{Cayrel} et~al.,}{{Cayrel}
  et~al.}{2004}]{Cayrel04}
{Cayrel} R.,  et~al., 2004, \mn@doi [\aap] {10.1051/0004-6361:20034074}, \href
  {http://adsabs.harvard.edu/abs/2004A%26A...416.1117C} {416, 1117}

\bibitem[\protect\citeauthoryear{{Centuri{\'o}n}, {Molaro}, {Vladilo},
  {P{\'e}roux}, {Levshakov}  \& {D'Odorico}}{{Centuri{\'o}n}
  et~al.}{2003}]{Centurion03}
{Centuri{\'o}n} M.,  {Molaro} P.,  {Vladilo} G.,  {P{\'e}roux} C.,  {Levshakov}
  S.~A.,   {D'Odorico} V.,  2003, \mn@doi [\aap] {10.1051/0004-6361:20030273},
  \href {http://adsabs.harvard.edu/abs/2003A%26A...403...55C} {403, 55}

\bibitem[\protect\citeauthoryear{{Cescutti}, {Romano}, {Matteucci}, {Chiappini}
   \& {Hirschi}}{{Cescutti} et~al.}{2015}]{Cescutti15}
{Cescutti} G.,  {Romano} D.,  {Matteucci} F.,  {Chiappini} C.,   {Hirschi} R.,
  2015, \mn@doi [\aap] {10.1051/0004-6361/201525698}, \href
  {http://adsabs.harvard.edu/abs/2015A%26A...577A.139C} {577, A139}

\bibitem[\protect\citeauthoryear{{Ceverino}, {Almeida}, {Tu{\~n}{\'o}n},
  {Dekel}, {Elmegreen}, {Elmegreen}  \& {Primack}}{{Ceverino}
  et~al.}{2016}]{Ceverino16}
{Ceverino} D.,  {Almeida} J.~S.,  {Tu{\~n}{\'o}n} C.~M.,  {Dekel} A.,
  {Elmegreen} B.~G.,  {Elmegreen} D.~M.,   {Primack} J.,  2016, \mn@doi
  [\mnras] {10.1093/mnras/stw064}, 457, 2605

\bibitem[\protect\citeauthoryear{{Chen}, {Hou}  \& {Wang}}{{Chen}
  et~al.}{2003}]{Chen03}
{Chen} L.,  {Hou} J.~L.,   {Wang} J.~J.,  2003, \mn@doi [\aj] {10.1086/367911},
  \href {http://adsabs.harvard.edu/abs/2003AJ....125.1397C} {125, 1397}

\bibitem[\protect\citeauthoryear{{Chevalier}, {Kirshner}  \&
  {Raymond}}{{Chevalier} et~al.}{1980}]{Chevalier80}
{Chevalier} R.~A.,  {Kirshner} R.~P.,   {Raymond} J.~C.,  1980, \mn@doi [\apj]
  {10.1086/157623}, \href {http://adsabs.harvard.edu/abs/1980ApJ...235..186C}
  {235, 186}

\bibitem[\protect\citeauthoryear{{Chevallard} \& {Charlot}}{{Chevallard} \&
  {Charlot}}{2016}]{Chevallard16}
{Chevallard} J.,  {Charlot} S.,  2016, \mn@doi [\mnras]
  {10.1093/mnras/stw1756}, \href
  {http://adsabs.harvard.edu/abs/2016MNRAS.462.1415C} {462, 1415}

\bibitem[\protect\citeauthoryear{{Chiappini}, {Gratton}  \& {R.}}{{Chiappini}
  et~al.}{1997}]{Chiappini97}
{Chiappini} C.,  {Gratton}  {R.} 1997, \apj, \href
  {http://adsabs.harvard.edu/abs/1997ApJ...477..765C} {477, 765}

\bibitem[\protect\citeauthoryear{{Chiappini}, {Matteucci}  \&
  {Romano}}{{Chiappini} et~al.}{2001}]{Chiappini01}
{Chiappini} C.,  {Matteucci} F.,   {Romano} D.,  2001, \mn@doi [\apj]
  {10.1086/321427}, \href {http://adsabs.harvard.edu/abs/2001ApJ...554.1044C}
  {554, 1044}

\bibitem[\protect\citeauthoryear{{Chiappini}, {Romano}  \&
  {Matteucci}}{{Chiappini} et~al.}{2003}]{Chiappini03}
{Chiappini} C.,  {Romano} D.,   {Matteucci} F.,  2003, \mn@doi [\mnras]
  {10.1046/j.1365-8711.2003.06154.x}, \href
  {http://adsabs.harvard.edu/abs/2003MNRAS.339...63C} {339, 63}

\bibitem[\protect\citeauthoryear{{Chiappini}, {Matteucci}  \&
  {Ballero}}{{Chiappini} et~al.}{2005}]{Chiappini05}
{Chiappini} C.,  {Matteucci} F.,   {Ballero} S.~K.,  2005, \mn@doi [\aap]
  {10.1051/0004-6361:20042292}, \href
  {http://adsabs.harvard.edu/abs/2005A%26A...437..429C} {437, 429}

\bibitem[\protect\citeauthoryear{{Chisholm}, {Tremonti}  \&
  {Leitherer}}{{Chisholm} et~al.}{2018}]{Chisholm18}
{Chisholm} J.,  {Tremonti} C.,   {Leitherer} C.,  2018, \mn@doi [\mnras]
  {10.1093/mnras/sty2380}, \href
  {http://adsabs.harvard.edu/abs/2018MNRAS.481.1690C} {481, 1690}

\bibitem[\protect\citeauthoryear{{Choi}, {Conroy}, {Moustakas}, {Graves},
  {Holden}, {Brodwin}, {Brown}  \& {van Dokkum}}{{Choi} et~al.}{2014}]{Choi14}
{Choi} J.,  {Conroy} C.,  {Moustakas} J.,  {Graves} G.~J.,  {Holden} B.~P.,
  {Brodwin} M.,  {Brown} M.~J.~I.,   {van Dokkum} P.~G.,  2014, \mn@doi [\apj]
  {10.1088/0004-637X/792/2/95}, \href
  {http://adsabs.harvard.edu/abs/2014ApJ...792...95C} {792, 95}

\bibitem[\protect\citeauthoryear{{Choi}, {Conroy}  \& {Byler}}{{Choi}
  et~al.}{2017}]{Choi17}
{Choi} J.,  {Conroy} C.,   {Byler} N.,  2017, \mn@doi [\apj]
  {10.3847/1538-4357/aa679f}, \href
  {http://adsabs.harvard.edu/abs/2017ApJ...838..159C} {838, 159}

\bibitem[\protect\citeauthoryear{{Christensen} et~al.,}{{Christensen}
  et~al.}{2012}]{Christensen12b}
{Christensen} L.,  et~al., 2012, \mn@doi [\mnras]
  {10.1111/j.1365-2966.2012.22007.x}, \href
  {http://adsabs.harvard.edu/abs/2012MNRAS.427.1973C} {427, 1973}

\bibitem[\protect\citeauthoryear{{Christensen}, {M{\o}ller}, {Fynbo}  \&
  {Zafar}}{{Christensen} et~al.}{2014}]{Christensen14}
{Christensen} L.,  {M{\o}ller} P.,  {Fynbo} J.~P.~U.,   {Zafar} T.,  2014,
  \mn@doi [\mnras] {10.1093/mnras/stu1726}, \href
  {http://adsabs.harvard.edu/abs/2014MNRAS.445..225C} {445, 225}

\bibitem[\protect\citeauthoryear{{Christensen}, {Dav{\'e}}, {Governato},
  {Pontzen}, {Brooks}, {Munshi}, {Quinn}  \& {Wadsley}}{{Christensen}
  et~al.}{2016}]{Christensen16}
{Christensen} C.~R.,  {Dav{\'e}} R.,  {Governato} F.,  {Pontzen} A.,  {Brooks}
  A.,  {Munshi} F.,  {Quinn} T.,   {Wadsley} J.,  2016, \mn@doi [\apj]
  {10.3847/0004-637X/824/1/57}, \href
  {http://adsabs.harvard.edu/abs/2016ApJ...824...57C} {824, 57}

\bibitem[\protect\citeauthoryear{{Chung}, {Rey}, {Sung}, {Yeom}, {Humphrey},
  {Yi}  \& {Kyeong}}{{Chung} et~al.}{2013}]{Chung13}
{Chung} J.,  {Rey} S.-C.,  {Sung} E.-C.,  {Yeom} B.-S.,  {Humphrey} A.,  {Yi}
  W.,   {Kyeong} J.,  2013, \mn@doi [\apjl] {10.1088/2041-8205/767/1/L15},
  \href {http://adsabs.harvard.edu/abs/2013ApJ...767L..15C} {767, L15}

\bibitem[\protect\citeauthoryear{{Cicone} et~al.,}{{Cicone}
  et~al.}{2014}]{Cicone14}
{Cicone} C.,  et~al., 2014, \mn@doi [\aap] {10.1051/0004-6361/201322464}, \href
  {http://adsabs.harvard.edu/abs/2014A%26A...562A..21C} {562, A21}

\bibitem[\protect\citeauthoryear{{Cid Fernandes}, {Stasi{\'n}ska},
  {Schlickmann}, {Mateus}, {Vale Asari}, {Schoenell}  \& {Sodr{\'e}}}{{Cid
  Fernandes} et~al.}{2010}]{Cid-Fernandes10a}
{Cid Fernandes} R.,  {Stasi{\'n}ska} G.,  {Schlickmann} M.~S.,  {Mateus} A.,
  {Vale Asari} N.,  {Schoenell} W.,   {Sodr{\'e}} L.,  2010, \mn@doi [\mnras]
  {10.1111/j.1365-2966.2009.16185.x}, \href
  {http://adsabs.harvard.edu/abs/2010MNRAS.403.1036C} {403, 1036}

\bibitem[\protect\citeauthoryear{{Cirasuolo} et~al.,}{{Cirasuolo}
  et~al.}{2012}]{Cirasuolo12}
{Cirasuolo} M.,  et~al., 2012, in Society of Photo-Optical Instrumentation
  Engineers (SPIE) Conference Series.  (\mn@eprint {} {1208.5780}),
  \mn@doi{10.1117/12.925871}

\bibitem[\protect\citeauthoryear{{Cohen} \& {Huang}}{{Cohen} \&
  {Huang}}{2010}]{Cohen10}
{Cohen} J.~G.,  {Huang} W.,  2010, \mn@doi [\apj]
  {10.1088/0004-637X/719/1/931}, \href
  {http://adsabs.harvard.edu/abs/2010ApJ...719..931C} {719, 931}

\bibitem[\protect\citeauthoryear{{Coil} et~al.,}{{Coil} et~al.}{2015}]{Coil15}
{Coil} A.~L.,  et~al., 2015, \mn@doi [\apj] {10.1088/0004-637X/801/1/35}, \href
  {http://adsabs.harvard.edu/abs/2015ApJ...801...35C} {801, 35}

\bibitem[\protect\citeauthoryear{{Cole}, {Lacey}, {Baugh}  \& {Frenk}}{{Cole}
  et~al.}{2000}]{Cole00}
{Cole} S.,  {Lacey} C.~G.,  {Baugh} C.~M.,   {Frenk} C.~S.,  2000, \mn@doi
  [\mnras] {10.1046/j.1365-8711.2000.03879.x}, \href
  {http://adsabs.harvard.edu/abs/2000MNRAS.319..168C} {319, 168}

\bibitem[\protect\citeauthoryear{{Collacchioni}, {Cora}, {Lagos}  \&
  {Vega-Mart{\'{\i}}nez}}{{Collacchioni} et~al.}{2018}]{Collacchioni18}
{Collacchioni} F.,  {Cora} S.~A.,  {Lagos} C.~D.~P.,   {Vega-Mart{\'{\i}}nez}
  C.~A.,  2018, \mn@doi [\mnras] {10.1093/mnras/sty2347}, \href
  {http://adsabs.harvard.edu/abs/2018MNRAS.481..954C} {481, 954}

\bibitem[\protect\citeauthoryear{{Conroy}}{{Conroy}}{2013}]{Conroy13}
{Conroy} C.,  2013, \mn@doi [\araa] {10.1146/annurev-astro-082812-141017},
  \href {http://adsabs.harvard.edu/abs/2013ARA%26A..51..393C} {51, 393}

\bibitem[\protect\citeauthoryear{{Conroy} \& {van Dokkum}}{{Conroy} \& {van
  Dokkum}}{2012}]{Conroy12}
{Conroy} C.,  {van Dokkum} P.~G.,  2012, \mn@doi [\apj]
  {10.1088/0004-637X/760/1/71}, \href
  {http://adsabs.harvard.edu/abs/2012ApJ...760...71C} {760, 71}

\bibitem[\protect\citeauthoryear{{Conroy}, {Gunn}  \& {White}}{{Conroy}
  et~al.}{2009}]{Conroy09}
{Conroy} C.,  {Gunn} J.~E.,   {White} M.,  2009, \mn@doi [\apj]
  {10.1088/0004-637X/699/1/486}, \href
  {http://adsabs.harvard.edu/abs/2009ApJ...699..486C} {699, 486}

\bibitem[\protect\citeauthoryear{{Conroy}, {Graves}  \& {van Dokkum}}{{Conroy}
  et~al.}{2014}]{Conroy14}
{Conroy} C.,  {Graves} G.~J.,   {van Dokkum} P.~G.,  2014, \mn@doi [\apj]
  {10.1088/0004-637X/780/1/33}, \href
  {http://adsabs.harvard.edu/abs/2014ApJ...780...33C} {780, 33}

\bibitem[\protect\citeauthoryear{{Contini}, {Treyer}, {Sullivan}  \&
  {Ellis}}{{Contini} et~al.}{2002}]{Contini02}
{Contini} T.,  {Treyer} M.~A.,  {Sullivan} M.,   {Ellis} R.~S.,  2002, \mn@doi
  [\mnras] {10.1046/j.1365-8711.2002.05042.x}, \href
  {http://adsabs.harvard.edu/abs/2002MNRAS.330...75C} {330, 75}

\bibitem[\protect\citeauthoryear{{Cooke}, {Pettini}  \& {Murphy}}{{Cooke}
  et~al.}{2012}]{Cooke12}
{Cooke} R.,  {Pettini} M.,   {Murphy} M.~T.,  2012, \mn@doi [\mnras]
  {10.1111/j.1365-2966.2012.21470.x}, \href
  {http://adsabs.harvard.edu/abs/2012MNRAS.425..347C} {425, 347}

\bibitem[\protect\citeauthoryear{{Cooke}, {Pettini}  \& {Jorgenson}}{{Cooke}
  et~al.}{2015}]{Cooke15}
{Cooke} R.~J.,  {Pettini} M.,   {Jorgenson} R.~A.,  2015, \mn@doi [\apj]
  {10.1088/0004-637X/800/1/12}, \href
  {http://adsabs.harvard.edu/abs/2015ApJ...800...12C} {800, 12}

\bibitem[\protect\citeauthoryear{{Cooke}, {Pettini}  \& {Steidel}}{{Cooke}
  et~al.}{2017}]{Cooke17}
{Cooke} R.~J.,  {Pettini} M.,   {Steidel} C.~C.,  2017, \mn@doi [\mnras]
  {10.1093/mnras/stx037}, \href
  {http://adsabs.harvard.edu/abs/2017MNRAS.467..802C} {467, 802}

\bibitem[\protect\citeauthoryear{{Cooksey}, {Prochaska}, {Chen}, {Mulchaey}  \&
  {Weiner}}{{Cooksey} et~al.}{2008}]{Cooksey08}
{Cooksey} K.~L.,  {Prochaska} J.~X.,  {Chen} H.-W.,  {Mulchaey} J.~S.,
  {Weiner} B.~J.,  2008, \mn@doi [\apj] {10.1086/528704}, \href
  {http://adsabs.harvard.edu/abs/2008ApJ...676..262C} {676, 262}

\bibitem[\protect\citeauthoryear{{Cooper}, {Tremonti}, {Newman}  \&
  {Zabludoff}}{{Cooper} et~al.}{2008}]{Cooper08b}
{Cooper} M.~C.,  {Tremonti} C.~A.,  {Newman} J.~A.,   {Zabludoff} A.~I.,  2008,
  \mn@doi [\mnras] {10.1111/j.1365-2966.2008.13714.x}, \href
  {http://adsabs.harvard.edu/abs/2008MNRAS.390..245C} {390, 245}

\bibitem[\protect\citeauthoryear{{Cortijo-Ferrero} et~al.,}{{Cortijo-Ferrero}
  et~al.}{2017}]{Cortijo-Ferrero17}
{Cortijo-Ferrero} C.,  et~al., 2017, \mn@doi [\mnras] {10.1093/mnras/stx383},
  \href {http://adsabs.harvard.edu/abs/2017MNRAS.467.3898C} {467, 3898}

\bibitem[\protect\citeauthoryear{{Costa}, {Sijacki}  \& {Haehnelt}}{{Costa}
  et~al.}{2015}]{Costa15}
{Costa} T.,  {Sijacki} D.,   {Haehnelt} M.~G.,  2015, \mn@doi [\mnras]
  {10.1093/mnrasl/slu193}, \href
  {http://adsabs.harvard.edu/abs/2015MNRAS.448L..30C} {448, L30}

\bibitem[\protect\citeauthoryear{{Cousin}, {Buat}, {Boissier}, {Bethermin},
  {Roehlly}  \& {G{\'e}nois}}{{Cousin} et~al.}{2016}]{Cousin16}
{Cousin} M.,  {Buat} V.,  {Boissier} S.,  {Bethermin} M.,  {Roehlly} Y.,
  {G{\'e}nois} M.,  2016, \mn@doi [\aap] {10.1051/0004-6361/201527734}, \href
  {http://adsabs.harvard.edu/abs/2016A%26A...589A.109C} {589, A109}

\bibitem[\protect\citeauthoryear{{Cowie} \& {Barger}}{{Cowie} \&
  {Barger}}{2008}]{Cowie08}
{Cowie} L.~L.,  {Barger} A.~J.,  2008, \mn@doi [\apj] {10.1086/591176}, \href
  {http://adsabs.harvard.edu/abs/2008ApJ...686...72C} {686, 72}

\bibitem[\protect\citeauthoryear{{Cowie}, {Songaila}, {Hu}  \& {Cohen}}{{Cowie}
  et~al.}{1996}]{Cowie96}
{Cowie} L.~L.,  {Songaila} A.,  {Hu} E.~M.,   {Cohen} J.~G.,  1996, \mn@doi
  [\aj] {10.1086/118058}, \href
  {http://adsabs.harvard.edu/abs/1996AJ....112..839C} {112, 839}

\bibitem[\protect\citeauthoryear{{Cowie}, {Barger}  \& {Songaila}}{{Cowie}
  et~al.}{2016}]{Cowie16}
{Cowie} L.~L.,  {Barger} A.~J.,   {Songaila} A.,  2016, \mn@doi [\apj]
  {10.3847/0004-637X/817/1/57}, \href
  {http://adsabs.harvard.edu/abs/2016ApJ...817...57C} {817, 57}

\bibitem[\protect\citeauthoryear{{Coziol}, {Reyes}, {Consid{\`e}re}, {Davoust}
  \& {Contini}}{{Coziol} et~al.}{1999}]{Coziol99}
{Coziol} R.,  {Reyes} R.~E.~C.,  {Consid{\`e}re} S.,  {Davoust} E.,   {Contini}
  T.,  1999, \aap, \href {http://adsabs.harvard.edu/abs/1999A%26A...345..733C}
  {345, 733}

\bibitem[\protect\citeauthoryear{{Crain} et~al.,}{{Crain}
  et~al.}{2015}]{Crain15}
{Crain} R.~A.,  et~al., 2015, \mn@doi [\mnras] {10.1093/mnras/stv725}, \href
  {http://adsabs.harvard.edu/abs/2015MNRAS.450.1937C} {450, 1937}

\bibitem[\protect\citeauthoryear{{Cresci}, {Mannucci}, {Maiolino}, {Marconi},
  {Gnerucci}  \& {Magrini}}{{Cresci} et~al.}{2010}]{Cresci10}
{Cresci} G.,  {Mannucci} F.,  {Maiolino} R.,  {Marconi} A.,  {Gnerucci} A.,
  {Magrini} L.,  2010, \mn@doi [\nat] {10.1038/nature09451}, \href
  {http://adsabs.harvard.edu/abs/2010Natur.467..811C} {467, 811}

\bibitem[\protect\citeauthoryear{{Cresci}, {Mannucci}, {Sommariva}, {Maiolino},
  {Marconi}  \& {Brusa}}{{Cresci} et~al.}{2012}]{Cresci12}
{Cresci} G.,  {Mannucci} F.,  {Sommariva} V.,  {Maiolino} R.,  {Marconi} A.,
  {Brusa} M.,  2012, \mn@doi [\mnras] {10.1111/j.1365-2966.2011.20299.x}, \href
  {http://adsabs.harvard.edu/abs/2012MNRAS.421..262C} {421, 262}

\bibitem[\protect\citeauthoryear{{Cresci}, {Vanzi}, {Telles}, {Lanzuisi},
  {Brusa}, {Mingozzi}, {Sauvage}  \& {Johnson}}{{Cresci}
  et~al.}{2017}]{Cresci17}
{Cresci} G.,  {Vanzi} L.,  {Telles} E.,  {Lanzuisi} G.,  {Brusa} M.,
  {Mingozzi} M.,  {Sauvage} M.,   {Johnson} K.,  2017, \mn@doi [\aap]
  {10.1051/0004-6361/201730876}, \href
  {http://adsabs.harvard.edu/abs/2017A%26A...604A.101C} {604, A101}

\bibitem[\protect\citeauthoryear{{Cresci}, {Mannucci}  \& {Curti}}{{Cresci}
  et~al.}{2018}]{Cresci19}
{Cresci} G.,  {Mannucci} F.,   {Curti} M.,  2018, preprint, \href
  {http://adsabs.harvard.edu/abs/2018arXiv181106015C} {} (\mn@eprint {arXiv}
  {1811.06015})

\bibitem[\protect\citeauthoryear{{Crighton}, {O'Meara}  \& {Murphy}}{{Crighton}
  et~al.}{2016}]{Crighton16}
{Crighton} N.~H.~M.,  {O'Meara} J.~M.,   {Murphy} M.~T.,  2016, \mn@doi
  [\mnras] {10.1093/mnrasl/slv191}, \href
  {http://adsabs.harvard.edu/abs/2016MNRAS.457L..44C} {457, L44}

\bibitem[\protect\citeauthoryear{{Croton} et~al.,}{{Croton}
  et~al.}{2006}]{Croton06}
{Croton} D.~J.,  et~al., 2006, \mn@doi [\mnras]
  {10.1111/j.1365-2966.2005.09675.x}, \href
  {http://adsabs.harvard.edu/abs/2006MNRAS.365...11C} {365, 11}

\bibitem[\protect\citeauthoryear{{Croxall} et~al.,}{{Croxall}
  et~al.}{2013}]{Croxall13}
{Croxall} K.~V.,  et~al., 2013, \mn@doi [\apj] {10.1088/0004-637X/777/2/96},
  \href {http://adsabs.harvard.edu/abs/2013ApJ...777...96C} {777, 96}

\bibitem[\protect\citeauthoryear{{Cullen}, {Cirasuolo}, {McLure}, {Dunlop}  \&
  {Bowler}}{{Cullen} et~al.}{2014}]{Cullen14}
{Cullen} F.,  {Cirasuolo} M.,  {McLure} R.~J.,  {Dunlop} J.~S.,   {Bowler}
  R.~A.~A.,  2014, \mn@doi [\mnras] {10.1093/mnras/stu443}, \href
  {http://adsabs.harvard.edu/abs/2014MNRAS.440.2300C} {440, 2300}

\bibitem[\protect\citeauthoryear{{Cullen}, {Cirasuolo}, {Kewley}, {McLure},
  {Dunlop}  \& {Bowler}}{{Cullen} et~al.}{2016}]{Cullen16}
{Cullen} F.,  {Cirasuolo} M.,  {Kewley} L.~J.,  {McLure} R.~J.,  {Dunlop}
  J.~S.,   {Bowler} R.~A.~A.,  2016, \mn@doi [\mnras] {10.1093/mnras/stw1181},
  \href {http://adsabs.harvard.edu/abs/2016MNRAS.460.3002C} {460, 3002}

\bibitem[\protect\citeauthoryear{{Cunha}, {Sellgren}, {Smith}, {Ramirez},
  {Blum}  \& {Terndrup}}{{Cunha} et~al.}{2007}]{Cunha07}
{Cunha} K.,  {Sellgren} K.,  {Smith} V.~V.,  {Ramirez} S.~V.,  {Blum} R.~D.,
  {Terndrup} D.~M.,  2007, \mn@doi [\apj] {10.1086/521813}, \href
  {http://adsabs.harvard.edu/abs/2007ApJ...669.1011C} {669, 1011}

\bibitem[\protect\citeauthoryear{{Curti}}{{Curti}}{2018}]{Curti18a}
{Curti} M.,  2018, in prep.

\bibitem[\protect\citeauthoryear{{Curti}}{{Curti}}{2019}]{Curti19a}
{Curti} M. e.~a.,  2019, in prep.

\bibitem[\protect\citeauthoryear{{Curti}, {Cresci}, {Mannucci}, {Marconi},
  {Maiolino}  \& {Esposito}}{{Curti} et~al.}{2017}]{Curti17}
{Curti} M.,  {Cresci} G.,  {Mannucci} F.,  {Marconi} A.,  {Maiolino} R.,
  {Esposito} S.,  2017, \mn@doi [\mnras] {10.1093/mnras/stw2766}, \href
  {http://adsabs.harvard.edu/abs/2017MNRAS.465.1384C} {465, 1384}

\bibitem[\protect\citeauthoryear{{Cyburt}, {Fields}, {Olive}  \&
  {Yeh}}{{Cyburt} et~al.}{2016}]{Cyburt16}
{Cyburt} R.~H.,  {Fields} B.~D.,  {Olive} K.~A.,   {Yeh} T.-H.,  2016, \mn@doi
  [Reviews of Modern Physics] {10.1103/RevModPhys.88.015004}, \href
  {http://adsabs.harvard.edu/abs/2016RvMP...88a5004C} {88, 015004}

\bibitem[\protect\citeauthoryear{{D'Eugenio}, {Colless}, {Groves}, {Bian}  \&
  {Barone}}{{D'Eugenio} et~al.}{2018}]{DEugenio18}
{D'Eugenio} F.,  {Colless} M.,  {Groves} B.,  {Bian} F.,   {Barone} T.~M.,
  2018, \mn@doi [\mnras] {10.1093/mnras/sty1424}, \href
  {http://cdsads.u-strasbg.fr/abs/2018MNRAS.479.1807D} {479, 1807}

\bibitem[\protect\citeauthoryear{{D'Odorico}, {Cristiani}, {Romano}, {Granato}
  \& {Danese}}{{D'Odorico} et~al.}{2004}]{DOdorico04}
{D'Odorico} V.,  {Cristiani} S.,  {Romano} D.,  {Granato} G.~L.,   {Danese} L.,
   2004, \mn@doi [\mnras] {10.1111/j.1365-2966.2004.07840.x}, \href
  {http://adsabs.harvard.edu/abs/2004MNRAS.351..976D} {351, 976}

\bibitem[\protect\citeauthoryear{{D'Odorico}, {Calura}, {Cristiani}  \&
  {Viel}}{{D'Odorico} et~al.}{2010}]{DOdorico10}
{D'Odorico} V.,  {Calura} F.,  {Cristiani} S.,   {Viel} M.,  2010, \mn@doi
  [\mnras] {10.1111/j.1365-2966.2009.15856.x}, \href
  {http://adsabs.harvard.edu/abs/2010MNRAS.401.2715D} {401, 2715}

\bibitem[\protect\citeauthoryear{{D'Odorico} et~al.,}{{D'Odorico}
  et~al.}{2013}]{DOdorico13}
{D'Odorico} V.,  et~al., 2013, \mn@doi [\mnras] {10.1093/mnras/stt1365}, 435,
  1198

\bibitem[\protect\citeauthoryear{{D'Odorico} et~al.,}{{D'Odorico}
  et~al.}{2016}]{DOdorico16}
{D'Odorico} V.,  et~al., 2016, \mn@doi [\mnras] {10.1093/mnras/stw2161}, \href
  {http://cdsads.u-strasbg.fr/abs/2016MNRAS.463.2690D} {463, 2690}

\bibitem[\protect\citeauthoryear{{Daddi} et~al.,}{{Daddi}
  et~al.}{2007}]{Daddi07}
{Daddi} E.,  et~al., 2007, \mn@doi [\apj] {10.1086/521818}, \href
  {http://adsabs.harvard.edu/abs/2007ApJ...670..156D} {670, 156}

\bibitem[\protect\citeauthoryear{{Daflon} \& {Cunha}}{{Daflon} \&
  {Cunha}}{2004}]{Daflon04}
{Daflon} S.,  {Cunha} K.,  2004, \mn@doi [\apj] {10.1086/425607}, \href
  {http://adsabs.harvard.edu/abs/2004ApJ...617.1115D} {617, 1115}

\bibitem[\protect\citeauthoryear{{Dalcanton}}{{Dalcanton}}{2007}]{Dalcanton07}
{Dalcanton} J.~J.,  2007, \mn@doi [\apj] {10.1086/508913}, \href
  {http://adsabs.harvard.edu/abs/2007ApJ...658..941D} {658, 941}

\bibitem[\protect\citeauthoryear{{Dalton}}{{Dalton}}{2016}]{Dalton16}
{Dalton} G.,  2016, in {Skillen} I.,  {Balcells} M.,   {Trager} S.,  eds,
  Astronomical Society of the Pacific Conference Series Vol. 507, Multi-Object
  Spectroscopy in the Next Decade: Big Questions, Large Surveys, and Wide
  Fields. p.~97

\bibitem[\protect\citeauthoryear{{Dav{\'e}}, {Finlator}  \&
  {Oppenheimer}}{{Dav{\'e}} et~al.}{2011}]{Dave11c}
{Dav{\'e}} R.,  {Finlator} K.,   {Oppenheimer} B.~D.,  2011, \mn@doi [\mnras]
  {10.1111/j.1365-2966.2011.19132.x}, \href
  {http://adsabs.harvard.edu/abs/2011MNRAS.416.1354D} {416, 1354}

\bibitem[\protect\citeauthoryear{{Dav{\'e}}, {Finlator}  \&
  {Oppenheimer}}{{Dav{\'e}} et~al.}{2012}]{Dave12}
{Dav{\'e}} R.,  {Finlator} K.,   {Oppenheimer} B.~D.,  2012, \mn@doi [\mnras]
  {10.1111/j.1365-2966.2011.20148.x}, \href
  {http://adsabs.harvard.edu/abs/2012MNRAS.421...98D} {421, 98}

\bibitem[\protect\citeauthoryear{{Dav{\'e}}, {Thompson}  \&
  {Hopkins}}{{Dav{\'e}} et~al.}{2016}]{Dave16}
{Dav{\'e}} R.,  {Thompson} R.,   {Hopkins} P.~F.,  2016, \mn@doi [\mnras]
  {10.1093/mnras/stw1862}, \href
  {http://adsabs.harvard.edu/abs/2016MNRAS.462.3265D} {462, 3265}

\bibitem[\protect\citeauthoryear{{Dav{\'e}}, {Rafieferantsoa}, {Thompson}  \&
  {Hopkins}}{{Dav{\'e}} et~al.}{2017}]{Dave17b}
{Dav{\'e}} R.,  {Rafieferantsoa} M.~H.,  {Thompson} R.~J.,   {Hopkins} P.~F.,
  2017, \mn@doi [\mnras] {10.1093/mnras/stx108}, \href
  {http://adsabs.harvard.edu/abs/2017MNRAS.467..115D} {467, 115}

\bibitem[\protect\citeauthoryear{{Davies}, {Mueller S{\'a}nchez}, {Genzel},
  {Tacconi}, {Hicks}, {Friedrich}  \& {Sternberg}}{{Davies}
  et~al.}{2007}]{Davies07}
{Davies} R.~I.,  {Mueller S{\'a}nchez} F.,  {Genzel} R.,  {Tacconi} L.~J.,
  {Hicks} E.~K.~S.,  {Friedrich} S.,   {Sternberg} A.,  2007, \mn@doi [\apj]
  {10.1086/523032}, \href {http://adsabs.harvard.edu/abs/2007ApJ...671.1388D}
  {671, 1388}

\bibitem[\protect\citeauthoryear{{Davies}, {Kudritzki}  \& {Figer}}{{Davies}
  et~al.}{2010}]{Davies10}
{Davies} B.,  {Kudritzki} R.-P.,   {Figer} D.~F.,  2010, \mn@doi [\mnras]
  {10.1111/j.1365-2966.2010.16965.x}, \href
  {http://adsabs.harvard.edu/abs/2010MNRAS.407.1203D} {407, 1203}

\bibitem[\protect\citeauthoryear{{Davies} et~al.,}{{Davies}
  et~al.}{2015}]{Davies15}
{Davies} B.,  et~al., 2015, The Messenger, \href
  {http://adsabs.harvard.edu/abs/2015Msngr.161...32D} {161, 32}

\bibitem[\protect\citeauthoryear{{Davies} et~al.,}{{Davies}
  et~al.}{2016}]{Davies16}
{Davies} R.~L.,  et~al., 2016, \mn@doi [\mnras] {10.1093/mnras/stw1754}, \href
  {http://adsabs.harvard.edu/abs/2016MNRAS.462.1616D} {462, 1616}

\bibitem[\protect\citeauthoryear{{Davies} et~al.,}{{Davies}
  et~al.}{2017a}]{Davies17b}
{Davies} R.~L.,  et~al., 2017a, \mn@doi [\mnras] {10.1093/mnras/stx1559}, \href
  {http://adsabs.harvard.edu/abs/2017MNRAS.470.4974D} {470, 4974}

\bibitem[\protect\citeauthoryear{{Davies} et~al.,}{{Davies}
  et~al.}{2017b}]{Davies17a}
{Davies} B.,  et~al., 2017b, \mn@doi [\apj] {10.3847/1538-4357/aa89ed}, \href
  {http://adsabs.harvard.edu/abs/2017ApJ...847..112D} {847, 112}

\bibitem[\protect\citeauthoryear{{Dayal} \& {Ferrara}}{{Dayal} \&
  {Ferrara}}{2018}]{Dayal18}
{Dayal} P.,  {Ferrara} A.,  2018, preprint, \href
  {http://adsabs.harvard.edu/abs/2018arXiv180909136D} {} (\mn@eprint {}
  {1809.09136})

\bibitem[\protect\citeauthoryear{{Dayal}, {Ferrara}  \& {Dunlop}}{{Dayal}
  et~al.}{2013}]{Dayal13}
{Dayal} P.,  {Ferrara} A.,   {Dunlop} J.~S.,  2013, \mn@doi [\mnras]
  {10.1093/mnras/stt083}, \href
  {http://adsabs.harvard.edu/abs/2013MNRAS.430.2891D} {430, 2891}

\bibitem[\protect\citeauthoryear{{De Cia}}{{De Cia}}{2018}]{DeCia18a}
{De Cia} A.,  2018, \mn@doi [\aap] {10.1051/0004-6361/201833034}, \href
  {http://adsabs.harvard.edu/abs/2018A%26A...613L...2D} {613, L2}

\bibitem[\protect\citeauthoryear{{De Cia}, {Ledoux}, {Savaglio}, {Schady}  \&
  {Vreeswijk}}{{De Cia} et~al.}{2013}]{De-Cia13}
{De Cia} A.,  {Ledoux} C.,  {Savaglio} S.,  {Schady} P.,   {Vreeswijk} P.~M.,
  2013, \mn@doi [\aap] {10.1051/0004-6361/201321834}, \href
  {http://adsabs.harvard.edu/abs/2013A%26A...560A..88D} {560, A88}

\bibitem[\protect\citeauthoryear{{De Cia}, {Ledoux}, {Mattsson}, {Petitjean},
  {Srianand}, {Gavignaud}  \& {Jenkins}}{{De Cia} et~al.}{2016}]{DeCia16}
{De Cia} A.,  {Ledoux} C.,  {Mattsson} L.,  {Petitjean} P.,  {Srianand} R.,
  {Gavignaud} I.,   {Jenkins} E.~B.,  2016, \mn@doi [\aap]
  {10.1051/0004-6361/201527895}, \href
  {http://adsabs.harvard.edu/abs/2016A%26A...596A..97D} {596, A97}

\bibitem[\protect\citeauthoryear{{De Cia}, {Ledoux}, {Petitjean}  \&
  {Savaglio}}{{De Cia} et~al.}{2018}]{DeCia18b}
{De Cia} A.,  {Ledoux} C.,  {Petitjean} P.,   {Savaglio} S.,  2018, \mn@doi
  [\aap] {10.1051/0004-6361/201731970}, \href
  {http://adsabs.harvard.edu/abs/2018A%26A...611A..76D} {611, A76}

\bibitem[\protect\citeauthoryear{{De Lucia}}{{De Lucia}}{2010}]{De-Lucia09}
{De Lucia} G.,  2010, \mn@doi [\pasa] {10.1071/AS09064}, \href
  {http://cdsads.u-strasbg.fr/abs/2010PASA...27..242D} {27, 242}

\bibitem[\protect\citeauthoryear{{De Lucia} \& {Blaizot}}{{De Lucia} \&
  {Blaizot}}{2007}]{De-Lucia07}
{De Lucia} G.,  {Blaizot} J.,  2007, \mn@doi [\mnras]
  {10.1111/j.1365-2966.2006.11287.x}, \href
  {http://adsabs.harvard.edu/abs/2007MNRAS.375....2D} {375, 2}

\bibitem[\protect\citeauthoryear{{De Lucia}, {Kauffmann}  \& {White}}{{De
  Lucia} et~al.}{2004}]{De-Lucia04}
{De Lucia} G.,  {Kauffmann} G.,   {White} S.~D.~M.,  2004, \mn@doi [\mnras]
  {10.1111/j.1365-2966.2004.07584.x}, \href
  {http://adsabs.harvard.edu/abs/2004MNRAS.349.1101D} {349, 1101}

\bibitem[\protect\citeauthoryear{{De Lucia}, {Fontanot}  \& {Hirschmann}}{{De
  Lucia} et~al.}{2017a}]{De-Lucia17}
{De Lucia} G.,  {Fontanot} F.,   {Hirschmann} M.,  2017a, \mn@doi [\mnras]
  {10.1093/mnrasl/slw242}, \href
  {http://adsabs.harvard.edu/abs/2017MNRAS.466L..88D} {466, L88}

\bibitem[\protect\citeauthoryear{{De Lucia}, {Fontanot}  \& {Hirschmann}}{{De
  Lucia} et~al.}{2017b}]{DeLucia17}
{De Lucia} G.,  {Fontanot} F.,   {Hirschmann} M.,  2017b, \mn@doi [\mnras]
  {10.1093/mnrasl/slw242}, \href
  {http://adsabs.harvard.edu/abs/2017MNRAS.466L..88D} {466, L88}

\bibitem[\protect\citeauthoryear{{De Masi}, {Vincenzo}, {Matteucci}, {Rosani},
  {Barbera}, {Pasquali}  \& {Spitoni}}{{De Masi} et~al.}{2018}]{De-Masi18}
{De Masi} C.,  {Vincenzo} F.,  {Matteucci} F.,  {Rosani} G.,  {Barbera} L.,
  {Pasquali} A.,   {Spitoni} E.,  2018, preprint, \href
  {http://adsabs.harvard.edu/abs/2018arXiv180506841D} {} (\mn@eprint {arXiv}
  {1805.06841})

\bibitem[\protect\citeauthoryear{{De Rosa}, {Decarli}, {Walter}, {Fan},
  {Jiang}, {Kurk}, {Pasquali}  \& {Rix}}{{De Rosa} et~al.}{2011}]{DeRosa11}
{De Rosa} G.,  {Decarli} R.,  {Walter} F.,  {Fan} X.,  {Jiang} L.,  {Kurk} J.,
  {Pasquali} A.,   {Rix} H.~W.,  2011, \mn@doi [\apj]
  {10.1088/0004-637X/739/2/56}, \href
  {http://adsabs.harvard.edu/abs/2011ApJ...739...56D} {739, 56}

\bibitem[\protect\citeauthoryear{{De Rosa} et~al.,}{{De Rosa}
  et~al.}{2014}]{DeRosa14}
{De Rosa} G.,  et~al., 2014, \mn@doi [\apj] {10.1088/0004-637X/790/2/145},
  \href {http://adsabs.harvard.edu/abs/2014ApJ...790..145D} {790, 145}

\bibitem[\protect\citeauthoryear{{De Rossi}, {Theuns}, {Font}  \&
  {McCarthy}}{{De Rossi} et~al.}{2015}]{De-Rossi15a}
{De Rossi} M.~E.,  {Theuns} T.,  {Font} A.~S.,   {McCarthy} I.~G.,  2015,
  \mn@doi [\mnras] {10.1093/mnras/stv1287}, \href
  {http://cdsads.u-strasbg.fr/abs/2015MNRAS.452..486D} {452, 486}

\bibitem[\protect\citeauthoryear{{De Rossi}, {Bower}, {Font}, {Schaye}  \&
  {Theuns}}{{De Rossi} et~al.}{2017}]{De-Rossi17}
{De Rossi} M.~E.,  {Bower} R.~G.,  {Font} A.~S.,  {Schaye} J.,   {Theuns} T.,
  2017, \mn@doi [\mnras] {10.1093/mnras/stx2158}, \href
  {http://adsabs.harvard.edu/abs/2017MNRAS.472.3354D} {472, 3354}

\bibitem[\protect\citeauthoryear{{De Rossi}, {Bower}, {Font}  \& {Schaye}}{{De
  Rossi} et~al.}{2018}]{De-Rossi18}
{De Rossi} M.~E.,  {Bower} R.~G.,  {Font} A.~S.,   {Schaye} T.,  2018, Boletin
  de la Asociacion Argentina de Astronomia La Plata Argentina, \href
  {http://adsabs.harvard.edu/abs/2018BAAA...60..121R} {60, 121}

\bibitem[\protect\citeauthoryear{{De Silva} et~al.,}{{De Silva}
  et~al.}{2015}]{De-Silva15}
{De Silva} G.~M.,  et~al., 2015, \mn@doi [\mnras] {10.1093/mnras/stv327}, 449,
  2604

\bibitem[\protect\citeauthoryear{{Deharveng}, {Pe{\~n}a}, {Caplan}  \&
  {Costero}}{{Deharveng} et~al.}{2000}]{Deharveng00}
{Deharveng} L.,  {Pe{\~n}a} M.,  {Caplan} J.,   {Costero} R.,  2000, \mn@doi
  [\mnras] {10.1046/j.1365-8711.2000.03030.x}, \href
  {http://adsabs.harvard.edu/abs/2000MNRAS.311..329D} {311, 329}

\bibitem[\protect\citeauthoryear{{Dekel} \& {Mandelker}}{{Dekel} \&
  {Mandelker}}{2014}]{Dekel14}
{Dekel} A.,  {Mandelker} N.,  2014, \mn@doi [\mnras] {10.1093/mnras/stu1427},
  \href {http://adsabs.harvard.edu/abs/2014MNRAS.444.2071D} {444, 2071}

\bibitem[\protect\citeauthoryear{{Dekel}, {Zolotov}, {Tweed}, {Cacciato},
  {Ceverino}  \& {Primack}}{{Dekel} et~al.}{2013}]{Dekel13}
{Dekel} A.,  {Zolotov} A.,  {Tweed} D.,  {Cacciato} M.,  {Ceverino} D.,
  {Primack} J.~R.,  2013, \mn@doi [\mnras] {10.1093/mnras/stt1338}, \href
  {http://adsabs.harvard.edu/abs/2013MNRAS.435..999D} {435, 999}

\bibitem[\protect\citeauthoryear{{Delahaye} \& {Pinsonneault}}{{Delahaye} \&
  {Pinsonneault}}{2006}]{Delahaye06}
{Delahaye} F.,  {Pinsonneault} M.~H.,  2006, \mn@doi [\apj] {10.1086/505260},
  \href {http://adsabs.harvard.edu/abs/2006ApJ...649..529D} {649, 529}

\bibitem[\protect\citeauthoryear{{Denicol{\'o}}, {Terlevich}  \&
  {Terlevich}}{{Denicol{\'o}} et~al.}{2002}]{Denicolo02}
{Denicol{\'o}} G.,  {Terlevich} R.,   {Terlevich} E.,  2002, \mn@doi [\mnras]
  {10.1046/j.1365-8711.2002.05041.x}, \href
  {http://adsabs.harvard.edu/abs/2002MNRAS.330...69D} {330, 69}

\bibitem[\protect\citeauthoryear{{Dessauges-Zavadsky}, {Prochaska},
  {D'Odorico}, {Calura}  \& {Matteucci}}{{Dessauges-Zavadsky}
  et~al.}{2006}]{Dessauges-Zavadsky06}
{Dessauges-Zavadsky} M.,  {Prochaska} J.~X.,  {D'Odorico} S.,  {Calura} F.,
  {Matteucci} F.,  2006, \mn@doi [\aap] {10.1051/0004-6361:20053200}, \href
  {http://adsabs.harvard.edu/abs/2006A%26A...445...93D} {445, 93}

\bibitem[\protect\citeauthoryear{{Dessauges-Zavadsky}, {D'Odorico}, {Schaerer},
  {Modigliani}, {Tapken}  \& {Vernet}}{{Dessauges-Zavadsky}
  et~al.}{2010}]{Dessauges-Zavadsky10}
{Dessauges-Zavadsky} M.,  {D'Odorico} S.,  {Schaerer} D.,  {Modigliani} A.,
  {Tapken} C.,   {Vernet} J.,  2010, \mn@doi [\aap]
  {10.1051/0004-6361/200913337}, \href
  {http://adsabs.harvard.edu/abs/2010A%26A...510A..26D} {510, A26+}

\bibitem[\protect\citeauthoryear{{Di Matteo}, {Pipino}, {Lehnert}, {Combes}  \&
  {Semelin}}{{Di Matteo} et~al.}{2009}]{DiMatteo09}
{Di Matteo} P.,  {Pipino} A.,  {Lehnert} M.~D.,  {Combes} F.,   {Semelin} B.,
  2009, \mn@doi [\aap] {10.1051/0004-6361/200911715}, \href
  {http://adsabs.harvard.edu/abs/2009A%26A...499..427D} {499, 427}

\bibitem[\protect\citeauthoryear{{D{\'{\i}}az}, {Castellanos}, {Terlevich}  \&
  {Luisa Garc{\'{\i}}a-Vargas}}{{D{\'{\i}}az} et~al.}{2000}]{Diaz00}
{D{\'{\i}}az} A.~I.,  {Castellanos} M.,  {Terlevich} E.,   {Luisa
  Garc{\'{\i}}a-Vargas} M.,  2000, \mn@doi [\mnras]
  {10.1046/j.1365-8711.2000.03737.x}, \href
  {http://adsabs.harvard.edu/abs/2000MNRAS.318..462D} {318, 462}

\bibitem[\protect\citeauthoryear{{Dietrich}, {Hamann}, {Shields}, {Constantin},
  {Heidt}, {J{\"a}ger}, {Vestergaard}  \& {Wagner}}{{Dietrich}
  et~al.}{2003a}]{Dietrich03b}
{Dietrich} M.,  {Hamann} F.,  {Shields} J.~C.,  {Constantin} A.,  {Heidt} J.,
  {J{\"a}ger} K.,  {Vestergaard} M.,   {Wagner} S.~J.,  2003a, \mn@doi [\apj]
  {10.1086/374662}, \href {http://adsabs.harvard.edu/abs/2003ApJ...589..722D}
  {589, 722}

\bibitem[\protect\citeauthoryear{{Dietrich}, {Hamann}, {Appenzeller}  \&
  {Vestergaard}}{{Dietrich} et~al.}{2003b}]{Dietrich03a}
{Dietrich} M.,  {Hamann} F.,  {Appenzeller} I.,   {Vestergaard} M.,  2003b,
  \mn@doi [\apj] {10.1086/378045}, \href
  {http://adsabs.harvard.edu/abs/2003ApJ...596..817D} {596, 817}

\bibitem[\protect\citeauthoryear{{Dinerstein}}{{Dinerstein}}{1990}]{Dinerstein90}
{Dinerstein} H.~L.,  1990, in {Thronson} Jr. H.~A.,  {Shull} J.~M.,  eds,
  Astrophysics and Space Science Library Vol. 161, The Interstellar Medium in
  Galaxies. pp 257--285, \mn@doi{10.1007/978-94-009-0595-5_10}

\bibitem[\protect\citeauthoryear{{Divoy} et~al.,}{{Divoy}
  et~al.}{2014}]{Divoy14}
{Divoy} C.,  et~al., 2014, \mn@doi [\aap] {10.1051/0004-6361/201423911}, \href
  {http://adsabs.harvard.edu/abs/2014A%26A...569A..64D} {569, A64}

\bibitem[\protect\citeauthoryear{{Dixon} \& {Furlanetto}}{{Dixon} \&
  {Furlanetto}}{2009}]{Dixon09}
{Dixon} K.~L.,  {Furlanetto} S.~R.,  2009, \mn@doi [\apj]
  {10.1088/0004-637X/706/2/970}, \href
  {http://adsabs.harvard.edu/abs/2009ApJ...706..970D} {706, 970}

\bibitem[\protect\citeauthoryear{{Dopita}, {Kewley}, {Heisler}  \&
  {Sutherland}}{{Dopita} et~al.}{2000}]{Dopita00}
{Dopita} M.~A.,  {Kewley} L.~J.,  {Heisler} C.~A.,   {Sutherland} R.~S.,  2000,
  \mn@doi [\apj] {10.1086/309538}, \href
  {http://adsabs.harvard.edu/abs/2000ApJ...542..224D} {542, 224}

\bibitem[\protect\citeauthoryear{{Dopita} et~al.,}{{Dopita}
  et~al.}{2006}]{Dopita06}
{Dopita} M.~A.,  et~al., 2006, \mn@doi [\apjs] {10.1086/508261}, \href
  {http://adsabs.harvard.edu/abs/2006ApJS..167..177D} {167, 177}

\bibitem[\protect\citeauthoryear{{Dopita}, {Sutherland}, {Nicholls}, {Kewley}
  \& {Vogt}}{{Dopita} et~al.}{2013}]{Dopita13}
{Dopita} M.~A.,  {Sutherland} R.~S.,  {Nicholls} D.~C.,  {Kewley} L.~J.,
  {Vogt} F.~P.~A.,  2013, \mn@doi [\apjs] {10.1088/0067-0049/208/1/10}, \href
  {http://adsabs.harvard.edu/abs/2013ApJS..208...10D} {208, 10}

\bibitem[\protect\citeauthoryear{{Dopita}, {Rich}, {Vogt}, {Kewley}, {Ho},
  {Basurah}, {Ali}  \& {Amer}}{{Dopita} et~al.}{2014}]{Dopita14}
{Dopita} M.~A.,  {Rich} J.,  {Vogt} F.~P.~A.,  {Kewley} L.~J.,  {Ho} I.-T.,
  {Basurah} H.~M.,  {Ali} A.,   {Amer} M.~A.,  2014, \mn@doi [\apss]
  {10.1007/s10509-013-1753-2}, \href
  {http://adsabs.harvard.edu/abs/2014Ap%26SS.350..741D} {350, 741}

\bibitem[\protect\citeauthoryear{{Dopita}, {Kewley}, {Sutherland}  \&
  {Nicholls}}{{Dopita} et~al.}{2016}]{Dopita16}
{Dopita} M.~A.,  {Kewley} L.~J.,  {Sutherland} R.~S.,   {Nicholls} D.~C.,
  2016, \mn@doi [\apss] {10.1007/s10509-016-2657-8}, \href
  {http://adsabs.harvard.edu/abs/2016Ap%26SS.361...61D} {361, 61}

\bibitem[\protect\citeauthoryear{{Dors}, {Cardaci}, {H{\"a}gele}  \&
  {Krabbe}}{{Dors} et~al.}{2014}]{Dors14}
{Dors} O.~L.,  {Cardaci} M.~V.,  {H{\"a}gele} G.~F.,   {Krabbe} {\^A}.~C.,
  2014, \mn@doi [\mnras] {10.1093/mnras/stu1218}, \href
  {http://adsabs.harvard.edu/abs/2014MNRAS.443.1291D} {443, 1291}

\bibitem[\protect\citeauthoryear{{Dors}, {Cardaci}, {H{\"a}gele}, {Rodrigues},
  {Grebel}, {Pilyugin}, {Freitas-Lemes}  \& {Krabbe}}{{Dors}
  et~al.}{2015}]{Dors15}
{Dors} O.~L.,  {Cardaci} M.~V.,  {H{\"a}gele} G.~F.,  {Rodrigues} I.,  {Grebel}
  E.~K.,  {Pilyugin} L.~S.,  {Freitas-Lemes} P.,   {Krabbe} A.~C.,  2015,
  \mn@doi [\mnras] {10.1093/mnras/stv1916}, \href
  {http://adsabs.harvard.edu/abs/2015MNRAS.453.4102D} {453, 4102}

\bibitem[\protect\citeauthoryear{{Dors}, {Arellano-C{\'o}rdova}, {Cardaci}  \&
  {H{\"a}gele}}{{Dors} et~al.}{2017}]{Dors17}
{Dors} Jr. O.~L.,  {Arellano-C{\'o}rdova} K.~Z.,  {Cardaci} M.~V.,
  {H{\"a}gele} G.~F.,  2017, \mn@doi [\mnras] {10.1093/mnrasl/slx036}, \href
  {http://adsabs.harvard.edu/abs/2017MNRAS.468L.113D} {468, L113}

\bibitem[\protect\citeauthoryear{{Draine}}{{Draine}}{2003}]{Draine03}
{Draine} B.~T.,  2003, \mn@doi [\araa]
  {10.1146/annurev.astro.41.011802.094840}, \href
  {http://adsabs.harvard.edu/abs/2003ARA%26A..41..241D} {41, 241}

\bibitem[\protect\citeauthoryear{{Dray} \& {Tout}}{{Dray} \&
  {Tout}}{2003}]{Dray03b}
{Dray} L.~M.,  {Tout} C.~A.,  2003, \mn@doi [\mnras]
  {10.1046/j.1365-8711.2003.06420.x}, \href
  {http://adsabs.harvard.edu/abs/2003MNRAS.341..299D} {341, 299}

\bibitem[\protect\citeauthoryear{{Dray}, {Tout}, {Karakas}  \&
  {Lattanzio}}{{Dray} et~al.}{2003}]{Dray03a}
{Dray} L.~M.,  {Tout} C.~A.,  {Karakas} A.~I.,   {Lattanzio} J.~C.,  2003,
  \mn@doi [\mnras] {10.1046/j.1365-8711.2003.06142.x}, \href
  {http://adsabs.harvard.edu/abs/2003MNRAS.338..973D} {338, 973}

\bibitem[\protect\citeauthoryear{{Dubois} et~al.,}{{Dubois}
  et~al.}{2014}]{Dubois14}
{Dubois} Y.,  et~al., 2014, \mn@doi [\mnras] {10.1093/mnras/stu1227}, \href
  {http://adsabs.harvard.edu/abs/2014MNRAS.444.1453D} {444, 1453}

\bibitem[\protect\citeauthoryear{{Dubois}, {Peirani}, {Pichon}, {Devriendt},
  {Gavazzi}, {Welker}  \& {Volonteri}}{{Dubois} et~al.}{2016}]{Dubois16}
{Dubois} Y.,  {Peirani} S.,  {Pichon} C.,  {Devriendt} J.,  {Gavazzi} R.,
  {Welker} C.,   {Volonteri} M.,  2016, \mn@doi [\mnras]
  {10.1093/mnras/stw2265}, \href
  {http://adsabs.harvard.edu/abs/2016MNRAS.463.3948D} {463, 3948}

\bibitem[\protect\citeauthoryear{{Dunne}, {Eales}  \& {Edmunds}}{{Dunne}
  et~al.}{2003}]{Dunne03}
{Dunne} L.,  {Eales} S.~A.,   {Edmunds} M.~G.,  2003, \mn@doi [\mnras]
  {10.1046/j.1365-8711.2003.06440.x}, \href
  {http://adsabs.harvard.edu/abs/2003MNRAS.341..589D} {341, 589}

\bibitem[\protect\citeauthoryear{{Dutton} et~al.,}{{Dutton}
  et~al.}{2011}]{Dutton10}
{Dutton} A.~A.,  et~al., 2011, \mn@doi [\mnras]
  {10.1111/j.1365-2966.2010.17555.x}, \href
  {http://cdsads.u-strasbg.fr/abs/2011MNRAS.410.1660D} {410, 1660}

\bibitem[\protect\citeauthoryear{{Dwek}}{{Dwek}}{1998}]{Dwek98}
{Dwek} E.,  1998, \mn@doi [\apj] {10.1086/305829}, \href
  {http://adsabs.harvard.edu/abs/1998ApJ...501..643D} {501, 643}

\bibitem[\protect\citeauthoryear{{Edmunds}}{{Edmunds}}{1990}]{Edmunds90}
{Edmunds} M.~G.,  1990, \mnras, \href
  {http://adsabs.harvard.edu/abs/1990MNRAS.246..678E} {246, 678}

\bibitem[\protect\citeauthoryear{{Edmunds} \& {Pagel}}{{Edmunds} \&
  {Pagel}}{1978}]{Edmunds78}
{Edmunds} M.~G.,  {Pagel} B.~E.~J.,  1978, \mn@doi [\mnras]
  {10.1093/mnras/185.1.77P}, \href
  {http://adsabs.harvard.edu/abs/1978MNRAS.185P..77E} {185, 77P}

\bibitem[\protect\citeauthoryear{{Edmunds} \& {Phillipps}}{{Edmunds} \&
  {Phillipps}}{1997}]{Edmunds97}
{Edmunds} M.~G.,  {Phillipps} S.,  1997, \mn@doi [\mnras]
  {10.1093/mnras/292.3.733}, \href
  {http://adsabs.harvard.edu/abs/1997MNRAS.292..733E} {292, 733}

\bibitem[\protect\citeauthoryear{{Eldridge} \& {Stanway}}{{Eldridge} \&
  {Stanway}}{2012}]{Eldridge12}
{Eldridge} J.~J.,  {Stanway} E.~R.,  2012, \mn@doi [\mnras]
  {10.1111/j.1365-2966.2011.19713.x}, \href
  {http://adsabs.harvard.edu/abs/2012MNRAS.419..479E} {419, 479}

\bibitem[\protect\citeauthoryear{{Ellison}, {Patton}, {Simard}  \&
  {McConnachie}}{{Ellison} et~al.}{2008a}]{Ellison08b}
{Ellison} S.~L.,  {Patton} D.~R.,  {Simard} L.,   {McConnachie} A.~W.,  2008a,
  \mn@doi [\aj] {10.1088/0004-6256/135/5/1877}, \href
  {http://adsabs.harvard.edu/abs/2008AJ....135.1877E} {135, 1877}

\bibitem[\protect\citeauthoryear{{Ellison}, {Patton}, {Simard}  \&
  {McConnachie}}{{Ellison} et~al.}{2008b}]{Ellison08a}
{Ellison} S.~L.,  {Patton} D.~R.,  {Simard} L.,   {McConnachie} A.~W.,  2008b,
  \mn@doi [\apjl] {10.1086/527296}, \href
  {http://adsabs.harvard.edu/abs/2008ApJ...672L.107E} {672, L107}

\bibitem[\protect\citeauthoryear{{Ellison}, {Simard}, {Cowan}, {Baldry},
  {Patton}  \& {McConnachie}}{{Ellison} et~al.}{2009}]{Ellison09}
{Ellison} S.~L.,  {Simard} L.,  {Cowan} N.~B.,  {Baldry} I.~K.,  {Patton}
  D.~R.,   {McConnachie} A.~W.,  2009, \mn@doi [\mnras]
  {10.1111/j.1365-2966.2009.14817.x}, \href
  {http://adsabs.harvard.edu/abs/2009MNRAS.396.1257E} {396, 1257}

\bibitem[\protect\citeauthoryear{{Ellison}, {Nair}, {Patton}, {Scudder},
  {Mendel}  \& {Simard}}{{Ellison} et~al.}{2011}]{Ellison11}
{Ellison} S.~L.,  {Nair} P.,  {Patton} D.~R.,  {Scudder} J.~M.,  {Mendel}
  J.~T.,   {Simard} L.,  2011, \mn@doi [\mnras]
  {10.1111/j.1365-2966.2011.19195.x}, 416, 2182

\bibitem[\protect\citeauthoryear{{Ellison}, {Mendel}, {Patton}  \&
  {Scudder}}{{Ellison} et~al.}{2013}]{Ellison13}
{Ellison} S.~L.,  {Mendel} J.~T.,  {Patton} D.~R.,   {Scudder} J.~M.,  2013,
  \mn@doi [\mnras] {10.1093/mnras/stt1562}, \href
  {http://adsabs.harvard.edu/abs/2013MNRAS.435.3627E} {435, 3627}

\bibitem[\protect\citeauthoryear{{Ellison}, {S{\'a}nchez}, {Ibarra-Medel},
  {Antonio}, {Mendel}  \& {Barrera-Ballesteros}}{{Ellison}
  et~al.}{2018a}]{Ellison18a}
{Ellison} S.~L.,  {S{\'a}nchez} S.~F.,  {Ibarra-Medel} H.,  {Antonio} B.,
  {Mendel} J.~T.,   {Barrera-Ballesteros} J.,  2018a, \mn@doi [\mnras]
  {10.1093/mnras/stx2882}, \href
  {http://adsabs.harvard.edu/abs/2018MNRAS.474.2039E} {474, 2039}

\bibitem[\protect\citeauthoryear{{Ellison}, {Catinella}  \&
  {Cortese}}{{Ellison} et~al.}{2018b}]{Ellison18b}
{Ellison} S.~L.,  {Catinella} B.,   {Cortese} L.,  2018b, \mn@doi [\mnras]
  {10.1093/mnras/sty1247}, \href
  {http://adsabs.harvard.edu/abs/2018MNRAS.478.3447E} {478, 3447}

\bibitem[\protect\citeauthoryear{{Engler}, {Lisker}  \& {Pillepich}}{{Engler}
  et~al.}{2018}]{Engler18}
{Engler} C.,  {Lisker} T.,   {Pillepich} A.,  2018, \mn@doi [Research Notes of
  the American Astronomical Society] {10.3847/2515-5172/aabcce}, \href
  {http://adsabs.harvard.edu/abs/2018RNAAS...2b...6E} {2, 6}

\bibitem[\protect\citeauthoryear{{Epinat} et~al.,}{{Epinat}
  et~al.}{2009}]{Epinat09a}
{Epinat} B.,  et~al., 2009, \mn@doi [\aap] {10.1051/0004-6361/200911995}, \href
  {http://adsabs.harvard.edu/abs/2009A%26A...504..789E} {504, 789}

\bibitem[\protect\citeauthoryear{{Erb}}{{Erb}}{2008}]{Erb08}
{Erb} D.~K.,  2008, \mn@doi [\apj] {10.1086/524727}, \href
  {http://adsabs.harvard.edu/abs/2008ApJ...674..151E} {674, 151}

\bibitem[\protect\citeauthoryear{{Erb}, {Shapley}, {Pettini}, {Steidel},
  {Reddy}  \& {Adelberger}}{{Erb} et~al.}{2006a}]{Erb06a}
{Erb} D.~K.,  {Shapley} A.~E.,  {Pettini} M.,  {Steidel} C.~C.,  {Reddy} N.~A.,
    {Adelberger} K.~L.,  2006a, \mn@doi [\apj] {10.1086/503623}, \href
  {http://adsabs.harvard.edu/abs/2006ApJ...644..813E} {644, 813}

\bibitem[\protect\citeauthoryear{{Erb}, {Steidel}, {Shapley}, {Pettini},
  {Reddy}  \& {Adelberger}}{{Erb} et~al.}{2006b}]{Erb06c}
{Erb} D.~K.,  {Steidel} C.~C.,  {Shapley} A.~E.,  {Pettini} M.,  {Reddy} N.~A.,
    {Adelberger} K.~L.,  2006b, \mn@doi [\apj] {10.1086/504891}, \href
  {http://adsabs.harvard.edu/abs/2006ApJ...646..107E} {646, 107}

\bibitem[\protect\citeauthoryear{{Erb}, {Pettini}, {Shapley}, {Steidel}, {Law}
  \& {Reddy}}{{Erb} et~al.}{2010}]{Erb10}
{Erb} D.~K.,  {Pettini} M.,  {Shapley} A.~E.,  {Steidel} C.~C.,  {Law} D.~R.,
  {Reddy} N.~A.,  2010, \mn@doi [\apj] {10.1088/0004-637X/719/2/1168}, \href
  {http://adsabs.harvard.edu/abs/2010ApJ...719.1168E} {719, 1168}

\bibitem[\protect\citeauthoryear{{Esteban} \& {Garc{\'{\i}}a-Rojas}}{{Esteban}
  \& {Garc{\'{\i}}a-Rojas}}{2018}]{Esteban18}
{Esteban} C.,  {Garc{\'{\i}}a-Rojas} J.,  2018, \mn@doi [\mnras]
  {10.1093/mnras/sty1168}, \href
  {http://adsabs.harvard.edu/abs/2018MNRAS.tmp.1105E} {}

\bibitem[\protect\citeauthoryear{{Esteban}, {Peimbert}, {Torres-Peimbert}  \&
  {Rodr{\'{\i}}guez}}{{Esteban} et~al.}{2002}]{Esteban02}
{Esteban} C.,  {Peimbert} M.,  {Torres-Peimbert} S.,   {Rodr{\'{\i}}guez} M.,
  2002, \mn@doi [\apj] {10.1086/344104}, \href
  {http://adsabs.harvard.edu/abs/2002ApJ...581..241E} {581, 241}

\bibitem[\protect\citeauthoryear{{Esteban}, {Peimbert}, {Garc{\'{\i}}a-Rojas},
  {Ruiz}, {Peimbert}  \& {Rodr{\'{\i}}guez}}{{Esteban}
  et~al.}{2004}]{Esteban04}
{Esteban} C.,  {Peimbert} M.,  {Garc{\'{\i}}a-Rojas} J.,  {Ruiz} M.~T.,
  {Peimbert} A.,   {Rodr{\'{\i}}guez} M.,  2004, \mn@doi [\mnras]
  {10.1111/j.1365-2966.2004.08313.x}, \href
  {http://adsabs.harvard.edu/abs/2004MNRAS.355..229E} {355, 229}

\bibitem[\protect\citeauthoryear{{Esteban}, {Garc{\'{\i}}a-Rojas}, {Peimbert},
  {Peimbert}, {Ruiz}, {Rodr{\'{\i}}guez}  \& {Carigi}}{{Esteban}
  et~al.}{2005}]{Esteban05}
{Esteban} C.,  {Garc{\'{\i}}a-Rojas} J.,  {Peimbert} M.,  {Peimbert} A.,
  {Ruiz} M.~T.,  {Rodr{\'{\i}}guez} M.,   {Carigi} L.,  2005, \mn@doi [\apjl]
  {10.1086/426889}, \href {http://adsabs.harvard.edu/abs/2005ApJ...618L..95E}
  {618, L95}

\bibitem[\protect\citeauthoryear{{Esteban}, {Bresolin}, {Peimbert},
  {Garc{\'{\i}}a-Rojas}, {Peimbert}  \& {Mesa-Delgado}}{{Esteban}
  et~al.}{2009}]{Esteban09}
{Esteban} C.,  {Bresolin} F.,  {Peimbert} M.,  {Garc{\'{\i}}a-Rojas} J.,
  {Peimbert} A.,   {Mesa-Delgado} A.,  2009, \mn@doi [\apj]
  {10.1088/0004-637X/700/1/654}, \href
  {http://adsabs.harvard.edu/abs/2009ApJ...700..654E} {700, 654}

\bibitem[\protect\citeauthoryear{{Esteban}, {Garc{\'{\i}}a-Rojas}, {Carigi},
  {Peimbert}, {Bresolin}, {L{\'o}pez-S{\'a}nchez}  \& {Mesa-Delgado}}{{Esteban}
  et~al.}{2014}]{Esteban14}
{Esteban} C.,  {Garc{\'{\i}}a-Rojas} J.,  {Carigi} L.,  {Peimbert} M.,
  {Bresolin} F.,  {L{\'o}pez-S{\'a}nchez} A.~R.,   {Mesa-Delgado} A.,  2014,
  \mn@doi [\mnras] {10.1093/mnras/stu1177}, \href
  {http://adsabs.harvard.edu/abs/2014MNRAS.443..624E} {443, 624}

\bibitem[\protect\citeauthoryear{{Fabbian}, {Nissen}, {Asplund}, {Pettini}  \&
  {Akerman}}{{Fabbian} et~al.}{2009}]{Fabbian09}
{Fabbian} D.,  {Nissen} P.~E.,  {Asplund} M.,  {Pettini} M.,   {Akerman} C.,
  2009, \mn@doi [\aap] {10.1051/0004-6361/200810095}, \href
  {http://adsabs.harvard.edu/abs/2009A%26A...500.1143F} {500, 1143}

\bibitem[\protect\citeauthoryear{{Faber}}{{Faber}}{1973}]{Faber73}
{Faber} S.~M.,  1973, \mn@doi [\apj] {10.1086/151912}, \href
  {http://adsabs.harvard.edu/abs/1973ApJ...179..731F} {179, 731}

\bibitem[\protect\citeauthoryear{{Faisst} et~al.,}{{Faisst}
  et~al.}{2016}]{Faisst16}
{Faisst} A.~L.,  et~al., 2016, \mn@doi [\apj] {10.3847/0004-637X/822/1/29},
  \href {http://adsabs.harvard.edu/abs/2016ApJ...822...29F} {822, 29}

\bibitem[\protect\citeauthoryear{{Fang} \& {Bryan}}{{Fang} \&
  {Bryan}}{2001}]{Fang01}
{Fang} T.,  {Bryan} G.~L.,  2001, \mn@doi [\apjl] {10.1086/324571}, \href
  {http://adsabs.harvard.edu/abs/2001ApJ...561L..31F} {561, L31}

\bibitem[\protect\citeauthoryear{{Fang}, {Buote}, {Humphrey}, {Canizares},
  {Zappacosta}, {Maiolino}, {Tagliaferri}  \& {Gastaldello}}{{Fang}
  et~al.}{2010}]{Fang10}
{Fang} T.,  {Buote} D.~A.,  {Humphrey} P.~J.,  {Canizares} C.~R.,  {Zappacosta}
  L.,  {Maiolino} R.,  {Tagliaferri} G.,   {Gastaldello} F.,  2010, \mn@doi
  [\apj] {10.1088/0004-637X/714/2/1715}, \href
  {http://adsabs.harvard.edu/abs/2010ApJ...714.1715F} {714, 1715}

\bibitem[\protect\citeauthoryear{{Fang}, {Garc{\'{\i}}a-Benito}, {Guerrero},
  {Liu}, {Yuan}, {Zhang}  \& {Zhang}}{{Fang} et~al.}{2015}]{Fang15}
{Fang} X.,  {Garc{\'{\i}}a-Benito} R.,  {Guerrero} M.~A.,  {Liu} X.,  {Yuan}
  H.,  {Zhang} Y.,   {Zhang} B.,  2015, \mn@doi [\apj]
  {10.1088/0004-637X/815/1/69}, \href
  {http://adsabs.harvard.edu/abs/2015ApJ...815...69F} {815, 69}

\bibitem[\protect\citeauthoryear{{Feldmann}}{{Feldmann}}{2015}]{Feldmann15}
{Feldmann} R.,  2015, \mn@doi [\mnras] {10.1093/mnras/stv552}, \href
  {http://adsabs.harvard.edu/abs/2015MNRAS.449.3274F} {449, 3274}

\bibitem[\protect\citeauthoryear{{Feltre}, {Charlot}  \& {Gutkin}}{{Feltre}
  et~al.}{2016}]{Feltre16}
{Feltre} A.,  {Charlot} S.,   {Gutkin} J.,  2016, \mn@doi [\mnras]
  {10.1093/mnras/stv2794}, \href
  {http://adsabs.harvard.edu/abs/2016MNRAS.456.3354F} {456, 3354}

\bibitem[\protect\citeauthoryear{{Ferland} et~al.,}{{Ferland}
  et~al.}{2013}]{Ferland13}
{Ferland} G.~J.,  et~al., 2013, \rmxaa, \href
  {http://adsabs.harvard.edu/abs/2013RMxAA..49..137F} {49, 137}

\bibitem[\protect\citeauthoryear{{Fern{\'a}ndez-Alvar}
  et~al.,}{{Fern{\'a}ndez-Alvar} et~al.}{2018}]{Fernandez-Alvar18}
{Fern{\'a}ndez-Alvar} E.,  et~al., 2018, \mn@doi [\apj]
  {10.3847/1538-4357/aa9ced}, \href
  {http://adsabs.harvard.edu/abs/2018ApJ...852...50F} {852, 50}

\bibitem[\protect\citeauthoryear{{Fern{\'a}ndez-Ontiveros}
  et~al.,}{{Fern{\'a}ndez-Ontiveros} et~al.}{2017}]{Fernandez-Ontiveros17}
{Fern{\'a}ndez-Ontiveros} J.~A.,  et~al., 2017, \mn@doi [PASA]
  {10.1017/pasa.2017.43}, \href
  {http://adsabs.harvard.edu/abs/2017PASA...34...53F} {}

\bibitem[\protect\citeauthoryear{{Ferrara}, {Scannapieco}  \&
  {Bergeron}}{{Ferrara} et~al.}{2005}]{Ferrara05}
{Ferrara} A.,  {Scannapieco} E.,   {Bergeron} J.,  2005, \mn@doi [\apjl]
  {10.1086/498845}, \href {http://adsabs.harvard.edu/abs/2005ApJ...634L..37F}
  {634, L37}

\bibitem[\protect\citeauthoryear{{Ferreras} et~al.,}{{Ferreras}
  et~al.}{2009}]{Ferreras09}
{Ferreras} I.,  et~al., 2009, \mn@doi [\apj] {10.1088/0004-637X/706/1/158},
  \href {http://adsabs.harvard.edu/abs/2009ApJ...706..158F} {706, 158}

\bibitem[\protect\citeauthoryear{{Finkelstein}, {Papovich}, {Giavalisco},
  {Reddy}, {Ferguson}, {Koekemoer}  \& {Dickinson}}{{Finkelstein}
  et~al.}{2010}]{Finkelstein09}
{Finkelstein} S.~L.,  {Papovich} C.,  {Giavalisco} M.,  {Reddy} N.~A.,
  {Ferguson} H.~C.,  {Koekemoer} A.~M.,   {Dickinson} M.,  2010, \mn@doi [\apj]
  {10.1088/0004-637X/719/2/1250}, \href
  {http://adsabs.harvard.edu/abs/2010ApJ...719.1250F} {719, 1250}

\bibitem[\protect\citeauthoryear{{Finkelstein} et~al.,}{{Finkelstein}
  et~al.}{2011a}]{Finkelstein11a}
{Finkelstein} S.~L.,  et~al., 2011a, \mn@doi [\apj]
  {10.1088/0004-637X/729/2/140}, \href
  {http://adsabs.harvard.edu/abs/2011ApJ...729..140F} {729, 140}

\bibitem[\protect\citeauthoryear{{Finkelstein}, {Cohen}, {Moustakas},
  {Malhotra}, {Rhoads}  \& {Papovich}}{{Finkelstein}
  et~al.}{2011b}]{Finkelstein11b}
{Finkelstein} S.~L.,  {Cohen} S.~H.,  {Moustakas} J.,  {Malhotra} S.,  {Rhoads}
  J.~E.,   {Papovich} C.,  2011b, \mn@doi [\apj] {10.1088/0004-637X/733/2/117},
  \href {http://adsabs.harvard.edu/abs/2011ApJ...733..117F} {733, 117}

\bibitem[\protect\citeauthoryear{{Finlator} \& {Dav{\'e}}}{{Finlator} \&
  {Dav{\'e}}}{2008}]{Finlator08}
{Finlator} K.,  {Dav{\'e}} R.,  2008, \mn@doi [\mnras]
  {10.1111/j.1365-2966.2008.12991.x}, \href
  {http://adsabs.harvard.edu/abs/2008MNRAS.385.2181F} {385, 2181}

\bibitem[\protect\citeauthoryear{{Fitzpatrick} \& {Graves}}{{Fitzpatrick} \&
  {Graves}}{2015}]{Fitzpatrick15}
{Fitzpatrick} P.~J.,  {Graves} G.~J.,  2015, \mn@doi [\mnras]
  {10.1093/mnras/stu2509}, \href
  {http://adsabs.harvard.edu/abs/2015MNRAS.447.1383F} {447, 1383}

\bibitem[\protect\citeauthoryear{{Fluetsch} et~al.,}{{Fluetsch}
  et~al.}{2018}]{Fluetsch18}
{Fluetsch} A.,  et~al., 2018, preprint, \href
  {http://adsabs.harvard.edu/abs/2018arXiv180505352F} {} (\mn@eprint {arXiv}
  {1805.05352})

\bibitem[\protect\citeauthoryear{{Fontanot}, {De Lucia}, {Hirschmann},
  {Bruzual}, {Charlot}  \& {Zibetti}}{{Fontanot} et~al.}{2017}]{Fontanot17b}
{Fontanot} F.,  {De Lucia} G.,  {Hirschmann} M.,  {Bruzual} G.,  {Charlot} S.,
   {Zibetti} S.,  2017, \mn@doi [\mnras] {10.1093/mnras/stw2612}, \href
  {http://adsabs.harvard.edu/abs/2017MNRAS.464.3812F} {464, 3812}

\bibitem[\protect\citeauthoryear{{Forbes}, {Krumholz}, {Burkert}  \&
  {Dekel}}{{Forbes} et~al.}{2014}]{Forbes14}
{Forbes} J.~C.,  {Krumholz} M.~R.,  {Burkert} A.,   {Dekel} A.,  2014, \mn@doi
  [\mnras] {10.1093/mnras/stu1142}, \href
  {http://adsabs.harvard.edu/abs/2014MNRAS.443..168F} {443, 168}

\bibitem[\protect\citeauthoryear{{F{\"o}rster Schreiber} et~al.,}{{F{\"o}rster
  Schreiber} et~al.}{2009}]{Forster-Schreiber09}
{F{\"o}rster Schreiber} N.~M.,  et~al., 2009, \mn@doi [\apj]
  {10.1088/0004-637X/706/2/1364}, \href
  {http://adsabs.harvard.edu/abs/2009ApJ...706.1364F} {706, 1364}

\bibitem[\protect\citeauthoryear{{F{\"o}rster Schreiber} et~al.,}{{F{\"o}rster
  Schreiber} et~al.}{2018}]{Forster-Schreiber18}
{F{\"o}rster Schreiber} N.~M.,  et~al., 2018, \mn@doi [\apjs]
  {10.3847/1538-4365/aadd49}, \href
  {http://adsabs.harvard.edu/abs/2018ApJS..238...21F} {238, 21}

\bibitem[\protect\citeauthoryear{{Fosbury} et~al.,}{{Fosbury}
  et~al.}{2003}]{Fosbury03}
{Fosbury} R.~A.~E.,  et~al., 2003, \mn@doi [\apj] {10.1086/378228}, \href
  {http://adsabs.harvard.edu/abs/2003ApJ...596..797F} {596, 797}

\bibitem[\protect\citeauthoryear{{Foster} et~al.,}{{Foster}
  et~al.}{2012}]{Foster12}
{Foster} C.,  et~al., 2012, \mn@doi [\aap] {10.1051/0004-6361/201220050}, \href
  {http://cdsads.u-strasbg.fr/abs/2012A%26A...547A..79F} {547, A79}

\bibitem[\protect\citeauthoryear{{Frebel} \& {Norris}}{{Frebel} \&
  {Norris}}{2015}]{Frebel15}
{Frebel} A.,  {Norris} J.~E.,  2015, \mn@doi [\araa]
  {10.1146/annurev-astro-082214-122423}, \href
  {http://adsabs.harvard.edu/abs/2015ARA%26A..53..631F} {53, 631}

\bibitem[\protect\citeauthoryear{{Frebel}, {Johnson}  \& {Bromm}}{{Frebel}
  et~al.}{2007}]{Frebel07}
{Frebel} A.,  {Johnson} J.~L.,   {Bromm} V.,  2007, \mn@doi [\mnras]
  {10.1111/j.1745-3933.2007.00344.x}, \href
  {http://adsabs.harvard.edu/abs/2007MNRAS.380L..40F} {380, L40}

\bibitem[\protect\citeauthoryear{{Freeman}}{{Freeman}}{2012}]{Freeman12}
{Freeman} K.~C.,  2012, in {Aoki} W.,  {Ishigaki} M.,  {Suda} T.,  {Tsujimoto}
  T.,   {Arimoto} N.,  eds,  Astronomical Society of the Pacific Conference
  Series Vol. 458, Galactic Archaeology: Near-Field Cosmology and the Formation
  of the Milky Way. p.~393

\bibitem[\protect\citeauthoryear{{Freudling}, {Corbin}  \&
  {Korista}}{{Freudling} et~al.}{2003}]{Freudling03}
{Freudling} W.,  {Corbin} M.~R.,   {Korista} K.~T.,  2003, \mn@doi [\apjl]
  {10.1086/375338}, \href {http://adsabs.harvard.edu/abs/2003ApJ...587L..67F}
  {587, L67}

\bibitem[\protect\citeauthoryear{{Friedmann} \& {Maoz}}{{Friedmann} \&
  {Maoz}}{2018}]{Friedmann18}
{Friedmann} M.,  {Maoz} D.,  2018, \mn@doi [\mnras] {10.1093/mnras/sty1664},
  \href {http://adsabs.harvard.edu/abs/2018MNRAS.479.3563F} {479, 3563}

\bibitem[\protect\citeauthoryear{{Friel}}{{Friel}}{1995}]{Friel95}
{Friel} E.~D.,  1995, \mn@doi [\araa] {10.1146/annurev.aa.33.090195.002121},
  \href {http://adsabs.harvard.edu/abs/1995ARA%26A..33..381F} {33, 381}

\bibitem[\protect\citeauthoryear{{Fu} \& {Stockton}}{{Fu} \&
  {Stockton}}{2009}]{Fu09}
{Fu} H.,  {Stockton} A.,  2009, \mn@doi [\apj] {10.1088/0004-637X/690/1/953},
  \href {http://adsabs.harvard.edu/abs/2009ApJ...690..953F} {690, 953}

\bibitem[\protect\citeauthoryear{{Fu} et~al.,}{{Fu} et~al.}{2013}]{Fu13}
{Fu} J.,  et~al., 2013, \mn@doi [\mnras] {10.1093/mnras/stt1117}, \href
  {http://adsabs.harvard.edu/abs/2013MNRAS.434.1531F} {434, 1531}

\bibitem[\protect\citeauthoryear{{Fumagalli}}{{Fumagalli}}{2014}]{Fumagalli14}
{Fumagalli} M.,  2014, \memsai, \href
  {http://adsabs.harvard.edu/abs/2014MmSAI..85..355F} {85, 355}

\bibitem[\protect\citeauthoryear{{Fumagalli}, {O'Meara}, {Prochaska},
  {Rafelski}  \& {Kanekar}}{{Fumagalli} et~al.}{2015}]{Fumagalli15}
{Fumagalli} M.,  {O'Meara} J.~M.,  {Prochaska} J.~X.,  {Rafelski} M.,
  {Kanekar} N.,  2015, \mn@doi [\mnras] {10.1093/mnras/stu2325}, \href
  {http://adsabs.harvard.edu/abs/2015MNRAS.446.3178F} {446, 3178}

\bibitem[\protect\citeauthoryear{{Fynbo} et~al.,}{{Fynbo}
  et~al.}{2006}]{Fynbo06a}
{Fynbo} J.~P.~U.,  et~al., 2006, \mn@doi [\aap] {10.1051/0004-6361:20065056},
  \href {http://adsabs.harvard.edu/abs/2006A%26A...451L..47F} {451, L47}

\bibitem[\protect\citeauthoryear{{Fynbo}, {Prochaska}, {Sommer-Larsen},
  {Dessauges-Zavadsky}  \& {M{\o}ller}}{{Fynbo} et~al.}{2008}]{Fynbo08}
{Fynbo} J.~P.~U.,  {Prochaska} J.~X.,  {Sommer-Larsen} J.,
  {Dessauges-Zavadsky} M.,   {M{\o}ller} P.,  2008, \mn@doi [\apj]
  {10.1086/589555}, \href {http://adsabs.harvard.edu/abs/2008ApJ...683..321F}
  {683, 321}

\bibitem[\protect\citeauthoryear{{Fynbo} et~al.,}{{Fynbo}
  et~al.}{2010}]{Fynbo10}
{Fynbo} J.~P.~U.,  et~al., 2010, \mn@doi [\mnras]
  {10.1111/j.1365-2966.2010.17294.x}, \href
  {http://adsabs.harvard.edu/abs/2010MNRAS.408.2128F} {408, 2128}

\bibitem[\protect\citeauthoryear{{Gabel}, {Arav}  \& {Kim}}{{Gabel}
  et~al.}{2006}]{Gabel06}
{Gabel} J.~R.,  {Arav} N.,   {Kim} T.-S.,  2006, \mn@doi [\apj]
  {10.1086/505070}, \href {http://adsabs.harvard.edu/abs/2006ApJ...646..742G}
  {646, 742}

\bibitem[\protect\citeauthoryear{{Gallazzi}, {Charlot}, {Brinchmann}, {White}
  \& {Tremonti}}{{Gallazzi} et~al.}{2005}]{Gallazzi05}
{Gallazzi} A.,  {Charlot} S.,  {Brinchmann} J.,  {White} S.~D.~M.,   {Tremonti}
  C.~A.,  2005, \mn@doi [\mnras] {10.1111/j.1365-2966.2005.09321.x}, \href
  {http://adsabs.harvard.edu/abs/2005MNRAS.362...41G} {362, 41}

\bibitem[\protect\citeauthoryear{{Gallazzi}, {Charlot}, {Brinchmann}  \&
  {White}}{{Gallazzi} et~al.}{2006}]{Gallazzi06}
{Gallazzi} A.,  {Charlot} S.,  {Brinchmann} J.,   {White} S.~D.~M.,  2006,
  \mn@doi [\mnras] {10.1111/j.1365-2966.2006.10548.x}, \href
  {http://adsabs.harvard.edu/abs/2006MNRAS.370.1106G} {370, 1106}

\bibitem[\protect\citeauthoryear{{Gallazzi}, {Brinchmann}, {Charlot}  \&
  {White}}{{Gallazzi} et~al.}{2008}]{Gallazzi08}
{Gallazzi} A.,  {Brinchmann} J.,  {Charlot} S.,   {White} S.~D.~M.,  2008,
  \mn@doi [\mnras] {10.1111/j.1365-2966.2007.12632.x}, \href
  {http://adsabs.harvard.edu/abs/2008MNRAS.383.1439G} {383, 1439}

\bibitem[\protect\citeauthoryear{{Gallazzi}, {Bell}, {Zibetti}, {Brinchmann}
  \& {Kelson}}{{Gallazzi} et~al.}{2014}]{Gallazzi14}
{Gallazzi} A.,  {Bell} E.~F.,  {Zibetti} S.,  {Brinchmann} J.,   {Kelson}
  D.~D.,  2014, \mn@doi [\apj] {10.1088/0004-637X/788/1/72}, \href
  {http://adsabs.harvard.edu/abs/2014ApJ...788...72G} {788, 72}

\bibitem[\protect\citeauthoryear{{Galliano}, {Galametz}  \& {Jones}}{{Galliano}
  et~al.}{2018}]{Galliano18}
{Galliano} F.,  {Galametz} M.,   {Jones} A.~P.,  2018, \mn@doi [\araa]
  {10.1146/annurev-astro-081817-051900}, 56, 673

\bibitem[\protect\citeauthoryear{{Ganguly}, {Masiero}, {Charlton}  \&
  {Sembach}}{{Ganguly} et~al.}{2003}]{Ganguly03}
{Ganguly} R.,  {Masiero} J.,  {Charlton} J.~C.,   {Sembach} K.~R.,  2003,
  \mn@doi [\apj] {10.1086/379057}, \href
  {http://adsabs.harvard.edu/abs/2003ApJ...598..922G} {598, 922}

\bibitem[\protect\citeauthoryear{{Ganguly}, {Sembach}, {Tripp}, {Savage}  \&
  {Wakker}}{{Ganguly} et~al.}{2006}]{Ganguly06}
{Ganguly} R.,  {Sembach} K.~R.,  {Tripp} T.~M.,  {Savage} B.~D.,   {Wakker}
  B.~P.,  2006, \mn@doi [\apj] {10.1086/504395}, \href
  {http://adsabs.harvard.edu/abs/2006ApJ...645..868G} {645, 868}

\bibitem[\protect\citeauthoryear{{Garc{\'{\i}}a-Rojas} \&
  {Esteban}}{{Garc{\'{\i}}a-Rojas} \& {Esteban}}{2007}]{Garcia-Rojas07}
{Garc{\'{\i}}a-Rojas} J.,  {Esteban} C.,  2007, \mn@doi [\apj]
  {10.1086/521871}, \href {http://adsabs.harvard.edu/abs/2007ApJ...670..457G}
  {670, 457}

\bibitem[\protect\citeauthoryear{{Garc{\'{\i}}a-Rojas}, {Pe{\~n}a}  \&
  {Peimbert}}{{Garc{\'{\i}}a-Rojas} et~al.}{2009}]{Garcia-Rojas09}
{Garc{\'{\i}}a-Rojas} J.,  {Pe{\~n}a} M.,   {Peimbert} A.,  2009, \mn@doi
  [\aap] {10.1051/0004-6361:200811185}, \href
  {http://adsabs.harvard.edu/abs/2009A%26A...496..139G} {496, 139}

\bibitem[\protect\citeauthoryear{{Garc{\'{\i}}a-Rojas}, {Pe{\~n}a}, {Morisset},
  {Delgado-Inglada}, {Mesa-Delgado}  \& {Ruiz}}{{Garc{\'{\i}}a-Rojas}
  et~al.}{2013}]{Garcia-Rojas13}
{Garc{\'{\i}}a-Rojas} J.,  {Pe{\~n}a} M.,  {Morisset} C.,  {Delgado-Inglada}
  G.,  {Mesa-Delgado} A.,   {Ruiz} M.~T.,  2013, \mn@doi [\aap]
  {10.1051/0004-6361/201322354}, \href
  {http://adsabs.harvard.edu/abs/2013A%26A...558A.122G} {558, A122}

\bibitem[\protect\citeauthoryear{{Garnett}}{{Garnett}}{1990}]{Garnett90}
{Garnett} D.~R.,  1990, \mn@doi [\apj] {10.1086/169324}, \href
  {http://adsabs.harvard.edu/abs/1990ApJ...363..142G} {363, 142}

\bibitem[\protect\citeauthoryear{{Garnett}}{{Garnett}}{1992}]{Garnett92}
{Garnett} D.~R.,  1992, \mn@doi [\aj] {10.1086/116146}, \href
  {http://adsabs.harvard.edu/abs/1992AJ....103.1330G} {103, 1330}

\bibitem[\protect\citeauthoryear{{Garnett}}{{Garnett}}{2002}]{Garnett02a}
{Garnett} D.~R.,  2002, \mn@doi [\apj] {10.1086/344301}, \href
  {http://adsabs.harvard.edu/abs/2002ApJ...581.1019G} {581, 1019}

\bibitem[\protect\citeauthoryear{{Garnett} \& {Shields}}{{Garnett} \&
  {Shields}}{1987}]{Garnett87}
{Garnett} D.~R.,  {Shields} G.~A.,  1987, \mn@doi [\apj] {10.1086/165257},
  \href {http://adsabs.harvard.edu/abs/1987ApJ...317...82G} {317, 82}

\bibitem[\protect\citeauthoryear{{Garnett}, {Skillman}, {Dufour}, {Peimbert},
  {Torres-Peimbert}, {Terlevich}, {Terlevich}  \& {Shields}}{{Garnett}
  et~al.}{1995}]{Garnett95}
{Garnett} D.~R.,  {Skillman} E.~D.,  {Dufour} R.~J.,  {Peimbert} M.,
  {Torres-Peimbert} S.,  {Terlevich} R.,  {Terlevich} E.,   {Shields} G.~A.,
  1995, \mn@doi [\apj] {10.1086/175503}, \href
  {http://adsabs.harvard.edu/abs/1995ApJ...443...64G} {443, 64}

\bibitem[\protect\citeauthoryear{{Garnett}, {Skillman}, {Dufour}  \&
  {Shields}}{{Garnett} et~al.}{1997}]{Garnett97}
{Garnett} D.~R.,  {Skillman} E.~D.,  {Dufour} R.~J.,   {Shields} G.~A.,  1997,
  \mn@doi [\apj] {10.1086/304058}, \href
  {http://adsabs.harvard.edu/abs/1997ApJ...481..174G} {481, 174}

\bibitem[\protect\citeauthoryear{{Garnett}, {Shields}, {Peimbert},
  {Torres-Peimbert}, {Skillman}, {Dufour}, {Terlevich}  \&
  {Terlevich}}{{Garnett} et~al.}{1999}]{Garnett99}
{Garnett} D.~R.,  {Shields} G.~A.,  {Peimbert} M.,  {Torres-Peimbert} S.,
  {Skillman} E.~D.,  {Dufour} R.~J.,  {Terlevich} E.,   {Terlevich} R.~J.,
  1999, \mn@doi [\apj] {10.1086/306860}, \href
  {http://adsabs.harvard.edu/abs/1999ApJ...513..168G} {513, 168}

\bibitem[\protect\citeauthoryear{{Gavazzi} \& {Scodeggio}}{{Gavazzi} \&
  {Scodeggio}}{1996}]{Gavazzi96}
{Gavazzi} G.,  {Scodeggio} M.,  1996, \aap, \href
  {http://adsabs.harvard.edu/abs/1996A%26A...312L..29G} {312, L29}

\bibitem[\protect\citeauthoryear{{Gazak} et~al.,}{{Gazak}
  et~al.}{2015}]{Gazak15}
{Gazak} J.~Z.,  et~al., 2015, \mn@doi [\apj] {10.1088/0004-637X/805/2/182},
  \href {http://adsabs.harvard.edu/abs/2015ApJ...805..182G} {805, 182}

\bibitem[\protect\citeauthoryear{{Genel}}{{Genel}}{2016}]{Genel16}
{Genel} S.,  2016, \mn@doi [\apj] {10.3847/0004-637X/822/2/107}, \href
  {http://adsabs.harvard.edu/abs/2016ApJ...822..107G} {822, 107}

\bibitem[\protect\citeauthoryear{{Genel} et~al.,}{{Genel}
  et~al.}{2014}]{Genel14}
{Genel} S.,  et~al., 2014, \mn@doi [\mnras] {10.1093/mnras/stu1654}, \href
  {http://adsabs.harvard.edu/abs/2014MNRAS.445..175G} {445, 175}

\bibitem[\protect\citeauthoryear{{Genzel} et~al.,}{{Genzel}
  et~al.}{2015}]{Genzel15}
{Genzel} R.,  et~al., 2015, \mn@doi [\apj] {10.1088/0004-637X/800/1/20}, \href
  {http://adsabs.harvard.edu/abs/2015ApJ...800...20G} {800, 20}

\bibitem[\protect\citeauthoryear{{Gibson}, {Pilkington}, {Brook}, {Stinson}  \&
  {Bailin}}{{Gibson} et~al.}{2013}]{Gibson13}
{Gibson} B.~K.,  {Pilkington} K.,  {Brook} C.~B.,  {Stinson} G.~S.,   {Bailin}
  J.,  2013, \mn@doi [\aap] {10.1051/0004-6361/201321239}, \href
  {http://adsabs.harvard.edu/abs/2013A%26A...554A..47G} {554, A47}

\bibitem[\protect\citeauthoryear{{Gilmore} et~al.,}{{Gilmore}
  et~al.}{2012}]{Gilmore12}
{Gilmore} G.,  et~al., 2012, The Messenger, \href
  {http://adsabs.harvard.edu/abs/2012Msngr.147...25G} {147, 25}

\bibitem[\protect\citeauthoryear{{Gioannini}, {Matteucci}, {Vladilo}  \&
  {Calura}}{{Gioannini} et~al.}{2017a}]{Gioannini17a}
{Gioannini} L.,  {Matteucci} F.,  {Vladilo} G.,   {Calura} F.,  2017a, \mn@doi
  [\mnras] {10.1093/mnras/stw2343}, \href
  {http://adsabs.harvard.edu/abs/2017MNRAS.464..985G} {464, 985}

\bibitem[\protect\citeauthoryear{{Gioannini}, {Matteucci}  \&
  {Calura}}{{Gioannini} et~al.}{2017b}]{Gioannini17b}
{Gioannini} L.,  {Matteucci} F.,   {Calura} F.,  2017b, \mn@doi [\mnras]
  {10.1093/mnras/stx1914}, \href
  {http://adsabs.harvard.edu/abs/2017MNRAS.471.4615G} {471, 4615}

\bibitem[\protect\citeauthoryear{{Goddard}, {Bresolin}, {Kennicutt},
  {Ryan-Weber}  \& {Rosales-Ortega}}{{Goddard} et~al.}{2011}]{Goddard11}
{Goddard} Q.~E.,  {Bresolin} F.,  {Kennicutt} R.~C.,  {Ryan-Weber} E.~V.,
  {Rosales-Ortega} F.~F.,  2011, \mn@doi [\mnras]
  {10.1111/j.1365-2966.2010.17990.x}, \href
  {http://adsabs.harvard.edu/abs/2011MNRAS.412.1246G} {412, 1246}

\bibitem[\protect\citeauthoryear{{Goddard} et~al.,}{{Goddard}
  et~al.}{2017}]{Goddard17a}
{Goddard} D.,  et~al., 2017, \mn@doi [\mnras] {10.1093/mnras/stw3371}, \href
  {http://adsabs.harvard.edu/abs/2017MNRAS.466.4731G} {466, 4731}

\bibitem[\protect\citeauthoryear{{Gonz{\'a}lez Delgado}, {Leitherer},
  {Heckman}, {Lowenthal}, {Ferguson}  \& {Robert}}{{Gonz{\'a}lez Delgado}
  et~al.}{1998}]{Gonzalez-Delgado98a}
{Gonz{\'a}lez Delgado} R.~M.,  {Leitherer} C.,  {Heckman} T.,  {Lowenthal}
  J.~D.,  {Ferguson} H.~C.,   {Robert} C.,  1998, \mn@doi [\apj]
  {10.1086/305321}, \href {http://adsabs.harvard.edu/abs/1998ApJ...495..698G}
  {495, 698}

\bibitem[\protect\citeauthoryear{{Gonz{\'a}lez Delgado} et~al.,}{{Gonz{\'a}lez
  Delgado} et~al.}{2014}]{Gonzalez-Delgado14}
{Gonz{\'a}lez Delgado} R.~M.,  et~al., 2014, \mn@doi [\apjl]
  {10.1088/2041-8205/791/1/L16}, \href
  {http://adsabs.harvard.edu/abs/2014ApJ...791L..16G} {791, L16}

\bibitem[\protect\citeauthoryear{{Gonz{\'a}lez Delgado} et~al.,}{{Gonz{\'a}lez
  Delgado} et~al.}{2015}]{Gonzalez-Delgado15}
{Gonz{\'a}lez Delgado} R.~M.,  et~al., 2015, \mn@doi [\aap]
  {10.1051/0004-6361/201525938}, \href
  {http://adsabs.harvard.edu/abs/2015A%26A...581A.103G} {581, A103}

\bibitem[\protect\citeauthoryear{{Governato} et~al.,}{{Governato}
  et~al.}{2009}]{Governato09}
{Governato} F.,  et~al., 2009, \mn@doi [\mnras]
  {10.1111/j.1365-2966.2009.15143.x}, \href
  {http://adsabs.harvard.edu/abs/2009MNRAS.398..312G} {398, 312}

\bibitem[\protect\citeauthoryear{{Grandi}}{{Grandi}}{1976}]{Grandi76}
{Grandi} S.~A.,  1976, \mn@doi [\apj] {10.1086/154424}, \href
  {http://adsabs.harvard.edu/abs/1976ApJ...206..658G} {206, 658}

\bibitem[\protect\citeauthoryear{{Grasshorn Gebhardt}, {Zeimann}, {Ciardullo},
  {Gronwall}, {Hagen}, {Bridge}, {Schneider}  \& {Trump}}{{Grasshorn Gebhardt}
  et~al.}{2016}]{Grasshorn-Gebhardt16}
{Grasshorn Gebhardt} H.~S.,  {Zeimann} G.~R.,  {Ciardullo} R.,  {Gronwall} C.,
  {Hagen} A.,  {Bridge} J.~S.,  {Schneider} D.~P.,   {Trump} J.~R.,  2016,
  \mn@doi [\apj] {10.3847/0004-637X/817/1/10}, \href
  {http://adsabs.harvard.edu/abs/2016ApJ...817...10G} {817, 10}

\bibitem[\protect\citeauthoryear{{Graves}, {Faber}  \& {Schiavon}}{{Graves}
  et~al.}{2009}]{Graves09a}
{Graves} G.~J.,  {Faber} S.~M.,   {Schiavon} R.~P.,  2009, \mn@doi [\apj]
  {10.1088/0004-637X/693/1/486}, \href
  {http://adsabs.harvard.edu/abs/2009ApJ...693..486G} {693, 486}

\bibitem[\protect\citeauthoryear{{Gray} \& {Scannapieco}}{{Gray} \&
  {Scannapieco}}{2017}]{Gray17}
{Gray} W.~J.,  {Scannapieco} E.,  2017, \mn@doi [\apj]
  {10.3847/1538-4357/aa9121}, \href
  {http://adsabs.harvard.edu/abs/2017ApJ...849..132G} {849, 132}

\bibitem[\protect\citeauthoryear{{Greene}, {Murphy}, {Graves}, {Gunn},
  {Raskutti}, {Comerford}  \& {Gebhardt}}{{Greene} et~al.}{2013}]{Greene13}
{Greene} J.~E.,  {Murphy} J.~D.,  {Graves} G.~J.,  {Gunn} J.~E.,  {Raskutti}
  S.,  {Comerford} J.~M.,   {Gebhardt} K.,  2013, \mn@doi [\apj]
  {10.1088/0004-637X/776/2/64}, \href
  {http://adsabs.harvard.edu/abs/2013ApJ...776...64G} {776, 64}

\bibitem[\protect\citeauthoryear{{Greggio} \& {Renzini}}{{Greggio} \&
  {Renzini}}{1983}]{Greggio83}
{Greggio} L.,  {Renzini} A.,  1983, \aap, \href
  {http://adsabs.harvard.edu/abs/1983A%26A...118..217G} {118, 217}

\bibitem[\protect\citeauthoryear{{Greggio} \& {Renzini}}{{Greggio} \&
  {Renzini}}{2011}]{Greggio11}
{Greggio} L.,  {Renzini} A.,  2011, {Stellar Populations. A User Guide from Low
  to High Redshift}

\bibitem[\protect\citeauthoryear{{Griffith}, {Martini}  \& {Conroy}}{{Griffith}
  et~al.}{2018}]{Griffith18}
{Griffith} E.,  {Martini} P.,   {Conroy} C.,  2018, preprint, \href
  {http://adsabs.harvard.edu/abs/2018arXiv180905114G} {} (\mn@eprint {arXiv}
  {1809.05114})

\bibitem[\protect\citeauthoryear{{Grisoni}, {Spitoni}, {Matteucci},
  {Recio-Blanco}, {de Laverny}, {Hayden}, {Mikolaitis}  \& {Worley}}{{Grisoni}
  et~al.}{2017}]{Grisoni17}
{Grisoni} V.,  {Spitoni} E.,  {Matteucci} F.,  {Recio-Blanco} A.,  {de Laverny}
  P.,  {Hayden} M.,  {Mikolaitis} {\^S}.,   {Worley} C.~C.,  2017, \mn@doi
  [\mnras] {10.1093/mnras/stx2201}, \href
  {http://adsabs.harvard.edu/abs/2017MNRAS.472.3637G} {472, 3637}

\bibitem[\protect\citeauthoryear{{Grisoni}, {Spitoni}  \&
  {Matteucci}}{{Grisoni} et~al.}{2018}]{Grisoni18}
{Grisoni} V.,  {Spitoni} E.,   {Matteucci} F.,  2018, \mn@doi [\mnras]
  {10.1093/mnras/sty2444}, \href
  {http://adsabs.harvard.edu/abs/2018MNRAS.481.2570G} {481, 2570}

\bibitem[\protect\citeauthoryear{{Gr{\o}nnow}, {Finlator}  \&
  {Christensen}}{{Gr{\o}nnow} et~al.}{2015}]{Gronnow15}
{Gr{\o}nnow} A.~E.,  {Finlator} K.,   {Christensen} L.,  2015, \mn@doi [\mnras]
  {10.1093/mnras/stv1232}, \href
  {http://adsabs.harvard.edu/abs/2015MNRAS.451.4005G} {451, 4005}

\bibitem[\protect\citeauthoryear{{Groves}, {Heckman}  \& {Kauffmann}}{{Groves}
  et~al.}{2006}]{Groves06}
{Groves} B.~A.,  {Heckman} T.~M.,   {Kauffmann} G.,  2006, \mn@doi [\mnras]
  {10.1111/j.1365-2966.2006.10812.x}, \href
  {http://adsabs.harvard.edu/abs/2006MNRAS.371.1559G} {371, 1559}

\bibitem[\protect\citeauthoryear{{Guidi}, {Scannapieco}, {Walcher}  \&
  {Gallazzi}}{{Guidi} et~al.}{2016}]{Guidi16}
{Guidi} G.,  {Scannapieco} C.,  {Walcher} J.,   {Gallazzi} A.,  2016, \mn@doi
  [\mnras] {10.1093/mnras/stw1790}, \href
  {http://adsabs.harvard.edu/abs/2016MNRAS.462.2046G} {462, 2046}

\bibitem[\protect\citeauthoryear{{Guidi} et~al.,}{{Guidi}
  et~al.}{2018}]{Guidi18}
{Guidi} G.,  et~al., 2018, \mn@doi [\mnras] {10.1093/mnras/sty1480}, \href
  {http://adsabs.harvard.edu/abs/2018MNRAS.479..917G} {479, 917}

\bibitem[\protect\citeauthoryear{{Guo} et~al.,}{{Guo} et~al.}{2016}]{Guo16b}
{Guo} Y.,  et~al., 2016, \mn@doi [\apj] {10.3847/0004-637X/822/2/103}, \href
  {http://adsabs.harvard.edu/abs/2016ApJ...822..103G} {822, 103}

\bibitem[\protect\citeauthoryear{{Gustafsson}, {Karlsson}, {Olsson},
  {Edvardsson}  \& {Ryde}}{{Gustafsson} et~al.}{1999}]{Gustafsson99}
{Gustafsson} B.,  {Karlsson} T.,  {Olsson} E.,  {Edvardsson} B.,   {Ryde} N.,
  1999, \aap, \href {http://adsabs.harvard.edu/abs/1999A%26A...342..426G} {342,
  426}

\bibitem[\protect\citeauthoryear{{Gutkin}, {Charlot}  \& {Bruzual}}{{Gutkin}
  et~al.}{2016}]{Gutkin16}
{Gutkin} J.,  {Charlot} S.,   {Bruzual} G.,  2016, \mn@doi [\mnras]
  {10.1093/mnras/stw1716}, \href
  {http://adsabs.harvard.edu/abs/2016MNRAS.462.1757G} {462, 1757}

\bibitem[\protect\citeauthoryear{{Haehnelt}, {Steinmetz}  \&
  {Rauch}}{{Haehnelt} et~al.}{1998}]{Haehnelt98}
{Haehnelt} M.~G.,  {Steinmetz} M.,   {Rauch} M.,  1998, \mn@doi [\apj]
  {10.1086/305323}, \href {http://adsabs.harvard.edu/abs/1998ApJ...495..647H}
  {495, 647}

\bibitem[\protect\citeauthoryear{{Hainline}, {Shapley}, {Kornei}, {Pettini},
  {Buckley-Geer}, {Allam}  \& {Tucker}}{{Hainline} et~al.}{2009}]{Hainline09}
{Hainline} K.~N.,  {Shapley} A.~E.,  {Kornei} K.~A.,  {Pettini} M.,
  {Buckley-Geer} E.,  {Allam} S.~S.,   {Tucker} D.~L.,  2009, \mn@doi [\apj]
  {10.1088/0004-637X/701/1/52}, \href
  {http://adsabs.harvard.edu/abs/2009ApJ...701...52H} {701, 52}

\bibitem[\protect\citeauthoryear{{Halliday} et~al.,}{{Halliday}
  et~al.}{2008}]{Halliday08}
{Halliday} C.,  et~al., 2008, \mn@doi [\aap] {10.1051/0004-6361:20078673},
  \href {http://adsabs.harvard.edu/abs/2008A%26A...479..417H} {479, 417}

\bibitem[\protect\citeauthoryear{{Hamann} \& {Ferland}}{{Hamann} \&
  {Ferland}}{1999}]{Hamann99}
{Hamann} F.,  {Ferland} G.,  1999, \mn@doi [\araa]
  {10.1146/annurev.astro.37.1.487}, \href
  {http://adsabs.harvard.edu/abs/1999ARA%26A..37..487H} {37, 487}

\bibitem[\protect\citeauthoryear{{Hamilton}}{{Hamilton}}{1985}]{Hamilton85}
{Hamilton} D.,  1985, \mn@doi [\apj] {10.1086/163537}, \href
  {http://adsabs.harvard.edu/abs/1985ApJ...297..371H} {297, 371}

\bibitem[\protect\citeauthoryear{{Harrison}, {Colless}, {Kuntschner}, {Couch},
  {de Propris}  \& {Pracy}}{{Harrison} et~al.}{2011}]{Harrison11}
{Harrison} C.~D.,  {Colless} M.,  {Kuntschner} H.,  {Couch} W.~J.,  {de
  Propris} R.,   {Pracy} M.~B.,  2011, \mn@doi [\mnras]
  {10.1111/j.1365-2966.2011.18195.x}, \href
  {http://adsabs.harvard.edu/abs/2011MNRAS.413.1036H} {413, 1036}

\bibitem[\protect\citeauthoryear{{Harwit} \& {Brisbin}}{{Harwit} \&
  {Brisbin}}{2015}]{Harwit15}
{Harwit} M.,  {Brisbin} D.,  2015, \mn@doi [\apj] {10.1088/0004-637X/800/2/91},
  \href {http://adsabs.harvard.edu/abs/2015ApJ...800...91H} {800, 91}

\bibitem[\protect\citeauthoryear{{Hashimoto} et~al.,}{{Hashimoto}
  et~al.}{2018a}]{Hashimoto18b}
{Hashimoto} T.,  et~al., 2018a, preprint, \href
  {http://adsabs.harvard.edu/abs/2018arXiv180600486H} {} (\mn@eprint {arXiv}
  {1806.00486})

\bibitem[\protect\citeauthoryear{{Hashimoto}, {Goto}  \& {Momose}}{{Hashimoto}
  et~al.}{2018b}]{Hashimoto18c}
{Hashimoto} T.,  {Goto} T.,   {Momose} R.,  2018b, \mn@doi [\mnras]
  {10.1093/mnras/sty113}, \href
  {http://adsabs.harvard.edu/abs/2018MNRAS.475.4424H} {475, 4424}

\bibitem[\protect\citeauthoryear{{Hashimoto} et~al.,}{{Hashimoto}
  et~al.}{2018c}]{Hashimoto18a}
{Hashimoto} T.,  et~al., 2018c, \mn@doi [\nat] {10.1038/s41586-018-0117-z},
  \href {http://adsabs.harvard.edu/abs/2018Natur.557..392H} {557, 392}

\bibitem[\protect\citeauthoryear{{Haurberg}, {Rosenberg}  \&
  {Salzer}}{{Haurberg} et~al.}{2013}]{Haurberg13}
{Haurberg} N.~C.,  {Rosenberg} J.,   {Salzer} J.~J.,  2013, \mn@doi [\apj]
  {10.1088/0004-637X/765/1/66}, \href
  {http://adsabs.harvard.edu/abs/2013ApJ...765...66H} {765, 66}

\bibitem[\protect\citeauthoryear{{Haurberg}, {Salzer}, {Cannon}  \&
  {Marshall}}{{Haurberg} et~al.}{2015}]{Haurberg15}
{Haurberg} N.~C.,  {Salzer} J.~J.,  {Cannon} J.~M.,   {Marshall} M.~V.,  2015,
  \mn@doi [\apj] {10.1088/0004-637X/800/2/121}, \href
  {http://adsabs.harvard.edu/abs/2015ApJ...800..121H} {800, 121}

\bibitem[\protect\citeauthoryear{{Hayashi} et~al.,}{{Hayashi}
  et~al.}{2015}]{Hayashi15}
{Hayashi} M.,  et~al., 2015, \mn@doi [\pasj] {10.1093/pasj/psv041}, \href
  {http://adsabs.harvard.edu/abs/2015PASJ...67...80H} {67, 80}

\bibitem[\protect\citeauthoryear{{Hayden} et~al.,}{{Hayden}
  et~al.}{2015}]{Hayden15}
{Hayden} M.~R.,  et~al., 2015, \mn@doi [\apj] {10.1088/0004-637X/808/2/132},
  \href {http://adsabs.harvard.edu/abs/2015ApJ...808..132H} {808, 132}

\bibitem[\protect\citeauthoryear{{Heckman}}{{Heckman}}{1980}]{Heckman80}
{Heckman} T.~M.,  1980, \aap, \href
  {http://adsabs.harvard.edu/abs/1980A%26A....87..152H} {87, 152}

\bibitem[\protect\citeauthoryear{{Heckman}}{{Heckman}}{2002}]{Heckman02}
{Heckman} T.~M.,  2002, in {Mulchaey} J.~S.,  {Stocke} J.~T.,  eds,
  Astronomical Society of the Pacific Conference Series Vol. 254, Extragalactic
  Gas at Low Redshift. p.~292 (\mn@eprint {} {astro-ph/0107438})

\bibitem[\protect\citeauthoryear{{Heckman} \& {Thompson}}{{Heckman} \&
  {Thompson}}{2017}]{Heckman17}
{Heckman} T.~M.,  {Thompson} T.~A.,  2017, preprint, \href
  {http://adsabs.harvard.edu/abs/2017arXiv170109062H} {} (\mn@eprint {arXiv}
  {1701.09062})

\bibitem[\protect\citeauthoryear{{Heckman}, {Robert}, {Leitherer}, {Garnett}
  \& {van der Rydt}}{{Heckman} et~al.}{1998}]{Heckman98}
{Heckman} T.~M.,  {Robert} C.,  {Leitherer} C.,  {Garnett} D.~R.,   {van der
  Rydt} F.,  1998, \mn@doi [\apj] {10.1086/306035}, \href
  {http://adsabs.harvard.edu/abs/1998ApJ...503..646H} {503, 646}

\bibitem[\protect\citeauthoryear{{Heckman} et~al.,}{{Heckman}
  et~al.}{2005}]{Heckman05}
{Heckman} T.~M.,  et~al., 2005, \mn@doi [\apjl] {10.1086/425979}, \href
  {http://adsabs.harvard.edu/abs/2005ApJ...619L..35H} {619, L35}

\bibitem[\protect\citeauthoryear{{Heckman}, {Alexandroff}, {Borthakur},
  {Overzier}  \& {Leitherer}}{{Heckman} et~al.}{2015}]{Heckman15}
{Heckman} T.~M.,  {Alexandroff} R.~M.,  {Borthakur} S.,  {Overzier} R.,
  {Leitherer} C.,  2015, \mn@doi [\apj] {10.1088/0004-637X/809/2/147}, \href
  {http://adsabs.harvard.edu/abs/2015ApJ...809..147H} {809, 147}

\bibitem[\protect\citeauthoryear{{Hendricks}, {Koch}, {Lanfranchi}, {Boeche},
  {Walker}, {Johnson}, {Pe{\~n}arrubia}  \& {Gilmore}}{{Hendricks}
  et~al.}{2014}]{Hendricks14}
{Hendricks} B.,  {Koch} A.,  {Lanfranchi} G.~A.,  {Boeche} C.,  {Walker} M.,
  {Johnson} C.~I.,  {Pe{\~n}arrubia} J.,   {Gilmore} G.,  2014, \mn@doi [\apj]
  {10.1088/0004-637X/785/2/102}, \href
  {http://adsabs.harvard.edu/abs/2014ApJ...785..102H} {785, 102}

\bibitem[\protect\citeauthoryear{{Henry} \& {Prochaska}}{{Henry} \&
  {Prochaska}}{2007}]{Henry07}
{Henry} R.~B.~C.,  {Prochaska} J.~X.,  2007, \mn@doi [\pasp] {10.1086/522038},
  \href {http://adsabs.harvard.edu/abs/2007PASP..119..962H} {119, 962}

\bibitem[\protect\citeauthoryear{{Henry}, {Edmunds}  \& {K{\"o}ppen}}{{Henry}
  et~al.}{2000}]{Henry00}
{Henry} R.~B.~C.,  {Edmunds} M.~G.,   {K{\"o}ppen} J.,  2000, \mn@doi [\apj]
  {10.1086/309471}, \href {http://adsabs.harvard.edu/abs/2000ApJ...541..660H}
  {541, 660}

\bibitem[\protect\citeauthoryear{{Henry}, {Kwitter}, {Jaskot}, {Balick},
  {Morrison}  \& {Milingo}}{{Henry} et~al.}{2010}]{Henry10}
{Henry} R.~B.~C.,  {Kwitter} K.~B.,  {Jaskot} A.~E.,  {Balick} B.,  {Morrison}
  M.~A.,   {Milingo} J.~B.,  2010, \mn@doi [\apj]
  {10.1088/0004-637X/724/1/748}, \href
  {http://adsabs.harvard.edu/abs/2010ApJ...724..748H} {724, 748}

\bibitem[\protect\citeauthoryear{{Henry}, {Martin}, {Finlator}  \&
  {Dressler}}{{Henry} et~al.}{2013a}]{Henry13a}
{Henry} A.,  {Martin} C.~L.,  {Finlator} K.,   {Dressler} A.,  2013a, \mn@doi
  [\apj] {10.1088/0004-637X/769/2/148}, \href
  {http://adsabs.harvard.edu/abs/2013ApJ...769..148H} {769, 148}

\bibitem[\protect\citeauthoryear{{Henry} et~al.,}{{Henry}
  et~al.}{2013b}]{Henry13b}
{Henry} A.,  et~al., 2013b, \mn@doi [\apjl] {10.1088/2041-8205/776/2/L27},
  \href {http://adsabs.harvard.edu/abs/2013ApJ...776L..27H} {776, L27}

\bibitem[\protect\citeauthoryear{{Hill} \& {DART Collaboration}}{{Hill} \&
  {DART Collaboration}}{2012}]{Hill12}
{Hill} V.,  {DART Collaboration} 2012, in {Aoki} W.,  {Ishigaki} M.,  {Suda}
  T.,  {Tsujimoto} T.,   {Arimoto} N.,  eds,  Astronomical Society of the
  Pacific Conference Series Vol. 458, Galactic Archaeology: Near-Field
  Cosmology and the Formation of the Milky Way. p.~297

\bibitem[\protect\citeauthoryear{{Hirschauer}, {Salzer}, {Janowiecki}  \&
  {Wegner}}{{Hirschauer} et~al.}{2018}]{Hirschauer18}
{Hirschauer} A.~S.,  {Salzer} J.~J.,  {Janowiecki} S.,   {Wegner} G.~A.,  2018,
  \mn@doi [\aj] {10.3847/1538-3881/aaa4ba}, \href
  {http://adsabs.harvard.edu/abs/2018AJ....155...82H} {155, 82}

\bibitem[\protect\citeauthoryear{{Hirschmann} et~al.,}{{Hirschmann}
  et~al.}{2013}]{Hirschmann13b}
{Hirschmann} M.,  et~al., 2013, \mn@doi [\mnras] {10.1093/mnras/stt1770}, \href
  {http://adsabs.harvard.edu/abs/2013MNRAS.436.2929H} {436, 2929}

\bibitem[\protect\citeauthoryear{{Hirschmann}, {De Lucia}  \&
  {Fontanot}}{{Hirschmann} et~al.}{2016}]{Hirschmann16}
{Hirschmann} M.,  {De Lucia} G.,   {Fontanot} F.,  2016, \mn@doi [\mnras]
  {10.1093/mnras/stw1318}, \href
  {http://adsabs.harvard.edu/abs/2016MNRAS.461.1760H} {461, 1760}

\bibitem[\protect\citeauthoryear{{Hirschmann}, {Charlot}, {Feltre}, {Naab},
  {Choi}, {Ostriker}  \& {Somerville}}{{Hirschmann}
  et~al.}{2017}]{Hirschmann17}
{Hirschmann} M.,  {Charlot} S.,  {Feltre} A.,  {Naab} T.,  {Choi} E.,
  {Ostriker} J.~P.,   {Somerville} R.~S.,  2017, \mn@doi [\mnras]
  {10.1093/mnras/stx2180}, \href
  {http://adsabs.harvard.edu/abs/2017MNRAS.472.2468H} {472, 2468}

\bibitem[\protect\citeauthoryear{{Hitomi Collaboration}}{{Hitomi
  Collaboration}}{2017}]{Hitomi17}
{Hitomi Collaboration} 2017, \mn@doi [\nat] {10.1038/nature24301}, \href
  {http://adsabs.harvard.edu/abs/2017Natur.551..478H} {551, 478}

\bibitem[\protect\citeauthoryear{{Ho}, {Kudritzki}, {Kewley}, {Zahid},
  {Dopita}, {Bresolin}  \& {Rupke}}{{Ho} et~al.}{2015}]{Ho15}
{Ho} I.-T.,  {Kudritzki} R.-P.,  {Kewley} L.~J.,  {Zahid} H.~J.,  {Dopita}
  M.~A.,  {Bresolin} F.,   {Rupke} D.~S.~N.,  2015, \mn@doi [\mnras]
  {10.1093/mnras/stv067}, \href
  {http://adsabs.harvard.edu/abs/2015MNRAS.448.2030H} {448, 2030}

\bibitem[\protect\citeauthoryear{{Ho} et~al.,}{{Ho} et~al.}{2017}]{Ho17}
{Ho} I.-T.,  et~al., 2017, \mn@doi [\apj] {10.3847/1538-4357/aa8460}, \href
  {http://adsabs.harvard.edu/abs/2017ApJ...846...39H} {846, 39}

\bibitem[\protect\citeauthoryear{{Ho} et~al.,}{{Ho} et~al.}{2018}]{Ho18}
{Ho} I.-T.,  et~al., 2018, \mn@doi [\aap] {10.1051/0004-6361/201833262}, \href
  {http://adsabs.harvard.edu/abs/2018A%26A...618A..64H} {618, A64}

\bibitem[\protect\citeauthoryear{{Holden} et~al.,}{{Holden}
  et~al.}{2016}]{Holden16}
{Holden} B.~P.,  et~al., 2016, \mn@doi [\apj] {10.3847/0004-637X/820/1/73},
  \href {http://adsabs.harvard.edu/abs/2016ApJ...820...73H} {820, 73}

\bibitem[\protect\citeauthoryear{{Hoopes} et~al.,}{{Hoopes}
  et~al.}{2007}]{Hoopes07}
{Hoopes} C.~G.,  et~al., 2007, \mn@doi [\apjs] {10.1086/516644}, \href
  {http://adsabs.harvard.edu/abs/2007ApJS..173..441H} {173, 441}

\bibitem[\protect\citeauthoryear{{Hopkins}, {Kere{\v s}}, {O{\~n}orbe},
  {Faucher-Gigu{\`e}re}, {Quataert}, {Murray}  \& {Bullock}}{{Hopkins}
  et~al.}{2014}]{Hopkins14}
{Hopkins} P.~F.,  {Kere{\v s}} D.,  {O{\~n}orbe} J.,  {Faucher-Gigu{\`e}re}
  C.-A.,  {Quataert} E.,  {Murray} N.,   {Bullock} J.~S.,  2014, \mn@doi
  [\mnras] {10.1093/mnras/stu1738}, \href
  {http://adsabs.harvard.edu/abs/2014MNRAS.445..581H} {445, 581}

\bibitem[\protect\citeauthoryear{{Hosek} Jr. et~al.,}{{Hosek}
  et~al.}{2014}]{Hosek14}
{Hosek} Jr. M.~W.,  et~al., 2014, \mn@doi [\apj] {10.1088/0004-637X/785/2/151},
  \href {http://adsabs.harvard.edu/abs/2014ApJ...785..151H} {785, 151}

\bibitem[\protect\citeauthoryear{{Hughes}, {Cortese}, {Boselli}, {Gavazzi}  \&
  {Davies}}{{Hughes} et~al.}{2013}]{Hughes13}
{Hughes} T.~M.,  {Cortese} L.,  {Boselli} A.,  {Gavazzi} G.,   {Davies} J.~I.,
  2013, \mn@doi [\aap] {10.1051/0004-6361/201218822}, \href
  {http://adsabs.harvard.edu/abs/2013A%26A...550A.115H} {550, A115}

\bibitem[\protect\citeauthoryear{{Hunt} et~al.,}{{Hunt} et~al.}{2012}]{Hunt12}
{Hunt} L.,  et~al., 2012, \mn@doi [\mnras] {10.1111/j.1365-2966.2012.21761.x},
  \href {http://adsabs.harvard.edu/abs/2012MNRAS.427..906H} {427, 906}

\bibitem[\protect\citeauthoryear{{Hunt}, {Dayal}, {Magrini}  \&
  {Ferrara}}{{Hunt} et~al.}{2016a}]{Hunt16a}
{Hunt} L.,  {Dayal} P.,  {Magrini} L.,   {Ferrara} A.,  2016a, \mn@doi [\mnras]
  {10.1093/mnras/stw1993}, \href
  {http://adsabs.harvard.edu/abs/2016MNRAS.463.2002H} {463, 2002}

\bibitem[\protect\citeauthoryear{{Hunt}, {Dayal}, {Magrini}  \&
  {Ferrara}}{{Hunt} et~al.}{2016b}]{Hunt16b}
{Hunt} L.,  {Dayal} P.,  {Magrini} L.,   {Ferrara} A.,  2016b, \mn@doi [\mnras]
  {10.1093/mnras/stw2091}, \href
  {http://adsabs.harvard.edu/abs/2016MNRAS.463.2020H} {463, 2020}

\bibitem[\protect\citeauthoryear{{Husemann}, {Wisotzki}, {Jahnke}  \&
  {S{\'a}nchez}}{{Husemann} et~al.}{2011}]{Husemann11}
{Husemann} B.,  {Wisotzki} L.,  {Jahnke} K.,   {S{\'a}nchez} S.~F.,  2011,
  \mn@doi [\aap] {10.1051/0004-6361/201117596}, \href
  {http://adsabs.harvard.edu/abs/2011A%26A...535A..72H} {535, A72}

\bibitem[\protect\citeauthoryear{{Inoue} et~al.,}{{Inoue}
  et~al.}{2016}]{Inoue16}
{Inoue} A.~K.,  et~al., 2016, \mn@doi [Science] {10.1126/science.aaf0714},
  \href {http://adsabs.harvard.edu/abs/2016Sci...352.1559I} {352, 1559}

\bibitem[\protect\citeauthoryear{{Iwamoto}, {Brachwitz}, {Nomoto}, {Kishimoto},
  {Umeda}, {Hix}  \& {Thielemann}}{{Iwamoto} et~al.}{1999}]{Iwamoto99}
{Iwamoto} K.,  {Brachwitz} F.,  {Nomoto} K.,  {Kishimoto} N.,  {Umeda} H.,
  {Hix} W.~R.,   {Thielemann} F.-K.,  1999, \mn@doi [\apjs] {10.1086/313278},
  \href {http://adsabs.harvard.edu/abs/1999ApJS..125..439I} {125, 439}

\bibitem[\protect\citeauthoryear{{Iwamuro}, {Kimura}, {Eto}, {Maihara},
  {Motohara}, {Yoshii}  \& {Doi}}{{Iwamuro} et~al.}{2004}]{Iwamuro04}
{Iwamuro} F.,  {Kimura} M.,  {Eto} S.,  {Maihara} T.,  {Motohara} K.,  {Yoshii}
  Y.,   {Doi} M.,  2004, \mn@doi [\apj] {10.1086/423610}, \href
  {http://adsabs.harvard.edu/abs/2004ApJ...614...69I} {614, 69}

\bibitem[\protect\citeauthoryear{{Izotov} \& {Thuan}}{{Izotov} \&
  {Thuan}}{1999}]{Izotov99}
{Izotov} Y.~I.,  {Thuan} T.~X.,  1999, \mn@doi [\apj] {10.1086/306708}, \href
  {http://adsabs.harvard.edu/abs/1999ApJ...511..639I} {511, 639}

\bibitem[\protect\citeauthoryear{{Izotov} \& {Thuan}}{{Izotov} \&
  {Thuan}}{2007}]{Izotov07a}
{Izotov} Y.~I.,  {Thuan} T.~X.,  2007, \mn@doi [\apj] {10.1086/519922}, \href
  {http://adsabs.harvard.edu/abs/2007ApJ...665.1115I} {665, 1115}

\bibitem[\protect\citeauthoryear{{Izotov}, {Stasi{\'n}ska}, {Meynet}, {Guseva}
  \& {Thuan}}{{Izotov} et~al.}{2006a}]{Izotov06a}
{Izotov} Y.~I.,  {Stasi{\'n}ska} G.,  {Meynet} G.,  {Guseva} N.~G.,   {Thuan}
  T.~X.,  2006a, \mn@doi [\aap] {10.1051/0004-6361:20053763}, \href
  {http://adsabs.harvard.edu/abs/2006A%26A...448..955I} {448, 955}

\bibitem[\protect\citeauthoryear{{Izotov}, {Schaerer}, {Blecha}, {Royer},
  {Guseva}  \& {North}}{{Izotov} et~al.}{2006b}]{Izotov06b}
{Izotov} Y.~I.,  {Schaerer} D.,  {Blecha} A.,  {Royer} F.,  {Guseva} N.~G.,
  {North} P.,  2006b, \mn@doi [\aap] {10.1051/0004-6361:20065622}, \href
  {http://adsabs.harvard.edu/abs/2006A%26A...459...71I} {459, 71}

\bibitem[\protect\citeauthoryear{{Izotov}, {Guseva}  \& {Thuan}}{{Izotov}
  et~al.}{2011}]{Izotov11a}
{Izotov} Y.~I.,  {Guseva} N.~G.,   {Thuan} T.~X.,  2011, \mn@doi [\apj]
  {10.1088/0004-637X/728/2/161}, \href
  {http://adsabs.harvard.edu/abs/2011ApJ...728..161I} {728, 161}

\bibitem[\protect\citeauthoryear{{Izotov}, {Thuan}  \& {Guseva}}{{Izotov}
  et~al.}{2012}]{Izotov12}
{Izotov} Y.~I.,  {Thuan} T.~X.,   {Guseva} N.~G.,  2012, \mn@doi [\aap]
  {10.1051/0004-6361/201219733}, \href
  {http://adsabs.harvard.edu/abs/2012A%26A...546A.122I} {546, A122}

\bibitem[\protect\citeauthoryear{{Izotov}, {Guseva}, {Fricke}  \&
  {Henkel}}{{Izotov} et~al.}{2015}]{Izotov15}
{Izotov} Y.~I.,  {Guseva} N.~G.,  {Fricke} K.~J.,   {Henkel} C.,  2015, \mn@doi
  [\mnras] {10.1093/mnras/stv1115}, \href
  {http://adsabs.harvard.edu/abs/2015MNRAS.451.2251I} {451, 2251}

\bibitem[\protect\citeauthoryear{{Izotov}, {Thuan}, {Guseva}  \&
  {Liss}}{{Izotov} et~al.}{2018a}]{Izotov18}
{Izotov} Y.~I.,  {Thuan} T.~X.,  {Guseva} N.~G.,   {Liss} S.~E.,  2018a,
  \mn@doi [\mnras] {10.1093/mnras/stx2478}, \href
  {http://adsabs.harvard.edu/abs/2018MNRAS.473.1956I} {473, 1956}

\bibitem[\protect\citeauthoryear{{Izotov}, {Thuan}, {Guseva}  \&
  {Liss}}{{Izotov} et~al.}{2018b}]{Izotov18a}
{Izotov} Y.~I.,  {Thuan} T.~X.,  {Guseva} N.~G.,   {Liss} S.~E.,  2018b,
  \mn@doi [\mnras] {10.1093/mnras/stx2478}, \href
  {http://adsabs.harvard.edu/abs/2018MNRAS.473.1956I} {473, 1956}

\bibitem[\protect\citeauthoryear{{Izotov}, {Worseck}, {Schaerer}, {Guseva},
  {Thuan}, {Fricke}  \& {Orlitov{\'a}}}{{Izotov} et~al.}{2018c}]{Izotov18b}
{Izotov} Y.~I.,  {Worseck} G.,  {Schaerer} D.,  {Guseva} N.~G.,  {Thuan} T.~X.,
   {Fricke} A. V.,   {Orlitov{\'a}} I.,  2018c, \mn@doi [\mnras]
  {10.1093/mnras/sty1378}, \href
  {http://adsabs.harvard.edu/abs/2018MNRAS.478.4851I} {478, 4851}

\bibitem[\protect\citeauthoryear{{James}, {Tsamis}, {Barlow}, {Westmoquette},
  {Walsh}, {Cuisinier}  \& {Exter}}{{James} et~al.}{2009}]{James09}
{James} B.~L.,  {Tsamis} Y.~G.,  {Barlow} M.~J.,  {Westmoquette} M.~S.,
  {Walsh} J.~R.,  {Cuisinier} F.,   {Exter} K.~M.,  2009, \mn@doi [\mnras]
  {10.1111/j.1365-2966.2009.15172.x}, \href
  {http://adsabs.harvard.edu/abs/2009MNRAS.398....2J} {398, 2}

\bibitem[\protect\citeauthoryear{{James}, {Tsamis}, {Barlow}, {Walsh}  \&
  {Westmoquette}}{{James} et~al.}{2013}]{James13}
{James} B.~L.,  {Tsamis} Y.~G.,  {Barlow} M.~J.,  {Walsh} J.~R.,
  {Westmoquette} M.~S.,  2013, \mn@doi [\mnras] {10.1093/mnras/sts004}, \href
  {http://adsabs.harvard.edu/abs/2013MNRAS.428...86J} {428, 86}

\bibitem[\protect\citeauthoryear{{James} et~al.,}{{James}
  et~al.}{2014a}]{James14a}
{James} B.~L.,  et~al., 2014a, \mn@doi [\mnras] {10.1093/mnras/stu287}, \href
  {http://adsabs.harvard.edu/abs/2014MNRAS.440.1794J} {440, 1794}

\bibitem[\protect\citeauthoryear{{James}, {Aloisi}, {Heckman}, {Sohn}  \&
  {Wolfe}}{{James} et~al.}{2014b}]{James14b}
{James} B.~L.,  {Aloisi} A.,  {Heckman} T.,  {Sohn} S.~T.,   {Wolfe} M.~A.,
  2014b, \mn@doi [\apj] {10.1088/0004-637X/795/2/109}, \href
  {http://adsabs.harvard.edu/abs/2014ApJ...795..109J} {795, 109}

\bibitem[\protect\citeauthoryear{{Jaskot} \& {Ravindranath}}{{Jaskot} \&
  {Ravindranath}}{2016}]{Jaskot16}
{Jaskot} A.~E.,  {Ravindranath} S.,  2016, \mn@doi [\apj]
  {10.3847/1538-4357/833/2/136}, \href
  {http://adsabs.harvard.edu/abs/2016ApJ...833..136J} {833, 136}

\bibitem[\protect\citeauthoryear{{Jenkins}}{{Jenkins}}{2009}]{Jenkins09}
{Jenkins} E.~B.,  2009, \mn@doi [\apj] {10.1088/0004-637X/700/2/1299}, \href
  {http://adsabs.harvard.edu/abs/2009ApJ...700.1299J} {700, 1299}

\bibitem[\protect\citeauthoryear{{Jenkins}}{{Jenkins}}{2014}]{Jenkins14}
{Jenkins} E.~B.,  2014, preprint (\mn@eprint {} {1402.4765})

\bibitem[\protect\citeauthoryear{{Jiang}, {Fan}, {Vestergaard}, {Kurk},
  {Walter}, {Kelly}  \& {Strauss}}{{Jiang} et~al.}{2007}]{Jiang07}
{Jiang} L.,  {Fan} X.,  {Vestergaard} M.,  {Kurk} J.~D.,  {Walter} F.,  {Kelly}
  B.~C.,   {Strauss} M.~A.,  2007, \mn@doi [\aj] {10.1086/520811}, \href
  {http://adsabs.harvard.edu/abs/2007AJ....134.1150J} {134, 1150}

\bibitem[\protect\citeauthoryear{{Jiang} et~al.,}{{Jiang}
  et~al.}{2018}]{Jiang18}
{Jiang} J.,  et~al., 2018, \mn@doi [\mnras] {10.1093/mnras/sty836}, \href
  {http://adsabs.harvard.edu/abs/2018MNRAS.477.3711J} {477, 3711}

\bibitem[\protect\citeauthoryear{{Jimmy}, {Tran}, {Saintonge}, {Accurso},
  {Brough}  \& {Oliva-Altamirano}}{{Jimmy} et~al.}{2015}]{Jimmy15}
{Jimmy} {Tran} K.-V.,  {Saintonge} A.,  {Accurso} G.,  {Brough} S.,
  {Oliva-Altamirano} P.,  2015, \mn@doi [\apj] {10.1088/0004-637X/812/2/98},
  \href {http://adsabs.harvard.edu/abs/2015ApJ...812...98J} {812, 98}

\bibitem[\protect\citeauthoryear{{Johnson}, {Rich}, {Kobayashi}, {Kunder}  \&
  {Koch}}{{Johnson} et~al.}{2014}]{Johnson14}
{Johnson} C.~I.,  {Rich} R.~M.,  {Kobayashi} C.,  {Kunder} A.,   {Koch} A.,
  2014, \mn@doi [\aj] {10.1088/0004-6256/148/4/67}, \href
  {http://adsabs.harvard.edu/abs/2014AJ....148...67J} {148, 67}

\bibitem[\protect\citeauthoryear{{Jones}, {Ellis}, {Richard}  \&
  {Jullo}}{{Jones} et~al.}{2013}]{Jones13}
{Jones} T.,  {Ellis} R.~S.,  {Richard} J.,   {Jullo} E.,  2013, \mn@doi [\apj]
  {10.1088/0004-637X/765/1/48}, \href
  {http://adsabs.harvard.edu/abs/2013ApJ...765...48J} {765, 48}

\bibitem[\protect\citeauthoryear{{Jones} et~al.,}{{Jones}
  et~al.}{2015a}]{Jones15a}
{Jones} T.,  et~al., 2015a, \mn@doi [\aj] {10.1088/0004-6256/149/3/107}, \href
  {http://adsabs.harvard.edu/abs/2015AJ....149..107J} {149, 107}

\bibitem[\protect\citeauthoryear{{Jones}, {Martin}  \& {Cooper}}{{Jones}
  et~al.}{2015b}]{Jones15b}
{Jones} T.,  {Martin} C.,   {Cooper} M.~C.,  2015b, \mn@doi [\apj]
  {10.1088/0004-637X/813/2/126}, \href
  {http://adsabs.harvard.edu/abs/2015ApJ...813..126J} {813, 126}

\bibitem[\protect\citeauthoryear{{Jorgenson}, {Murphy}  \&
  {Thompson}}{{Jorgenson} et~al.}{2013}]{Jorgenson13}
{Jorgenson} R.~A.,  {Murphy} M.~T.,   {Thompson} R.,  2013, \mn@doi [\mnras]
  {10.1093/mnras/stt1309}, \href
  {http://adsabs.harvard.edu/abs/2013MNRAS.435..482J} {435, 482}

\bibitem[\protect\citeauthoryear{{Juarez}, {Maiolino}, {Mujica}, {Pedani},
  {Marinoni}, {Nagao}, {Marconi}  \& {Oliva}}{{Juarez} et~al.}{2009}]{Juarez09}
{Juarez} Y.,  {Maiolino} R.,  {Mujica} R.,  {Pedani} M.,  {Marinoni} S.,
  {Nagao} T.,  {Marconi} A.,   {Oliva} E.,  2009, \mn@doi [\aap]
  {10.1051/0004-6361:200811415}, \href
  {http://adsabs.harvard.edu/abs/2009A%26A...494L..25J} {494, L25}

\bibitem[\protect\citeauthoryear{{Juneau} et~al.,}{{Juneau}
  et~al.}{2014}]{Juneau14}
{Juneau} S.,  et~al., 2014, \mn@doi [\apj] {10.1088/0004-637X/788/1/88}, \href
  {http://adsabs.harvard.edu/abs/2014ApJ...788...88J} {788, 88}

\bibitem[\protect\citeauthoryear{{Kaasinen}, {Kewley}, {Bian}, {Groves},
  {Kashino}, {Silverman}  \& {Kartaltepe}}{{Kaasinen}
  et~al.}{2018}]{Kaasinen18}
{Kaasinen} M.,  {Kewley} L.,  {Bian} F.,  {Groves} B.,  {Kashino} D.,
  {Silverman} J.,   {Kartaltepe} J.,  2018, \mn@doi [\mnras]
  {10.1093/mnras/sty1012}, \href
  {http://adsabs.harvard.edu/abs/2018MNRAS.tmp..962K} {}

\bibitem[\protect\citeauthoryear{{Kacprzak} et~al.,}{{Kacprzak}
  et~al.}{2015}]{Kacprzak15}
{Kacprzak} G.~G.,  et~al., 2015, \mn@doi [\apjl] {10.1088/2041-8205/802/2/L26},
  \href {http://adsabs.harvard.edu/abs/2015ApJ...802L..26K} {802, L26}

\bibitem[\protect\citeauthoryear{{Kacprzak} et~al.,}{{Kacprzak}
  et~al.}{2016}]{Kacprzak16}
{Kacprzak} G.~G.,  et~al., 2016, \mn@doi [\apjl] {10.3847/2041-8205/826/1/L11},
  \href {http://adsabs.harvard.edu/abs/2016ApJ...826L..11K} {826, L11}

\bibitem[\protect\citeauthoryear{{Kanekar}, {Neeleman}, {Prochaska}  \&
  {Ghosh}}{{Kanekar} et~al.}{2018}]{Kanekar18}
{Kanekar} N.,  {Neeleman} M.,  {Prochaska} J.~X.,   {Ghosh} T.,  2018, \mn@doi
  [\mnras] {10.1093/mnrasl/slx162}, \href
  {http://adsabs.harvard.edu/abs/2018MNRAS.473L..54K} {473, L54}

\bibitem[\protect\citeauthoryear{{Kashino}, {Renzini}, {Silverman}  \&
  {Daddi}}{{Kashino} et~al.}{2016}]{Kashino16}
{Kashino} D.,  {Renzini} A.,  {Silverman} J.~D.,   {Daddi} E.,  2016, \mn@doi
  [\apjl] {10.3847/2041-8205/823/2/L24}, \href
  {http://adsabs.harvard.edu/abs/2016ApJ...823L..24K} {823, L24}

\bibitem[\protect\citeauthoryear{{Kashino} et~al.,}{{Kashino}
  et~al.}{2017}]{Kashino17a}
{Kashino} D.,  et~al., 2017, \mn@doi [\apj] {10.3847/1538-4357/835/1/88}, \href
  {http://adsabs.harvard.edu/abs/2017ApJ...835...88K} {835, 88}

\bibitem[\protect\citeauthoryear{{Katz}, {Sijacki}  \& {Haehnelt}}{{Katz}
  et~al.}{2015}]{Katz15}
{Katz} H.,  {Sijacki} D.,   {Haehnelt} M.~G.,  2015, \mn@doi [\mnras]
  {10.1093/mnras/stv1048}, \href
  {http://adsabs.harvard.edu/abs/2015MNRAS.451.2352K} {451, 2352}

\bibitem[\protect\citeauthoryear{{Katz}, {Kimm}, {Sijacki}  \&
  {Haehnelt}}{{Katz} et~al.}{2017}]{Katz17}
{Katz} H.,  {Kimm} T.,  {Sijacki} D.,   {Haehnelt} M.~G.,  2017, \mn@doi
  [\mnras] {10.1093/mnras/stx608}, \href
  {http://adsabs.harvard.edu/abs/2017MNRAS.468.4831K} {468, 4831}

\bibitem[\protect\citeauthoryear{{Kauffmann}, {White}  \&
  {Guiderdoni}}{{Kauffmann} et~al.}{1993}]{Kauffmann93}
{Kauffmann} G.,  {White} S.~D.~M.,   {Guiderdoni} B.,  1993, \mn@doi [\mnras]
  {10.1093/mnras/264.1.201}, \href
  {http://adsabs.harvard.edu/abs/1993MNRAS.264..201K} {264, 201}

\bibitem[\protect\citeauthoryear{{Kauffmann} et~al.,}{{Kauffmann}
  et~al.}{2003}]{Kauffmann03c}
{Kauffmann} G.,  et~al., 2003, \mn@doi [\mnras]
  {10.1111/j.1365-2966.2003.07154.x}, \href
  {http://adsabs.harvard.edu/abs/2003MNRAS.346.1055K} {346, 1055}

\bibitem[\protect\citeauthoryear{{Kauffmann}, {White}, {Heckman}, {M{\'e}nard},
  {Brinchmann}, {Charlot}, {Tremonti}  \& {Brinkmann}}{{Kauffmann}
  et~al.}{2004}]{Kauffmann04}
{Kauffmann} G.,  {White} S.~D.~M.,  {Heckman} T.~M.,  {M{\'e}nard} B.,
  {Brinchmann} J.,  {Charlot} S.,  {Tremonti} C.,   {Brinkmann} J.,  2004,
  \mn@doi [\mnras] {10.1111/j.1365-2966.2004.08117.x}, \href
  {http://adsabs.harvard.edu/abs/2004MNRAS.353..713K} {353, 713}

\bibitem[\protect\citeauthoryear{{Kelson}, {Illingworth}, {Franx}  \& {van
  Dokkum}}{{Kelson} et~al.}{2006}]{Kelson06}
{Kelson} D.~D.,  {Illingworth} G.~D.,  {Franx} M.,   {van Dokkum} P.~G.,  2006,
  \mn@doi [\apj] {10.1086/507832}, \href
  {http://adsabs.harvard.edu/abs/2006ApJ...653..159K} {653, 159}

\bibitem[\protect\citeauthoryear{{Kennicutt} \& {Evans}}{{Kennicutt} \&
  {Evans}}{2012}]{Kennicutt12}
{Kennicutt} R.~C.,  {Evans} N.~J.,  2012, \mn@doi [\araa]
  {10.1146/annurev-astro-081811-125610}, \href
  {http://adsabs.harvard.edu/abs/2012ARA%26A..50..531K} {50, 531}

\bibitem[\protect\citeauthoryear{{Kennicutt}, {Bresolin}  \&
  {Garnett}}{{Kennicutt} et~al.}{2003}]{Kennicutt03}
{Kennicutt} Jr. R.~C.,  {Bresolin} F.,   {Garnett} D.~R.,  2003, \mn@doi [\apj]
  {10.1086/375398}, \href {http://adsabs.harvard.edu/abs/2003ApJ...591..801K}
  {591, 801}

\bibitem[\protect\citeauthoryear{{Kere{\v s}}, {Katz}, {Weinberg}  \&
  {Dav{\'e}}}{{Kere{\v s}} et~al.}{2005}]{Keres05}
{Kere{\v s}} D.,  {Katz} N.,  {Weinberg} D.~H.,   {Dav{\'e}} R.,  2005, \mn@doi
  [\mnras] {10.1111/j.1365-2966.2005.09451.x}, \href
  {http://adsabs.harvard.edu/abs/2005MNRAS.363....2K} {363, 2}

\bibitem[\protect\citeauthoryear{{Kewley} \& {Dopita}}{{Kewley} \&
  {Dopita}}{2002}]{Kewley02a}
{Kewley} L.~J.,  {Dopita} M.~A.,  2002, \mn@doi [\apjs] {10.1086/341326}, \href
  {http://adsabs.harvard.edu/abs/2002ApJS..142...35K} {142, 35}

\bibitem[\protect\citeauthoryear{{Kewley} \& {Ellison}}{{Kewley} \&
  {Ellison}}{2008}]{Kewley08}
{Kewley} L.~J.,  {Ellison} S.~L.,  2008, \mn@doi [\apj] {10.1086/587500}, \href
  {http://adsabs.harvard.edu/abs/2008ApJ...681.1183K} {681, 1183}

\bibitem[\protect\citeauthoryear{{Kewley}, {Heisler}, {Dopita}  \&
  {Lumsden}}{{Kewley} et~al.}{2001a}]{Kewley01a}
{Kewley} L.~J.,  {Heisler} C.~A.,  {Dopita} M.~A.,   {Lumsden} S.,  2001a,
  \mn@doi [\apjs] {10.1086/318944}, \href
  {http://adsabs.harvard.edu/abs/2001ApJS..132...37K} {132, 37}

\bibitem[\protect\citeauthoryear{{Kewley}, {Dopita}, {Sutherland}, {Heisler}
  \& {Trevena}}{{Kewley} et~al.}{2001b}]{Kewley01b}
{Kewley} L.~J.,  {Dopita} M.~A.,  {Sutherland} R.~S.,  {Heisler} C.~A.,
  {Trevena} J.,  2001b, \mn@doi [\apj] {10.1086/321545}, \href
  {http://adsabs.harvard.edu/abs/2001ApJ...556..121K} {556, 121}

\bibitem[\protect\citeauthoryear{{Kewley}, {Geller}  \& {Barton}}{{Kewley}
  et~al.}{2006a}]{Kewley06}
{Kewley} L.~J.,  {Geller} M.~J.,   {Barton} E.~J.,  2006a, \mn@doi [\aj]
  {10.1086/500295}, \href {http://adsabs.harvard.edu/abs/2006AJ....131.2004K}
  {131, 2004}

\bibitem[\protect\citeauthoryear{{Kewley}, {Groves}, {Kauffmann}  \&
  {Heckman}}{{Kewley} et~al.}{2006b}]{Kewley06b}
{Kewley} L.~J.,  {Groves} B.,  {Kauffmann} G.,   {Heckman} T.,  2006b, \mn@doi
  [\mnras] {10.1111/j.1365-2966.2006.10859.x}, \href
  {http://adsabs.harvard.edu/abs/2006MNRAS.372..961K} {372, 961}

\bibitem[\protect\citeauthoryear{{Kewley}, {Dopita}, {Leitherer}, {Dav{\'e}},
  {Yuan}, {Allen}, {Groves}  \& {Sutherland}}{{Kewley}
  et~al.}{2013a}]{Kewley13b}
{Kewley} L.~J.,  {Dopita} M.~A.,  {Leitherer} C.,  {Dav{\'e}} R.,  {Yuan} T.,
  {Allen} M.,  {Groves} B.,   {Sutherland} R.,  2013a, \mn@doi [\apj]
  {10.1088/0004-637X/774/2/100}, \href
  {http://adsabs.harvard.edu/abs/2013ApJ...774..100K} {774, 100}

\bibitem[\protect\citeauthoryear{{Kewley}, {Maier}, {Yabe}, {Ohta}, {Akiyama},
  {Dopita}  \& {Yuan}}{{Kewley} et~al.}{2013b}]{Kewley13a}
{Kewley} L.~J.,  {Maier} C.,  {Yabe} K.,  {Ohta} K.,  {Akiyama} M.,  {Dopita}
  M.~A.,   {Yuan} T.,  2013b, \mn@doi [\apjl] {10.1088/2041-8205/774/1/L10},
  \href {http://adsabs.harvard.edu/abs/2013ApJ...774L..10K} {774, L10}

\bibitem[\protect\citeauthoryear{{Kewley}, {Zahid}, {Geller}, {Dopita}, {Hwang}
   \& {Fabricant}}{{Kewley} et~al.}{2015}]{Kewley15}
{Kewley} L.~J.,  {Zahid} H.~J.,  {Geller} M.~J.,  {Dopita} M.~A.,  {Hwang}
  H.~S.,   {Fabricant} D.,  2015, \mn@doi [\apjl]
  {10.1088/2041-8205/812/2/L20}, \href
  {http://adsabs.harvard.edu/abs/2015ApJ...812L..20K} {812, L20}

\bibitem[\protect\citeauthoryear{{Kinman} \& {Davidson}}{{Kinman} \&
  {Davidson}}{1981}]{Kinman81}
{Kinman} T.~D.,  {Davidson} K.,  1981, \mn@doi [\apj] {10.1086/158575}, \href
  {http://adsabs.harvard.edu/abs/1981ApJ...243..127K} {243, 127}

\bibitem[\protect\citeauthoryear{{Kirby}, {Cohen}, {Smith}, {Majewski}, {Sohn}
  \& {Guhathakurta}}{{Kirby} et~al.}{2011}]{Kirby11}
{Kirby} E.~N.,  {Cohen} J.~G.,  {Smith} G.~H.,  {Majewski} S.~R.,  {Sohn}
  S.~T.,   {Guhathakurta} P.,  2011, \mn@doi [\apj]
  {10.1088/0004-637X/727/2/79}, \href
  {http://adsabs.harvard.edu/abs/2011ApJ...727...79K} {727, 79}

\bibitem[\protect\citeauthoryear{{Kirby}, {Cohen}, {Guhathakurta}, {Cheng},
  {Bullock}  \& {Gallazzi}}{{Kirby} et~al.}{2013}]{Kirby13}
{Kirby} E.~N.,  {Cohen} J.~G.,  {Guhathakurta} P.,  {Cheng} L.,  {Bullock}
  J.~S.,   {Gallazzi} A.,  2013, \mn@doi [\apj] {10.1088/0004-637X/779/2/102},
  \href {http://adsabs.harvard.edu/abs/2013ApJ...779..102K} {779, 102}

\bibitem[\protect\citeauthoryear{{Kobayashi}}{{Kobayashi}}{2004}]{Kobayashi04}
{Kobayashi} C.,  2004, \mn@doi [\mnras] {10.1111/j.1365-2966.2004.07258.x},
  \href {http://adsabs.harvard.edu/abs/2004MNRAS.347..740K} {347, 740}

\bibitem[\protect\citeauthoryear{{Kobulnicky} \& {Kewley}}{{Kobulnicky} \&
  {Kewley}}{2004}]{Kobulnicky04}
{Kobulnicky} H.~A.,  {Kewley} L.~J.,  2004, \mn@doi [\apj] {10.1086/425299},
  \href {http://adsabs.harvard.edu/abs/2004ApJ...617..240K} {617, 240}

\bibitem[\protect\citeauthoryear{{Kobulnicky} \& {Koo}}{{Kobulnicky} \&
  {Koo}}{2000}]{Kobulnicky00}
{Kobulnicky} H.~A.,  {Koo} D.~C.,  2000, \mn@doi [\apj] {10.1086/317866}, \href
  {http://adsabs.harvard.edu/abs/2000ApJ...545..712K} {545, 712}

\bibitem[\protect\citeauthoryear{{Kobulnicky} \& {Skillman}}{{Kobulnicky} \&
  {Skillman}}{1998}]{Kobulnicky98}
{Kobulnicky} H.~A.,  {Skillman} E.~D.,  1998, \mn@doi [\apj] {10.1086/305491},
  \href {http://adsabs.harvard.edu/abs/1998ApJ...497..601K} {497, 601}

\bibitem[\protect\citeauthoryear{{Kobulnicky}, {Skillman}, {Roy}, {Walsh}  \&
  {Rosa}}{{Kobulnicky} et~al.}{1997}]{Kobulnicky97}
{Kobulnicky} H.~A.,  {Skillman} E.~D.,  {Roy} J.-R.,  {Walsh} J.~R.,   {Rosa}
  M.~R.,  1997, \mn@doi [\apj] {10.1086/303742}, \href
  {http://adsabs.harvard.edu/abs/1997ApJ...477..679K} {477, 679}

\bibitem[\protect\citeauthoryear{{Kobulnicky}, {Kennicutt}  \&
  {Pizagno}}{{Kobulnicky} et~al.}{1999}]{Kobulnicky99}
{Kobulnicky} H.~A.,  {Kennicutt} Jr. R.~C.,   {Pizagno} J.~L.,  1999, \mn@doi
  [\apj] {10.1086/306987}, \href
  {http://adsabs.harvard.edu/abs/1999ApJ...514..544K} {514, 544}

\bibitem[\protect\citeauthoryear{{Kobulnicky} et~al.,}{{Kobulnicky}
  et~al.}{2003}]{Kobulnicky03}
{Kobulnicky} H.~A.,  et~al., 2003, \mn@doi [\apj] {10.1086/379360}, \href
  {http://adsabs.harvard.edu/abs/2003ApJ...599.1006K} {599, 1006}

\bibitem[\protect\citeauthoryear{{Kojima}, {Ouchi}, {Nakajima}, {Shibuya},
  {Harikane}  \& {Ono}}{{Kojima} et~al.}{2017}]{Kojima17}
{Kojima} T.,  {Ouchi} M.,  {Nakajima} K.,  {Shibuya} T.,  {Harikane} Y.,
  {Ono} Y.,  2017, \mn@doi [\pasj] {10.1093/pasj/psx017}, \href
  {http://adsabs.harvard.edu/abs/2017PASJ...69...44K} {69, 44}

\bibitem[\protect\citeauthoryear{{Koleva}, {Prugniel}, {De Rijcke}  \&
  {Zeilinger}}{{Koleva} et~al.}{2011}]{Koleva11}
{Koleva} M.,  {Prugniel} P.,  {De Rijcke} S.,   {Zeilinger} W.~W.,  2011,
  \mn@doi [\mnras] {10.1111/j.1365-2966.2011.19057.x}, \href
  {http://adsabs.harvard.edu/abs/2011MNRAS.417.1643K} {417, 1643}

\bibitem[\protect\citeauthoryear{{K{\"o}ppen} \& {Hensler}}{{K{\"o}ppen} \&
  {Hensler}}{2005}]{Koppen05}
{K{\"o}ppen} J.,  {Hensler} G.,  2005, \mn@doi [\aap]
  {10.1051/0004-6361:20042266}, \href
  {http://adsabs.harvard.edu/abs/2005A%26A...434..531K} {434, 531}

\bibitem[\protect\citeauthoryear{{K{\"o}ppen}, {Weidner}  \&
  {Kroupa}}{{K{\"o}ppen} et~al.}{2007}]{Koppen07}
{K{\"o}ppen} J.,  {Weidner} C.,   {Kroupa} P.,  2007, \mn@doi [\mnras]
  {10.1111/j.1365-2966.2006.11328.x}, \href
  {http://adsabs.harvard.edu/abs/2007MNRAS.375..673K} {375, 673}

\bibitem[\protect\citeauthoryear{{Kreckel}, {Croxall}, {Groves}, {van de
  Weygaert}  \& {Pogge}}{{Kreckel} et~al.}{2015}]{Kreckel15}
{Kreckel} K.,  {Croxall} K.,  {Groves} B.,  {van de Weygaert} R.,   {Pogge}
  R.~W.,  2015, \mn@doi [\apjl] {10.1088/2041-8205/798/1/L15}, \href
  {http://adsabs.harvard.edu/abs/2015ApJ...798L..15K} {798, L15}

\bibitem[\protect\citeauthoryear{{Kriek} et~al.,}{{Kriek}
  et~al.}{2007}]{Kriek07}
{Kriek} M.,  et~al., 2007, \mn@doi [\apj] {10.1086/520789}, \href
  {http://adsabs.harvard.edu/abs/2007ApJ...669..776K} {669, 776}

\bibitem[\protect\citeauthoryear{{Kriek} et~al.,}{{Kriek}
  et~al.}{2016}]{Kriek16}
{Kriek} M.,  et~al., 2016, \mn@doi [\nat] {10.1038/nature20570}, \href
  {http://adsabs.harvard.edu/abs/2016Natur.540..248K} {540, 248}

\bibitem[\protect\citeauthoryear{{Krogager}, {Fynbo}, {M{\o}ller}, {Ledoux},
  {Noterdaeme}, {Christensen}, {Milvang-Jensen}  \& {Sparre}}{{Krogager}
  et~al.}{2012}]{Krogager12}
{Krogager} J.-K.,  {Fynbo} J.~P.~U.,  {M{\o}ller} P.,  {Ledoux} C.,
  {Noterdaeme} P.,  {Christensen} L.,  {Milvang-Jensen} B.,   {Sparre} M.,
  2012, \mn@doi [\mnras] {10.1111/j.1745-3933.2012.01272.x}, \href
  {http://adsabs.harvard.edu/abs/2012MNRAS.424L...1K} {424, L1}

\bibitem[\protect\citeauthoryear{{Krogager}, {M{\o}ller}, {Fynbo}  \&
  {Noterdaeme}}{{Krogager} et~al.}{2017}]{Krogager17}
{Krogager} J.-K.,  {M{\o}ller} P.,  {Fynbo} J.~P.~U.,   {Noterdaeme} P.,  2017,
  \mn@doi [\mnras] {10.1093/mnras/stx1011}, \href
  {http://adsabs.harvard.edu/abs/2017MNRAS.469.2959K} {469, 2959}

\bibitem[\protect\citeauthoryear{{Kroupa}, {Tout}  \& {Gilmore}}{{Kroupa}
  et~al.}{1993}]{Kroupa93}
{Kroupa} P.,  {Tout} C.~A.,   {Gilmore} G.,  1993, \mn@doi [\mnras]
  {10.1093/mnras/262.3.545}, \href
  {http://adsabs.harvard.edu/abs/1993MNRAS.262..545K} {262, 545}

\bibitem[\protect\citeauthoryear{{Kr{\"u}hler} et~al.,}{{Kr{\"u}hler}
  et~al.}{2015}]{Kruhler15}
{Kr{\"u}hler} T.,  et~al., 2015, \mn@doi [\aap] {10.1051/0004-6361/201425561},
  \href {http://adsabs.harvard.edu/abs/2015A%26A...581A.125K} {581, A125}

\bibitem[\protect\citeauthoryear{{Kudritzki}, {Urbaneja}, {Bresolin},
  {Przybilla}, {Gieren}  \& {Pietrzy{\'n}ski}}{{Kudritzki}
  et~al.}{2008}]{Kudritzki08}
{Kudritzki} R.-P.,  {Urbaneja} M.~A.,  {Bresolin} F.,  {Przybilla} N.,
  {Gieren} W.,   {Pietrzy{\'n}ski} G.,  2008, \mn@doi [\apj] {10.1086/588647},
  \href {http://adsabs.harvard.edu/abs/2008ApJ...681..269K} {681, 269}

\bibitem[\protect\citeauthoryear{{Kudritzki}, {Urbaneja}, {Gazak}, {Bresolin},
  {Przybilla}, {Gieren}  \& {Pietrzy{\'n}ski}}{{Kudritzki}
  et~al.}{2012}]{Kudritzki12}
{Kudritzki} R.-P.,  {Urbaneja} M.~A.,  {Gazak} Z.,  {Bresolin} F.,  {Przybilla}
  N.,  {Gieren} W.,   {Pietrzy{\'n}ski} G.,  2012, \mn@doi [\apj]
  {10.1088/0004-637X/747/1/15}, \href
  {http://adsabs.harvard.edu/abs/2012ApJ...747...15K} {747, 15}

\bibitem[\protect\citeauthoryear{{Kudritzki}, {Ho}, {Schruba}, {Burkert},
  {Zahid}, {Bresolin}  \& {Dima}}{{Kudritzki} et~al.}{2015}]{Kudritzki15}
{Kudritzki} R.-P.,  {Ho} I.-T.,  {Schruba} A.,  {Burkert} A.,  {Zahid} H.~J.,
  {Bresolin} F.,   {Dima} G.~I.,  2015, \mn@doi [\mnras]
  {10.1093/mnras/stv522}, \href
  {http://adsabs.harvard.edu/abs/2015MNRAS.450..342K} {450, 342}

\bibitem[\protect\citeauthoryear{{Kudritzki}, {Castro}, {Urbaneja}, {Ho},
  {Bresolin}, {Gieren}, {Pietrzy{\'n}ski}  \& {Przybilla}}{{Kudritzki}
  et~al.}{2016}]{Kudritzki16}
{Kudritzki} R.~P.,  {Castro} N.,  {Urbaneja} M.~A.,  {Ho} I.-T.,  {Bresolin}
  F.,  {Gieren} W.,  {Pietrzy{\'n}ski} G.,   {Przybilla} N.,  2016, \mn@doi
  [\apj] {10.3847/0004-637X/829/2/70}, \href
  {http://adsabs.harvard.edu/abs/2016ApJ...829...70K} {829, 70}

\bibitem[\protect\citeauthoryear{{Kulas} et~al.,}{{Kulas}
  et~al.}{2013}]{Kulas13}
{Kulas} K.~R.,  et~al., 2013, \mn@doi [\apj] {10.1088/0004-637X/774/2/130},
  \href {http://adsabs.harvard.edu/abs/2013ApJ...774..130K} {774, 130}

\bibitem[\protect\citeauthoryear{{Kumari}, {James}, {Irwin}, {Amor{\'{\i}}n}
  \& {P{\'e}rez-Montero}}{{Kumari} et~al.}{2018}]{Kumari18}
{Kumari} N.,  {James} B.~L.,  {Irwin} M.~J.,  {Amor{\'{\i}}n} R.,
  {P{\'e}rez-Montero} E.,  2018, \mn@doi [\mnras] {10.1093/mnras/sty402}, \href
  {http://adsabs.harvard.edu/abs/2018MNRAS.476.3793K} {476, 3793}

\bibitem[\protect\citeauthoryear{{Kuntschner}, {Lucey}, {Smith}, {Hudson}  \&
  {Davies}}{{Kuntschner} et~al.}{2001}]{Kuntschner01}
{Kuntschner} H.,  {Lucey} J.~R.,  {Smith} R.~J.,  {Hudson} M.~J.,   {Davies}
  R.~L.,  2001, \mn@doi [\mnras] {10.1046/j.1365-8711.2001.04263.x}, \href
  {http://adsabs.harvard.edu/abs/2001MNRAS.323..615K} {323, 615}

\bibitem[\protect\citeauthoryear{{La Barbera}, {Ferreras}, {de Carvalho},
  {Bruzual}, {Charlot}, {Pasquali}  \& {Merlin}}{{La Barbera}
  et~al.}{2012}]{La-Barbera12}
{La Barbera} F.,  {Ferreras} I.,  {de Carvalho} R.~R.,  {Bruzual} G.,
  {Charlot} S.,  {Pasquali} A.,   {Merlin} E.,  2012, \mn@doi [\mnras]
  {10.1111/j.1365-2966.2012.21848.x}, \href
  {http://adsabs.harvard.edu/abs/2012MNRAS.426.2300L} {426, 2300}

\bibitem[\protect\citeauthoryear{{La Barbera}, {Vazdekis}, {Ferreras},
  {Pasquali}, {Allende Prieto}, {Rock}, {Aguado}  \& {Peletier}}{{La Barbera}
  et~al.}{2016}]{La-Barbera16}
{La Barbera} F.,  {Vazdekis} A.,  {Ferreras} I.,  {Pasquali} A.,  {Allende
  Prieto} C.,  {Rock} B.,  {Aguado} D.~S.,   {Peletier} R.~F.,  2016, \mn@doi
  [\mnras] {10.1093/mnras/stw2407}, \href
  {http://adsabs.harvard.edu/abs/2017MNRAS.464.3597L} {464, 3597}

\bibitem[\protect\citeauthoryear{{Lacerda} et~al.,}{{Lacerda}
  et~al.}{2018}]{Lacerda18}
{Lacerda} E.~A.~D.,  et~al., 2018, \mn@doi [\mnras] {10.1093/mnras/stx3022},
  \href {http://adsabs.harvard.edu/abs/2018MNRAS.474.3727L} {474, 3727}

\bibitem[\protect\citeauthoryear{{Lacey} \& {Cole}}{{Lacey} \&
  {Cole}}{1993}]{Lacey93}
{Lacey} C.,  {Cole} S.,  1993, \mn@doi [\mnras] {10.1093/mnras/262.3.627},
  \href {http://adsabs.harvard.edu/abs/1993MNRAS.262..627L} {262, 627}

\bibitem[\protect\citeauthoryear{{Lagos}, {Demarco}, {Papaderos}, {Telles},
  {Nigoche-Netro}, {Humphrey}, {Roche}  \& {Gomes}}{{Lagos}
  et~al.}{2016a}]{Lagos16}
{Lagos} P.,  {Demarco} R.,  {Papaderos} P.,  {Telles} E.,  {Nigoche-Netro} A.,
  {Humphrey} A.,  {Roche} N.,   {Gomes} J.~M.,  2016a, \mn@doi [\mnras]
  {10.1093/mnras/stv2702}, \href
  {http://adsabs.harvard.edu/abs/2016MNRAS.456.1549L} {456, 1549}

\bibitem[\protect\citeauthoryear{{Lagos} et~al.,}{{Lagos}
  et~al.}{2016b}]{Lagos16b}
{Lagos} C.~d.~P.,  et~al., 2016b, \mn@doi [\mnras] {10.1093/mnras/stw717},
  \href {http://adsabs.harvard.edu/abs/2016MNRAS.459.2632D} {459, 2632}

\bibitem[\protect\citeauthoryear{{Lamareille}}{{Lamareille}}{2010}]{Lamareille10}
{Lamareille} F.,  2010, \mn@doi [\aap] {10.1051/0004-6361/200913168}, \href
  {http://adsabs.harvard.edu/abs/2010A%26A...509A..53L} {509, A53}

\bibitem[\protect\citeauthoryear{{Lamareille}, {Mouhcine}, {Contini}, {Lewis}
  \& {Maddox}}{{Lamareille} et~al.}{2004}]{Lamareille04}
{Lamareille} F.,  {Mouhcine} M.,  {Contini} T.,  {Lewis} I.,   {Maddox} S.,
  2004, \mn@doi [\mnras] {10.1111/j.1365-2966.2004.07697.x}, \href
  {http://adsabs.harvard.edu/abs/2004MNRAS.350..396L} {350, 396}

\bibitem[\protect\citeauthoryear{{Lara-L{\'o}pez} et~al.,}{{Lara-L{\'o}pez}
  et~al.}{2010}]{Lara-Lopez10b}
{Lara-L{\'o}pez} M.~A.,  et~al., 2010, \mn@doi [\aap]
  {10.1051/0004-6361/201014803}, \href
  {http://adsabs.harvard.edu/abs/2010A%26A...521L..53L} {521, L53}

\bibitem[\protect\citeauthoryear{{Lara-L{\'o}pez} et~al.,}{{Lara-L{\'o}pez}
  et~al.}{2013}]{Lara-Lopez13}
{Lara-L{\'o}pez} M.~A.,  et~al., 2013, \mn@doi [\mnras]
  {10.1093/mnrasl/slt054}, \href
  {http://adsabs.harvard.edu/abs/2013MNRAS.433L..35L} {433, L35}

\bibitem[\protect\citeauthoryear{{Lardo}, {Davies}, {Kudritzki}, {Gazak},
  {Evans}, {Patrick}, {Bergemann}  \& {Plez}}{{Lardo} et~al.}{2015}]{Lardo15}
{Lardo} C.,  {Davies} B.,  {Kudritzki} R.-P.,  {Gazak} J.~Z.,  {Evans} C.~J.,
  {Patrick} L.~R.,  {Bergemann} M.,   {Plez} B.,  2015, \mn@doi [\apj]
  {10.1088/0004-637X/812/2/160}, \href
  {http://adsabs.harvard.edu/abs/2015ApJ...812..160L} {812, 160}

\bibitem[\protect\citeauthoryear{{Larson}, {Tinsley}  \& {Caldwell}}{{Larson}
  et~al.}{1980}]{Larson80}
{Larson} R.~B.,  {Tinsley} B.~M.,   {Caldwell} C.~N.,  1980, \mn@doi [\apj]
  {10.1086/157917}, \href {http://adsabs.harvard.edu/abs/1980ApJ...237..692L}
  {237, 692}

\bibitem[\protect\citeauthoryear{{Laskar}, {Berger}  \& {Chary}}{{Laskar}
  et~al.}{2011}]{Laskar11}
{Laskar} T.,  {Berger} E.,   {Chary} R.-R.,  2011, \mn@doi [\apj]
  {10.1088/0004-637X/739/1/1}, \href
  {http://adsabs.harvard.edu/abs/2011ApJ...739....1L} {739, 1}

\bibitem[\protect\citeauthoryear{{Law}, {Steidel}, {Erb}, {Pettini}, {Reddy},
  {Shapley}, {Adelberger}  \& {Simenc}}{{Law} et~al.}{2007}]{Law07}
{Law} D.~R.,  {Steidel} C.~C.,  {Erb} D.~K.,  {Pettini} M.,  {Reddy} N.~A.,
  {Shapley} A.~E.,  {Adelberger} K.~L.,   {Simenc} D.~J.,  2007, \apj, \href
  {http://adsabs.harvard.edu/abs/2007ApJ...656....1L} {656, 1}

\bibitem[\protect\citeauthoryear{{Law}, {Wright}, {Ellis}, {Erb}, {Nesvadba},
  {Steidel}  \& {Swinbank}}{{Law} et~al.}{2009}]{Law09}
{Law} D.~R.,  {Wright} S.~A.,  {Ellis} R.~S.,  {Erb} D.~K.,  {Nesvadba} N.,
  {Steidel} C.~C.,   {Swinbank} M.,  2009, in astro2010: The Astronomy and
  Astrophysics Decadal Survey.  (\mn@eprint {arXiv} {0902.2567})

\bibitem[\protect\citeauthoryear{{Le Brun}, {Bergeron}, {Boisse}  \&
  {Deharveng}}{{Le Brun} et~al.}{1997}]{Le-Brun97}
{Le Brun} V.,  {Bergeron} J.,  {Boisse} P.,   {Deharveng} J.~M.,  1997, \aap,
  \href {http://adsabs.harvard.edu/abs/1997A%26A...321..733L} {321, 733}

\bibitem[\protect\citeauthoryear{{Ledoux}, {Petitjean}, {Fynbo}, {M{\o}ller}
  \& {Srianand}}{{Ledoux} et~al.}{2006}]{Ledoux06}
{Ledoux} C.,  {Petitjean} P.,  {Fynbo} J.~P.~U.,  {M{\o}ller} P.,   {Srianand}
  R.,  2006, \mn@doi [\aap] {10.1051/0004-6361:20054242}, \href
  {http://adsabs.harvard.edu/abs/2006A%26A...457...71L} {457, 71}

\bibitem[\protect\citeauthoryear{{Ledoux}, {Vreeswijk}, {Smette}, {Fox},
  {Petitjean}, {Ellison}, {Fynbo}  \& {Savaglio}}{{Ledoux}
  et~al.}{2009}]{Ledoux09}
{Ledoux} C.,  {Vreeswijk} P.~M.,  {Smette} A.,  {Fox} A.~J.,  {Petitjean} P.,
  {Ellison} S.~L.,  {Fynbo} J.~P.~U.,   {Savaglio} S.,  2009, \mn@doi [\aap]
  {10.1051/0004-6361/200811572}, \href
  {http://adsabs.harvard.edu/abs/2009A%26A...506..661L} {506, 661}

\bibitem[\protect\citeauthoryear{{Lee}, {Skillman}, {Cannon}, {Jackson},
  {Gehrz}, {Polomski}  \& {Woodward}}{{Lee} et~al.}{2006}]{Lee06}
{Lee} H.,  {Skillman} E.~D.,  {Cannon} J.~M.,  {Jackson} D.~C.,  {Gehrz} R.~D.,
   {Polomski} E.~F.,   {Woodward} C.~E.,  2006, \mn@doi [\apj]
  {10.1086/505573}, \href {http://adsabs.harvard.edu/abs/2006ApJ...647..970L}
  {647, 970}

\bibitem[\protect\citeauthoryear{{Leethochawalit}, {Jones}, {Ellis}, {Stark},
  {Richard}, {Zitrin}  \& {Auger}}{{Leethochawalit}
  et~al.}{2016}]{Leethochawalit16}
{Leethochawalit} N.,  {Jones} T.~A.,  {Ellis} R.~S.,  {Stark} D.~P.,  {Richard}
  J.,  {Zitrin} A.,   {Auger} M.,  2016, \mn@doi [\apj]
  {10.3847/0004-637X/820/2/84}, \href
  {http://adsabs.harvard.edu/abs/2016ApJ...820...84L} {820, 84}

\bibitem[\protect\citeauthoryear{{Leethochawalit}, {Kirby}, {Moran}, {Ellis}
  \& {Treu}}{{Leethochawalit} et~al.}{2018}]{Leethochawalit18}
{Leethochawalit} N.,  {Kirby} E.~N.,  {Moran} S.~M.,  {Ellis} R.~S.,   {Treu}
  T.,  2018, \mn@doi [\apj] {10.3847/1538-4357/aab26a}, \href
  {http://adsabs.harvard.edu/abs/2018ApJ...856...15L} {856, 15}

\bibitem[\protect\citeauthoryear{{Lehner}, {O'Meara}, {Howk}, {Prochaska}  \&
  {Fumagalli}}{{Lehner} et~al.}{2016}]{Lehner16}
{Lehner} N.,  {O'Meara} J.~M.,  {Howk} J.~C.,  {Prochaska} J.~X.,   {Fumagalli}
  M.,  2016, \mn@doi [\apj] {10.3847/1538-4357/833/2/283}, \href
  {http://adsabs.harvard.edu/abs/2016ApJ...833..283L} {833, 283}

\bibitem[\protect\citeauthoryear{{Lehnert} \& {Heckman}}{{Lehnert} \&
  {Heckman}}{1996}]{Lehnert96}
{Lehnert} M.~D.,  {Heckman} T.~M.,  1996, \mn@doi [\apj] {10.1086/177180},
  \href {http://adsabs.harvard.edu/abs/1996ApJ...462..651L} {462, 651}

\bibitem[\protect\citeauthoryear{{Lehnert}, {Nesvadba}, {Tiran}, {Di Matteo},
  {van Driel}, {Douglas}, {Chemin}  \& {Bournaud}}{{Lehnert}
  et~al.}{2009}]{Lehnert09}
{Lehnert} M.~D.,  {Nesvadba} N.~P.~H.,  {Tiran} L.~L.,  {Di Matteo} P.,  {van
  Driel} W.,  {Douglas} L.~S.,  {Chemin} L.,   {Bournaud} F.,  2009, \mn@doi
  [\apj] {10.1088/0004-637X/699/2/1660}, \href
  {http://adsabs.harvard.edu/abs/2009ApJ...699.1660L} {699, 1660}

\bibitem[\protect\citeauthoryear{{Leibundgut}}{{Leibundgut}}{2001}]{Leibundgut01}
{Leibundgut} B.,  2001, \mn@doi [\araa] {10.1146/annurev.astro.39.1.67}, \href
  {http://adsabs.harvard.edu/abs/2001ARA%26A..39...67L} {39, 67}

\bibitem[\protect\citeauthoryear{{Leitherer}}{{Leitherer}}{2011}]{Leitherer11b}
{Leitherer} C.,  2011, in {Treyer} M.,  {Wyder} T.,  {Neill} J.,  {Seibert} M.,
    {Lee} J.,  eds,  Astronomical Society of the Pacific Conference Series Vol.
  440, UP2010: Have Observations Revealed a Variable Upper End of the Initial
  Mass Function?. p.~309 (\mn@eprint {arXiv} {1009.0245})

\bibitem[\protect\citeauthoryear{{Leitherer}}{{Leitherer}}{2014}]{Leitherer14b}
{Leitherer} C.,  2014, in Astronomical Society of India Conference Series.
  (\mn@eprint {arXiv} {1312.2464})

\bibitem[\protect\citeauthoryear{{Leitherer} \& {Heckman}}{{Leitherer} \&
  {Heckman}}{1995}]{Leitherer95}
{Leitherer} C.,  {Heckman} T.~M.,  1995, \mn@doi [\apjs] {10.1086/192112},
  \href {http://adsabs.harvard.edu/abs/1995ApJS...96....9L} {96, 9}

\bibitem[\protect\citeauthoryear{{Leitherer}, {Le{\~a}o}, {Heckman}, {Lennon},
  {Pettini}  \& {Robert}}{{Leitherer} et~al.}{2001}]{Leitherer01}
{Leitherer} C.,  {Le{\~a}o} J.~R.~S.,  {Heckman} T.~M.,  {Lennon} D.~J.,
  {Pettini} M.,   {Robert} C.,  2001, \mn@doi [\apj] {10.1086/319814}, \href
  {http://adsabs.harvard.edu/abs/2001ApJ...550..724L} {550, 724}

\bibitem[\protect\citeauthoryear{{Leitherer}, {Ortiz Ot{\'a}lvaro}, {Bresolin},
  {Kudritzki}, {Lo Faro}, {Pauldrach}, {Pettini}  \& {Rix}}{{Leitherer}
  et~al.}{2010}]{Leitherer10}
{Leitherer} C.,  {Ortiz Ot{\'a}lvaro} P.~A.,  {Bresolin} F.,  {Kudritzki}
  R.-P.,  {Lo Faro} B.,  {Pauldrach} A.~W.~A.,  {Pettini} M.,   {Rix} S.~A.,
  2010, \mn@doi [\apjs] {10.1088/0067-0049/189/2/309}, \href
  {http://adsabs.harvard.edu/abs/2010ApJS..189..309L} {189, 309}

\bibitem[\protect\citeauthoryear{{Leitherer}, {Tremonti}, {Heckman}  \&
  {Calzetti}}{{Leitherer} et~al.}{2011}]{Leitherer11}
{Leitherer} C.,  {Tremonti} C.~A.,  {Heckman} T.~M.,   {Calzetti} D.,  2011,
  \mn@doi [\aj] {10.1088/0004-6256/141/2/37}, \href
  {http://adsabs.harvard.edu/abs/2011AJ....141...37L} {141, 37}

\bibitem[\protect\citeauthoryear{{Leitherer}, {Ekstr{\"o}m}, {Meynet},
  {Schaerer}, {Agienko}  \& {Levesque}}{{Leitherer} et~al.}{2014}]{Leitherer14}
{Leitherer} C.,  {Ekstr{\"o}m} S.,  {Meynet} G.,  {Schaerer} D.,  {Agienko}
  K.~B.,   {Levesque} E.~M.,  2014, \mn@doi [\apjs]
  {10.1088/0067-0049/212/1/14}, \href
  {http://adsabs.harvard.edu/abs/2014ApJS..212...14L} {212, 14}

\bibitem[\protect\citeauthoryear{{Lemasle} et~al.,}{{Lemasle}
  et~al.}{2014}]{Lemasle14}
{Lemasle} B.,  et~al., 2014, \mn@doi [\aap] {10.1051/0004-6361/201423919},
  \href {http://adsabs.harvard.edu/abs/2014A%26A...572A..88L} {572, A88}

\bibitem[\protect\citeauthoryear{{L{\'e}pine} et~al.,}{{L{\'e}pine}
  et~al.}{2011}]{Lepine11}
{L{\'e}pine} J.~R.~D.,  et~al., 2011, \mn@doi [\mnras]
  {10.1111/j.1365-2966.2011.19314.x}, \href
  {http://adsabs.harvard.edu/abs/2011MNRAS.417..698L} {417, 698}

\bibitem[\protect\citeauthoryear{{Lequeux}, {Peimbert}, {Rayo}, {Serrano}  \&
  {Torres-Peimbert}}{{Lequeux} et~al.}{1979}]{Lequeux79}
{Lequeux} J.,  {Peimbert} M.,  {Rayo} J.~F.,  {Serrano} A.,   {Torres-Peimbert}
  S.,  1979, \aap, \href {http://adsabs.harvard.edu/abs/1979A%26A....80..155L}
  {80, 155}

\bibitem[\protect\citeauthoryear{{Lester}, {Dinerstein}, {Werner}, {Watson}  \&
  {Genzel}}{{Lester} et~al.}{1983}]{Lester83}
{Lester} D.~F.,  {Dinerstein} H.~L.,  {Werner} M.~W.,  {Watson} D.~M.,
  {Genzel} R.~L.,  1983, \mn@doi [\apj] {10.1086/161229}, \href
  {http://adsabs.harvard.edu/abs/1983ApJ...271..618L} {271, 618}

\bibitem[\protect\citeauthoryear{{Lester}, {Dinerstein}, {Werner}, {Watson},
  {Genzel}  \& {Storey}}{{Lester} et~al.}{1987}]{Lester87}
{Lester} D.~F.,  {Dinerstein} H.~L.,  {Werner} M.~W.,  {Watson} D.~M.,
  {Genzel} R.,   {Storey} J.~W.~V.,  1987, \mn@doi [\apj] {10.1086/165575},
  \href {http://adsabs.harvard.edu/abs/1987ApJ...320..573L} {320, 573}

\bibitem[\protect\citeauthoryear{{Levesque}, {Leitherer}, {Ekstrom}, {Meynet}
  \& {Schaerer}}{{Levesque} et~al.}{2012}]{Levesque12a}
{Levesque} E.~M.,  {Leitherer} C.,  {Ekstrom} S.,  {Meynet} G.,   {Schaerer}
  D.,  2012, \mn@doi [\apj] {10.1088/0004-637X/751/1/67}, \href
  {http://adsabs.harvard.edu/abs/2012ApJ...751...67L} {751, 67}

\bibitem[\protect\citeauthoryear{{Li}, {Bresolin}  \& {Kennicutt}}{{Li}
  et~al.}{2013}]{Li13}
{Li} Y.,  {Bresolin} F.,   {Kennicutt} Jr. R.~C.,  2013, \mn@doi [\apj]
  {10.1088/0004-637X/766/1/17}, \href
  {http://adsabs.harvard.edu/abs/2013ApJ...766...17L} {766, 17}

\bibitem[\protect\citeauthoryear{{Li} et~al.,}{{Li} et~al.}{2018}]{Li18}
{Li} H.,  et~al., 2018, \mn@doi [\mnras] {10.1093/mnras/sty334}, \href
  {http://adsabs.harvard.edu/abs/2018MNRAS.476.1765L} {476, 1765}

\bibitem[\protect\citeauthoryear{{Lian}, {Li}, {Yan}  \& {Kong}}{{Lian}
  et~al.}{2015}]{Lian15}
{Lian} J.~H.,  {Li} J.~R.,  {Yan} W.,   {Kong} X.,  2015, \mn@doi [\mnras]
  {10.1093/mnras/stu2184}, \href
  {http://adsabs.harvard.edu/abs/2015MNRAS.446.1449L} {446, 1449}

\bibitem[\protect\citeauthoryear{{Lian}, {Thomas}, {Maraston}, {Goddard},
  {Comparat}, {Gonzalez-Perez}  \& {Ventura}}{{Lian} et~al.}{2018a}]{Lian18a}
{Lian} J.,  {Thomas} D.,  {Maraston} C.,  {Goddard} D.,  {Comparat} J.,
  {Gonzalez-Perez} V.,   {Ventura} P.,  2018a, \mn@doi [\mnras]
  {10.1093/mnras/stx2829}, \href
  {http://adsabs.harvard.edu/abs/2018MNRAS.474.1143L} {474, 1143}

\bibitem[\protect\citeauthoryear{{Lian} et~al.,}{{Lian}
  et~al.}{2018b}]{Lian18b}
{Lian} J.,  et~al., 2018b, \mn@doi [\mnras] {10.1093/mnras/sty425}, \href
  {http://adsabs.harvard.edu/abs/2018MNRAS.476.3883L} {476, 3883}

\bibitem[\protect\citeauthoryear{{Lian}, {Thomas}  \& {Maraston}}{{Lian}
  et~al.}{2018c}]{Lian18c}
{Lian} J.,  {Thomas} D.,   {Maraston} C.,  2018c, \mn@doi [\mnras]
  {10.1093/mnras/sty2506}, \href
  {http://adsabs.harvard.edu/abs/2018MNRAS.481.4000L} {481, 4000}

\bibitem[\protect\citeauthoryear{{Liang}, {Hammer}, {Flores}, {Elbaz},
  {Marcillac}  \& {Cesarsky}}{{Liang} et~al.}{2004}]{Liang04}
{Liang} Y.~C.,  {Hammer} F.,  {Flores} H.,  {Elbaz} D.,  {Marcillac} D.,
  {Cesarsky} C.~J.,  2004, \mn@doi [\aap] {10.1051/0004-6361:20035740}, \href
  {http://adsabs.harvard.edu/abs/2004A%26A...423..867L} {423, 867}

\bibitem[\protect\citeauthoryear{{Lilly} \& {Carollo}}{{Lilly} \&
  {Carollo}}{2016}]{Lilly16}
{Lilly} S.~J.,  {Carollo} C.~M.,  2016, \mn@doi [\apj]
  {10.3847/0004-637X/833/1/1}, \href
  {http://adsabs.harvard.edu/abs/2016ApJ...833....1L} {833, 1}

\bibitem[\protect\citeauthoryear{{Lilly}, {Carollo}  \& {Stockton}}{{Lilly}
  et~al.}{2003}]{Lilly03}
{Lilly} S.~J.,  {Carollo} C.~M.,   {Stockton} A.~N.,  2003, \mn@doi [\apj]
  {10.1086/378389}, \href {http://adsabs.harvard.edu/abs/2003ApJ...597..730L}
  {597, 730}

\bibitem[\protect\citeauthoryear{{Lilly}, {Carollo}, {Pipino}, {Renzini}  \&
  {Peng}}{{Lilly} et~al.}{2013}]{Lilly13}
{Lilly} S.~J.,  {Carollo} C.~M.,  {Pipino} A.,  {Renzini} A.,   {Peng} Y.,
  2013, \mn@doi [\apj] {10.1088/0004-637X/772/2/119}, \href
  {http://adsabs.harvard.edu/abs/2013ApJ...772..119L} {772, 119}

\bibitem[\protect\citeauthoryear{{Lindgren}, {Heiter}  \&
  {Seifahrt}}{{Lindgren} et~al.}{2016}]{Lindgren16}
{Lindgren} S.,  {Heiter} U.,   {Seifahrt} A.,  2016, \mn@doi [\aap]
  {10.1051/0004-6361/201526602}, \href
  {http://adsabs.harvard.edu/abs/2016A%26A...586A.100L} {586, A100}

\bibitem[\protect\citeauthoryear{{Liu}}{{Liu}}{2002}]{Liu02}
{Liu} X.-W.,  2002, in {Henney} W.~J.,  {Franco} J.,   {Martos} M.,  eds,
  Revista Mexicana de Astronomia y Astrofisica Conference Series Vol. 12,
  Revista Mexicana de Astronomia y Astrofisica Conference Series. pp 70--76

\bibitem[\protect\citeauthoryear{{Liu}}{{Liu}}{2003}]{Liu03}
{Liu} X.-W.,  2003, in {Kwok} S.,  {Dopita} M.,   {Sutherland} R.,  eds,  IAU
  Symposium Vol. 209, Planetary Nebulae: Their Evolution and Role in the
  Universe. p.~339

\bibitem[\protect\citeauthoryear{{Liu}, {Luo}, {Barlow}, {Danziger}  \&
  {Storey}}{{Liu} et~al.}{2001}]{Liu01}
{Liu} X.-W.,  {Luo} S.-G.,  {Barlow} M.~J.,  {Danziger} I.~J.,   {Storey}
  P.~J.,  2001, \mn@doi [\mnras] {10.1046/j.1365-8711.2001.04676.x}, \href
  {http://adsabs.harvard.edu/abs/2001MNRAS.327..141L} {327, 141}

\bibitem[\protect\citeauthoryear{{Liu}, {Shapley}, {Coil}, {Brinchmann}  \&
  {Ma}}{{Liu} et~al.}{2008}]{Liu08a}
{Liu} X.,  {Shapley} A.~E.,  {Coil} A.~L.,  {Brinchmann} J.,   {Ma} C.-P.,
  2008, \mn@doi [\apj] {10.1086/529030}, \href
  {http://adsabs.harvard.edu/abs/2008ApJ...678..758L} {678, 758}

\bibitem[\protect\citeauthoryear{{Liu}, {Wang}  \& {Mao}}{{Liu}
  et~al.}{2012}]{Liu12}
{Liu} J.,  {Wang} Q.~D.,   {Mao} S.,  2012, \mn@doi [\mnras]
  {10.1111/j.1365-2966.2011.20263.x}, \href
  {http://adsabs.harvard.edu/abs/2012MNRAS.420.3389L} {420, 3389}

\bibitem[\protect\citeauthoryear{{Lofthouse}, {Houghton}  \&
  {Kaviraj}}{{Lofthouse} et~al.}{2017}]{Lofthouse17}
{Lofthouse} E.~K.,  {Houghton} R.~C.~W.,   {Kaviraj} S.,  2017, \mn@doi
  [\mnras] {10.1093/mnras/stx1627}, \href
  {http://adsabs.harvard.edu/abs/2017MNRAS.471.2311L} {471, 2311}

\bibitem[\protect\citeauthoryear{{Lonoce} et~al.,}{{Lonoce}
  et~al.}{2015}]{Lonoce15}
{Lonoce} I.,  et~al., 2015, \mn@doi [\mnras] {10.1093/mnras/stv2150}, \href
  {http://adsabs.harvard.edu/abs/2015MNRAS.454.3912L} {454, 3912}

\bibitem[\protect\citeauthoryear{{L{\'o}pez-S{\'a}nchez}, {Esteban},
  {Garc{\'{\i}}a-Rojas}, {Peimbert}  \&
  {Rodr{\'{\i}}guez}}{{L{\'o}pez-S{\'a}nchez} et~al.}{2007}]{Lopez-Sanchez07}
{L{\'o}pez-S{\'a}nchez} {\'A}.~R.,  {Esteban} C.,  {Garc{\'{\i}}a-Rojas} J.,
  {Peimbert} M.,   {Rodr{\'{\i}}guez} M.,  2007, \mn@doi [\apj]
  {10.1086/510112}, \href {http://adsabs.harvard.edu/abs/2007ApJ...656..168L}
  {656, 168}

\bibitem[\protect\citeauthoryear{{L{\'o}pez-S{\'a}nchez}, {Dopita}, {Kewley},
  {Zahid}, {Nicholls}  \& {Scharw{\"a}chter}}{{L{\'o}pez-S{\'a}nchez}
  et~al.}{2012}]{Lopez-Sanchez12}
{L{\'o}pez-S{\'a}nchez} {\'A}.~R.,  {Dopita} M.~A.,  {Kewley} L.~J.,  {Zahid}
  H.~J.,  {Nicholls} D.~C.,   {Scharw{\"a}chter} J.,  2012, \mn@doi [\mnras]
  {10.1111/j.1365-2966.2012.21145.x}, \href
  {http://adsabs.harvard.edu/abs/2012MNRAS.426.2630L} {426, 2630}

\bibitem[\protect\citeauthoryear{{Lu}, {Mo}  \& {Lu}}{{Lu}
  et~al.}{2015}]{Lu15c}
{Lu} Z.,  {Mo} H.~J.,   {Lu} Y.,  2015, \mn@doi [\mnras]
  {10.1093/mnras/stv671}, \href
  {http://adsabs.harvard.edu/abs/2015MNRAS.450..606L} {450, 606}

\bibitem[\protect\citeauthoryear{{Luck}, {Kovtyukh}  \& {Andrievsky}}{{Luck}
  et~al.}{2006}]{Luck06}
{Luck} R.~E.,  {Kovtyukh} V.~V.,   {Andrievsky} S.~M.,  2006, \mn@doi [\aj]
  {10.1086/505687}, \href {http://adsabs.harvard.edu/abs/2006AJ....132..902L}
  {132, 902}

\bibitem[\protect\citeauthoryear{{Luck}, {Andrievsky}, {Kovtyukh}, {Gieren}  \&
  {Graczyk}}{{Luck} et~al.}{2011}]{Luck11}
{Luck} R.~E.,  {Andrievsky} S.~M.,  {Kovtyukh} V.~V.,  {Gieren} W.,   {Graczyk}
  D.,  2011, \mn@doi [\aj] {10.1088/0004-6256/142/2/51}, \href
  {http://adsabs.harvard.edu/abs/2011AJ....142...51L} {142, 51}

\bibitem[\protect\citeauthoryear{{Ludwig}, {Greene}, {Barth}  \& {Ho}}{{Ludwig}
  et~al.}{2012}]{Ludwig12}
{Ludwig} R.~R.,  {Greene} J.~E.,  {Barth} A.~J.,   {Ho} L.~C.,  2012, \mn@doi
  [\apj] {10.1088/0004-637X/756/1/51}, \href
  {http://adsabs.harvard.edu/abs/2012ApJ...756...51L} {756, 51}

\bibitem[\protect\citeauthoryear{{Luridiana}, {Morisset}  \&
  {Shaw}}{{Luridiana} et~al.}{2012}]{Luridiana12}
{Luridiana} V.,  {Morisset} C.,   {Shaw} R.~A.,  2012, in IAU Symposium. pp
  422--423, \mn@doi{10.1017/S1743921312011738}

\bibitem[\protect\citeauthoryear{{Luridiana}, {Morisset}  \&
  {Shaw}}{{Luridiana} et~al.}{2015}]{Luridiana15}
{Luridiana} V.,  {Morisset} C.,   {Shaw} R.~A.,  2015, \mn@doi [\aap]
  {10.1051/0004-6361/201323152}, \href
  {http://adsabs.harvard.edu/abs/2015A%26A...573A..42L} {573, A42}

\bibitem[\protect\citeauthoryear{{Ly}, {Rigby}, {Cooper}  \& {Yan}}{{Ly}
  et~al.}{2015}]{Ly15}
{Ly} C.,  {Rigby} J.~R.,  {Cooper} M.,   {Yan} R.,  2015, \mn@doi [\apj]
  {10.1088/0004-637X/805/1/45}, \href
  {http://adsabs.harvard.edu/abs/2015ApJ...805...45L} {805, 45}

\bibitem[\protect\citeauthoryear{Ly, Malkan, Rigby  \& Nagao}{Ly
  et~al.}{2016}]{Ly16}
Ly C.,  Malkan M.~A.,  Rigby J.~R.,   Nagao T.,  2016, The Astrophysical
  Journal, 828, 67

\bibitem[\protect\citeauthoryear{{Ma}, {Hopkins}, {Faucher-Gigu{\`e}re},
  {Zolman}, {Muratov}, {Kere{\v s}}  \& {Quataert}}{{Ma} et~al.}{2016}]{Ma16}
{Ma} X.,  {Hopkins} P.~F.,  {Faucher-Gigu{\`e}re} C.-A.,  {Zolman} N.,
  {Muratov} A.~L.,  {Kere{\v s}} D.,   {Quataert} E.,  2016, \mn@doi [\mnras]
  {10.1093/mnras/stv2659}, \href
  {http://adsabs.harvard.edu/abs/2016MNRAS.456.2140M} {456, 2140}

\bibitem[\protect\citeauthoryear{{Ma}, {Hopkins}, {Feldmann}, {Torrey},
  {Faucher-Gigu{\`e}re}  \& {Kere{\v s}}}{{Ma} et~al.}{2017}]{Ma17a}
{Ma} X.,  {Hopkins} P.~F.,  {Feldmann} R.,  {Torrey} P.,  {Faucher-Gigu{\`e}re}
  C.-A.,   {Kere{\v s}} D.,  2017, \mn@doi [\mnras] {10.1093/mnras/stx034},
  \href {http://adsabs.harvard.edu/abs/2017MNRAS.466.4780M} {466, 4780}

\bibitem[\protect\citeauthoryear{{Maciel} \& {Quireza}}{{Maciel} \&
  {Quireza}}{1999}]{Maciel99}
{Maciel} W.~J.,  {Quireza} C.,  1999, \aap, \href
  {http://adsabs.harvard.edu/abs/1999A%26A...345..629M} {345, 629}

\bibitem[\protect\citeauthoryear{{Maciel}, {Costa}  \& {Uchida}}{{Maciel}
  et~al.}{2003}]{Maciel03}
{Maciel} W.~J.,  {Costa} R.~D.~D.,   {Uchida} M.~M.~M.,  2003, \mn@doi [\aap]
  {10.1051/0004-6361:20021530}, \href
  {http://adsabs.harvard.edu/abs/2003A%26A...397..667M} {397, 667}

\bibitem[\protect\citeauthoryear{{Madau} \& {Dickinson}}{{Madau} \&
  {Dickinson}}{2014}]{Madau14}
{Madau} P.,  {Dickinson} M.,  2014, \mn@doi [\araa]
  {10.1146/annurev-astro-081811-125615}, \href
  {http://adsabs.harvard.edu/abs/2014ARA%26A..52..415M} {52, 415}

\bibitem[\protect\citeauthoryear{{Madau} \& {Shull}}{{Madau} \&
  {Shull}}{1996}]{Madau96a}
{Madau} P.,  {Shull} J.~M.,  1996, \mn@doi [\apj] {10.1086/176751}, \href
  {http://adsabs.harvard.edu/abs/1996ApJ...457..551M} {457, 551}

\bibitem[\protect\citeauthoryear{{Madau}, {Pozzetti}  \& {Dickinson}}{{Madau}
  et~al.}{1998}]{Madau98a}
{Madau} P.,  {Pozzetti} L.,   {Dickinson} M.,  1998, \mn@doi [\apj]
  {10.1086/305523}, \href {http://adsabs.harvard.edu/abs/1998ApJ...498..106M}
  {498, 106}

\bibitem[\protect\citeauthoryear{{Magrini}, {Sestito}, {Randich}  \&
  {Galli}}{{Magrini} et~al.}{2009}]{Magrini09a}
{Magrini} L.,  {Sestito} P.,  {Randich} S.,   {Galli} D.,  2009, \mn@doi [\aap]
  {10.1051/0004-6361:200810634}, \href
  {http://adsabs.harvard.edu/abs/2009A%26A...494...95M} {494, 95}

\bibitem[\protect\citeauthoryear{{Magrini}, {Sommariva}, {Cresci}, {Sani},
  {Galametz}, {Mannucci}, {Petropoulou}  \& {Fumana}}{{Magrini}
  et~al.}{2012}]{Magrini12a}
{Magrini} L.,  {Sommariva} V.,  {Cresci} G.,  {Sani} E.,  {Galametz} A.,
  {Mannucci} F.,  {Petropoulou} V.,   {Fumana} M.,  2012, \mn@doi [\mnras]
  {10.1111/j.1365-2966.2012.21460.x}, \href
  {http://adsabs.harvard.edu/abs/2012MNRAS.426.1195M} {426, 1195}

\bibitem[\protect\citeauthoryear{{Magrini}, {Coccato}, {Stanghellini},
  {Casasola}  \& {Galli}}{{Magrini} et~al.}{2016}]{Magrini16}
{Magrini} L.,  {Coccato} L.,  {Stanghellini} L.,  {Casasola} V.,   {Galli} D.,
  2016, \mn@doi [\aap] {10.1051/0004-6361/201527799}, \href
  {http://adsabs.harvard.edu/abs/2016A%26A...588A..91M} {588, A91}

\bibitem[\protect\citeauthoryear{{Magrini} et~al.,}{{Magrini}
  et~al.}{2017}]{Magrini17b}
{Magrini} L.,  et~al., 2017, \mn@doi [\aap] {10.1051/0004-6361/201630294},
  \href {http://adsabs.harvard.edu/abs/2017A%26A...603A...2M} {603, A2}

\bibitem[\protect\citeauthoryear{{Maier}, {Meisenheimer}  \&
  {Hippelein}}{{Maier} et~al.}{2004}]{Maier04}
{Maier} C.,  {Meisenheimer} K.,   {Hippelein} H.,  2004, \mn@doi [\aap]
  {10.1051/0004-6361:20035795}, \href
  {http://adsabs.harvard.edu/abs/2004A%26A...418..475M} {418, 475}

\bibitem[\protect\citeauthoryear{{Maier}, {Lilly}, {Carollo}, {Stockton}  \&
  {Brodwin}}{{Maier} et~al.}{2005}]{Maier05}
{Maier} C.,  {Lilly} S.~J.,  {Carollo} C.~M.,  {Stockton} A.,   {Brodwin} M.,
  2005, \mn@doi [\apj] {10.1086/497091}, \href
  {http://adsabs.harvard.edu/abs/2005ApJ...634..849M} {634, 849}

\bibitem[\protect\citeauthoryear{{Maier}, {Lilly}, {Carollo}, {Meisenheimer},
  {Hippelein}  \& {Stockton}}{{Maier} et~al.}{2006}]{Maier06}
{Maier} C.,  {Lilly} S.~J.,  {Carollo} C.~M.,  {Meisenheimer} K.,  {Hippelein}
  H.,   {Stockton} A.,  2006, \mn@doi [\apj] {10.1086/499518}, \href
  {http://adsabs.harvard.edu/abs/2006ApJ...639..858M} {639, 858}

\bibitem[\protect\citeauthoryear{{Maier}, {Lilly}, {Ziegler}, {Contini},
  {P{\'e}rez Montero}, {Peng}  \& {Balestra}}{{Maier} et~al.}{2014}]{Maier14a}
{Maier} C.,  {Lilly} S.~J.,  {Ziegler} B.~L.,  {Contini} T.,  {P{\'e}rez
  Montero} E.,  {Peng} Y.,   {Balestra} I.,  2014, \mn@doi [\apj]
  {10.1088/0004-637X/792/1/3}, \href
  {http://adsabs.harvard.edu/abs/2014ApJ...792....3M} {792, 3}

\bibitem[\protect\citeauthoryear{{Maier}, {Lilly}  \& {Ziegler}}{{Maier}
  et~al.}{2015a}]{Maier15a}
{Maier} C.,  {Lilly} S.~J.,   {Ziegler} B.~L.,  2015a, in {Ziegler} B.~L.,
  {Combes} F.,  {Dannerbauer} H.,   {Verdugo} M.,  eds,  IAU Symposium Vol.
  309, Galaxies in 3D across the Universe. pp 281--282 (\mn@eprint {arXiv}
  {1408.5896}), \mn@doi{10.1017/S1743921314009867}

\bibitem[\protect\citeauthoryear{{Maier}, {Ziegler}, {Lilly}, {Contini},
  {P{\'e}rez-Montero}, {Lamareille}, {Bolzonella}  \& {Le Floc'h}}{{Maier}
  et~al.}{2015b}]{Maier15b}
{Maier} C.,  {Ziegler} B.~L.,  {Lilly} S.~J.,  {Contini} T.,
  {P{\'e}rez-Montero} E.,  {Lamareille} F.,  {Bolzonella} M.,   {Le Floc'h} E.,
   2015b, \mn@doi [\aap] {10.1051/0004-6361/201425224}, \href
  {http://adsabs.harvard.edu/abs/2015A%26A...577A..14M} {577, A14}

\bibitem[\protect\citeauthoryear{{Maier} et~al.,}{{Maier}
  et~al.}{2016}]{Maier16}
{Maier} C.,  et~al., 2016, \mn@doi [\aap] {10.1051/0004-6361/201628223}, \href
  {http://adsabs.harvard.edu/abs/2016A%26A...590A.108M} {590, A108}

\bibitem[\protect\citeauthoryear{{Maier}, {Ziegler}, {Haines}  \&
  {Smith}}{{Maier} et~al.}{2018}]{Maier18}
{Maier} C.,  {Ziegler} B.~L.,  {Haines} C.~P.,   {Smith} G.~P.,  2018,
  preprint, \href {http://adsabs.harvard.edu/abs/2018arXiv180907675M} {}
  (\mn@eprint {arXiv} {1809.07675})

\bibitem[\protect\citeauthoryear{{Maiolino}, {Juarez}, {Mujica}, {Nagar}  \&
  {Oliva}}{{Maiolino} et~al.}{2003}]{Maiolino03}
{Maiolino} R.,  {Juarez} Y.,  {Mujica} R.,  {Nagar} N.~M.,   {Oliva} E.,  2003,
  \mn@doi [\apjl] {10.1086/379600}, \href
  {http://adsabs.harvard.edu/abs/2003ApJ...596L.155M} {596, L155}

\bibitem[\protect\citeauthoryear{{Maiolino} et~al.,}{{Maiolino}
  et~al.}{2008}]{Maiolino08}
{Maiolino} R.,  et~al., 2008, \mn@doi [\aap] {10.1051/0004-6361:200809678},
  \href {http://adsabs.harvard.edu/abs/2008A%26A...488..463M} {488, 463}

\bibitem[\protect\citeauthoryear{{Maiolino} et~al.,}{{Maiolino}
  et~al.}{2015}]{Maiolino15}
{Maiolino} R.,  et~al., 2015, \mn@doi [\mnras] {10.1093/mnras/stv1194}, \href
  {http://adsabs.harvard.edu/abs/2015MNRAS.452...54M} {452, 54}

\bibitem[\protect\citeauthoryear{{Majewski} et~al.,}{{Majewski}
  et~al.}{2017}]{Majewski17}
{Majewski} S.~R.,  et~al., 2017, \mn@doi [\aj] {10.3847/1538-3881/aa784d}, 154,
  94

\bibitem[\protect\citeauthoryear{{Mannucci}, {Maoz}, {Sharon}, {Botticella},
  {Della Valle}, {Gal-Yam}  \& {Panagia}}{{Mannucci} et~al.}{2008}]{Mannucci08}
{Mannucci} F.,  {Maoz} D.,  {Sharon} K.,  {Botticella} M.~T.,  {Della Valle}
  M.,  {Gal-Yam} A.,   {Panagia} N.,  2008, \mn@doi [\mnras]
  {10.1111/j.1365-2966.2007.12603.x}, \href
  {http://adsabs.harvard.edu/abs/2008MNRAS.383.1121M} {383, 1121}

\bibitem[\protect\citeauthoryear{{Mannucci} et~al.,}{{Mannucci}
  et~al.}{2009}]{Mannucci09b}
{Mannucci} F.,  et~al., 2009, \mn@doi [\mnras]
  {10.1111/j.1365-2966.2009.15185.x}, \href
  {http://adsabs.harvard.edu/abs/2009MNRAS.398.1915M} {398, 1915}

\bibitem[\protect\citeauthoryear{{Mannucci}, {Cresci}, {Maiolino}, {Marconi}
  \& {Gnerucci}}{{Mannucci} et~al.}{2010}]{Mannucci10}
{Mannucci} F.,  {Cresci} G.,  {Maiolino} R.,  {Marconi} A.,   {Gnerucci} A.,
  2010, \mn@doi [\mnras] {10.1111/j.1365-2966.2010.17291.x}, \href
  {http://adsabs.harvard.edu/abs/2010MNRAS.408.2115M} {408, 2115}

\bibitem[\protect\citeauthoryear{{Mannucci}, {Salvaterra}  \&
  {Campisi}}{{Mannucci} et~al.}{2011}]{Mannucci11a}
{Mannucci} F.,  {Salvaterra} R.,   {Campisi} M.~A.,  2011, \mn@doi [\mnras]
  {10.1111/j.1365-2966.2011.18459.x}, \href
  {http://adsabs.harvard.edu/abs/2011MNRAS.414.1263M} {414, 1263}

\bibitem[\protect\citeauthoryear{{Maoz} \& {Mannucci}}{{Maoz} \&
  {Mannucci}}{2012}]{Maoz12a}
{Maoz} D.,  {Mannucci} F.,  2012, \mn@doi [\pasa] {10.1071/AS11052}, \href
  {http://adsabs.harvard.edu/abs/2012PASA...29..447M} {29, 447}

\bibitem[\protect\citeauthoryear{{Maoz}, {Mannucci}  \& {Nelemans}}{{Maoz}
  et~al.}{2014}]{Maoz14}
{Maoz} D.,  {Mannucci} F.,   {Nelemans} G.,  2014, \mn@doi [\araa]
  {10.1146/annurev-astro-082812-141031}, \href
  {http://adsabs.harvard.edu/abs/2014ARA%26A..52..107M} {52, 107}

\bibitem[\protect\citeauthoryear{{Maraston}, {Nieves Colmen{\'a}rez}, {Bender}
  \& {Thomas}}{{Maraston} et~al.}{2009}]{Maraston09}
{Maraston} C.,  {Nieves Colmen{\'a}rez} L.,  {Bender} R.,   {Thomas} D.,  2009,
  \mn@doi [\aap] {10.1051/0004-6361:20066907}, \href
  {http://adsabs.harvard.edu/abs/2009A%26A...493..425M} {493, 425}

\bibitem[\protect\citeauthoryear{{Maraston}, {Pforr}, {Renzini}, {Daddi},
  {Dickinson}, {Cimatti}  \& {Tonini}}{{Maraston} et~al.}{2010}]{Maraston10}
{Maraston} C.,  {Pforr} J.,  {Renzini} A.,  {Daddi} E.,  {Dickinson} M.,
  {Cimatti} A.,   {Tonini} C.,  2010, \mn@doi [\mnras]
  {10.1111/j.1365-2966.2010.16973.x}, \href
  {http://adsabs.harvard.edu/abs/2010MNRAS.407..830M} {407, 830}

\bibitem[\protect\citeauthoryear{{Marino} et~al.,}{{Marino}
  et~al.}{2013}]{Marino13}
{Marino} R.~A.,  et~al., 2013, \mn@doi [\aap] {10.1051/0004-6361/201321956},
  \href {http://adsabs.harvard.edu/abs/2013A%26A...559A.114M} {559, A114}

\bibitem[\protect\citeauthoryear{{Martel}, {Kawata}  \& {Ellison}}{{Martel}
  et~al.}{2013}]{Martel13}
{Martel} H.,  {Kawata} D.,   {Ellison} S.~L.,  2013, \mn@doi [\mnras]
  {10.1093/mnras/stt354}, \href
  {http://adsabs.harvard.edu/abs/2013MNRAS.431.2560M} {431, 2560}

\bibitem[\protect\citeauthoryear{{Martel}, {Carles}, {Robichaud}, {Ellison}  \&
  {Williamson}}{{Martel} et~al.}{2018}]{Martel18}
{Martel} H.,  {Carles} C.,  {Robichaud} F.,  {Ellison} S.~L.,   {Williamson}
  D.~J.,  2018, \mn@doi [\mnras] {10.1093/mnras/sty932}, \href
  {http://adsabs.harvard.edu/abs/2018MNRAS.477.5367M} {477, 5367}

\bibitem[\protect\citeauthoryear{{Martin}, {Shapley}, {Coil}, {Kornei},
  {Bundy}, {Weiner}, {Noeske}  \& {Schiminovich}}{{Martin}
  et~al.}{2012}]{Martin12b}
{Martin} C.~L.,  {Shapley} A.~E.,  {Coil} A.~L.,  {Kornei} K.~A.,  {Bundy} K.,
  {Weiner} B.~J.,  {Noeske} K.~G.,   {Schiminovich} D.,  2012, \mn@doi [\apj]
  {10.1088/0004-637X/760/2/127}, \href
  {http://adsabs.harvard.edu/abs/2012ApJ...760..127M} {760, 127}

\bibitem[\protect\citeauthoryear{{Maseda} et~al.,}{{Maseda}
  et~al.}{2014}]{Maseda14}
{Maseda} M.~V.,  et~al., 2014, \mn@doi [\apj] {10.1088/0004-637X/791/1/17},
  \href {http://adsabs.harvard.edu/abs/2014ApJ...791...17M} {791, 17}

\bibitem[\protect\citeauthoryear{{Masters} et~al.,}{{Masters}
  et~al.}{2014}]{Masters14}
{Masters} D.,  et~al., 2014, \mn@doi [\apj] {10.1088/0004-637X/785/2/153},
  \href {http://adsabs.harvard.edu/abs/2014ApJ...785..153M} {785, 153}

\bibitem[\protect\citeauthoryear{{Masters}, {Faisst}  \& {Capak}}{{Masters}
  et~al.}{2016}]{Masters16}
{Masters} D.,  {Faisst} A.,   {Capak} P.,  2016, \mn@doi [\apj]
  {10.3847/0004-637X/828/1/18}, \href
  {http://adsabs.harvard.edu/abs/2016ApJ...828...18M} {828, 18}

\bibitem[\protect\citeauthoryear{{Matsuoka}, {Nagao}, {Maiolino}, {Marconi}  \&
  {Taniguchi}}{{Matsuoka} et~al.}{2009}]{Matsuoka09}
{Matsuoka} K.,  {Nagao} T.,  {Maiolino} R.,  {Marconi} A.,   {Taniguchi} Y.,
  2009, \mn@doi [\aap] {10.1051/0004-6361/200811478}, \href
  {http://adsabs.harvard.edu/abs/2009A%26A...503..721M} {503, 721}

\bibitem[\protect\citeauthoryear{{Matsuoka}, {Nagao}, {Marconi}, {Maiolino}  \&
  {Taniguchi}}{{Matsuoka} et~al.}{2011a}]{Matsuoka11a}
{Matsuoka} K.,  {Nagao} T.,  {Marconi} A.,  {Maiolino} R.,   {Taniguchi} Y.,
  2011a, \mn@doi [\aap] {10.1051/0004-6361/201015584}, \href
  {http://adsabs.harvard.edu/abs/2011A%26A...527A.100M} {527, A100}

\bibitem[\protect\citeauthoryear{{Matsuoka}, {Nagao}, {Maiolino}, {Marconi}  \&
  {Taniguchi}}{{Matsuoka} et~al.}{2011b}]{Matsuoka11b}
{Matsuoka} K.,  {Nagao} T.,  {Maiolino} R.,  {Marconi} A.,   {Taniguchi} Y.,
  2011b, \mn@doi [\aap] {10.1051/0004-6361/201117641}, \href
  {http://adsabs.harvard.edu/abs/2011A%26A...532L..10M} {532, L10}

\bibitem[\protect\citeauthoryear{{Matsuoka}, {Nagao}, {Marconi}, {Maiolino},
  {Mannucci}, {Cresci}, {Terao}  \& {Ikeda}}{{Matsuoka}
  et~al.}{2018}]{Matsuoka18}
{Matsuoka} K.,  {Nagao} T.,  {Marconi} A.,  {Maiolino} R.,  {Mannucci} F.,
  {Cresci} G.,  {Terao} K.,   {Ikeda} H.,  2018, \mn@doi [\aap]
  {10.1051/0004-6361/201833418}, \href
  {http://adsabs.harvard.edu/abs/2018A%26A...616L...4M} {616, L4}

\bibitem[\protect\citeauthoryear{{Matteucci}}{{Matteucci}}{1986}]{Matteucci86b}
{Matteucci} F.,  1986, \mn@doi [\mnras] {10.1093/mnras/221.4.911}, \href
  {http://adsabs.harvard.edu/abs/1986MNRAS.221..911M} {221, 911}

\bibitem[\protect\citeauthoryear{{Matteucci}}{{Matteucci}}{1994}]{Matteucci94}
{Matteucci} F.,  1994, \aap, \href
  {http://adsabs.harvard.edu/abs/1994A%26A...288...57M} {288, 57}

\bibitem[\protect\citeauthoryear{{Matteucci}}{{Matteucci}}{2001}]{Matteucci01b}
{Matteucci} F.,  2001, {The chemical evolution of the Galaxy}.
 Astrophysics and Space Science Library Vol. 253, Kluwer

\bibitem[\protect\citeauthoryear{{Matteucci}}{{Matteucci}}{2008}]{Matteucci08a}
{Matteucci} F.,  2008, in IAU Symposium. pp 391--400,
  \mn@doi{10.1017/S1743921308020747}

\bibitem[\protect\citeauthoryear{{Matteucci}}{{Matteucci}}{2012}]{Matteucci12}
{Matteucci} F.,  2012, {Chemical Evolution of Galaxies},
  \mn@doi{10.1007/978-3-642-22491-1.
}

\bibitem[\protect\citeauthoryear{{Matteucci} \& {Brocato}}{{Matteucci} \&
  {Brocato}}{1990}]{Matteucci90}
{Matteucci} F.,  {Brocato} E.,  1990, \mn@doi [\apj] {10.1086/169508}, \href
  {http://adsabs.harvard.edu/abs/1990ApJ...365..539M} {365, 539}

\bibitem[\protect\citeauthoryear{{Matteucci} \& {Chiosi}}{{Matteucci} \&
  {Chiosi}}{1983}]{Matteucci83}
{Matteucci} F.,  {Chiosi} C.,  1983, \aap, \href
  {http://adsabs.harvard.edu/abs/1983A%26A...123..121M} {123, 121}

\bibitem[\protect\citeauthoryear{{Matteucci} \& {Francois}}{{Matteucci} \&
  {Francois}}{1989}]{Matteucci89}
{Matteucci} F.,  {Francois} P.,  1989, \mn@doi [\mnras]
  {10.1093/mnras/239.3.885}, \href
  {http://adsabs.harvard.edu/abs/1989MNRAS.239..885M} {239, 885}

\bibitem[\protect\citeauthoryear{{Matteucci} \& {Greggio}}{{Matteucci} \&
  {Greggio}}{1986}]{Matteucci86a}
{Matteucci} F.,  {Greggio} L.,  1986, \aap, \href
  {http://adsabs.harvard.edu/abs/1986A%26A...154..279M} {154, 279}

\bibitem[\protect\citeauthoryear{{Matteucci} \& {Tornambe}}{{Matteucci} \&
  {Tornambe}}{1987}]{Matteucci87}
{Matteucci} F.,  {Tornambe} A.,  1987, \aap, \href
  {http://adsabs.harvard.edu/abs/1987A%26A...185...51M} {185, 51}

\bibitem[\protect\citeauthoryear{{Matteucci}, {Ponzone}  \&
  {Gibson}}{{Matteucci} et~al.}{1998}]{Matteucci98}
{Matteucci} F.,  {Ponzone} R.,   {Gibson} B.~K.,  1998, \aap, \href
  {http://adsabs.harvard.edu/abs/1998A%26A...335..855M} {335, 855}

\bibitem[\protect\citeauthoryear{{Matteucci}, {Romano}, {Arcones}, {Korobkin}
  \& {Rosswog}}{{Matteucci} et~al.}{2014}]{Matteucci14}
{Matteucci} F.,  {Romano} D.,  {Arcones} A.,  {Korobkin} O.,   {Rosswog} S.,
  2014, \mn@doi [\mnras] {10.1093/mnras/stt2350}, \href
  {http://adsabs.harvard.edu/abs/2014MNRAS.438.2177M} {438, 2177}

\bibitem[\protect\citeauthoryear{{Matthee} \& {Schaye}}{{Matthee} \&
  {Schaye}}{2018}]{Matthee18}
{Matthee} J.,  {Schaye} J.,  2018, \mn@doi [\mnras] {10.1093/mnrasl/sly093},
  \href {http://adsabs.harvard.edu/abs/2018MNRAS.479L..34M} {479, L34}

\bibitem[\protect\citeauthoryear{{Mazzucchelli} et~al.,}{{Mazzucchelli}
  et~al.}{2017}]{Mazzucchelli17}
{Mazzucchelli} C.,  et~al., 2017, \mn@doi [\apj] {10.3847/1538-4357/aa9185},
  \href {http://adsabs.harvard.edu/abs/2017ApJ...849...91M} {849, 91}

\bibitem[\protect\citeauthoryear{{McAlpine} et~al.,}{{McAlpine}
  et~al.}{2016}]{McAlpine16}
{McAlpine} S.,  et~al., 2016, \mn@doi [Astronomy and Computing]
  {10.1016/j.ascom.2016.02.004}, \href
  {http://adsabs.harvard.edu/abs/2016A%26C....15...72M} {15, 72}

\bibitem[\protect\citeauthoryear{{McCall}, {Rybski}  \& {Shields}}{{McCall}
  et~al.}{1985}]{McCall85}
{McCall} M.~L.,  {Rybski} P.~M.,   {Shields} G.~A.,  1985, \mn@doi [\apjs]
  {10.1086/190994}, \href {http://adsabs.harvard.edu/abs/1985ApJS...57....1M}
  {57, 1}

\bibitem[\protect\citeauthoryear{{McClure} \& {van den Bergh}}{{McClure} \&
  {van den Bergh}}{1968}]{McClure68}
{McClure} R.~D.,  {van den Bergh} S.,  1968, \mn@doi [\aj] {10.1086/110760},
  \href {http://adsabs.harvard.edu/abs/1968AJ.....73.1008M} {73, 1008}

\bibitem[\protect\citeauthoryear{{McGaugh}}{{McGaugh}}{1991}]{McGaugh91}
{McGaugh} S.~S.,  1991, \mn@doi [\apj] {10.1086/170569}, \href
  {http://adsabs.harvard.edu/abs/1991ApJ...380..140M} {380, 140}

\bibitem[\protect\citeauthoryear{{Mehlert} et~al.,}{{Mehlert}
  et~al.}{2002}]{Mehlert02}
{Mehlert} D.,  et~al., 2002, \mn@doi [\aap] {10.1051/0004-6361:20021052}, \href
  {http://adsabs.harvard.edu/abs/2002A%26A...393..809M} {393, 809}

\bibitem[\protect\citeauthoryear{{Meiksin}}{{Meiksin}}{2009}]{Meiksin09}
{Meiksin} A.~A.,  2009, \mn@doi [Reviews of Modern Physics]
  {10.1103/RevModPhys.81.1405}, \href
  {http://adsabs.harvard.edu/abs/2009RvMP...81.1405M} {81, 1405}

\bibitem[\protect\citeauthoryear{{Mernier} et~al.,}{{Mernier}
  et~al.}{2016}]{Mernier16}
{Mernier} F.,  et~al., 2016, \mn@doi [\aap] {10.1051/0004-6361/201628765},
  \href {http://adsabs.harvard.edu/abs/2016A%26A...595A.126M} {595, A126}

\bibitem[\protect\citeauthoryear{{Mernier} et~al.,}{{Mernier}
  et~al.}{2017}]{Mernier17}
{Mernier} F.,  et~al., 2017, \mn@doi [\aap] {10.1051/0004-6361/201630075},
  \href {http://adsabs.harvard.edu/abs/2017A%26A...603A..80M} {603, A80}

\bibitem[\protect\citeauthoryear{{Mernier} et~al.,}{{Mernier}
  et~al.}{2018}]{Mernier18}
{Mernier} F.,  et~al., 2018, \mn@doi [\mnras] {10.1093/mnrasl/sly134}, \href
  {http://adsabs.harvard.edu/abs/2018MNRAS.480L..95M} {480, L95}

\bibitem[\protect\citeauthoryear{{Micali}, {Matteucci}  \& {Romano}}{{Micali}
  et~al.}{2013}]{Micali13}
{Micali} A.,  {Matteucci} F.,   {Romano} D.,  2013, \mn@doi [\mnras]
  {10.1093/mnras/stt1681}, \href
  {http://adsabs.harvard.edu/abs/2013MNRAS.436.1648M} {436, 1648}

\bibitem[\protect\citeauthoryear{{Michel-Dansac}, {Lambas}, {Alonso}  \&
  {Tissera}}{{Michel-Dansac} et~al.}{2008}]{Michel-Dansac08}
{Michel-Dansac} L.,  {Lambas} D.~G.,  {Alonso} M.~S.,   {Tissera} P.,  2008,
  \mn@doi [\mnras] {10.1111/j.1745-3933.2008.00466.x}, \href
  {http://adsabs.harvard.edu/abs/2008MNRAS.386L..82M} {386, L82}

\bibitem[\protect\citeauthoryear{{Mitchell} et~al.,}{{Mitchell}
  et~al.}{2018}]{Mitchell18}
{Mitchell} P.~D.,  et~al., 2018, \mn@doi [\mnras] {10.1093/mnras/stx2770},
  \href {http://adsabs.harvard.edu/abs/2018MNRAS.474..492M} {474, 492}

\bibitem[\protect\citeauthoryear{{Molendi}, {Eckert}, {De Grandi}, {Ettori},
  {Gastaldello}, {Ghizzardi}, {Pratt}  \& {Rossetti}}{{Molendi}
  et~al.}{2016}]{Molendi16}
{Molendi} S.,  {Eckert} D.,  {De Grandi} S.,  {Ettori} S.,  {Gastaldello} F.,
  {Ghizzardi} S.,  {Pratt} G.~W.,   {Rossetti} M.,  2016, \mn@doi [\aap]
  {10.1051/0004-6361/201527356}, \href
  {http://adsabs.harvard.edu/abs/2016A%26A...586A..32M} {586, A32}

\bibitem[\protect\citeauthoryear{{Moll{\'a}}, {Ferrini}  \&
  {D{\'{\i}}az}}{{Moll{\'a}} et~al.}{1997}]{Molla97}
{Moll{\'a}} M.,  {Ferrini} F.,   {D{\'{\i}}az} A.~I.,  1997, \mn@doi [\apj]
  {10.1086/303550}, \href {http://adsabs.harvard.edu/abs/1997ApJ...475..519M}
  {475, 519}

\bibitem[\protect\citeauthoryear{{Moll{\'a}}, {Cavichia}, {Gavil{\'a}n}  \&
  {Gibson}}{{Moll{\'a}} et~al.}{2015}]{Molla15}
{Moll{\'a}} M.,  {Cavichia} O.,  {Gavil{\'a}n} M.,   {Gibson} B.~K.,  2015,
  \mn@doi [\mnras] {10.1093/mnras/stv1102}, \href
  {http://adsabs.harvard.edu/abs/2015MNRAS.451.3693M} {451, 3693}

\bibitem[\protect\citeauthoryear{{M{\o}ller}, {Fynbo}, {Ledoux}  \&
  {Nilsson}}{{M{\o}ller} et~al.}{2013}]{Moller13}
{M{\o}ller} P.,  {Fynbo} J.~P.~U.,  {Ledoux} C.,   {Nilsson} K.~K.,  2013,
  \mn@doi [\mnras] {10.1093/mnras/stt067}, \href
  {http://adsabs.harvard.edu/abs/2013MNRAS.430.2680M} {430, 2680}

\bibitem[\protect\citeauthoryear{{Monaco}, {Bellazzini}, {Bonifacio},
  {Ferraro}, {Marconi}, {Pancino}, {Sbordone}  \& {Zaggia}}{{Monaco}
  et~al.}{2005}]{Monaco05}
{Monaco} L.,  {Bellazzini} M.,  {Bonifacio} P.,  {Ferraro} F.~R.,  {Marconi}
  G.,  {Pancino} E.,  {Sbordone} L.,   {Zaggia} S.,  2005, \mn@doi [\aap]
  {10.1051/0004-6361:20053333}, \href
  {http://adsabs.harvard.edu/abs/2005A%26A...441..141M} {441, 141}

\bibitem[\protect\citeauthoryear{{Monreal-Ibero}, {Walsh}  \&
  {V{\'{\i}}lchez}}{{Monreal-Ibero} et~al.}{2012}]{Monreal-Ibero12}
{Monreal-Ibero} A.,  {Walsh} J.~R.,   {V{\'{\i}}lchez} J.~M.,  2012, \mn@doi
  [\aap] {10.1051/0004-6361/201219543}, \href
  {http://adsabs.harvard.edu/abs/2012A%26A...544A..60M} {544, A60}

\bibitem[\protect\citeauthoryear{{Moorwood}, {Salinari}, {Furniss}, {Jennings}
  \& {King}}{{Moorwood} et~al.}{1980a}]{Moorwood80b}
{Moorwood} A.~F.~M.,  {Salinari} P.,  {Furniss} I.,  {Jennings} R.~E.,   {King}
  K.~J.,  1980a, \aap, \href
  {http://adsabs.harvard.edu/abs/1980A%26A....90..304M} {90, 304}

\bibitem[\protect\citeauthoryear{{Moorwood}, {Baluteau}, {Anderegg}, {Coron},
  {Biraud}  \& {Fitton}}{{Moorwood} et~al.}{1980b}]{Moorwood80a}
{Moorwood} A.~F.~M.,  {Baluteau} J.~P.,  {Anderegg} M.,  {Coron} N.,  {Biraud}
  Y.,   {Fitton} B.,  1980b, \mn@doi [\apj] {10.1086/158013}, \href
  {http://adsabs.harvard.edu/abs/1980ApJ...238..565M} {238, 565}

\bibitem[\protect\citeauthoryear{{Morales-Luis}, {S{\'a}nchez Almeida},
  {Aguerri}  \& {Mu{\~n}oz-Tu{\~n}{\'o}n}}{{Morales-Luis}
  et~al.}{2011}]{Morales-Luis11}
{Morales-Luis} A.~B.,  {S{\'a}nchez Almeida} J.,  {Aguerri} J.~A.~L.,
  {Mu{\~n}oz-Tu{\~n}{\'o}n} C.,  2011, \mn@doi [\apj]
  {10.1088/0004-637X/743/1/77}, \href
  {http://adsabs.harvard.edu/abs/2011ApJ...743...77M} {743, 77}

\bibitem[\protect\citeauthoryear{{Moran} et~al.,}{{Moran}
  et~al.}{2012}]{Moran12}
{Moran} S.~M.,  et~al., 2012, \mn@doi [\apj] {10.1088/0004-637X/745/1/66},
  \href {http://adsabs.harvard.edu/abs/2012ApJ...745...66M} {745, 66}

\bibitem[\protect\citeauthoryear{{Morelli}, {Corsini}, {Pizzella}, {Dalla
  Bont{\`a}}, {Coccato}  \& {M{\'e}ndez-Abreu}}{{Morelli}
  et~al.}{2015}]{Morelli15}
{Morelli} L.,  {Corsini} E.~M.,  {Pizzella} A.,  {Dalla Bont{\`a}} E.,
  {Coccato} L.,   {M{\'e}ndez-Abreu} J.,  2015, \mn@doi [\mnras]
  {10.1093/mnras/stv1357}, \href
  {http://adsabs.harvard.edu/abs/2015MNRAS.452.1128M} {452, 1128}

\bibitem[\protect\citeauthoryear{{Mortlock} et~al.,}{{Mortlock}
  et~al.}{2011}]{Mortlock11}
{Mortlock} D.~J.,  et~al., 2011, \mn@doi [\nat] {10.1038/nature10159}, \href
  {http://adsabs.harvard.edu/abs/2011Natur.474..616M} {474, 616}

\bibitem[\protect\citeauthoryear{{Mott}, {Spitoni}  \& {Matteucci}}{{Mott}
  et~al.}{2013}]{Mott13}
{Mott} A.,  {Spitoni} E.,   {Matteucci} F.,  2013, \mn@doi [\mnras]
  {10.1093/mnras/stt1495}, \href
  {http://adsabs.harvard.edu/abs/2013MNRAS.435.2918M} {435, 2918}

\bibitem[\protect\citeauthoryear{{Mouhcine}, {Baldry}  \& {Bamford}}{{Mouhcine}
  et~al.}{2007}]{Mouhcine07}
{Mouhcine} M.,  {Baldry} I.~K.,   {Bamford} S.~P.,  2007, \mn@doi [\mnras]
  {10.1111/j.1365-2966.2007.12405.x}, \href
  {http://adsabs.harvard.edu/abs/2007MNRAS.382..801M} {382, 801}

\bibitem[\protect\citeauthoryear{{Mouhcine}, {Kriwattanawong}  \&
  {James}}{{Mouhcine} et~al.}{2011}]{Mouhcine11}
{Mouhcine} M.,  {Kriwattanawong} W.,   {James} P.~A.,  2011, \mn@doi [\mnras]
  {10.1111/j.1365-2966.2010.17993.x}, \href
  {http://adsabs.harvard.edu/abs/2011MNRAS.412.1295M} {412, 1295}

\bibitem[\protect\citeauthoryear{{Mould}}{{Mould}}{1984}]{Mould84}
{Mould} J.~R.,  1984, \mn@doi [\pasp] {10.1086/131418}, \href
  {http://adsabs.harvard.edu/abs/1984PASP...96..773M} {96, 773}

\bibitem[\protect\citeauthoryear{{Mould}, {Kristian}  \& {Da Costa}}{{Mould}
  et~al.}{1983}]{Mould83}
{Mould} J.~R.,  {Kristian} J.,   {Da Costa} G.~S.,  1983, \mn@doi [\apj]
  {10.1086/161141}, \href {http://adsabs.harvard.edu/abs/1983ApJ...270..471M}
  {270, 471}

\bibitem[\protect\citeauthoryear{{Moustakas}, {Kennicutt}, {Tremonti}, {Dale},
  {Smith}  \& {Calzetti}}{{Moustakas} et~al.}{2010}]{Moustakas10}
{Moustakas} J.,  {Kennicutt} R.~C.,  {Tremonti} C.~A.,  {Dale} D.~A.,  {Smith}
  J.,   {Calzetti} D.,  2010, \mn@doi [\apjs] {10.1088/0067-0049/190/2/233},
  \href {http://adsabs.harvard.edu/abs/2010ApJS..190..233M} {190, 233}

\bibitem[\protect\citeauthoryear{{Moustakas} et~al.,}{{Moustakas}
  et~al.}{2011}]{Moustakas11}
{Moustakas} J.,  et~al., 2011, preprint, \href
  {http://adsabs.harvard.edu/abs/2011arXiv1112.3300M} {} (\mn@eprint {}
  {1112.3300})

\bibitem[\protect\citeauthoryear{{Mushotzky} \& {Loewenstein}}{{Mushotzky} \&
  {Loewenstein}}{1997}]{Mushotzky97}
{Mushotzky} R.~F.,  {Loewenstein} M.,  1997, \mn@doi [\apjl] {10.1086/310651},
  \href {http://adsabs.harvard.edu/abs/1997ApJ...481L..63M} {481, L63}

\bibitem[\protect\citeauthoryear{{Mushotzky}, {Loewenstein}, {Arnaud},
  {Tamura}, {Fukazawa}, {Matsushita}, {Kikuchi}  \& {Hatsukade}}{{Mushotzky}
  et~al.}{1996}]{Mushotzky96}
{Mushotzky} R.,  {Loewenstein} M.,  {Arnaud} K.~A.,  {Tamura} T.,  {Fukazawa}
  Y.,  {Matsushita} K.,  {Kikuchi} K.,   {Hatsukade} I.,  1996, \mn@doi [\apj]
  {10.1086/177541}, \href {http://adsabs.harvard.edu/abs/1996ApJ...466..686M}
  {466, 686}

\bibitem[\protect\citeauthoryear{{Muzzin} et~al.,}{{Muzzin}
  et~al.}{2013}]{Muzzin13}
{Muzzin} A.,  et~al., 2013, \mn@doi [\apj] {10.1088/0004-637X/777/1/18}, \href
  {http://adsabs.harvard.edu/abs/2013ApJ...777...18M} {777, 18}

\bibitem[\protect\citeauthoryear{{Naab} \& {Ostriker}}{{Naab} \&
  {Ostriker}}{2006}]{Naab06}
{Naab} T.,  {Ostriker} J.~P.,  2006, \mn@doi [\mnras]
  {10.1111/j.1365-2966.2005.09807.x}, \href
  {http://adsabs.harvard.edu/abs/2006MNRAS.366..899N} {366, 899}

\bibitem[\protect\citeauthoryear{{Naab} \& {Ostriker}}{{Naab} \&
  {Ostriker}}{2017}]{Naab17}
{Naab} T.,  {Ostriker} J.~P.,  2017, \mn@doi [\araa]
  {10.1146/annurev-astro-081913-040019}, \href
  {http://adsabs.harvard.edu/abs/2017ARA%26A..55...59N} {55, 59}

\bibitem[\protect\citeauthoryear{{Naab}, {Johansson}, {Ostriker}  \&
  {Efstathiou}}{{Naab} et~al.}{2007}]{Naab07}
{Naab} T.,  {Johansson} P.~H.,  {Ostriker} J.~P.,   {Efstathiou} G.,  2007,
  \mn@doi [\apj] {10.1086/510841}, \href
  {http://adsabs.harvard.edu/abs/2007ApJ...658..710N} {658, 710}

\bibitem[\protect\citeauthoryear{{Naab}, {Johansson}  \& {Ostriker}}{{Naab}
  et~al.}{2009}]{Naab09}
{Naab} T.,  {Johansson} P.~H.,   {Ostriker} J.~P.,  2009, \mn@doi [\apjl]
  {10.1088/0004-637X/699/2/L178}, \href
  {http://adsabs.harvard.edu/abs/2009ApJ...699L.178N} {699, L178}

\bibitem[\protect\citeauthoryear{{Nagao}, {Marconi}  \& {Maiolino}}{{Nagao}
  et~al.}{2006a}]{Nagao06b}
{Nagao} T.,  {Marconi} A.,   {Maiolino} R.,  2006a, \mn@doi [\aap]
  {10.1051/0004-6361:20054024}, \href
  {http://adsabs.harvard.edu/abs/2006A%26A...447..157N} {447, 157}

\bibitem[\protect\citeauthoryear{{Nagao}, {Maiolino}  \& {Marconi}}{{Nagao}
  et~al.}{2006b}]{Nagao06c}
{Nagao} T.,  {Maiolino} R.,   {Marconi} A.,  2006b, \mn@doi [\aap]
  {10.1051/0004-6361:20054127}, \href
  {http://adsabs.harvard.edu/abs/2006A%26A...447..863N} {447, 863}

\bibitem[\protect\citeauthoryear{{Nagao}, {Maiolino}  \& {Marconi}}{{Nagao}
  et~al.}{2006c}]{Nagao06}
{Nagao} T.,  {Maiolino} R.,   {Marconi} A.,  2006c, \mn@doi [\aap]
  {10.1051/0004-6361:20065216}, \href
  {http://adsabs.harvard.edu/abs/2006A%26A...459...85N} {459, 85}

\bibitem[\protect\citeauthoryear{{Nagao}, {Maiolino}, {Marconi}  \&
  {Matsuhara}}{{Nagao} et~al.}{2011}]{Nagao11}
{Nagao} T.,  {Maiolino} R.,  {Marconi} A.,   {Matsuhara} H.,  2011, \mn@doi
  [\aap] {10.1051/0004-6361/201015471}, \href
  {http://adsabs.harvard.edu/abs/2011A%26A...526A.149N} {526, A149}

\bibitem[\protect\citeauthoryear{{Nagao}, {Maiolino}, {De Breuck}, {Caselli},
  {Hatsukade}  \& {Saigo}}{{Nagao} et~al.}{2012}]{Nagao12}
{Nagao} T.,  {Maiolino} R.,  {De Breuck} C.,  {Caselli} P.,  {Hatsukade} B.,
  {Saigo} K.,  2012, \mn@doi [\aap] {10.1051/0004-6361/201219518}, \href
  {http://adsabs.harvard.edu/abs/2012A%26A...542L..34N} {542, L34}

\bibitem[\protect\citeauthoryear{{Nakajima} \& {Ouchi}}{{Nakajima} \&
  {Ouchi}}{2014}]{Nakajima14}
{Nakajima} K.,  {Ouchi} M.,  2014, \mn@doi [\mnras] {10.1093/mnras/stu902},
  \href {http://adsabs.harvard.edu/abs/2014MNRAS.442..900N} {442, 900}

\bibitem[\protect\citeauthoryear{{Nakajima} et~al.,}{{Nakajima}
  et~al.}{2012}]{Nakajima12}
{Nakajima} K.,  et~al., 2012, \mn@doi [\apj] {10.1088/0004-637X/745/1/12},
  \href {http://adsabs.harvard.edu/abs/2012ApJ...745...12N} {745, 12}

\bibitem[\protect\citeauthoryear{{Nakajima}, {Ouchi}, {Shimasaku}, {Hashimoto},
  {Ono}  \& {Lee}}{{Nakajima} et~al.}{2013}]{Nakajima13}
{Nakajima} K.,  {Ouchi} M.,  {Shimasaku} K.,  {Hashimoto} T.,  {Ono} Y.,
  {Lee} J.~C.,  2013, \mn@doi [\apj] {10.1088/0004-637X/769/1/3}, \href
  {http://adsabs.harvard.edu/abs/2013ApJ...769....3N} {769, 3}

\bibitem[\protect\citeauthoryear{{Nakajima}, {Ellis}, {Iwata}, {Inoue},
  {Kusakabe}, {Ouchi}  \& {Robertson}}{{Nakajima} et~al.}{2016}]{Nakajima16}
{Nakajima} K.,  {Ellis} R.~S.,  {Iwata} I.,  {Inoue} A.~K.,  {Kusakabe} H.,
  {Ouchi} M.,   {Robertson} B.~E.,  2016, \mn@doi [\apjl]
  {10.3847/2041-8205/831/1/L9}, \href
  {http://adsabs.harvard.edu/abs/2016ApJ...831L...9N} {831, L9}

\bibitem[\protect\citeauthoryear{{Nakajima} et~al.,}{{Nakajima}
  et~al.}{2018}]{Nakajima18}
{Nakajima} K.,  et~al., 2018, \mn@doi [\aap] {10.1051/0004-6361/201731935},
  \href {http://adsabs.harvard.edu/abs/2018A%26A...612A..94N} {612, A94}

\bibitem[\protect\citeauthoryear{{Nardini}, {Wang}, {Fabbiano}, {Elvis},
  {Pellegrini}, {Risaliti}, {Karovska}  \& {Zezas}}{{Nardini}
  et~al.}{2013}]{Nardini13}
{Nardini} E.,  {Wang} J.,  {Fabbiano} G.,  {Elvis} M.,  {Pellegrini} S.,
  {Risaliti} G.,  {Karovska} M.,   {Zezas} A.,  2013, \mn@doi [\apj]
  {10.1088/0004-637X/765/2/141}, 765, 141

\bibitem[\protect\citeauthoryear{{Neeleman}, {Wolfe}, {Prochaska}  \&
  {Rafelski}}{{Neeleman} et~al.}{2013}]{Neeleman13}
{Neeleman} M.,  {Wolfe} A.~M.,  {Prochaska} J.~X.,   {Rafelski} M.,  2013,
  \mn@doi [\apj] {10.1088/0004-637X/769/1/54}, \href
  {http://adsabs.harvard.edu/abs/2013ApJ...769...54N} {769, 54}

\bibitem[\protect\citeauthoryear{{Nelan}, {Smith}, {Hudson}, {Wegner}, {Lucey},
  {Moore}, {Quinney}  \& {Suntzeff}}{{Nelan} et~al.}{2005}]{Nelan05}
{Nelan} J.~E.,  {Smith} R.~J.,  {Hudson} M.~J.,  {Wegner} G.~A.,  {Lucey}
  J.~R.,  {Moore} S.~A.~W.,  {Quinney} S.~J.,   {Suntzeff} N.~B.,  2005,
  \mn@doi [\apj] {10.1086/431962}, \href
  {http://adsabs.harvard.edu/abs/2005ApJ...632..137N} {632, 137}

\bibitem[\protect\citeauthoryear{{Newman} et~al.,}{{Newman}
  et~al.}{2014}]{Newman14}
{Newman} S.~F.,  et~al., 2014, \mn@doi [\apj] {10.1088/0004-637X/781/1/21},
  \href {http://adsabs.harvard.edu/abs/2014ApJ...781...21N} {781, 21}

\bibitem[\protect\citeauthoryear{{Nicastro}, {Krongold}, {Mathur}  \&
  {Elvis}}{{Nicastro} et~al.}{2017}]{Nicastro17}
{Nicastro} F.,  {Krongold} Y.,  {Mathur} S.,   {Elvis} M.,  2017, \mn@doi
  [Astronomische Nachrichten] {10.1002/asna.201713343}, \href
  {http://adsabs.harvard.edu/abs/2017AN....338..281N} {338, 281}

\bibitem[\protect\citeauthoryear{{Nicastro} et~al.,}{{Nicastro}
  et~al.}{2018}]{Nicastro18}
{Nicastro} F.,  et~al., 2018, \mn@doi [\nat] {10.1038/s41586-018-0204-1}, \href
  {http://adsabs.harvard.edu/abs/2018Natur.558..406N} {558, 406}

\bibitem[\protect\citeauthoryear{{Nicholls}, {Dopita}  \&
  {Sutherland}}{{Nicholls} et~al.}{2012}]{Nicholls12}
{Nicholls} D.~C.,  {Dopita} M.~A.,   {Sutherland} R.~S.,  2012, \mn@doi [\apj]
  {10.1088/0004-637X/752/2/148}, \href
  {http://adsabs.harvard.edu/abs/2012ApJ...752..148N} {752, 148}

\bibitem[\protect\citeauthoryear{{Nicholls}, {Sutherland}, {Dopita}, {Kewley}
  \& {Groves}}{{Nicholls} et~al.}{2017}]{Nicholls17}
{Nicholls} D.~C.,  {Sutherland} R.~S.,  {Dopita} M.~A.,  {Kewley} L.~J.,
  {Groves} B.~A.,  2017, \mn@doi [\mnras] {10.1093/mnras/stw3235}, \href
  {http://adsabs.harvard.edu/abs/2017MNRAS.466.4403N} {466, 4403}

\bibitem[\protect\citeauthoryear{{Nieva} \& {Przybilla}}{{Nieva} \&
  {Przybilla}}{2012}]{Nieva12}
{Nieva} M.-F.,  {Przybilla} N.,  2012, \mn@doi [\aap]
  {10.1051/0004-6361/201118158}, \href
  {http://adsabs.harvard.edu/abs/2012A%26A...539A.143N} {539, A143}

\bibitem[\protect\citeauthoryear{{Niino}}{{Niino}}{2012}]{Niino12b}
{Niino} Y.,  2012, \mn@doi [\apj] {10.1088/0004-637X/761/2/126}, \href
  {http://adsabs.harvard.edu/abs/2012ApJ...761..126N} {761, 126}

\bibitem[\protect\citeauthoryear{{Nissen} \& {Schuster}}{{Nissen} \&
  {Schuster}}{2010}]{Nissen10}
{Nissen} P.~E.,  {Schuster} W.~J.,  2010, \mn@doi [\aap]
  {10.1051/0004-6361/200913877}, \href
  {http://adsabs.harvard.edu/abs/2010A%26A...511L..10N} {511, L10}

\bibitem[\protect\citeauthoryear{{Nissen}, {Chen}, {Carigi}, {Schuster}  \&
  {Zhao}}{{Nissen} et~al.}{2014}]{Nissen14}
{Nissen} P.~E.,  {Chen} Y.~Q.,  {Carigi} L.,  {Schuster} W.~J.,   {Zhao} G.,
  2014, \mn@doi [\aap] {10.1051/0004-6361/201424184}, \href
  {http://adsabs.harvard.edu/abs/2014A%26A...568A..25N} {568, A25}

\bibitem[\protect\citeauthoryear{{Nomoto}, {Kobayashi}  \& {Tominaga}}{{Nomoto}
  et~al.}{2013}]{Nomoto13}
{Nomoto} K.,  {Kobayashi} C.,   {Tominaga} N.,  2013, \mn@doi [\araa]
  {10.1146/annurev-astro-082812-140956}, \href
  {http://adsabs.harvard.edu/abs/2013ARA%26A..51..457N} {51, 457}

\bibitem[\protect\citeauthoryear{{Noterdaeme}, {Ledoux}, {Petitjean}  \&
  {Srianand}}{{Noterdaeme} et~al.}{2008}]{Noterdaeme08}
{Noterdaeme} P.,  {Ledoux} C.,  {Petitjean} P.,   {Srianand} R.,  2008, \mn@doi
  [\aap] {10.1051/0004-6361:20078780}, \href
  {http://adsabs.harvard.edu/abs/2008A%26A...481..327N} {481, 327}

\bibitem[\protect\citeauthoryear{{Noterdaeme} et~al.,}{{Noterdaeme}
  et~al.}{2012}]{Noterdaeme12}
{Noterdaeme} P.,  et~al., 2012, \mn@doi [\aap] {10.1051/0004-6361/201118691},
  \href {http://adsabs.harvard.edu/abs/2012A%26A...540A..63N} {540, A63}

\bibitem[\protect\citeauthoryear{{Oey} et~al.,}{{Oey} et~al.}{2007}]{Oey07}
{Oey} M.~S.,  et~al., 2007, \mn@doi [\apj] {10.1086/517867}, \href
  {http://adsabs.harvard.edu/abs/2007ApJ...661..801O} {661, 801}

\bibitem[\protect\citeauthoryear{{Olave-Rojas}, {Torres-Flores}, {Carrasco},
  {Mendes de Oliveira}, {de Mello}  \& {Scarano}}{{Olave-Rojas}
  et~al.}{2015}]{Olave-Rojas15}
{Olave-Rojas} D.,  {Torres-Flores} S.,  {Carrasco} E.~R.,  {Mendes de Oliveira}
  C.,  {de Mello} D.~F.,   {Scarano} S.,  2015, \mn@doi [\mnras]
  {10.1093/mnras/stv1798}, \href
  {http://adsabs.harvard.edu/abs/2015MNRAS.453.2808O} {453, 2808}

\bibitem[\protect\citeauthoryear{{{\"O}nehag}, {Heiter}, {Gustafsson},
  {Piskunov}, {Plez}  \& {Reiners}}{{{\"O}nehag} et~al.}{2012}]{Onehag12}
{{\"O}nehag} A.,  {Heiter} U.,  {Gustafsson} B.,  {Piskunov} N.,  {Plez} B.,
  {Reiners} A.,  2012, \mn@doi [\aap] {10.1051/0004-6361/201118101}, \href
  {http://adsabs.harvard.edu/abs/2012A%26A...542A..33O} {542, A33}

\bibitem[\protect\citeauthoryear{{Onodera} et~al.,}{{Onodera}
  et~al.}{2015}]{Onodera15}
{Onodera} M.,  et~al., 2015, \mn@doi [\apj] {10.1088/0004-637X/808/2/161},
  \href {http://adsabs.harvard.edu/abs/2015ApJ...808..161O} {808, 161}

\bibitem[\protect\citeauthoryear{{Onodera} et~al.,}{{Onodera}
  et~al.}{2016}]{Onodera16}
{Onodera} M.,  et~al., 2016, \mn@doi [\apj] {10.3847/0004-637X/822/1/42}, \href
  {http://adsabs.harvard.edu/abs/2016ApJ...822...42O} {822, 42}

\bibitem[\protect\citeauthoryear{{Oppenheimer} \& {Dav{\'e}}}{{Oppenheimer} \&
  {Dav{\'e}}}{2008}]{Oppenheimer08}
{Oppenheimer} B.~D.,  {Dav{\'e}} R.,  2008, \mn@doi [\mnras]
  {10.1111/j.1365-2966.2008.13280.x}, \href
  {http://adsabs.harvard.edu/abs/2008MNRAS.387..577O} {387, 577}

\bibitem[\protect\citeauthoryear{{Oppenheimer}, {Dav{\'e}}, {Kere{\v s}},
  {Fardal}, {Katz}, {Kollmeier}  \& {Weinberg}}{{Oppenheimer}
  et~al.}{2010}]{Oppenheimer10}
{Oppenheimer} B.~D.,  {Dav{\'e}} R.,  {Kere{\v s}} D.,  {Fardal} M.,  {Katz}
  N.,  {Kollmeier} J.~A.,   {Weinberg} D.~H.,  2010, \mn@doi [\mnras]
  {10.1111/j.1365-2966.2010.16872.x}, \href
  {http://adsabs.harvard.edu/abs/2010MNRAS.406.2325O} {406, 2325}

\bibitem[\protect\citeauthoryear{{Oser}, {Ostriker}, {Naab}, {Johansson}  \&
  {Burkert}}{{Oser} et~al.}{2010}]{Oser10}
{Oser} L.,  {Ostriker} J.~P.,  {Naab} T.,  {Johansson} P.~H.,   {Burkert} A.,
  2010, \mn@doi [\apj] {10.1088/0004-637X/725/2/2312}, \href
  {http://adsabs.harvard.edu/abs/2010ApJ...725.2312O} {725, 2312}

\bibitem[\protect\citeauthoryear{{Pagel}}{{Pagel}}{2002}]{Pagel02}
{Pagel} B.~E.~J.,  2002, in {Fusco-Femiano} R.,  {Matteucci} F.,  eds,
  Astronomical Society of the Pacific Conference Series Vol. 253, Chemical
  Enrichment of Intracluster and Intergalactic Medium. p.~489 (\mn@eprint {}
  {astro-ph/0107215})

\bibitem[\protect\citeauthoryear{{Pagel} \& {Edmunds}}{{Pagel} \&
  {Edmunds}}{1981}]{Pagel81}
{Pagel} B.~E.~J.,  {Edmunds} M.~G.,  1981, \mn@doi [\araa]
  {10.1146/annurev.aa.19.090181.000453}, \href
  {http://adsabs.harvard.edu/abs/1981ARA%26A..19...77P} {19, 77}

\bibitem[\protect\citeauthoryear{{Pallottini}, {Ferrara}, {Gallerani},
  {Salvadori}  \& {D'Odorico}}{{Pallottini} et~al.}{2014}]{Pallottini14a}
{Pallottini} A.,  {Ferrara} A.,  {Gallerani} S.,  {Salvadori} S.,   {D'Odorico}
  V.,  2014, \mn@doi [\mnras] {10.1093/mnras/stu451}, \href
  {http://adsabs.harvard.edu/abs/2014MNRAS.440.2498P} {440, 2498}

\bibitem[\protect\citeauthoryear{{Pallottini}, {Ferrara}, {Gallerani},
  {Vallini}, {Maiolino}  \& {Salvadori}}{{Pallottini}
  et~al.}{2017}]{Pallottini17a}
{Pallottini} A.,  {Ferrara} A.,  {Gallerani} S.,  {Vallini} L.,  {Maiolino} R.,
    {Salvadori} S.,  2017, \mn@doi [\mnras] {10.1093/mnras/stw2847}, \href
  {http://adsabs.harvard.edu/abs/2017MNRAS.465.2540P} {465, 2540}

\bibitem[\protect\citeauthoryear{{Panter}, {Jimenez}, {Heavens}  \&
  {Charlot}}{{Panter} et~al.}{2008}]{Panter08}
{Panter} B.,  {Jimenez} R.,  {Heavens} A.~F.,   {Charlot} S.,  2008, \mn@doi
  [\mnras] {10.1111/j.1365-2966.2008.13981.x}, \href
  {http://adsabs.harvard.edu/abs/2008MNRAS.391.1117P} {391, 1117}

\bibitem[\protect\citeauthoryear{{Papastergis}, {Cattaneo}, {Huang},
  {Giovanelli}  \& {Haynes}}{{Papastergis} et~al.}{2012}]{Papastergis12}
{Papastergis} E.,  {Cattaneo} A.,  {Huang} S.,  {Giovanelli} R.,   {Haynes}
  M.~P.,  2012, \mn@doi [\apj] {10.1088/0004-637X/759/2/138}, \href
  {http://adsabs.harvard.edu/abs/2012ApJ...759..138P} {759, 138}

\bibitem[\protect\citeauthoryear{{Pasquali}, {Gallazzi}, {Fontanot}, {van den
  Bosch}, {De Lucia}, {Mo}  \& {Yang}}{{Pasquali} et~al.}{2010}]{Pasquali10}
{Pasquali} A.,  {Gallazzi} A.,  {Fontanot} F.,  {van den Bosch} F.~C.,  {De
  Lucia} G.,  {Mo} H.~J.,   {Yang} X.,  2010, \mn@doi [\mnras]
  {10.1111/j.1365-2966.2010.17074.x}, \href
  {http://adsabs.harvard.edu/abs/2010MNRAS.407..937P} {407, 937}

\bibitem[\protect\citeauthoryear{{Pasquali}, {Gallazzi}  \& {van den
  Bosch}}{{Pasquali} et~al.}{2012}]{Pasquali12}
{Pasquali} A.,  {Gallazzi} A.,   {van den Bosch} F.~C.,  2012, \mn@doi [\mnras]
  {10.1111/j.1365-2966.2012.21454.x}, \href
  {http://adsabs.harvard.edu/abs/2012MNRAS.425..273P} {425, 273}

\bibitem[\protect\citeauthoryear{{Patr{\'{\i}}cio}, {Christensen}, {Rhodin},
  {Ca{\~n}ameras}  \& {Lara-L{\'o}pez}}{{Patr{\'{\i}}cio}
  et~al.}{2018}]{Patricio18}
{Patr{\'{\i}}cio} V.,  {Christensen} L.,  {Rhodin} H.,  {Ca{\~n}ameras} R.,
  {Lara-L{\'o}pez} M.~A.,  2018, preprint, \href
  {http://adsabs.harvard.edu/abs/2018arXiv180903612P} {} (\mn@eprint {arXiv}
  {1809.03612})

\bibitem[\protect\citeauthoryear{{Patrick}, {Evans}, {Davies}, {Kudritzki},
  {H{\'e}nault-Brunet}, {Bastian}, {Lapenna}  \& {Bergemann}}{{Patrick}
  et~al.}{2016}]{Patrick16}
{Patrick} L.~R.,  {Evans} C.~J.,  {Davies} B.,  {Kudritzki} R.-P.,
  {H{\'e}nault-Brunet} V.,  {Bastian} N.,  {Lapenna} E.,   {Bergemann} M.,
  2016, \mn@doi [\mnras] {10.1093/mnras/stw561}, \href
  {http://adsabs.harvard.edu/abs/2016MNRAS.458.3968P} {458, 3968}

\bibitem[\protect\citeauthoryear{{Pe{\~n}a-Guerrero}, {Leitherer}, {de Mink},
  {Wofford}  \& {Kewley}}{{Pe{\~n}a-Guerrero} et~al.}{2017}]{Pena-Guerrero17}
{Pe{\~n}a-Guerrero} M.~A.,  {Leitherer} C.,  {de Mink} S.,  {Wofford} A.,
  {Kewley} L.,  2017, \mn@doi [\apj] {10.3847/1538-4357/aa88bf}, \href
  {http://adsabs.harvard.edu/abs/2017ApJ...847..107P} {847, 107}

\bibitem[\protect\citeauthoryear{{Peeples} \& {Shankar}}{{Peeples} \&
  {Shankar}}{2011}]{Peeples11}
{Peeples} M.~S.,  {Shankar} F.,  2011, \mn@doi [\mnras]
  {10.1111/j.1365-2966.2011.19456.x}, \href
  {http://adsabs.harvard.edu/abs/2011MNRAS.417.2962P} {417, 2962}

\bibitem[\protect\citeauthoryear{{Peeples}, {Pogge}  \& {Stanek}}{{Peeples}
  et~al.}{2008}]{Peeples08}
{Peeples} M.~S.,  {Pogge} R.~W.,   {Stanek} K.~Z.,  2008, \mn@doi [\apj]
  {10.1086/591492}, \href {http://adsabs.harvard.edu/abs/2008ApJ...685..904P}
  {685, 904}

\bibitem[\protect\citeauthoryear{{Peeples}, {Pogge}  \& {Stanek}}{{Peeples}
  et~al.}{2009}]{Peeples09}
{Peeples} M.~S.,  {Pogge} R.~W.,   {Stanek} K.~Z.,  2009, \mn@doi [\apj]
  {10.1088/0004-637X/695/1/259}, \href
  {http://adsabs.harvard.edu/abs/2009ApJ...695..259P} {695, 259}

\bibitem[\protect\citeauthoryear{{Peeples}, {Werk}, {Tumlinson}, {Oppenheimer},
  {Prochaska}, {Katz}  \& {Weinberg}}{{Peeples} et~al.}{2014}]{Peeples14}
{Peeples} M.~S.,  {Werk} J.~K.,  {Tumlinson} J.,  {Oppenheimer} B.~D.,
  {Prochaska} J.~X.,  {Katz} N.,   {Weinberg} D.~H.,  2014, \mn@doi [\apj]
  {10.1088/0004-637X/786/1/54}, \href
  {http://adsabs.harvard.edu/abs/2014ApJ...786...54P} {786, 54}

\bibitem[\protect\citeauthoryear{{Pei} \& {Fall}}{{Pei} \&
  {Fall}}{1995}]{Pei95}
{Pei} Y.~C.,  {Fall} S.~M.,  1995, \mn@doi [\apj] {10.1086/176466}, \href
  {http://adsabs.harvard.edu/abs/1995ApJ...454...69P} {454, 69}

\bibitem[\protect\citeauthoryear{{Peimbert}}{{Peimbert}}{1967}]{Peimbert67}
{Peimbert} M.,  1967, \mn@doi [\apj] {10.1086/149385}, \href
  {http://adsabs.harvard.edu/abs/1967ApJ...150..825P} {150, 825}

\bibitem[\protect\citeauthoryear{{Peimbert}}{{Peimbert}}{2003}]{Peimbert03}
{Peimbert} A.,  2003, \mn@doi [\apj] {10.1086/345793}, \href
  {http://adsabs.harvard.edu/abs/2003ApJ...584..735P} {584, 735}

\bibitem[\protect\citeauthoryear{{Peimbert} \& {Peimbert}}{{Peimbert} \&
  {Peimbert}}{2014}]{Peimbert14a}
{Peimbert} M.,  {Peimbert} A.,  2014, in Revista Mexicana de Astronomia y
  Astrofisica Conference Series. pp 137--137 (\mn@eprint {arXiv} {1310.0089}),
  \mn@doi{10.1088/0004-637X/778/2/89}

\bibitem[\protect\citeauthoryear{{Peimbert} \& {Spinrad}}{{Peimbert} \&
  {Spinrad}}{1970}]{Peimbert70}
{Peimbert} M.,  {Spinrad} H.,  1970, \aap, \href
  {http://adsabs.harvard.edu/abs/1970A%26A.....7..311P} {7, 311}

\bibitem[\protect\citeauthoryear{{Peimbert}, {Peimbert}, {Ruiz}  \&
  {Esteban}}{{Peimbert} et~al.}{2004}]{Peimbert04}
{Peimbert} M.,  {Peimbert} A.,  {Ruiz} M.~T.,   {Esteban} C.,  2004, \mn@doi
  [\apjs] {10.1086/381090}, \href
  {http://adsabs.harvard.edu/abs/2004ApJS..150..431P} {150, 431}

\bibitem[\protect\citeauthoryear{{Peimbert}, {Peimbert}  \& {Ruiz}}{{Peimbert}
  et~al.}{2005}]{Peimbert05}
{Peimbert} A.,  {Peimbert} M.,   {Ruiz} M.~T.,  2005, \mn@doi [\apj]
  {10.1086/444557}, \href {http://adsabs.harvard.edu/abs/2005ApJ...634.1056P}
  {634, 1056}

\bibitem[\protect\citeauthoryear{{Peimbert}, {Peimbert}, {Esteban},
  {Garc{\'{\i}}a-Rojas}, {Bresolin}, {Carigi}, {Ruiz}  \&
  {L{\'o}pez-S{\'a}nchez}}{{Peimbert} et~al.}{2007}]{Peimbert07}
{Peimbert} M.,  {Peimbert} A.,  {Esteban} C.,  {Garc{\'{\i}}a-Rojas} J.,
  {Bresolin} F.,  {Carigi} L.,  {Ruiz} M.~T.,   {L{\'o}pez-S{\'a}nchez} A.~R.,
  2007, in {Guzm{\'a}n} R.,  ed.,  Revista Mexicana de Astronomia y Astrofisica
  Conference Series Vol. 29, Revista Mexicana de Astronomia y Astrofisica
  Conference Series. pp 72--79 (\mn@eprint {} {astro-ph/0608440})

\bibitem[\protect\citeauthoryear{{Peimbert}, {Pe{\~n}a-Guerrero}  \&
  {Peimbert}}{{Peimbert} et~al.}{2012}]{Peimbert12}
{Peimbert} A.,  {Pe{\~n}a-Guerrero} M.~A.,   {Peimbert} M.,  2012, \mn@doi
  [\apj] {10.1088/0004-637X/753/1/39}, \href
  {http://adsabs.harvard.edu/abs/2012ApJ...753...39P} {753, 39}

\bibitem[\protect\citeauthoryear{{Peimbert}, {Peimbert}, {Delgado-Inglada},
  {Garc{\'{\i}}a-Rojas}  \& {Pe{\~n}a}}{{Peimbert} et~al.}{2014}]{Peimbert14b}
{Peimbert} A.,  {Peimbert} M.,  {Delgado-Inglada} G.,  {Garc{\'{\i}}a-Rojas}
  J.,   {Pe{\~n}a} M.,  2014, \rmxaa, \href
  {http://adsabs.harvard.edu/abs/2014RMxAA..50..329P} {50, 329}

\bibitem[\protect\citeauthoryear{{Peimbert}, {Peimbert}  \&
  {Delgado-Inglada}}{{Peimbert} et~al.}{2017}]{Peimbert17}
{Peimbert} M.,  {Peimbert} A.,   {Delgado-Inglada} G.,  2017, \mn@doi [\pasp]
  {10.1088/1538-3873/aa72c3}, \href
  {http://adsabs.harvard.edu/abs/2017PASP..129h2001P} {129, 082001}

\bibitem[\protect\citeauthoryear{{Peng} \& {Maiolino}}{{Peng} \&
  {Maiolino}}{2014a}]{Peng14}
{Peng} Y.-j.,  {Maiolino} R.,  2014a, \mn@doi [\mnras] {10.1093/mnras/stt2175},
  \href {http://adsabs.harvard.edu/abs/2014MNRAS.438..262P} {438, 262}

\bibitem[\protect\citeauthoryear{{Peng} \& {Maiolino}}{{Peng} \&
  {Maiolino}}{2014b}]{Peng14b}
{Peng} Y.-j.,  {Maiolino} R.,  2014b, \mn@doi [\mnras] {10.1093/mnras/stu1288},
  \href {http://adsabs.harvard.edu/abs/2014MNRAS.443.3643P} {443, 3643}

\bibitem[\protect\citeauthoryear{{Peng} et~al.,}{{Peng} et~al.}{2010}]{Peng10}
{Peng} Y.-j.,  et~al., 2010, \mn@doi [\apj] {10.1088/0004-637X/721/1/193},
  \href {http://adsabs.harvard.edu/abs/2010ApJ...721..193P} {721, 193}

\bibitem[\protect\citeauthoryear{{Peng}, {Lilly}, {Renzini}  \&
  {Carollo}}{{Peng} et~al.}{2012}]{Peng12}
{Peng} Y.-j.,  {Lilly} S.~J.,  {Renzini} A.,   {Carollo} M.,  2012, \mn@doi
  [\apj] {10.1088/0004-637X/757/1/4}, \href
  {http://adsabs.harvard.edu/abs/2012ApJ...757....4P} {757, 4}

\bibitem[\protect\citeauthoryear{{Peng}, {Maiolino}  \& {Cochrane}}{{Peng}
  et~al.}{2015}]{Peng15}
{Peng} Y.,  {Maiolino} R.,   {Cochrane} R.,  2015, \mn@doi [\nat]
  {10.1038/nature14439}, \href
  {http://adsabs.harvard.edu/abs/2015Natur.521..192P} {521, 192}

\bibitem[\protect\citeauthoryear{{Pentericci} et~al.,}{{Pentericci}
  et~al.}{2016}]{Pentericci16}
{Pentericci} L.,  et~al., 2016, \mn@doi [\apjl] {10.3847/2041-8205/829/1/L11},
  \href {http://adsabs.harvard.edu/abs/2016ApJ...829L..11P} {829, L11}

\bibitem[\protect\citeauthoryear{{Pereira-Santaella}, {Rigopoulou}, {Farrah},
  {Lebouteiller}  \& {Li}}{{Pereira-Santaella}
  et~al.}{2017}]{Pereira-Santaella17}
{Pereira-Santaella} M.,  {Rigopoulou} D.,  {Farrah} D.,  {Lebouteiller} V.,
  {Li} J.,  2017, \mn@doi [\mnras] {10.1093/mnras/stx1284}, 470, 1218

\bibitem[\protect\citeauthoryear{{P{\'e}rez-Gonz{\'a}lez}
  et~al.,}{{P{\'e}rez-Gonz{\'a}lez} et~al.}{2008}]{Perez-Gonzalez08}
{P{\'e}rez-Gonz{\'a}lez} P.~G.,  et~al., 2008, \mn@doi [\apj] {10.1086/523690},
  \href {http://adsabs.harvard.edu/abs/2008ApJ...675..234P} {675, 234}

\bibitem[\protect\citeauthoryear{{Perez-Martinez}}{{Perez-Martinez}}{2014}]{Perez-Martinez14}
{Perez-Martinez} J.~M.,  2014, preprint, \href
  {http://adsabs.harvard.edu/abs/2014arXiv1412.3853P} {} (\mn@eprint {}
  {1412.3853})

\bibitem[\protect\citeauthoryear{{P{\'e}rez-Montero}}{{P{\'e}rez-Montero}}{2014}]{Perez-Montero14}
{P{\'e}rez-Montero} E.,  2014, \mn@doi [\mnras] {10.1093/mnras/stu753}, \href
  {http://adsabs.harvard.edu/abs/2014MNRAS.441.2663P} {441, 2663}

\bibitem[\protect\citeauthoryear{{P{\'e}rez-Montero}}{{P{\'e}rez-Montero}}{2017}]{Perez-Montero17b}
{P{\'e}rez-Montero} E.,  2017, \pasp, \href
  {http://adsabs.harvard.edu/abs/2017PASP..129d3001P} {129, 043001}

\bibitem[\protect\citeauthoryear{{P{\'e}rez-Montero} \&
  {Amor{\'{\i}}n}}{{P{\'e}rez-Montero} \&
  {Amor{\'{\i}}n}}{2017}]{Perez-Montero17}
{P{\'e}rez-Montero} E.,  {Amor{\'{\i}}n} R.,  2017, \mn@doi [\mnras]
  {10.1093/mnras/stx186}, \href
  {http://adsabs.harvard.edu/abs/2017MNRAS.467.1287P} {467, 1287}

\bibitem[\protect\citeauthoryear{{P{\'e}rez-Montero} \&
  {Contini}}{{P{\'e}rez-Montero} \& {Contini}}{2009}]{Perez-Montero09}
{P{\'e}rez-Montero} E.,  {Contini} T.,  2009, \mn@doi [\mnras]
  {10.1111/j.1365-2966.2009.15145.x}, \href
  {http://adsabs.harvard.edu/abs/2009MNRAS.398..949P} {398, 949}

\bibitem[\protect\citeauthoryear{{P{\'e}rez-Montero} \&
  {D{\'{\i}}az}}{{P{\'e}rez-Montero} \& {D{\'{\i}}az}}{2003}]{Perez-Montero03}
{P{\'e}rez-Montero} E.,  {D{\'{\i}}az} A.~I.,  2003, \mn@doi [\mnras]
  {10.1046/j.1365-2966.2003.07064.x}, \href
  {http://adsabs.harvard.edu/abs/2003MNRAS.346..105P} {346, 105}

\bibitem[\protect\citeauthoryear{{P{\'e}rez-Montero} \&
  {D{\'{\i}}az}}{{P{\'e}rez-Montero} \& {D{\'{\i}}az}}{2005}]{Perez-Montero05}
{P{\'e}rez-Montero} E.,  {D{\'{\i}}az} A.~I.,  2005, \mn@doi [\mnras]
  {10.1111/j.1365-2966.2005.09263.x}, \href
  {http://adsabs.harvard.edu/abs/2005MNRAS.361.1063P} {361, 1063}

\bibitem[\protect\citeauthoryear{{P{\'e}rez-Montero}, {H{\"a}gele}, {Contini}
  \& {D{\'{\i}}az}}{{P{\'e}rez-Montero} et~al.}{2007}]{Perez-Montero07}
{P{\'e}rez-Montero} E.,  {H{\"a}gele} G.~F.,  {Contini} T.,   {D{\'{\i}}az}
  {\'A}.~I.,  2007, \mn@doi [\mnras] {10.1111/j.1365-2966.2007.12213.x}, \href
  {http://adsabs.harvard.edu/abs/2007MNRAS.381..125P} {381, 125}

\bibitem[\protect\citeauthoryear{{P{\'e}rez-Montero}
  et~al.,}{{P{\'e}rez-Montero} et~al.}{2011}]{Perez-Montero11}
{P{\'e}rez-Montero} E.,  et~al., 2011, \mn@doi [\aap]
  {10.1051/0004-6361/201116582}, \href
  {http://adsabs.harvard.edu/abs/2011A%26A...532A.141P} {532, A141}

\bibitem[\protect\citeauthoryear{{P{\'e}rez-Montero}
  et~al.,}{{P{\'e}rez-Montero} et~al.}{2013}]{Perez-Montero13}
{P{\'e}rez-Montero} E.,  et~al., 2013, \mn@doi [\aap]
  {10.1051/0004-6361/201220070}, \href
  {http://adsabs.harvard.edu/abs/2013A%26A...549A..25P} {549, A25}

\bibitem[\protect\citeauthoryear{{P{\'e}rez-Montero}
  et~al.,}{{P{\'e}rez-Montero} et~al.}{2016}]{Perez-Montero16}
{P{\'e}rez-Montero} E.,  et~al., 2016, \mn@doi [\aap]
  {10.1051/0004-6361/201628601}, \href
  {http://adsabs.harvard.edu/abs/2016A%26A...595A..62P} {595, A62}

\bibitem[\protect\citeauthoryear{{Perez}, {Michel-Dansac}  \&
  {Tissera}}{{Perez} et~al.}{2011}]{Perez11}
{Perez} J.,  {Michel-Dansac} L.,   {Tissera} P.~B.,  2011, \mn@doi [\mnras]
  {10.1111/j.1365-2966.2011.19300.x}, \href
  {http://adsabs.harvard.edu/abs/2011MNRAS.417..580P} {417, 580}

\bibitem[\protect\citeauthoryear{{P{\'e}rez}, {Hoyos}, {D{\'{\i}}az}, {Koo}  \&
  {Willmer}}{{P{\'e}rez} et~al.}{2016}]{Perez16}
{P{\'e}rez} J.~M.,  {Hoyos} C.,  {D{\'{\i}}az} {\'A}.~I.,  {Koo} D.~C.,
  {Willmer} C.~N.~A.,  2016, \mn@doi [\mnras] {10.1093/mnras/stv1949}, \href
  {http://adsabs.harvard.edu/abs/2016MNRAS.455.3359P} {455, 3359}

\bibitem[\protect\citeauthoryear{{Perna} et~al.,}{{Perna}
  et~al.}{2018}]{Perna18}
{Perna} M.,  et~al., 2018, \mn@doi [\aap] {10.1051/0004-6361/201732387}, \href
  {http://adsabs.harvard.edu/abs/2018A%26A...618A..36P} {618, A36}

\bibitem[\protect\citeauthoryear{{P{\'e}roux}, {Bouch{\'e}}, {Kulkarni}, {York}
   \& {Vladilo}}{{P{\'e}roux} et~al.}{2011}]{Peroux11}
{P{\'e}roux} C.,  {Bouch{\'e}} N.,  {Kulkarni} V.~P.,  {York} D.~G.,
  {Vladilo} G.,  2011, \mn@doi [\mnras] {10.1111/j.1365-2966.2010.17598.x},
  \href {http://adsabs.harvard.edu/abs/2011MNRAS.410.2237P} {410, 2237}

\bibitem[\protect\citeauthoryear{{Petropoulou}, {V{\'{\i}}lchez},
  {Iglesias-P{\'a}ramo}, {Papaderos}, {Magrini}, {Cedr{\'e}s}  \&
  {Reverte}}{{Petropoulou} et~al.}{2011}]{Petropoulou11}
{Petropoulou} V.,  {V{\'{\i}}lchez} J.,  {Iglesias-P{\'a}ramo} J.,  {Papaderos}
  P.,  {Magrini} L.,  {Cedr{\'e}s} B.,   {Reverte} D.,  2011, \mn@doi [\apj]
  {10.1088/0004-637X/734/1/32}, \href
  {http://adsabs.harvard.edu/abs/2011ApJ...734...32P} {734, 32}

\bibitem[\protect\citeauthoryear{{Petropoulou}, {V{\'{\i}}lchez}  \&
  {Iglesias-P{\'a}ramo}}{{Petropoulou} et~al.}{2012}]{Petropoulou12}
{Petropoulou} V.,  {V{\'{\i}}lchez} J.,   {Iglesias-P{\'a}ramo} J.,  2012,
  \mn@doi [\apj] {10.1088/0004-637X/749/2/133}, \href
  {http://adsabs.harvard.edu/abs/2012ApJ...749..133P} {749, 133}

\bibitem[\protect\citeauthoryear{{Pettini}}{{Pettini}}{2004}]{Pettini04b}
{Pettini} M.,  2004, in {Esteban} C.,  {Garc{\'{\i}}a L{\'o}pez} R.,  {Herrero}
  A.,   {S{\'a}nchez} F.,  eds, Cosmochemistry. The melting pot of the
  elements. pp 257--298 (\mn@eprint {} {astro-ph/0303272})

\bibitem[\protect\citeauthoryear{{Pettini}}{{Pettini}}{2006}]{Pettini06}
{Pettini} M.,  2006, in {Le Brun} V.,  {Mazure} A.,  {Arnouts} S.,
  {Burgarella} D.,  eds, The Fabulous Destiny of Galaxies: Bridging Past and
  Present. p.~319 (\mn@eprint {} {astro-ph/0603066})

\bibitem[\protect\citeauthoryear{{Pettini} \& {Pagel}}{{Pettini} \&
  {Pagel}}{2004}]{Pettini04}
{Pettini} M.,  {Pagel} B.~E.~J.,  2004, \mn@doi [\mnras]
  {10.1111/j.1365-2966.2004.07591.x}, \href
  {http://adsabs.harvard.edu/abs/2004MNRAS.348L..59P} {348, L59}

\bibitem[\protect\citeauthoryear{{Pettini}, {Ellison}, {Steidel}  \&
  {Bowen}}{{Pettini} et~al.}{1999}]{Pettini99}
{Pettini} M.,  {Ellison} S.~L.,  {Steidel} C.~C.,   {Bowen} D.~V.,  1999,
  \mn@doi [\apj] {10.1086/306635}, \href
  {http://adsabs.harvard.edu/abs/1999ApJ...510..576P} {510, 576}

\bibitem[\protect\citeauthoryear{{Pettini}, {Steidel}, {Adelberger},
  {Dickinson}  \& {Giavalisco}}{{Pettini} et~al.}{2000}]{Pettini00}
{Pettini} M.,  {Steidel} C.~C.,  {Adelberger} K.~L.,  {Dickinson} M.,
  {Giavalisco} M.,  2000, \mn@doi [\apj] {10.1086/308176}, \href
  {http://adsabs.harvard.edu/abs/2000ApJ...528...96P} {528, 96}

\bibitem[\protect\citeauthoryear{{Pettini}, {Shapley}, {Steidel}, {Cuby},
  {Dickinson}, {Moorwood}, {Adelberger}  \& {Giavalisco}}{{Pettini}
  et~al.}{2001}]{Pettini01}
{Pettini} M.,  {Shapley} A.~E.,  {Steidel} C.~C.,  {Cuby} J.-G.,  {Dickinson}
  M.,  {Moorwood} A.~F.~M.,  {Adelberger} K.~L.,   {Giavalisco} M.,  2001,
  \mn@doi [\apj] {10.1086/321403}, \href
  {http://adsabs.harvard.edu/abs/2001ApJ...554..981P} {554, 981}

\bibitem[\protect\citeauthoryear{{Pettini}, {Ellison}, {Bergeron}  \&
  {Petitjean}}{{Pettini} et~al.}{2002a}]{Pettini02c}
{Pettini} M.,  {Ellison} S.~L.,  {Bergeron} J.,   {Petitjean} P.,  2002a,
  \mn@doi [\aap] {10.1051/0004-6361:20020809}, \href
  {http://adsabs.harvard.edu/abs/2002A%26A...391...21P} {391, 21}

\bibitem[\protect\citeauthoryear{{Pettini}, {Rix}, {Steidel}, {Adelberger},
  {Hunt}  \& {Shapley}}{{Pettini} et~al.}{2002b}]{Pettini02a}
{Pettini} M.,  {Rix} S.~A.,  {Steidel} C.~C.,  {Adelberger} K.~L.,  {Hunt}
  M.~P.,   {Shapley} A.~E.,  2002b, \mn@doi [\apj] {10.1086/339355}, \href
  {http://adsabs.harvard.edu/abs/2002ApJ...569..742P} {569, 742}

\bibitem[\protect\citeauthoryear{{Pettini}, {Zych}, {Steidel}  \&
  {Chaffee}}{{Pettini} et~al.}{2008}]{Pettini08}
{Pettini} M.,  {Zych} B.~J.,  {Steidel} C.~C.,   {Chaffee} F.~H.,  2008,
  \mn@doi [\mnras] {10.1111/j.1365-2966.2008.12951.x}, \href
  {http://adsabs.harvard.edu/abs/2008MNRAS.385.2011P} {385, 2011}

\bibitem[\protect\citeauthoryear{{Pian} et~al.,}{{Pian} et~al.}{2017}]{Pian17}
{Pian} E.,  et~al., 2017, \mn@doi [\nat] {10.1038/nature24298}, \href
  {http://adsabs.harvard.edu/abs/2017Natur.551...67P} {551, 67}

\bibitem[\protect\citeauthoryear{{Pilbratt} et~al.,}{{Pilbratt}
  et~al.}{2010}]{Pilbratt10}
{Pilbratt} G.~L.,  et~al., 2010, \mn@doi [\aap] {10.1051/0004-6361/201014759},
  \href {http://adsabs.harvard.edu/abs/2010A%26A...518L...1P} {518, L1}

\bibitem[\protect\citeauthoryear{{Pilkington} et~al.,}{{Pilkington}
  et~al.}{2012}]{Pilkington12}
{Pilkington} K.,  et~al., 2012, \mn@doi [\aap] {10.1051/0004-6361/201117466},
  \href {http://adsabs.harvard.edu/abs/2012A%26A...540A..56P} {540, A56}

\bibitem[\protect\citeauthoryear{{Pillepich} et~al.,}{{Pillepich}
  et~al.}{2018}]{Pillepich18}
{Pillepich} A.,  et~al., 2018, \mn@doi [\mnras] {10.1093/mnras/stx2656}, \href
  {http://adsabs.harvard.edu/abs/2018MNRAS.473.4077P} {473, 4077}

\bibitem[\protect\citeauthoryear{{Pilyugin}}{{Pilyugin}}{2007}]{Pilyugin07a}
{Pilyugin} L.~S.,  2007, \mn@doi [\mnras] {10.1111/j.1365-2966.2006.11333.x},
  \href {http://adsabs.harvard.edu/abs/2007MNRAS.375..685P} {375, 685}

\bibitem[\protect\citeauthoryear{{Pilyugin} \& {Grebel}}{{Pilyugin} \&
  {Grebel}}{2016}]{Pilyugin16}
{Pilyugin} L.~S.,  {Grebel} E.~K.,  2016, \mn@doi [\mnras]
  {10.1093/mnras/stw238}, \href
  {http://adsabs.harvard.edu/abs/2016MNRAS.457.3678P} {457, 3678}

\bibitem[\protect\citeauthoryear{{Pilyugin} \& {Thuan}}{{Pilyugin} \&
  {Thuan}}{2005}]{Pilyugin05b}
{Pilyugin} L.~S.,  {Thuan} T.~X.,  2005, \mn@doi [\apj] {10.1086/432408}, \href
  {http://adsabs.harvard.edu/abs/2005ApJ...631..231P} {631, 231}

\bibitem[\protect\citeauthoryear{{Pilyugin}, {Thuan}  \&
  {V{\'{\i}}lchez}}{{Pilyugin} et~al.}{2006a}]{Pilyugin06a}
{Pilyugin} L.~S.,  {Thuan} T.~X.,   {V{\'{\i}}lchez} J.~M.,  2006a, \mn@doi
  [\mnras] {10.1111/j.1365-2966.2006.10033.x}, \href
  {http://adsabs.harvard.edu/abs/2006MNRAS.367.1139P} {367, 1139}

\bibitem[\protect\citeauthoryear{{Pilyugin}, {V{\'{\i}}lchez}  \&
  {Thuan}}{{Pilyugin} et~al.}{2006b}]{Pilyugin06b}
{Pilyugin} L.~S.,  {V{\'{\i}}lchez} J.~M.,   {Thuan} T.~X.,  2006b, \mn@doi
  [\mnras] {10.1111/j.1365-2966.2006.10618.x}, \href
  {http://adsabs.harvard.edu/abs/2006MNRAS.370.1928P} {370, 1928}

\bibitem[\protect\citeauthoryear{{Pilyugin}, {Mattsson}, {V{\'{\i}}lchez}  \&
  {Cedr{\'e}s}}{{Pilyugin} et~al.}{2009}]{Pilyugin09}
{Pilyugin} L.~S.,  {Mattsson} L.,  {V{\'{\i}}lchez} J.~M.,   {Cedr{\'e}s} B.,
  2009, \mn@doi [\mnras] {10.1111/j.1365-2966.2009.15182.x}, \href
  {http://adsabs.harvard.edu/abs/2009MNRAS.398..485P} {398, 485}

\bibitem[\protect\citeauthoryear{{Pilyugin}, {V{\'{\i}}lchez}, {Cedr{\'e}s}  \&
  {Thuan}}{{Pilyugin} et~al.}{2010a}]{Pilyugin10a}
{Pilyugin} L.~S.,  {V{\'{\i}}lchez} J.~M.,  {Cedr{\'e}s} B.,   {Thuan} T.~X.,
  2010a, \mn@doi [\mnras] {10.1111/j.1365-2966.2009.16166.x}, \href
  {http://adsabs.harvard.edu/abs/2010MNRAS.403..896P} {403, 896}

\bibitem[\protect\citeauthoryear{{Pilyugin}, {V{\'{\i}}lchez}  \&
  {Thuan}}{{Pilyugin} et~al.}{2010b}]{Pilyugin10b}
{Pilyugin} L.~S.,  {V{\'{\i}}lchez} J.~M.,   {Thuan} T.~X.,  2010b, \mn@doi
  [\apj] {10.1088/0004-637X/720/2/1738}, \href
  {http://adsabs.harvard.edu/abs/2010ApJ...720.1738P} {720, 1738}

\bibitem[\protect\citeauthoryear{{Pilyugin}, {V{\'{\i}}lchez}, {Mattsson}  \&
  {Thuan}}{{Pilyugin} et~al.}{2012}]{Pilyugin12}
{Pilyugin} L.~S.,  {V{\'{\i}}lchez} J.~M.,  {Mattsson} L.,   {Thuan} T.~X.,
  2012, \mn@doi [\mnras] {10.1111/j.1365-2966.2012.20420.x}, \href
  {http://adsabs.harvard.edu/abs/2012MNRAS.421.1624P} {421, 1624}

\bibitem[\protect\citeauthoryear{{Pilyugin}, {Lara-L{\'o}pez}, {Grebel},
  {Kehrig}, {Zinchenko}, {L{\'o}pez-S{\'a}nchez}, {V{\'{\i}}lchez}  \&
  {Mattsson}}{{Pilyugin} et~al.}{2013}]{Pilyugin13}
{Pilyugin} L.~S.,  {Lara-L{\'o}pez} M.~A.,  {Grebel} E.~K.,  {Kehrig} C.,
  {Zinchenko} I.~A.,  {L{\'o}pez-S{\'a}nchez} {\'A}.~R.,  {V{\'{\i}}lchez}
  J.~M.,   {Mattsson} L.,  2013, \mn@doi [\mnras] {10.1093/mnras/stt539}, \href
  {http://adsabs.harvard.edu/abs/2013MNRAS.432.1217P} {432, 1217}

\bibitem[\protect\citeauthoryear{{Pilyugin}, {Grebel}  \&
  {Zinchenko}}{{Pilyugin} et~al.}{2015}]{Pilyugin15}
{Pilyugin} L.~S.,  {Grebel} E.~K.,   {Zinchenko} I.~A.,  2015, preprint
  (\mn@eprint {} {1505.00337})

\bibitem[\protect\citeauthoryear{{Pilyugin}, {Grebel}, {Zinchenko}, {Nefedyev}
  \& {Mattsson}}{{Pilyugin} et~al.}{2017}]{Pilyugin17a}
{Pilyugin} L.~S.,  {Grebel} E.~K.,  {Zinchenko} I.~A.,  {Nefedyev} Y.~A.,
  {Mattsson} L.,  2017, \mn@doi [\mnras] {10.1093/mnras/stw2831}, \href
  {http://adsabs.harvard.edu/abs/2017MNRAS.465.1358P} {465, 1358}

\bibitem[\protect\citeauthoryear{{Pinto}, {Kaastra}, {Costantini}  \& {de
  Vries}}{{Pinto} et~al.}{2013}]{Pinto13}
{Pinto} C.,  {Kaastra} J.~S.,  {Costantini} E.,   {de Vries} C.,  2013, \mn@doi
  [\aap] {10.1051/0004-6361/201220481}, \href
  {http://adsabs.harvard.edu/abs/2013A%26A...551A..25P} {551, A25}

\bibitem[\protect\citeauthoryear{{Pipino} \& {Matteucci}}{{Pipino} \&
  {Matteucci}}{2004}]{Pipino04}
{Pipino} A.,  {Matteucci} F.,  2004, \mn@doi [\mnras]
  {10.1111/j.1365-2966.2004.07268.x}, \href
  {http://adsabs.harvard.edu/abs/2004MNRAS.347..968P} {347, 968}

\bibitem[\protect\citeauthoryear{{Pipino}, {Matteucci}  \&
  {Chiappini}}{{Pipino} et~al.}{2006}]{Pipino06}
{Pipino} A.,  {Matteucci} F.,   {Chiappini} C.,  2006, \mn@doi [\apj]
  {10.1086/499033}, \href {http://adsabs.harvard.edu/abs/2006ApJ...638..739P}
  {638, 739}

\bibitem[\protect\citeauthoryear{{Pipino}, {D'Ercole}  \& {Matteucci}}{{Pipino}
  et~al.}{2008}]{Pipino08}
{Pipino} A.,  {D'Ercole} A.,   {Matteucci} F.,  2008, \mn@doi [\aap]
  {10.1051/0004-6361:20078121}, \href
  {http://adsabs.harvard.edu/abs/2008A%26A...484..679P} {484, 679}

\bibitem[\protect\citeauthoryear{{Pipino}, {D'Ercole}, {Chiappini}  \&
  {Matteucci}}{{Pipino} et~al.}{2010}]{Pipino10}
{Pipino} A.,  {D'Ercole} A.,  {Chiappini} C.,   {Matteucci} F.,  2010, \mn@doi
  [\mnras] {10.1111/j.1365-2966.2010.17007.x}, \href
  {http://adsabs.harvard.edu/abs/2010MNRAS.407.1347P} {407, 1347}

\bibitem[\protect\citeauthoryear{{Pipino}, {Fan}, {Matteucci}, {Calura},
  {Silva}, {Granato}  \& {Maiolino}}{{Pipino} et~al.}{2011}]{Pipino11}
{Pipino} A.,  {Fan} X.~L.,  {Matteucci} F.,  {Calura} F.,  {Silva} L.,
  {Granato} G.,   {Maiolino} R.,  2011, \mn@doi [\aap]
  {10.1051/0004-6361/201014843}, \href
  {http://adsabs.harvard.edu/abs/2011A%26A...525A..61P} {525, A61}

\bibitem[\protect\citeauthoryear{{Pipino}, {Lilly}  \& {Carollo}}{{Pipino}
  et~al.}{2014}]{Pipino14}
{Pipino} A.,  {Lilly} S.~J.,   {Carollo} C.~M.,  2014, \mn@doi [\mnras]
  {10.1093/mnras/stu579}, \href
  {http://adsabs.harvard.edu/abs/2014MNRAS.441.1444P} {441, 1444}

\bibitem[\protect\citeauthoryear{{Piranomonte} et~al.,}{{Piranomonte}
  et~al.}{2015}]{Piranomonte15}
{Piranomonte} S.,  et~al., 2015, \mn@doi [\mnras] {10.1093/mnras/stv1569},
  \href {http://adsabs.harvard.edu/abs/2015MNRAS.452.3293P} {452, 3293}

\bibitem[\protect\citeauthoryear{{Poetrodjojo} et~al.,}{{Poetrodjojo}
  et~al.}{2018}]{Poetrodjojo18}
{Poetrodjojo} H.,  et~al., 2018, \mn@doi [\mnras] {10.1093/mnras/sty1782},
  \href {http://adsabs.harvard.edu/abs/2018MNRAS.479.5235P} {479, 5235}

\bibitem[\protect\citeauthoryear{{Pontzen} et~al.,}{{Pontzen}
  et~al.}{2008}]{Pontzen08}
{Pontzen} A.,  et~al., 2008, \mn@doi [\mnras]
  {10.1111/j.1365-2966.2008.13782.x}, \href
  {http://adsabs.harvard.edu/abs/2008MNRAS.390.1349P} {390, 1349}

\bibitem[\protect\citeauthoryear{{Porter}, {Somerville}, {Primack}, {Croton},
  {Covington}, {Graves}  \& {Faber}}{{Porter} et~al.}{2014}]{Porter14}
{Porter} L.~A.,  {Somerville} R.~S.,  {Primack} J.~R.,  {Croton} D.~J.,
  {Covington} M.~D.,  {Graves} G.~J.,   {Faber} S.~M.,  2014, \mn@doi [\mnras]
  {10.1093/mnras/stu1701}, \href
  {http://adsabs.harvard.edu/abs/2014MNRAS.445.3092P} {445, 3092}

\bibitem[\protect\citeauthoryear{{Poudel}, {Kulkarni}, {Morrison},
  {P{\'e}roux}, {Som}, {Rahmani}  \& {Quiret}}{{Poudel}
  et~al.}{2018}]{Poudel18}
{Poudel} S.,  {Kulkarni} V.~P.,  {Morrison} S.,  {P{\'e}roux} C.,  {Som} D.,
  {Rahmani} H.,   {Quiret} S.,  2018, \mn@doi [\mnras] {10.1093/mnras/stx2607},
  \href {http://adsabs.harvard.edu/abs/2018MNRAS.473.3559P} {473, 3559}

\bibitem[\protect\citeauthoryear{{Prantzos}}{{Prantzos}}{2008}]{Prantzos08}
{Prantzos} N.,  2008, in {Charbonnel} C.,  {Zahn} J.-P.,  eds,  EAS
  Publications Series Vol. 32, EAS Publications Series. pp 311--356 (\mn@eprint
  {} {0709.0833}), \mn@doi{10.1051/eas:0832009}

\bibitem[\protect\citeauthoryear{{Prantzos}}{{Prantzos}}{2009}]{Prantzos09}
{Prantzos} N.,  2009, in {Andersen} J.,  {Nordstr{\"o}ara} {m} B.,
  {Bland-Hawthorn} J.,  eds,  IAU Symposium Vol. 254, The Galaxy Disk in
  Cosmological Context. pp 381--392 (\mn@eprint {} {0809.2507}),
  \mn@doi{10.1017/S1743921308027841}

\bibitem[\protect\citeauthoryear{{Prantzos} \& {Boissier}}{{Prantzos} \&
  {Boissier}}{2000}]{Prantzos00}
{Prantzos} N.,  {Boissier} S.,  2000, \mn@doi [\mnras]
  {10.1046/j.1365-8711.2000.03228.x}, \href
  {http://adsabs.harvard.edu/abs/2000MNRAS.313..338P} {313, 338}

\bibitem[\protect\citeauthoryear{{Prochaska} \& {Wolfe}}{{Prochaska} \&
  {Wolfe}}{1999}]{Prochaska99}
{Prochaska} J.~X.,  {Wolfe} A.~M.,  1999, \mn@doi [\apjs] {10.1086/313200},
  \href {http://adsabs.harvard.edu/abs/1999ApJS..121..369P} {121, 369}

\bibitem[\protect\citeauthoryear{{Prochaska} \& {Wolfe}}{{Prochaska} \&
  {Wolfe}}{2009}]{Prochaska09}
{Prochaska} J.~X.,  {Wolfe} A.~M.,  2009, \mn@doi [\apj]
  {10.1088/0004-637X/696/2/1543}, \href
  {http://adsabs.harvard.edu/abs/2009ApJ...696.1543P} {696, 1543}

\bibitem[\protect\citeauthoryear{{Prochaska}, {Howk}  \& {Wolfe}}{{Prochaska}
  et~al.}{2003a}]{Prochaska03}
{Prochaska} J.~X.,  {Howk} J.~C.,   {Wolfe} A.~M.,  2003a, \mn@doi [\nat]
  {10.1038/nature01524}, \href
  {http://adsabs.harvard.edu/abs/2003Natur.423...57P} {423, 57}

\bibitem[\protect\citeauthoryear{{Prochaska}, {Gawiser}, {Wolfe}, {Castro}  \&
  {Djorgovski}}{{Prochaska} et~al.}{2003b}]{Prochaska03b}
{Prochaska} J.~X.,  {Gawiser} E.,  {Wolfe} A.~M.,  {Castro} S.,   {Djorgovski}
  S.~G.,  2003b, \mn@doi [\apjl] {10.1086/378945}, 595, L9

\bibitem[\protect\citeauthoryear{{Prochaska} et~al.,}{{Prochaska}
  et~al.}{2004}]{Prochaska04}
{Prochaska} J.~X.,  et~al., 2004, \mn@doi [\apj] {10.1086/421988}, \href
  {http://adsabs.harvard.edu/abs/2004ApJ...611..200P} {611, 200}

\bibitem[\protect\citeauthoryear{{Prochaska}, {Chen}, {Dessauges-Zavadsky}  \&
  {Bloom}}{{Prochaska} et~al.}{2007}]{Prochaska07}
{Prochaska} J.~X.,  {Chen} H.-W.,  {Dessauges-Zavadsky} M.,   {Bloom} J.~S.,
  2007, \mn@doi [\apj] {10.1086/520042}, 666, 267

\bibitem[\protect\citeauthoryear{{Prochaska}, {O'Meara}  \&
  {Worseck}}{{Prochaska} et~al.}{2010}]{Prochaska10}
{Prochaska} J.~X.,  {O'Meara} J.~M.,   {Worseck} G.,  2010, \mn@doi [\apj]
  {10.1088/0004-637X/718/1/392}, \href
  {http://adsabs.harvard.edu/abs/2010ApJ...718..392P} {718, 392}

\bibitem[\protect\citeauthoryear{{Prochaska} et~al.,}{{Prochaska}
  et~al.}{2013}]{Prochaska13}
{Prochaska} J.~X.,  et~al., 2013, \mn@doi [\apj] {10.1088/0004-637X/776/2/136},
  \href {http://adsabs.harvard.edu/abs/2013ApJ...776..136P} {776, 136}

\bibitem[\protect\citeauthoryear{{Prochaska}, {Lau}  \& {Hennawi}}{{Prochaska}
  et~al.}{2014}]{Prochaska14}
{Prochaska} J.~X.,  {Lau} M.~W.,   {Hennawi} J.~F.,  2014, \mn@doi [\apj]
  {10.1088/0004-637X/796/2/140}, \href
  {http://adsabs.harvard.edu/abs/2014ApJ...796..140P} {796, 140}

\bibitem[\protect\citeauthoryear{{Prochaska} et~al.,}{{Prochaska}
  et~al.}{2017}]{Prochaska17}
{Prochaska} J.~X.,  et~al., 2017, \mn@doi [\apj] {10.3847/1538-4357/aa6007},
  \href {http://adsabs.harvard.edu/abs/2017ApJ...837..169P} {837, 169}

\bibitem[\protect\citeauthoryear{{Puglisi} et~al.,}{{Puglisi}
  et~al.}{2017}]{Puglisi17}
{Puglisi} A.,  et~al., 2017, \mn@doi [\apjl] {10.3847/2041-8213/aa66c9}, \href
  {http://adsabs.harvard.edu/abs/2017ApJ...838L..18P} {838, L18}

\bibitem[\protect\citeauthoryear{{Queyrel} et~al.,}{{Queyrel}
  et~al.}{2012}]{Queyrel12}
{Queyrel} J.,  et~al., 2012, \mn@doi [\aap] {10.1051/0004-6361/201117718},
  \href {http://adsabs.harvard.edu/abs/2012A%26A...539A..93Q} {539, A93}

\bibitem[\protect\citeauthoryear{{Quider}, {Pettini}, {Shapley}  \&
  {Steidel}}{{Quider} et~al.}{2009}]{Quider09}
{Quider} A.~M.,  {Pettini} M.,  {Shapley} A.~E.,   {Steidel} C.~C.,  2009,
  \mn@doi [\mnras] {10.1111/j.1365-2966.2009.15234.x}, \href
  {http://adsabs.harvard.edu/abs/2009MNRAS.398.1263Q} {398, 1263}

\bibitem[\protect\citeauthoryear{{Rafelski}, {Wolfe}, {Prochaska}, {Neeleman}
  \& {Mendez}}{{Rafelski} et~al.}{2012}]{Rafelski12}
{Rafelski} M.,  {Wolfe} A.~M.,  {Prochaska} J.~X.,  {Neeleman} M.,   {Mendez}
  A.~J.,  2012, \mn@doi [\apj] {10.1088/0004-637X/755/2/89}, \href
  {http://adsabs.harvard.edu/abs/2012ApJ...755...89R} {755, 89}

\bibitem[\protect\citeauthoryear{{Rafelski}, {Neeleman}, {Fumagalli}, {Wolfe}
  \& {Prochaska}}{{Rafelski} et~al.}{2014}]{Rafelski14}
{Rafelski} M.,  {Neeleman} M.,  {Fumagalli} M.,  {Wolfe} A.~M.,   {Prochaska}
  J.~X.,  2014, \mn@doi [\apjl] {10.1088/2041-8205/782/2/L29}, \href
  {http://adsabs.harvard.edu/abs/2014ApJ...782L..29R} {782, L29}

\bibitem[\protect\citeauthoryear{{Ranalli}, {Comastri}, {Origlia}  \&
  {Maiolino}}{{Ranalli} et~al.}{2008}]{Ranalli08}
{Ranalli} P.,  {Comastri} A.,  {Origlia} L.,   {Maiolino} R.,  2008, \mn@doi
  [\mnras] {10.1111/j.1365-2966.2008.13128.x}, \href
  {http://adsabs.harvard.edu/abs/2008MNRAS.386.1464R} {386, 1464}

\bibitem[\protect\citeauthoryear{{Rauch}}{{Rauch}}{1998}]{Rauch98}
{Rauch} M.,  1998, \mn@doi [\araa] {10.1146/annurev.astro.36.1.267}, \href
  {http://adsabs.harvard.edu/abs/1998ARA%26A..36..267R} {36, 267}

\bibitem[\protect\citeauthoryear{{Rauscher} \& {Patk{\'o}s}}{{Rauscher} \&
  {Patk{\'o}s}}{2011}]{Rauscher11}
{Rauscher} T.,  {Patk{\'o}s} A.,  2011, {Origin of the Chemical Elements}.
p.~611

\bibitem[\protect\citeauthoryear{{Recchi}, {Matteucci}  \& {D'Ercole}}{{Recchi}
  et~al.}{2001}]{Recchi01}
{Recchi} S.,  {Matteucci} F.,   {D'Ercole} A.,  2001, \mn@doi [\mnras]
  {10.1046/j.1365-8711.2001.04189.x}, \href
  {http://adsabs.harvard.edu/abs/2001MNRAS.322..800R} {322, 800}

\bibitem[\protect\citeauthoryear{{Reddy}, {Lambert}  \& {Allende
  Prieto}}{{Reddy} et~al.}{2006}]{Reddy06}
{Reddy} B.~E.,  {Lambert} D.~L.,   {Allende Prieto} C.,  2006, \mn@doi [\mnras]
  {10.1111/j.1365-2966.2006.10148.x}, \href
  {http://adsabs.harvard.edu/abs/2006MNRAS.367.1329R} {367, 1329}

\bibitem[\protect\citeauthoryear{{Reichard}, {Heckman}, {Rudnick},
  {Brinchmann}, {Kauffmann}  \& {Wild}}{{Reichard} et~al.}{2009}]{Reichard09}
{Reichard} T.~A.,  {Heckman} T.~M.,  {Rudnick} G.,  {Brinchmann} J.,
  {Kauffmann} G.,   {Wild} V.,  2009, \mn@doi [\apj]
  {10.1088/0004-637X/691/2/1005}, \href
  {http://adsabs.harvard.edu/abs/2009ApJ...691.1005R} {691, 1005}

\bibitem[\protect\citeauthoryear{{Renzini} \& {Andreon}}{{Renzini} \&
  {Andreon}}{2014}]{Renzini14}
{Renzini} A.,  {Andreon} S.,  2014, \mn@doi [\mnras] {10.1093/mnras/stu1689},
  \href {http://adsabs.harvard.edu/abs/2014MNRAS.444.3581R} {444, 3581}

\bibitem[\protect\citeauthoryear{{Rhodin}, {Christensen}, {M{\o}ller}, {Zafar}
  \& {Fynbo}}{{Rhodin} et~al.}{2018}]{Rhodin18}
{Rhodin} N.~H.~P.,  {Christensen} L.,  {M{\o}ller} P.,  {Zafar} T.,   {Fynbo}
  J.~P.~U.,  2018, preprint, \href
  {http://adsabs.harvard.edu/abs/2018arXiv180701755R} {} (\mn@eprint {arXiv}
  {1807.01755})

\bibitem[\protect\citeauthoryear{{Rich}, {Dopita}, {Kewley}  \& {Rupke}}{{Rich}
  et~al.}{2010}]{Rich10}
{Rich} J.~A.,  {Dopita} M.~A.,  {Kewley} L.~J.,   {Rupke} D.~S.~N.,  2010,
  \mn@doi [\apj] {10.1088/0004-637X/721/1/505}, \href
  {http://adsabs.harvard.edu/abs/2010ApJ...721..505R} {721, 505}

\bibitem[\protect\citeauthoryear{{Rich}, {Origlia}  \& {Valenti}}{{Rich}
  et~al.}{2012a}]{Rich12b}
{Rich} R.~M.,  {Origlia} L.,   {Valenti} E.,  2012a, \mn@doi [\apj]
  {10.1088/0004-637X/746/1/59}, \href
  {http://adsabs.harvard.edu/abs/2012ApJ...746...59R} {746, 59}

\bibitem[\protect\citeauthoryear{{Rich}, {Torrey}, {Kewley}, {Dopita}  \&
  {Rupke}}{{Rich} et~al.}{2012b}]{Rich12a}
{Rich} J.~A.,  {Torrey} P.,  {Kewley} L.~J.,  {Dopita} M.~A.,   {Rupke}
  D.~S.~N.,  2012b, \mn@doi [\apj] {10.1088/0004-637X/753/1/5}, \href
  {http://adsabs.harvard.edu/abs/2012ApJ...753....5R} {753, 5}

\bibitem[\protect\citeauthoryear{{Richard}, {Jones}, {Ellis}, {Stark},
  {Livermore}  \& {Swinbank}}{{Richard} et~al.}{2011}]{Richard11}
{Richard} J.,  {Jones} T.,  {Ellis} R.,  {Stark} D.~P.,  {Livermore} R.,
  {Swinbank} M.,  2011, \mn@doi [\mnras] {10.1111/j.1365-2966.2010.18161.x},
  \href {http://adsabs.harvard.edu/abs/2011MNRAS.413..643R} {413, 643}

\bibitem[\protect\citeauthoryear{{Rix}, {Pettini}, {Leitherer}, {Bresolin},
  {Kudritzki}  \& {Steidel}}{{Rix} et~al.}{2004}]{Rix04}
{Rix} S.~A.,  {Pettini} M.,  {Leitherer} C.,  {Bresolin} F.,  {Kudritzki} R.,
  {Steidel} C.~C.,  2004, \mn@doi [\apj] {10.1086/424031}, \href
  {http://adsabs.harvard.edu/abs/2004ApJ...615...98R} {615, 98}

\bibitem[\protect\citeauthoryear{{Rodrigues}, {Puech}, {Hammer}, {Rothberg}  \&
  {Flores}}{{Rodrigues} et~al.}{2012}]{Rodrigues12}
{Rodrigues} M.,  {Puech} M.,  {Hammer} F.,  {Rothberg} B.,   {Flores} H.,
  2012, \mn@doi [\mnras] {10.1111/j.1365-2966.2012.20518.x}, \href
  {http://adsabs.harvard.edu/abs/2012MNRAS.421.2888R} {421, 2888}

\bibitem[\protect\citeauthoryear{{Rodr{\'{\i}}guez-Puebla}, {Primack},
  {Behroozi}  \& {Faber}}{{Rodr{\'{\i}}guez-Puebla}
  et~al.}{2016}]{Rodriguez-Puebla16}
{Rodr{\'{\i}}guez-Puebla} A.,  {Primack} J.~R.,  {Behroozi} P.,   {Faber}
  S.~M.,  2016, \mn@doi [\mnras] {10.1093/mnras/stv2513}, \href
  {http://adsabs.harvard.edu/abs/2016MNRAS.455.2592R} {455, 2592}

\bibitem[\protect\citeauthoryear{{Roelfsema} et~al.,}{{Roelfsema}
  et~al.}{2018}]{Roelfsema18}
{Roelfsema} P.~R.,  et~al., 2018, \mn@doi [\pasa] {10.1017/pasa.2018.15}, 35,
  e030

\bibitem[\protect\citeauthoryear{{Roig}, {Blanton}  \& {Yan}}{{Roig}
  et~al.}{2015}]{Roig15}
{Roig} B.,  {Blanton} M.~R.,   {Yan} R.,  2015, \mn@doi [\apj]
  {10.1088/0004-637X/808/1/26}, \href
  {http://adsabs.harvard.edu/abs/2015ApJ...808...26R} {808, 26}

\bibitem[\protect\citeauthoryear{{Rojas-Arriagada} et~al.,}{{Rojas-Arriagada}
  et~al.}{2017}]{Rojas-Arriagada17}
{Rojas-Arriagada} A.,  et~al., 2017, \mn@doi [\aap]
  {10.1051/0004-6361/201629160}, \href
  {http://adsabs.harvard.edu/abs/2017A%26A...601A.140R} {601, A140}

\bibitem[\protect\citeauthoryear{{Romano}, {Silva}, {Matteucci}  \&
  {Danese}}{{Romano} et~al.}{2002}]{Romano02}
{Romano} D.,  {Silva} L.,  {Matteucci} F.,   {Danese} L.,  2002, \mn@doi
  [\mnras] {10.1046/j.1365-8711.2002.05534.x}, \href
  {http://adsabs.harvard.edu/abs/2002MNRAS.334..444R} {334, 444}

\bibitem[\protect\citeauthoryear{{Romano}, {Karakas}, {Tosi}  \&
  {Matteucci}}{{Romano} et~al.}{2010}]{Romano10}
{Romano} D.,  {Karakas} A.~I.,  {Tosi} M.,   {Matteucci} F.,  2010, \mn@doi
  [\aap] {10.1051/0004-6361/201014483}, \href
  {http://adsabs.harvard.edu/abs/2010A%26A...522A..32R} {522, A32}

\bibitem[\protect\citeauthoryear{{Roseboom} et~al.,}{{Roseboom}
  et~al.}{2012}]{Roseboom12}
{Roseboom} I.~G.,  et~al., 2012, \mn@doi [\mnras]
  {10.1111/j.1365-2966.2012.21777.x}, \href
  {http://adsabs.harvard.edu/abs/2012MNRAS.426.1782R} {426, 1782}

\bibitem[\protect\citeauthoryear{{Rubin}, {Simpson}, {Erickson}  \&
  {Haas}}{{Rubin} et~al.}{1988}]{Rubin88}
{Rubin} R.~H.,  {Simpson} J.~P.,  {Erickson} E.~F.,   {Haas} M.~R.,  1988,
  \mn@doi [\apj] {10.1086/166200}, \href
  {http://adsabs.harvard.edu/abs/1988ApJ...327..377R} {327, 377}

\bibitem[\protect\citeauthoryear{{Rudolph}, {Fich}, {Bell}, {Norsen},
  {Simpson}, {Haas}  \& {Erickson}}{{Rudolph} et~al.}{2006}]{Rudolph06}
{Rudolph} A.~L.,  {Fich} M.,  {Bell} G.~R.,  {Norsen} T.,  {Simpson} J.~P.,
  {Haas} M.~R.,   {Erickson} E.~F.,  2006, \mn@doi [\apjs] {10.1086/498869},
  \href {http://adsabs.harvard.edu/abs/2006ApJS..162..346R} {162, 346}

\bibitem[\protect\citeauthoryear{{Rupke}, {Veilleux}  \& {Baker}}{{Rupke}
  et~al.}{2008}]{Rupke08}
{Rupke} D.~S.~N.,  {Veilleux} S.,   {Baker} A.~J.,  2008, \mn@doi [\apj]
  {10.1086/522363}, \href {http://adsabs.harvard.edu/abs/2008ApJ...674..172R}
  {674, 172}

\bibitem[\protect\citeauthoryear{{Rupke}, {Kewley}  \& {Barnes}}{{Rupke}
  et~al.}{2010a}]{Rupke10a}
{Rupke} D.~S.~N.,  {Kewley} L.~J.,   {Barnes} J.~E.,  2010a, \mn@doi [\apjl]
  {10.1088/2041-8205/710/2/L156}, \href
  {http://adsabs.harvard.edu/abs/2010ApJ...710L.156R} {710, L156}

\bibitem[\protect\citeauthoryear{{Rupke}, {Kewley}  \& {Chien}}{{Rupke}
  et~al.}{2010b}]{Rupke10b}
{Rupke} D.~S.~N.,  {Kewley} L.~J.,   {Chien} L.-H.,  2010b, \mn@doi [\apj]
  {10.1088/0004-637X/723/2/1255}, \href
  {http://adsabs.harvard.edu/abs/2010ApJ...723.1255R} {723, 1255}

\bibitem[\protect\citeauthoryear{{Ryde}, {Schultheis}, {Grieco}, {Matteucci},
  {Rich}  \& {Uttenthaler}}{{Ryde} et~al.}{2016}]{Ryde16}
{Ryde} N.,  {Schultheis} M.,  {Grieco} V.,  {Matteucci} F.,  {Rich} R.~M.,
  {Uttenthaler} S.,  2016, \mn@doi [\aj] {10.3847/0004-6256/151/1/1}, \href
  {http://adsabs.harvard.edu/abs/2016AJ....151....1R} {151, 1}

\bibitem[\protect\citeauthoryear{{Salim}, {Lee}, {Ly}, {Brinchmann},
  {Dav{\'e}}, {Dickinson}, {Salzer}  \& {Charlot}}{{Salim}
  et~al.}{2014}]{Salim14}
{Salim} S.,  {Lee} J.~C.,  {Ly} C.,  {Brinchmann} J.,  {Dav{\'e}} R.,
  {Dickinson} M.,  {Salzer} J.~J.,   {Charlot} S.,  2014, \mn@doi [\apj]
  {10.1088/0004-637X/797/2/126}, \href
  {http://adsabs.harvard.edu/abs/2014ApJ...797..126S} {797, 126}

\bibitem[\protect\citeauthoryear{{Salim}, {Lee}, {Dav{\'e}}  \&
  {Dickinson}}{{Salim} et~al.}{2015}]{Salim15}
{Salim} S.,  {Lee} J.~C.,  {Dav{\'e}} R.,   {Dickinson} M.,  2015, \mn@doi
  [\apj] {10.1088/0004-637X/808/1/25}, \href
  {http://adsabs.harvard.edu/abs/2015ApJ...808...25S} {808, 25}

\bibitem[\protect\citeauthoryear{{S{\'a}nchez Almeida} \& {Dalla
  Vecchia}}{{S{\'a}nchez Almeida} \& {Dalla
  Vecchia}}{2018}]{Sanchez-Almeida18b}
{S{\'a}nchez Almeida} J.,  {Dalla Vecchia} C.,  2018, \mn@doi [\apj]
  {10.3847/1538-4357/aac086}, \href
  {http://adsabs.harvard.edu/abs/2018ApJ...859..109S} {859, 109}

\bibitem[\protect\citeauthoryear{{S{\'a}nchez Almeida},
  {Mu{\~n}oz-Tu{\~n}{\'o}n}, {Elmegreen}, {Elmegreen}  \&
  {M{\'e}ndez-Abreu}}{{S{\'a}nchez Almeida} et~al.}{2013}]{Sanchez-Almeida13}
{S{\'a}nchez Almeida} J.,  {Mu{\~n}oz-Tu{\~n}{\'o}n} C.,  {Elmegreen} D.~M.,
  {Elmegreen} B.~G.,   {M{\'e}ndez-Abreu} J.,  2013, \mn@doi [\apj]
  {10.1088/0004-637X/767/1/74}, \href
  {http://adsabs.harvard.edu/abs/2013ApJ...767...74S} {767, 74}

\bibitem[\protect\citeauthoryear{{S{\'a}nchez Almeida}, {Morales-Luis},
  {Mu{\~n}oz-Tu{\~n}{\'o}n}, {Elmegreen}, {Elmegreen}  \&
  {M{\'e}ndez-Abreu}}{{S{\'a}nchez Almeida} et~al.}{2014}]{Sanchez-Almeida14}
{S{\'a}nchez Almeida} J.,  {Morales-Luis} A.~B.,  {Mu{\~n}oz-Tu{\~n}{\'o}n} C.,
   {Elmegreen} D.~M.,  {Elmegreen} B.~G.,   {M{\'e}ndez-Abreu} J.,  2014,
  \mn@doi [\apj] {10.1088/0004-637X/783/1/45}, \href
  {http://adsabs.harvard.edu/abs/2014ApJ...783...45S} {783, 45}

\bibitem[\protect\citeauthoryear{{S{\'a}nchez Almeida} et~al.,}{{S{\'a}nchez
  Almeida} et~al.}{2015}]{Sanchez-Almeida15}
{S{\'a}nchez Almeida} J.,  et~al., 2015, \mn@doi [\apjl]
  {10.1088/2041-8205/810/2/L15}, \href
  {http://adsabs.harvard.edu/abs/2015ApJ...810L..15S} {810, L15}

\bibitem[\protect\citeauthoryear{{S{\'a}nchez Almeida}, {P{\'e}rez-Montero},
  {Morales-Luis}, {Mu{\~n}oz-Tu{\~n}{\'o}n}, {Garc{\'{\i}}a-Benito}, {Nuza}  \&
  {Kitaura}}{{S{\'a}nchez Almeida} et~al.}{2016}]{Sanchez-Almeida16}
{S{\'a}nchez Almeida} J.,  {P{\'e}rez-Montero} E.,  {Morales-Luis} A.~B.,
  {Mu{\~n}oz-Tu{\~n}{\'o}n} C.,  {Garc{\'{\i}}a-Benito} R.,  {Nuza} S.~E.,
  {Kitaura} F.~S.,  2016, \mn@doi [\apj] {10.3847/0004-637X/819/2/110}, \href
  {http://adsabs.harvard.edu/abs/2016ApJ...819..110S} {819, 110}

\bibitem[\protect\citeauthoryear{{S{\'a}nchez Almeida}, {Caon},
  {Mu{\~n}oz-Tu{\~n}{\'o}n}, {Filho}  \& {Cervi{\~n}o}}{{S{\'a}nchez Almeida}
  et~al.}{2018}]{Sanchez-Almeida18a}
{S{\'a}nchez Almeida} J.,  {Caon} N.,  {Mu{\~n}oz-Tu{\~n}{\'o}n} C.,  {Filho}
  M.,   {Cervi{\~n}o} M.,  2018, \mn@doi [\mnras] {10.1093/mnras/sty510}, \href
  {http://adsabs.harvard.edu/abs/2018MNRAS.476.4765S} {476, 4765}

\bibitem[\protect\citeauthoryear{{S{\'a}nchez-Bl{\'a}zquez}, {Gorgas}  \&
  {Cardiel}}{{S{\'a}nchez-Bl{\'a}zquez} et~al.}{2006}]{Sanchez-Blazquez06}
{S{\'a}nchez-Bl{\'a}zquez} P.,  {Gorgas} J.,   {Cardiel} N.,  2006, \mn@doi
  [\aap] {10.1051/0004-6361:20064846}, \href
  {http://adsabs.harvard.edu/abs/2006A%26A...457..823S} {457, 823}

\bibitem[\protect\citeauthoryear{{S{\'a}nchez-Bl{\'a}zquez}
  et~al.,}{{S{\'a}nchez-Bl{\'a}zquez} et~al.}{2014}]{Sanchez-Blazquez14}
{S{\'a}nchez-Bl{\'a}zquez} P.,  et~al., 2014, \mn@doi [\aap]
  {10.1051/0004-6361/201423635}, \href
  {http://adsabs.harvard.edu/abs/2014A%26A...570A...6S} {570, A6}

\bibitem[\protect\citeauthoryear{{S{\'a}nchez-Menguiano}
  et~al.,}{{S{\'a}nchez-Menguiano} et~al.}{2016}]{Sanchez-Menguiano16}
{S{\'a}nchez-Menguiano} L.,  et~al., 2016, \mn@doi [\aap]
  {10.1051/0004-6361/201527450}, \href
  {http://adsabs.harvard.edu/abs/2016A%26A...587A..70S} {587, A70}

\bibitem[\protect\citeauthoryear{{S{\'a}nchez-Menguiano}
  et~al.,}{{S{\'a}nchez-Menguiano} et~al.}{2017a}]{Sanchez-Menguiano17b}
{S{\'a}nchez-Menguiano} L.,  et~al., 2017a, preprint, \href
  {http://adsabs.harvard.edu/abs/2017arXiv171001188S} {} (\mn@eprint {arXiv}
  {1710.01188})

\bibitem[\protect\citeauthoryear{{S{\'a}nchez-Menguiano}
  et~al.,}{{S{\'a}nchez-Menguiano} et~al.}{2017b}]{Sanchez-Menguiano17a}
{S{\'a}nchez-Menguiano} L.,  et~al., 2017b, \mn@doi [\aap]
  {10.1051/0004-6361/201630062}, \href
  {http://adsabs.harvard.edu/abs/2017A%26A...603A.113S} {603, A113}

\bibitem[\protect\citeauthoryear{{S{\'a}nchez} et~al.,}{{S{\'a}nchez}
  et~al.}{2012}]{Sanchez12a}
{S{\'a}nchez} S.~F.,  et~al., 2012, \mn@doi [\aap]
  {10.1051/0004-6361/201117353}, \href
  {http://adsabs.harvard.edu/abs/2012A%26A...538A...8S} {538, A8}

\bibitem[\protect\citeauthoryear{{Sanchez} et~al.,}{{Sanchez}
  et~al.}{2013a}]{Sanchez13b}
{Sanchez} S.~F.,  et~al., 2013a, preprint, \href
  {http://adsabs.harvard.edu/abs/2013arXiv1311.7052S} {} (\mn@eprint {}
  {1311.7052})

\bibitem[\protect\citeauthoryear{{S{\'a}nchez} et~al.,}{{S{\'a}nchez}
  et~al.}{2013b}]{Sanchez13a}
{S{\'a}nchez} S.~F.,  et~al., 2013b, \mn@doi [\aap]
  {10.1051/0004-6361/201220669}, \href
  {http://adsabs.harvard.edu/abs/2013A%26A...554A..58S} {554, A58}

\bibitem[\protect\citeauthoryear{{S{\'a}nchez} et~al.,}{{S{\'a}nchez}
  et~al.}{2014}]{Sanchez14}
{S{\'a}nchez} S.~F.,  et~al., 2014, \mn@doi [\aap]
  {10.1051/0004-6361/201322343}, \href
  {http://adsabs.harvard.edu/abs/2014A%26A...563A..49S} {563, A49}

\bibitem[\protect\citeauthoryear{{S{\'a}nchez} et~al.,}{{S{\'a}nchez}
  et~al.}{2017}]{Sanchez17}
{S{\'a}nchez} S.~F.,  et~al., 2017, \mn@doi [\mnras] {10.1093/mnras/stx808},
  \href {http://adsabs.harvard.edu/abs/2017MNRAS.469.2121S} {469, 2121}

\bibitem[\protect\citeauthoryear{{Sandage}}{{Sandage}}{1972}]{Sandage72}
{Sandage} A.,  1972, \mn@doi [\apj] {10.1086/151606}, \href
  {http://adsabs.harvard.edu/abs/1972ApJ...176...21S} {176, 21}

\bibitem[\protect\citeauthoryear{{Sanders} et~al.,}{{Sanders}
  et~al.}{2015}]{Sanders15}
{Sanders} R.~L.,  et~al., 2015, \mn@doi [\apj] {10.1088/0004-637X/799/2/138},
  \href {http://adsabs.harvard.edu/abs/2015ApJ...799..138S} {799, 138}

\bibitem[\protect\citeauthoryear{{Sanders} et~al.,}{{Sanders}
  et~al.}{2016a}]{Sanders16}
{Sanders} J.~S.,  et~al., 2016a, \mn@doi [\mnras] {10.1093/mnras/stv2972},
  \href {http://adsabs.harvard.edu/abs/2016MNRAS.457...82S} {457, 82}

\bibitem[\protect\citeauthoryear{{Sanders} et~al.,}{{Sanders}
  et~al.}{2016b}]{Sanders16a}
{Sanders} R.~L.,  et~al., 2016b, \mn@doi [\apj] {10.3847/0004-637X/816/1/23},
  \href {http://adsabs.harvard.edu/abs/2016ApJ...816...23S} {816, 23}

\bibitem[\protect\citeauthoryear{{Sanders} et~al.,}{{Sanders}
  et~al.}{2016c}]{Sanders16b}
{Sanders} R.~L.,  et~al., 2016c, \mn@doi [\apjl] {10.3847/2041-8205/825/2/L23},
  \href {http://adsabs.harvard.edu/abs/2016ApJ...825L..23S} {825, L23}

\bibitem[\protect\citeauthoryear{{Sanders}, {Shapley}, {Zhang}  \&
  {Yan}}{{Sanders} et~al.}{2017}]{Sanders17a}
{Sanders} R.~L.,  {Shapley} A.~E.,  {Zhang} K.,   {Yan} R.,  2017, \mn@doi
  [\apj] {10.3847/1538-4357/aa93e4}, \href
  {http://adsabs.harvard.edu/abs/2017ApJ...850..136S} {850, 136}

\bibitem[\protect\citeauthoryear{{Sanders} et~al.,}{{Sanders}
  et~al.}{2018}]{Sanders18}
{Sanders} R.~L.,  et~al., 2018, \mn@doi [\apj] {10.3847/1538-4357/aabcbd},
  \href {http://adsabs.harvard.edu/abs/2018ApJ...858...99S} {858, 99}

\bibitem[\protect\citeauthoryear{{Santini} et~al.,}{{Santini}
  et~al.}{2015}]{Santini15}
{Santini} P.,  et~al., 2015, \mn@doi [\apj] {10.1088/0004-637X/801/2/97}, \href
  {http://adsabs.harvard.edu/abs/2015ApJ...801...97S} {801, 97}

\bibitem[\protect\citeauthoryear{{Sarzi} et~al.,}{{Sarzi}
  et~al.}{2010}]{Sarzi10}
{Sarzi} M.,  et~al., 2010, \mn@doi [\mnras] {10.1111/j.1365-2966.2009.16039.x},
  \href {http://adsabs.harvard.edu/abs/2010MNRAS.402.2187S} {402, 2187}

\bibitem[\protect\citeauthoryear{{Sarzi}, {Spiniello}, {Barbera},
  {Krajnovi{\'c}}  \& {Bosch}}{{Sarzi} et~al.}{2018}]{Sarzi18}
{Sarzi} M.,  {Spiniello} C.,  {Barbera} F.~L.,  {Krajnovi{\'c}} D.,   {Bosch}
  R.~v.~d.,  2018, \mn@doi [\mnras] {10.1093/mnras/sty1092}, \href
  {http://adsabs.harvard.edu/abs/2018MNRAS.tmp.1090S} {}

\bibitem[\protect\citeauthoryear{{Sato}, {Tokoi}, {Matsushita}, {Ishisaki},
  {Yamasaki}, {Ishida}  \& {Ohashi}}{{Sato} et~al.}{2007}]{Sato07}
{Sato} K.,  {Tokoi} K.,  {Matsushita} K.,  {Ishisaki} Y.,  {Yamasaki} N.~Y.,
  {Ishida} M.,   {Ohashi} T.,  2007, \mn@doi [\apjl] {10.1086/522031}, \href
  {http://adsabs.harvard.edu/abs/2007ApJ...667L..41S} {667, L41}

\bibitem[\protect\citeauthoryear{{Savage} \& {Sembach}}{{Savage} \&
  {Sembach}}{1996}]{Savage96}
{Savage} B.~D.,  {Sembach} K.~R.,  1996, \mn@doi [\araa]
  {10.1146/annurev.astro.34.1.279}, \href
  {http://adsabs.harvard.edu/abs/1996ARA%26A..34..279S} {34, 279}

\bibitem[\protect\citeauthoryear{{Savaglio} et~al.,}{{Savaglio}
  et~al.}{2004}]{Savaglio04}
{Savaglio} S.,  et~al., 2004, \mn@doi [\apj] {10.1086/380903}, \href
  {http://adsabs.harvard.edu/abs/2004ApJ...602...51S} {602, 51}

\bibitem[\protect\citeauthoryear{{Savaglio} et~al.,}{{Savaglio}
  et~al.}{2005}]{Savaglio05}
{Savaglio} S.,  et~al., 2005, \mn@doi [\apj] {10.1086/497331}, \href
  {http://adsabs.harvard.edu/abs/2005ApJ...635..260S} {635, 260}

\bibitem[\protect\citeauthoryear{{Sbordone}, {Bonifacio}, {Buonanno},
  {Marconi}, {Monaco}  \& {Zaggia}}{{Sbordone} et~al.}{2007}]{Sbordone07}
{Sbordone} L.,  {Bonifacio} P.,  {Buonanno} R.,  {Marconi} G.,  {Monaco} L.,
  {Zaggia} S.,  2007, \mn@doi [\aap] {10.1051/0004-6361:20066385}, \href
  {http://adsabs.harvard.edu/abs/2007A%26A...465..815S} {465, 815}

\bibitem[\protect\citeauthoryear{{Scannapieco}, {Gadotti}, {Jonsson}  \&
  {White}}{{Scannapieco} et~al.}{2010}]{Scannapieco10}
{Scannapieco} C.,  {Gadotti} D.~A.,  {Jonsson} P.,   {White} S.~D.~M.,  2010,
  \mn@doi [\mnras] {10.1111/j.1745-3933.2010.00900.x}, 407, L41

\bibitem[\protect\citeauthoryear{{Schaerer}}{{Schaerer}}{2003}]{Schaerer03}
{Schaerer} D.,  2003, \mn@doi [\aap] {10.1051/0004-6361:20021525}, \href
  {http://adsabs.harvard.edu/abs/2003A%26A...397..527S} {397, 527}

\bibitem[\protect\citeauthoryear{{Schaye} et~al.,}{{Schaye}
  et~al.}{2015}]{Schaye15}
{Schaye} J.,  et~al., 2015, \mn@doi [\mnras] {10.1093/mnras/stu2058}, \href
  {http://adsabs.harvard.edu/abs/2015MNRAS.446..521S} {446, 521}

\bibitem[\protect\citeauthoryear{{Schiavon}}{{Schiavon}}{2007}]{Schiavon07}
{Schiavon} R.~P.,  2007, \mn@doi [\apjs] {10.1086/511753}, \href
  {http://adsabs.harvard.edu/abs/2007ApJS..171..146S} {171, 146}

\bibitem[\protect\citeauthoryear{{Schneider}, {Omukai}, {Bianchi}  \&
  {Valiante}}{{Schneider} et~al.}{2012a}]{Schneider12a}
{Schneider} R.,  {Omukai} K.,  {Bianchi} S.,   {Valiante} R.,  2012a, \mn@doi
  [\mnras] {10.1111/j.1365-2966.2011.19818.x}, \href
  {http://adsabs.harvard.edu/abs/2012MNRAS.419.1566S} {419, 1566}

\bibitem[\protect\citeauthoryear{{Schneider}, {Omukai}, {Limongi}, {Ferrara},
  {Salvaterra}, {Chieffi}  \& {Bianchi}}{{Schneider}
  et~al.}{2012b}]{Schneider12b}
{Schneider} R.,  {Omukai} K.,  {Limongi} M.,  {Ferrara} A.,  {Salvaterra} R.,
  {Chieffi} A.,   {Bianchi} S.,  2012b, \mn@doi [\mnras]
  {10.1111/j.1745-3933.2012.01257.x}, \href
  {http://adsabs.harvard.edu/abs/2012MNRAS.423L..60S} {423, L60}

\bibitem[\protect\citeauthoryear{{Sch{\"o}nrich} \& {Binney}}{{Sch{\"o}nrich}
  \& {Binney}}{2009a}]{Schonrich09b}
{Sch{\"o}nrich} R.,  {Binney} J.,  2009a, \mn@doi [\mnras]
  {10.1111/j.1365-2966.2009.14750.x}, \href
  {http://adsabs.harvard.edu/abs/2009MNRAS.396..203S} {396, 203}

\bibitem[\protect\citeauthoryear{{Sch{\"o}nrich} \& {Binney}}{{Sch{\"o}nrich}
  \& {Binney}}{2009b}]{Schonrich09a}
{Sch{\"o}nrich} R.,  {Binney} J.,  2009b, \mn@doi [\mnras]
  {10.1111/j.1365-2966.2009.15365.x}, \href
  {http://adsabs.harvard.edu/abs/2009MNRAS.399.1145S} {399, 1145}

\bibitem[\protect\citeauthoryear{{Sch{\"o}nrich} \& {McMillan}}{{Sch{\"o}nrich}
  \& {McMillan}}{2017}]{Schonrich17}
{Sch{\"o}nrich} R.,  {McMillan} P.~J.,  2017, \mn@doi [\mnras]
  {10.1093/mnras/stx093}, \href
  {http://adsabs.harvard.edu/abs/2017MNRAS.467.1154S} {467, 1154}

\bibitem[\protect\citeauthoryear{{Schultheis} et~al.,}{{Schultheis}
  et~al.}{2017}]{Schultheis17}
{Schultheis} M.,  et~al., 2017, \mn@doi [\aap] {10.1051/0004-6361/201630154},
  \href {http://adsabs.harvard.edu/abs/2017A%26A...600A..14S} {600, A14}

\bibitem[\protect\citeauthoryear{{Searle}}{{Searle}}{1971}]{Searle71}
{Searle} L.,  1971, \mn@doi [\apj] {10.1086/151090}, \href
  {http://adsabs.harvard.edu/abs/1971ApJ...168..327S} {168, 327}

\bibitem[\protect\citeauthoryear{{Searle} \& {Sargent}}{{Searle} \&
  {Sargent}}{1972}]{Searle72}
{Searle} L.,  {Sargent} W.~L.~W.,  1972, \mn@doi [\apj] {10.1086/151398}, \href
  {http://adsabs.harvard.edu/abs/1972ApJ...173...25S} {173, 25}

\bibitem[\protect\citeauthoryear{{Segers}, {Crain}, {Schaye}, {Bower},
  {Furlong}, {Schaller}  \& {Theuns}}{{Segers} et~al.}{2016}]{Segers16}
{Segers} M.~C.,  {Crain} R.~A.,  {Schaye} J.,  {Bower} R.~G.,  {Furlong} M.,
  {Schaller} M.,   {Theuns} T.,  2016, \mn@doi [\mnras]
  {10.1093/mnras/stv2562}, \href
  {http://adsabs.harvard.edu/abs/2016MNRAS.456.1235S} {456, 1235}

\bibitem[\protect\citeauthoryear{{Senchyna} \& {Stark}}{{Senchyna} \&
  {Stark}}{2018}]{Senchyna18}
{Senchyna} P.,  {Stark} D.~P.,  2018, preprint, \href
  {http://adsabs.harvard.edu/abs/2018arXiv180600551S} {} (\mn@eprint {arXiv}
  {1806.00551})

\bibitem[\protect\citeauthoryear{{Sestito}, {Bragaglia}, {Randich}, {Carretta},
  {Prisinzano}  \& {Tosi}}{{Sestito} et~al.}{2006}]{Sestito06}
{Sestito} P.,  {Bragaglia} A.,  {Randich} S.,  {Carretta} E.,  {Prisinzano} L.,
    {Tosi} M.,  2006, \mn@doi [\aap] {10.1051/0004-6361:20065175}, \href
  {http://adsabs.harvard.edu/abs/2006A%26A...458..121S} {458, 121}

\bibitem[\protect\citeauthoryear{{Shapley}, {Steidel}, {Pettini}  \&
  {Adelberger}}{{Shapley} et~al.}{2003}]{Shapley03}
{Shapley} A.~E.,  {Steidel} C.~C.,  {Pettini} M.,   {Adelberger} K.~L.,  2003,
  \mn@doi [\apj] {10.1086/373922}, \href
  {http://adsabs.harvard.edu/abs/2003ApJ...588...65S} {588, 65}

\bibitem[\protect\citeauthoryear{{Shapley}, {Coil}, {Ma}  \& {Bundy}}{{Shapley}
  et~al.}{2005}]{Shapley05a}
{Shapley} A.~E.,  {Coil} A.~L.,  {Ma} C.-P.,   {Bundy} K.,  2005, \mn@doi
  [\apj] {10.1086/497630}, \href
  {http://adsabs.harvard.edu/abs/2005ApJ...635.1006S} {635, 1006}

\bibitem[\protect\citeauthoryear{{Shapley} et~al.,}{{Shapley}
  et~al.}{2015}]{Shapley15}
{Shapley} A.~E.,  et~al., 2015, \mn@doi [\apj] {10.1088/0004-637X/801/2/88},
  \href {http://adsabs.harvard.edu/abs/2015ApJ...801...88S} {801, 88}

\bibitem[\protect\citeauthoryear{{Sheth}, {Jimenez}, {Panter}  \&
  {Heavens}}{{Sheth} et~al.}{2006}]{Sheth06}
{Sheth} R.~K.,  {Jimenez} R.,  {Panter} B.,   {Heavens} A.~F.,  2006, \mn@doi
  [\apjl] {10.1086/508683}, \href
  {http://adsabs.harvard.edu/abs/2006ApJ...650L..25S} {650, L25}

\bibitem[\protect\citeauthoryear{{Sheth} et~al.,}{{Sheth}
  et~al.}{2008}]{Sheth08}
{Sheth} K.,  et~al., 2008, \mn@doi [\apj] {10.1086/524980}, \href
  {http://adsabs.harvard.edu/abs/2008ApJ...675.1141S} {675, 1141}

\bibitem[\protect\citeauthoryear{{Shields}, {Skillman}  \&
  {Kennicutt}}{{Shields} et~al.}{1991}]{Shields91}
{Shields} G.~A.,  {Skillman} E.~D.,   {Kennicutt} Jr. R.~C.,  1991, \mn@doi
  [\apj] {10.1086/169872}, \href
  {http://adsabs.harvard.edu/abs/1991ApJ...371...82S} {371, 82}

\bibitem[\protect\citeauthoryear{{Shimakawa}, {Kodama}, {Tadaki}, {Hayashi},
  {Koyama}  \& {Tanaka}}{{Shimakawa} et~al.}{2015a}]{Shimakawa15a}
{Shimakawa} R.,  {Kodama} T.,  {Tadaki} K.-i.,  {Hayashi} M.,  {Koyama} Y.,
  {Tanaka} I.,  2015a, \mn@doi [\mnras] {10.1093/mnras/stv051}, \href
  {http://adsabs.harvard.edu/abs/2015MNRAS.448..666S} {448, 666}

\bibitem[\protect\citeauthoryear{{Shimakawa} et~al.,}{{Shimakawa}
  et~al.}{2015b}]{Shimakawa15b}
{Shimakawa} R.,  et~al., 2015b, \mn@doi [\mnras] {10.1093/mnras/stv915}, \href
  {http://adsabs.harvard.edu/abs/2015MNRAS.451.1284S} {451, 1284}

\bibitem[\protect\citeauthoryear{{Shin}, {Woo}, {Nagao}  \& {Kim}}{{Shin}
  et~al.}{2013}]{Shin13}
{Shin} J.,  {Woo} J.-H.,  {Nagao} T.,   {Kim} S.~C.,  2013, \mn@doi [\apj]
  {10.1088/0004-637X/763/1/58}, \href
  {http://adsabs.harvard.edu/abs/2013ApJ...763...58S} {763, 58}

\bibitem[\protect\citeauthoryear{{Shin}, {nagao}  \& {Woo}}{{Shin}
  et~al.}{2017}]{Shin17}
{Shin} J.,  {nagao} T.,   {Woo} J.-H.,  2017, \mn@doi [\apj]
  {10.3847/1538-4357/835/1/24}, \href
  {http://adsabs.harvard.edu/abs/2017ApJ...835...24S} {835, 24}

\bibitem[\protect\citeauthoryear{{Shirazi}, {Vegetti}, {Nesvadba}, {Allam},
  {Brinchmann}  \& {Tucker}}{{Shirazi} et~al.}{2014}]{Shirazi14}
{Shirazi} M.,  {Vegetti} S.,  {Nesvadba} N.,  {Allam} S.,  {Brinchmann} J.,
  {Tucker} D.,  2014, \mn@doi [\mnras] {10.1093/mnras/stu316}, \href
  {http://adsabs.harvard.edu/abs/2014MNRAS.440.2201S} {440, 2201}

\bibitem[\protect\citeauthoryear{{Shivaei} et~al.,}{{Shivaei}
  et~al.}{2018}]{Shivaei18}
{Shivaei} I.,  et~al., 2018, \mn@doi [\apj] {10.3847/1538-4357/aaad62}, \href
  {http://adsabs.harvard.edu/abs/2018ApJ...855...42S} {855, 42}

\bibitem[\protect\citeauthoryear{{Shull} \& {McKee}}{{Shull} \&
  {McKee}}{1979}]{Shull79}
{Shull} J.~M.,  {McKee} C.~F.,  1979, \mn@doi [\apj] {10.1086/156712}, \href
  {http://adsabs.harvard.edu/abs/1979ApJ...227..131S} {227, 131}

\bibitem[\protect\citeauthoryear{{Shull}, {Danforth}  \& {Tilton}}{{Shull}
  et~al.}{2014}]{Shull14}
{Shull} J.~M.,  {Danforth} C.~W.,   {Tilton} E.~M.,  2014, \mn@doi [\apj]
  {10.1088/0004-637X/796/1/49}, \href
  {http://adsabs.harvard.edu/abs/2014ApJ...796...49S} {796, 49}

\bibitem[\protect\citeauthoryear{{Silk} \& {Mamon}}{{Silk} \&
  {Mamon}}{2012}]{Silk12}
{Silk} J.,  {Mamon} G.~A.,  2012, \mn@doi [Research in Astronomy and
  Astrophysics] {10.1088/1674-4527/12/8/004}, \href
  {http://adsabs.harvard.edu/abs/2012RAA....12..917S} {12, 917}

\bibitem[\protect\citeauthoryear{{Silk}, {Di Cintio}  \& {Dvorkin}}{{Silk}
  et~al.}{2014}]{Silk14}
{Silk} J.,  {Di Cintio} A.,   {Dvorkin} I.,  2014, in Proceedings of the
  International School of Physics 'Enrico Fermi' Course 186 'New Horizons for
  Observational Cosmology' Vol. 186, p. 137-187. pp 137--187 (\mn@eprint
  {arXiv} {1312.0107}), \mn@doi{10.3254/978-1-61499-476-3-137}

\bibitem[\protect\citeauthoryear{{Simcoe} et~al.,}{{Simcoe}
  et~al.}{2011}]{Simcoe11b}
{Simcoe} R.~A.,  et~al., 2011, \mn@doi [\apj] {10.1088/0004-637X/743/1/21},
  \href {http://adsabs.harvard.edu/abs/2011ApJ...743...21S} {743, 21}

\bibitem[\protect\citeauthoryear{{Simionescu}, {Werner}, {Forman}, {Miller},
  {Takei}, {B{\"o}hringer}, {Churazov}  \& {Nulsen}}{{Simionescu}
  et~al.}{2010}]{Simionescu10}
{Simionescu} A.,  {Werner} N.,  {Forman} W.~R.,  {Miller} E.~D.,  {Takei} Y.,
  {B{\"o}hringer} H.,  {Churazov} E.,   {Nulsen} P.~E.~J.,  2010, \mn@doi
  [\mnras] {10.1111/j.1365-2966.2010.16450.x}, \href
  {http://adsabs.harvard.edu/abs/2010MNRAS.405...91S} {405, 91}

\bibitem[\protect\citeauthoryear{{Simionescu}, {Werner}, {Urban}, {Allen},
  {Ichinohe}  \& {Zhuravleva}}{{Simionescu} et~al.}{2015}]{Simionescu15}
{Simionescu} A.,  {Werner} N.,  {Urban} O.,  {Allen} S.~W.,  {Ichinohe} Y.,
  {Zhuravleva} I.,  2015, \mn@doi [\apjl] {10.1088/2041-8205/811/2/L25}, \href
  {http://adsabs.harvard.edu/abs/2015ApJ...811L..25S} {811, L25}

\bibitem[\protect\citeauthoryear{{Simionescu} et~al.,}{{Simionescu}
  et~al.}{2018}]{Simionescu18}
{Simionescu} A.,  et~al., 2018, preprint, \href
  {http://adsabs.harvard.edu/abs/2018arXiv180600932S} {} (\mn@eprint {arXiv}
  {1806.00932})

\bibitem[\protect\citeauthoryear{{Simon} \& {Hamann}}{{Simon} \&
  {Hamann}}{2010}]{Simon10a}
{Simon} L.~E.,  {Hamann} F.,  2010, \mn@doi [\mnras]
  {10.1111/j.1365-2966.2010.17306.x}, \href
  {http://adsabs.harvard.edu/abs/2010MNRAS.409..269S} {409, 269}

\bibitem[\protect\citeauthoryear{{Sim{\'o}n-D{\'{\i}}az}}{{Sim{\'o}n-D{\'{\i}}az}}{2010}]{Simon-Diaz10}
{Sim{\'o}n-D{\'{\i}}az} S.,  2010, \mn@doi [\aap]
  {10.1051/0004-6361/200913120}, \href
  {http://adsabs.harvard.edu/abs/2010A%26A...510A..22S} {510, A22}

\bibitem[\protect\citeauthoryear{{Sim{\'o}n-D{\'{\i}}az} \&
  {Stasi{\'n}ska}}{{Sim{\'o}n-D{\'{\i}}az} \&
  {Stasi{\'n}ska}}{2011}]{Simon-Diaz11}
{Sim{\'o}n-D{\'{\i}}az} S.,  {Stasi{\'n}ska} G.,  2011, \mn@doi [\aap]
  {10.1051/0004-6361/201015512}, \href
  {http://adsabs.harvard.edu/abs/2011A%26A...526A..48S} {526, A48}

\bibitem[\protect\citeauthoryear{{Singh} et~al.,}{{Singh}
  et~al.}{2013}]{Singh13}
{Singh} R.,  et~al., 2013, \mn@doi [\aap] {10.1051/0004-6361/201322062}, \href
  {http://adsabs.harvard.edu/abs/2013A%26A...558A..43S} {558, A43}

\bibitem[\protect\citeauthoryear{{Skillman}}{{Skillman}}{1989}]{Skillman89c}
{Skillman} E.~D.,  1989, \mn@doi [\apj] {10.1086/168179}, \href
  {http://adsabs.harvard.edu/abs/1989ApJ...347..883S} {347, 883}

\bibitem[\protect\citeauthoryear{{Skillman}, {Melnick}, {Terlevich}  \&
  {Moles}}{{Skillman} et~al.}{1988}]{Skillman88}
{Skillman} E.~D.,  {Melnick} J.,  {Terlevich} R.,   {Moles} M.,  1988, \aap,
  \href {http://adsabs.harvard.edu/abs/1988A%26A...196...31S} {196, 31}

\bibitem[\protect\citeauthoryear{{Skillman}, {Kennicutt}  \&
  {Hodge}}{{Skillman} et~al.}{1989}]{Skillman89b}
{Skillman} E.~D.,  {Kennicutt} R.~C.,   {Hodge} P.~W.,  1989, \mn@doi [\apj]
  {10.1086/168178}, \href {http://adsabs.harvard.edu/abs/1989ApJ...347..875S}
  {347, 875}

\bibitem[\protect\citeauthoryear{{Skillman}, {Kennicutt}, {Shields}  \&
  {Zaritsky}}{{Skillman} et~al.}{1996}]{Skillman96}
{Skillman} E.~D.,  {Kennicutt} Jr. R.~C.,  {Shields} G.~A.,   {Zaritsky} D.,
  1996, \mn@doi [\apj] {10.1086/177138}, \href
  {http://adsabs.harvard.edu/abs/1996ApJ...462..147S} {462, 147}

\bibitem[\protect\citeauthoryear{{Somerville} \& {Dav{\'e}}}{{Somerville} \&
  {Dav{\'e}}}{2015}]{Somerville15}
{Somerville} R.~S.,  {Dav{\'e}} R.,  2015, \mn@doi [\araa]
  {10.1146/annurev-astro-082812-140951}, \href
  {http://adsabs.harvard.edu/abs/2015ARA%26A..53...51S} {53, 51}

\bibitem[\protect\citeauthoryear{{Somerville}, {Hopkins}, {Cox}, {Robertson}
  \& {Hernquist}}{{Somerville} et~al.}{2008}]{Somerville08}
{Somerville} R.~S.,  {Hopkins} P.~F.,  {Cox} T.~J.,  {Robertson} B.~E.,
  {Hernquist} L.,  2008, \mn@doi [\mnras] {10.1111/j.1365-2966.2008.13805.x},
  \href {http://adsabs.harvard.edu/abs/2008MNRAS.391..481S} {391, 481}

\bibitem[\protect\citeauthoryear{{Somerville}, {Gilmore}, {Primack}  \&
  {Dom{\'{\i}}nguez}}{{Somerville} et~al.}{2012}]{Somerville12}
{Somerville} R.~S.,  {Gilmore} R.~C.,  {Primack} J.~R.,   {Dom{\'{\i}}nguez}
  A.,  2012, \mn@doi [\mnras] {10.1111/j.1365-2966.2012.20490.x}, \href
  {http://adsabs.harvard.edu/abs/2012MNRAS.423.1992S} {423, 1992}

\bibitem[\protect\citeauthoryear{{Sommariva}, {Mannucci}, {Cresci}, {Maiolino},
  {Marconi}, {Nagao}, {Baroni}  \& {Grazian}}{{Sommariva}
  et~al.}{2012}]{Sommariva12}
{Sommariva} V.,  {Mannucci} F.,  {Cresci} G.,  {Maiolino} R.,  {Marconi} A.,
  {Nagao} T.,  {Baroni} A.,   {Grazian} A.,  2012, \mn@doi [\aap]
  {10.1051/0004-6361/201118134}, \href
  {http://adsabs.harvard.edu/abs/2012A%26A...539A.136S} {539, A136}

\bibitem[\protect\citeauthoryear{{Song} et~al.,}{{Song} et~al.}{2014}]{Song14}
{Song} M.,  et~al., 2014, \mn@doi [\apj] {10.1088/0004-637X/791/1/3}, \href
  {http://adsabs.harvard.edu/abs/2014ApJ...791....3S} {791, 3}

\bibitem[\protect\citeauthoryear{{Spite} et~al.,}{{Spite}
  et~al.}{2005}]{Spite05}
{Spite} M.,  et~al., 2005, \mn@doi [\aap] {10.1051/0004-6361:20041274}, \href
  {http://adsabs.harvard.edu/abs/2005A%26A...430..655S} {430, 655}

\bibitem[\protect\citeauthoryear{{Spitoni}}{{Spitoni}}{2015}]{Spitoni15b}
{Spitoni} E.,  2015, \mn@doi [\mnras] {10.1093/mnras/stv1008}, \href
  {http://adsabs.harvard.edu/abs/2015MNRAS.451.1090S} {451, 1090}

\bibitem[\protect\citeauthoryear{{Spitoni} \& {Matteucci}}{{Spitoni} \&
  {Matteucci}}{2011}]{Spitoni11}
{Spitoni} E.,  {Matteucci} F.,  2011, \mn@doi [\aap]
  {10.1051/0004-6361/201015749}, \href
  {http://adsabs.harvard.edu/abs/2011A%26A...531A..72S} {531, A72}

\bibitem[\protect\citeauthoryear{{Spitoni}, {Matteucci}  \&
  {Marcon-Uchida}}{{Spitoni} et~al.}{2013}]{Spitoni13}
{Spitoni} E.,  {Matteucci} F.,   {Marcon-Uchida} M.~M.,  2013, \mn@doi [\aap]
  {10.1051/0004-6361/201220401}, \href
  {http://adsabs.harvard.edu/abs/2013A%26A...551A.123S} {551, A123}

\bibitem[\protect\citeauthoryear{{Spitoni}, {Romano}, {Matteucci}  \&
  {Ciotti}}{{Spitoni} et~al.}{2015}]{Spitoni15a}
{Spitoni} E.,  {Romano} D.,  {Matteucci} F.,   {Ciotti} L.,  2015, \mn@doi
  [\apj] {10.1088/0004-637X/802/2/129}, \href
  {http://adsabs.harvard.edu/abs/2015ApJ...802..129S} {802, 129}

\bibitem[\protect\citeauthoryear{{Spitoni}, {Vincenzo}  \&
  {Matteucci}}{{Spitoni} et~al.}{2017}]{Spitoni17}
{Spitoni} E.,  {Vincenzo} F.,   {Matteucci} F.,  2017, \mn@doi [\aap]
  {10.1051/0004-6361/201629745}, \href
  {http://adsabs.harvard.edu/abs/2017A%26A...599A...6S} {599, A6}

\bibitem[\protect\citeauthoryear{{Spolaor}, {Kobayashi}, {Forbes}, {Couch}  \&
  {Hau}}{{Spolaor} et~al.}{2010}]{Spolaor10}
{Spolaor} M.,  {Kobayashi} C.,  {Forbes} D.~A.,  {Couch} W.~J.,   {Hau}
  G.~K.~T.,  2010, \mn@doi [\mnras] {10.1111/j.1365-2966.2010.17080.x}, \href
  {http://adsabs.harvard.edu/abs/2010MNRAS.408..272S} {408, 272}

\bibitem[\protect\citeauthoryear{{Springel} et~al.,}{{Springel}
  et~al.}{2018}]{Springel18}
{Springel} V.,  et~al., 2018, \mn@doi [\mnras] {10.1093/mnras/stx3304}, \href
  {http://adsabs.harvard.edu/abs/2018MNRAS.475..676S} {475, 676}

\bibitem[\protect\citeauthoryear{{Stanghellini} \& {Haywood}}{{Stanghellini} \&
  {Haywood}}{2010}]{Stanghellini10}
{Stanghellini} L.,  {Haywood} M.,  2010, \mn@doi [\apj]
  {10.1088/0004-637X/714/2/1096}, \href
  {http://adsabs.harvard.edu/abs/2010ApJ...714.1096S} {714, 1096}

\bibitem[\protect\citeauthoryear{{Stanghellini} \& {Haywood}}{{Stanghellini} \&
  {Haywood}}{2018}]{Stanghellini18}
{Stanghellini} L.,  {Haywood} M.,  2018, \mn@doi [\apj]
  {10.3847/1538-4357/aacaf8}, \href
  {http://adsabs.harvard.edu/abs/2018ApJ...862...45S} {862, 45}

\bibitem[\protect\citeauthoryear{{Stanghellini}, {Magrini}, {Casasola}  \&
  {Villaver}}{{Stanghellini} et~al.}{2014}]{Stanghellini14}
{Stanghellini} L.,  {Magrini} L.,  {Casasola} V.,   {Villaver} E.,  2014,
  \mn@doi [\aap] {10.1051/0004-6361/201423423}, \href
  {http://adsabs.harvard.edu/abs/2014A%26A...567A..88S} {567, A88}

\bibitem[\protect\citeauthoryear{{Stanway}, {Eldridge}  \& {Becker}}{{Stanway}
  et~al.}{2016}]{Stanway16}
{Stanway} E.~R.,  {Eldridge} J.~J.,   {Becker} G.~D.,  2016, \mn@doi [\mnras]
  {10.1093/mnras/stv2661}, \href
  {http://adsabs.harvard.edu/abs/2016MNRAS.456..485S} {456, 485}

\bibitem[\protect\citeauthoryear{{Stark} et~al.,}{{Stark}
  et~al.}{2013}]{Stark13b}
{Stark} D.~P.,  et~al., 2013, \mn@doi [\mnras] {10.1093/mnras/stt1624}, \href
  {http://adsabs.harvard.edu/abs/2013MNRAS.436.1040S} {436, 1040}

\bibitem[\protect\citeauthoryear{{Stark} et~al.,}{{Stark}
  et~al.}{2017}]{Stark17}
{Stark} D.~P.,  et~al., 2017, \mn@doi [\mnras] {10.1093/mnras/stw2233}, \href
  {http://adsabs.harvard.edu/abs/2017MNRAS.464..469S} {464, 469}

\bibitem[\protect\citeauthoryear{{Starkenburg} et~al.,}{{Starkenburg}
  et~al.}{2013}]{Starkenburg13}
{Starkenburg} E.,  et~al., 2013, \mn@doi [\aap] {10.1051/0004-6361/201220349},
  \href {http://adsabs.harvard.edu/abs/2013A%26A...549A..88S} {549, A88}

\bibitem[\protect\citeauthoryear{{Stasi{\'n}ska}}{{Stasi{\'n}ska}}{2002}]{Stasinska02}
{Stasi{\'n}ska} G.,  2002, in {Henney} W.~J.,  {Franco} J.,   {Martos} M.,
  eds,  Revista Mexicana de Astronomia y Astrofisica Conference Series Vol. 12,
  Revista Mexicana de Astronomia y Astrofisica Conference Series. pp 62--69
  (\mn@eprint {} {astro-ph/0102403})

\bibitem[\protect\citeauthoryear{{Stasi{\'n}ska}}{{Stasi{\'n}ska}}{2006}]{Stasinska06a}
{Stasi{\'n}ska} G.,  2006, \mn@doi [\aap] {10.1051/0004-6361:20065516}, \href
  {http://adsabs.harvard.edu/abs/2006A%26A...454L.127S} {454, L127}

\bibitem[\protect\citeauthoryear{{Stasi{\'n}ska} \& {Szczerba}}{{Stasi{\'n}ska}
  \& {Szczerba}}{2001}]{Stasinska01}
{Stasi{\'n}ska} G.,  {Szczerba} R.,  2001, \mn@doi [\aap]
  {10.1051/0004-6361:20011403}, \href
  {http://adsabs.harvard.edu/abs/2001A%26A...379.1024S} {379, 1024}

\bibitem[\protect\citeauthoryear{{Stasi{\'n}ska}, {Cid Fernandes}, {Mateus},
  {Sodr{\'e}}  \& {Asari}}{{Stasi{\'n}ska} et~al.}{2006}]{Stasinska06b}
{Stasi{\'n}ska} G.,  {Cid Fernandes} R.,  {Mateus} A.,  {Sodr{\'e}} L.,
  {Asari} N.~V.,  2006, \mn@doi [\mnras] {10.1111/j.1365-2966.2006.10732.x},
  \href {http://adsabs.harvard.edu/abs/2006MNRAS.371..972S} {371, 972}

\bibitem[\protect\citeauthoryear{{Stasi{\'n}ska}, {Tenorio-Tagle},
  {Rodr{\'{\i}}guez}  \& {Henney}}{{Stasi{\'n}ska} et~al.}{2007}]{Stasinska07b}
{Stasi{\'n}ska} G.,  {Tenorio-Tagle} G.,  {Rodr{\'{\i}}guez} M.,   {Henney}
  W.~J.,  2007, \mn@doi [\aap] {10.1051/0004-6361:20065675}, \href
  {http://adsabs.harvard.edu/abs/2007A%26A...471..193S} {471, 193}

\bibitem[\protect\citeauthoryear{{Steidel}, {Shapley}, {Pettini}, {Adelberger},
  {Erb}, {Reddy}  \& {Hunt}}{{Steidel} et~al.}{2004}]{Steidel04}
{Steidel} C.~C.,  {Shapley} A.~E.,  {Pettini} M.,  {Adelberger} K.~L.,  {Erb}
  D.~K.,  {Reddy} N.~A.,   {Hunt} M.~P.,  2004, \mn@doi [\apj]
  {10.1086/381960}, \href {http://adsabs.harvard.edu/abs/2004ApJ...604..534S}
  {604, 534}

\bibitem[\protect\citeauthoryear{{Steidel}, {Erb}, {Shapley}, {Pettini},
  {Reddy}, {Bogosavljevi{\'c}}, {Rudie}  \& {Rakic}}{{Steidel}
  et~al.}{2010}]{Steidel10}
{Steidel} C.~C.,  {Erb} D.~K.,  {Shapley} A.~E.,  {Pettini} M.,  {Reddy} N.,
  {Bogosavljevi{\'c}} M.,  {Rudie} G.~C.,   {Rakic} O.,  2010, \mn@doi [\apj]
  {10.1088/0004-637X/717/1/289}, \href
  {http://adsabs.harvard.edu/abs/2010ApJ...717..289S} {717, 289}

\bibitem[\protect\citeauthoryear{{Steidel} et~al.,}{{Steidel}
  et~al.}{2014}]{Steidel14}
{Steidel} C.~C.,  et~al., 2014, \mn@doi [\apj] {10.1088/0004-637X/795/2/165},
  \href {http://adsabs.harvard.edu/abs/2014ApJ...795..165S} {795, 165}

\bibitem[\protect\citeauthoryear{{Steidel}, {Strom}, {Pettini}, {Rudie},
  {Reddy}  \& {Trainor}}{{Steidel} et~al.}{2016}]{Steidel16}
{Steidel} C.~C.,  {Strom} A.~L.,  {Pettini} M.,  {Rudie} G.~C.,  {Reddy} N.~A.,
    {Trainor} R.~F.,  2016, \mn@doi [\apj] {10.3847/0004-637X/826/2/159}, \href
  {http://adsabs.harvard.edu/abs/2016ApJ...826..159S} {826, 159}

\bibitem[\protect\citeauthoryear{{Steinmetz} et~al.,}{{Steinmetz}
  et~al.}{2006}]{Steinmetz06}
{Steinmetz} M.,  et~al., 2006, \mn@doi [\aj] {10.1086/506564}, 132, 1645

\bibitem[\protect\citeauthoryear{{Stern} \& {Laor}}{{Stern} \&
  {Laor}}{2013}]{Stern13}
{Stern} J.,  {Laor} A.,  2013, \mn@doi [\mnras] {10.1093/mnras/stt211}, \href
  {http://adsabs.harvard.edu/abs/2013MNRAS.431..836S} {431, 836}

\bibitem[\protect\citeauthoryear{{Stott} et~al.,}{{Stott}
  et~al.}{2013}]{Stott13}
{Stott} J.~P.,  et~al., 2013, \mn@doi [\mnras] {10.1093/mnras/stt1641}, \href
  {http://adsabs.harvard.edu/abs/2013MNRAS.436.1130S} {436, 1130}

\bibitem[\protect\citeauthoryear{{Stott} et~al.,}{{Stott}
  et~al.}{2014}]{Stott14}
{Stott} J.~P.,  et~al., 2014, \mn@doi [\mnras] {10.1093/mnras/stu1343}, \href
  {http://adsabs.harvard.edu/abs/2014MNRAS.443.2695S} {443, 2695}

\bibitem[\protect\citeauthoryear{{Strom}, {Steidel}, {Rudie}, {Trainor}  \&
  {Pettini}}{{Strom} et~al.}{2017a}]{Strom17b}
{Strom} A.~L.,  {Steidel} C.~C.,  {Rudie} G.~C.,  {Trainor} R.~F.,   {Pettini}
  M.,  2017a, preprint, \href
  {http://adsabs.harvard.edu/abs/2017arXiv171108820S} {} (\mn@eprint {arXiv}
  {1711.08820})

\bibitem[\protect\citeauthoryear{{Strom}, {Steidel}, {Rudie}, {Trainor},
  {Pettini}  \& {Reddy}}{{Strom} et~al.}{2017b}]{Strom17a}
{Strom} A.~L.,  {Steidel} C.~C.,  {Rudie} G.~C.,  {Trainor} R.~F.,  {Pettini}
  M.,   {Reddy} N.~A.,  2017b, \mn@doi [\apj] {10.3847/1538-4357/836/2/164},
  \href {http://adsabs.harvard.edu/abs/2017ApJ...836..164S} {836, 164}

\bibitem[\protect\citeauthoryear{{Sutherland} \& {Dopita}}{{Sutherland} \&
  {Dopita}}{1993}]{Sutherland93}
{Sutherland} R.~S.,  {Dopita} M.~A.,  1993, \mn@doi [\apjs] {10.1086/191823},
  \href {http://adsabs.harvard.edu/abs/1993ApJS...88..253S} {88, 253}

\bibitem[\protect\citeauthoryear{{Suzuki} et~al.,}{{Suzuki}
  et~al.}{2017}]{Suzuki17}
{Suzuki} T.~L.,  et~al., 2017, \mn@doi [\apj] {10.3847/1538-4357/aa8df3}, \href
  {http://adsabs.harvard.edu/abs/2017ApJ...849...39S} {849, 39}

\bibitem[\protect\citeauthoryear{{Swinbank}, {Sobral}, {Smail}, {Geach},
  {Best}, {McCarthy}, {Crain}  \& {Theuns}}{{Swinbank}
  et~al.}{2012}]{Swinbank12}
{Swinbank} A.~M.,  {Sobral} D.,  {Smail} I.,  {Geach} J.~E.,  {Best} P.~N.,
  {McCarthy} I.~G.,  {Crain} R.~A.,   {Theuns} T.,  2012, \mn@doi [\mnras]
  {10.1111/j.1365-2966.2012.21774.x}, \href
  {http://adsabs.harvard.edu/abs/2012MNRAS.426..935S} {426, 935}

\bibitem[\protect\citeauthoryear{{Sybilska} et~al.,}{{Sybilska}
  et~al.}{2017}]{Sybilska17}
{Sybilska} A.,  et~al., 2017, \mn@doi [\mnras] {10.1093/mnras/stx1138}, \href
  {http://adsabs.harvard.edu/abs/2017MNRAS.470..815S} {470, 815}

\bibitem[\protect\citeauthoryear{{Talent}}{{Talent}}{1981}]{Talent81}
{Talent} D.~L.,  1981, PhD thesis, Rice Univ., Houston, TX.

\bibitem[\protect\citeauthoryear{{Tamura} et~al.,}{{Tamura}
  et~al.}{2016}]{Tamura16}
{Tamura} N.,  et~al., 2016, in Ground-based and Airborne Instrumentation for
  Astronomy VI. p. 99081M (\mn@eprint {arXiv} {1608.01075}),
  \mn@doi{10.1117/12.2232103}

\bibitem[\protect\citeauthoryear{{Tamura} et~al.,}{{Tamura}
  et~al.}{2018}]{Tamura18}
{Tamura} Y.,  et~al., 2018, preprint, \href
  {http://adsabs.harvard.edu/abs/2018arXiv180604132T} {} (\mn@eprint {arXiv}
  {1806.04132})

\bibitem[\protect\citeauthoryear{{Tautvai{\v s}ien{e}}, {Drazdauskas},
  {Bragaglia}, {Randich}  \& {{\v Z}enovien{e}}}{{Tautvai{\v s}ien{e}}
  et~al.}{2016}]{Tautvaisiene16}
{Tautvai{\v s}ien{e}} G.,  {Drazdauskas} A.,  {Bragaglia} A.,  {Randich} S.,
  {{\v Z}enovien{e}} R.,  2016, \mn@doi [\aap] {10.1051/0004-6361/201629273},
  \href {http://adsabs.harvard.edu/abs/2016A%26A...595A..16T} {595, A16}

\bibitem[\protect\citeauthoryear{{Telford}, {Dalcanton}, {Skillman}  \&
  {Conroy}}{{Telford} et~al.}{2016}]{Telford16}
{Telford} O.~G.,  {Dalcanton} J.~J.,  {Skillman} E.~D.,   {Conroy} C.,  2016,
  \mn@doi [\apj] {10.3847/0004-637X/827/1/35}, \href
  {http://adsabs.harvard.edu/abs/2016ApJ...827...35T} {827, 35}

\bibitem[\protect\citeauthoryear{{Telford}, {Werk}, {Dalcanton}  \&
  {Williams}}{{Telford} et~al.}{2018}]{Telford18}
{Telford} O.~G.,  {Werk} J.~K.,  {Dalcanton} J.~J.,   {Williams} B.~F.,  2018,
  preprint, \href {http://adsabs.harvard.edu/abs/2018arXiv181102589T} {}
  (\mn@eprint {arXiv} {1811.02589})

\bibitem[\protect\citeauthoryear{{Teplitz} et~al.,}{{Teplitz}
  et~al.}{2000}]{Teplitz00a}
{Teplitz} H.~I.,  et~al., 2000, \mn@doi [\apjl] {10.1086/312595}, \href
  {http://adsabs.harvard.edu/abs/2000ApJ...533L..65T} {533, L65}

\bibitem[\protect\citeauthoryear{{Thomas}, {Greggio}  \& {Bender}}{{Thomas}
  et~al.}{1999}]{Thomas99a}
{Thomas} D.,  {Greggio} L.,   {Bender} R.,  1999, \mn@doi [\mnras]
  {10.1046/j.1365-8711.1999.02138.x}, \href
  {http://adsabs.harvard.edu/abs/1999MNRAS.302..537T} {302, 537}

\bibitem[\protect\citeauthoryear{{Thomas}, {Maraston}  \& {Bender}}{{Thomas}
  et~al.}{2003}]{Thomas03a}
{Thomas} D.,  {Maraston} C.,   {Bender} R.,  2003, \mn@doi [\mnras]
  {10.1046/j.1365-8711.2003.06248.x}, \href
  {http://adsabs.harvard.edu/abs/2003MNRAS.339..897T} {339, 897}

\bibitem[\protect\citeauthoryear{{Thomas}, {Maraston}, {Bender}  \& {Mendes de
  Oliveira}}{{Thomas} et~al.}{2005}]{Thomas05}
{Thomas} D.,  {Maraston} C.,  {Bender} R.,   {Mendes de Oliveira} C.,  2005,
  \mn@doi [\apj] {10.1086/426932}, \href
  {http://adsabs.harvard.edu/abs/2005ApJ...621..673T} {621, 673}

\bibitem[\protect\citeauthoryear{{Thomas}, {Maraston}, {Schawinski}, {Sarzi}
  \& {Silk}}{{Thomas} et~al.}{2010a}]{Thomas10a}
{Thomas} D.,  {Maraston} C.,  {Schawinski} K.,  {Sarzi} M.,   {Silk} J.,
  2010a, \mn@doi [\mnras] {10.1111/j.1365-2966.2010.16427.x}, \href
  {http://adsabs.harvard.edu/abs/2010MNRAS.404.1775T} {404, 1775}

\bibitem[\protect\citeauthoryear{{Thomas}, {Maraston}, {Schawinski}, {Sarzi}
  \& {Silk}}{{Thomas} et~al.}{2010b}]{Thomas10}
{Thomas} D.,  {Maraston} C.,  {Schawinski} K.,  {Sarzi} M.,   {Silk} J.,
  2010b, \mn@doi [\mnras] {10.1111/j.1365-2966.2010.16427.x}, \href
  {http://adsabs.harvard.edu/abs/2010MNRAS.404.1775T} {404, 1775}

\bibitem[\protect\citeauthoryear{{Thomas}, {Maraston}  \& {Johansson}}{{Thomas}
  et~al.}{2011}]{Thomas11d}
{Thomas} D.,  {Maraston} C.,   {Johansson} J.,  2011, \mn@doi [\mnras]
  {10.1111/j.1365-2966.2010.18049.x}, \href
  {http://adsabs.harvard.edu/abs/2011MNRAS.412.2183T} {412, 2183}

\bibitem[\protect\citeauthoryear{{Thuan}, {Izotov}  \& {Lipovetsky}}{{Thuan}
  et~al.}{1995}]{Thuan95}
{Thuan} T.~X.,  {Izotov} Y.~I.,   {Lipovetsky} V.~A.,  1995, \mn@doi [\apj]
  {10.1086/175676}, \href {http://adsabs.harvard.edu/abs/1995ApJ...445..108T}
  {445, 108}

\bibitem[\protect\citeauthoryear{{Tinsley}}{{Tinsley}}{1974}]{Tinsley74}
{Tinsley} B.~M.,  1974, \mn@doi [\apj] {10.1086/153099}, \href
  {http://adsabs.harvard.edu/abs/1974ApJ...192..629T} {192, 629}

\bibitem[\protect\citeauthoryear{{Tinsley}}{{Tinsley}}{1978}]{Tinsley78}
{Tinsley} B.~M.,  1978, \mn@doi [\apj] {10.1086/156116}, \href
  {http://adsabs.harvard.edu/abs/1978ApJ...222...14T} {222, 14}

\bibitem[\protect\citeauthoryear{{Tinsley}}{{Tinsley}}{1980}]{Tinsley80a}
{Tinsley} B.~M.,  1980, FCP, \href
  {http://adsabs.harvard.edu/abs/1980FCPh....5..287T} {5, 287}

\bibitem[\protect\citeauthoryear{{Tissera}, {Pedrosa}, {Sillero}  \&
  {Vilchez}}{{Tissera} et~al.}{2016a}]{Tissera16a}
{Tissera} P.~B.,  {Pedrosa} S.~E.,  {Sillero} E.,   {Vilchez} J.~M.,  2016a,
  \mn@doi [\mnras] {10.1093/mnras/stv2736}, \href
  {http://adsabs.harvard.edu/abs/2016MNRAS.456.2982T} {456, 2982}

\bibitem[\protect\citeauthoryear{{Tissera}, {Machado}, {Sanchez-Blazquez},
  {Pedrosa}, {S{\'a}nchez}, {Snaith}  \& {Vilchez}}{{Tissera}
  et~al.}{2016b}]{Tissera16b}
{Tissera} P.~B.,  {Machado} R.~E.~G.,  {Sanchez-Blazquez} P.,  {Pedrosa} S.~E.,
   {S{\'a}nchez} S.~F.,  {Snaith} O.,   {Vilchez} J.,  2016b, \mn@doi [\aap]
  {10.1051/0004-6361/201628188}, \href
  {http://adsabs.harvard.edu/abs/2016A%26A...592A..93T} {592, A93}

\bibitem[\protect\citeauthoryear{{Tissera}, {Machado}, {Vilchez}, {Pedrosa},
  {Sanchez-Blazquez}  \& {Varela}}{{Tissera} et~al.}{2017}]{Tissera17}
{Tissera} P.~B.,  {Machado} R.~E.~G.,  {Vilchez} J.~M.,  {Pedrosa} S.~E.,
  {Sanchez-Blazquez} P.,   {Varela} S.,  2017, \mn@doi [\aap]
  {10.1051/0004-6361/201628915}, \href
  {http://adsabs.harvard.edu/abs/2017A%26A...604A.118T} {604, A118}

\bibitem[\protect\citeauthoryear{{Tissera}, {Rosas-Guevara}, {Bower}, {Crain},
  {Lagos}, {Schaller}, {Schaye}  \& {Theuns}}{{Tissera}
  et~al.}{2018}]{Tissera18}
{Tissera} P.~B.,  {Rosas-Guevara} Y.,  {Bower} R.~G.,  {Crain} R.~A.,  {Lagos}
  C.~d.~P.,  {Schaller} M.,  {Schaye} J.,   {Theuns} T.,  2018, preprint, \href
  {http://adsabs.harvard.edu/abs/2018arXiv180604575T} {} (\mn@eprint {arXiv}
  {1806.04575})

\bibitem[\protect\citeauthoryear{{Tolstoy}, {Hill}  \& {Tosi}}{{Tolstoy}
  et~al.}{2009}]{Tolstoy09}
{Tolstoy} E.,  {Hill} V.,   {Tosi} M.,  2009, \mn@doi [\araa]
  {10.1146/annurev-astro-082708-101650}, \href
  {http://adsabs.harvard.edu/abs/2009ARA%26A..47..371T} {47, 371}

\bibitem[\protect\citeauthoryear{{Toribio San Cipriano},
  {Dom{\'{\i}}nguez-Guzm{\'a}n}, {Esteban}, {Garc{\'{\i}}a-Rojas},
  {Mesa-Delgado}, {Bresolin}, {Rodr{\'{\i}}guez}  \&
  {Sim{\'o}n-D{\'{\i}}az}}{{Toribio San Cipriano}
  et~al.}{2017}]{Toribio-San-Cipriano17}
{Toribio San Cipriano} L.,  {Dom{\'{\i}}nguez-Guzm{\'a}n} G.,  {Esteban} C.,
  {Garc{\'{\i}}a-Rojas} J.,  {Mesa-Delgado} A.,  {Bresolin} F.,
  {Rodr{\'{\i}}guez} M.,   {Sim{\'o}n-D{\'{\i}}az} S.,  2017, \mn@doi [\mnras]
  {10.1093/mnras/stx328}, \href
  {http://adsabs.harvard.edu/abs/2017MNRAS.467.3759T} {467, 3759}

\bibitem[\protect\citeauthoryear{{Torres-Flores}, {Scarano}, {Mendes de
  Oliveira}, {de Mello}, {Amram}  \& {Plana}}{{Torres-Flores}
  et~al.}{2014a}]{Torres-Flores13}
{Torres-Flores} S.,  {Scarano} S.,  {Mendes de Oliveira} C.,  {de Mello} D.~F.,
   {Amram} P.,   {Plana} H.,  2014a, \mn@doi [\mnras] {10.1093/mnras/stt2340},
  \href {http://adsabs.harvard.edu/abs/2014MNRAS.438.1894T} {438, 1894}

\bibitem[\protect\citeauthoryear{{Torres-Flores}, {Scarano}, {Mendes de
  Oliveira}, {de Mello}, {Amram}  \& {Plana}}{{Torres-Flores}
  et~al.}{2014b}]{Torres-Flores14}
{Torres-Flores} S.,  {Scarano} S.,  {Mendes de Oliveira} C.,  {de Mello} D.~F.,
   {Amram} P.,   {Plana} H.,  2014b, \mn@doi [\mnras] {10.1093/mnras/stt2340},
  \href {http://adsabs.harvard.edu/abs/2014MNRAS.438.1894T} {438, 1894}

\bibitem[\protect\citeauthoryear{{Torres-Papaqui}, {Coziol}, {Ortega-Minakata}
  \& {Neri-Larios}}{{Torres-Papaqui} et~al.}{2012}]{Torres-Papaqui12}
{Torres-Papaqui} J.~P.,  {Coziol} R.,  {Ortega-Minakata} R.~A.,   {Neri-Larios}
  D.~M.,  2012, \mn@doi [\apj] {10.1088/0004-637X/754/2/144}, \href
  {http://adsabs.harvard.edu/abs/2012ApJ...754..144T} {754, 144}

\bibitem[\protect\citeauthoryear{{Torrey}, {Cox}, {Kewley}  \&
  {Hernquist}}{{Torrey} et~al.}{2012}]{Torrey12}
{Torrey} P.,  {Cox} T.~J.,  {Kewley} L.,   {Hernquist} L.,  2012, \mn@doi
  [\apj] {10.1088/0004-637X/746/1/108}, 746, 108

\bibitem[\protect\citeauthoryear{{Torrey}, {Vogelsberger}, {Genel}, {Sijacki},
  {Springel}  \& {Hernquist}}{{Torrey} et~al.}{2014}]{Torrey14}
{Torrey} P.,  {Vogelsberger} M.,  {Genel} S.,  {Sijacki} D.,  {Springel} V.,
  {Hernquist} L.,  2014, \mn@doi [\mnras] {10.1093/mnras/stt2295}, \href
  {http://adsabs.harvard.edu/abs/2014MNRAS.438.1985T} {438, 1985}

\bibitem[\protect\citeauthoryear{{Torrey} et~al.,}{{Torrey}
  et~al.}{2017}]{Torrey17}
{Torrey} P.,  et~al., 2017, preprint (\mn@eprint {} {1711.05261})

\bibitem[\protect\citeauthoryear{{Torrey} et~al.,}{{Torrey}
  et~al.}{2018}]{Torrey18}
{Torrey} P.,  et~al., 2018, \mn@doi [\mnras] {10.1093/mnrasl/sly031}, \href
  {http://adsabs.harvard.edu/abs/2018MNRAS.477L..16T} {477, L16}

\bibitem[\protect\citeauthoryear{{Tortora}, {Napolitano}, {Cardone},
  {Capaccioli}, {Jetzer}  \& {Molinaro}}{{Tortora} et~al.}{2010}]{Tortora10}
{Tortora} C.,  {Napolitano} N.~R.,  {Cardone} V.~F.,  {Capaccioli} M.,
  {Jetzer} P.,   {Molinaro} R.,  2010, \mn@doi [\mnras]
  {10.1111/j.1365-2966.2010.16938.x}, \href
  {http://adsabs.harvard.edu/abs/2010MNRAS.407..144T} {407, 144}

\bibitem[\protect\citeauthoryear{{Trager}, {Worthey}, {Faber}, {Burstein}  \&
  {Gonz{\'a}lez}}{{Trager} et~al.}{1998}]{Trager98}
{Trager} S.~C.,  {Worthey} G.,  {Faber} S.~M.,  {Burstein} D.,   {Gonz{\'a}lez}
  J.~J.,  1998, \mn@doi [\apjs] {10.1086/313099}, \href
  {http://adsabs.harvard.edu/abs/1998ApJS..116....1T} {116, 1}

\bibitem[\protect\citeauthoryear{{Trager}, {Faber}, {Worthey}  \&
  {Gonz{\'a}lez}}{{Trager} et~al.}{2000a}]{Trager00a}
{Trager} S.~C.,  {Faber} S.~M.,  {Worthey} G.,   {Gonz{\'a}lez} J.~J.,  2000a,
  \mn@doi [\aj] {10.1086/301299}, \href
  {http://adsabs.harvard.edu/abs/2000AJ....119.1645T} {119, 1645}

\bibitem[\protect\citeauthoryear{{Trager}, {Faber}, {Worthey}  \&
  {Gonz{\'a}lez}}{{Trager} et~al.}{2000b}]{Trager00b}
{Trager} S.~C.,  {Faber} S.~M.,  {Worthey} G.,   {Gonz{\'a}lez} J.~J.,  2000b,
  \mn@doi [\aj] {10.1086/301442}, \href
  {http://adsabs.harvard.edu/abs/2000AJ....120..165T} {120, 165}

\bibitem[\protect\citeauthoryear{{Trainor}, {Strom}, {Steidel}  \&
  {Rudie}}{{Trainor} et~al.}{2016}]{Trainor16}
{Trainor} R.~F.,  {Strom} A.~L.,  {Steidel} C.~C.,   {Rudie} G.~C.,  2016,
  \mn@doi [\apj] {10.3847/0004-637X/832/2/171}, 832, 171

\bibitem[\protect\citeauthoryear{{Tremonti} et~al.,}{{Tremonti}
  et~al.}{2004}]{Tremonti04}
{Tremonti} C.~A.,  et~al., 2004, \mn@doi [\apj] {10.1086/423264}, \href
  {http://adsabs.harvard.edu/abs/2004ApJ...613..898T} {613, 898}

\bibitem[\protect\citeauthoryear{{Trenti}, {Perna}  \& {Jimenez}}{{Trenti}
  et~al.}{2015}]{Trenti15}
{Trenti} M.,  {Perna} R.,   {Jimenez} R.,  2015, \mn@doi [\apj]
  {10.1088/0004-637X/802/2/103}, \href
  {http://adsabs.harvard.edu/abs/2015ApJ...802..103T} {802, 103}

\bibitem[\protect\citeauthoryear{{Tresse}, {Rola}, {Hammer}, {Stasi{\'n}ska},
  {Le Fevre}, {Lilly}  \& {Crampton}}{{Tresse} et~al.}{1996}]{Tresse96}
{Tresse} L.,  {Rola} C.,  {Hammer} F.,  {Stasi{\'n}ska} G.,  {Le Fevre} O.,
  {Lilly} S.~J.,   {Crampton} D.,  1996, \mn@doi [\mnras]
  {10.1093/mnras/281.3.847}, \href
  {http://adsabs.harvard.edu/abs/1996MNRAS.281..847T} {281, 847}

\bibitem[\protect\citeauthoryear{{Tripp} \& {Savage}}{{Tripp} \&
  {Savage}}{2000}]{Tripp00}
{Tripp} T.~M.,  {Savage} B.~D.,  2000, \mn@doi [\apj] {10.1086/309506}, \href
  {http://adsabs.harvard.edu/abs/2000ApJ...542...42T} {542, 42}

\bibitem[\protect\citeauthoryear{{Troncoso} et~al.,}{{Troncoso}
  et~al.}{2014}]{Troncoso14}
{Troncoso} P.,  et~al., 2014, \mn@doi [\aap] {10.1051/0004-6361/201322099},
  \href {http://adsabs.harvard.edu/abs/2014A%26A...563A..58T} {563, A58}

\bibitem[\protect\citeauthoryear{{Trussler}, {Maiolino}, {Maraston}, {Thomas},
  {Goddard}, {Lian}  \& {Peng}}{{Trussler} et~al.}{2018}]{Trussler18}
{Trussler} J.,  {Maiolino} R.,  {Maraston} C.,  {Thomas} D.,  {Goddard} D.,
  {Lian} J.,   {Peng} Y.,  2018, preprint, \href
  {http://adsabs.harvard.edu/abs/2018arXiv2478591M} {} (\mn@eprint {arXiv}
  {1811.2478591})

\bibitem[\protect\citeauthoryear{{Tsamis} \& {P{\'e}quignot}}{{Tsamis} \&
  {P{\'e}quignot}}{2005}]{Tsamis05}
{Tsamis} Y.~G.,  {P{\'e}quignot} D.,  2005, \mn@doi [\mnras]
  {10.1111/j.1365-2966.2005.09595.x}, \href
  {http://adsabs.harvard.edu/abs/2005MNRAS.364..687T} {364, 687}

\bibitem[\protect\citeauthoryear{{Tsamis}, {Barlow}, {Liu}, {Danziger}  \&
  {Storey}}{{Tsamis} et~al.}{2003}]{Tsamis03}
{Tsamis} Y.~G.,  {Barlow} M.~J.,  {Liu} X.-W.,  {Danziger} I.~J.,   {Storey}
  P.~J.,  2003, \mn@doi [\mnras] {10.1046/j.1365-8711.2003.06081.x}, \href
  {http://adsabs.harvard.edu/abs/2003MNRAS.338..687T} {338, 687}

\bibitem[\protect\citeauthoryear{{Tumlinson} et~al.,}{{Tumlinson}
  et~al.}{2011}]{Tumlinson11}
{Tumlinson} J.,  et~al., 2011, \mn@doi [Science] {10.1126/science.1209840},
  334, 948

\bibitem[\protect\citeauthoryear{{Tumlinson}, {Peeples}  \& {Werk}}{{Tumlinson}
  et~al.}{2017}]{Tumlinson17}
{Tumlinson} J.,  {Peeples} M.~S.,   {Werk} J.~K.,  2017, \mn@doi [\araa]
  {10.1146/annurev-astro-091916-055240}, \href
  {http://adsabs.harvard.edu/abs/2017ARA%26A..55..389T} {55, 389}

\bibitem[\protect\citeauthoryear{{Vale Asari}, {Stasi{\'n}ska}, {Morisset}  \&
  {Cid Fernandes}}{{Vale Asari} et~al.}{2016}]{Vale-Asari16}
{Vale Asari} N.,  {Stasi{\'n}ska} G.,  {Morisset} C.,   {Cid Fernandes} R.,
  2016, \mn@doi [\mnras] {10.1093/mnras/stw971}, \href
  {http://adsabs.harvard.edu/abs/2016MNRAS.460.1739V} {460, 1739}

\bibitem[\protect\citeauthoryear{{Valentino} et~al.,}{{Valentino}
  et~al.}{2015}]{Valentino15}
{Valentino} F.,  et~al., 2015, \mn@doi [\apj] {10.1088/0004-637X/801/2/132},
  \href {http://adsabs.harvard.edu/abs/2015ApJ...801..132V} {801, 132}

\bibitem[\protect\citeauthoryear{{Valiante}, {Schneider}, {Bianchi}  \&
  {Andersen}}{{Valiante} et~al.}{2009}]{Valiante09}
{Valiante} R.,  {Schneider} R.,  {Bianchi} S.,   {Andersen} A.~C.,  2009,
  \mn@doi [\mnras] {10.1111/j.1365-2966.2009.15076.x}, \href
  {http://adsabs.harvard.edu/abs/2009MNRAS.397.1661V} {397, 1661}

\bibitem[\protect\citeauthoryear{{Vangioni}, {Dvorkin}, {Olive}, {Dubois},
  {Molaro}, {Petitjean}, {Silk}  \& {Kimm}}{{Vangioni}
  et~al.}{2018}]{Vangioni18}
{Vangioni} E.,  {Dvorkin} I.,  {Olive} K.~A.,  {Dubois} Y.,  {Molaro} P.,
  {Petitjean} P.,  {Silk} J.,   {Kimm} T.,  2018, \mn@doi [\mnras]
  {10.1093/mnras/sty559}, \href
  {http://adsabs.harvard.edu/abs/2018MNRAS.477...56V} {477, 56}

\bibitem[\protect\citeauthoryear{{Veilleux} \& {Osterbrock}}{{Veilleux} \&
  {Osterbrock}}{1987}]{Veilleux87}
{Veilleux} S.,  {Osterbrock} D.~E.,  1987, \mn@doi [\apjs] {10.1086/191166},
  \href {http://adsabs.harvard.edu/abs/1987ApJS...63..295V} {63, 295}

\bibitem[\protect\citeauthoryear{{Veilleux}, {Teng}, {Rupke}, {Maiolino}  \&
  {Sturm}}{{Veilleux} et~al.}{2014}]{Veilleux14}
{Veilleux} S.,  {Teng} S.~H.,  {Rupke} D.~S.~N.,  {Maiolino} R.,   {Sturm} E.,
  2014, \mn@doi [\apj] {10.1088/0004-637X/790/2/116}, \href
  {http://adsabs.harvard.edu/abs/2014ApJ...790..116V} {790, 116}

\bibitem[\protect\citeauthoryear{{Venturi}, {Marconi}, {Mingozzi}, {Carniani},
  {Cresci}, {Risaliti}  \& {Mannucci}}{{Venturi} et~al.}{2017}]{Venturi17}
{Venturi} G.,  {Marconi} A.,  {Mingozzi} M.,  {Carniani} S.,  {Cresci} G.,
  {Risaliti} G.,   {Mannucci} F.,  2017, \mn@doi [Frontiers in Astronomy and
  Space Sciences] {10.3389/fspas.2017.00046}, \href
  {http://adsabs.harvard.edu/abs/2017FrASS...4...46V} {4, 46}

\bibitem[\protect\citeauthoryear{{Venturi} et~al.,}{{Venturi}
  et~al.}{2018}]{Venturi18}
{Venturi} G.,  et~al., 2018, preprint, \href
  {http://adsabs.harvard.edu/abs/2018arXiv180901206V} {} (\mn@eprint {arXiv}
  {1809.01206})

\bibitem[\protect\citeauthoryear{{Vergani} et~al.,}{{Vergani}
  et~al.}{2017}]{Vergani17}
{Vergani} S.~D.,  et~al., 2017, \mn@doi [\aap] {10.1051/0004-6361/201629759},
  \href {http://adsabs.harvard.edu/abs/2017A%26A...599A.120V} {599, A120}

\bibitem[\protect\citeauthoryear{{Verner} \& {Peterson}}{{Verner} \&
  {Peterson}}{2004}]{Verner04a}
{Verner} E.~M.,  {Peterson} B.~A.,  2004, \mn@doi [\apjl] {10.1086/422389},
  \href {http://adsabs.harvard.edu/abs/2004ApJ...608L..85V} {608, L85}

\bibitem[\protect\citeauthoryear{{Verner}, {Bruhweiler}, {Verner}, {Johansson},
  {Kallman}  \& {Gull}}{{Verner} et~al.}{2004}]{Verner04b}
{Verner} E.,  {Bruhweiler} F.,  {Verner} D.,  {Johansson} S.,  {Kallman} T.,
  {Gull} T.,  2004, \mn@doi [\apj] {10.1086/422303}, \href
  {http://adsabs.harvard.edu/abs/2004ApJ...611..780V} {611, 780}

\bibitem[\protect\citeauthoryear{{Vila-Costas} \& {Edmunds}}{{Vila-Costas} \&
  {Edmunds}}{1992}]{Vila-Costas92}
{Vila-Costas} M.~B.,  {Edmunds} M.~G.,  1992, \mn@doi [\mnras]
  {10.1093/mnras/259.1.121}, \href
  {http://adsabs.harvard.edu/abs/1992MNRAS.259..121V} {259, 121}

\bibitem[\protect\citeauthoryear{{Vila Costas} \& {Edmunds}}{{Vila Costas} \&
  {Edmunds}}{1993}]{Vila-Costas93}
{Vila Costas} M.~B.,  {Edmunds} M.~G.,  1993, \mn@doi [\mnras]
  {10.1093/mnras/265.1.199}, \href
  {http://adsabs.harvard.edu/abs/1993MNRAS.265..199V} {265, 199}

\bibitem[\protect\citeauthoryear{{Villar-Mart{\'{\i}}n}, {Cervi{\~n}o}  \&
  {Gonz{\'a}lez Delgado}}{{Villar-Mart{\'{\i}}n}
  et~al.}{2004}]{Villar-Martin04}
{Villar-Mart{\'{\i}}n} M.,  {Cervi{\~n}o} M.,   {Gonz{\'a}lez Delgado} R.~M.,
  2004, \mn@doi [\mnras] {10.1111/j.1365-2966.2004.08395.x}, \href
  {http://adsabs.harvard.edu/abs/2004MNRAS.355.1132V} {355, 1132}

\bibitem[\protect\citeauthoryear{{Vincenzo} \& {Kobayashi}}{{Vincenzo} \&
  {Kobayashi}}{2018}]{Vincenzo18}
{Vincenzo} F.,  {Kobayashi} C.,  2018, \mn@doi [\mnras]
  {10.1093/mnras/sty1047}, \href
  {http://adsabs.harvard.edu/abs/2018MNRAS.478..155V} {478, 155}

\bibitem[\protect\citeauthoryear{{Vincenzo}, {Matteucci}, {Belfiore}  \&
  {Maiolino}}{{Vincenzo} et~al.}{2016a}]{Vincenzo16b}
{Vincenzo} F.,  {Matteucci} F.,  {Belfiore} F.,   {Maiolino} R.,  2016a,
  \mn@doi [\mnras] {10.1093/mnras/stv2598}, \href
  {http://adsabs.harvard.edu/abs/2016MNRAS.455.4183V} {455, 4183}

\bibitem[\protect\citeauthoryear{{Vincenzo}, {Belfiore}, {Maiolino},
  {Matteucci}  \& {Ventura}}{{Vincenzo} et~al.}{2016b}]{Vincenzo16}
{Vincenzo} F.,  {Belfiore} F.,  {Maiolino} R.,  {Matteucci} F.,   {Ventura} P.,
   2016b, \mn@doi [\mnras] {10.1093/mnras/stw532}, \href
  {http://adsabs.harvard.edu/abs/2016MNRAS.458.3466V} {458, 3466}

\bibitem[\protect\citeauthoryear{{Vladilo}, {Abate}, {Yin}, {Cescutti}  \&
  {Matteucci}}{{Vladilo} et~al.}{2011}]{Vladilo11}
{Vladilo} G.,  {Abate} C.,  {Yin} J.,  {Cescutti} G.,   {Matteucci} F.,  2011,
  \mn@doi [\aap] {10.1051/0004-6361/201016330}, \href
  {http://adsabs.harvard.edu/abs/2011A%26A...530A..33V} {530, A33}

\bibitem[\protect\citeauthoryear{{Vogelsberger} et~al.,}{{Vogelsberger}
  et~al.}{2014}]{Vogelsberger14}
{Vogelsberger} M.,  et~al., 2014, \mn@doi [\mnras] {10.1093/mnras/stu1536},
  \href {http://adsabs.harvard.edu/abs/2014MNRAS.444.1518V} {444, 1518}

\bibitem[\protect\citeauthoryear{{Vogt}, {Dopita}, {Kewley}, {Sutherland},
  {Scharw{\"a}chter}, {Basurah}, {Ali}  \& {Amer}}{{Vogt}
  et~al.}{2014}]{Vogt14}
{Vogt} F.~P.~A.,  {Dopita} M.~A.,  {Kewley} L.~J.,  {Sutherland} R.~S.,
  {Scharw{\"a}chter} J.,  {Basurah} H.~M.,  {Ali} A.,   {Amer} M.~A.,  2014,
  \mn@doi [\apj] {10.1088/0004-637X/793/2/127}, \href
  {http://adsabs.harvard.edu/abs/2014ApJ...793..127V} {793, 127}

\bibitem[\protect\citeauthoryear{{Vreeswijk} et~al.,}{{Vreeswijk}
  et~al.}{2014}]{Vreeswijk14}
{Vreeswijk} P.~M.,  et~al., 2014, \mn@doi [\apj] {10.1088/0004-637X/797/1/24},
  \href {http://adsabs.harvard.edu/abs/2014ApJ...797...24V} {797, 24}

\bibitem[\protect\citeauthoryear{{Walborn}, {Lennon}, {Haser}, {Kudritzki}  \&
  {Voels}}{{Walborn} et~al.}{1995}]{Walborn95}
{Walborn} N.~R.,  {Lennon} D.~J.,  {Haser} S.~M.,  {Kudritzki} R.-P.,   {Voels}
  S.~A.,  1995, \mn@doi [\pasp] {10.1086/133524}, \href
  {http://adsabs.harvard.edu/abs/1995PASP..107..104W} {107, 104}

\bibitem[\protect\citeauthoryear{{Walcher}, {Coelho}, {Gallazzi}  \&
  {Charlot}}{{Walcher} et~al.}{2009}]{Walcher09}
{Walcher} C.~J.,  {Coelho} P.,  {Gallazzi} A.,   {Charlot} S.,  2009, \mn@doi
  [\mnras] {10.1111/j.1745-3933.2009.00705.x}, \href
  {http://adsabs.harvard.edu/abs/2009MNRAS.398L..44W} {398, L44}

\bibitem[\protect\citeauthoryear{{Walsh} \& {Roy}}{{Walsh} \&
  {Roy}}{1989}]{Walsh89}
{Walsh} J.~R.,  {Roy} J.-R.,  1989, \mn@doi [\mnras] {10.1093/mnras/239.2.297},
  \href {http://adsabs.harvard.edu/abs/1989MNRAS.239..297W} {239, 297}

\bibitem[\protect\citeauthoryear{{Wang}, {Zhou}, {Yuan}  \& {Wang}}{{Wang}
  et~al.}{2012}]{Wang12}
{Wang} H.,  {Zhou} H.,  {Yuan} W.,   {Wang} T.,  2012, \mn@doi [\apjl]
  {10.1088/2041-8205/751/2/L23}, \href
  {http://adsabs.harvard.edu/abs/2012ApJ...751L..23W} {751, L23}

\bibitem[\protect\citeauthoryear{{Wang} et~al.,}{{Wang} et~al.}{2017}]{Wang17}
{Wang} X.,  et~al., 2017, \mn@doi [\apj] {10.3847/1538-4357/aa603c}, \href
  {http://adsabs.harvard.edu/abs/2017ApJ...837...89W} {837, 89}

\bibitem[\protect\citeauthoryear{{Wang} et~al.,}{{Wang} et~al.}{2018}]{Wang18}
{Wang} X.,  et~al., 2018, preprint, \href
  {http://adsabs.harvard.edu/abs/2018arXiv180808800W} {} (\mn@eprint {arXiv}
  {1808.08800})

\bibitem[\protect\citeauthoryear{{Weiner} et~al.,}{{Weiner}
  et~al.}{2009}]{Weiner09}
{Weiner} B.~J.,  et~al., 2009, \mn@doi [\apj] {10.1088/0004-637X/692/1/187},
  692, 187

\bibitem[\protect\citeauthoryear{{Werk}, {Putman}, {Meurer}  \&
  {Santiago-Figueroa}}{{Werk} et~al.}{2011}]{Werk11}
{Werk} J.~K.,  {Putman} M.~E.,  {Meurer} G.~R.,   {Santiago-Figueroa} N.,
  2011, \mn@doi [\apj] {10.1088/0004-637X/735/2/71}, \href
  {http://adsabs.harvard.edu/abs/2011ApJ...735...71W} {735, 71}

\bibitem[\protect\citeauthoryear{{Werk} et~al.,}{{Werk} et~al.}{2014}]{Werk14}
{Werk} J.~K.,  et~al., 2014, \mn@doi [\apj] {10.1088/0004-637X/792/1/8}, \href
  {http://adsabs.harvard.edu/abs/2014ApJ...792....8W} {792, 8}

\bibitem[\protect\citeauthoryear{{Werner} et~al.,}{{Werner}
  et~al.}{2004}]{Werner04}
{Werner} M.~W.,  et~al., 2004, \mn@doi [\apjs] {10.1086/422992}, \href
  {http://adsabs.harvard.edu/abs/2004ApJS..154....1W} {154, 1}

\bibitem[\protect\citeauthoryear{{Westmoquette}, {James}, {Monreal-Ibero}  \&
  {Walsh}}{{Westmoquette} et~al.}{2013}]{Westmoquette13}
{Westmoquette} M.~S.,  {James} B.,  {Monreal-Ibero} A.,   {Walsh} J.~R.,  2013,
  \mn@doi [\aap] {10.1051/0004-6361/201220580}, \href
  {http://adsabs.harvard.edu/abs/2013A%26A...550A..88W} {550, A88}

\bibitem[\protect\citeauthoryear{{Wilkinson}, {Maraston}, {Goddard}, {Thomas}
  \& {Parikh}}{{Wilkinson} et~al.}{2017}]{Wilkinson17}
{Wilkinson} D.~M.,  {Maraston} C.,  {Goddard} D.,  {Thomas} D.,   {Parikh} T.,
  2017, \mn@doi [\mnras] {10.1093/mnras/stx2215}, \href
  {http://adsabs.harvard.edu/abs/2017MNRAS.472.4297W} {472, 4297}

\bibitem[\protect\citeauthoryear{{Williams}, {Maiolino}, {Santini}, {Marconi},
  {Cresci}, {Mannucci}  \& {Lutz}}{{Williams} et~al.}{2014}]{Williams14}
{Williams} R.~J.,  {Maiolino} R.,  {Santini} P.,  {Marconi} A.,  {Cresci} G.,
  {Mannucci} F.,   {Lutz} D.,  2014, \mn@doi [\mnras] {10.1093/mnras/stu1422},
  \href {http://adsabs.harvard.edu/abs/2014MNRAS.443.3780W} {443, 3780}

\bibitem[\protect\citeauthoryear{{Wiseman}, {Schady}, {Bolmer}, {Kr{\"u}hler},
  {Yates}, {Greiner}  \& {Fynbo}}{{Wiseman} et~al.}{2017}]{Wiseman17}
{Wiseman} P.,  {Schady} P.,  {Bolmer} J.,  {Kr{\"u}hler} T.,  {Yates} R.~M.,
  {Greiner} J.,   {Fynbo} J.~P.~U.,  2017, \mn@doi [\aap]
  {10.1051/0004-6361/201629228}, \href
  {http://adsabs.harvard.edu/abs/2017A%26A...599A..24W} {599, A24}

\bibitem[\protect\citeauthoryear{{Wolfe}, {Turnshek}, {Smith}  \&
  {Cohen}}{{Wolfe} et~al.}{1986}]{Wolfe86}
{Wolfe} A.~M.,  {Turnshek} D.~A.,  {Smith} H.~E.,   {Cohen} R.~D.,  1986,
  \mn@doi [\apjs] {10.1086/191114}, \href
  {http://adsabs.harvard.edu/abs/1986ApJS...61..249W} {61, 249}

\bibitem[\protect\citeauthoryear{{Wolfe}, {Gawiser}  \& {Prochaska}}{{Wolfe}
  et~al.}{2005}]{Wolfe05}
{Wolfe} A.~M.,  {Gawiser} E.,   {Prochaska} J.~X.,  2005, \mn@doi [\araa]
  {10.1146/annurev.astro.42.053102.133950}, \href
  {http://adsabs.harvard.edu/abs/2005ARA%26A..43..861W} {43, 861}

\bibitem[\protect\citeauthoryear{{Worthey}}{{Worthey}}{1994}]{Worthey94b}
{Worthey} G.,  1994, \mn@doi [\apjs] {10.1086/192096}, \href
  {http://adsabs.harvard.edu/abs/1994ApJS...95..107W} {95, 107}

\bibitem[\protect\citeauthoryear{{Worthey} \& {Ottaviani}}{{Worthey} \&
  {Ottaviani}}{1997}]{Worthey97}
{Worthey} G.,  {Ottaviani} D.~L.,  1997, \mn@doi [\apjs] {10.1086/313021},
  \href {http://adsabs.harvard.edu/abs/1997ApJS..111..377W} {111, 377}

\bibitem[\protect\citeauthoryear{{Worthey}, {Faber}, {Gonzalez}  \&
  {Burstein}}{{Worthey} et~al.}{1994}]{Worthey94a}
{Worthey} G.,  {Faber} S.~M.,  {Gonzalez} J.~J.,   {Burstein} D.,  1994,
  \mn@doi [\apjs] {10.1086/192087}, \href
  {http://adsabs.harvard.edu/abs/1994ApJS...94..687W} {94, 687}

\bibitem[\protect\citeauthoryear{{Wright}, {Larkin}, {Law}, {Steidel},
  {Shapley}  \& {Erb}}{{Wright} et~al.}{2009}]{Wright09}
{Wright} S.~A.,  {Larkin} J.~E.,  {Law} D.~R.,  {Steidel} C.~C.,  {Shapley}
  A.~E.,   {Erb} D.~K.,  2009, \mn@doi [\apj] {10.1088/0004-637X/699/1/421},
  \href {http://adsabs.harvard.edu/abs/2009ApJ...699..421W} {699, 421}

\bibitem[\protect\citeauthoryear{{Wright}, {Larkin}, {Graham}  \&
  {Ma}}{{Wright} et~al.}{2010}]{Wright10}
{Wright} S.~A.,  {Larkin} J.~E.,  {Graham} J.~R.,   {Ma} C.,  2010, \mn@doi
  [\apj] {10.1088/0004-637X/711/2/1291}, \href
  {http://adsabs.harvard.edu/abs/2010ApJ...711.1291W} {711, 1291}

\bibitem[\protect\citeauthoryear{{Wu}, {Kudritzki}, {Tully}  \& {Neill}}{{Wu}
  et~al.}{2015}]{Wu15}
{Wu} P.-F.,  {Kudritzki} R.-P.,  {Tully} R.~B.,   {Neill} J.~D.,  2015, \mn@doi
  [\apj] {10.1088/0004-637X/810/2/151}, \href
  {http://adsabs.harvard.edu/abs/2015ApJ...810..151W} {810, 151}

\bibitem[\protect\citeauthoryear{{Wu}, {Zhang}, {Zhao}  \& {Zhang}}{{Wu}
  et~al.}{2016}]{Wu16}
{Wu} Y.-Z.,  {Zhang} S.-N.,  {Zhao} Y.-H.,   {Zhang} W.,  2016, \mn@doi
  [\mnras] {10.1093/mnras/stw113}, \href
  {http://adsabs.harvard.edu/abs/2016MNRAS.457.2929W} {457, 2929}

\bibitem[\protect\citeauthoryear{{Wu}, {Zahid}, {Hwang}  \& {Geller}}{{Wu}
  et~al.}{2017}]{Wu17}
{Wu} P.-F.,  {Zahid} H.~J.,  {Hwang} H.~S.,   {Geller} M.~J.,  2017, \mn@doi
  [\mnras] {10.1093/mnras/stx597}, \href
  {http://adsabs.harvard.edu/abs/2017MNRAS.468.1881W} {468, 1881}

\bibitem[\protect\citeauthoryear{{Wuyts}, {Rigby}, {Sharon}  \&
  {Gladders}}{{Wuyts} et~al.}{2012}]{Wuyts12}
{Wuyts} E.,  {Rigby} J.~R.,  {Sharon} K.,   {Gladders} M.~D.,  2012, \mn@doi
  [\apj] {10.1088/0004-637X/755/1/73}, \href
  {http://adsabs.harvard.edu/abs/2012ApJ...755...73W} {755, 73}

\bibitem[\protect\citeauthoryear{{Wuyts}, {Rigby}, {Gladders}  \&
  {Sharon}}{{Wuyts} et~al.}{2014a}]{Wuyts14a}
{Wuyts} E.,  {Rigby} J.~R.,  {Gladders} M.~D.,   {Sharon} K.,  2014a, \mn@doi
  [\apj] {10.1088/0004-637X/781/2/61}, \href
  {http://adsabs.harvard.edu/abs/2014ApJ...781...61W} {781, 61}

\bibitem[\protect\citeauthoryear{{Wuyts} et~al.,}{{Wuyts}
  et~al.}{2014b}]{Wuyts14b}
{Wuyts} E.,  et~al., 2014b, \mn@doi [\apjl] {10.1088/2041-8205/789/2/L40},
  \href {http://adsabs.harvard.edu/abs/2014ApJ...789L..40W} {789, L40}

\bibitem[\protect\citeauthoryear{{Wuyts} et~al.,}{{Wuyts}
  et~al.}{2016}]{Wuyts16}
{Wuyts} E.,  et~al., 2016, \mn@doi [\apj] {10.3847/0004-637X/827/1/74}, \href
  {http://adsabs.harvard.edu/abs/2016ApJ...827...74W} {827, 74}

\bibitem[\protect\citeauthoryear{{Xia} et~al.,}{{Xia} et~al.}{2012}]{Xia12}
{Xia} L.,  et~al., 2012, \mn@doi [\aj] {10.1088/0004-6256/144/1/28}, \href
  {http://adsabs.harvard.edu/abs/2012AJ....144...28X} {144, 28}

\bibitem[\protect\citeauthoryear{{Xiao}, {Stanway}  \& {Eldridge}}{{Xiao}
  et~al.}{2018}]{Xiao18}
{Xiao} L.,  {Stanway} E.~R.,   {Eldridge} J.~J.,  2018, \mn@doi [\mnras]
  {10.1093/mnras/sty646}, \href
  {http://adsabs.harvard.edu/abs/2018MNRAS.477..904X} {477, 904}

\bibitem[\protect\citeauthoryear{{Xu}, {Bian}, {Shen}, {Zuo}, {Fan}  \&
  {Zhu}}{{Xu} et~al.}{2018}]{Xu18}
{Xu} F.,  {Bian} F.,  {Shen} Y.,  {Zuo} W.,  {Fan} X.,   {Zhu} Z.,  2018,
  \mn@doi [\mnras] {10.1093/mnras/sty1763}

\bibitem[\protect\citeauthoryear{{Yabe} et~al.,}{{Yabe} et~al.}{2012}]{Yabe12}
{Yabe} K.,  et~al., 2012, \mn@doi [\pasj] {10.1093/pasj/64.3.60}, \href
  {http://adsabs.harvard.edu/abs/2012PASJ...64...60Y} {64, 60}

\bibitem[\protect\citeauthoryear{{Yabe} et~al.,}{{Yabe} et~al.}{2014}]{Yabe14a}
{Yabe} K.,  et~al., 2014, \mn@doi [\mnras] {10.1093/mnras/stt2185}, \href
  {http://adsabs.harvard.edu/abs/2014MNRAS.437.3647Y} {437, 3647}

\bibitem[\protect\citeauthoryear{{Yabe} et~al.,}{{Yabe} et~al.}{2015a}]{Yabe15}
{Yabe} K.,  et~al., 2015a, \mn@doi [\pasj] {10.1093/pasj/psv079}, \href
  {http://adsabs.harvard.edu/abs/2015PASJ...67..102Y} {67, 102}

\bibitem[\protect\citeauthoryear{{Yabe}, {Ohta}, {Akiyama}, {Iwamuro},
  {Tamura}, {Yuma}, {Dalton}  \& {Lewis}}{{Yabe} et~al.}{2015b}]{Yabe15a}
{Yabe} K.,  {Ohta} K.,  {Akiyama} M.,  {Iwamuro} F.,  {Tamura} N.,  {Yuma} S.,
  {Dalton} G.,   {Lewis} I.,  2015b, \mn@doi [\apj]
  {10.1088/0004-637X/798/1/45}, 798, 45

\bibitem[\protect\citeauthoryear{{Yanny} et~al.,}{{Yanny}
  et~al.}{2009}]{Yanny09}
{Yanny} B.,  et~al., 2009, \mn@doi [\aj] {10.1088/0004-6256/137/5/4377}, \href
  {http://adsabs.harvard.edu/abs/2009AJ....137.4377Y} {137, 4377}

\bibitem[\protect\citeauthoryear{{Yates} \& {Kauffmann}}{{Yates} \&
  {Kauffmann}}{2014}]{Yates14}
{Yates} R.~M.,  {Kauffmann} G.,  2014, \mn@doi [\mnras] {10.1093/mnras/stu233},
  \href {http://adsabs.harvard.edu/abs/2014MNRAS.tmp..399Y} {}

\bibitem[\protect\citeauthoryear{{Yates}, {Kauffmann}  \& {Guo}}{{Yates}
  et~al.}{2012}]{Yates12}
{Yates} R.~M.,  {Kauffmann} G.,   {Guo} Q.,  2012, \mn@doi [\mnras]
  {10.1111/j.1365-2966.2012.20595.x}, \href
  {http://adsabs.harvard.edu/abs/2012MNRAS.422..215Y} {422, 215}

\bibitem[\protect\citeauthoryear{{Yuan} \& {Kewley}}{{Yuan} \&
  {Kewley}}{2009}]{Yuan09}
{Yuan} T.-T.,  {Kewley} L.~J.,  2009, \mn@doi [\apjl]
  {10.1088/0004-637X/699/2/L161}, \href
  {http://adsabs.harvard.edu/abs/2009ApJ...699L.161Y} {699, L161}

\bibitem[\protect\citeauthoryear{{Yuan}, {Kewley}, {Swinbank}, {Richard}  \&
  {Livermore}}{{Yuan} et~al.}{2011}]{Yuan11}
{Yuan} T.-T.,  {Kewley} L.~J.,  {Swinbank} A.~M.,  {Richard} J.,   {Livermore}
  R.~C.,  2011, \mn@doi [\apjl] {10.1088/2041-8205/732/1/L14}, \href
  {http://adsabs.harvard.edu/abs/2011ApJ...732L..14Y} {732, L14}

\bibitem[\protect\citeauthoryear{{Yuan}, {Kewley}  \& {Richard}}{{Yuan}
  et~al.}{2013}]{Yuan13}
{Yuan} T.-T.,  {Kewley} L.~J.,   {Richard} J.,  2013, \mn@doi [\apj]
  {10.1088/0004-637X/763/1/9}, \href
  {http://adsabs.harvard.edu/abs/2013ApJ...763....9Y} {763, 9}

\bibitem[\protect\citeauthoryear{{Zafar}, {Centuri{\'o}n}, {P{\'e}roux},
  {Molaro}, {D'Odorico}, {Vladilo}  \& {Popping}}{{Zafar}
  et~al.}{2014}]{Zafar14}
{Zafar} T.,  {Centuri{\'o}n} M.,  {P{\'e}roux} C.,  {Molaro} P.,  {D'Odorico}
  V.,  {Vladilo} G.,   {Popping} A.,  2014, \mn@doi [\mnras]
  {10.1093/mnras/stu1473}, \href
  {http://adsabs.harvard.edu/abs/2014MNRAS.444..744Z} {444, 744}

\bibitem[\protect\citeauthoryear{{Zahedy}, {Chen}, {Johnson}, {Pierce},
  {Rauch}, {Huang}, {Weiner}  \& {Gauthier}}{{Zahedy} et~al.}{2018}]{Zahedy18}
{Zahedy} F.~S.,  {Chen} H.-W.,  {Johnson} S.~D.,  {Pierce} R.~M.,  {Rauch} M.,
  {Huang} Y.-H.,  {Weiner} B.~D.,   {Gauthier} J.-R.,  2018, preprint, \href
  {http://adsabs.harvard.edu/abs/2018arXiv180905115Z} {} (\mn@eprint {arXiv}
  {1809.05115})

\bibitem[\protect\citeauthoryear{{Zahid}, {Kewley}  \& {Bresolin}}{{Zahid}
  et~al.}{2011}]{Zahid11}
{Zahid} H.~J.,  {Kewley} L.~J.,   {Bresolin} F.,  2011, \mn@doi [\apj]
  {10.1088/0004-637X/730/2/137}, \href
  {http://adsabs.harvard.edu/abs/2011ApJ...730..137Z} {730, 137}

\bibitem[\protect\citeauthoryear{{Zahid}, {Dima}, {Kewley}, {Erb}  \&
  {Dav{\'e}}}{{Zahid} et~al.}{2012}]{Zahid12b}
{Zahid} H.~J.,  {Dima} G.~I.,  {Kewley} L.~J.,  {Erb} D.~K.,   {Dav{\'e}} R.,
  2012, \mn@doi [\apj] {10.1088/0004-637X/757/1/54}, \href
  {http://adsabs.harvard.edu/abs/2012ApJ...757...54Z} {757, 54}

\bibitem[\protect\citeauthoryear{{Zahid}, {Geller}, {Kewley}, {Hwang},
  {Fabricant}  \& {Kurtz}}{{Zahid} et~al.}{2013}]{Zahid13a}
{Zahid} H.~J.,  {Geller} M.~J.,  {Kewley} L.~J.,  {Hwang} H.~S.,  {Fabricant}
  D.~G.,   {Kurtz} M.~J.,  2013, \mn@doi [\apjl] {10.1088/2041-8205/771/2/L19},
  \href {http://adsabs.harvard.edu/abs/2013ApJ...771L..19Z} {771, L19}

\bibitem[\protect\citeauthoryear{{Zahid}, {Torrey}, {Vogelsberger},
  {Hernquist}, {Kewley}  \& {Dav{\'e}}}{{Zahid} et~al.}{2014a}]{Zahid14}
{Zahid} H.~J.,  {Torrey} P.,  {Vogelsberger} M.,  {Hernquist} L.,  {Kewley} L.,
    {Dav{\'e}} R.,  2014a, \mn@doi [\apss] {10.1007/s10509-013-1666-0}, \href
  {http://adsabs.harvard.edu/abs/2014Ap%26SS.349..873Z} {349, 873}

\bibitem[\protect\citeauthoryear{{Zahid}, {Dima}, {Kudritzki}, {Kewley},
  {Geller}, {Hwang}, {Silverman}  \& {Kashino}}{{Zahid}
  et~al.}{2014b}]{Zahid14a}
{Zahid} H.~J.,  {Dima} G.~I.,  {Kudritzki} R.-P.,  {Kewley} L.~J.,  {Geller}
  M.~J.,  {Hwang} H.~S.,  {Silverman} J.~D.,   {Kashino} D.,  2014b, \mn@doi
  [\apj] {10.1088/0004-637X/791/2/130}, \href
  {http://adsabs.harvard.edu/abs/2014ApJ...791..130Z} {791, 130}

\bibitem[\protect\citeauthoryear{{Zahid} et~al.,}{{Zahid}
  et~al.}{2014c}]{Zahid14b}
{Zahid} H.~J.,  et~al., 2014c, \mn@doi [\apj] {10.1088/0004-637X/792/1/75},
  \href {http://adsabs.harvard.edu/abs/2014ApJ...792...75Z} {792, 75}

\bibitem[\protect\citeauthoryear{{Zahid}, {Kudritzki}, {Conroy}, {Andrews}  \&
  {Ho}}{{Zahid} et~al.}{2017}]{Zahid17}
{Zahid} H.~J.,  {Kudritzki} R.-P.,  {Conroy} C.,  {Andrews} B.,   {Ho} I.-T.,
  2017, \mn@doi [\apj] {10.3847/1538-4357/aa88ae}, \href
  {http://adsabs.harvard.edu/abs/2017ApJ...847...18Z} {847, 18}

\bibitem[\protect\citeauthoryear{{Zappacosta}, {Nicastro}, {Maiolino},
  {Tagliaferri}, {Buote}, {Fang}, {Humphrey}  \& {Gastaldello}}{{Zappacosta}
  et~al.}{2010}]{Zappacosta10}
{Zappacosta} L.,  {Nicastro} F.,  {Maiolino} R.,  {Tagliaferri} G.,  {Buote}
  D.~A.,  {Fang} T.,  {Humphrey} P.~J.,   {Gastaldello} F.,  2010, \mn@doi
  [\apj] {10.1088/0004-637X/717/1/74}, \href
  {http://adsabs.harvard.edu/abs/2010ApJ...717...74Z} {717, 74}

\bibitem[\protect\citeauthoryear{{Zappacosta}, {Nicastro}, {Krongold}  \&
  {Maiolino}}{{Zappacosta} et~al.}{2012}]{Zappacosta12}
{Zappacosta} L.,  {Nicastro} F.,  {Krongold} Y.,   {Maiolino} R.,  2012,
  \mn@doi [\apj] {10.1088/0004-637X/753/2/137}, \href
  {http://adsabs.harvard.edu/abs/2012ApJ...753..137Z} {753, 137}

\bibitem[\protect\citeauthoryear{{Zaritsky}, {Kennicutt}  \&
  {Huchra}}{{Zaritsky} et~al.}{1994}]{Zaritsky94}
{Zaritsky} D.,  {Kennicutt} Jr. R.~C.,   {Huchra} J.~P.,  1994, \mn@doi [\apj]
  {10.1086/173544}, \href {http://adsabs.harvard.edu/abs/1994ApJ...420...87Z}
  {420, 87}

\bibitem[\protect\citeauthoryear{{Zepf} \& {Silk}}{{Zepf} \&
  {Silk}}{1996}]{Zepf96}
{Zepf} S.~E.,  {Silk} J.,  1996, \mn@doi [\apj] {10.1086/177496}, \href
  {http://adsabs.harvard.edu/abs/1996ApJ...466..114Z} {466, 114}

\bibitem[\protect\citeauthoryear{{Zetterlund}, {Levesque}, {Leitherer}  \&
  {Danforth}}{{Zetterlund} et~al.}{2015}]{Zetterlund15}
{Zetterlund} E.,  {Levesque} E.~M.,  {Leitherer} C.,   {Danforth} C.~W.,  2015,
  \mn@doi [\apj] {10.1088/0004-637X/805/2/151}, \href
  {http://adsabs.harvard.edu/abs/2015ApJ...805..151Z} {805, 151}

\bibitem[\protect\citeauthoryear{{Zhang}, {Wang}, {Ji}, {Smith}, {Foster}  \&
  {Zhou}}{{Zhang} et~al.}{2014}]{Zhang14a}
{Zhang} S.,  {Wang} Q.~D.,  {Ji} L.,  {Smith} R.~K.,  {Foster} A.~R.,   {Zhou}
  X.,  2014, \mn@doi [\apj] {10.1088/0004-637X/794/1/61}, \href
  {http://adsabs.harvard.edu/abs/2014ApJ...794...61Z} {794, 61}

\bibitem[\protect\citeauthoryear{{Zhang} et~al.,}{{Zhang}
  et~al.}{2017}]{Zhang17}
{Zhang} K.,  et~al., 2017, \mn@doi [\mnras] {10.1093/mnras/stw3308}, \href
  {http://adsabs.harvard.edu/abs/2017MNRAS.466.3217Z} {466, 3217}

\bibitem[\protect\citeauthoryear{{Zhang}, {Romano}, {Ivison}, {Papadopoulos}
  \& {Matteucci}}{{Zhang} et~al.}{2018a}]{Zhang18a}
{Zhang} Z.-Y.,  {Romano} D.,  {Ivison} R.~J.,  {Papadopoulos} P.~P.,
  {Matteucci} F.,  2018a, \mn@doi [\nat] {10.1038/s41586-018-0196-x}, \href
  {http://adsabs.harvard.edu/abs/2018Natur.558..260Z} {558, 260}

\bibitem[\protect\citeauthoryear{{Zhang} et~al.,}{{Zhang}
  et~al.}{2018b}]{Zhang18b}
{Zhang} H.-X.,  et~al., 2018b, \mn@doi [\apj] {10.3847/1538-4357/aab88a}, \href
  {http://adsabs.harvard.edu/abs/2018ApJ...858...37Z} {858, 37}

\bibitem[\protect\citeauthoryear{{Zinchenko}, {Pilyugin}, {Grebel},
  {S{\'a}nchez}  \& {V{\'{\i}}lchez}}{{Zinchenko} et~al.}{2016}]{Zinchenko16}
{Zinchenko} I.~A.,  {Pilyugin} L.~S.,  {Grebel} E.~K.,  {S{\'a}nchez} S.~F.,
  {V{\'{\i}}lchez} J.~M.,  2016, \mn@doi [\mnras] {10.1093/mnras/stw1857},
  \href {http://adsabs.harvard.edu/abs/2016MNRAS.462.2715Z} {462, 2715}

\bibitem[\protect\citeauthoryear{{Zoccali} et~al.,}{{Zoccali}
  et~al.}{2017}]{Zoccali17}
{Zoccali} M.,  et~al., 2017, \mn@doi [\aap] {10.1051/0004-6361/201629805},
  \href {http://adsabs.harvard.edu/abs/2017A%26A...599A..12Z} {599, A12}

\bibitem[\protect\citeauthoryear{{Zoldan}, {De Lucia}, {Xie}, {Fontanot}  \&
  {Hirschmann}}{{Zoldan} et~al.}{2017}]{Zoldan17}
{Zoldan} A.,  {De Lucia} G.,  {Xie} L.,  {Fontanot} F.,   {Hirschmann} M.,
  2017, \mn@doi [\mnras] {10.1093/mnras/stw2901}, \href
  {http://adsabs.harvard.edu/abs/2017MNRAS.465.2236Z} {465, 2236}

\bibitem[\protect\citeauthoryear{{de Bennassuti}, {Salvadori}, {Schneider},
  {Valiante}  \& {Omukai}}{{de Bennassuti} et~al.}{2017}]{deBennassuti17}
{de Bennassuti} M.,  {Salvadori} S.,  {Schneider} R.,  {Valiante} R.,
  {Omukai} K.,  2017, \mn@doi [\mnras] {10.1093/mnras/stw2687}, \href
  {http://adsabs.harvard.edu/abs/2017MNRAS.465..926D} {465, 926}

\bibitem[\protect\citeauthoryear{{de Jong} et~al.,}{{de Jong}
  et~al.}{2016}]{De-Jong16}
{de Jong} R.~S.,  et~al., 2016, in Ground-based and Airborne Instrumentation
  for Astronomy VI. p. 99081O, \mn@doi{10.1117/12.2232832}

\bibitem[\protect\citeauthoryear{{de Plaa}}{{de Plaa}}{2013}]{De-Plaa13}
{de Plaa} J.,  2013, \mn@doi [Astronomische Nachrichten]
  {10.1002/asna.201211870}, \href
  {http://adsabs.harvard.edu/abs/2013AN....334..416D} {334, 416}

\bibitem[\protect\citeauthoryear{{de Plaa}, {Werner}, {Bleeker}, {Vink},
  {Kaastra}  \& {M{\'e}ndez}}{{de Plaa} et~al.}{2007}]{de-Plaa07}
{de Plaa} J.,  {Werner} N.,  {Bleeker} J.~A.~M.,  {Vink} J.,  {Kaastra} J.~S.,
   {M{\'e}ndez} M.,  2007, \mn@doi [\aap] {10.1051/0004-6361:20066382}, \href
  {http://adsabs.harvard.edu/abs/2007A%26A...465..345D} {465, 345}

\bibitem[\protect\citeauthoryear{{de Plaa} et~al.,}{{de Plaa}
  et~al.}{2017}]{de-Plaa17}
{de Plaa} J.,  et~al., 2017, \mn@doi [\aap] {10.1051/0004-6361/201629926},
  \href {http://adsabs.harvard.edu/abs/2017A%26A...607A..98D} {607, A98}

\bibitem[\protect\citeauthoryear{{de los Reyes} et~al.,}{{de los Reyes}
  et~al.}{2015}]{de-los-Reyes15}
{de los Reyes} M.~A.,  et~al., 2015, \mn@doi [\aj]
  {10.1088/0004-6256/149/2/79}, \href
  {http://adsabs.harvard.edu/abs/2015AJ....149...79D} {149, 79}

\bibitem[\protect\citeauthoryear{{van Dokkum} et~al.,}{{van Dokkum}
  et~al.}{2010}]{vanDokkum10}
{van Dokkum} P.~G.,  et~al., 2010, \mn@doi [\apj]
  {10.1088/0004-637X/709/2/1018}, \href
  {http://adsabs.harvard.edu/abs/2010ApJ...709.1018V} {709, 1018}

\bibitem[\protect\citeauthoryear{{van Zee} \& {Haynes}}{{van Zee} \&
  {Haynes}}{2006}]{van-Zee06}
{van Zee} L.,  {Haynes} M.~P.,  2006, \mn@doi [\apj] {10.1086/498017}, \href
  {http://adsabs.harvard.edu/abs/2006ApJ...636..214V} {636, 214}

\bibitem[\protect\citeauthoryear{{van Zee}, {Salzer}  \& {Haynes}}{{van Zee}
  et~al.}{1998}]{van-Zee98a}
{van Zee} L.,  {Salzer} J.~J.,   {Haynes} M.~P.,  1998, \mn@doi [\apjl]
  {10.1086/311263}, \href {http://adsabs.harvard.edu/abs/1998ApJ...497L...1V}
  {497, L1}

\makeatother
\end{thebibliography}

\end{document}